\documentclass[rmp,aps,nofootinbib,twocolumn,showpacs]{revtex4-1}

\usepackage[latin2]{inputenc}
\usepackage[T1]{fontenc}
\usepackage{amsthm}
\usepackage{amssymb}
\usepackage{graphicx}
\usepackage{epsfig}
\usepackage{multirow}

\begin{document}
\title{Radioactive decays at limits of nuclear stability}

\author{M.~Pf\"utzner}
\email{pfutzner@fuw.edu.pl}
\affiliation{Faculty of Physics, University of Warsaw, Ho\.za 69, PL-00-681
Warszawa, Poland}

\author{L.V.~Grigorenko}
\affiliation{Flerov Laboratory of Nuclear Reactions, JINR, RU-141980, Dubna,
Russia}

\author{M.~Karny}
\affiliation{Faculty of Physics, University of Warsaw, Ho\.za 69, PL-00-681
Warszawa, Poland}

\author{K.~Riisager}
\affiliation{Department of Physics and Astronomy, Aarhus University, DK-8000
Aarhus C, Denmark}

\begin{abstract}
The last decades brought an impressive progress in synthesizing and studying
properties of
nuclides located very far from the beta stability line.
Among the most fundamental properties of such exotic nuclides,
usually established first, is the half-life, possible radioactive decay modes,
and their
relative probabilities. When approaching limits of nuclear stability, new decay
modes set in.
First, beta decays become accompanied by emission of nucleons from highly
excited states of
daughter nuclei. Second, when the nucleon separation energy becomes negative,
nucleons start
to be emitted from the ground state. Here, we present a review of the decay
modes occurring
close to the limits of stability. The experimental methods used to produce,
identify and detect
new species and their radiation are discussed. The current theoretical
understanding of these
decay processes is overviewed. The theoretical description of the most recently
discovered and
most complex radioactive process --- the two-proton radioactivity --- is
discussed in
more detail.
\end{abstract}


\date{\today}


\maketitle

\tableofcontents


\newpage
\section{INTRODUCTION}
\label{sec:I}


We consider the atomic nucleus as a quantum object composed of $A$ nucleons
(mass number): $Z$ protons (atomic number) and $N$ neutrons, held together
mainly by strong nuclear forces. A neutral atom with the specified numbers
$A$ and $Z$ is called a nuclide. When using this term, however, we focus
on the nuclear component of the atom. Such a system is stable only for certain
combinations of numbers $Z$ and $N$. Presently, 256 stable nuclides are known.
Systems different from stable configurations undergo spontaneous, radioactive
decays until the stability is reached. A nucleus of such an unstable nuclide
is considered as a well defined object if its half-life is much longer than
$10^{-21}$~s
which is a characteristic timescale for processes governed by strong
interaction.
These nuclides are bound
by nuclear forces and/or by Coulomb and centrifugal barriers. The number of
unstable nuclides synthesized in laboratories is constantly growing, and up to
now more than 3000 were identified. In this review, we concentrate on
radioactive processes observed for nuclides located at the limits of the nuclear
chart.
The emphasis is
given on new decay processes and features of classical decay modes which do not
take place among nuclides close to stability. We will refrain, however, from
discussing the heavy frontier
of the nuclear chart. The quest for the superheavy elements was reviewed by
\textcite{Hofmann:2000} and more recently by \textcite{Oganessian:2007}
and by \textcite{Hofmann:2009a}.

\subsection{Radioactivity and Nuclides}
\label{sec:IB}

\begin{figure*}[th]
\centerline{
\includegraphics[width=0.78\textwidth]{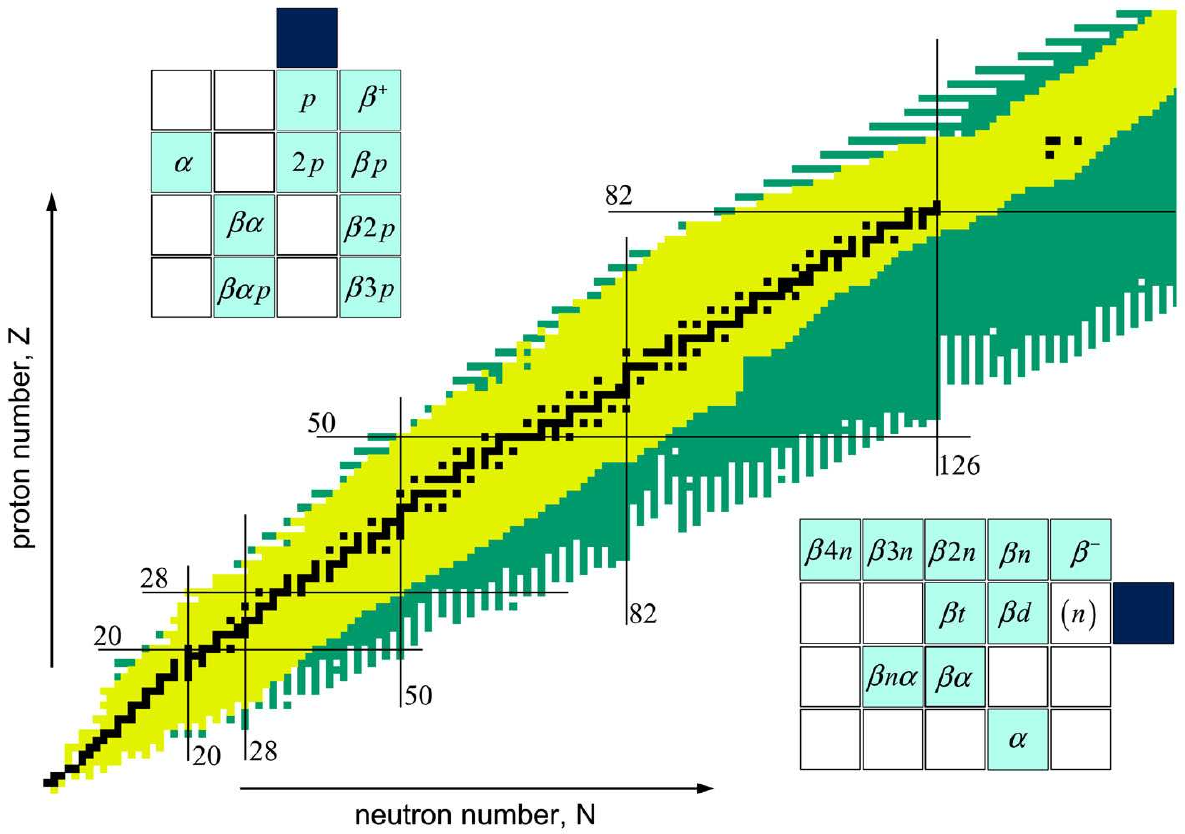}
}
\caption{(Color online) The chart of nuclei. The stable nuclides and the
radioactive ones which were experimentally identified are shown by black and
light/yellow,
respectively. The nuclides predicted to have positive nucleon separation energy
according to
the FRDM mass model \cite{Moller:1997}, but not yet observed, are shown in
dark/green.
The lines indicate
position of magic numbers corresponding to the closed neutron and proton shells
(the numbers smaller than 20 are not shown).
The insets show the location on the chart of the decay products of the parent
nucleus which is
indicated by a dark square. The observed
decay channels of the proton-rich and the neutron-rich nuclei are shown on the
left and on the right inset, respectively.}
\label{fig:IB_Chart}
\end{figure*}

The notion of radioactivity is useful in making distinction between emission of
rays or particles by a highly unstable system (for example undergoing a nuclear
reaction) from radiation emitted spontaneously by a system whose nuclear and
atomic degrees of freedom are close to equilibrium.
Such distinction, however, has to be arbitrary and usually a characteristic time
scale is used as a criterion. Throughout this review we adopt
the following definition. Radioactivity is a process of emission of
particles by an atomic nucleus
which occurs with characteristic time (half-life) much longer than the K-shell
vacancy half-life in a carbon atom, which amounts to about $2 \times 10^{-14}$~s
\cite{Bambynek:1972}. A relativistic particle travels in the
time of $10^{-14}$~s a distance of a few micrometers which is close to the
measurement limit in a nuclear emulsion.
In addition, this value coincides with a decay width, defined as $\Gamma =
\ln 2 \, \hbar/T_{1/2}$, of about 0.03~eV which is roughly the thermal
energy at room temperature. Thus, nuclear processes much slower
than filling the K-vacancy, whose duration in principle can be measured
directly,
and with the width much smaller than the thermal energy at room temperature
will be called radioactive. This definition applies to both nuclear ground
states and to long-lived excited nuclear states (isomers).

The definition of a nuclide relates to the definition of radioactivity. A
nuclide is a neutral atom, specified by the numbers $A$ and $Z$ of its nucleus,
which is either stable or lives long enough to be classified as radioactive. We
say that a nuclide does not "exist" if its nucleus decays too fast to be called
radioactive. All existing nuclides are represented on a chart of nuclides
spanned by the atomic number $Z$ and neutron number $N$
(Figure~\ref{fig:IB_Chart}). In the last three decades their number was growing
almost steadily from about 2200 in 1981 to about 3000 in 2006
\cite{Pfennig:2008}, giving an average of about 30 new nuclides identified per
year. Due to vigorous growth of nuclear facilities (Sec.~\ref{sec:IIIB}) this
trend is expected to continue in next decades.

The domain of processes occurring on a time scale shorter than radioactivity
is referred to as the resonant regime. The resonant phenomena are characterized
by features having a directly measurable width in the energy spectra. Typically,
widths of order meV is taken as the lower end of the resonant regime.
Thus, the distinction between radioactive and resonant phenomena is that the
former have a characteristic time which can be measured directly while the
latter have a characteristic energy width which can be measured directly.

\begin{figure*}[tbh]
\centerline{
\includegraphics[width=0.75\textwidth]{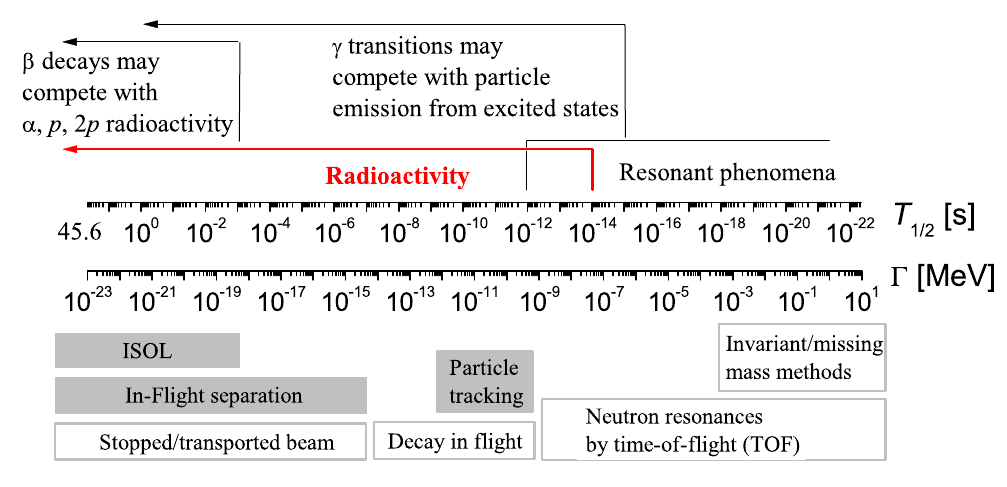}
}
\caption[Energy ranges, time scales, and experimental methods for decay
phenomena.]{Characteristic time scales and decay widths of radioactive decays.
Ranges of application of
selected experimental techniques are sketched.}
\label{fig:IB1:timescale}
\end{figure*}

Characteristic time scales for different radioactive decays are illustrated
schematically in Figure~\ref{fig:IB1:timescale}.

\subsection{Brief history}
\label{sec:IA}

The discovery of radioactivity by \textcite{Becquerel:1896} and the subsequent
discovery of new radioactive elements, polonium \cite{Curie:1898} and radium
\cite{Curie:1898b}
initiated a scientific breakthrough into the world of subatomic structure of
matter. Through early works on $\alpha$- and $\beta$-rays, distinguished and
named by \textcite{Rutherford:1899} and on $\gamma$-radiation, discovered by
\textcite{Villard:1900}, a manifold of nuclear phenomena started to emerge. The
nature of the new rays was soon clarified \cite{Rutherford:1908},
\cite{Kaufmann:1902}, \cite{Rutherford:1914}. Finally, two other milestones: the
discovery of the atomic nucleus \cite{Rutherford:1911} and of the neutron
\cite{Chadwick:1932} gave birth to a new discipline --- nuclear physics. A
brilliant and detailed account of these early years was given by
\textcite{Pais:1986}.

The full understanding of decay processes required quantum mechanics, as an
illustrious explanation of $\alpha$-decay by \textcite{Gamow:1928} and
\textcite{Gurney:1928} demonstrated. Soon afterwards,
early ideas of quantum field theory led Fermi to formulate the first theory of
$\beta$-decay \cite{Fermi:1934}. A novel type of radioactivity --- the
$\beta ^+$ decay --- was discovered by \textcite{Curie:1934}.
It was also the first instant of a radioactive nuclide synthesized in a
laboratory
in contrast to all previously known radioactive substances of natural origin.
\textcite{Wick:1934}, and independently
\textcite{Bethe:1934}, realized that a process of orbital electron capture (EC)
by an atomic nucleus is possible. Thus, the EC decay was the first type of
radioactivity predicted theoretically. It was confirmed by an observation of
K-capture by \textcite{Alvarez:1937}.
Spontaneous fission, discovered by \textcite{Flerov:1940}, completed the list of
"classical" radioactive decay modes. Ever since, they have played a crucial role
in learning about nuclear properties, in tagging reaction channels, and in
identifying new nuclides.

The beginning of the modern era of radioactivity can be marked by the work of
\textcite{Goldansky:1960} who in a systematical study considered
properties of very neutron deficient nuclei and discussed
possible decay modes like $\beta$-delayed proton emission and
proton radioactivity. He was the first to point out the possibility
of the two-proton radioactivity and to describe its key features.
The first emission of a proton following $\beta$ decay was reported by
\textcite{Karnaukhov:1963}. The first emission from an identified precursor was
observed
from an excited state of $^{25}$Al populated in the $\beta ^+$ decay of
$^{25}$Si \cite{Burton:1963}.
Such a variant of $\beta$ decay --- the $\beta$-delayed emission of particles
from states of a daughter nucleus --- become a field of study in its own and
includes various decay channels with emission of protons, neutrons, $\alpha$
particles,
deuterons, and tritons, see Figure\ref{fig:IB_Chart}.
This is the subject of Sec.~\ref{sec:IV}.
The first direct emission of a proton from a nuclear state was observed
to proceed from an isomeric state in $^{53}$Co \cite{Jackson:1970}.
The ground-state proton radioactivity was
observed for the first time in $^{151}$Lu \cite{Hofmann:1982} and
in $^{147}$Tm \cite{Klepper:1982}. Since then, almost fifty
proton emitters, including emission from isomeric states, were identified. This
field is
covered in Sec.~\ref{sec:V}. The prediction of the
two-proton radioactivity (\emph{2p}) had to wait much longer for experimental
confirmation. Such a decay mode was observed first for $^{45}$Fe by
\textcite{Pfutzner:2002} and in an independent experiment by
\textcite{Giovinazzo:2002}. This freshly opened field of nuclear spectroscopy is
covered in Sec.~\ref{sec:VII}.

For completeness, one should mention the cluster radioactivity, in which a
nuclear fragment heavier than the $\alpha$-particle but lighter than fission
fragments, is emitted. This decay mode was discovered by \textcite{Rose:1984}
who identified $^{14}$C ions emitted by $^{223}$Ra.
Later, many similar decay channels were observed
\cite{Poenaru:2002,Bonetti:2007}.
In all of them, however, the dominant decay mode is $\alpha$ decay, the cluster
emission
being less probable by a factor of at least $10^{9}$. In addition, this rare
decay mode can be established only in long-living nuclei, not far from the
nuclear stability. Thus, it is beyond the scope of this paper.

\subsection{Links and connections}
\label{sec:IC}

In general, information on nuclei come from different and complementary
experimental approaches.
Radioactive decay always was one of them and still remains a major tool in
nuclear
physics as well as in numerous branches of physics where nuclear degrees of
freedom are relevant.
An important attribute of the decay is the characteristic half-life. The very
delay between
the moment of production and the decay event, offers a filtering possibility,
thus helping to
increase the signal-to-background ratio. The values of the half-life provide a
way
to identify processes and offer tests to nuclear models. Another important facet
of spontaneous
decays are selection rules, which reflect conservation laws obeyed by underlying
interactions.
Effectively, they provide different filtering mechanisms, essential for the
planning of measurements
and for interpretation of their results.

The quest for superheavy elements may exemplify the role of radioactive
decays \cite{Hofmann:2000}. The emission of $\alpha$ particles by synthesized
heavy nuclides provides the very proof of their existence and a clean way for
their identification.
Moreover, the $\alpha$ decay is a powerful marker helping to establish the first
chemical properties of a few-atom sample of a new element, like in the case of
$^{283}$Cn \cite{Eichler:2007}.

Radioactivity plays a significant role in fundamental research, as can be
illustrated by historical examples of the neutrino hypothesis by
\textcite{Pauli:1930} and
the
discovery of parity non-conservation in the $\beta$~decay of $^{60}$Co
\cite{Wu:1957}.
A recent review of tests of the standard electroweak model by the beta
decay \cite{Severijns:2006} gives a comprehensive summary of this field.

However, since the present paper is limited to radioactive studies at the
limits of stability,
below we mention briefly a few research areas where they are particularly
important.

\subsubsection{Nuclear structure}
\label{sec:IC1}

With increasing scope of studies on the chart of nuclei new patterns emerge
and novel features are predicted.
An observed anomaly in masses of neutron-rich sodium isotopes
\cite{Thibault:1975} is
an early example which led to the notion of an "island of inversion" in vicinity
of $^{32}$Mg resulting from changes in sequence of single-particle
orbitals \cite{Nummela:2001,Yordanov:2007}. Recently, a similar phenomenon was
claimed to
occur around $^{62}$Ti \cite{Tarasov:2009,Flanagan:2009}. It was realized that
the classical shell
gaps --- the cornerstones of microscopic description of nuclei --- do migrate in
areas distant from stability. A detailed survey of the present situation
is given by \textcite{Sorlin:2008}. A new feature of quenching
of the shell-gaps due to influence of continuum states was predicted close
to the neutron drip line \cite{Dobaczewski:1994}.
A comprehensive review of novel aspects of nuclear structure emerging far from
the beta
stability is given by \textcite{Dobaczewski:2007}.

The experimental reach for most exotic nuclides suffers from very low production
yields. Radioactive decays allow not merely to identify a system of interest
but often offer the only practical source of structural information.
The first half-life determination for the
doubly magic $^{78}$Ni was done with 11 atoms \cite{Hosmer:2005}.
The first insight into the structure of $^{45}$Fe resulted from decays of 75
atoms \cite{Miernik:2007b}.
The first information on excited states in $^{70}$Ni was deduced from a few tens
of counts in a singles gamma spectrum \cite{Grzywacz:1998}. Here, the very clean
selection was provided by a delayed decay of a microsecond isomer.
Such first information, in turn, becomes essential in later, more advanced
experiments, allowing for coincidence measurements or for selection of events in
high-background conditions, for example, by decay tagging
\cite{Seweryniak:1997,Jenkins:2000}.

Recently, the insights gained on nuclear structure near the proton drip-line by
means of radioactive decay studies were reviewed by \textcite{Blank:2008}.

\subsubsection{Nuclear astrophysics}
\label{sec:IC2}

Several subjects of modern astrophysics are related to properties of atomic
nuclei.
They include questions on the origin of the elements, physics of compact
objects like neutron stars or white dwarfs, studies of stellar explosions
like supernovae, x-ray bursts, and many others. Although the most frequently
requested
nuclear data in astrophysics are reaction cross sections for some key
processes, the radioactive decays provide indispensable information
for many applications. Nuclei very far from the beta stability
play a particularly important role in nucleosynthesis of elements
heavier than iron, as pointed out in the seminal paper of
\textcite{Burbidge:1957}.
The rapid neutron capture (\emph{r}-process), responsible for about half of
the abundance of elements above iron, including all of uranium and thorium,
passes through regions of very neutron-rich nuclei, mostly far beyond the reach
of present experiments \cite{Kratz:2007}.
In turn, a number of neutron-deficient isotopes was produced in the rapid
proton capture (\emph{rp} - process) which involves very neutron-deficient
areas on chart of nuclei \cite{Schatz:1998}. Both processes
are expected to occur in explosive stellar conditions, details of which are
still
under debate. Theoretical reconstruction of these processes is additionally
hindered by the lack of relevant nuclear data. To those of special interest
belong masses, half-lives, branching ratios, and beta-delayed particle
emission probabilities. In principle, these values can be determined from
radioactive decay studies. Presently, however, most of the nuclei important in
this
context are either difficult or impossible to synthesize
in reasonable quantities. The theoretical models or empirical
parameterizations have to be relied on, instead.
Thus, the motivation to extend experimental decay studies to the limits of
nuclear
stability is twofold. They deliver data directly needed
by astrophysical calculations, and they help to test and improve
models of nuclear structure, increasing reliability of theoretical
extrapolations.

A thorough and detailed discussion of the present status of the interplay
between nuclear structure and astrophysics, with the emphasis on stellar
evolution and nucleosynthesis, can be found in \textcite{Langanke:2003} and
in \textcite{Grawe:2007}. Both these reviews reveal the necessity of
exploring exotic nuclei and their decay properties in order to fully
understand astrophysical aspects of our Universe.

\subsubsection{Open quantum systems}
\label{sec:IC3}

Several fields of modern physics face the goal of describing quantum many-body
systems which
are not isolated from its quantal environments. Examples of such open quantum
systems
include quantum dots, droplets of neutral atoms, microwave cavities, or weakly
bound
nuclei very far from the beta stability \cite{Okolowicz:2003}.
Nuclear physics at the limits of stability appears as a particularly
promising testing ground of new concepts. When the nuclear binding energy
decreases, the
conceptual separation of well localized, bound states from the continuum
scattering states
becomes artificial and, in fact, is hampering the correct description of various
features, like neutron halos, Thomas-Ehrman shifts, or clustering phenomena
\cite{Dobaczewski:2007, Michel:2010}.

The quest to formulate a unified description of nuclear structure and nuclear
reactions
resulted in extensions of nuclear shell model, like the Shell Model Embedded in
the Continuum \cite{Bennaceur:1998,Okolowicz:2003}, or more recently the Gamow
Shell Model
\cite{Michel:2002,Michel:2009}.
New spectroscopic data on very exotic nuclei, and on complex decay modes, like
two-proton
radioactivity, or beta-delayed multi-particle emission would stimulate
further developments in this field. Thus, radioactive decays at the limits of
nuclear stability may be instrumental in improving our fundamental understanding
of many-body quantum systems.

\subsection{Outline}
\label{sec:IE}

In Section~\ref{sec:II} we discuss the limits of nuclear stability
using the concept of the drip-line based on nucleon
separation energy. The experimental situation in accessing
both the proton and the neutron drip-line is briefly presented.
The experimental techniques pertaining to radioactivity studies
far form stability are reviewed in Section~\ref{sec:III}.
Various reactions used to produce exotic nuclides and the
main methods of their separation are shortly described. Finally,
selected detection techniques of special importance for
measurements of radioactive decays are presented.
The following Sections~\ref{sec:IV}--\ref{sec:VII} are
devoted to the main radioactive decay modes at the limits of
stability: $\beta$-delayed particle emission, proton radioactivity,
$\alpha$ emission, and two-proton radioactivity. The latter decay mode
is treated in considerably more detailed way, as it is the least known
and its understanding is still in a status of development.
In Section~\ref{sec:VIII}
the prospects of neutron radioactivity are examined. The main conclusions
of the paper are shortly summarized in the final Section~\ref{sec:IX}.

Throughout this work we use the system of units in which $\hbar = c = 1$.


\newpage
\section{LIMITS OF STABILITY}
\label{sec:II}


The limits of the nuclear world are determined by the nuclear binding energies.
The limits relevant to this review are often characterized by the
\emph{drip lines} which separate bound systems from the unbound ones. Although
different definitions can be encountered in the literature, we adhere to
the simplest and most common one which is based on the single-nucleon separation
energy. The proton- and the neutron separation energy of a nuclide with numbers
$N$
and $Z$ are given by:
\begin{eqnarray}
    S_p(N,Z) = B(N,Z)-B(N,Z-1) \\
    S_n(N,Z) = B(N,Z)-B(N-1,Z).
\end{eqnarray}
The $B(N,Z)$ is the binding energy of the nuclide related to its mass $M(N,Z)$ :
\begin{equation}
    M(N,Z) = Z \, M_H + N \, m_n - B(N,Z),
\end{equation}
where $M_H$ and $m_n$ are masses of the hydrogen atom and the neutron,
respectively.

\begin{figure}[th]
\centerline{
\includegraphics[width=0.48\textwidth]{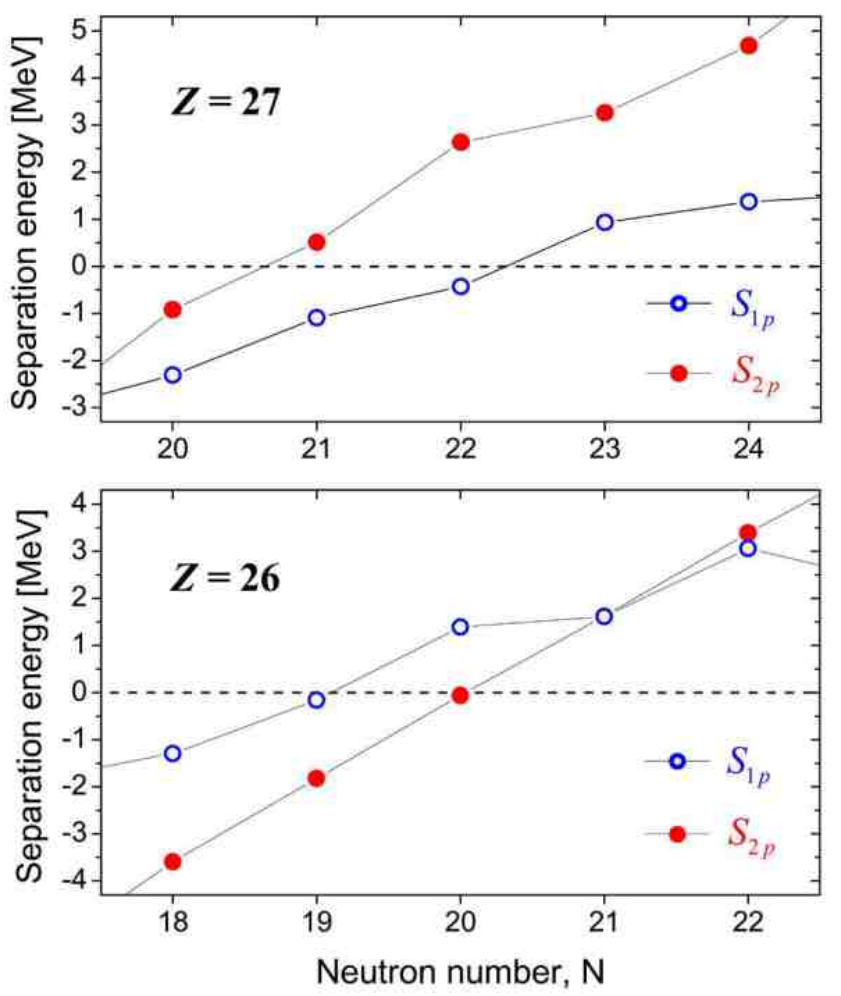}
}
\caption{(Color online) The proton- and two-proton separation energies of iron
and
cobalt isotopes as predicted by the FRDM mass model \cite{Moller:1997}. }
\label{fig:II_psep}
\end{figure}

When we move along the line of isotopes with the given atomic number $Z$,
starting
from stability towards neutron-deficient nuclides, the proton separation energy
$S_p$
decreases and at certain location it becomes negative. The proton drip-line is
defined as the border between the last proton-bound isotope and the first
one with the negative value of the $S_p$. The typical situation, according to
the
predictions of a particular mass model \cite{Moller:1997} for the isotopes of
iron
and cobalt is presented in Figure~\ref{fig:II_psep}. It follows from this model
that
the proton drip-line for iron should lie between $^{45}_{26}$Fe$_{19}$ and
$^{46}_{26}$Fe$_{20}$,
while in case of cobalt it is located between $^{49}_{27}$Co$_{22}$ and
$^{50}_{27}$Co$_{23}$.
Generally, the proton drip-line for odd $Z$ isotopes is closer to stability
than in case of the neighboring even-$Z$ which results from the proton pairing
energy.

\begin{figure}[th]
\centerline{
\includegraphics[width=0.48\textwidth]{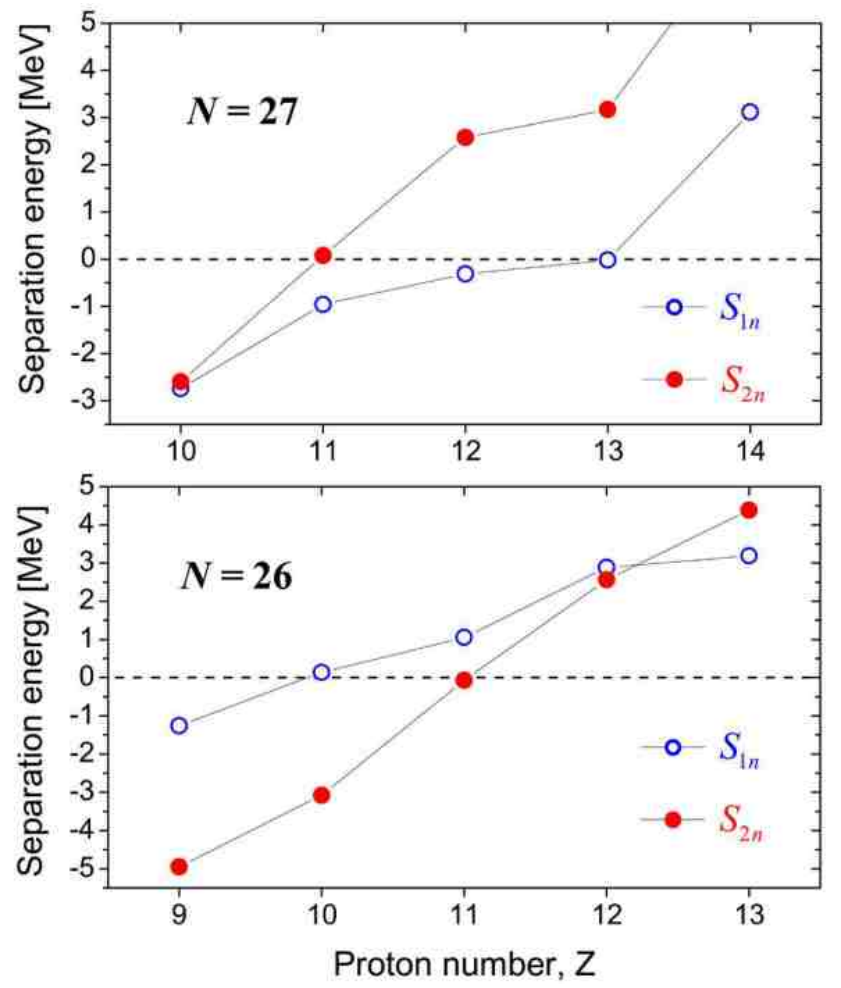}
}
\caption{(Color online) The neutron- and two-neutron separation energies for the
$N=26$ and $N=27$ isotones, as predicted by the FRDM mass model
\cite{Moller:1997}. }
\label{fig:II_nsep}
\end{figure}

In the fully analogous way, the neutron drip-line for a given neutron number $N$
is
defined as a border between the last neutron bound isotone, when counting from
stability, and the first one for which the neutron separation energy $S_n$ is
negative. The predicted separation energies for the $N=26$ and $N=27$ isotones
are shown in Figure~\ref{fig:II_nsep}. Thus, for $N=26$ the neutron drip-line
is expected to lie between $^{35}_{9}$F$_{26}$ and $^{36}_{10}$Ne$_{26}$.
Similarly to the proton case, the neutron drip-line for the odd-$N$ is closer
to stability than for the neighboring even-$N$ which
reflects the neutron pairing energy.

The drip lines as defined above are very useful in identifying and discussing
limits of stability, but to some extend they are arbitrary and they do not
provide the unambiguous demarkation of nuclear stability. This can be seen by
inspecting
the two-nucleon separation energies:
\begin{eqnarray}
    S_{2p}(N,Z) = B(N,Z)-B(N,Z-2) \\
    S_{2n}(N,Z) = B(N,Z)-B(N-2,Z).
\end{eqnarray}
Due to the pairing interaction, in case of an even number of nucleons
the two-nucleon separation energy can be smaller than the single-nucleon value.
For the cases discussed above, it is illustrated in
Figures~\ref{fig:II_psep} and \ref{fig:II_nsep}. Although the proton separation
energy in $^{46}_{26}$Fe$_{20}$ is positive, this nuclide is expected to be
slightly two-proton unbound. Similar situation is observed in the $N=26$
isotones --- the two-neutron instability develops first, before the neutron
drip-line is reached. The additional complication, which is essential on
the neutron-deficient side, comes from the fact that the exact position of a
nuclide
with respect to the drip-line cannot determine alone its dominant decay
mode. This is caused by the Coulomb and centrifugal barriers which hamper
emission of nucleons. Only when the nucleon penetration probability through the
barrier,
depending on the energy and the angular momentum of the initial state, is large
enough, the particle radioactivity can compete with $\beta$ decay.
Thus, although the mentioned $^{46}_{26}$Fe$_{20}$ may be two-proton unbound,
it is known to decay by $\beta^+$ transition. In turn, $^{45}_{26}$Fe$_{19}$
is sufficiently two-proton unbound to decay predominantly by the \emph{2p}
radioactivity, although the $\beta^+$ channel has a substantial
branching \cite{Miernik:2009}. The exact position of this nucleus with respect
to the proton drip-line turns out to be irrelevant for its radioactive decay.
In case of odd-$Z$ nuclides, the proton radioactivity can win the competition
with $\beta$ decay only when its proton separation energy is sufficiently
negative.
Thus, the observation of the proton radioactivity proves that the nuclide is
located
beyond the proton drip-line but the exact position of this line cannot be
determined from decay data alone. This can be achieved only by precise
mass measurements of nuclides in the region of interest.

We note, that on the neutron-deficient edge of the chart of nuclides above
tungsten
the dominant decay mode is the $\alpha$ emission which happens to proceed
faster than $\beta$ decay. Thus, beyond the proton
drip-line in this region, the proton radioactivity competes actually
with $\alpha$ decay.

\begin{figure}[th]
\centerline{
\includegraphics[width=0.48\textwidth]{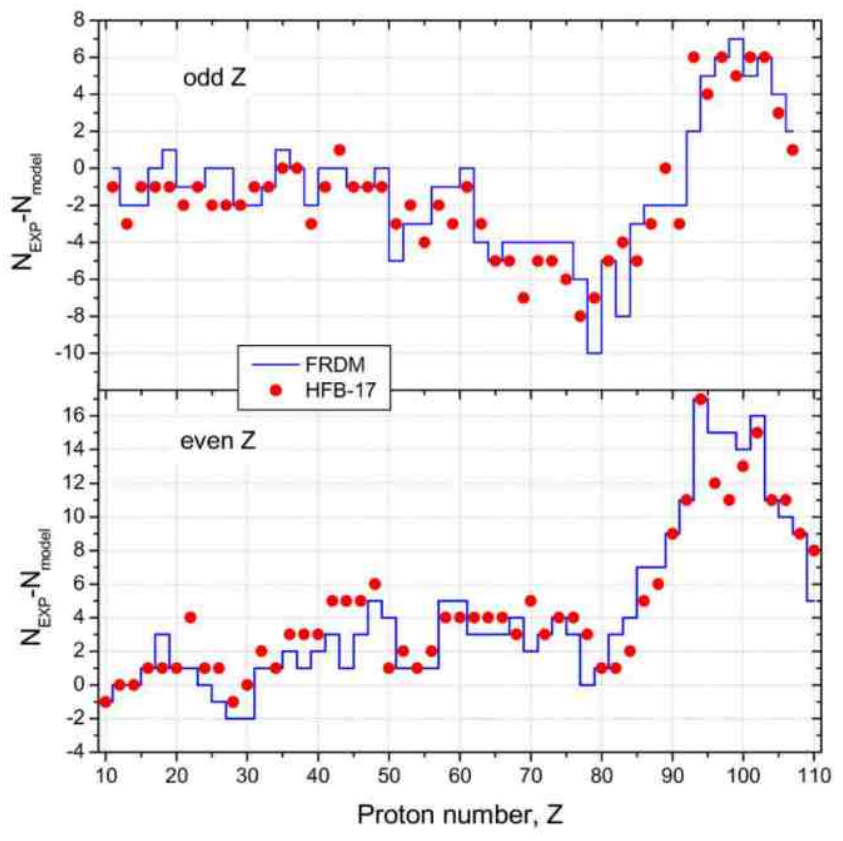}
}
\caption{(Color online) The difference between the neutron number of the
lightest experimentally observed isotope for a given atomic number $Z$ and the
corresponding prediction for the last isotope before the proton drip-line
according to the FRDM mass model \cite{Moller:1997}(line) and the HFB-17 model
\cite{Goriely:2009}
(circles). The results for the even $Z$ and the odd $Z$ are shown in the
bottom and in the top, respectively. The experimental values were taken from
\textcite{Magill:2009} with corrections contained in \textcite{Baumann:2007}. }
\label{fig:II_pdrip}
\end{figure}

At the neutron drip-line the situation is different because the unbound
neutrons are not affected by the Coulomb barrier. The influence of the
centrifugal
potential alone is much weaker as it decreases with radius effectively
as $1/r^2$ in contrast to the $1/r$ dependence of the Coulomb potential.
In consequence, the effect of the centrifugal barrier is expected to be
observable only in rare cases of very low decay energies and large angular
momentum. The resulting possible neutron and two-neutron radioactivity
is examined in more detail in Sec.~\ref{sec:VIII}. For practical purposes,
any system with negative neutron- or two-neutron separation energy can be
expected to live too short to be qualified as radioactive. Therefore, the
limits of stability on the neutron-rich side could in principle be
established rather precisely by inspecting which is the lightest isotone
still undergoing radioactive decay. The problem, however, is that
it is very difficult to reach experimentally the neutron drip-line
for $N > 28$.

\begin{figure}[th]
\centerline{
\includegraphics[width=0.48\textwidth]{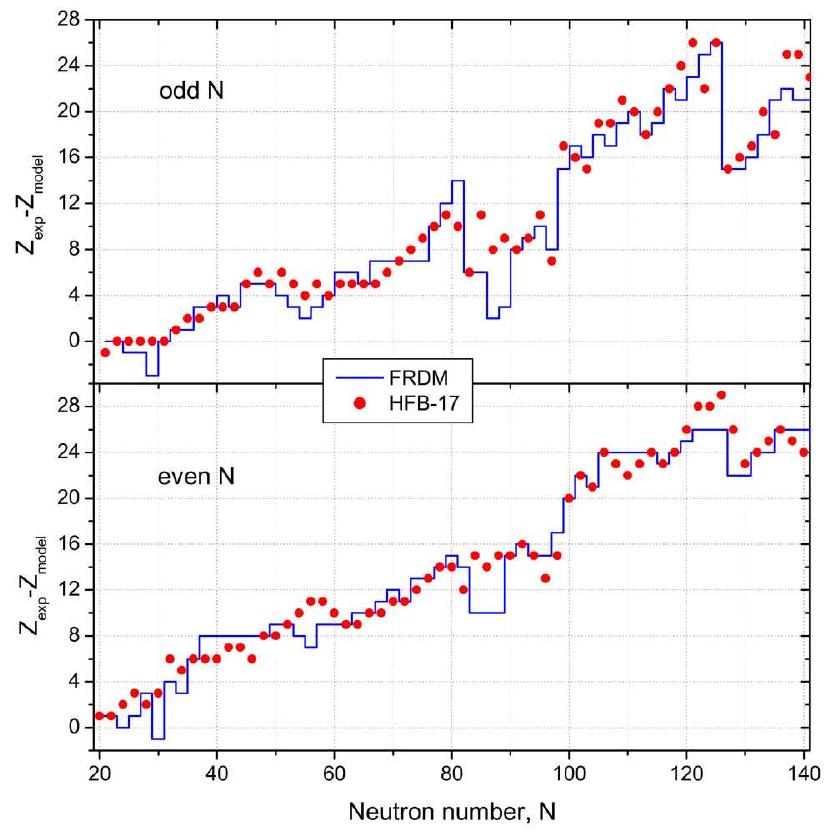}
}
\caption{(Color online) The difference between the proton number of the
lightest observed isotone for a given neutron number $N$ and the prediction
of the last stable isotone before the neutron drip-line according to the
two theoretical mass models. The results for the even $N$ and the odd $N$ are
shown in the bottom and in the top, respectively. The plot details are the
same as in Figure~\ref{fig:II_pdrip}. }
\label{fig:II_ndrip}
\end{figure}

The general picture is presented in Figure~\ref{fig:IB_Chart} where all
nuclides identified experimentally until now are superimposed on the
plot of all nuclides predicted to have positive proton
and neutron separation energy by the Finite Range Droplet Model (FRDM)
developed by \textcite{Moller:1995,Moller:1997}. This model is a successful
representant of a class of macroscopic-microscopic mass formulae,
combining the macroscopic liquid-drop parametrization with the
microscopic shell and pairing corrections. The prediction reveals
a few characteristic features, like the even-odd staggering
for the neutron deficient isotopes and for the proton deficient
isotones, or the strong influence of the $N=82$ and $N=126$ shells
on the neutron-rich side of the chart. To illustrate theoretical
uncertainties we compare predictions of the FRDM model with results of
the HFB-17 model \cite{Goriely:2009} which represents a class of fully
microscopic approaches based on Hartree-Fock-Bogoliubov formalism and
Skyrme forces \cite{Goriely:2010}.
Figure~\ref{fig:II_pdrip} presents the current experimental situation
and the comparison of the two models. It shows differences between
the neutron number of the lightest observed isotope and the predicted
values for the last proton-stable isotope before the proton drip-line.
We see that both predictions agree well with each other --- they follow
the same pattern and they differ by a few units at most.
The negative values indicate those observed nuclides which are located
beyond the predicted proton-drip line, in most cases they are proton
emitters. The large group of such nuclides, seen for
odd-$Z$ values between 50 and 90 illustrates the strong impact of
the Coulomb barrier on the heavy nuclei. On the other hand for almost
all even-$Z$ elements, there are predicted bound isotopes which remain
to be observed. A distinguished peak of positive values for the $Z>90$
results from experimental difficulties to produce proton-rich nuclei
in this region.

The analogous information for the neutron-rich side is given in
Figure~\ref{fig:II_ndrip} where the differences between the proton number
of the lightest observed isotone and the predicted values for the last
neutron-stable isotone before the neutron drip-line are plotted.
Again, both models are consistent with each other. The largest
difference is seen above the $N=80$ where the FRDM model
seems to exhibit large variations due to the neutron shell closure --- an
effect not pronounced in the HFB-17 model. In contrast to the proton-rich
side, however, almost all values are positive and they increase rapidly
with the increasing neutron number. This reflects the fact that except
for the light nuclei, the neutron drip-line is far from the body of
presently observed nuclides, as is dramatically evident also in
Figure~\ref{fig:IB_Chart}. In fact, the drip-line has been determined
experimentally and unambiguously for even $N$ only up to $N=20$
and for odd $N$ up to $N=27$ \cite{Thoennessen:2004}.

The significant expansion of the body of observed nuclides, especially on
the neutron-rich side of the chart, is expected only when the next generation
of radioactive beam facilities will come into operation, see Sec.~\ref{sec:III}.

The current status of the knowledge on atomic masses and of the global
mass models can be found in \textcite{Lunney:2003}. The detailed discussion
of the present experimental knowledge of the limits of nuclear stability
was presented by \textcite{Thoennessen:2004}.


\newpage
\section{EXPERIMENTAL TECHNIQUES}
\label{sec:III}


Experimental studies of nuclei at the limits of stability belong to the
front-line of physical research. A view on experimental techniques,
given in this section, provides a general perspective on the advanced methods
of present-day low-energy nuclear physics. First, reactions used to produce
radioactive nuclides will be mentioned followed by a short description of the
main
methods of their extraction and separation. Then, selected aspects of modern
detection systems will be reviewed with an emphasis on recording manifestations
of radioactivity.

\subsection{Production}
\label{sec:IIIA}

The methods of production of nuclides far from beta stability are almost
exclusively based
on nuclear reactions involving stable nuclides or their ions.
In a simplified view, a new nucleus
is formed either by fusion of two other nuclei (projectile and target),
by exchange of nucleons between the projectile and the target nuclei (transfer),
or in reaction leading to removal of nucleons either from the target or from the
projectile nucleus (fragmentation, spallation, fission).

In principle, a radioactive nucleus produced in one of these reaction, having
sufficiently
long half-life, can be used as a projectile to initiate a secondary reaction in
which
nuclei even further from stability are formed. This is the idea of radioactive
beams
which has been driving many experimental developments
\cite{Tanihata:2008,ISOLbeams:2011}.
It is anticipated that reactions
induced by radioactive beams will play a major role in the future expansion of
the chart of nuclei. In addition, the secondary reactions induced by radioactive
projectiles represent one of the main methods to produce radioactive nuclides
with the
shortest half-lives, in the nanosecond range or shorter. For completeness, we
note
that in some cases a radioactive target may be used for the production of exotic
nuclei. For example, in the recent discovery of a new element with the atomic
number $Z=117$, the radioactive target of $^{249}$Bk was used
\cite{Oganessian:2010}.

Each production method has its own characteristics and a typical application
range. In the
following we mention briefly various reactions which are being used and we
direct the reader
to papers providing more detailed and broader presentations, as well as
references to relevant technical contributions.
Some aspects of reactions used to produce exotic nuclei were
discussed by \textcite{Geissel:1995} and more recently by
\textcite{Schmidt:2002}.

\subsubsection{Fusion-evaporation}
\label{sec:IIIA1}

In a central collision at low energy two nuclei can join together
(\emph{fusion}) to
form a single heavier nucleus. In the second step, the resulting
compound nucleus releases its excitation energy by
emission of nucleons (\emph{evaporation}) and radiation.
The reaction cross section is very sensitive to the initial energy in the
projectile-target system, which must be close to the Coulomb barrier.
If the energy is too low, the probability of barrier penetration
drops dramatically, if it is too large, other channels
start to dominate.
This is the key reaction in the synthesis of superheavy
elements \cite{Schmidt:1991,Hofmann:2009}.
However, since the final nucleus tends to be located on the neutron-deficient
side of stability, the fusion-evaporation is successfully used to
produce very neutron-deficient systems. In fact, most of research on proton
radioactivity employs this reaction \cite{Ferreira:2007, Woods:1997}.
The most commonly used tool
for optimizing experimental conditions and for prediction of cross-sections
is the statistical code \textsc{HIVAP} \cite{Reisdorf:1981, Reisdorf:1992},
but other statistical codes like CASCADE \cite{Puhlhofer:1977} or PACE
\cite{Gavron:1980}
are also being used.
The potential of the fusion-evaporation reaction can be illustrated by an
attempt to reach $\alpha$-emitters above $^{100}$Sn \cite{Korgul:2008}
and by an investigation to produce the lightest isotopes of bismuth and
polonium \cite{Andreyev:2005}.

\subsubsection{Multi-nucleon transfer}
\label{sec:IIIA2}

Transfer reactions belong to the category of binary processes where instead
of a fused system of two heavy ions a projectile-like and a target-like nucleus
appear in the final state. This happens if the collision is not central.
If it is also deep-inelastic (damped), a few nucleons can be
exchanged between reaction partners leading to radioactive products.
Although a part of the energy of the relative motion goes
into the excitation of the final fragments, which is released by evaporation
of light particles, still residual nuclei far from stability can be formed.
Such multi-nucleon transfer reactions at Coulomb barrier energies has been
used to produce unstable nuclides, including neutron rich
ones \cite{Broda:2006}. The method is mainly used in combination
with in-beam $\gamma$-ray spectroscopy and isomeric spectroscopy in various
regions of the chart on nuclei \cite{Cocks:2000, Montanari:2011}. The main
advances on
this field in the last decade and the summary of theoretical understanding
of the reaction mechanism are given in the recent review by
\textcite{Corradi:2009}.
The current limits of nuclear stability cannot be reached by multi-nucleon
transfer between
stable projectile and target, but the importance of this reaction is increasing
with developments of radioactive beams. The transfer reactions are considered
also
as a tool to produce new isotopes in the region of superheavy nuclei
\cite{Zagrebaev:2008}.

\subsubsection{Fragmentation}
\label{sec:IIIA3}

When the collision energy of two heavy nuclei is large compared to the Fermi
energy of nucleons, the probability that nucleons will be exchanged between the
reaction partners
becomes very small. Instead, violent interactions occur in the overlapping zone
of the
projectile and the target (\emph{participants}), while their parts outside this
zone
(\emph{spectators}) emanate as the projectile-, and target-like prefragments,
respectively. After this abrasion phase, the cooling of prefragments  by
evaporation
of particles, by radiation, or by fission proceeds and the final fragments are
formed.
If the excitation energy of the prefragment is large, which happens in more
central
collisions, the multifragmentation takes place, i.e. the break-up into many
intermediate-mass fragments.
In the so called \emph{limiting fragmentation} regime, for projectile energies
above 100 MeV/nucleon, a characteristic feature is observed that the total
reaction cross section weakly depends on projectile energy and
can be approximated by a simple geometric formula
\begin{equation}
    \sigma_R = \pi \, r^{2}_{0} \, (A^{1/3}_T + A^{1/3}_P - c),
\end{equation}
where $A_T$ and $A_P$ are the mass numbers of the target and the projectile,
respectively,
the radius parameter is $r_0 = 1.1$~fm, and a correction for nuclear
transparency
is introduced by a parameter $c \cong 2$ \cite{Kox:1987}.

In the present context, the fragmentation of the projectile plays a special
role.
When high energy projectile ions collide with target nuclei, the projectile-like
fragments surviving the abrasion phase continue moving with almost no change of
velocity.
Thus, the resulting unstable nuclei form a secondary beam which can be
transported
and filtered by means of ion-optical devices. This method is very fast and
universal, since practically any nucleus with numbers $N$ and $Z$ smaller than
those of the projectile can be produced. These features make fragmentation one
of the key reactions for radioactive beam facilities.
The method of projectile fragmentation was pioneered at Bevalac facility at
the Lawrence Berkeley Laboratory \cite{Symons:1979,Westfall:1979}.
Later the systematic studies of fragmentation cross sections as a function
of energy, projectile, and target were carried out by \textcite{Webber:1990}.
Recently, comprehensive
studies of projectile-like fragmentation are being carried out at SIS-FRS system
at
GSI Darmstadt \cite{Benlliure:2008,Henzlova:2008}. The most advanced theoretical
description of the fragmentation is currently achieved in a modern version of
the abrasion-ablation model and is implemented as the Monte-Carlo
code ABRABLA \cite{Gaimard:1991}.
The mechanism of prefragment excitation is understood in this model as a result
of random
creation of holes in the nucleonic Fermi distribution \cite{Schmidt:1993}.
The evaporation stage is modeled with the code ABLA \cite{Kelic:2008}.
The simpler, analytical version COFRA \cite{Benlliure:1999, Benlliure:2000}
is applicable to the very-neutron rich fragments (cold fragmentation).
For practical estimates, the empirical parametrization of the fragmentation
cross sections is given by a simple analytical model EPAX \cite{Summerer:2000}.

\subsubsection{Spallation}
\label{sec:IIIA4}

\begin{figure*}[th]
\centerline{
\includegraphics[width=0.99\textwidth]{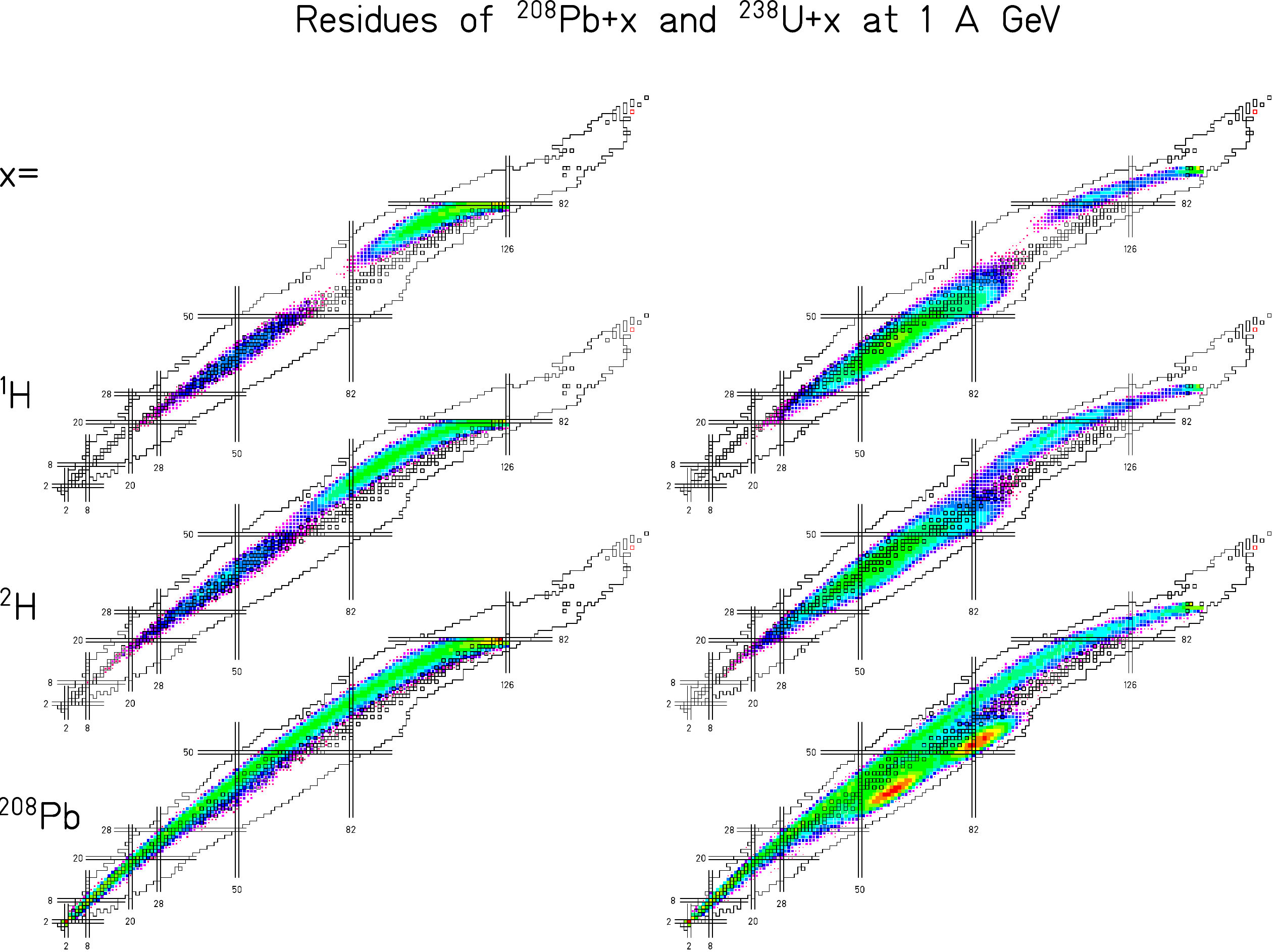}
}
\caption{(Color online) Systematic overview on calculated isotopic production
cross sections in different reactions. For clarity only values above 100~$\mu$b
are shown. From \textcite{Schmidt:2002}.}
\label{fig:IIIA_KHS}
\end{figure*}

If in a high energy collision (above 100 MeV/nucleon) one of the reaction
partners
is a light ion, like proton, deuteron, or triton, the process is referred
to as spallation.
From the perspective of production of exotic nuclei, the main difference
from projectile fragmentation is that in case of spallation usually the target
nucleus being the heavier partner is the source of radioactive nuclei.
In addition, the mechanism of its primary excitation is different.
The first step of the spallation is usually described as a series of collisions
between nucleons in the target nucleus, induced by the projectile,
which forms the base of intranuclear cascade models (INC). Such a cascade
of collisions leads to a highly excited system which, in the
second phase, deexcites in the same way as the
hot target-like prefragment. In consequence, the target nucleus
is destroyed, and in analogy to the fragmentation, practically
any nucleus with numbers $N$ and $Z$ smaller than those of the target can be
produced. Due to this universality, the spallation is the second main production
process considered for radioactive beam facilities. In contrast to
fragmentation,
however, the radioactive products have to be extracted from the target material.

The example of a modern version of an INC approach with a discussion
of the physics involved is given
by \textcite{Boudard:2002}. For the second stage of the reaction, various
versions of statistical evaporation codes are being used \cite{Gentil:2008}.
The detailed experimental studies of the spallation are conveniently performed
in the inverse kinematics, where a light target (hydrogen, deuterium)
is bombarded by heavy nuclei accelerated to relativistic energy.
Such studies are performed with use of the fragmentation facility which
reflects a symmetry between these two reactions.
This method was recently used at GSI Darmstadt in a comprehensive study of
residual
fragments produced by the spallation of $^{238}$U by
protons \cite{Taieb:2003,Ricciardi:2006} and by
deuterons \cite{Casarejos:2006,Pereira:2007}. A similar
work on spallation of $^{136}$Xe and $^{56}$Fe by protons was reported
by \textcite{Napolitani:2007} and by \textcite{Gentil:2008}, respectively.

\subsubsection{Fission}
\label{sec:IIA5}

Since heavier stable nuclei are more neutron-rich than the lighter ones, the
process of fission of a heavy nucleus is a source of neutron-rich medium mass
nuclei. In addition, fission is one of the important decay
channels of excited heavy nuclei. Thus, it plays a role as a direct
source of exotic nuclei and as a process interfering with other reactions
used for this purpose. Applications of fission to generate neutron-rich
nuclei differ in methods used to excite a fissile nucleus and in the range
of excitation energies imparted. On the low-excitation end is
the spontaneous fission and the thermal-neutron induced
fission \cite{Wahl:1988,Rochman:2004}. To this class belongs also fission
resulting from electromagnetic excitation (photofission) \cite{Cetina:2002}.
The photons inducing fission may be produced directly, e.g. by converting
an intense electron beam into bremsstrahlung \cite{Diamond:1999} or they
can be virtual, resulting from a fast motion of a fissile system relative
to a high-$Z$ target \cite{Bertulani:1986}. The higher excitation energies
are achieved by bombarding fissile targets with beams of fast neutrons or
charged particles. Low energy proton-induced fission is a frequent
choice because of the relative technical simplicity. The main aspects of
this method are discussed by \textcite{Penttila:2010} who recently developed
a novel method to measure the particle-induced fission yields.
The high energy reactions induced by light or heavy ions lead to high
excitation energies and subsequent fission becomes one of the main deexcitation
channels influencing the outcome of the spallation and fragmentation
reactions, respectively. High energy reactions in inverse kinematics,
where a heavy fissile nucleus is the projectile, has been proven to be
exceptionally fruitful. The pioneering experiments with relativistic $^{238}$U
beam at GSI Darmstadt revealed the properties and advantages of this
approach \cite{Bernas:1994,Bernas:1997}. When the target nucleus has a large $Z$
number,
the excitation of a fissile projectile-like fragment has a nuclear
contribution (fragmentation) and an electromagnetic one (photofission).
This situation was systematically investigated in the reaction
of 1 GeV/nucleon $^{238}$U impinging on a lead target \cite{Enqvist:1999}.
A similar technique was applied in a broad campaign dedicated to the
comprehensive study of spallation of heavy nuclei. The contribution of
fission was investigated in detail for the spallation of $^{238}$U by
hydrogen \cite{Bernas:2003} and by deuterium \cite{Pereira:2007}, as well as
of $^{208}$Pb by hydrogen \cite{FernandezDominguez:2005}.

\begin{figure}[t]
\centerline{
\includegraphics[width=0.5\textwidth]{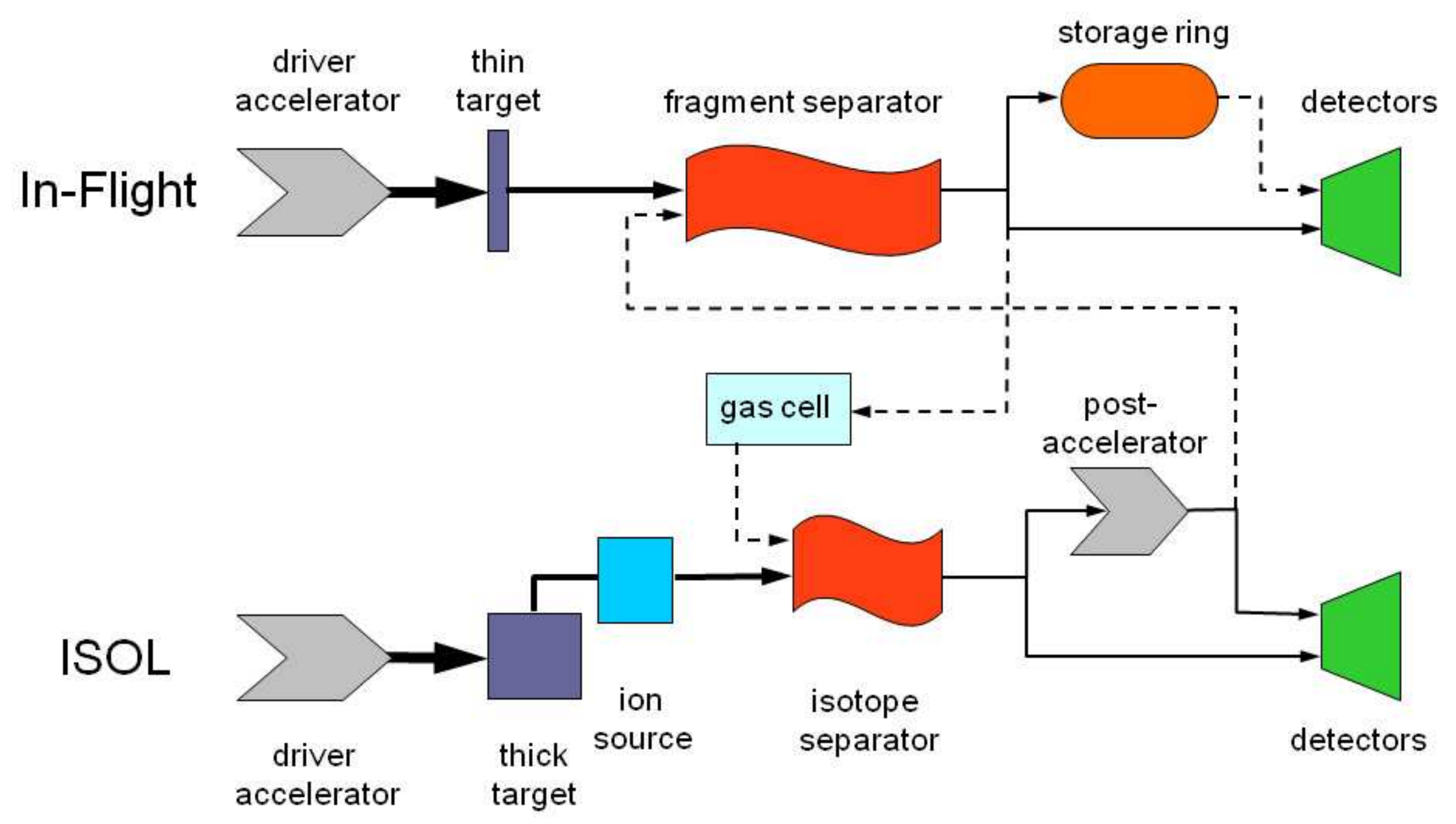}
}
\caption{(Color online) A general scheme of the two main methods used to extract
and separate radioactive nuclei. The dashed lines indicate connections which are
considered in planning future facilities.}
\label{fig:IIIB_scheme}
\end{figure}

The emerging general picture of the production of nuclei in both the spallation
and the fragmentation reactions with inclusion of the fission channel was
discussed by \textcite{Schmidt:2002}. It is illustrated in
Figure~\ref{fig:IIIA_KHS}
which shows the production cross sections for beams of $^{208}$Pb and $^{238}$U
impinging at 1~$A$~GeV on hydrogen, deuterium, and lead targets. Although
results of model calculations are shown in Figure~\ref{fig:IIIA_KHS}, they
represent very well features observed in experiments.

Recently, in the new-generation RIBF facility at RIKEN Nishina Center,
the in-flight fission of a 345 MeV/nucleon $^{238}$U beam has been
used to produce 45 new neutron-rich isotopes \cite{Ohnishi:2010}.

\subsection{Separation}
\label{sec:IIIB}

In reactions used to produce radioactive nuclei always a large number of
different products is formed and some method of selection is necessary to filter
nuclei of interest from an unwanted background. Very generally, one can
classify all separation methods which are used into two distinct classes.
The main difference is the target thickness relative to the range of products
in the target material. If the relative target thickness is small, the products
emerge from the target with significant kinetic energy and can be promptly
manipulated by ion-optical devices. The separation techniques of this type
are called \emph{In-Flight} and the filtering devices are called recoil or
fragment separators. On the other hand, if the production target is relatively
thick, such that the products are stopped in its volume, the nuclei of interest
have to be extracted from the target for further filtering. For historical
reasons such approach is referred to as \emph{ISOL} (which stands for Isotope
Separator On Line) technique. The principle of operation of In-Flight and ISOL
type of facility is shown schematically in Figure~\ref{fig:IIIB_scheme}.
In the following we discuss briefly only basic features of both these
methods and provide selected examples of laboratories in which they are
implemented,
referring the reader to technical papers with detailed information.
A comparison of the two separation methods with a discussion of some future
prospects, including the hybrid combination, was made by
\textcite{Tanihata:2008}.

\subsubsection{In-Flight}
\label{sec:IIIB1}

The key feature of the In-Flight methods is that the kinetic energy of the
reaction
product is large enough to escape from the relatively thin production target.
This method is applicable to reactions induced by heavy ions like
fusion-evaporation,
multi-nucleon transfer and projectile fragmentation or fission in inverse
kinematics.
The products emerging from the target enter an ion-optical system of magnetic
and
electric fields where they are separated from unwanted contaminants and
delivered to
the final experimental station. Usually, the main selection is applied to the
mass-over-charge ratio $A/q$ of particles by means of a uniform magnetic field,
according to
the formula relating the magnetic field $B$ with the momentum $p$ of a particle
having
charge $q$ and moving in this field along a circular trajectory of the radius
$\rho$:
\begin{equation}
    B \, \rho = \frac{p}{q} = u \, \beta \, \gamma \, \frac{A}{q},
\label{eq:IIB1}
\end{equation}
where $u$ is the atomic mass unit,
$\beta$ is the particle velocity, and
$\gamma$ is the Lorentz factor ($\gamma = 1/\sqrt{1-\beta ^2}$).
Additionally, in some separators the crossed magnetic and electric fields are
used to select the velocity of particles (Wien filter).
The In-Flight method is fast as the typical time of flight
through the separator is of the order of a microsecond or shorter. Another
important
feature is the lack of chemical sensitivity.

At the low-energy end, the fusion reaction requires projectiles with the energy
of the order of 10~MeV/nucleon (Coulomb barrier). The target thickness is
usually
of the order of 1~mg/cm$^2$. The average energy of the
reaction product results simply from the momentum conservation in a system of
collision partners. For this reason, the filtering devices in this case
are called \emph{recoil} separators. The energy of the recoiling products
is usually too small to allow for the in-flight identification of ions.
The general properties of recoil separators were reviewed by
\textcite{Davids:2003}.

\begin{table*}
\caption{The leading laboratories employing the In-Flight method to produce and
study the
          radioactive decays very far from the stability. In the lower part of
          the table
          the facilities under construction are listed.}
\vspace{0.5\baselineskip}
\begin{tabular}{lllccll}
  \hline
  \hline
  Country & Laboratory  & Driver      &  Beams   & Max beam     & Separator &
  Reference        \\
          &             & accelerator &          & energy [$A \,$MeV]      &
                 &     \\
  \hline
  Finland & Jyv\"{a}skyl\"{a}& cyclotron &  Ne--Kr & $\simeq10$ & RITU      &
  \textcite{Leino:1995} \\
  Germany & GSI   & linac          & H--U      & 11        &  SHIP      &
  \textcite{Muenzenberg:1979} \\
  USA     & ORNL  & tandem         & H--U      & $\simeq10$&  RMS       &
  \textcite{Gross:2000}   \\
  USA     & ANL   & tandem + linac & H--U      & 17        &  FMA       &
  \textcite{Davids:1992}  \\
  Russia  & FLNR  & cyclotron    & Li--Ar      & 50        &  ACCULINNA &
  \textcite{Rodin:2003}   \\
  China   & HIRFL & cyclotron    & C--U        & 60        &  RIBLL     &
  \textcite{Sun:2003}\\
  France  & GANIL & 2 cyclotrons & C--U        & 95        &  LISE      &
  \textcite{Mueller:1991} \\
  USA     & NSCL  & 2 cyclotrons & O--U        & 170       &  A1900     &
  \textcite{Morrissey:2003}\\
  Japan   & RIBF  & 4 cyclotrons & H--U        & 350       &  BigRIPS   &
  \textcite{Sakurai:2008}  \\
  Germany & GSI   & synchrotron  & H--U        & 1000      &  FRS       &
  \textcite{Geissel:1992} \\
  \hline
  USA     & FRIB  & linac        & H--U        & 500       &            &
  \textcite{Thoennessen:2010}  \\
  Germany & FAIR  & synchrotron  & H--U        & 1500      &  Super-FRS &
  \textcite{Winkler:2008} \\
  \hline
  \hline
\end{tabular}
\label{tab:IIIB1}
\end{table*}

By increasing the energy and shifting to the projectile fragmentation regime
thicker targets can be used. At the energy of about 50~$A\cdot$MeV the typical
targets have a few hundred mg/cm$^2$ thickness, while at 1000~$A\cdot$MeV
they reach the thickness of a few g/cm$^2$. At the larger projectile energies
the kinematical focusing helps to achieve larger acceptance by the
electro-magnetic
\emph{fragment} separator. Additional filtering, according to the atomic number
$Z$ of the particle,
is achieved by means of an energy degrader mounted in the middle of the
fragment separator. The high energy of the reaction products allows for
their full in-flight identification by means of time-of-flight and energy-loss
measurements for individual ions. The resulting extreme sensitivity is one of
the
most important advantages of the medium- and high-energy In-Flight technique.
Indeed, this was the
key factor allowing the first observation of doubly-magic nuclei
$^{100}$Sn \cite{Schneider:1994,Lewitowicz:1994}, $^{78}$Ni
\cite{Engelmann:1995},
and $^{48}$Ni \cite{Blank:2000}, as well as the discovery of the two-proton
radioactivity \cite{Pfutzner:2002,Giovinazzo:2002}.

Although the transmission, separation, and in-flight identification of reaction
products are easier at large projectile energy, the maximum beam intensity which
can
be achieved decreases with increasing energy. Moreover, the quality of the beam
of
fragmentation products is rather poor. The energy spread of the fragments
resulting from both the reaction kinematics and the energy-loss straggling in
layers
of matter hinders efficient stopping in the final detectors. Hence, the optimal
conditions for production of a given nucleus result from a compromise between
different factors.

The special role of the high-energy In-Flight method comes from a relatively
simple
way of delivering beams of radioactive nuclei. Such radioactive beams may be
injected
into more sophisticated devices like storage rings \cite{Nolden:2008} or
ion traps \cite{Kluge:2008}, or can be used to induce secondary reactions
leading
to a significant expansion of the nuclear physics field of research.

The leading facilities using the In-Flight method to study decays of very exotic
nuclides
together with their brief characteristics and the corresponding reference to the
detailed information are collected in Table~\ref{tab:IIIB1}.

\subsubsection{ISOL}
\label{sec:IIIB2}

In an ISOL-type facility the nuclei of interest are
produced in a relatively thick target irradiated by a primary
beam from a driver accelerator. If the products recoil from the target they
are stopped by means of a catcher or a gas cell, otherwise they diffuse
out of the target material. Subsequently they are transferred
to the ion source and extracted, mostly as $1^+$ ions, by means of an HV
potential of the
order of 50 kV. Because of the constant charge of the extracted ions, the
following
separation in a uniform magnetic field corresponds to the mass separation, see
Eq.~\ref{eq:IIB1}.
In the first realizations of this technique, such mass separated, very low
energy ions
were deposited on a thin catcher foil in front of a detection system.
In the modern variant of this technique the mass separated ions are
postaccelerated and
a high quality beam is formed allowing better manipulation of the ions and
inducing secondary reactions \cite{ISOLbeams:2011}.

\begin{table*}
\caption{The main facilities based on the ISOL method for the radioactive decay
studies
         far from the stability. In the lower part of the table the facilities
         under
         construction or planning are listed.}
\resizebox{\textwidth}{!}{
\begin{tabular}{llllccll}
  \hline
  \hline
  Country & Laboratory & Facility & \multicolumn{2}{c}{Driver}& Postaccelerator,
        &  Reference \\
          &            &          & accelerator &    beam     & energy [$A
          \,$MeV]         &            \\
  \hline
  Finland & Jyv\"{a}skyl\"{a}& IGISOL & cyclotron &  H--Xe, $130 q^2/A$ MeV  &
       & \textcite{Aysto:2001} \\
  Belgium & Louvain-La-Neuve& LISOL   & cyclotron & H--Ni, 10 AMeV  &         &
  \textcite{Kudryavtsev:2008} \\
  Italy   & INFN-LNS   & EXCYT    & cyclotron   & $A<48$, 80 AMeV & tandem, 8 &
  \textcite{Cuttone:2008}  \\
  USA     & ORNL       & HRIBF   & cyclotron   & $p$, $d$,
                                       $\alpha$, 42--85 MeV  & tandem,
                                       $\simeq10$ & \textcite{Stracener:2003}\\
  France  & GANIL      & SPIRAL  & 2 cyclotrons & H--U, 95 AMeV  & cyclotron, 25
  & \textcite{Villari:2003}\\
  Canada  & TRIUMF     & ISAC    & cyclotron    & $p$, 500 MeV & linac, 11   &
  \textcite{Shotter:2003}\\
  Switzerland& CERN    & REX-ISOLDE& synchrotron & $p$, 1.4 GeV & linac, 3   &
  \textcite{Voulot:2008}\\
    \hline
  Italy   & LNL        & SPES     & cyclotron    & $p$, 70 MeV & linac, 11   &
  \textcite{Cinausero:2009}\\
  France  & GANIL  & SPIRAL-2 & linac  &$d$, 40 MeV; HI, 14.5 A MeV & cyclotron,
  25 &\textcite{Lewitowicz:2008}\\
  Canada  & TRIUMF     & ARIEL    & e-linac     &  $e$, 50 MeV & linac, 11   &
  http://www.triumf.ca/ariel\\
  \multicolumn{2}{c}{to be decided}
                       & EURISOL &  linac &  $p$, 1 GeV     & linac, 150     &
                       \textcite{Blumenfeld:2009}\\
  \hline
  \hline
\end{tabular}}
\label{tab:IIIB2}
\end{table*}

Although all kinds of nuclear reactions can (and are) employed in this method,
the most
important are spallation induced by protons, and fission of target nuclei
induced
by protons, light ions or neutrons. The latter can come either from a reactor or
from a beam of deuterons hitting a neutron converter in front of the target.
An interesting concept is to induce fission by bremsstrahlung
photons originating from a very high intensity electron beam hitting the
high-$Z$ converter \cite{CheikhMhamed:2008}.
The main point is that the delivered intensities of light beams, like protons,
deuterons, or electrons, can be significantly larger than maximal intensities
of heavy ion beams. This, in combination with thick targets which can be used,
results in the high yields of radioactive nuclei which is the main
advantage of the ISOL method.
On the other hand, the transfer processes occurring in the target and in the ion
source
take time of the order of milliseconds (see Figure~\ref{fig:IB1:timescale})
which
imposes limits on the half-lives which
can be accessed by this method. In addition, some of these processes exhibit
chemical
sensitivity which for example hinders extraction of refractory elements in
some implementations of this technique.

In the last decades, a remarkable progress in ion-source techniques and in
manipulating
low-energy ions has been achieved \cite{Lecesne:2008,Wenander:2008}.
A spectacular example is the application of
resonant laser ionization allowing extremely efficient and clean extraction
of selected elements from the source \cite{Cheal:2010}.

A selection of facilities employing the ISOL method with short characteristics
and
the reference to the corresponding technical information is presented in
Table~\ref{tab:IIIB2}.

\begin{table}
\begin{tabular}{l p{200pt}}
\hline
\hline
  \multicolumn{2}{l}{Some abbreviations used in Table~\ref{tab:IIIB1} and
  \ref{tab:IIIB2}}\\
        &  \\
  ANL   &  Argonne National Laboratory, USA\\
  FAIR  &  Facility for Antiproton and Ion Research, Darmstadt, Germany\\
  FLNR  &  Flerov Laboratory of Nuclear Reactions at Joint Insitute for Nuclear
  Research, Dubna, Russia\\
  FRIB  &  Facility for Rare Isotope Beams at Michigan State University, East
  Lansing, USA\\
  GANIL &  Grand Acc\'{e}l\'{e}rateur National d'Ions Lourds, Caen, France\\
  GSI   &  Helmholtzzentrum f\"{u}r Schwerionenforschung GmbH, Darmstadt,
  Germany\\
  HRIBF &  Holifield Radioactive Ion Beam Facility at ORNL, USA \\
  HRIFL &  Heavy Ion Research Facility in Lanzhou, China \\
  INFN-LNS & Istituto Nazionale di Fisica Nucleare, Laboratori Nazionali del
  Sud, Catania, Italy \\
  LNL   &  Laboratori Nazionali di Legnaro, Legnaro, Italy \\
  NSCL  &  National Superconducting Cyclotron Laboratory by Michigan State
  University, East Lansing, USA\\
  ORNL  &  Oak Ridge National Laboratory, Oak Ridge, USA\\
  RIBF  &  Radioactive Isotope Beam Factory at RIKEN laboratory, Wako, Saitama,
  Japan \\
\hline
  BigRIPS & Big RIKEN Projectile Fragment Separator at RIBF \\
  EXCYT   & EXotics with CYclotron and Tandem at INFN-LNS \\
  FMA   &  Fragment Mass Analyser at ANL\\
  FRS   &  FRagment Separator at GSI Darmstadt\\
  IGISOL&  Ion Guide and Isotope Separator On-Line\\
  LISE  &  Ligne d'Ions Super Epluch\'{e}s at GANIL \\
  LISOL &  Leuven Isotope Separator On-Line \\
  RIBLL &  Radioactive Ion Beam Line in Lanzhou at HRIFL\\
  RITU  &  Recoil Ion Transport Unit \\
  RMS   &  Recoil Mass Spectrometer at HRIBF\\
  SHIP  &  Separator for Heavy Ion reaction Products at GSI Darmstadt \\
  SPES  & Selective Production of Exotic Species at LNL \\
  SPIRAL  & Syst\`{e}eme de Production d'Ions Radioactifs avec
  Acc\'{e}l\'{e}ration en Ligne at GANIL \\
\hline
\hline
\end{tabular}
\end{table}

\subsubsection{Future facilities}

In the lower part of both Table~\ref{tab:IIIB1} and Table~\ref{tab:IIIB2} the
future facilities
which are being constructed or planned are listed. A new idea which is
considered in these recent developments is to combine advantages of both the
In-Flight and the ISOL
methods into hybrid solutions \cite{Tanihata:2008}.
Sufficient postacceleration of the ISOL secondary beam may enable
taking advantage of instrumentation developed for the In-Flight technique.
Such an option is discussed in the EURISOL design study \cite{Blumenfeld:2009}.
On the other hand, the fast In-Flight fragment beam may be stopped in a gas cell
and extracted at low energy with help of the ISOL techniques \cite{Facina:2008}.
An example of such a solution will be realized in the currently constructed FRIB
facility
\cite{Thoennessen:2010}.
The possible connections between the two main approaches to extract and separate
radioactive
nuclei are marked in Figure~\ref{fig:IIIB_scheme} by dashed lines.

\subsection{Detection}
\label{sec:IIIC}

The detection of a radioactive decay requires detection of particle(s) emitted
in the process.
The large majority of nuclides decay by $\beta$ transitions where primarily an
electron or a positron
is emitted (the presence of neutrinos can be safely neglected in this context)
and in the second step
electromagnetic radiation follows in form of $\gamma$ radiation if the daughter
nucleus was formed
in an excited state and/or of characteristic $X$-rays if the final atom was
excited. This secondary
electromagnetic deexcitations may proceed by emission of Auger electrons. When
we move away from
the stability line, however, the mass differences between isobars (and thus the
beta decay energy
windows) increase and the particle unbound states become populated. On the
neutron-deficient side
this leads to the $\beta$-delayed charged particle emission, mainly of protons.
Beyond the proton-drip
line the direct emission of protons comes into play. That is why the detection
of particles
like $p$ and $\alpha$ plays the central role in the radioactivity studies at the
neutron-deficient limit of stability. In turn, on the neutron-rich side emission
of
$\beta$-delayed neutrons becomes important which makes the neutron spectroscopy
necessary.
However, also charged particles, like $d$, $t$, and $\alpha$, are emitted
following $\beta$
decay of very neutron-rich nuclei.
In the following we sketch the modern methods used to
detect charged particles and neutrons. Finally, we discuss briefly the digital
signal processing
techniques which represent a new development in the systems of nuclear data
acquisition.

The technique of $\gamma$ spectroscopy is the large subject in its own, as it is
the main source of
detailed data on the nuclear excited states. Here we refrain from discussing its
development
referring the reader to the rich literature on this topic
\cite{Lee:2003,Gelletly:2006,Farnea:2010}.

\subsubsection{Charged particles}
\label{sec:IIIC1}

The most common devices for detecting charge particles are based on silicon
detectors
which record the energy deposit of a charge particle passing through its
material.
To increase the sensitivity of the measurement and to provide an additional
information about the
particle, stacks of two or more detectors are frequently used. In such a
\emph{telescope} the energy loss information from thin transmission detectors is
combined
with the total kinetic energy from the final thick detector. A particle
telescope may
combine a thin gas chamber, acting as a transmission counter with a thick
silicon
detector \cite{Moltz:1994,Axelsson:1998}. Additional information on the location
of the
particle's hit can be extracted from a position sensitive detectors where the
signal is read
from two ends of a resistive electrode. A significant advance in detection
technique was
introduced with a concept of a silicon strip detector. Particularly successful
was the
development of a double sided silicon strip detector (DSSSD) with perpendicular
sets of strip electrodes on its both sides \cite{Sellin:1992}. The achieved
granularity
provides not only a simple measure of the position but allows to establish
a position correlations between subsequent events, like the implantation of an
ion and its
decay by particle emission in condition of the high total counting rate.
In addition, it reduces the effect of energy summing between $\beta$-delayed
particles
and electrons which deposit much less energy in a strip (a pixel) area
\cite{Buscher:2008}.
Most of results on $\beta$-delayed proton emission
(Sec.~\ref{sec:IV}) and on proton radioactivity (Sec.~\ref{sec:V}) were obtained
with
help of such detectors. An example of a recent improvement is a novel design of
a
large-area DSSSD with an ultra-thin dead layer \cite{Tengblad:2004}. The modern
detection
set-ups which require granularity but also a large angular coverage are usually
constructed as arrays of silicon detectors. They may consist of a large number
of
simple Si diodes \cite{Fraile:2003}, a box of DSSSD detector \cite{Adimi:2010},
or
a combination of DSSSD detectors with a gaseous multiwire proportional chamber
and
germanium detectors \cite{Page:2003}. In the decay studies at projectile
fragment
separators (the In-Flight method) the large area of the final focal plane
and the large range distributions of selected ions have to be taken into
account.
To meet the latter challenge, the stacks of many DSSSD detectors are used.
The example solution used for the $\beta$-decay studies of $^{100}$Sn at the
FRS separator (GSI Darmstadt) consisted of three large-area DSSSD detectors,
providing in total $3\times60\times40=7200$ pixels,
sandwiched between two sets of ten single-side silicon strip detector
\cite{Eppinger:2009}.
In another development for the FRS separator, a set of two rows consisting of
three
5~cm$\times$5~cm DSSSD detectors, each with 256 (16$\times$16 strips) pixels
was developed \cite{Kumar:2009}. Even more ambitious Advanced Implantation
Detector
Array (AIDA), to be used at the FAIR facility, will comprise twenty four $8
\times 8$~cm$^2$
DSSSD detectors with $128 \times 128$ strips \cite{Davinson:2010}.

\begin{figure}[th]
\centerline{
\includegraphics[width=0.49\textwidth]{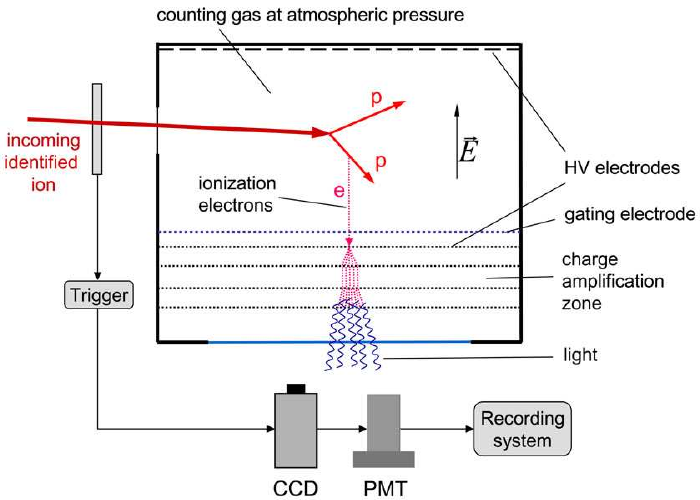}
}
\caption{(Color online) A schematic view of the Optical Time Projection Chamber
(OTPC). For each recorded event, the data consist of a 2D image taken by a CCD 
camera in a given exposure time and the total light intensity detected by a 
photomultiplier (PMT) as a function of time, sampled by a digital oscilloscope. 
The gating electrode is used to block the charge induced by incoming ions.}
\label{fig:IIIC_OTPC}
\end{figure}

\begin{figure}[th]
\centerline{
\includegraphics[width=0.44\textwidth]{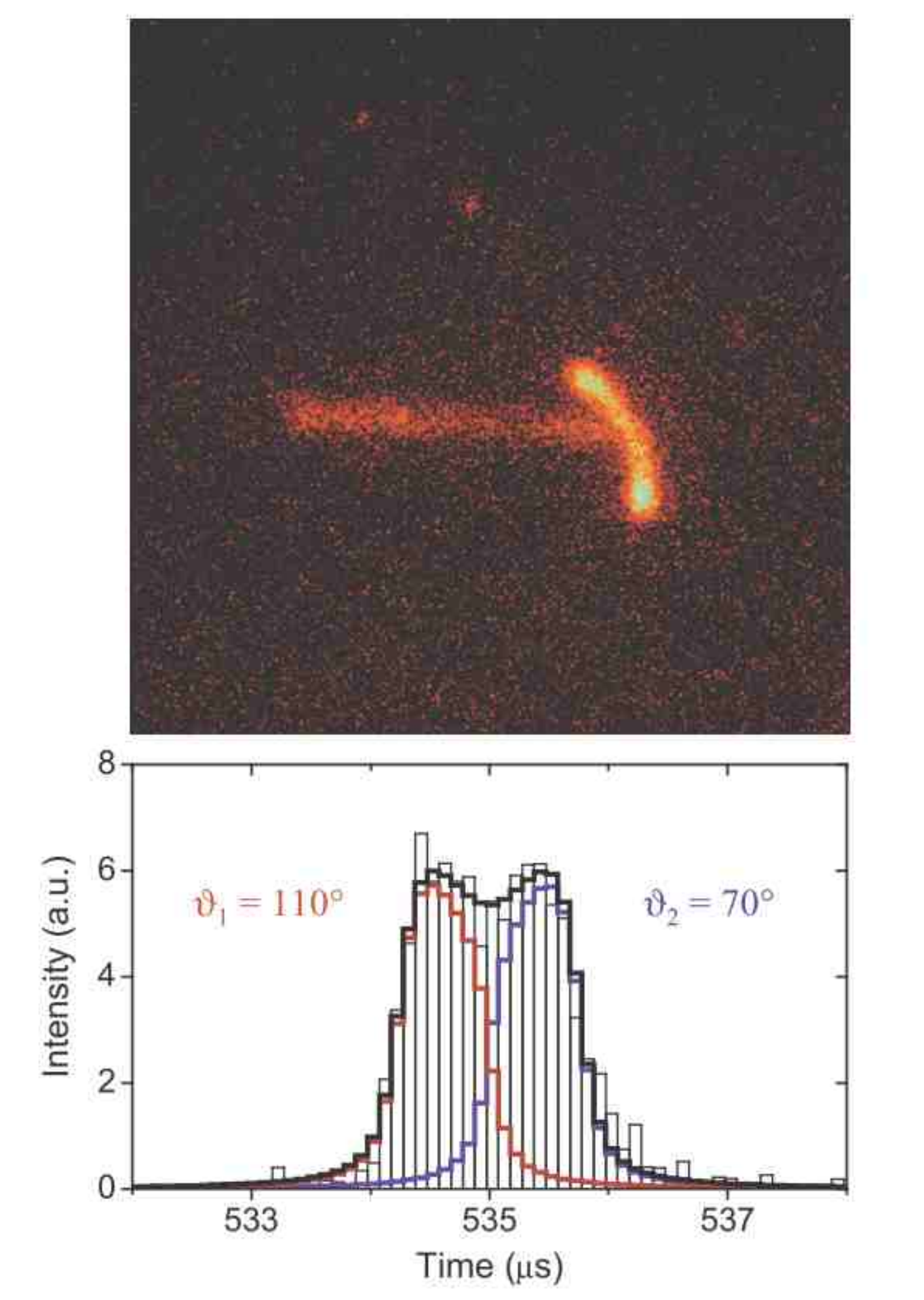}
}
\caption{(Color online) An example of a registered two-proton decay event of
$^{45}$Fe.
Top: an image recorded by the CCD camera in a 25 ms exposure. A track of a
$^{45}$Fe
ion entering the chamber from left is seen. The two bright, short tracks are
protons of
approximately 0.6 MeV, emitted 535 $\mu$s after the implantation.
Bottom: a part of the time profile of the total light intensity measured by the
PMT (histogram)
showing in detail the 2p emission. Lines show results of the reconstruction
procedure
yielding the emission angles $\vartheta$ with respect to the axis normal to the
image.
From \textcite{Miernik:2007b}.}
\label{fig:IIIC_event2p}
\end{figure}

The implantation of a nucleus which undergoes a multiparticle decay into a
silicon detector
has a serious limitation that only the total decay energy can be measured.
The energy sharing between products and their momentum correlations
cannot be accessed. Such problem appeared in case of the two-proton
radioactivity.
The first observation of this decay mode was accomplished by implanting ions of
$^{45}$Fe
into a stack of silicon detectors \cite{Pfutzner:2002} and into a DSSSD
detector \cite{Giovinazzo:2002}. The information on the decay time and on the
total decay energy were sufficient to claim the new type of radioactivity, but
for
the detailed study of this process a novel experimental approach was necessary.
To meet this challenge, two new developments were undertaken, both based on a
principle of the time projection chamber (TPC). The key idea is that such a
gaseous
ionization chamber can record tracks of charged particles, allowing their
reconstruction in three dimensions. The radioactive ion stopped inside the
active volume and the subsequently emitted particles ionize the counting gas.
The primary ionization electrons drift in a uniform electric field towards
the charge amplification section producing the two-dimensional representation
of the particles' tracks. The drift time contains the position information
along the electric field direction. In one solution, the amplified ionization
charges are collected electronically by means of an anode plate with two sets
of orthogonal strips \cite{Blank:2010}. This detector rendered the first direct
evidence for the two protons emitted in the decay of $^{45}$Fe
\cite{Giovinazzo:2007}.
In the second design, the idea of an optical readout \cite{Charpak:1988} was
implemented.
It is based on the observation that light is emitted in the final stage of
charge amplification. In the Optical Time Projection Chamber (OTPC) this light
is
collected by a CCD camera and by a photomultiplier (PMT) \cite{Miernik:2007c}.
The construction of this detector is shown schematically in
Figure~\ref{fig:IIIC_OTPC}.
The application of the OTPC yielded spectacular results, including the detailed
proton-proton correlation picture for the \emph{2p} decay of $^{45}$Fe
\cite{Miernik:2007b}
and the first observation of the $\beta$-delayed three-proton emission
channel \cite{Miernik:2007}. An example event of the two-proton radioactivity
of $^{45}$Fe is shown in Figure~\ref{fig:IIIC_event2p}.

In case of very short decay half-lives, in the subnanosecond range, the
implantation
technique generally cannot be used. A short-lived precursor decays in-flight
very close to the place of its production. The identification of the nucleus and
its
properties can be deduced from the detection and tracking of all decay products.
This approach was successfully applied to the study the \emph{2p} decay of
$^{19}$Mg \cite{Mukha:2007} which also exemplifies advantages of radioactive
beams.
Ions of $^{19}$Mg were produced in a secondary target by a neutron knock-out
reaction
from a beam of $^{20}$Mg delivered by the GSI FRS separator \cite{Geissel:1992}.
The tracking of emitted protons by means of silicon microstrip detectors
\cite{Stanoiu:2008}
allowed to establish the longitudinal distribution of decay vertexes and to
determine
the half-life of $^{19}$Mg to be $4.0(15)$~ps. At the same time the information
on
correlations between emitted protons was collected. Since the beam impinging
on the secondary target contains usually a mixture of different ions ("cocktail"
beam)
other reactions can be addressed simultaneously. For example, in the measurement
of $^{19}$Mg, the data on proton and two-proton decays from $^{15}$F, $^{16}$Ne,
and
$^{19}$Na were obtained \cite{Mukha:2010}. Similar technique has been applied to
study two-proton emission form excited states of $^{17}$Ne
\cite{Chromik:2002,Zerguerras:2004}.
The tracking method of the in-flight decay products is expected to provide
information
on several \emph{2p} emitters among light nuclei, see Sec.~\ref{sec:VII}.

\subsubsection{Neutrons}
\label{sec:IIIC2}

\begin{figure}[th]
\centerline{
\includegraphics[width=0.49\textwidth]{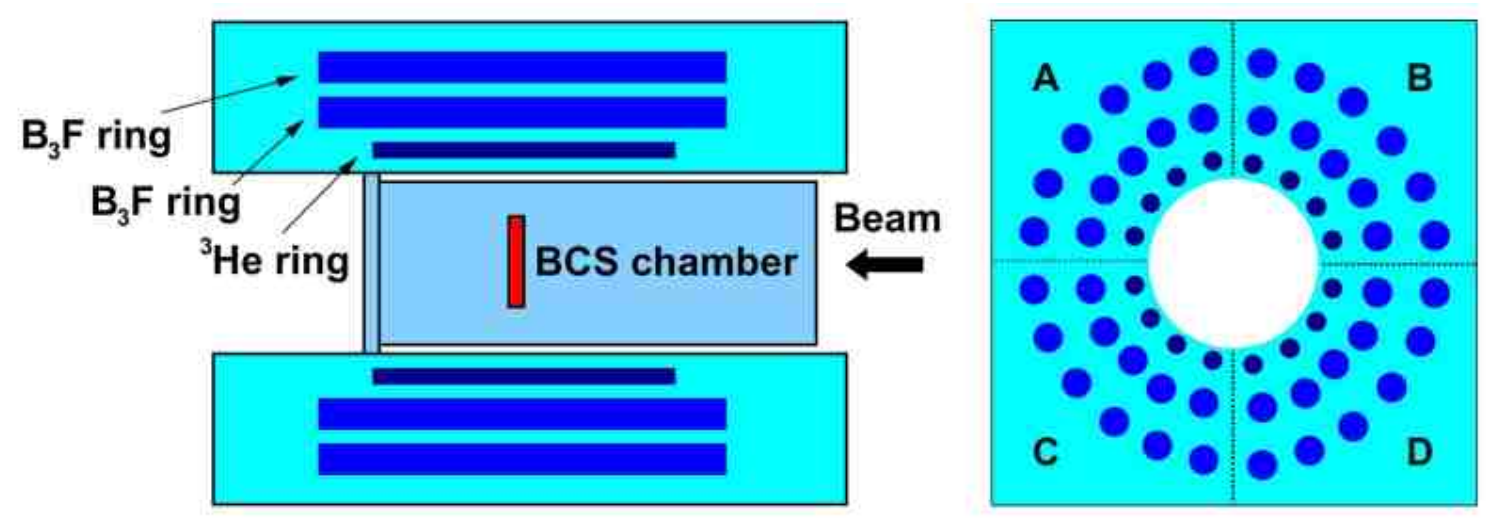}
}
\caption{(Color online) Schematic drawings of the NERO detector. Left: side view
showing the Beta Counting Station  (BCS) chamber located inside of NERO with the
DSSSD at the central position. Right: backside showing the cylindrical cavity to
house the BCS and the three concentric rings of gas-filled proportional
counters. The labels A, B, C and D designate the four quadrants. From
\textcite{Pereira:2010}.}
\label{fig:IIIC_NERO}
\end{figure}

Currently, two different methods for neutron detection are used in nuclear
spectroscopy.
The first one is based on thermal-neutron induced reactions, like
$^{3}$He(n,p)$^{3}$H,
$^{6}$Li(n,$\alpha$)$^{3}$H, or $^{10}$B(n,$\alpha$)$^{7}$Li, leading to charged
particles
which can be easily detected. The neutrons emitted by a radioactive source have
to be first
thermalized and this is achieved by means of a moderator --- usually a large
block of
polyethylene surrounding the source. In the moderator cylindrical cavities,
arranged in concentric rings, are drilled in which proportional counters
are mounted \cite{Mehren:1996}. In these counters which are filled with $^{3}$He
or BF$_3$ gas,
neutron-capture reactions take place and are detected. Such construction allows
to cover
a large solid angle, approaching $4 \, \pi$ and the large total efficiency of up
to 30\% can
be achieved for a broad neutron energy-range from meV to tens of MeV and almost
independent
on the neutron energy due to thermalization. Since the information on energy is
lost,
such detector is used primarily for counting which makes it well suited for
determination of
branching ratios for various neutron emission channels. Another disadvantage of
the moderation
is that a neutron is detected up to about 100~$\mu$s after the emission. Such a
delay
reduces the total counting rate which can be accepted. An example of a modern
version of such a
$4 \, \pi$ neutron counter is the NERO detector, recently built at the NSCL
laboratory \cite{Pereira:2010}. Its layout is shown in
Figure~\ref{fig:IIIC_NERO}.

A different, and to a large degree a complementary solution, employs
scintillation detectors
in which interactions of fast neutrons are detected, predominantly by recording
elastic
proton recoils. The neutron energy can be then determined by means of the
time-of-flight (TOF) after
a trigger signal, given for example by a $\beta$ particle. The panels containing
liquid or
plastic scintillators are mounted at some distance from the radioactive source,
which
usually reduces the solid angle which can be achieved. The efficiency depends on
the
neutron energy and exhibits a low-energy threshold at about few hundred keV. In
addition,
such detector is sensitive also to $\gamma$ radiation and the pulse-shape
analysis
has to be performed for the n--$\gamma$ discrimination \cite{Skeppstedt:1999}.
Such an approach to the neutron TOF spectroscopy can be exemplified by the
detector
TONNERRE developed at the GANIL laboratory \cite{Buta:2000}.

\subsubsection{Signal processing}
\label{sec:IIIC3}

In the conventional approach, signals from detectors of nuclear radiation are
preamplified
and then processed in analogue-electronics modules like shapers,
amplifiers, discriminators, etc., to be finally converted to the digital form in
analogue-to-digital (ADC) and time-to-digital (TDC) converters, and stored in
the
electronic memory of the data acquisition system. With the increasing number of
channels which have to be read, resulting from pixelization of the detectors
(strips,
segments, pads) the amount of necessary electronic units is growing and
thus magnifying the complexity and the cost of the instrumentation. In addition,
by storing only the values of energy and time for an event, the information on
the pulse shape is lost, which is very disadvantageous in some applications.

\begin{figure}[th]
\centerline{
\includegraphics[width=0.48\textwidth]{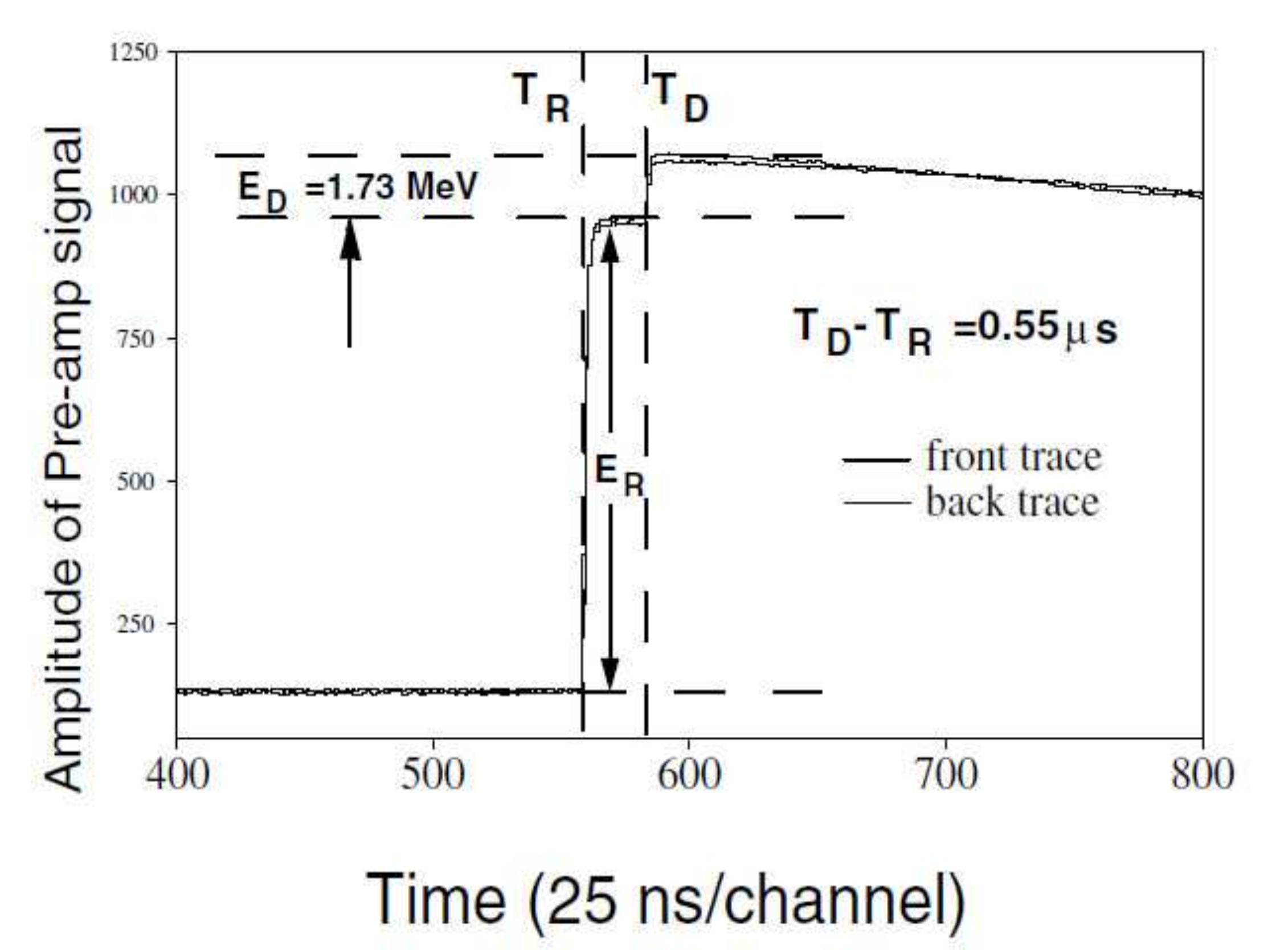}
}
\caption{Part of the preamplifier signal waveforms recorded by the front and
         the back strip of the 65-$\mu$m DSSSD during the $^{145}$Tm experiment.
         The recoil depositing about 14 MeV energy is followed after 0.55 $\mu$s
         by the 1.73 MeV signal. From \textcite{Karny:2003}.}
\label{fig:III-catcher}
\end{figure}

A possible solution to these problems is offered by the technique of the digital
signal
processing (DSP) which since recently is taking over the conventional data
acquisition systems in nuclear spectroscopy. Its basic principle is that the
output
of a preamplifier is digitized first and all further manipulations are performed
by numerical algorithms acting on this digital representation of the signal.
These algorithms replace all functions of analogue electronics and additionally
offer a choice of much more complex and flexible operations on the pulse.
Originally, the introduction of the DSP methods was motivated by needs of
segmented X-ray and $\gamma$-ray arrays and this sector is the main
recipient of this technology
\cite{Cromaz:2008,Crespi:2009,Starosta:2009,Pietri:2007}.
One commercial development --- the Digital Gamma Finder module (DGF-4C) by
XIA LLC \cite{Hubbard:1999}--- proved to be particularly successful also
in the domain of particle spectroscopy. Some applications of the DGF electronics
in various decay studies of exotic nuclei were reviewed by
\textcite{Grzywacz:2003}.
The trends and new products from this developer were presented by
\textcite{Warburton:2006}.

A good illustration of new possibilities provided by the DSP is the measurement
of
very short-lived proton radioactivity in $^{145}$Tm \cite{Karny:2003}.
The technical challenge is to detect a low-energy proton ($\sim 1.5$~MeV)
emitted very shortly ($\sim 1 \, \mu$s) after stopping of the parent nucleus
which deposits up to about 35~MeV in the implantation DSSSD detector. Such
sequence of events
cannot be resolved easily when signals are passed through the analogue
amplifiers.
The solution offered by the DGF electronics is to store the whole waveforms
of the signals from the silicon strips and to analyze their shapes off-line.
In addition, the special triggering mode was implemented to the DGF board
which allowed to store only those events in which the pile-up of two pulses
was detected. This feature leads to the large increase of sensitivity,
which is especially important when many different ions are coming to the
final detector and the decay investigated is rare. This technique was the
key factor leading to the discovery of the fine structure in the
decay of $^{145}$Tm (Sec.~\ref{sec:V}). An example of the recorded waveform
representing a low-energy proton signal superimposed on the large implantation
signal of the $^{145}$Tm ion is shown in Figure~\ref{fig:III-catcher}.
The same method was instrumental in the observation of the superallowed
$\alpha$-decay
chain from $^{109}$Xe \cite{Darby:2010} which is discussed in Sec.~\ref{sec:VI}.

Another development which is recently gaining importance, especially in the
domain
of detectors with high granularity, is the technology of so called
application-specific
integrated circuit (ASIC). It is based on a highly integrated circuit which is
customized for a specific use rather than for a general-purpose application.
Usually, one integrated circuit (IC) chip features several independent channels,
each capable of handling energy and timing of a single detector element (pixel
or strip).
In fact, the large silicon array AIDA \cite{Davinson:2010} as well as one of the
TPC detectors developed to study $2p$ radioactivity \cite{Blank:2010}, see
Sec.~\ref{sec:IIIC1}, employ ASIC-type chips in their read-out electronic
system. An example of multi-channel
IC for the detection of nuclear radiation is described by \textcite{Engel:2011}.
In combination with an array of Si-strip detectors \cite{Wallace:2007} it was
used in a study of $^{6}$Be \cite{Grigorenko:2009c}, presented in
Sec.~\ref{sec:VIID1}.


\newpage
\section{BETA DELAYED PARTICLE EMISSION}
\label{sec:IV}


\subsection{Beta decay, general observations}
\label{sec:IVA}

\subsubsection{The beta strength}

The weak interactions and in particular their low-energy manifestation
in nuclear beta decay are by now well understood
\cite{Behrens:1982,Grotz:1990}, see \cite{Severijns:2006} for a recent
survey of weak interaction tests in nuclear physics. The decay rate
for allowed $\beta^-$ or $\beta^+$ decays can be transformed to give
the known expression for the $ft$-value
\begin{equation}
   ft = \frac{K}{g_V^2 B_F + g_A^2 B_{GT}} ,
\end{equation}
where $t$ is the partial halflife of the transition,
$K/g_V^2 = 6144.2(1.6)$ s and $g_A/g_V = -1.2694(28)$
\cite{Towner:2010}, and $B_F$ and $B_{GT}$ are the reduced matrix
elements squared for the Fermi and Gamow-Teller parts. (Note that some
authors define $B_{GT}$ to include the factor $(g_A/g_V)^2$.) Nuclear
electron capture will also contribute, but is mainly noticeable for
low decay energies and in heavier elements. The phase space factor can
be approximated roughly by $f = (1+Q/m_ec^2)^5/30$ in terms of the
decay energy ($Q$-value) for $\beta^{\pm}$-decays, more accurate determinations
can
be found through \cite{Behrens:1982,Wilkinson:1995}. Due to the
parabolic mass surface the $Q$-values increase as one moves away from
beta-stability which by itself enhances beta decay rates. Empirically
the beta halflives fall off approximately exponentially away from
stability \cite{Zhang:2007}.

There are only few beta decays where most of the beta strength is
energetically accessible in the decay, and the detailed distribution
in general plays a crucial role. The Fermi strength is concentrated
around the Isobaric Analogue State (IAS). The summed strength
fulfils a sum rule $\sum B_F^+
- \sum B_F^- = Z-N$ that involves Fermi transitions in ``both
directions'', this is relevant e.g.\ for the odd-odd $N=Z$ nuclei
from $^{34}$Cl to $^{98}$In where most ground states have $T=1$ and
can be fed as well as decay by Fermi transitions.

The Gamow-Teller strength obeys the Ikeda sum rule:
\begin{equation} \label{eq:IVA2}
   \sum B_{GT}^-  -  \sum B_{GT}^+  = 3 (N-Z) ,
\end{equation}
and much of this strength is collected in the several MeV wide
so-called Gamow-Teller Giant Resonance (GTGR), although mixing with
higher lying configurations is important and removes a sizeable part
of the strength to higher energies \cite{Ichimura:2006}; this is often
referred to as the quenching of the GT strength, the key point being
that the quenching factor, although depending on what approximations
are used when calculating the GT strength, is varying slowly as a
function of $N$ and $Z$. Note that the summed value of $B_{GT}$, even
including quenching, is larger than $B_F$. More details on the
strength distribution will be given in the following two subsections.

For the lightest nuclei the $Q$-values are for a given mass number
slightly higher on the proton-rich side than on the neutron-rich side
of beta-stability and experimentally halflives are systematically
shorter on the proton-rich side. The asymmetry is enhanced by
the contribution from the IAS transition in
nuclei with $N<Z$. For masses above 100 the situation has changed:
experimentally the halflives for nuclei more than 3--4 nucleons away
from the beta-stability line are systematically shorter on the
neutron-rich side than on the proton-rich side. For these nuclei the
$\beta^+$ decay increases isospin and the systematic difference can
be understood from eq.\ (\ref{eq:IVA2}) since the summed $\beta^-$
strength is significantly higher than the $\beta^+$ strength.

Several approaches have been employed to reproduce and predict the
beta decay halflives in large parts of the nuclear chart
\cite{Honma:1996,Nakata:1997,Borzov:2000,Moller:2003}.  The increase
in computing power has allowed the use of increasingly sophisticated
microscopic models \cite{Brown:2001,Borzov:2006,Borzov:2008}, even ab
initio methods \cite{Navratil:2007,Pervin:2007} for the very light
nuclei, and more extended RPA \cite{Toivanen:2010} and shell-model
calculations are underway (G. Mart\'{\i}nez-Pinedo, private communication,
2010).

\subsubsection{Delayed particles}

\begin{figure}
\centerline{
\includegraphics[width=0.5\textwidth]{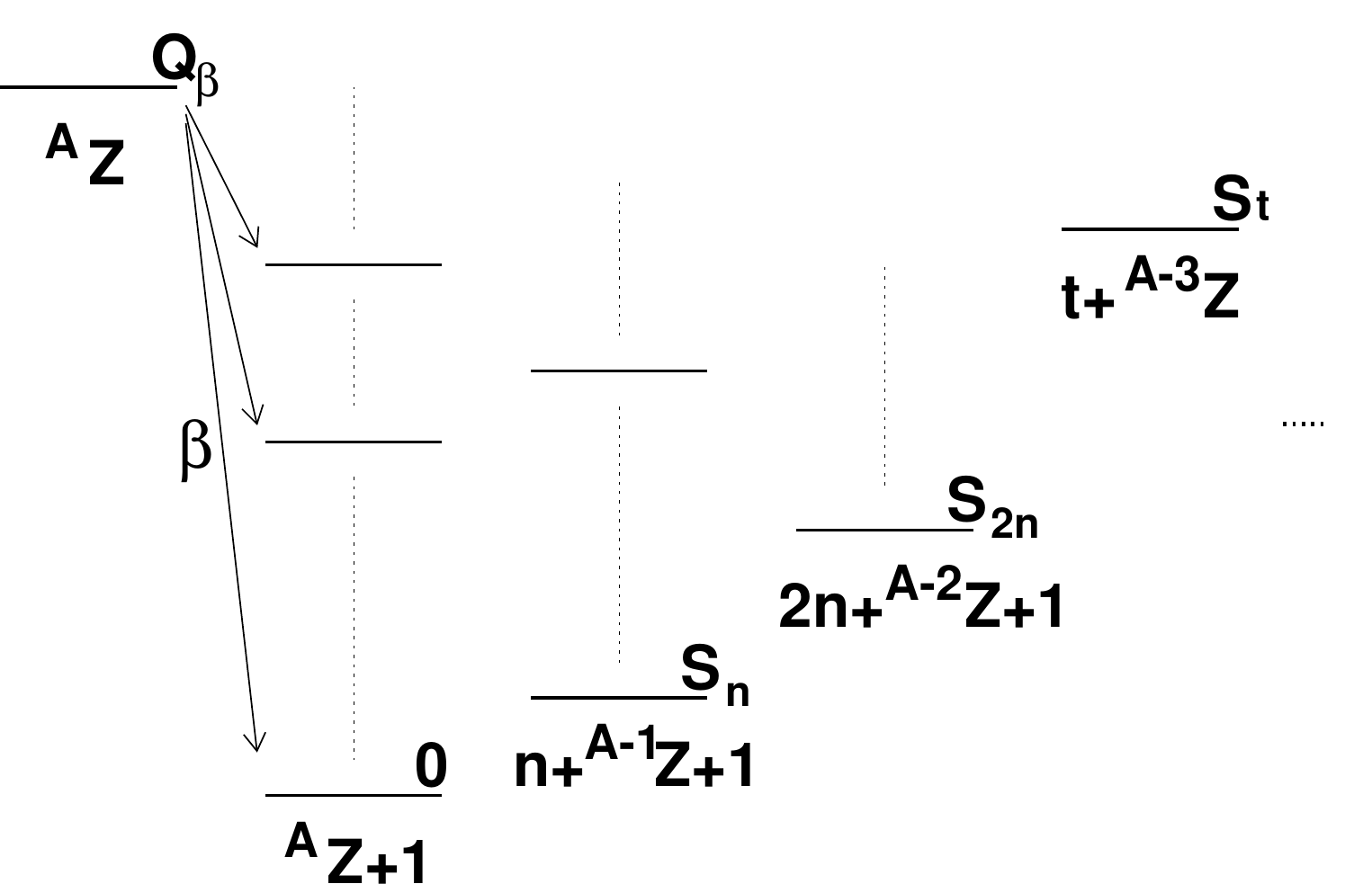}
}
 \caption{\label{fig:iva1} Some of the possible energetically allowed
   beta-decay channels for a neutron rich nucleus. The precursor,
   $^AZ$, beta decays to the emitter, $^A(Z+1)$, that has particle
   unbound excited states.  All energies are given relative to the
   ground state of the emitter.}
\end{figure}

Defining as usual the relative probability $P_S$ for a given
beta-delayed decay mode $S$ (1p, 2p, 1n, 2n, $\alpha$ etc), as the
fraction of all decays that results in a final state containing $S$,
one can find the average number of emitted neutrons as $P_n = \sum_i i
P_{in}$ with $P_p$ defined in a similar way. The energetics for the
different channels is sketched in figure \ref{fig:iva1}. The figure
implicitly assumes that decays take place through states in the
emitter and that multi-particle emission happen sequentially;
as will be argued below these assumptions may not hold for all cases.
The $Q$-value for delayed emission of neutrons explicitly depends on the
neutron separation energies, but also the $Q$-value for some other modes
can be rewritten \cite{Jonson:2001} to show that they depend on the
neutron separation energies of the precursor nucleus:
\begin{equation} \label{eq:IVA3}
 Q_{\beta d} = 3.007 \mathrm{MeV} - S_{2n} , \, \, \,
 Q_{\beta t} = 9.264 \mathrm{MeV} - S_{3n} .
\end{equation}
Most beta-delayed decay modes will therefore be enhanced at the
driplines since multi-nucleon separation energies will be low there:
the ``dripline'' for emission of two or more neutrons will lie very
close to the one neutron dripline. We shall return in section
\ref{sec:IVD} in more detail to the particle emission processes, but
can note already now that the probability for a delayed multi-particle
emission may depend on the emission mechanism (simultaneous or
sequential emission) as well as on the energy available.

As we approach the driplines, the enhanced role played by beta-delayed
particle emission implies that the physics problems investigated via
beta decay will overlap more and more with the ones investigated via
reaction studies, but the selection rules for beta decay will provide
a spin selectivity that often is useful. We shall focus here on the
general features of the beta decay processes and specific challenges
met in decay studies, but shall give as well selected examples of
structure questions that have been studied.

A specific example of this is the population in beta decay of excited
states that enter in the astrophysical \emph{rp}-process \cite{Wormer:1994}.
An even more direct need for beta-decay data in astrophysics arises in
processes where weak interactions play a role, either directly as beta
decay rates or indirectly where neutrino interactions are important,
see \cite{Langanke:2003,Borzov:2006,Arnould:2007,Grawe:2007}.

Quite apart from the general interest in the coupling to continuum
degrees of freedom, cf.\ section \ref{sec:IC3}, beta decay processes
may provide specific information on isospin mixing that is expected to
be enhanced for continuum states at low energy, see e.g.\
\cite{Garrido:2007,Mitchell:2010,Michel:2010a}.

The experimental considerations were covered in general in section
\ref{sec:III}, but a few specific comments may be relevant. Since beta
halflives are longer than about one ms, essentially all experiments
make use of a stopped beam. This gives a source distribution that
at in-flight facilities
may have a considerable spatial extent. Experiments in storage rings
\cite{Litvinov:2011} or ion and atom traps \cite{Severijns:2006} have
been undertaken in several cases, but still present practical problems
in particular concerning efficient detection of all decay products.
Complementary experiments at different type of facilities may overcome
the disadvantages for a specific production method, one example being
the study of $^{32}$Ar \cite{Bhattacharya:2008} where a
high-resolution spectrum obtained at ISOLDE/CERN was combined with an
absolute intensity determination carried out at NSCL/MSU. ISOL
facilities often have problems for very short halflives and in
determining absolute activities, whereas in-flight facilities
frequently employ composite beams (so-called cocktail beams) where
special procedures may be needed in order to correct for background
from decay of non-relevant isotopes, see e.g.\
\cite{Dossat:2007,KurtukianNieto:2008}. As will be seen below many
results from the past decade come from in-flight facilities, often
through implantation of the radioactive ion into a Si detector.

Beta-delayed particle emission has been the subject of
several earlier reviews, both more general ones
\cite{Jonson:1996,Jonson:2001} and specific ones for proton-rich
nuclei \cite{Hardy:1988,Blank:2008}, neutron-rich nuclei
\cite{Hansen:1988} and heavy nuclei \cite{Hall:1992}. Since
more detailed accounts can be found there the treatment here, in
sections \ref{sec:IVB}, \ref{sec:IVC} and \ref{sec:IVE}, will be
somewhat brief. The remaining section \ref{sec:IVD} deals with
beta-delayed multi-particle emission and naturally has more
interconnections to other parts of the present paper.

\subsection{$\beta^+$ delayed emission of one particle}
\label{sec:IVB}

\subsubsection{Occurrence of particle emission}

\begin{figure}
\centerline{
\includegraphics[width=0.46\textwidth]{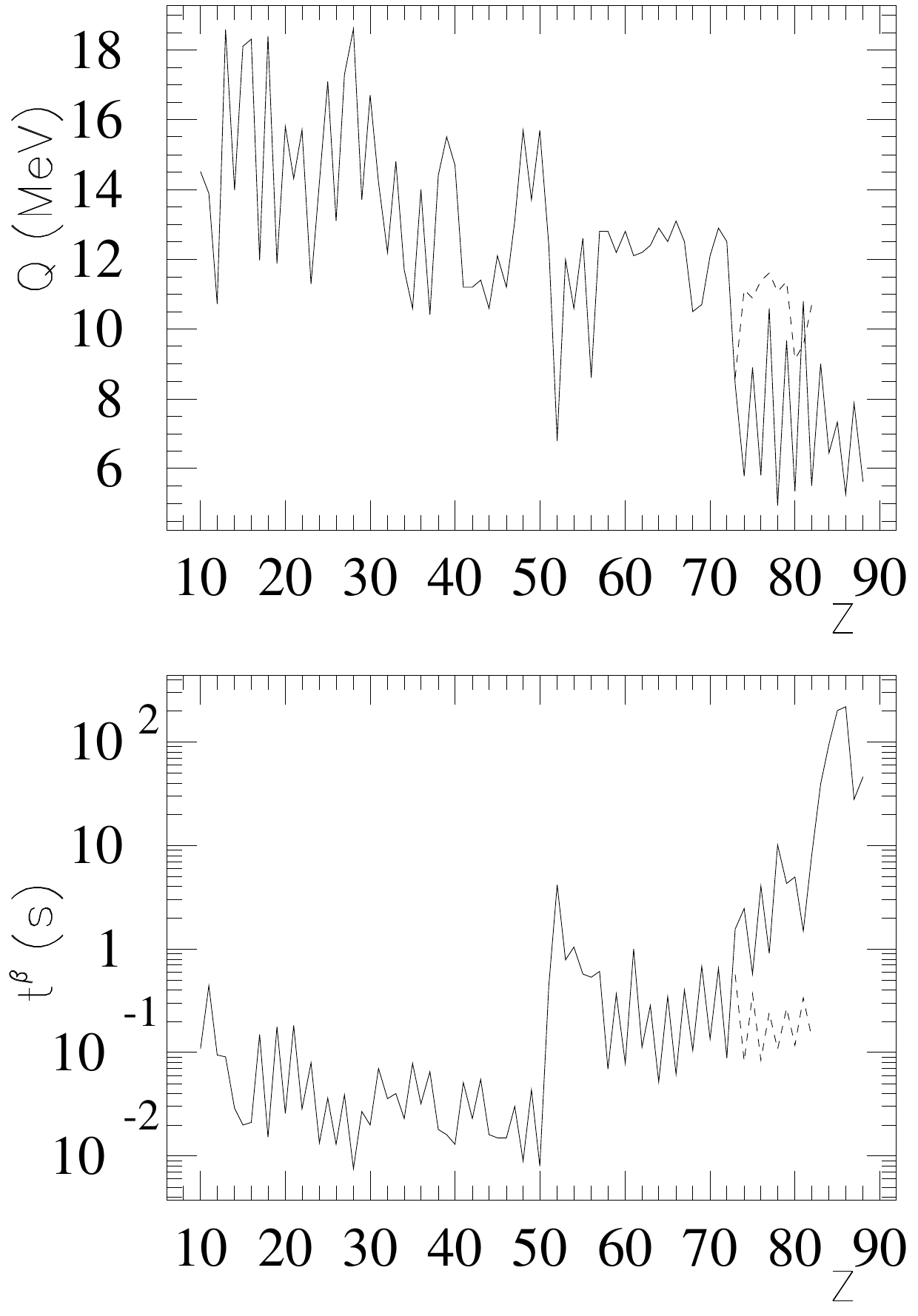}
}
 \caption{ \label{fig:IVB1}
 $Q_{EC}$-value (upper figure) and beta-decay halflife (lower
 figure) as a function of proton number $Z$ for the lightest nucleus
 for each element where beta-decay remains the dominating decay mode.
 Dashed lines (for $73 \leq Z \leq 82$)
 are the values at the proton dripline. Experimental values from
 \protect\cite{Audi:2003,Audi:2003a,Achouri:2006,Dossat:2007}
 are completed by estimates based on \protect\cite{Moller:1997}.}
\end{figure}

Figure \ref{fig:IVB1} shows $Q_{EC}$ and the beta-decay halflives for
the most proton-rich nuclei where beta-decay still plays a role. There
is considerable scatter in the values, but also clear effects of the
proton shells at $Z=50$, enhanced by the fact that Fermi transitions
contribute below this value and not above, and $Z=82$ as well as the
neutron shell at 82 corresponding to $Z=72$: below this the competing
decay mode is proton emission, above alpha decay takes over inside the
proton dripline. For beta decays along the proton dripline (dashed
lines in the figure) one still finds that the $Q$-values decrease
towards 10 MeV in the heavier nuclei. The scatter indicates that local
nuclear structure still plays an important role in these decays.
Even though protons and neutrons in many cases still are within the same
major shell, forbidden decays will play an important role for the
heavier nuclei.

\begin{figure}
\centerline{
\includegraphics[width=0.46\textwidth]{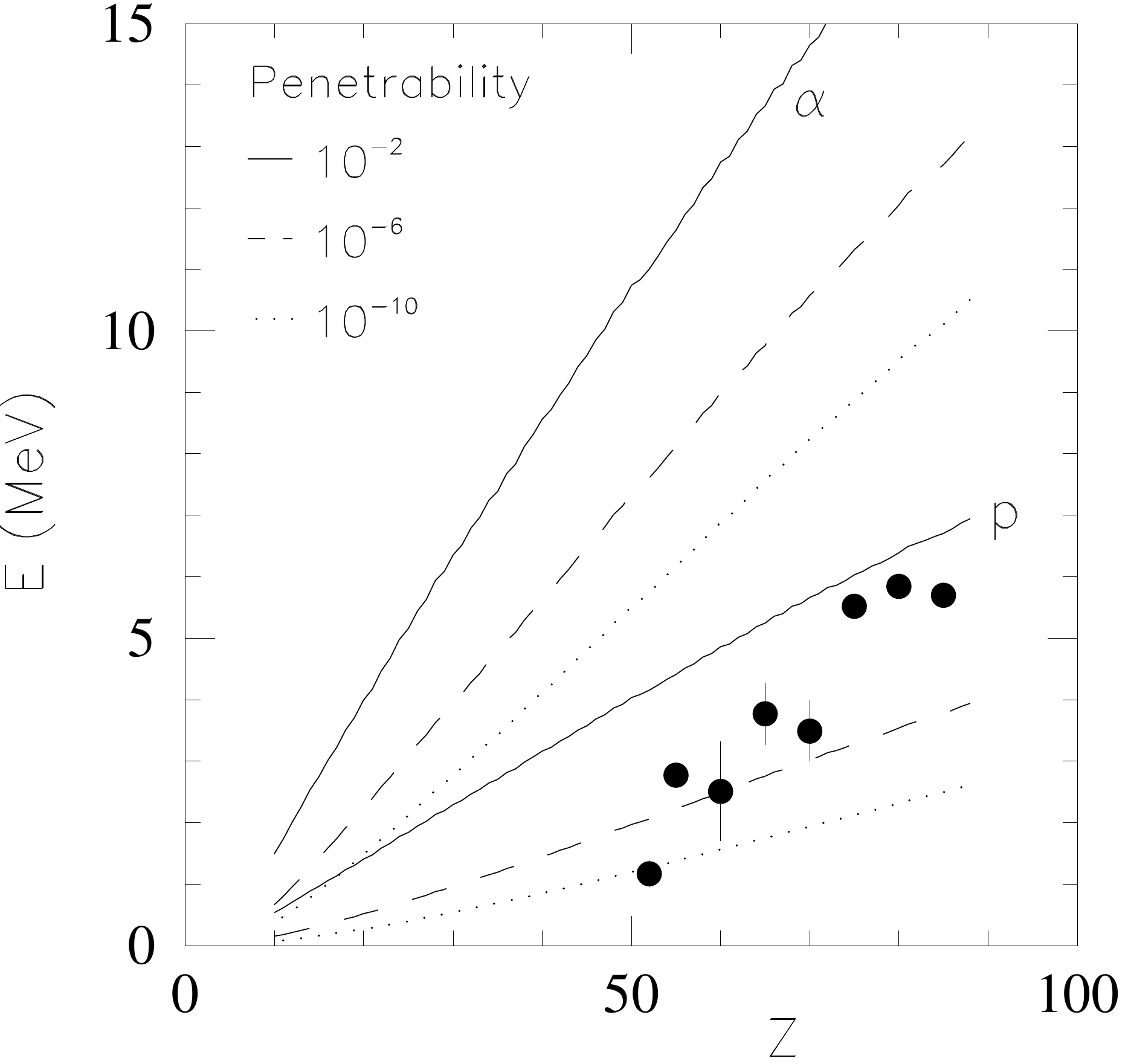}
}
 \caption{ \label{fig:IVB2} The centre-of-mass energy that gives an
   s-wave penetrability of $10^{-2}$ (full lines), $10^{-6}$ (dashed
   lines) or $10^{-10}$ (dotted lines) for a beta-delayed proton or
   alpha particle are shown versus the charge $Z$ for the precursors
   shown in figure \protect\ref{fig:IVB1}. Gamma emission can be
   expected to compete for penetrabilities below $10^{-6}$ (cf.\
   figure \protect\ref{fig:IB1:timescale}). For illustration the
   emitter $Q_{\alpha}$ value \cite{Audi:2003a} is
   shown for a few beta-decaying nuclei, see the text.}
\end{figure}

The Coulomb barrier plays a dominating role in beta-delayed particle
emission in proton-rich nuclei, as illustrated in figure
\ref{fig:IVB2}, and essentially limits the emitted particles to
protons. Delayed alpha emission is energetically allowed for many
proton-rich nuclei and may seem energetically favoured for nuclei
above $Z=50$ where the beta-daughter often has a positive
$Q_{\alpha} $ value (see figure \ref{fig:IVB2}). However, this decay
mode is mainly important in light nuclei: the only
nucleus above mass 20 where the $\beta\alpha$ branching ratio gets
above 1\% is $^{110}$I. The $\beta\alpha$ process has in heavier
nuclei mostly been observed just above closed shells similar to what
is seen for ground state alpha decays. The competing process to
delayed proton emission is therefore delayed gamma emission.
The retardation from the Coulomb barrier will also be significant
for protons, but the staggering of the proton dripline of course
implies that there will be nuclei with sizable $P_p$ for most even
$Z$-values. Turning to nuclei that lie within the ``odd-$Z$ dripline'' it
appears, with our present incomplete experimental knowledge, that
beta-delayed proton emission with $P_p$ above 1\% with few exceptions
occur in nuclei that are at most one or two nucleons away from the line.
One may thus regard significant beta-delayed proton emission
as a dripline phenomenon. To give one
example, the nucleus $^{167}$Ir has a ground state and an isomer that
both decay by proton emission, alpha particle emission and beta decay
\cite{Davids:1997}, but beta-delayed particle emission has not been
reported even though the proton separation energy is below 2 MeV in
the daughter nucleus $^{167}$Os (however, such events may be harder to
see with the tagging technique employed in the experiment).

The competition between proton and gamma emission can lead to the
occurrence of gamma-delayed proton emission. The angular momentum
barrier for outgoing protons seems to make this happen frequently in
high-spin physics \cite{Rudolph:2002}, but it may also happen after
beta decays where angular momentum barriers are smaller. It has been
suggested to take place in the decays of $^{20}$Na
\cite{Clifford:1989,Kirsebom:2010} and $^{31}$Ar \cite{Wrede:2009},
but may be expected also in other nuclei. Gamma emission preceding
particle emission is well-known in the light nuclei and recent
dedicated reaction experiments have now succeeded to observe it even
for cases where one or both of the unbound states are broad, namely
for the $4^+$ to $2^+$ transition in $^8$Be \cite{Datar:2005} and several
transitions in $^{12}$C \cite{Kirsebom:2009}.

\subsubsection{Fermi decays}
\begin{figure}
\centerline{
\includegraphics[width=0.46\textwidth]{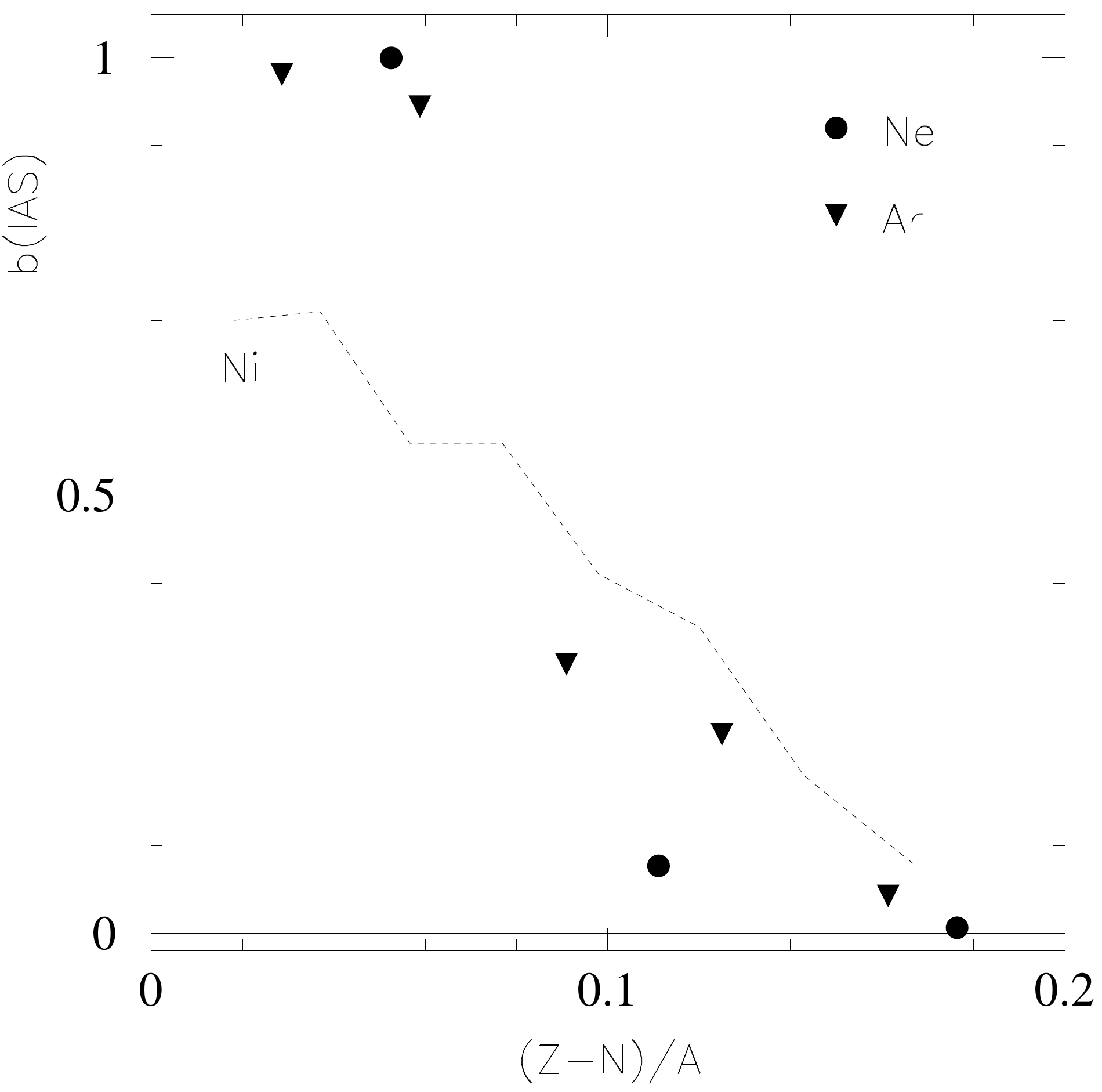}
}
 \caption{ \label{fig:IVB3}
  The total branching ratio of beta-decays to the IAS is shown
   as a function of relative proton excess for the light Ne and Ar
   isotopes.
   The dashed line gives the estimated branching ratio for Fermi
   beta-decays to the IAS for the Ni isotopes. Only the Fermi part of
   the transition is included, the partial halflife is assumed to
   scale inversely with $Z-N$ and total halflives are taken from
   \protect\cite{Dossat:2007}.}
\end{figure}

For $Z$ up to 50 the dripline nuclei have $N \leq Z$ so the Fermi
strength contribute to beta decay.  The approximate model-independence
of $B_F$ makes the IAS transition interesting even though it, as shown
in figure \ref{fig:IVB3}, only dominates the decays close to $N=Z$
where the IAS is at low excitation energy. The decay rate for the
transition to the IAS is proportional to $(Z-N)f_{\beta}(\Delta E_C)$
where the Coulomb energy shift $\Delta E_C$ depends only slowly on
mass number for a given set of isotopes \cite{Antony:1997}. Even
though this strength increases with $Z-N$ the branching ratio to the
IAS will decrease. Furthermore since the IAS will be situated at
higher and higher excitation energy its decay will become more
fragmented and there will for the most proton-rich nuclei such as
$^{17}$Ne and $^{31}$Ar not any longer be a dominant IAS peak in the
final state spectrum.

A first approximation of the wavefunction of the IAS will be $| A >$,
the (normalized) state obtained by letting the isospin lowering
operator work on the parent state. However, the two will not be
exactly identical, a fact often referred to as isospin symmetry
breaking. There are two aspects of this: the isospin of the IAS will
not be pure (isospin mixing) and the radial wavefunctions may differ
slightly (imperfect overlap, here isospin is in principle
conserved). The magnitude of the combined effect has been an important
issue in precision determinations of the Cabibbo-Kobayashi-Maskawa
quark-mixing matrix element $V_{ud}$
\cite{Towner:2010}. Models predict increases from around 0.2\% for
$^{10}$C to close to 2\% for $^{74}$Rb \cite{Grinyer:2010} and it has
been measured to 2.0(4)\% in $^{32}$Ar \cite{Bhattacharya:2008} and
can be expected to be of similar magnitude along the dripline.
Isospin mixing is known from reaction studies to be small in the sense
that the spreading width of an IAS typically is in the range 10 to 100
keV, see \cite{Harney:1986,Mitchell:2010} and references therein.
Note that many of the present beta-delayed proton experiments may not
be able to resolve the IAS from close-by levels that it mixes with;
the average level spacing can be estimated from the mirror systems
where proton scattering on even-even nuclei give about 10 keV for
nuclei at mass 50 \cite{Bilpuch:1976}.

To discuss the Fermi strength distribution in more detail one can
start from the simple relation
\begin{equation} \label{eq:IVB1}
   < A | H | A > = \sum_i  E_i |< i | A >|^2   ,
\end{equation}
where $H$ is the Hamiltonian of the system and $|i>$ a complete set of
states. Although $|A>$ is not an eigenstate of the system, the
expectation value on the left is the quantity that formally enters in
the isobaric multiplet mass quation (IMME). However, the IMME is very
resilient \cite{Benenson:1979,Bentley:2007} and will hold in many
cases even though isospin symmetry breaking may be significant in the
intermediate multiplet members. The Fermi strength that is spread out
via isospin mixing will obviously remain close to the IAS, the
important consequence of eq. (\ref{eq:IVB1}) being that this also
holds on average for the Fermi strength spread out due to imperfect
overlap between $|A>$ and the IAS. The redistribution of Fermi
strength has been checked experimentally in a few decays, e.g.\
$^{20}$Na \cite{Clifford:1989}, and it would be interesting to have
thorough studies in more nuclei where the effects are expected to be
large such as when continuum effects become important.

The particle emission from the IAS will in most cases be isospin
forbidden \cite{Auerbach:1983,Brown:1990} and the width will
consequently be so narrow that gamma decays may have a substantial
branching ratio. This is well established in light nuclei and must be
kept in mind when detailed investigations of the Fermi strength become
possible also in heavier nuclei.

\subsubsection{Gamow-Teller decays}

The GTGR will for proton-rich nuclei
lie above the IAS, but can be reached in
beta decay e.g.\ for the lightest Ar \cite{Borge:1989} and Ca
\cite{Trinder:1997} isotopes allowing for experimental tests of the
predicted strength distribution.  The GTGR is predicted to be
accessible even for $N=Z$ nuclei above mass 64 \cite{Hamamoto:1993}.
The experimental knowledge is still limited, but present data appear
consistent with shell-model calculations \cite{Dossat:2007}. For
nuclei with $N>Z$ the systematics of the Gamow-Teller strength is
given in \cite{Langanke:2000,Batist:2010}.

There is a special interest in the nuclear structure around the doubly
magic nucleus $^{100}$Sn. This is the last particle stable $N=Z$
nucleus, the halflife has now been measured \cite{Bazin:2008} for all
of them. For nuclei with $Z \leq 50$ and $50 \leq N$ allowed beta
decay will mainly proceed via the $\pi g_{9/2} \rightarrow \nu g_{7/2}$
transition and for nuclei approaching $^{100}$Sn all of the strength
again appears to be accessible in beta decay. A comprehensive overview
was given recently by \textcite{Batist:2010}.

\subsubsection{Selected spectroscopic tools}

This subsection will present a few physics phenomena that can be employed
to extract more detailed information on the states entering in
beta-delayed particle emission, namely recoil shifts, interference
between levels and decays where individual levels are not resolved.

To experimentally distinguish Fermi and Gamow-Teller transitions one
may be guided by spin selection rules, but in general have to resort
to beta recoil effects \cite{Holstein:1974}. The beta-neutrino angular
correlation will give a significantly larger recoil shift in Fermi
transitions than in Gamow-Teller transitions and can be studied either
as a function of beta-particle angle \cite{Clifford:1989} or through
measurement of the peak shape \cite{Schardt:1993}. The size of the
shift scales inversely with the mass number and is
therefore easier to measure for light nuclei. It will depend on the
spin sequences in the decay and has e.g.\ been used to determine the
spin of $^{31}$Ar \cite{Thaysen:1999}.

The level density of nuclei increases with excitation energy and with
mass number. As it increases the local structure changes from rather
regular to essentially chaotic, a transition well-studied
theoretically but experimentally less understood in nuclei
\cite{Weidenmuller:2009}. In many nuclei around mass 100 the
beta-delayed proton spectra will be dominated by unresolved isolated
resonances and fluctuation analysis is needed to extract information
on the average spectral properties \cite{Hansen:1990} (see also
\textcite{Giovinazzo:2000} for later work around mass 70). The larger
windows for beta-delayed protons in lighter nuclei close to the
driplines will enable these studies to be continued to cases with
different level density.
In the decays with highest $Q_{EC}$ values one may reach excitation
energies close to where Ericsson fluctuations have been observed in
nuclear reactions, i.e.\ the region where the level widths are larger
than the average level distance. It will be experimentally challenging
to look for such fluctuations in beta decay.

Another aspect of spectroscopy at high level density is that
``complete spectroscopy'' will be very challenging to achieve, see the
discussion in \textcite{Hansen:1990}. A way of overcoming
this challenge is the total absorption technique \cite{Rubio:2005,Janas:2005},
where the aim is to measure the total emitted energy (apart from the
emitted beta and neutrino particles) rather than the individual
protons and gamma rays. This of course also holds for decays of
neutron-rich nuclei where it, as demonstrated recently
\cite{Algora:2010}, is essential for a correct understanding of the
decay heat in nuclear reactors.

A final effect that can influence decay spectra significantly is
interference due to overlapping levels of the same spin and
parity. This will occur not only at high excitation energy, but also
for otherwise well-resolved states whose tails overlap. Interference
will be clearly prominent in light nuclei where broad states occur
frequently, but it will certainly be an issue also for the broad
states that can appear for heavier nuclei due to width
collectivization once the level density is so large that levels start
to overlap, see \cite{Zelevinsky:1996,Celardo:2008} and references
therein. Interference effects needs a more careful theoretical
treatment, e.g.\ via the R-matrix formalism.
The effects are often easy to identify once the
statistics is sufficient and may range from slight distortions, as in
the beta-delayed proton spectrum from $^{33}$Ar \cite{Schardt:1993},
to considerable spectral modifications, as in the beta-delayed alpha
spectrum from $^{18}$N \cite{Buchmann:2007}. However, interference
effects are not always easily recognizable, as seen in the
beta-delayed alpha-decays of $^8$B \cite{Barker:1989}, $^{12}$N (see
subsection \ref{sec:IVD}) and to some extent also $^{16}$N
\cite{Buchmann:2009}, and when statistics is insufficient spectral
features arising from broad and interfering levels are easily
misinterpreted as new weak transitions as demonstrated e.g.\ for the
beta-delayed proton spectrum from $^{17}$Ne \cite{Borge:1988}. This
underlines the care that must be taken when interpreting decay
spectra.

\subsection{$\beta^-$ delayed emission of one particle}
\label{sec:IVC}

\subsubsection{Occurrence of particle emission}
\begin{figure}
\centerline{
\includegraphics[width=0.46\textwidth]{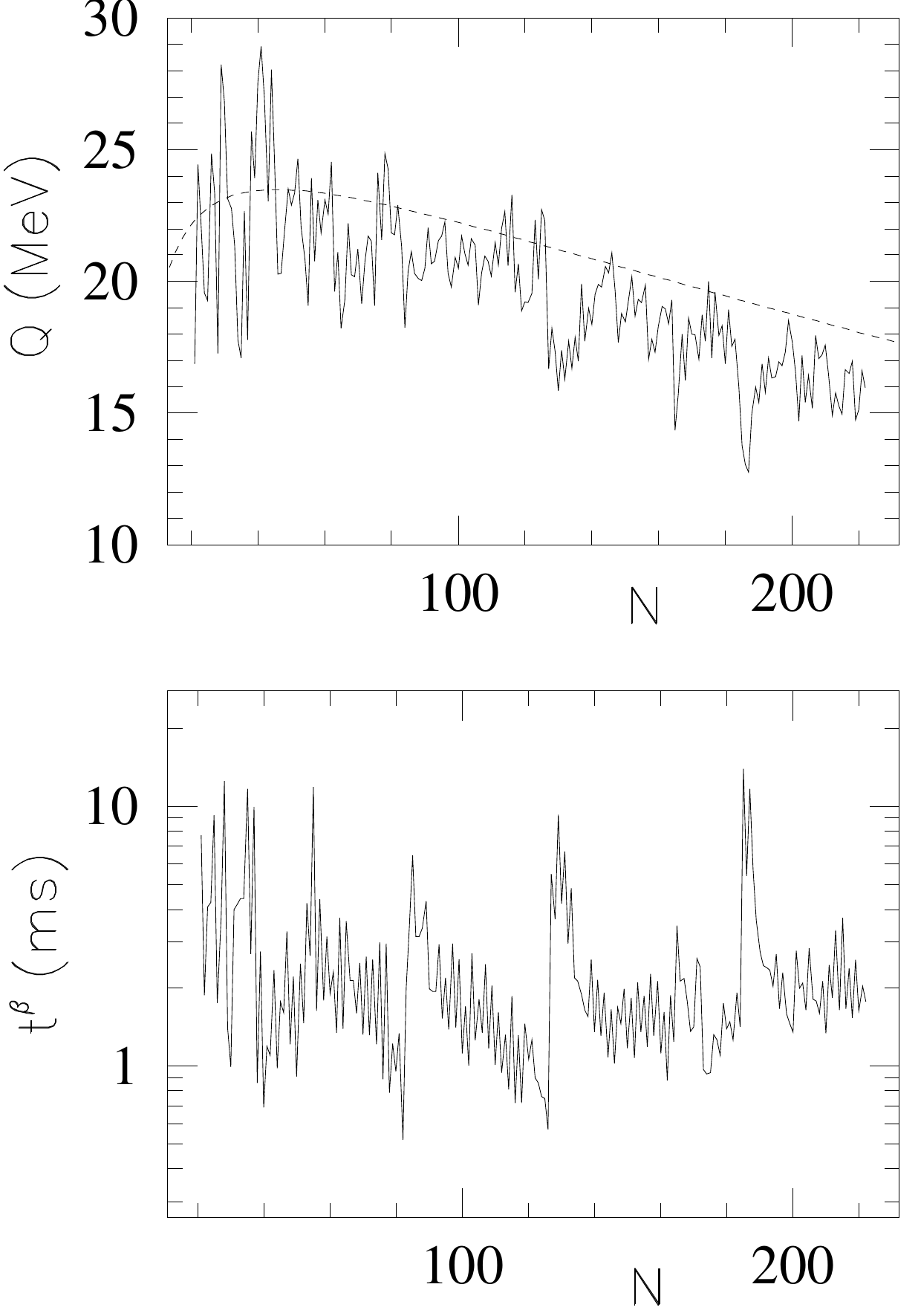}
}
 \caption{ \label{fig:IVC1}
 $Q_{\beta}$-value (upper figure) and beta-decay halflife (lower
 figure) as a function of neutron number $N$ for the lightest particle
 stable nucleus for a given $N$. The experimental dripline position is
 used for $N$ up to 30, all other values are taken from
 \protect\cite{Moller:1997,Moller:2003}. The dashed line gives the $Q$-values
 from an estimate based on the Weizs\"{a}cker mass formula.}
\end{figure}

Figure \ref{fig:IVC1} shows $Q_{\beta}$ and the beta-decay halflives
along the neutron dripline. The values are theoretical estimates and
will depend on the theoretical model chosen, in particular on how the
model predicts the nuclear shell structure \cite{Sorlin:2008} evolves.
However, the following general observations are most likely robust.
(As shown by \textcite{Moller:1997} their theoretical halflives agree
better with experimental value the larger the $Q$-value is.)

The halflives for nuclei at the neutron dripline vary somewhat for the
experimentally known ones with $N<30$, but are likely to mainly be in
the range 1--3 ms once we get above $N$ about 40. Deviations will be
due to changes in the $Q_{\beta}$-value rather than other structure
effects. The $Q$-values may be affected by shell structure, but
decrease slowly towards higher masses. This overall trend is seen
already from the simplest possible liquid drop formula, as also
indicated in the figure. An even smoother dependence of halflive with
nucleon number is found in recent work based on the density functional
approach \cite{Borzov:2008}, but it is clear that both $Q$-values and
halflives vary much less for neutron-rich nuclei than for the
proton-rich ones.

\begin{figure*}
\centerline{
\includegraphics[width=0.95\textwidth]{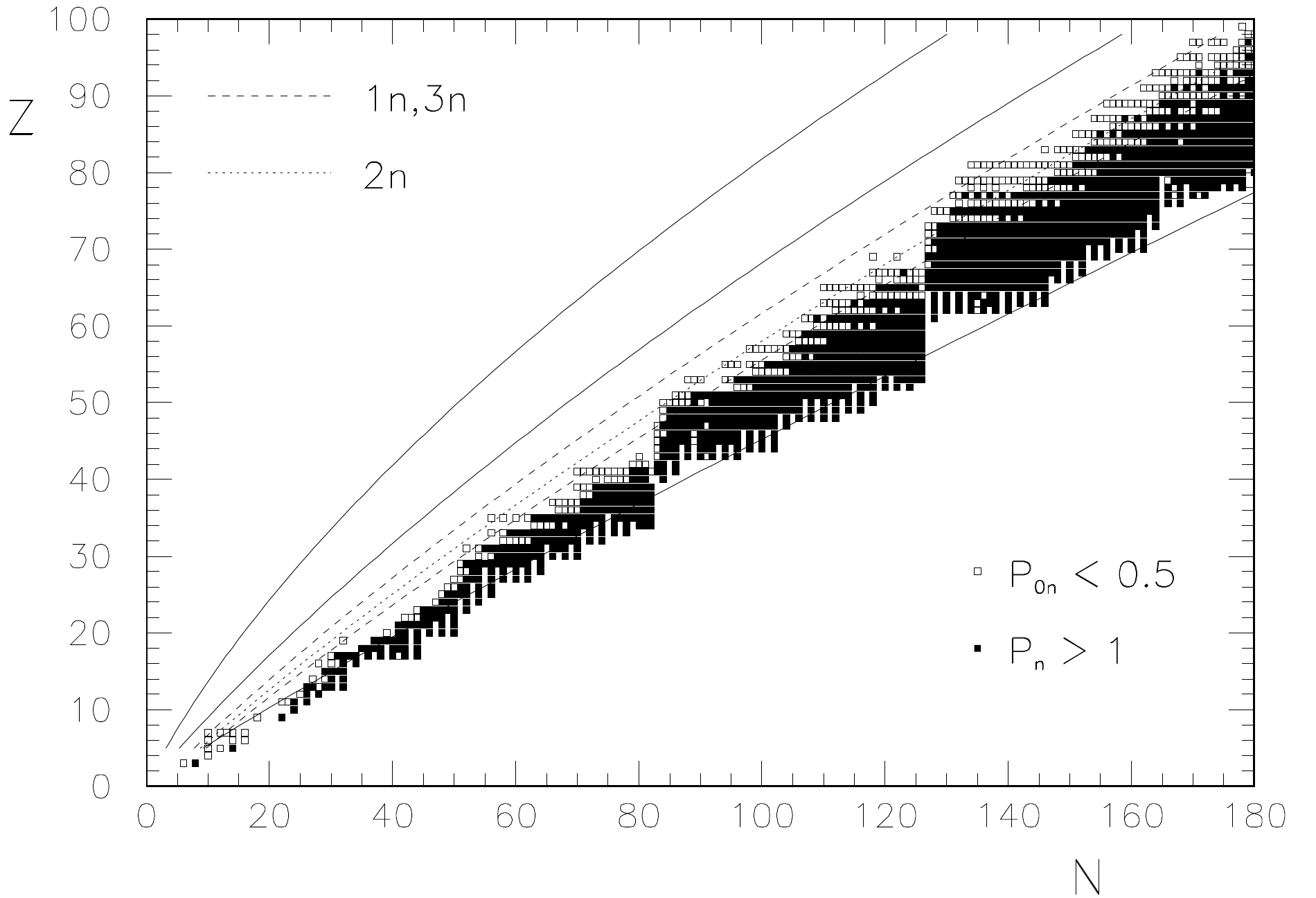}
}
 \caption{ \label{fig:IVC2}
   Nuclei with large beta-delayed neutron-emission probability are
   marked with an open square if the probability for emitting one or
   more neutrons is larger than 50\% and with a filled square if the
   average number of emitted neutrons is larger than one.  The $P_n$
   values are taken from experiment
   \protect\cite{Audi:2003,Borge:1997,Yoneda:2003} for $N<20$ and from
   \protect\textcite{Moller:2003} otherwise. The full lines indicate
   the line of beta-stability and the two driplines estimated from the
   Weizs\"{a}cker mass formula and the broken lines the corresponding
   estimates for where beta-delayed one-, two- and three-neutron emission
   becomes energetically allowed.}
\end{figure*}

Neutron emission will take place once it is energetically allowed and
beta-delayed neutron emission will therefore be an important feature
for neutron-rich nuclei. The extent is illustrated in figure
\ref{fig:IVC2}: not only are beta-delayed multi-neutron decays
energetically allowed shortly after beta-delayed one-neutron decay,
the estimated beta strength distribution will soon give more than one
emitted neutron on average per decay. As an example, for neutron
dripline nuclei around mass 180 one expects more than 10 neutrons
emitted in the decay chain towards beta-stability. Experimentally, we
have today mainly reached this extended region of high $P_n$-values
for the light nuclei.

Similar to what is observed for proton-rich nuclei the Coulomb barrier
will limit significant beta-delayed alpha emission to the very light
nuclei. However, the delayed emission of hydrogen isotopes, in
particular deuterons and tritons, may also occur with small
probabilities. Since their Q-values are limited, as seen from eq.\
(\ref{eq:IVA3}), the deuteron emission will be suppressed by three
orders of magnitude at mass 100 whereas triton emission may still be
possible to see up to mass 200. Their physics relevance will be
discussed shortly. The major difference to the situation for
proton-rich nuclei is therefore the prominent beta-delayed one-neutron
and multi-neutron emission.

The excitation energy of the Gamow-Teller Giant Resonance, that
carries the main part of the Gamow-Teller strength, is known to
decrease linearly with respect to the Isobaric Analogue State as a
function of $(N-Z)/A$ \cite{Osterfeld:1992,Langanke:2000}. It was
pointed out by \textcite{Sagawa:1993} that the GTGR for light
neutron-rich nuclei (oxygen or below) even could move below the initial
state so that a major part of the strength can be accessed in
beta-decay. For heavier nuclei, decays will take place to the tail of
the GTGR.

\subsubsection{Decays in different mass regions}

To illustrate the present stage of the field, this section will
present experimental results from several currently investigated mass
regions, starting with the lightest nuclei which is where the neutron
dripline is reached and halo structures \cite{Jensen:2004} have been
studied.

Neutron halo nuclei must have low neutron separation energy and have a
``clustered'' structure in the sense that the halo neutrons should
decouple from the core to a large extent.  It is obvious from eq.\
(\ref{eq:IVA3}) that beta-delayed deuteron emission will be
energetically favored in two-neutron halo nuclei, furthermore the
component where the two halo neutrons decay to a deuteron (with the
core as spectator) will give an important contribution to this decay
mode. In fact, most theoretical calculations of the $\beta \, d$ decay
only includes decays of the halo neutrons directly to continuum
deuteron states. The early work on this decay mode is reviewed in
\textcite{Nilsson:2000}. The decay has so far only been seen in $^6$He
and $^{11}$Li and the first experiments at ISOLDE have now been
extended at other laboratories both for $^6$He
\cite{Anthony:2002,Raabe:2009} and $^{11}$Li \cite{Raabe:2008}. The
branching ratio is for $^6$He now determined to be $1.65(10) \cdot
10^{-6}$ above a centre-of-mass energy of 525 keV. This very low value
is understood to be due to cancellation in the matrix elements between
contributions from small and large radii. The latest calculations
\cite{Tursonov:2006,Tursonov:2006E} reproduce both shape and intensity
of the deuteron distribution, but it is not yet clear whether the
theoretical and experimental maximum intensity positions agree, so
measurements at lower energy would still be valuable. For $^{11}$Li
the branching ratio is $1.30(13) \cdot 10^{-4}$ above a centre-of-mass
energy of 200 keV and the spectrum is again rather featureless
\cite{Raabe:2008}. The most recent theoretical calculations
\cite{Baye:2006} give a qualitative agreement with data, but a real
test of the theoretical understanding seems only possible once
experimental data on the $^9$Li+d interaction at low energy are
available.

\begin{figure*}
 \includegraphics[width=0.65\textwidth]{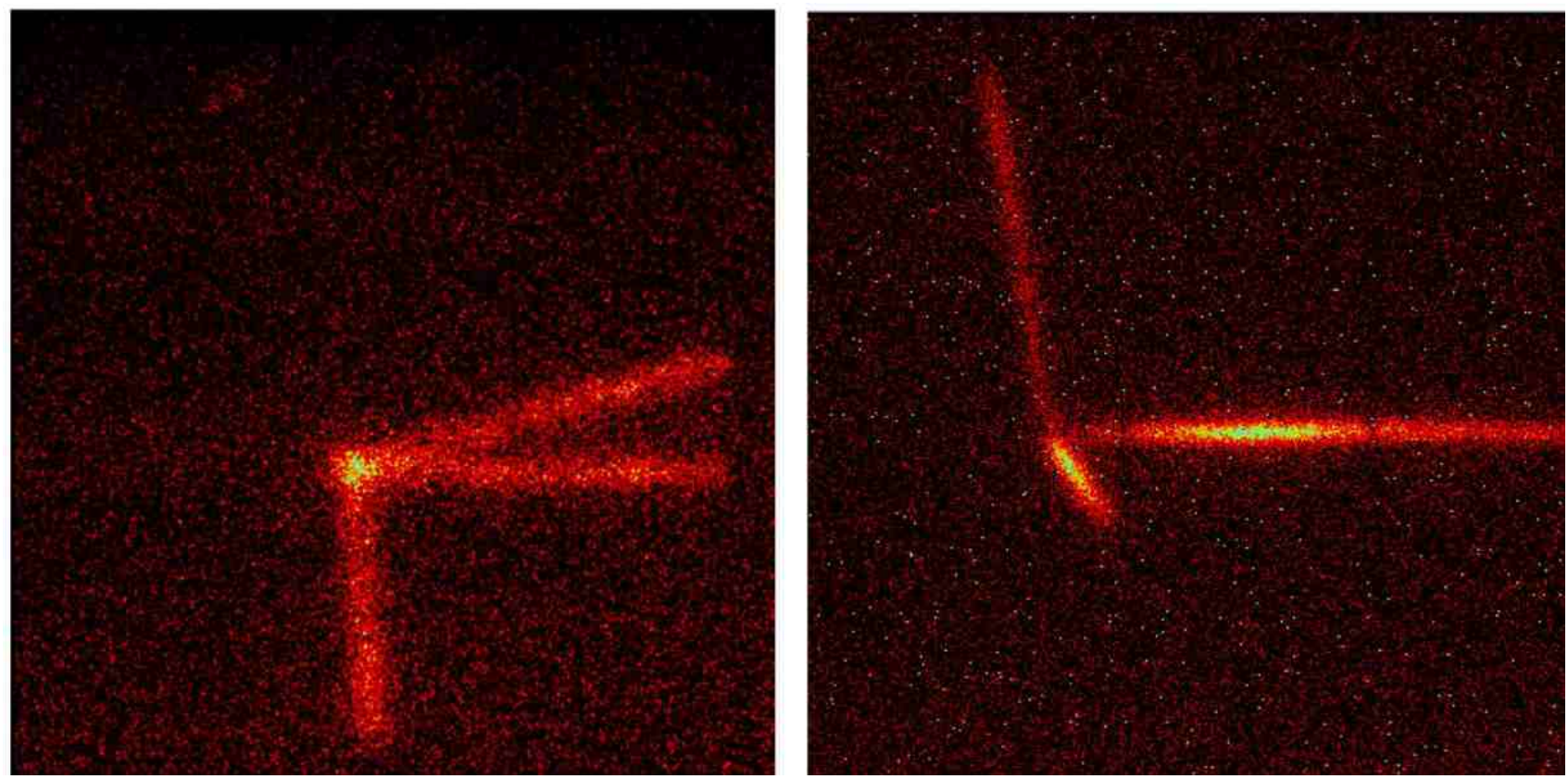}
 \caption{ \label{fig:IVD1}
   Beta-delayed multi-particle decays recorded with the optical time
   projection chamber described in section \protect\ref{sec:IIIC1}.
   The left panel shows beta-delayed three-proton emission from
   $^{45}$Fe (from \protect\cite{Miernik:2007}) recorded so the
   incoming track is not visible, the right panel the track of a $^8$He
   ion entering from the right that after beta-decay breaks up into a
   triton (long weak track), an alpha particle and an invisible neutron.}
\end{figure*}

The beta-delayed triton emission is again favoured at the neutron
dripline and has been observed clearly in $^8$He and $^{11}$Li and at
the $10^{-4}$ level in $^{14}$Be \cite{Jeppesen:2002}, but its
relation to the structure of the emitting nucleus is less well
understood. Recent experiments on $^{11}$Li \cite{Madurga:2009} and
$^8$He \cite{Mianowski:2010} have confirmed the decay mode with new
experimental procedures, see figure \ref{fig:IVD1}, but it seems that
more experimental data is needed before one can determine e.g.\
whether the triton decays proceed through states in the daughter
nucleus or, as the deuterons, directly to the continuum. In the latter
case the decay mode may depend on three-nucleon correlations in the
decaying nucleus.

The decays of $A=9$ nuclei lead mainly to final states with two alpha
particles and a nucleon. Complete kinematics decay studies have been
performed on these nuclei during the last decade and have resulted in
the discovery of new decay branches and in spin determination of
several intermediate levels \cite{Prezado:2003,Prezado:2005}. Strong
Gamow-Teller branches in the mirror decays of $^9$Li and $^9$C go to
states at around 12 MeV excitation energy, but the deduced $B_{GT}$
values are large and a factor 4--5 larger for $^9$Li than for $^9$C
which is not possible to understand from conventional theory
\cite{Millener:2005,Kanada-Enyo:2010}. The reason for this is still
unknown, the experimental strength \cite{Prezado:2003} may perhaps
involve more than one level, and a proper theoretical investigation of
the three-body continuum may help resolve the puzzle.

It would also be valuable to have calculations of the decays of the
halo nuclei $^{11}$Li and $^{14}$Be that take the continuum into
account explicitly. A more complete decay scheme for $^{11}$Li is now
available both at low \cite{Mattoon:2009} and high \cite{Madurga:2008}
excitation energies in the daughter, but the deduced strength is still
significantly less than that predicted by recent theory
\cite{Kanada-Enyo:2010}, in particular it has still not been possible
to experimentally check the above mentioned prediction of the GTGR
being placed below the initial state. A similar situation seems to be present
for $^{14}$Be where the experimental decay strength distribution
\cite{Jeppesen:2002} at high excitation energy is significantly lower
than theoretical predictions. A better determination of the
beta-delayed neutron branches could alleviate the problem, but may not
suffice to solve it.

Much less is known about the decay of heavier dripline nuclei, but at
least major beta-delayed neutron and/or gamma lines are known out to
$^{17}$B \cite{Raimann:1996}, $^{19}$C \cite{Ozawa:1995}, $^{22}$N
\cite{Sumithrarachchi:2010} and $^{24}$O \cite{Reed:1999} and
halflives and $P_n$ values are known for the heavier B, C and N
isotopes \cite{Yoneda:2003}. In the region above oxygen the halflives
are not known
for the most neutron-rich isotopes of any element. The major decay
branches are established for nuclei at a similar distance from the
line of stability, e.g.\ for $^{29}$Ne \cite{Tripathi:2006} and
$^{33}$Na \cite{Nummela:2001,Radivojevic:2002}, but the dripline from
here on is significantly further out (see section \ref{sec:II}).

Among the different physics questions that have been investigated in
the heavier neutron-rich nuclei can be mentioned the stability of the
$N=28$ shell that has been probed by extensive halflife measurements
\cite{Grevy:2004} as well as the $N=32$ and possible $N=34$ subshells
probed in decays of Sc and Ti isotopes \cite{Crawford:2010}. The
observed isotopic anomalies in some meteorites is known to depend on
decay properties of very neutron-rich nuclei and motivated new
measurements on the heavy Ar \cite{Weissman:2003} and Sc-Co
\cite{Sorlin:2003} isotopes.

Recent experiments \cite{Hosmer:2005,Hosmer:2010,Winger:2009} have
succeeded in determining halflives and $P_n$ values for $^{78}$Ni and
nuclei around it. Apart from the interest in settling the properties
of this doubly magic nucleus the information is also needed to
fine-tune calculations of the astrophysical \emph{r}-process in this mass
range where there is sensitivity in particular to the halflife of
$^{78}$Ni itself \cite{Hosmer:2010}. At higher masses the nuclei
participating in the \emph{r}-process have been reached experimentally at
$N=82$, see \cite{Pfeiffer:2001,Langanke:2003} for more details.

\begin{figure*}
\centerline{
\includegraphics[width=0.63\textwidth]{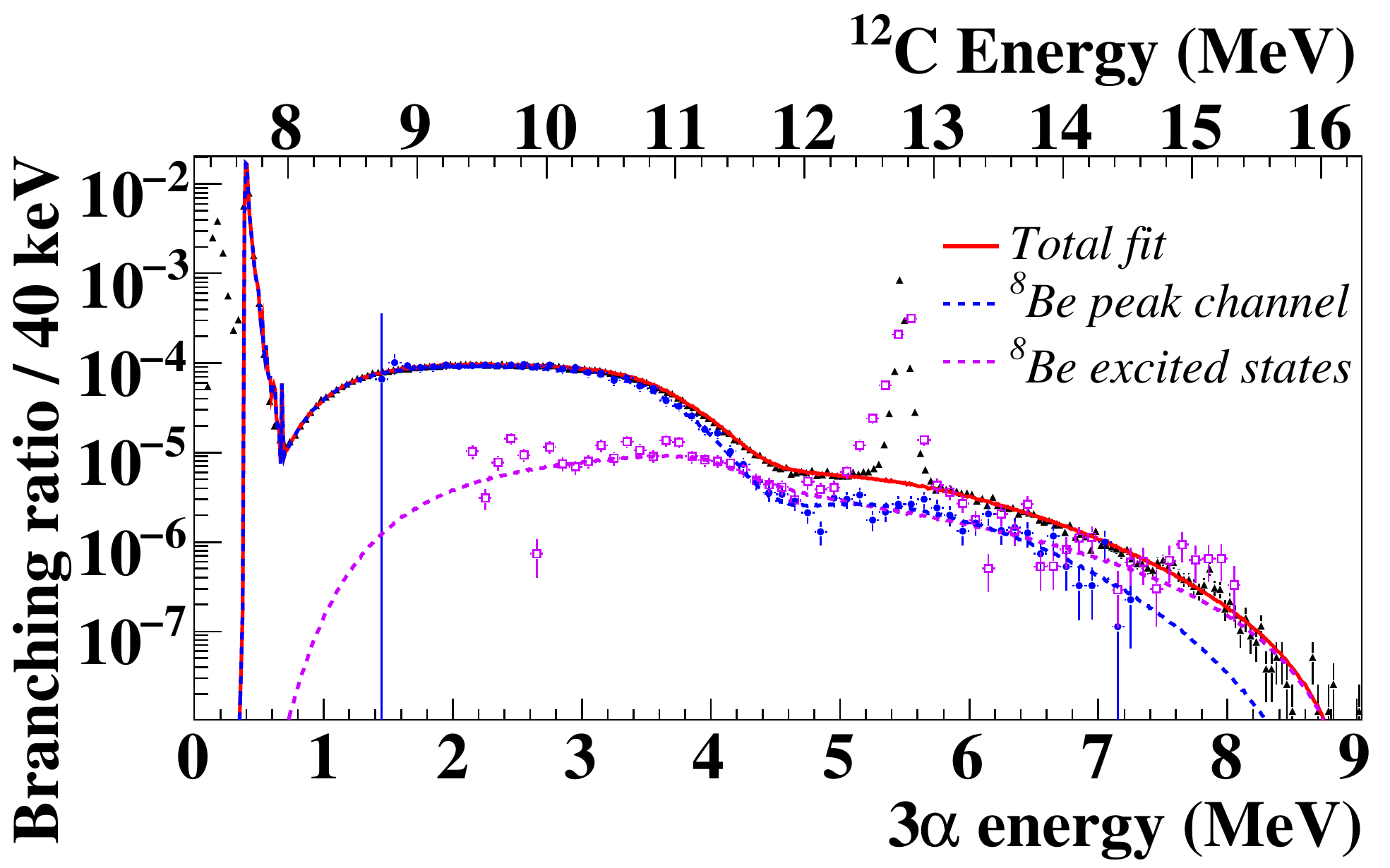}
}
 \caption{ \label{fig:IVD2}
   (Color online) The branching ratio for beta-delayed alpha decay of $^{12}$N
   (filled triangles) is shown as a function of the total energy
   \protect\cite{Hyldegaard:2010}. The solid line is a fit to the
   feeding to $0^+$ and $2^+$ states and does not include the
   contribution to the $1^+$ state at 12.7 MeV. The filled circles
   (open squares) give the contribution from decays that do (do not)
   proceed through the $^8$Be ground state, the dashed lines are the
   corresponding fits. See the text for details.}
\end{figure*}

\subsubsection{Selected spectroscopic tools}

Beta-recoil effects may play a role for beta-delayed neutrons similar
to that discussed above for beta-delayed protons, but has only been
explored in a few cases such as $^9$Li \cite{Nyman:1990}.  If the
nucleus recoiling from neutron emission emits a gamma ray the latter
will also be Doppler broadened provided the gamma-emitting state is
sufficiently short-lived. This has been used to improve the decay
scheme of $^{11}$Li \cite{Fynbo:2004,Mattoon:2009}, and is a valuable
way to cross-check results from the sometimes complex neutron spectra
and neutron-gamma coincidence measurements, see \cite{Hirayama:2005}
and references therein for $^{11}$Li.  Analogous neutron-gamma
experiments have already been performed e.g.\ for $^{21}$N
\cite{Li:2009}, $^{33}$Mg and $^{35}$Al \cite{Angelique:2006}.

The analysis of decays through regions of high level density proceeds
similarly to the case for the proton rich nuclei, except that the
experimental challenges are higher due to the neutron detection.
Fluctuation analysis will again be an important tool in order to
extract reliable interpretations from beta-delayed neutron spectra
\cite{Hardy:1978}.
For lighter nuclei where the level density is smaller one should in
principle in the analysis of neutron spectra worry about exact lineshapes,
interference effects etc.\ as was the case for the corresponding
delayed proton spectra. However, most experiments presently determine
neutron energies through time-of-flight and assume (at least
implicitly) that their resolution will smear out such effects so that
peaks in the spectra can be fitted with Gaussians. This practice could
lead to wrong assignments.

\subsection{Beta delayed emission of several light particles}
\label{sec:IVD}

Apart from decays through $^8$Be and states in $^{12}$C above the
triple-$\alpha$ threshold (and a few weak transitions involving an
$\alpha$ particle and a nucleon such as occurring in the decay of
$^{17}$Ne \cite{Chow:2002}) beta-delayed emission of several light
particles involves only nucleons.  The first beta-delayed
multi-nucleon decays, $\beta$2n and $\beta$3n, were discovered about
30 years ago \cite{Azuma:1979,Azuma:1980}. The $\beta$2p process
followed shortly after \cite{Cable:1983}, but the $\beta$3p process
was only observed a few years ago in $^{45}$Fe
\cite{Miernik:2007} (see figure \ref{fig:IVD1}) and only recently also
reported in $^{43}$Cr \cite{Pomorski:2011}.

As shown in figure
\ref{fig:IVC2} beta-delayed multi-neutron emission will become
dominant in the decays of very neutron-rich nuclei, whereas the other
processes only occur with small to moderate intensity (with the
exception of the $A=8,9$ decays). Somewhat ironically, the
multi-neutron process are the least studied ones, partly for
experimental reasons due to the difficulty of neutron detection,
partly due to the quite few cases of beta-delayed multi-neutron
emission known today. The one case, $^{17}$B, where beta-delayed four
neutron emission has been reported \cite{Dufour:1988} needs to be
confirmed since other multi-neutron branches reported in the same work
has since been shown to be too large \cite{Bergmann:1999}.

\begin{figure*}
\centerline{
\includegraphics[width=0.63\textwidth]{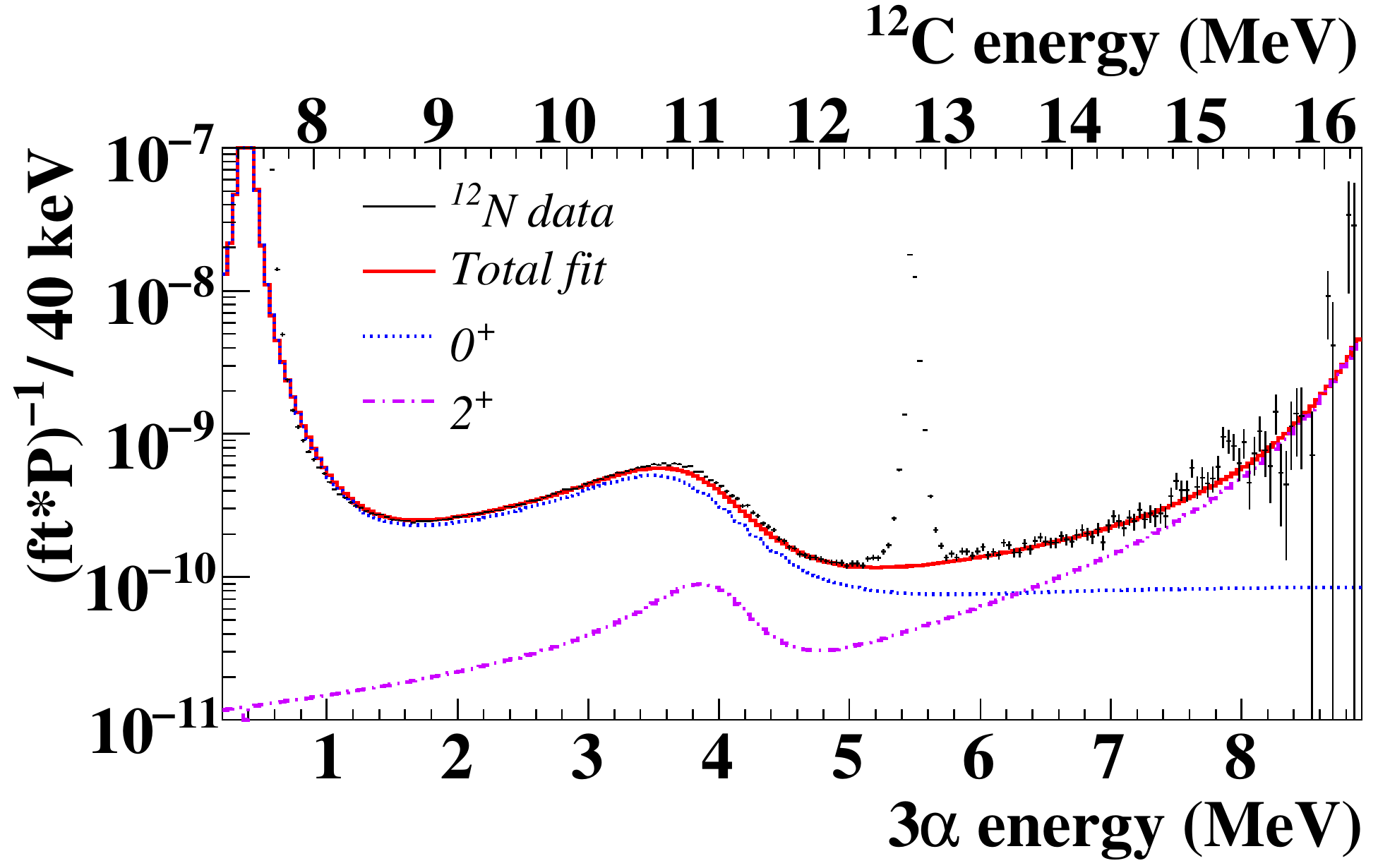}
}
 \caption{ \label{fig:IVD3}
   (Color online) The beta-delayed alpha decay data for $^{12}$N shown in figure
   \protect\ref{fig:IVD2} are displayed corrected for the beta-decay
   phase space factor and alpha particle penetrability factors (from
   \cite{Hyldegaard:2010a}). The total fit is divided into
   contributions from $0^+$ and $2^+$ states in $^{12}$C. Note the
   clear interference between $0^+$ states at low energy and the
   enhanced decay rate at high energy.}
\end{figure*}

The question of the particle emission mechanism is of prime
importance.  Although, as discussed in more detail in section
\ref{sec:VII}, calculations of multi-particle final states in
principle are becoming feasible now, it is still of interest to know
whether simpler decay mechanisms, such as sequential decay, can
describe a process or whether break-up directly into multi-particle
continuum states takes place.  Currently the only
experimental information on beta-delayed multi-neutron emission
comes from single neutron spectra
\cite{Azuma:1979}. There is more knowledge on $\beta$2p decays, as
recently reviewed in \cite{Blank:2008}. The most thoroughly studied
case, that of $^{31}$Ar \cite{Fynbo:2000}, seems to display only
sequential emission and other $\beta$2p cases are consistent with
this. Turning to other beta-delayed multi-particles modes,
the $\beta\alpha$p decays from $^{17}$Ne \cite{Chow:2002} could
also be analyzed assuming only sequential decays. For the light
nuclei, a recent study \cite{Madurga:2008} of $^{11}$Li indicated that
most of the decays in the three-body (n$+\alpha+^6$He) and the
five-body ($2\alpha+3$n) channels are sequential and to a large extent
proceeds through various He isotopes; a smaller direct break-up
component is, however, still possible. Finally, for the $A=9$ decays
mentioned above in section \ref{sec:IVC} most decay branches can be
approximately described in a sequential picture, but there are
indications \cite{Prezado:2005} that there is a direct break-up
component from the 5/2$^-$ level at 2.43 MeV in $^9$Be. The
hypothetical sequential branches through $^5$He$_{gs}$ and
$^8$Be($2^+$), which give energy distributions that for the
higher-lying broader states can be distinguished and seem to be
observed, would give overlapping energy distributions for the 2.43 MeV
level. The sequential picture anyway does not really make sense in
this situation as discussed in section \ref{sec:VII}. For the specific
case of the 2.43 MeV level the break-up mechanism has been
investigated in three-body calculations \cite{Alvarez:2008} where the
experimental energy and angular distributions could be reproduced.

The analysis of such data often makes use of the R-matrix formalism
since this allows for level parameters to be fitted to experiment. The
adaptation of the formalism to beta decay is described e.g.\ in
\textcite{Barker:1988}. It is a
priori applicable only for two-body decays, but has been employed in
practice also for sequential decays due to the lack of better
approaches. \textcite{Robson:1975} has shown how to formally make
sense of extensions of R-matrix to multi-particle situations, but this
has not been implemented in data analysis. One can therefore not rely
fully on results derived from present R-matrix fits.

It would be interesting to have more detailed data on the decay
mechanism for $\beta$3p decays. On a longer timescale it is without
doubt the multi-neutron detection capabilities that constitute the key
challenge for future progress in this field.

\subsubsection{The case of $A=12$}

The complications that may arise in beta-delayed decays can be
illustrated with the case of $^{12}$N (and $^{12}$B) whose decay into
the 3$\alpha$ continuum has recently been studied in detail
\cite{Diget:2009,Hyldegaard:2009,Hyldegaard:2010} motivated by the
importance of this continuum for the astrophysical triple-$\alpha$
process \cite{Fynbo:2005}. The decay goes through narrow $1^+$ states
as well as through several $0^+$ and $2^+$ states that all couple
strongly to the continuum. The experimental spectrum is shown in
figure \ref{fig:IVD2} as a function of the excitation energy in
$^{12}$C.  One can experimentally identify decays that proceed through
the narrow ground state in $^8$Be, these decays are also marked in the
figure and obviously correspond to sequential break-up. Apart from the
Hoyle state at 7.65 MeV all other states in the fit are broad. The
resulting interference may be easier to see in figure \ref{fig:IVD3}
where the data have been corrected for the beta phase-space factor and
the alpha particle penetrabilities. The interference actually also
involves the upper tail of the Hoyle state (its ``ghost'', cf.\
\textcite{Barker:1962}) that owes its narrow width to a small value
for the penetrability.

The increase in strength at higher energies seen for the $2^+$ states
is very hard to reconcile with the sum rule, eq.\ (\ref{eq:IVA2}), if
the decays are assumed to proceed through levels in $^{12}$C. This was
taken \cite{Hyldegaard:2010} as an indication that some of the
$^{12}$N decays proceed directly to continuum states. As mentioned
when discussing figure \ref{fig:iva1} the standard assumption is that
beta decays proceed through states in the emitter rather than directly
into continuum states. Corresponding transitions directly to the
continuum have been known for a long time for strong and
electromagnetic processes (direct reactions and direct radiative
capture, respectively), but are not generally recognized to occur also
in weak decays. However, as mentioned in section \ref{sec:IVC} it is
the most natural explanation also for $\beta$d decays.

\subsection{Beta delayed fission}
\label{sec:IVE}

Beta-delayed emission of particles heavier than alpha particles has
only been seen as beta-delayed fission. An overview of this phenomenon
with references to the early work can be found through
\textcite{Hall:1992,Kuznetsov:1999,Shaughnessy:2002}.
The probability for such decays depends both on
the beta strength at high excitation energy and on the fission
barriers \cite{Moller:2009}, and the decay mode may therefore provide
experimental information on fission in regions with high $Q_{\beta}$
values that is hard to obtain otherwise. Such information will enhance
our understanding of the fission process and can help to determine
better the role of fission in the \emph{r}-process \cite{Martinez:2007}.

Much of the recent activity has been driven by the continuous
developments in radioactive beam production capabilities and has
focussed on EC delayed fission in the light mass region. Experiments
have been carried out on $^{194}$At at GSI and on $^{180}$Tl at ISOLDE
\cite{Andreyev:2010}. The latter experiment showed a
surprising asymmetry in the mass distribution of the fragments. More
detailed information should become available in the coming years.


\newpage
\section{SINGLE-PROTON RADIOACTIVITY}
\label{sec:V}


\subsection{Introduction}
\label{sec:VA}

The proton radioactivity is the process occurring in odd-$Z$ nuclei located
beyond the proton-drip line. Due to the potential barrier (Coulomb and
centrifugal)
the emission of a proton from an unbound nucleus successfully competes
with other forms of decays ($\beta^+$, $\alpha$) only when the $Q_p$ value for
the decay is sufficiently large,
see Eq.~\ref{eq:VA1-Q}. The proton radioactivity was discovered  by
\textcite{Jackson:1970} who observed protons emitted from an isomeric state in
$^{53}$Co
at the excitation energy of 3.2~MeV.
First observation of the ground-state proton radioactivity was reported 12 years
later
by \textcite{Hofmann:1982} for $^{151}$Lu and by \textcite{Klepper:1982} for
$^{147}$Tm.
Presently, more then 40 proton emitters (from $^{109}_{53}$I to
$^{185}_{83}$Bi),
including emission from long-lived isomeric states, have been
established experimentally. Six of them
($^{131}$Eu,$^{141m}$Ho,$^{141gs}$Ho,$^{144}$Tm,
$^{145}$Tm,$^{146}$Tm) have transitions (so called fine structure) to the
excited
states in the respective daughter nuclei.

The importance of proton radioactivity follows from the fact that the knowledge
of the decay energy and the half-life (width), combined with the relatively
simple
model of the potential barrier penetration, yields information on the nuclear
wave function.
Thus, relatively simple observables provide constrains on nuclear models for
exotic nuclei, located beyond the proton drip-line. Since nuclear structure
information
is usually interpreted with help of the shell-model, it is convenient to divide
proton
emitters into two groups: those of the combined seniority one or two ($s\le 2$)
and
others with the combined seniority larger then two ($s>2$). The combined
seniority
is defined as the number of unpaired nucleons (protons and neutrons).
In the first case, apart from the odd-proton, and possibly an odd-neutron,
no proton and neutron pairs are broken.
Such decays are typical for ground states and for low lying isomers. In the
$s=1$ case
the odd proton can be pictured as moving in a single-particle orbital in the
nuclear potential
of the even-even daughter nucleus, while the $s=2$ case corresponds to even-odd
daughter
with an odd neutron acting as a spectator. The majority of known proton emitters
belong to this category, they will be discussed in Sec.~\ref{sec:VB}.
The situation of $s>2$ corresponds to the proton emission from highly excited
isomers having multiparticle character, which involves breaking additional pairs
of protons or neutrons. In Sec.~\ref{sec:VC} we discuss a few known cases.

In addition to the information extracted directly from proton emission
observables,
nuclear structure information has been gathered by using
the emitted protons as a tag in the recoil decay tagging (RDT) studies
\cite{Seweryniak:2001,Yu:1998,Seweryniak:2007}.

An overview of nuclear structure studies at the proton drip-line by means
of proton radioactivity was given recently by \textcite{Blank:2008}.
More detailed discussion of proton radioactivity
was given by \textcite{Woods:1997} and \textcite{Hofmann:1995}.
The work of \textcite{Sonzogni:2002} contains a compilation of results
on proton emitters known in 2001.

\subsubsection{Fundamentals}
\label{sec:VA1}

The necessary condition for a nuclide to decay by proton radioactivity
is a positive decay energy $Q_p$ defined as the difference between binding
energies (Eq.~\ref{eq:VA1-Q}) of the parent and the daughter atoms:
\begin{equation}
    Q_p = B(N,Z-1) - B(N,Z) = - S_p.
\label{eq:VA1-Q}
\end{equation}
To separate the contribution from the atomic electrons, the decay energy is
expressed
in the form:
\begin{equation}
    Q_p = Q_p^{nuc}-ES,
\label{eq:VA1-Q2}
\end{equation}
where $Q_p^{nuc}$ is the \emph{nuclear} part of the decay energy.
It is determined by the nuclear masses:
\begin{equation}
    Q_p^{nuc} = M^{nuc}(N,Z) - M^{nuc}(N,Z-1) - m_p \, ,
\label{eq:VA1-Q3}
\end{equation}
where $m_p$ is the proton mass.
The $ES$ is the electron screening correction defined as the difference between
total electron binding energies in the parent and the daughter nuclides:
\begin{equation}
    ES = Be(N,Z) - Be(N,Z-1).
\label{eq:VA1-ES}
\end{equation}
In the above the electron binding energy in the hydrogen atom has been
neglected.
The value of the screening correction $ES$ can be calculated
from the tabulated electron binding energies \cite{Huang:1976},
or estimated by a simple formula:
\begin{eqnarray}
\label{eq:ES-param}
 ES=0.49 + 0.0144 \cdot Z^{1.6} \; {\rm keV}.
\end{eqnarray}
The accuracy of this parametrization is better than 0.5\% for
$42<Z<75$ and drops to 1.6\% for Z=83. No influence of the neutron
number (isotopic effect) is taken here into account.

The decay energy is shared between the proton $E_p$ and the recoiling atom.
Therefore, the measured kinetic energy of the emitted proton is given by:
\begin{eqnarray} \label{eq:Ep}
E_p & = &
\frac{(M(N,Z-1)+m_e)}{m_p+M(N,Z-1)+m_e} \, Q_p \, .
\end{eqnarray}
In addition, the angular momentum as well as parity
conservation laws have to be satisfied:
\begin{eqnarray}
\label{eq:zzmp}
 \vec{I_i} & = & \vec{I_f}+ (\vec{l}+\vec{s})\\
 \pi_i \cdot \pi_f & = & (-1)^l
\end{eqnarray}
where, $\vec{I_i}$ and $\vec{I_f}$ are spins of the initial and final nuclear
states respectively,
$\vec{l}$ is the angular momentum of the emitted proton, $\vec{s}$ is the spin
of the proton,
$\pi_i$, $\pi_f$ are parities of the initial and final states, respectively.

\subsubsection{Probability of proton emission}

\begin{figure}
\centerline{
\includegraphics[width=0.37\textwidth]{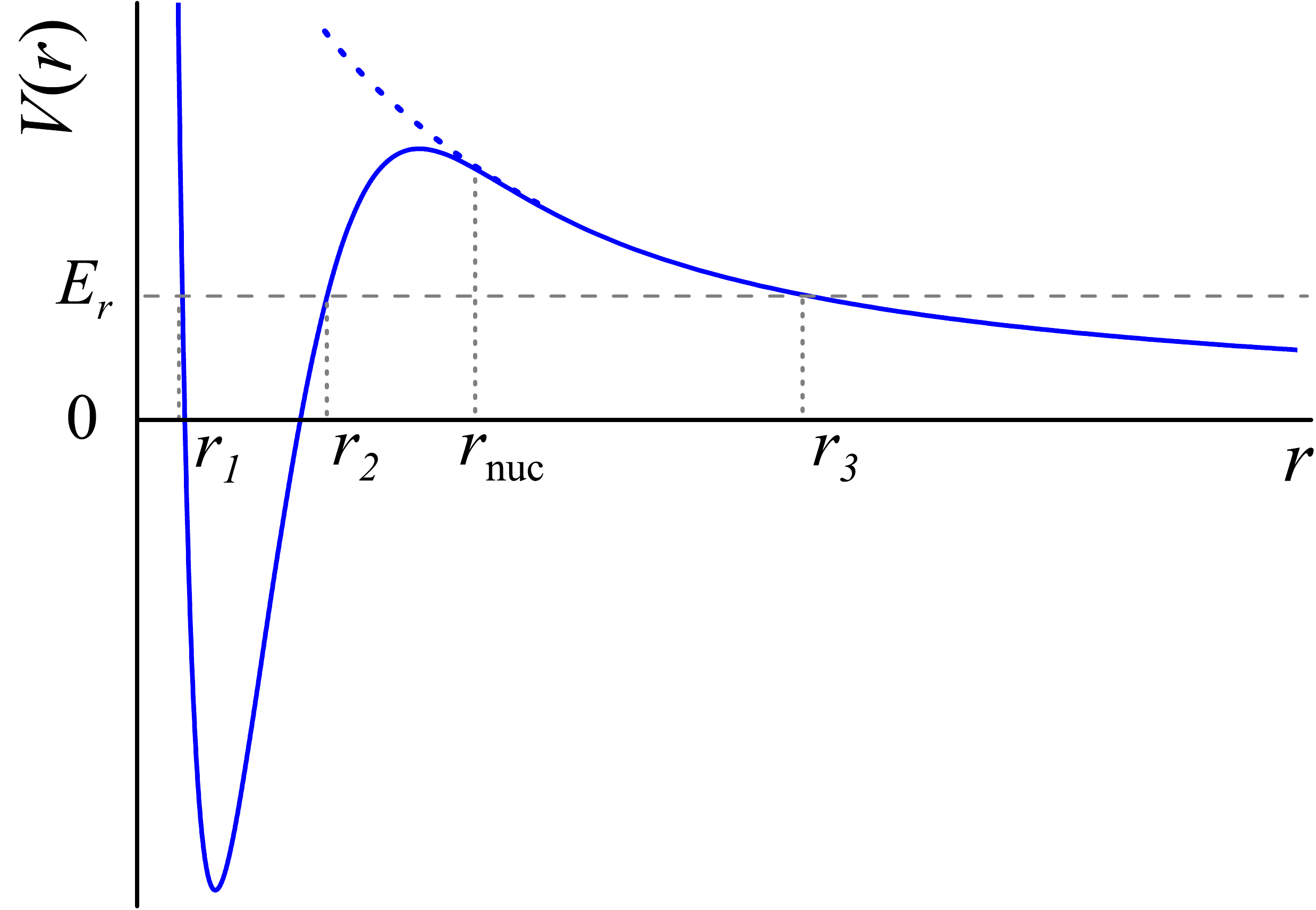}
}
\caption{Schematic view of the radial part of the nucleus-proton potential. The
classical turning points
$r_i$ for a particle with energy $E_r$ are marked. The nuclear contribution
turns
to zero around
$r=r_{\text{nuc}}$. }
\label{fig:VA-pot-bar-schem}
\end{figure}

Relatively simple calculations of the proton emission lifetimes
are based on the result obtained by \textcite{Gurvitz:1987} who analyzed
the decay widths and shifts of quasistationary states in the quantum
mechanical two-potential approach. By investigating the quasiclassical
limit they provided simple formulae which are similar to, but more general
than, those achieved in the framework of WKB approximations \cite{Brink:1983}.
In this approach the width of the proton-emitting state is given by:
\begin{eqnarray}
\Gamma_p&=& S_p \, \frac{N}{4\mu}\exp [-2\int_{r_2}^{r_3}{k(r)\, dr}] \\
 k(r)&=&\left( 2 \, \mu |Q_p^{nuc}-V(r)| \right)^{1/2},
\label{eq:VA2_Gamma}
\end{eqnarray}
where  the normalization factor $N$ has to satisfy the equation:
\begin{equation}
N \, \int_{r_1}^{r_2}{\frac{dr}{2 \, k(r)}} = 1.
\end{equation}
$S_p$ is the spectroscopic factor described later and $\mu$ is the reduced mass
of the proton and the daughter nucleus. Integration limits $r_i$
are the classical turning points, defined by $V(r_i)=Q_p^{nuc}$,
where $V(r)$ is the radial part of the nucleus-proton potential,
see Fig.~\ref{fig:VA-pot-bar-schem}.

To simplify calculations, some authors replace the factor $\frac{N}{4\mu}$ by
the so called frequency of assaults factor $\nu$ \cite{Hofmann:1996} calculated
for
the case of an $s$-wave proton leaving the square well plus Coulomb potential
\cite{Bethe:1937}
\begin{eqnarray}\label{eq:nu}
   \nu = \frac{\sqrt{2} \, \pi^2}{\mu^{\frac{3}{2}} \, R_c^3\sqrt{(Z-1)e^2/R_c
   -Q_p^{nuc}}},
\end{eqnarray}
where $R_c$ is the channel radius $R_c = r_{nuc} \cong r_0 (A-1)^{1/3}$ with
$r_0=1.21$~fm.
For example, in the decay of $^{151}$Lu $\nu$ equals $4.1 \cdot 10^{21}$
s$^{-1}$.

Then, the proton emission decay constant is calculated as:
\begin{equation}
\lambda_p= S_p \, \nu \, \exp [-2 \, \int_{r_2}^{r_3}{k(r)dr}]
\end{equation}

The potential $V(r)$ is taken as a superposition of nuclear ($V_N$),
coulomb ($V_C$), centrifugal $V_L$, and spin-orbit $V_{ls}$ terms.
The nuclear part is usually described by the Woods-Saxon form with various
parametrization:
\begin{equation}
V_N(r) = -V_r \frac{1}{1+\exp{\frac{r-R}{a}}} \,.
\end{equation}
Comparison between different parametrizations was done in the work of
\textcite{Ferreira:2002}.
Although not supported by the work of \textcite{Ferreira:2002}, the most
frequently used
parametrization is that of \textcite{Becchetti:1969}.
Detailed potential descriptions and calculations for some of the emitters
can be found in \textcite{Aberg:1997,Buck:1992,Hofmann:1995}.

\begin{figure}
\centerline{
\includegraphics[width=0.4\textwidth]{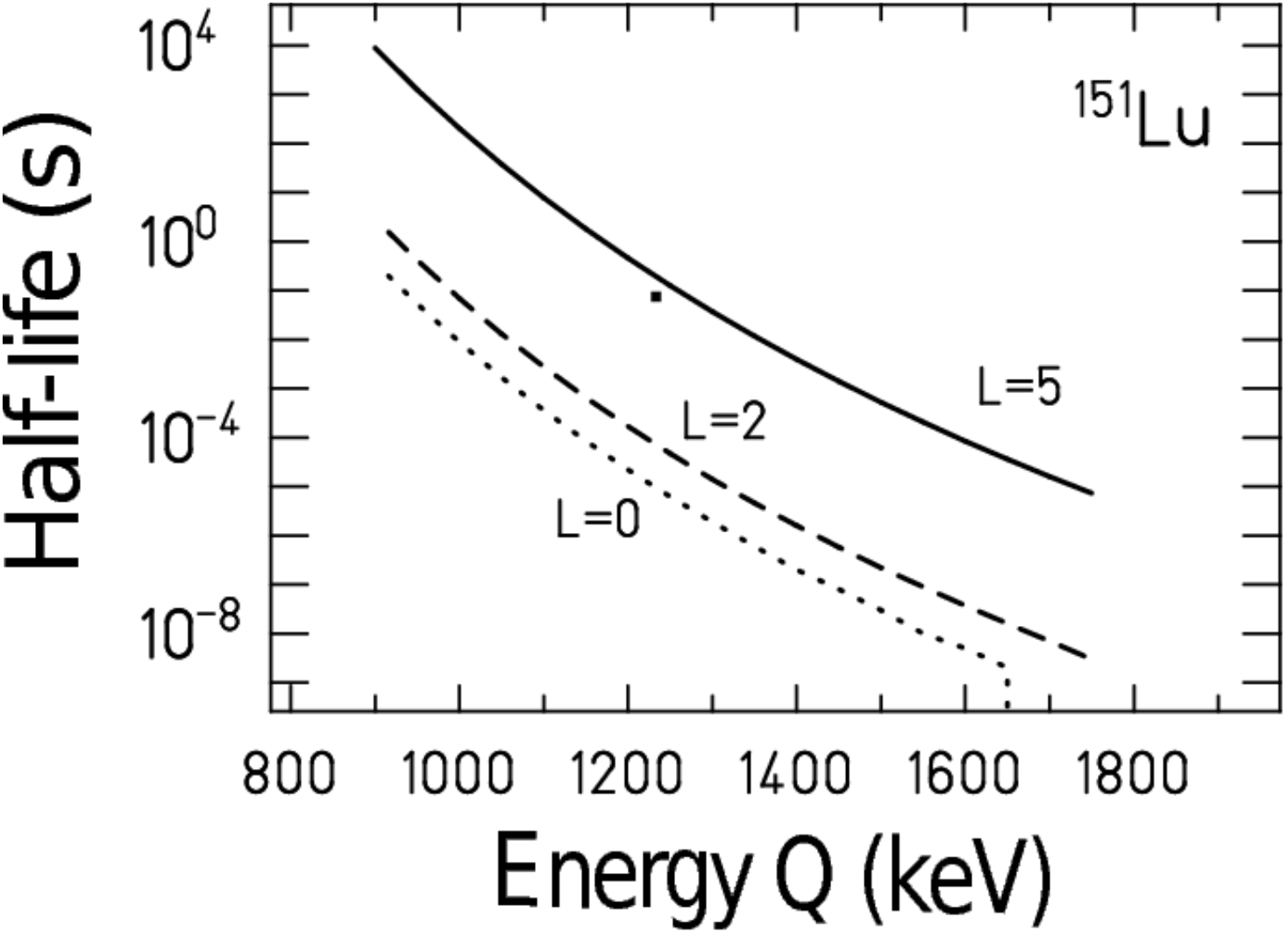}
}
\caption{The half-life for the proton emission as a function of nuclear decay
energy $Q_{p}^{nuc}$ and the orbital angular momentum carried away by the
proton.
Calculations were done for the case of $^{151}$Lu with S$_p=0.54$. The measured
values of the decay energy and the half-life, indicated by the black square,
suggest the transition with $L=5$. }
\label{fig:VA-T12}
\end{figure}

As an example, the half-life for the proton emission from $^{151}$Lu as a
function
of the decay energy, for three values of the orbital angular momentum is shown
in Fig.~\ref{fig:VA-T12}.
The characteristic strong dependence on the available energy and the angular
momentum
is clearly seen.

\subsubsection{Spectroscopic factor $S_p$}
\label{sec:VA3}

The spectroscopic factor $S_p$ is a measure of the single-particle purity
of the initial wave function.
Within the BCS theory the spectroscopic factor is given by
 $S_p^{th} = u_j^2$, where the vacancy factor $u^2$ is the probability that
the spherical shell-model orbital with $(n,l,j)$ quantum numbers is empty
in the daughter nucleus.
For some proton emitters the factors $u_j^2$ can be found in the
work of \textcite{Aberg:1997}.
This theoretical value is compared with the experimental value $S_p^{exp}$
derived as the
ratio of the measured partial decay constant and the calculated one
assuming $S_p = 1$. The agreement between the two values indicates that
the correct assumption about the initial wave function has been taken.
For example, in the case of $^{151}$Lu $S_p^{exp}=0.5$ and $S_p^{th}=0.54$
assuming the proton was emitted from the $[\pi h_{11/2}]11/2^-$ state.
The good agreement supports such an interpretation.
In contrast, for the case of $^{145}$Tm $S_p^{exp}=0.48$ and $S_p^{th}=0.65$,
which suggests that the spherical potential used in the calculation may
not be applicable. Indeed, the deformed nature of $^{145}$Tm
was confirmed experimentally \cite{Karny:2003,Seweryniak:2007a} and
theoretically \cite{Arumugam:2008}.

\begin{table*}
\caption{Summary of literature data for $s=1$ proton emitters.
Empty place means no data available. For isotopes with ground and isomeric state
emission a combined literature is given. In 'Reference' column additional letter
E - refers to experimental papers, while letter T - points to the theoretical
papers where properties of the referred nuclei  are explicitly calculated.}
\label{tab:VB2}
\begin{tabular}{|c|c|c|c|c|c|l|}
\hline \hline
Emitter &Cross sec.& E$_p$(MeV) & Q$_p^{nucl}$(MeV) & T$_{1/2}$  & Ang.mom. &
References \\
\hline \hline
$^{109}_{53}$I & 40$\mu$b &0.8126(40)&0.829(4) &93.5(3)$\mu$s & l=2 &
\textcite{Faesterman:1984} (E)\\
               &          &          &         &              &     &
               \textcite{Gillitzer:1987} (E)\\
               &          &          &         &              &     &
               \textcite{Heine:1991} (E)\\
               &          &          &         &              &     &
               \textcite{Barmore:2000} (T)\\
               &          &          &         &              &     &
               \textcite{Sellin:1993} (E)\\
               &          &          &         &              &     &
               \textcite{Mazzocchi:2007} (E)\\
\hline
$^{112}_{55}$Cs & 0.5$\mu$b&0.807(7)& 0.815(7)& 0.5(1)ms      & l=2 &
\textcite{Page:1994} (E)\\
               &          &          &         &              &     &
               \textcite{Ferreira:2001} (T) \\
\hline
$^{113}_{55}$Cs & 30$\mu$b&& 0.9771(37)& 16.7(7)$\mu$s        & l=2 &
\textcite{Faesterman:1984} (E)\\
               &          &          &         &              &     &
               \textcite{Gillitzer:1987} (E)\\
               &          &          &         &              &     &
               \textcite{Page:1994} (E)\\
               &          &          &         &              &     &
               \textcite{Batchelder:1998} (E)\\
               &          &          &         &              &     &
               \textcite{Maglione:1998} (T)\\
               &          &          &         &              &     &
               \textcite{Barmore:2000} (T)\\
\hline
$^{117}_{57}$La & 240nb&0.806(5)&0.813(5) &24(3)ms            & l=2 &
\textcite{Soramel:2001} (E)\\
               &          &          &         &              &     &
               \textcite{Mahmud:2001} (E)\\
\hline
$^{121}_{59}$Pr  &0.3nb& 0.882(10)& 0.900(10)& $10^{+6}_{-3}$ ms &  l=4 or 5  &
\textcite{Bogdanov:1973} (E)\\
               &          &          &         &                 &            &
               \textcite{Robinson:2005} (E)\\
\hline
$^{130}_{63}$Eu  &9nb &1.020(15)& 1.028(15)& $0.90^{+0.49}_{-0.29}$ ms &  l=2 &
\textcite{Mahmud:2002} (E)\\
               &          &          &         &                       &      &
               \textcite{Davids:2004a} (E)\\
\hline
$^{131}_{63}$Eu &90nb& 0.932(7)& &17.8(19)ms                 & l=2 &
\textcite{Davids:1998} (E)\\
                &    & 811(7)\footnote{transitions to the excited states}&  &  &
                l=2 & \textcite{Sonzogni:1999} (E)\\
               &          &          &         &              &     &
               \textcite{Maglione:1999} (T)\\
               &          &          &         &              &     &
               \textcite{Davids:2000} (T)\\
               &          &          &         &              &     &
               \textcite{Kruppa:2000} (T)\\
               &          &          &         &              &     &
               \textcite{Maglione:2000} (T)\\
               &          &          &         &              &     &
               \textcite{Esbensen:2001} (T)\\
               &          &          &         &              &     &
               \textcite{Ferreira:2002} (T)\\
               &          &          &         &              &     &
               \textcite{Ferreira:2005} (T)\\
\hline
$^{135}_{65}$Tb  &6nb& 1.179(7) &1.188(7) & $0.94^{+0.33}_{-0.22}$ms&  l=5 &
\textcite{Woods:2004} (E)\\
\hline
$^{140}_{~67}$Ho  &13nb& 1086(10)& & 6(3)ms                   & l=5 &
\textcite{Rykaczewski:1999} (E)\\
               &          &          &         &              &     &
               \textcite{Maglione:2002} (T)\\
               &          &          &         &              &     &
               \textcite{Ferreira:2001} (T)\\
\hline
$^{141}_{~67}$Ho  &1.4$\mu$b& 1169(8)   &       & 4.1(1)      & l=5 &
\textcite{Davids:1998} (E)\\
               &          & 968(8)$^a$  &       &             & l=3 &
               \textcite{Rykaczewski:1999} (E)\\
$^{141m}_{~67}$Ho &       & 1235(9)     &       &7.4(3)$\mu$s & l=0 &
\textcite{Maglione:1999} (T)\\
               &          & 1031(11)$^a$&       &7.4(3)$\mu$s & l=2 &
               \textcite{Barmore:2000} (T)\\
               &          &          &         &              &     &
               \textcite{Esbensen:2001} (T)\\
               &          &          &         &              &     &
               \textcite{Seweryniak:2001} (E)\\
               &          &          &         &              &     &
               \textcite{Karny:2008} (E)\\
               &          &          &         &              &     &
               \textcite{Arumugam:2009} (T)\\
\hline
$^{144}_{~69}$Tm  &10nb& 1700(16) & & $1.9^{+1.2}_{-0.5}\mu$s & l=5 &
\textcite{Grzywacz:2005} (E)\\
  &  &  1430(25)\footnote{evidence of the transition to the excited state based
  on two counts}&  &
             & l=2$^b$ & \textcite{Bingham:2005} (E)\\
\hline \hline
\end{tabular}
\end{table*}

\begin{table*}
\caption{Summary of literature data for $s=1$ proton emitters. Continuation of
Table \ref{tab:VB2}.}
\label{tab:VB2m}
\begin{tabular}{|c|c|c|c|c|c|l|}
\hline \hline
Emitter &Cross sec.& E$_p$(MeV) & Q$_p^{nucl}$(MeV) & T$_{1/2}$  & Ang.mom. &
References \\
\hline \hline
$^{145}_{~69}$Tm  &0.5$\mu$b& 1728(10),  &    & 3.1(2)$\mu$s  & l=5 &
\textcite{Batchelder:1998} (E)\\
              &             & 1398(10)   &    &               & l=3 &
              \textcite{Rykaczewski:2001} (E)\\
              &          &          &         &               &     &
              \textcite{Karny:2003} (E)\\
              &          &          &         &               &     &
              \textcite{Seweryniak:2005} (E)\\
              &          &          &         &               &     &
              \textcite{Seweryniak:2007a} (E)\\
              &          &          &         &               &     &
              \textcite{Arumugam:2008} (T)\\
\hline
$^{146}_{~69}$Tm  &1$\mu$b& 1191(1) &         & 66ms        & l=5  &
\textcite{Livingston:1993} (E)\\
              &        & 1016(4)$^a$&         &             & l=3  &
              \textcite{Rykaczewski:2002} (E)\\
$^{146m}_{~~~69}$Tm&   & 889(8)$^a$ &         & 200ms       & l=3  &
\textcite{Rykaczewski:2001} (E)\\
              &        &1120(1)$^a$ &         &             & l=5  &
              \textcite{Rykaczewski:2001b} (E)\\
              &          &          &         &              &     &
              \textcite{Ginter:2003} (E)\\
              &          &          &         &              &     &
              \textcite{Seweryniak:2005} (E)\\
              &          &          &         &              &     &
              \textcite{Tantawy:2006} (E)\\
\hline
$^{147}_{~69}$Tm  &30$\mu$b& &1071.4(3.3) & 560(40)ms       & l=5 &
\textcite{Klepper:1982} (E)\\
$^{147m}_{~~~69}$Tm & &1110.8(3.9)&1139.3(5.3)&360(40)$\mu$s& l=2 &
\textcite{Larsson:1983} (E)\\
             &          &          &         &              &     &
             \textcite{Sellin:1993} (E)\\
             &          &          &         &              &     &
             \textcite{Toth:1993} (E)\\
             &          &          &         &              &     &
             \textcite{Seweryniak:1997} (E)\\
             &          &          &         &              &     &
             \textcite{Seweryniak:2005} (E)\\
\hline
$^{150}_{~71}$Lu  & 3$\mu$b&1261(4)& &46(5)ms               & l=5 &
\textcite{Sellin:1993} (E)\\
$^{150m}_{~~~71}$Lu & & 1286(6)& & $39^{+8}_{-6}\mu$s       & l=2 &
\textcite{Woods:1993} (E)\\
             &          &          &         &              &     &
             \textcite{Ginter:2000} (E)\\
             &          &          &         &              &     &
             \textcite{Ferreira:2001} (T)\\
             &          &          &         &              &     &
             \textcite{Maglione:2002} (T)\\
             &          &          &         &              &     &
             \textcite{Ginter:2003} (E)\\
             &          &          &         &              &     &
             \textcite{Robinson:2003}(E)\\
\hline
$^{151}_{~71}$Lu  & 70$\mu$b&1232.9(2.0) &   & 80(2)ms      & l=5 &
\textcite{Hofmann:1982} (E)\\
$^{151m}_{~~~71}$Lu  &  & 1310(10) &         & 16(1)$\mu$s  & l=2 &
\textcite{Sellin:1993} (E)\\
             &          &          &         &              &     &
             \textcite{Yu:1998} (E)\\
             &          &          &         &              &     &
             \textcite{Bingham:1999} (E)\\
             &          &          &         &              &     &
             \textcite{Ferreira:2000} (T)\\
\hline
$^{155}_{~73}$Ta\footnote{see the discussion in \cite{Page:2007} about
contradictory results from the work of
\cite{Uusitalo:1999}}&  &1444(15)  &  &$2.9^{+1.5}_{-1.1}$ms& l=5 &
\textcite{Uusitalo:1999} (E)\\
             &          &          &         &              &     &
             \textcite{Page:2007} (E)\\
             &          &          &         &              &     &
             \textcite{Joss:2006} (E)\\
\hline
$^{156}_{~73}$Ta  & & 1007(5)& & 144(24)ms                  & l=2 &
\textcite{Page:1992} (E)\\
$^{156m}_{~~~73}$Ta &50nb& 1108(8)&  & 375(54)ms            & l=5 &
\textcite{Livingston:1993} (E)\\
             &          &          &         &              &     &
             \textcite{Page:1996} (E)\\
\hline
$^{157}_{~73}$Ta  & 20nb&927(7)& & $12.1^{+3.1}_{-2.3}$ms   & l=0 &
\textcite{Page:1996} (E)\\
             &          &          &         &              &     &
             \textcite{Irvine:1997} (E)\\
\hline
$^{159}_{~75}$Re  & &1805(20)& &21(4)$\mu$s                 & l=5 &
\textcite{Joss:2006} (E)\\
             &          &          &         &              &     &
             \textcite{Page:2007} (E)\\
\hline
$^{160}_{~75}$Re  & 1$\mu$b&1261(6)& &0.79(16)ms            & l=2 &
\textcite{Page:1992} (E)\\
             &          &          &         &              &     &
             \textcite{Page:1996} (E)\\
             &          &          &         &              &     &
             \textcite{Hagino:2001} (T)\\
\hline
$^{161}_{~75}$Re  & 150nb&1192(2)& & 440(1)$\mu$s           & l=0 &
\textcite{Irvine:1997} (E)\\
$^{161m}_{~75}$Re  & & 1315(7)& & 14.7(3)ms                 & l=5 &
\textcite{Lagergren:2006} (E)\\
             &          &          &         &              &     &
             \textcite{Hagino:2001} (T)\\
             &          &          &         &              &     &
             \textcite{Arumugam:2007} (T)\\
\hline \hline
\end{tabular}
\end{table*}

\begin{table*}
\caption{Summary of literature data for $s=1$ proton emitters. Continuation of
Table \ref{tab:VB2}.}
\label{tab:VB2mm}
\begin{tabular}{|c|c|c|c|c|c|l|}
\hline \hline
Emitter &Cross sec.& E$_p$(MeV) & Q$_p^{nucl}$(MeV) & T$_{1/2}$  & Ang.mom. &
References \\
\hline \hline
$^{164m}_{~~~77}$Ir  & n/a& 1807(14)& & $58^{+46}_{-18}\mu$s&  l=5 &
\textcite{Kettunen:2001} (E)\\
             &          &          &         &              &     &
             \textcite{Mahmud:2002} (E)\\
\hline
$^{165m}_{~77}$Ir  & 0.2$\mu$b&1707(7)& & 0.30(6)ms         &  l=5 &
\textcite{Davids:1997} (E)\\
\hline
$^{166}_{~77}$Ir  & 6$\mu$b&1145(8)& & 10.5(2.2)ms          &  l=2 &
\textcite{Davids:1997} (E)\\
$^{166m}_{~~~77}$Ir & & 1316(8)& &15.1(9)                   &  l=5 &  \\
\hline
$^{167}_{~77}$Ir  & 10$\mu$b&1064(6)& &35.2(20)ms           &  l=0 &
\textcite{Davids:1997} (E)\\
$^{167m}_{~~~77}$Ir  && 1238(7)& & 30.0(6)ms                &  l=5 &
\textcite{Davids:2000} (T) \\
             &          &          &         &              &      &
             \textcite{Scholey:2005} (E)\\
\hline
$^{170}_{~79}$Au & &1463(12) & &$286^{+50}_{-40}\mu$s       & l=2 &
\textcite{Mahmud:2002} (E)\\
$^{170m}_{~~~79}$Au & & 1735(9)& &$617^{+50}_{-40}\mu$ms    & l=5 &
\textcite{Kettunen:2004} (E)\\
\hline
$^{171}_{~79}$Au &  2$\mu$b&1437(12)& &$22^{+3}_{-2}\mu$s   & l=0 &
\textcite{Davids:1997} (E)\\
$^{171m}_{~~~79}$Au & & 1694(6)& &1.09(3)ms                 & l=5 &
\textcite{Poli:1999} (E)\\
             &          &          &         &              &     &
             \textcite{Back:2003} (E)\\
             &          &          &         &              &     &
             \textcite{Kettunen:2004} (E)\\
\hline
$^{176}_{~81}$Tl  & &1258(18)& & $5.2^{+3.0}_{-1.4}$ms      & l=0 &
\textcite{Kettunen:2004} (E)\\
\hline
$^{177}_{~81}$Tl  & 30nb&1156(20)& &18(5)ms                 & l=0 &
\textcite{Poli:1999} (E)\\
$^{177m}_{~~~81}$Tl  && 1958(10)& & $230(40)\mu$s           & l=5 &
\textcite{Davids:2001b} (E)\\
             &          &          &         &              &     &
             \textcite{Kettunen:2004} (E)\\
\hline
$^{185}_{~83}$Bi & 60nb\footnote{6-10nb for 3n evaporation channel
\citetext{Andreyev:2004}}&1594(9)& &$60(4)\mu$s               & l=0 &
\textcite{Davids:1996} (E)\\
             &          &          &         &              &     &
             \textcite{Poli:2001} (E)\\
             &          &          &         &              &     &
             \textcite{Andreyev:2004} (E)\\
             &          &          &         &              &     &
             \textcite{Andreyev:2005} (E)\\
             &          &          &         &              &     &
             \textcite{Arumugam:2007} (T)\\
\hline \hline
\end{tabular}
\end{table*}

\subsubsection{Models of proton emission}
\label{sec:VA4}

The usefulness as well as limitations of simple models, introduced in previous
sections,
may be illustrated with the example of $^{145}$Tm.
In this nucleus two proton transitions from the same state have been
observed \cite{Karny:2003}. The proton energies are
$E_p=1.73$~MeV and $E_p=1.40$~MeV, while the corresponding partial half-lives
are $3.4 \, \mu$s and $32 \, \mu$s. The first transition is interpreted as
a decay to the $0^+$ ground state of $^{144}$Er, while the
second goes to the first excited $2^+$ state in this nucleus.
In the frame of the spherical quasi-classical approach we may
assume that the first transition originates from the $\pi h_{11/2}$
orbital ($l=5$) while the second from the $\pi f_{7/2}$ ($l=3$)
component of the initial wave function. The calculations, including
theoretical spectroscopic factors ($u^2(h_{11/2}) =0.647$ and
$u^2(f_{7/2})=0.985$) yield the partial half-lives of
2.29$\mu$s and $1.28\mu$s, for the two transitions, respectively.
Thus, by comparing with the experimental values, we may conclude that
the emitter wave function consists of 67\%$=\frac{2.29}{3.4}\cdot 100$\%
of the $l=5$, $\pi h_{11/2} \otimes 0^+$ and 3.7\% of $l=3$,
 $\pi f_{7/2}\otimes 2^+$ components. The remaining 29\% of the wave function
 does not
participate in proton emission and therefore can not be determined within
this simple model. The more elaborate coupled channel
model which takes into account the dynamic deformation
\cite{Karny:2003,Hagino:2001}
yields values of  56\% for $\pi h_{11/2}\otimes 0^+$ state
and $\approx$ 3\% for $\pi f_{7/2} \otimes 2^+$ state.

Although simple models are useful for the first order
approximations, they cannot be expected to yield correct results
for highly deformed nuclides.
For those  cases more elaborate theoretical approaches have to be applied.
Examples of such approaches are given by
 \textcite{Aberg:1997,Esbensen:2001,Kruppa:2004,Fiorin:2003}.
Calculations with tri-axially deformed potential are considered
by \textcite{Davids:2004,Kruppa:2004,Arumugam:2009}.
An extended description of the theoretical models
used for proton emission can be found in chapters 4,5, and 6 of the Lecture
notes in Physics by \textcite{Delion:2010}, as well as in
\textcite{Delion:2006a}.

\subsection{Seniority $s\le2$ proton emitters}
\label{sec:VB}

Seniority $s \le 2$ proton emitters are the "classical" proton emitters in which
an
unpaired proton leaves the nucleus from the ground state or the isomeric state.
The isomeric state has to be low enough to allow only for unpaired particle
excitation.
The half-life of those emitters span from
$T_{1/2}=1.9^{+1.2}_{-0.5} \,\mu$s for $^{144}$Tm to 0.560(40)s for $^{147}$Tm.
They were
all but one produced in fusion-evaporation reactions with the exit channel
containing
a proton and
from 1 to 6 neutrons. $^{185}$Bi was also produced in the 3n evaporation channel
\cite{Andreyev:2004}.
Typical cross sections range from 0.3 nb (1p, 6n) for $^{121}$Pr
to $70$ $\mu$b (1p, 2n) for $^{151}$Lu. Combined properties of "classical"
proton
emitters are presented in table \ref{tab:VB2}.

\subsubsection{Odd-mass, $s=1$ proton emitters}
\label{sec:VB1}

As mentioned above, due to the strong dependence of the half-life on the
$Q_p$-value and
on the angular momentum of the emitted proton (see figure \ref{fig:VA-T12}),
proton emission is a valuable tool for
nuclear structure study  beyond the proton drip line. Measured proton energy and
decay half-life
in most of the cases directly point to the configuration of the decaying
orbital.
 This is especially true for
the odd-mass seniority $s=1$ cases where the emitting nuclei can be described as
even-even $0^+$ core coupled to the unpaired proton. With the $0^+$ ground state
of
the daughter nuclei, establishing the angular momentum leaves only two
possibilities
for the total angular momentum $j=l\pm1/2$ of the emitting state.
This value can then be used
to calculate the components of the nuclear wave function of the emitter.
Depending on the nuclear shape either spherical single-particle orbitals or
Nilsson type deformed orbitals can be used. Evolution of proton emitting states
starts
with $^{109}$I ($Z=53$, $l=2$) proton radioactivity. In this region just above
$Z=50$
the $\pi d_{5/2}$ and  $\pi g_{7/2}$ orbitals are close to each other,
nevertheless due to the lower $l$ of the former configuration
($\pi d_{5/2}$) most of the emitted protons carry away
two units of angular momentum.
There is though one case discussed in the literature namely $^{121}$Pr,
for which \textcite{Robinson:2005} suggests two possible configurations
$3/2^+$[422] from the spherical $\pi g_{7/2}$ orbital and $3/2^-$[541] from
the spherical $\pi h_{11/2}$ orbital. While in case of a high deformation
the $3/2^+$[422] state can contain admixtures of the spherical $\pi d_{3/2}$
and $\pi d_{5/2}$ orbitals,
making the $l=2$ proton transition possible, the negative parity $3/2^-$[541]
state
cannot contain an $l=2$ contribution. In the
\textcite{Delion:2006a} systematics $l=2$ is assigned to the proton emission
from $^{121}$Pr
making it "compatible" with other proton emitters in this region.

Although for proton emitters above $Z=64$ three orbitals ($d_{3/2}$, $h_{11/2}$
and $s_{1/2}$) should be considered as proton emitters, an interesting
phenomenon
occurs for $^{135}$Tb ($Z=65$) and for $^{141}$Ho ($Z=67$) due to the high
deformation.
The $7/2^-$ ground states of both  emitters are dominated by the [523] component
from the $h_{11/2}$ orbital, but the proton emission is driven by a small
admixture of the
$f_{7/2}$ orbital to the wave function. The small $l=3$ component (<2\% in case
of $^{141}$Ho) wins over the larger $l=5$ component (>78\% in case of
$^{141}$Ho)
\cite{Karny:2008}.
The presence of the $1\pi f_{7/2}$ component in the wave function of nuclei in
this region  is also confirmed by the analysis of $^{145}$Tm fine structure
data.
The 9.6\% proton branching to the first excited $2^+$ level in $^{144}$Er can
only
be explained by the presence of the $l=3$ component in the wave function
\cite{Karny:2003}.
The small energy difference between $h_{11/2}$ and $d_{3/2}$ related levels
manifests itself through the presence of $l=2$ proton emitting low lying isomers
in $^{147m}$Tm and $^{151m}$Lu.

Odd-mass proton emitters above $Z=72$ are characterized by the presence of
two proton emitting states $1/2^+$ ground state and the $11/2^-$ isomer.
There are two cases $^{159}$Re~\cite{Joss:2006} and $^{155}$Ta~\cite{Page:2007}
where  only one $l=5$ proton
emission has been observed.
In the case of $^{159}$Re, the expected half-life for $l=0$ emission is
below $1 \mu$s, which was beyond the capability of applied experimental
technique.
The $^{155}$Ta has been observed as a second generation
decay after $^{159}$Re ion implantation and its subsequent $\alpha$ decay
to $^{155}$Ta.
Although the expected half-life of $l=0$ proton emission from $^{155}$Ta is
long enough to be observed, the combination of the low production cross
section for the $1/2^+$ state with a small detection efficiency for
the second generation decays, may explain the non-observation of
the $l=0$ proton channel. Further studies have still to confirm that
observed decays in both cases do originate from the $11/2^-$ isomeric states.
The heaviest known proton emitter $^{185}$Bi ($Z=83$) decays with the $l=0$
proton
emission from the $1/2^+[400]$ intruder state pushed by the deformation above
the
$Z=82$ shell.

\begin{table}[t]
\begin{center}
\caption{Proton energies $E_p$ and branching ratios $I_p^{exp}$ measured for
proton emission channels from $^{146}$Tm
together with the calculated values based on the particle-core vibration
coupling model
\cite{Hagino:2001,Tantawy:2006}.
E$_f$ denote excitation energy of the final state in $^{145}$Er.
The spin and parity for states in $^{145}$Er is $1/2^+,\ 3/2^+,\ 11/2^-, and\
13/2^-$ for
the excitation energy of 0, 175, 253, and 484~keV, respectively.
All energies are in keV.} \label{tab:VB1}
\begin{tabular}{cccccc}
\hline\hline
$E_p$ & $I_p^{exp}$ (\%) & Wave function composition & $I_p^{cal}$ & $\Delta l$
& E$_f$\\
\hline
&&  ground state &&&\\
&&   $I^\pi=5^-$, $T_{1/2}=68(5)$ms&&&\\
\hline
 $938(4)$ & 13.8(9) &   2\% $\pi s_{1/2}  \otimes \nu h_{11/2}\otimes 0^+ $
& (15)\footnote{Value based on the experimental intensity ratios not predicted
by the particle-core vibration coupling model}  &0 &  253\\
$1016(4)$ & 18.3(11)&   4\% $\pi f_{7/2}  \otimes \nu s_{1/2} \otimes 2^+$  & 15
  &3 &  175  \\
          &         &  41\% $\pi h_{11/2} \otimes \nu s_{1/2} \otimes 2^+$  &
          0.003&5 & 175\\
$1191(1)$ & 68.1(19)&  53\% $\pi h_{11/2} \otimes \nu s_{1/2} \otimes 0^+$  & 70
  &5 &  0\\
\hline
&&  isomeric state &&\\
&&  $I^\pi=10^+$, $T_{1/2}=198(3)$ms&&\\
\hline
$889(8)$  & 1.0(4)  & 2.5\% $\pi f_{7/2} \otimes \nu  h_{11/2}\otimes 2^+$ & 1.2
 &3  &  484\\
           &        &  41\% $\pi h_{11/2}\otimes \nu  h_{11/2}\otimes 2^+$ &
           0.04 &5  &  484\\
$1120(1)$ & 100(1)  &  55\% $\pi h_{11/2}\otimes \nu  h_{11/2}\otimes 0^+$ &
98.6 &5  &  253\\
           &        & 0.1\% $\pi h_{9/2} \otimes \nu  h_{11/2}\otimes 0^+$ & 0.2
            &5  &  253\\
           &        & 0.4\% $\pi (l>5)   \otimes \nu  h_{11/2}$            &
             &   & \\
\hline \hline
\end{tabular}
\end{center}
\end{table}

It is worth noting that the $s=1$ $^{141}$Ho proton emitter is among the most
extensively
studied  and understood isotopes beyond the proton-drip line.
We know proton emission from both ground and isomeric states in $^{141}$Ho to
the $0^+$ ground state
as well as first excited $2^+$ state in
$^{140}$Dy~\cite{Davids:1998,Rykaczewski:1999,Karny:2008}.
In these decays three different angular momenta ($l=0$, 2, and 3) are involved.
Proton emission was also used in a recoil decay tagging study of this isotope
allowing  observation and interpretation of rotational bands up to
$I^{\pi}=35/2^{-}$ \cite{Seweryniak:2001}.
Theoretical works focused on $^{141}$Ho include
\textcite{Barmore:2000,Davids:2000,Kadmensky:2000,
Esbensen:2001,Fiorin:2003,Davids:2004,Kruppa:2004,Arumugam:2009}

\begin{table*}
\caption{Properties of $s>2$ proton emitters.}
\label{tab:VC1}
\begin{tabular}{|c|c|c|c|c|c|l|}
\hline \hline
Emitter &E$^*$(MeV)& E$_p$(MeV)  & T$_{1/2}$ & Configuration & Ang.mom. &
References \\
\hline \hline
$^{53m}$Co & 3.197(29)&1.57(3)& 242(15)~ms &$[\pi f^{-1}_{7/2} \otimes \nu
f^{-2}_{7/2}]19/2^-$ & l=9 &\textcite{Jackson:1970} \\
           &          &       &            &
                            &     &\textcite{Cerny:1970}   \\
$^{54m}$Ni &6.457(1)& 1.28(5)& 152(4)~ns&$[\pi (f^{-1}_{7/2}p_{3/2}) \otimes \nu
f^{-2}_{7/2}]10^+$ & l=5 &\textcite{Rudolph:2008} \\
$^{94m}$Ag &5.780(30)& 0.79(3)\footnote{A second transition of 1.01~MeV reported
by \textcite{Mukha:2005b} has not been confirmed by the work of
\textcite{Cerny:2009} and therefore is not included in the table.} & 0.39(4)~s &
$[\pi g_{9/2}^{-3} \otimes \nu g_{9/2}^{-3}] 21^+$ &l=4&
\textcite{Mukha:2005b} \\
           &          &       &           &        &  & \textcite{Cerny:2009} \\
\hline
\end{tabular}
\end{table*}

\subsubsection{Even-mass, $s=2$ proton emitters}
\label{sec:VB2}

There are 17 even-mass $s=2$ proton emitters known. The lightest known
is $^{112}_{55}$Cs$_{57}$ and $^{176}_{81}$Tl$_{95}$ is the heaviest.
In these odd-odd nuclides the wave function compositions results from the
interaction
of the unpaired proton and neutron. For nuclei with the neutron number $N < 82$
the valence neutron has no significant influence on the proton emission, i.e.\
proton emission from the odd-odd emitters follows the pattern of the odd-even
emitters of the same element. For example, $^{150}$Lu has two states decaying
via proton
emission:
the ground state which emits $l=5$ proton and an isomeric state emitting proton
with $l=2$.
The same pattern is found in $^{151}$Lu, where the ground- and isomeric states
decay
with $l=5$ and $l=2$ proton emission, respectively.
The situation changes when neutrons start filling $\nu f_{7/2}$
orbital above $N=82$. In these cases an attractive interaction due to the tensor
force
between $\pi d_{3/2}$ and $\nu f_{7/2}$ orbitals pushes the former above the
$\pi s_{1/2}$ orbital leading to $l=2$ proton emission from the ground state.
In other words, the $s_{1/2}$ ground state in odd-mass emitters is replaced by
the $d_{3/2}$ ground state in odd-odd emitters of the same element.
The $l=2$ emission from the even-mass isotopes has been observed for
$^{156}_{73}$Ta$_{83}$, $^{160}_{75}$Re$_{85}$, $^{166}_{77}$Ir$_{89}$,
$^{170}_{79}$Au$_{91}$.
The $^{176}_{81}$Tl$_{95}$ emits an $l=0$ proton from its ground state
just like its neighbor $^{177}$Tl.

The $^{146}$Tm, $s=2$, proton emitter is the richest proton emitter known.
There are 5 proton transition known in this case~\cite{Tantawy:2006}. Three
transitions
are coming from the ground state and two were assigned to its isomeric state.
Table~\ref{tab:VB1} shows experimental results obtained for these
transitions together with the calculated wave functions components,
based on the work of \textcite{Tantawy:2006} and \textcite{Hagino:2001}.

The ground state emits protons with $l=0$, $3$ and $5$, while the
emission from the isomeric state has mainly the $l=5$ component.
It is worth noting that in the cited calculation $l=0$ emission
is due to the $\pi s_{1/2}  \otimes \nu h_{11/2}\otimes 0^+$  $2\%$ component,
which is
the isospin symmetric to the dominant ($53\%$) $\pi h_{11/2} \otimes \nu s_{1/2}
\otimes 0^+$.

\subsection{Seniority $s>2$ proton emitters}
\label{sec:VC}

In this category there are 3 proton emitters to be mentioned:
$^{53m}$Co~\cite{Jackson:1970},
$^{54m}$Ni~\cite{Rudolph:2008} and $^{94m}$Ag~\cite{Mukha:2005b}.
They are all high spin, high excitation isomers with a multi-particle
configuration
of the wave function. The $^{53m}$Co was the first proton emitter discovered.
T$_{1/2}=242(15)$~ms.
Its wave function is best described by the $[\pi f^{-1}_{7/2} \otimes \nu
f^{-2}_{7/2}]19/2^-$ configuration. The transition goes to the $0^+$ ground
state of $^{52}$Fe, thus the proton
carries 9 units of angular momentum \cite{Jackson:1970}.
The $^{54m}$Ni is the first and so far the only
proton emitter produced in fragmentation reaction.
The angular momentum of the emitted proton
equals 5, although the $\pi h_{11/2}$ orbital is not present in the proposed
configuration of the
emitting state. Both cases can not be described by the core plus proton
model used in case of $s \le 2$ proton emitters. In the case of $^{94m}$Ag $l=4$
emission is assumed to originate from $\pi g_{9/2}$ orbital.
The $21^+$ isomeric state is created by  three proton holes on the $g_{9/2}$
orbital coupled to three neutron holes on the $g_{9/2}$ orbital.
Table~\ref{tab:VC1} shows the combined information on
these high spin proton emitters.

\subsection{Outlook}

The wealth of nuclear data established by proton radioactivity studies is
impressive and indicates that this field of research is mature and the
applied experimental techniques are well advanced. They appear, however, to be
still not sufficient to address potential proton emitters with atomic numbers
below 50.
The low production rates and short half-lives, expected for these nuclei,
present
a challenge to the experimentalists. Observation of proton radioactivity in
nuclei with $Z<50$ and establishing their properties will be important
for calculations of the astrophysical \emph{rp}-process. Of special
interest are nuclei around the waiting points, like $^{68}$Se (see a
recent article of \textcite{Rogers:2011}), and the region just
below  $^{100}$Sn, at the expected end of the \emph{rp}-process path.

The prospects for experimental studies of proton emission in the region
between $N=82$ and $Z=82$ were discussed recently by \textcite{Page:2011}.


\newpage
\section{ALPHA DECAYS}
\label{sec:VI}


Emission of $\alpha$ particles belongs to the oldest known (together with
$\beta$-decay) types of radioactivity. Its first theoretical description
by \textcite{Gamow:1928} and independently by \textcite{Gurney:1928} was
one of the early triumphs of quantum mechanics applied for the first time
to the atomic nucleus. In particular the empirical law of \textcite{Geiger:1912}
could be successfully explained. Presently, the
calculations of the $\alpha$-decay lifetimes are performed in analogy to
proton radioactivity by using Eq.~\ref{eq:VA2_Gamma} \cite{Gurvitz:1987}
where the proton spectroscopic factor $S_p$ is replaced by the $\alpha$
preformation factor $S_{\alpha}$. The latter measures the probability
that the $\alpha$ particle is formed inside the mother nucleus.
Combining the shell model with the cluster model proved to be successful
in calculating the absolute $\alpha$ decay width of $^{212}$Po
\cite{Varga:1992}.
The result, $\Gamma=1.45\cdot10^{-15}$MeV, agrees very well with the
experimental
value of $1.5\cdot10^{-15}$MeV.
The large body of experimental and theoretical findings about $\alpha$
decay mode is covered extensively in a number of books and
reviews \cite{Rasmussen:1966,Roeckl:1996,Delion:2010}.
For a compilation of even-even $\alpha$-decay data
see \textcite{Akovali:1998}. An extended version of the Geiger-Nutall rule has
been recently proposed by \textcite{Qi:2009,Qi:2009a}. Since this paper is
devoted to
the decays at the limits of stability, here we focus mainly on the latest
studies of $\alpha$ decay close to the proton drip-line.

One of the regions which attracts attention is located above $^{100}$Sn, where
due to the proximity of $N=50$ and $Z=50$ shell closures, the energy available
for $\alpha$ decay
is large enough to overcome the Coulomb barrier. This results in an island of
$\alpha$ radioactivity for $52 \le Z \le 56$ and
the neutron number $N$ up to 60.
Apart from the energy factor one has to note that the nuclei in this region
are among the heaviest with
protons and neutrons occupying the same type of single-particle orbitals.
 For these nuclei the active single-particle
orbitals are $g_{7/2}$ and $d_{5/2}$ which differ in excitation energy by only a
few hundred keV. In the case of protons and neutrons occupying the same
orbitals,
their spatial overlap is maximized leading to the maximal preformation factor.
For this reason alpha decays $^{104}$Te $\rightarrow$ $^{100}$Sn
and $^{106}$Te $\rightarrow$ $^{102}$Sn are expected to be the best
examples of the superallowed alpha decay \cite{Macfarlane:1965,Roeckl:1995}.

\begin{figure}
\centerline{
\includegraphics[width=0.47\textwidth]{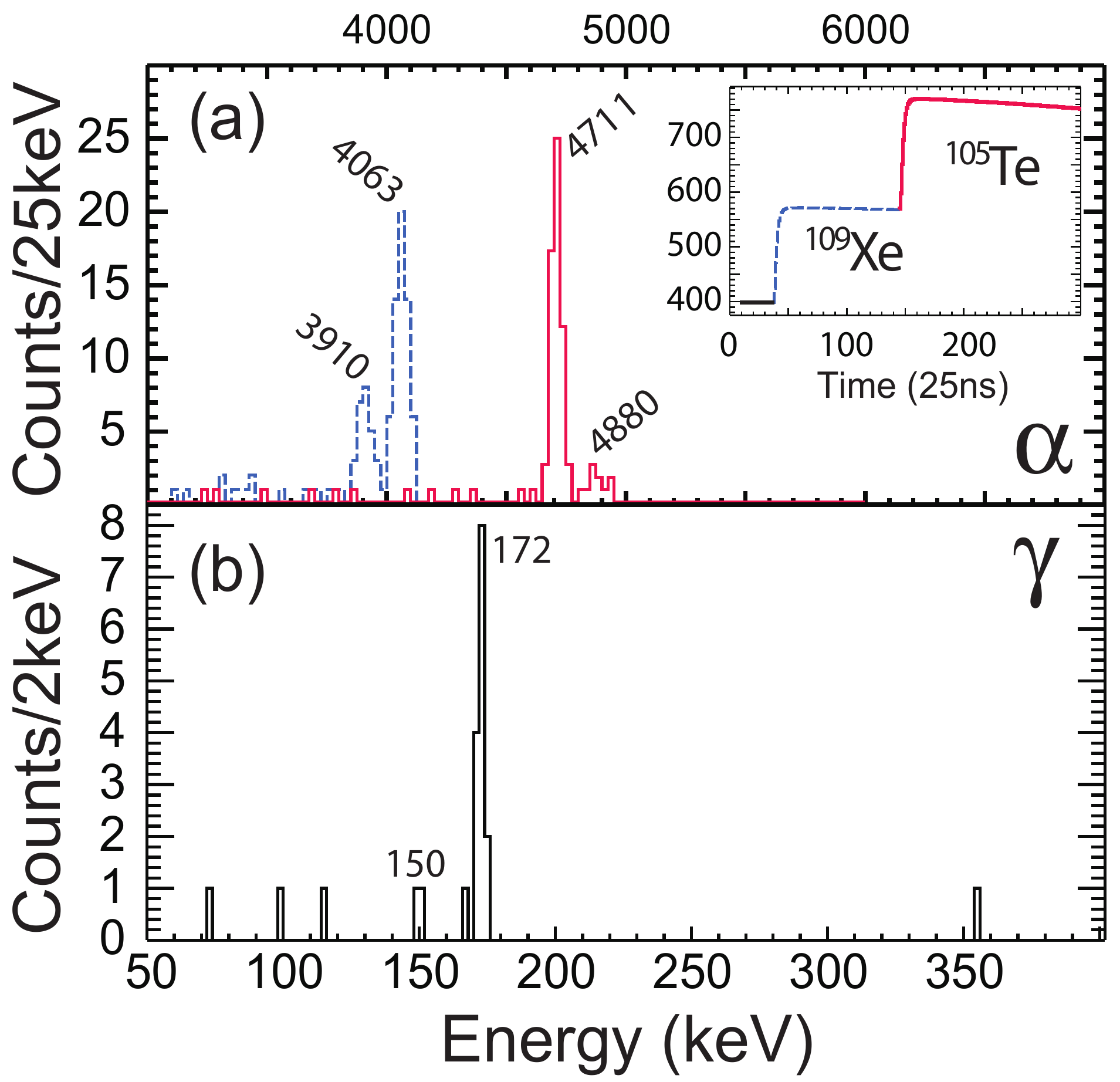}
}
\caption{(Color online) (a) Energy spectrum of the first ($^{109}$Xe) and the
second ($^{105}$Te)
$\alpha$ pulses obtained from the $\alpha-\alpha$ pileup traces (inset).
The lines at 3910(10) and 4063(4) keV are assigned to the $^{109}$Xe$\rightarrow
^{105}$Te
transitions, while the lines at 4711(3) and 4880(20) keV are assigned to the
$^{105}$Xe$\rightarrow ^{101}$Sn decay.
(b) $\gamma$ spectrum in coincidence with the analyzed $\alpha-\alpha$ traces.
From \cite{Darby:2010}.}
\label{fig:VI_Darby}
\end{figure}

While the search for the superallowed $^{104}$Te $\rightarrow$ $^{100}$Sn decay
is
still an ongoing effort, the successful
measurements of $^{105}$Te $\rightarrow$ $^{101}$Sn have been reported \cite
{Seweryniak:2006,Liddick:2006}.
In the work of \textcite{Seweryniak:2006} the decay of $^{105}$Te was measured
directly.
Ions of $^{105}$Te were produced in the fusion-evaporation reaction of
a $^{58}$Ni beam impinging
on a $^{50}$Cr target. The products were separated by means of the Fragment Mass
Analyzer of Argonne National Laboratory \cite{Davids:1992} and implanted into a
DSSSD detector.
Alpha decays events were measured and correlated with the implanted $^{105}$Te
ions.
Thirteen counts were identified as representing the decay $^{105}$Te
$\rightarrow$ $^{101}$Sn.
As a result, an $\alpha$ decay energy $E_{\alpha}=4720(50)$~keV, corresponding
to
Q$_{\alpha}=4900(50)$, and a half-life of $T_{1/2}=0.7_{-0.17}^{+0.25}$ $\mu$s
were established.
In the different experiment of \textcite{Liddick:2006} the $^{58}$Ni beam
impinging on
a $^{54}$Fe target was used. With the Recoil Mass Separator (RMS) of the HRIBF
facility in Oak Ridge, the $^{109}$Xe $\rightarrow$ $^{105}$Te $\rightarrow$
$^{101}$Sn
decay chain was analyzed. The relatively long half-life of $^{109}$Xe ($\sim
13$~ms) helped
to overcome the inevitable losses due to the finite flight time through the RMS
($\sim 2 \, \mu$s)
in case $^{105}$Te was studied directly.
Pulses of correlated $\alpha -\alpha $ decays were analyzed with help of digital
electronics programmed to trigger and collect only the signals of high energy
decay events.
Analysis revealed two branches of $^{109}$Xe $\alpha$ decay with energies
$E_{\alpha 1} = 3918(9)$~keV and $E_{\alpha 2}=4062(7)$~keV, followed by
the $\alpha$ decay of $^{105}$Te with the energy $E_{\alpha}=4703(5)$~keV.
The half-lives of $13(2)$~ms and $0.62(7)\mu$s were determined for $^{109}$Xe
and $^{105}$Te,
respectively. The preformation factors derived from these experiments were found
to
be larger by about a factor of three from the values in the well studied region
of
the doubly magic $^{208}$Pb \cite{Liddick:2006, Mohr:2007}.

In the next experiment at the HRIBF laboratory, using a similar technique, the
fine structure in the $\alpha$ decay of $^{105}$Te was found \cite{Darby:2010}.
The setup used by \textcite{Liddick:2006} was additionally equipped with four
germanium clover detectors placed around the DSSSD implantation detector.
Double pulses of $\alpha$-$\alpha$ events from the
$^{109}$Xe $\rightarrow$ $^{105}$Te $\rightarrow$ $^{101}$Sn decay chain were
stored and analyzed. The result is shown in Fig.~\ref{fig:VI_Darby}.
The observed $\alpha$ lines at 4711~keV and 4880~keV are assigned to the
decays of $^{105}$Te leading to the first excited and to the ground state of
$^{101}$Sn,
respectively. The $\gamma$ transition between these two states (172~keV),
coincident with double $\alpha$ pulses was also detected
(Fig.~\ref{fig:VI_Darby}(b)).
This result confirmed the previous evidence for the first excited state
in $^{101}$Sn obtained by \textcite{Seweryniak:2007} who employed the
recoil-decay tagging method (RDT) \cite{Paul:1995} by combining $\gamma$
spectroscopy with $\beta$-delayed proton detection.

The stronger $\alpha$ line in the decay of $^{105}$Te, at 4711~keV
with the intensity of 89(4)\%, is interpreted as corresponding to the decay
with no change of the orbital angular momentum ($\Delta l = 0$),
while the line at 4880~keV is assigned to the $\Delta l = 2$ channel
\cite{Darby:2010}.
In addition, the strong $\Delta l = 0$ transition goes to the first excited
state in $^{101}$Sn in contrast to the $\alpha$ decay of $^{107}$Te where
the strong $\Delta l = 0$ decay connects nuclear ground states.
Thus, the level inversion occurs between $^{101}$Sn and $^{103}$Sn.
The $5/2^+$ ground state in $^{103}$Sn becomes the first excited state
in $^{101}$Sn, while the $7/2^+$ excited state
in $^{103}$Sn becomes the ground state of $^{101}$Sn.
This  phenomenon is interpreted as a result of the interplay between the
pairing on the $\nu g_{7/2}$ orbital ($V^{pair}(g_{7/2})=1.4$~MeV) being much
stronger than for the $\nu d_{5/2}$ orbital  ($V^{pair}(d_{5/2})=0.56$~MeV),
and a small (0.17~MeV) energy difference between these two orbitals
\cite{Darby:2010}.
This interpretation contradicts the conclusions
of \textcite{Seweryniak:2007} who assigned spin and parity $5/2^+$ to
the ground state of $^{101}$Sn.
We note that both experiments agree on their common experimental finding
but differ in theoretical interpretation. Further experiments are
required to firmly establish the $d_{5/2}-g_{7/2}$ order in $^{101}$Sn. For
example,
an observation (or exclusion) of the Gamow-Teller $\beta$-decay between the
ground state
of $^{101}$Sn and the $9/2^+$ ground state of $^{101}$In should settle the
controversy.


In the region of very neutron-deficient lead isotopes, recent $\alpha$ decay
studies
provided information on the shape coexistence in $^{186}$Pb
\cite{Andreyev:2000}.
The states of $^{186}$Pb were populated in the $\alpha$ decay
of $^{190}$Po, produced in the fusion-evaporation reaction of $^{52}$Cr beam
impinging on a $^{142}$NdF$_{3}$ target. The products were separated by means of
the SHIP velocity filter \cite{Muenzenberg:1979} at GSI Darmstadt and implanted
into a position
sensitive silicon detector, backed by a germanium clover detector for X-ray
measurements.
A set of silicon detectors was mounted for measurement of conversion electrons.
In addition to the ground-state-to-ground-state decay, two other channels were
observed in coincidence with conversion electrons.
Due to the similar half-life all three alpha transitions were assigned to the
decay of the $^{190}$Po ground state. Analysis of coincidences between $\alpha$
particles,
electrons, X- and  $\gamma$-rays suggested that the spin of the three final
states is $0^+$.
The analysis of the preformation factor lead to the
conclusion that the presence of the three $0^+$ states in $^{186}$Pb 
similar energy,
is a manifestation of shape coexistence where the ground state is spherical, the
first excited
$0^+$ state at 532~keV is oblate and second excited $0^+$ state at 650~keV is
prolate.

It is worth noting that in recent years $\alpha$ decay served as a tagging
signal
in recoil decay tagging studies of heavy nuclei providing valuable nuclear
structure
information. Recent highlights from the RITU spectrometer at University of
Jyv\"askyl\"a
were presented by \textcite{Julin:2010}. The RDT experiments with the
FMA separator  coupled to germanium detector arrays  were reported by
\textcite{Carpenter:1997}, \textcite{Reiter:1999}, and
by \textcite{Seweryniak:1998,Seweryniak:2005a}. Finally, it should be mentioned
that
most of the discoveries of new elements rely on $\alpha$ decay
\cite{Hofmann:2009a}.


\newpage
\section{TWO-PROTON RADIOACTIVITY}
\label{sec:VII}


\subsection{Introduction}
\label{sec:VIIA}

The two-proton radioactivity ($2p$) is the most recently observed type of decay
and thus the least known. The experimental studies are still in the early stage.
The detailed understanding of its mechanism requires novel theoretical
approaches which in certain aspects are still under development. A very early
look can be found in the book by \textcite{Baz:1972}. The current experimental
and theoretical status of the $2p$ decay were summarized recently by
\textcite{Blank:2008a} and by \textcite{Grigorenko:2009b} with focus on specific
theoretical methods. Because of its exceptional status, we discuss here this
decay mode in more detail, emphasizing the major qualitative features of the
phenomenon. The illustrations are provided mainly by the examples of $^{6}$Be,
$^{19}$Mg, and $^{45}$Fe. These nuclei belong to $p$, $s$-$d$, and $p$-$f$
shells respectively and their lifetimes span about 18 orders of the magnitude,
providing support for universality of the currently achieved understanding of
the two-proton decay.

\begin{figure*}[tbh]
\centerline{
\includegraphics[width=0.3\textwidth]{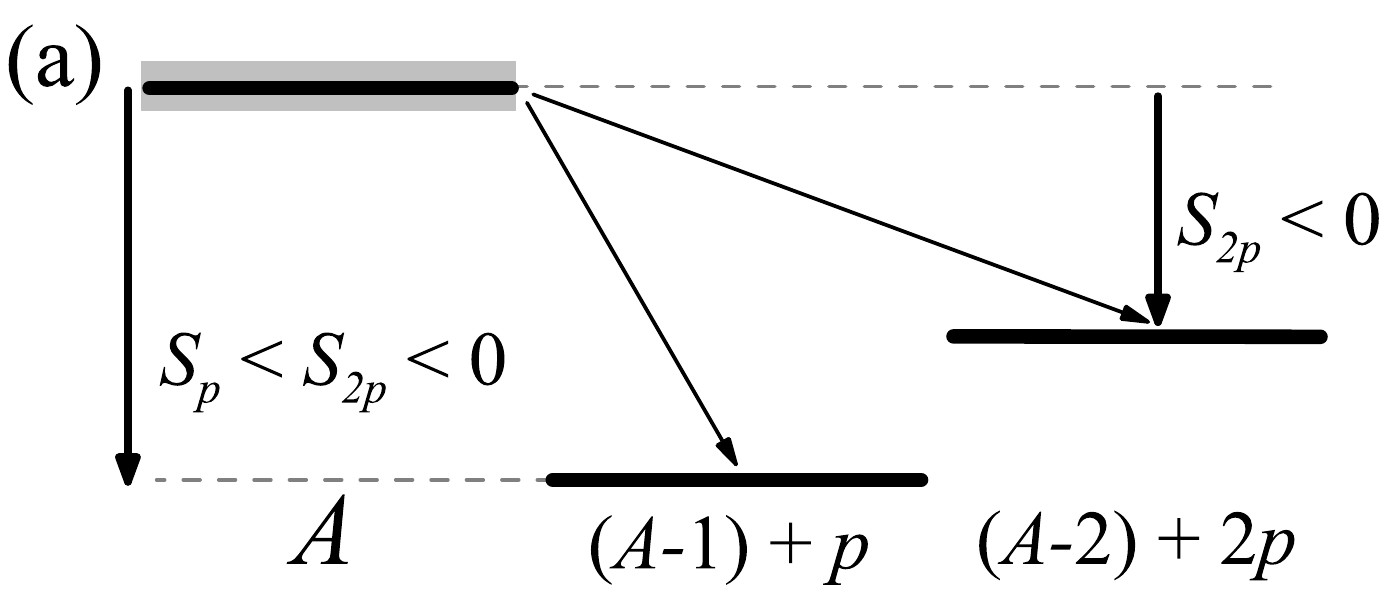}
\includegraphics[width=0.3\textwidth]{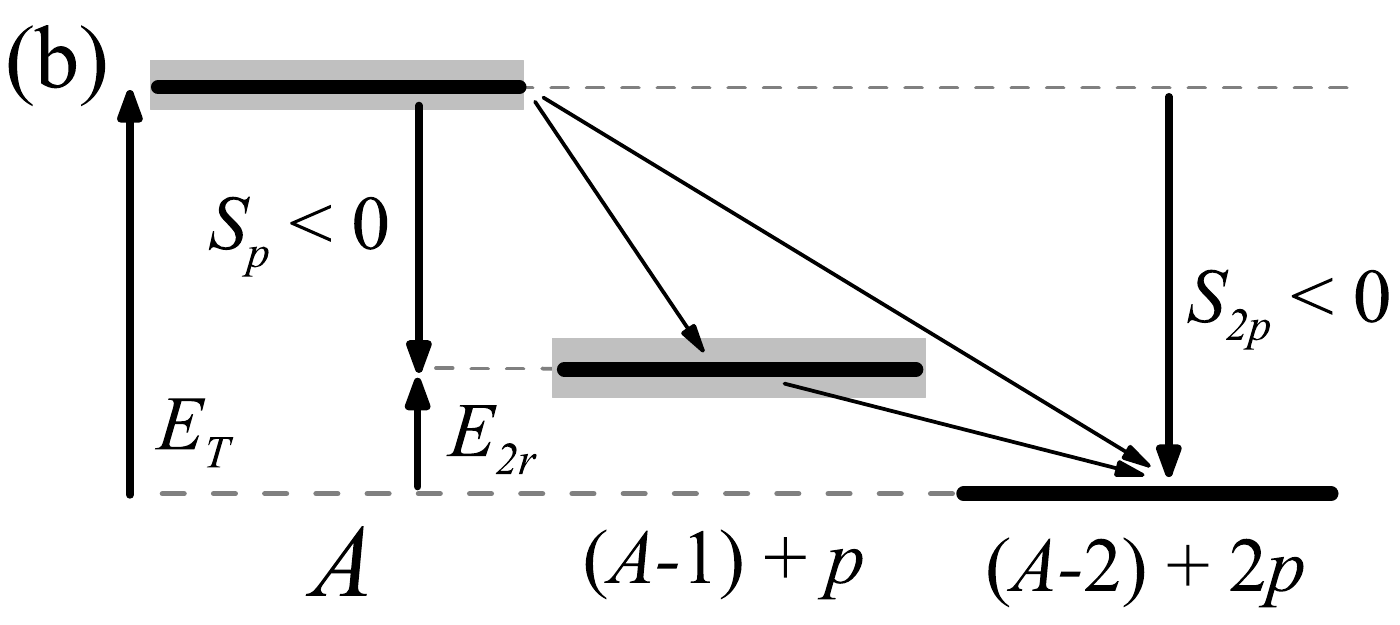}
\includegraphics[width=0.3\textwidth]{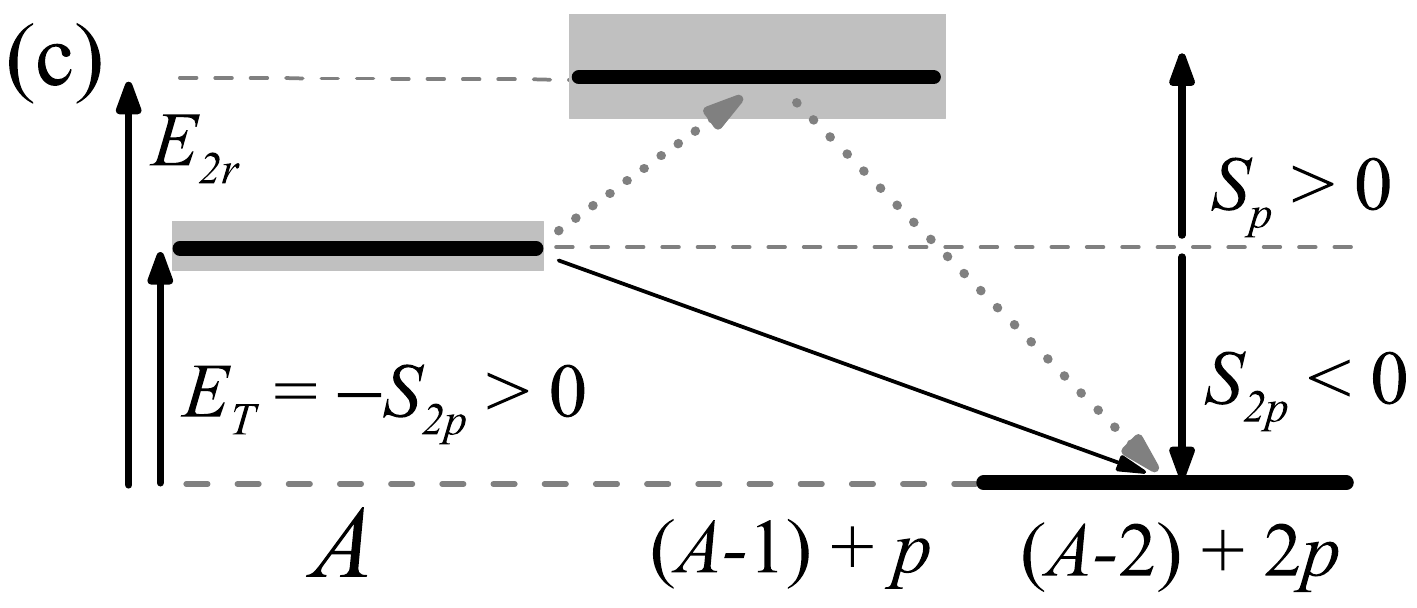}
}
\centerline{
\includegraphics[width=0.3\textwidth]{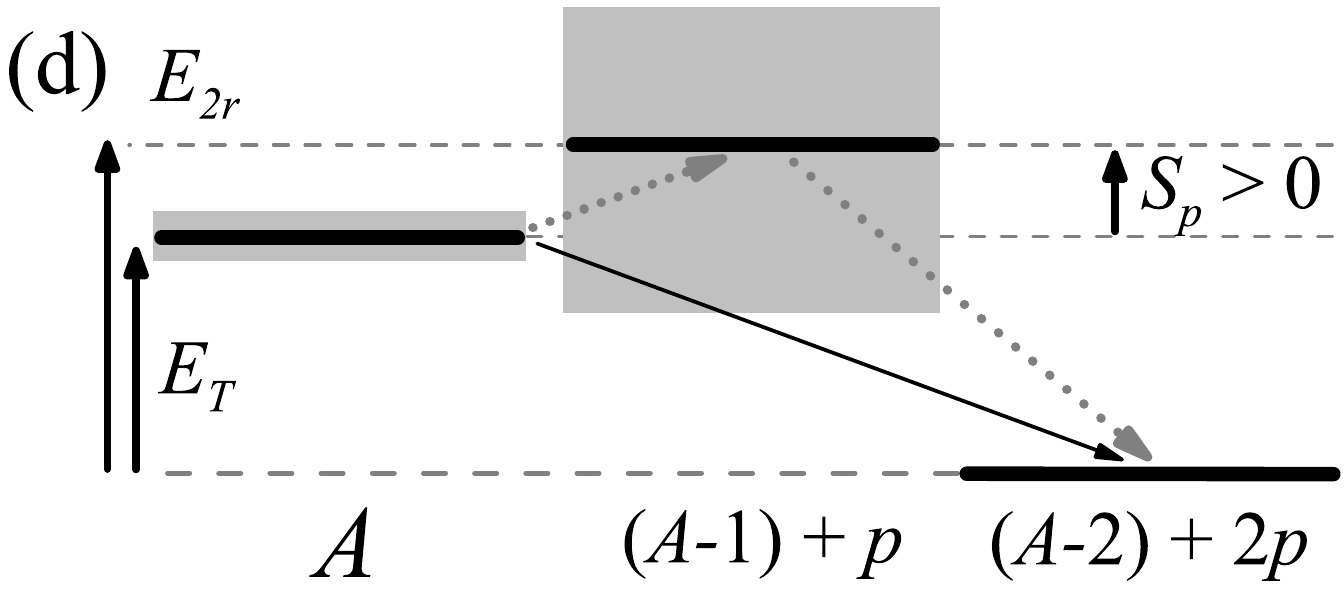}
\includegraphics[width=0.3\textwidth]{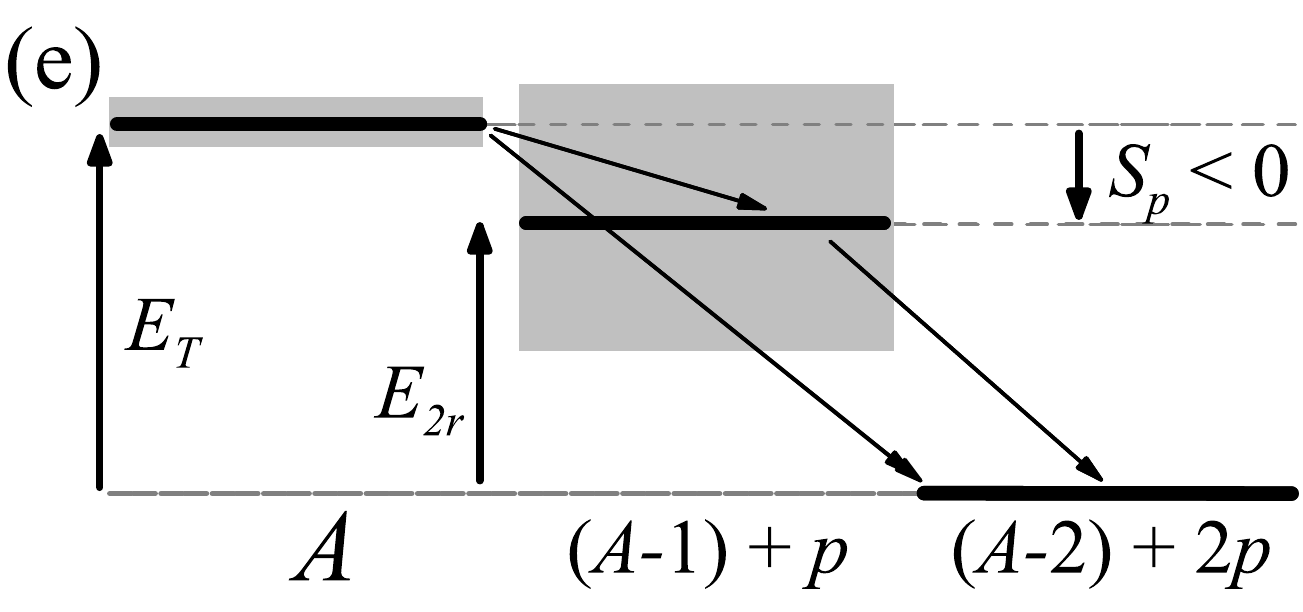}
}
\caption[Energy conditions for different modes of the two-proton
emission.]{Energy conditions for different modes of the
two-proton emission: (a) typical situation for decays of excited states (both
$1p$ and $2p$ decays are possible), (b) sequential decay via narrow intermediate
resonance, (c) true $2p$ decay. The cases (d) and (e) represent ``democratic''
decays. The gray dotted arrows in (c) and (d) indicate the ``decay path''
through the states available only as virtual excitations.}
\label{fig:VIIA-scheme}
\end{figure*}

The emission of two protons from a nuclear state is in principle possible in
various decay scheme situations which are sketched in
Fig.~\ref{fig:VIIA-scheme}. We introduce here the following notation: $E_T$
is the system energy relative to the nearest three-body breakup threshold, while
$E_{2r}$ is the lowest two-body resonance energy relative to this threshold.
The $2p$ decay in the pure form, which we will call the \emph{true 2p decay} (or
true three-body decay) is represented in Fig.~\ref{fig:VIIA-scheme}(c). In this
case sequential emission of protons is energetically prohibited and all
final-state fragments are emitted simultaneously. Such a situation is common
among even-$Z$ nuclei at the proton-drip-line and results from pairing
interactions, see Sec.~\ref{sec:II}.
The decay dynamics of true $2p$ decay is not reducible to the conventional
two-body
dynamics and should be addressed by the methods of few-body physics.

A somewhat special situation, represented in Fig.~\ref{fig:VIIA-scheme}(d),
occurs when the ground state of the subsystem is so broad that the emission
of the first proton becomes energetically possible (although $E_{2r}>E_T$) which
opens a way for a sequential transition. Similarly, the decay may formally
proceed in a
sequential manner ($E_{2r}<E_T$) but the ground state of the subsystem is so
broad that no
strong correlation between outgoing fragments at given resonance energy can be
formed, see Fig.~\ref{fig:VIIA-scheme}(e). We refer to such scenarios as
\emph{democratic} decays and discuss them briefly in Sec.~\ref{sec:VIID1}.

The three-body character of the $2p$ radioactivity places it in the broader
context of nuclear processes exhibiting essential many-body features. This
includes studies of the broad states in continuum and excitation modes, like the
soft dipole mode \cite{Aumann:2005}. Another topic, pursued actively in the last
decades, is the phenomenon of two-neutron halo \cite{Zhukov:1993} with its
\emph{Borromean property} that none of the three two-body subsystems is bound.
The $2p$ decay can be seen as an analogue of the two-neutron halo, requiring
similar ingredients in the proper many-body description of its properties. The
illustration of this point is provided by the isobaric mirror partners $^{6}$He
and
$^{6}$Be: the first is the ``classical'' Borromean halo nucleus and the second
is the lightest true $2p$ emitter. The crucial difference, however, comes from
the fact that the $2p$ decays involve charged particles in the continuum which
significantly complicates the theoretical description. Another example:
$^{17}$Ne is a Borromean two-proton halo nucleus, while the first excited state
of
$^{17}$Ne and the less bound $^{16}$Ne are true $2p$ emitters.

All ground-state two-proton emitters studied experimentally up to now
are collected in Table~\ref{tab:VIIC3-2p}.

\subsubsection{Two-proton correlations}
\label{sec:VIIB}

The two-body decay of a resonance is characterized only by the energy and the
width of the state. The three-body decay is much more ``rich'' as complex
information about momentum correlations becomes available.

For decays with three particles in the final state there are nine degrees of
freedom (spins are not counted). Three of them describe the center-of-mass
motion and three describe the Euler rotation of the decay plane. Therefore,
for a fixed decay energy $E_T$ there are two parameters representing the
complete correlation picture. It is convenient to choose the energy distribution
parameter $\varepsilon$ and an angle $\theta_k$ between the Jacobi momenta
$\mathbf{k}_{x}$ and $\mathbf{k}_{y}$:
\begin{eqnarray}
E_T =E_x+E_y=\frac{(A_1+A_2)k^2_x}{2MA_1 A_2} +
\frac{(A_1+A_2+A_3)k^2_y}{2M(A_1+A_2) A_3} \, , \nonumber  \\
{\bf k}_x  =  \frac{A_2 {\bf k}_1-A_1 {\bf k}_2 }{A_1+A_2} \, ,  \,\;
{\bf k}_y  =  \frac{A_3 ({\bf k}_1+{\bf k}_2)-(A_1+A_2) {\bf k}_3}
{A_1+A_2+A_3},\nonumber \\
\varepsilon = E_x/E_T \quad ,\quad \cos(\theta_k)=(\mathbf{k}_{x} \cdot
\mathbf{k}_{y}) /(k_x\,k_y) \, , \quad \quad \quad
\label{eq:VIIB-var}
\end{eqnarray}
where $M$ is ``scaling'' nucleon mass, $M_{\text{nucleus}}=M(A_1+A_2+A_3)$. The
Jacobi momenta for two-proton emitters (protons are indistinguishable) can be
defined in two ``irreducible'' Jacobi systems, called ``T'' and ``Y'', see Fig.\
\ref{fig:VIIB-coord-syst}. In the ``T'' Jacobi system, the core is the particle
$A_3$ and the parameter $\varepsilon$ describes the energy distribution between
the
two protons. In the ``Y'' Jacobi system, the core is the particle $A_2$ and
$\varepsilon$ corresponds to the core and proton subsystem. The Jacobi momentum
${\bf k}_x$ is the momentum of particle 1 in the c.m.\ of particles 1 and 2,
${\bf k}_y$ is the c.m.\ momentum of particles 1 and 2 in the c.m.\ of the whole
system (particles 1, 2, and 3). Distributions constructed in the different
Jacobi systems are just different representations of the same physical picture.

A more general (5-dimensional) correlation pattern becomes available for systems
with total spin $J>1/2$. Manifestation of such correlations requires existence
of a selected direction in space and spin alignment, which is naturally achieved
for short-lived states populated in nuclear reactions. The only
example of such detailed studies is so far the \emph{two-neutron} decay of broad
states in $^{5}$H \cite{Golovkov:2005}

\subsubsection{Historical note}
\label{sec:VIIC}

The possibility of a true two-proton emission was mentioned for the first time
by \textcite{Zeldovich:1960}. This work comprises the dripline prediction for
light systems. After predictions about existence of $^{13}$O and $^{20}$Mg
isotopes Zeldovich notes: ``The existence of $^{12}$O, $^{16}$Ne, and $^{19}$Mg
is not excluded [...] These nuclei could appear to be unstable with respect to
emission of two protons simultaneously.'' The explicit and detailed statement of
the two-proton radioactivity phenomenon was given by V.I.\ Goldansky a bit later
\cite{Goldansky:1960}\footnote{Zeldovich and Goldansky lived next door to each
other. The problem is known to be a subject of many of their informal
discussions (which are acknowledged in the paper of Goldansky). Later, on
occasion of the priority discussion raised by some people, Zeldovich rejected
any credits for the idea. Zeldovich was famous for providing in his
works insights important for later development of physics in a very compact form
and without attempt of further elaboration.}. While the proton and cluster
radioactivity are quite straightforward generalizations of
$\alpha$-radioactivity, the few-body decays are qualitatively different and
required ingenuity to foresee. The pioneering work of Goldansky contained
several important insights which we illustrate by the following citation
\cite{Goldansky:1960}:

``Thus the simplest approach to the theory of two-proton decay would consist in
using the product of two usual barrier factors, that is, in introducing an
exponential factor of the type
\begin{equation}
w(\varepsilon)=\exp \left\{\frac{-2\pi(Z-2) \alpha \sqrt{M}}{\sqrt{E_{T}}}
\left[\frac{1}{\sqrt{\varepsilon}}+\frac{1}{\sqrt{1-\varepsilon}} \right]
\right\},
\label{eq:VIIB-w-goldan}
\end{equation}
where $E_{T}$ is the sum of the energies of the two protons (energy of emitted
diproton), $\varepsilon$ and $(1-\varepsilon)$ are the fractions of energy
referring to each of the
protons.

It can easily  be seen that the total barrier factor $w(x)$ is maximum for
$\varepsilon=0.5$, i.e., when the proton energies are equal. It will be noted
that the value in the exponent is just the same as for the sub-barrier emission
of a diproton with the energy $E_{T}$ as a whole.''\footnote{In Eq.\
(\ref{eq:VIIB-w-goldan}) we have modified the notation of Goldansky to make it
consistent with the notation of this work.}

The general character of the energy distribution predicted by Eq.\
(\ref{eq:VIIB-w-goldan}) has proven to be correct and is now confirmed also
experimentally. The idea of emission of a ``diproton particle'' turned to be an
attractive concept but finally appeared to be misleading.

\begin{table}[t]
\caption{Ground-state $2p$  emitters investigated experimentally. The indicated
half-life
         corresponds to the partial value for the $2p$ decay. }
\begin{ruledtabular}
\begin{tabular}[c]{ccccc}
$^NZ$  & $E$ keV & $\Gamma$ or $T_{1/2}$ & Reference \\
\hline
$^{6}$Be   & 1371(5)   & 92(6) keV & \cite{Whaling:1966} \\
$^{12}$O   & 1820(120) & 400(250)\footnotemark[1] keV & \cite{KeKelis:1978}  \\
           & 1790(40)  & 580(200)\footnotemark[1] keV & \cite{Kryger:1995}   \\
           & 1800(400) & 600(500)\footnotemark[1] keV & \cite{Suzuki:2009}   \\
$^{16}$Ne  & 1350(80)  & 200(100)\footnotemark[1] keV & \cite{KeKelis:1978} \\
           & 1400(20)  &  110(40)\footnotemark[1] keV & \cite{Woodward:1983} \\
           & 1350(80)  &  <200 keV                    & \cite{Mukha:2008} \\
$^{19}$Mg  & 750(50)   & 4.0(15) ps              & \cite{Mukha:2007} \\
$^{45}$Fe  & 1100(100) & $3.2^{+2.6}_{-1.0}$ ms  & \cite{Pfutzner:2002} \\
           & 1140(50)  & $5.7^{+2.7}_{-1.4}$ ms  & \cite{Giovinazzo:2002} \\
           & 1154(16)  & $2.8^{+1.0}_{-0.7}$ ms  & \cite{Dossat:2005} \\
           &           & $3.7^{+0.4}_{-0.4}$ ms  & \cite{Miernik:2007b} \\
$^{48}$Ni  & 1350(20)  & $8.4^{+12.8}_{-7.0}$ ms \footnotemark[2]  &
\cite{Dossat:2005} \\
           &           & $3.0^{+2.2}_{-1.2}$ ms  & \cite{Pomorski:2011b} \\
$^{54}$Zn  & 1480(20)  & $3.2^{+1.8}_{-7.8}$ ms  & \cite{Blank:2005}
\end{tabular}
\end{ruledtabular}
\label{tab:VIIC3-2p}
\footnotetext[1]{According to theoretical calculations much smaller widths are
expected \cite{Barker:1999,Barker:2001,Grigorenko:2002}.}
\footnotetext[2]{Only one decay event observed.}
\end{table}

Later, significant theoretical work was devoted to identifying the best
candidates
for the observation of the $2p$ radioactivity. Due to the extreme sensitivity of
the
$2p$ decay probability to the width of the Coulomb barrier, the decay energy of
a candidate must fall into a rather narrow window \cite{Nazarewicz:1996}. The
resulting $2p$ partial half-life should be long enough for an efficient
separation in the spectrometer (typically a fraction of a microsecond) but short
enough to compete with the $\beta^+$ decay channel ($\sim 10$ ms). Thus accurate
mass predictions for nuclei beyond the drip line were necessary. One of the most
exact methods was the application of the isobaric multiplet mass equation (IMME)
\cite{Benenson:1979} combined with the experimentally measured mass of the
neutron-rich member of the multiplet. Coefficients of the IMME can be calculated
within the shell model or deduced from the Coulomb-energy systematics. Both
approaches were undertaken \cite{Brown:1991,Ormand:1996,Cole:1996} and the
choice of the best candidates was narrowed down to three cases: $^{45}$Fe,
$^{48}$Ni, and $^{54}$Zn. These predictions played an essential role in
motivating
experimental search for the $2p$ radioactivity.

\subsection{Experimental results}
\label{sec:VIID}

While the discovery of the true $2p$ radioactivity awaited more than 40 years
from its prediction, interesting information about related phenomena has been
accumulated. In particular, this include $2p$ democratic decays and the $2p$
emission from excited nuclear states populated in $\beta$ decay and in
reactions. In the following we overview the main steps in the experimental
progress.

\subsubsection{Democratic decays}
\label{sec:VIID1}

\begin{figure}[tb]
\centerline{
\includegraphics[width=0.43\textwidth]{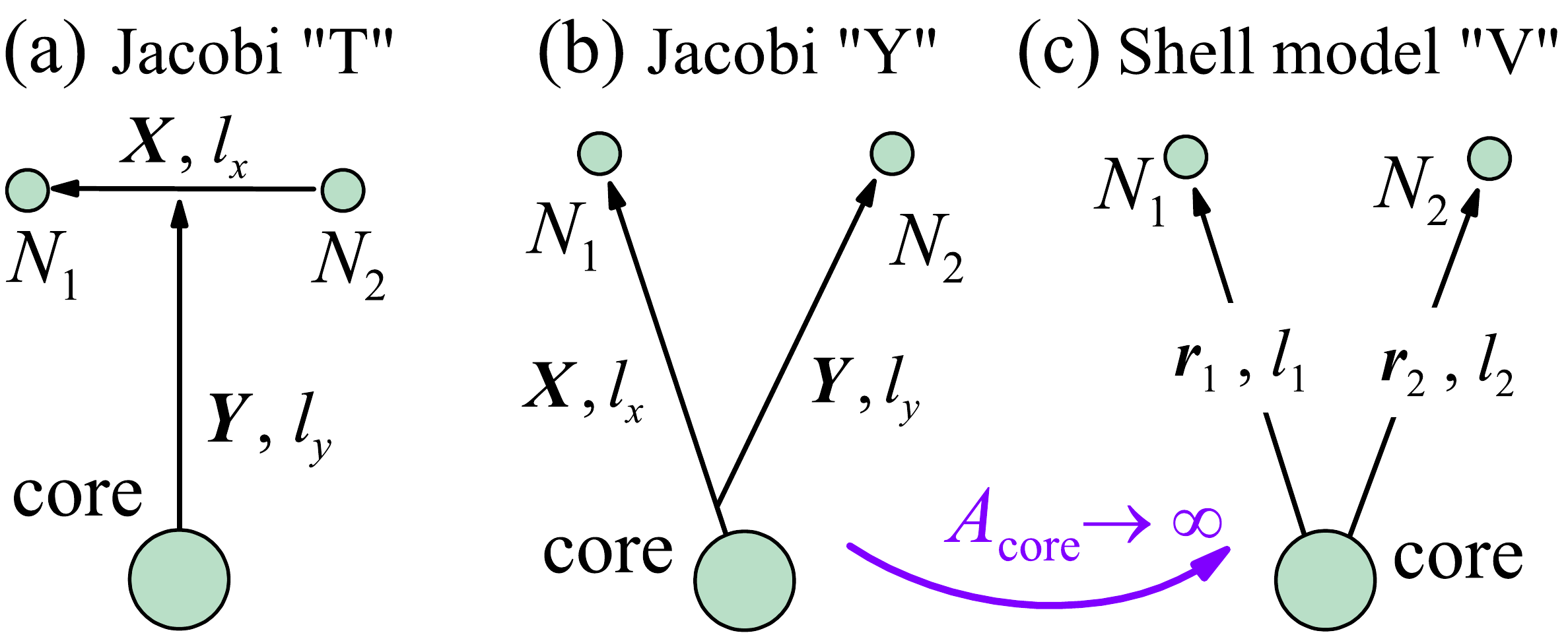}
}
\caption[Coordinate systems for two-nucleon plus core problem.]{(Color online)
Coordinate systems for two-nucleon plus core problem. In the Jacobi ``T'' system
(a), the ``diproton'' and the core are explicitly in configurations with
definite angular momenta $l_x$ and $l_y$. For a heavy core the Jacobi ``Y''
system (b) is close to the single-particle ``V''-system (c) used typically in
many-body approaches.}
\label{fig:VIIB-coord-syst}
\end{figure}

Since it is relatively easier to reach experimentally light proton-drip line
nuclei, the first attempts to search for $2p$ emission phenomena started with
the light $2p$ unbound systems: $^{6}$Be \cite{Geesaman:1977,Bochkarev:1984},
$^{12}$O \cite{KeKelis:1978,Kryger:1995}, and $^{16}$Ne \cite{KeKelis:1978}, see
also Table \ref{tab:VIIC3-2p}. Due to a very low Coulomb barrier these cases
have
half-lives much too short to classify as being radioactive, more appropriately
they should be referred to as $2p$ resonances.

Already \textcite{Geesaman:1977} could not fit the results on $^{6}$Be using
simple decay scenarios (phase volume, diproton decay, simultaneous emission of
$p$-wave protons) and concluded that a full three-body computation is necessary
to understand the measured spectra. Later the interpretation of correlations
observed in the decay of the $^{6}$Be ground state have led to the concept of
``democratic decay'' \cite{Bochkarev:1989}. In such a decay there is no strong
energy focusing of the particles and they are smoothly distributed in the
momentum space. The system is demonstrating a kind of equal rights among
different parts of the kinematical space\footnote{Now the term \emph{democratic
decay} has become accepted for description of the mentioned class of phenomena.
There exist, however, an anecdotic story that when it was used for the first
time at some conference at Soviet Union in the end of 80-s the authors were
heavily criticized because ``there can not be any democracy in nuclear
physics''.}. The democratic decay is not a phenomenon on its own but rather a
name for the experimental fingerprint of a true three-body decay in light
systems
with the relatively small Coulomb force or in two-neutron emitters.

The study of opening angle between protons emitted from $^{12}$O was motivated
by the search for diproton correlation. The measured spectrum, however, was
found to be consistent with the sequential emission via intermediate $^{11}$N
state. Later it was found that indeed, the ground state energy of $^{11}$N is
below that of $^{12}$O and the decay of the latter belongs to the class shown in
Fig.~\ref{fig:VIIA-scheme}(e) \cite{Azhari:1998}.

\begin{figure}
\centerline{
\includegraphics[width=0.48\textwidth]{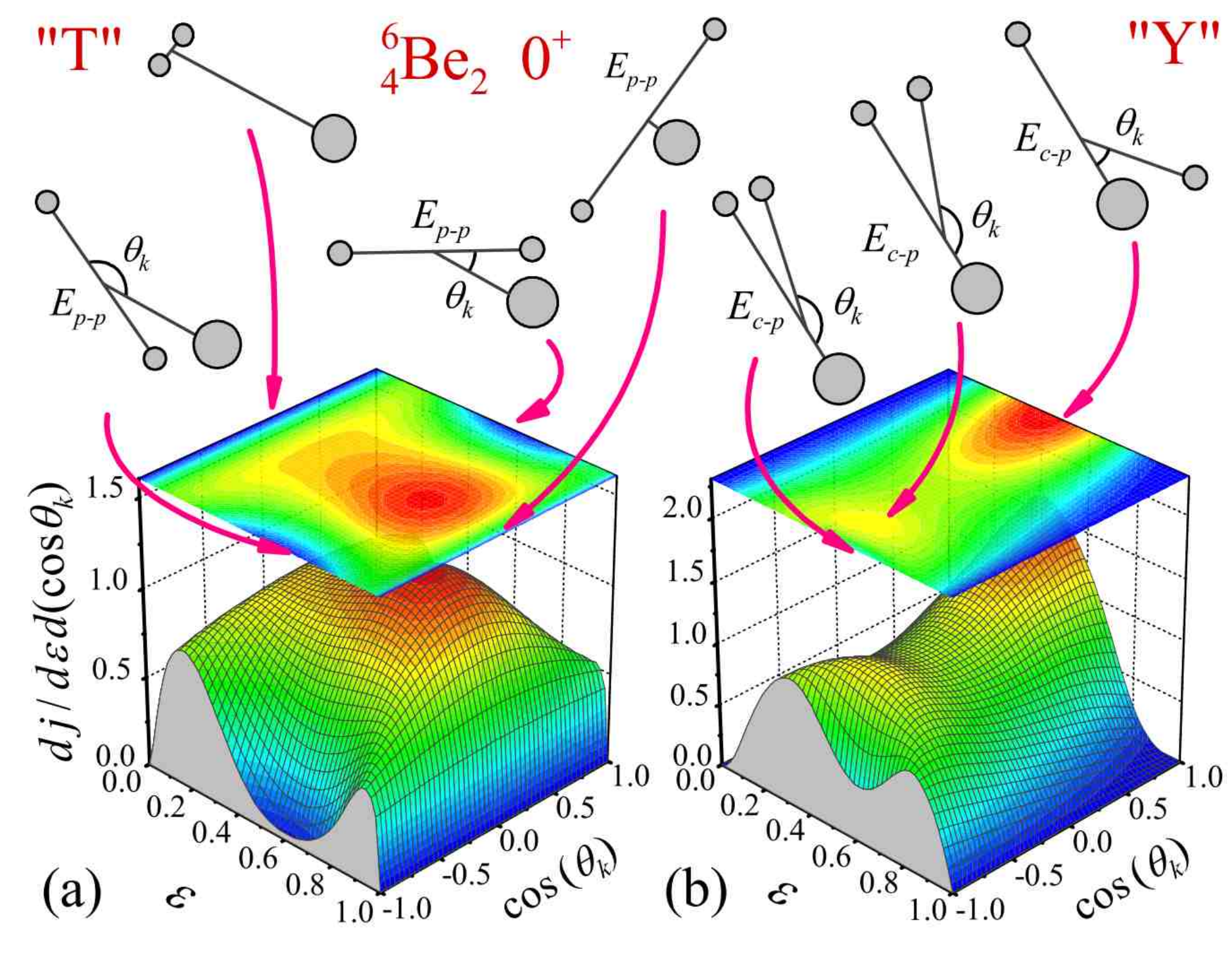}
}
\centerline{
\includegraphics[width=0.235\textwidth]{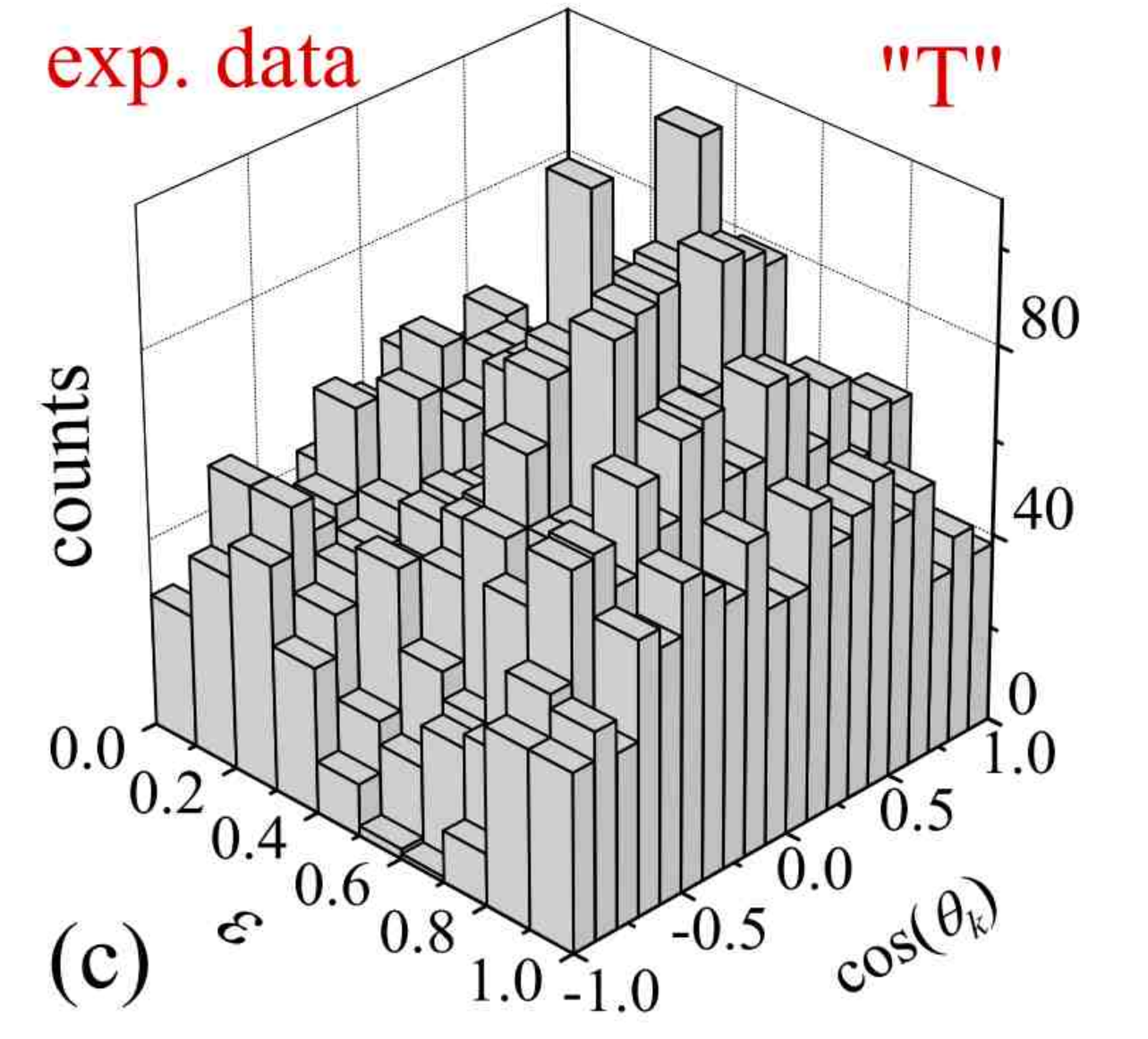}
\includegraphics[width=0.235\textwidth]{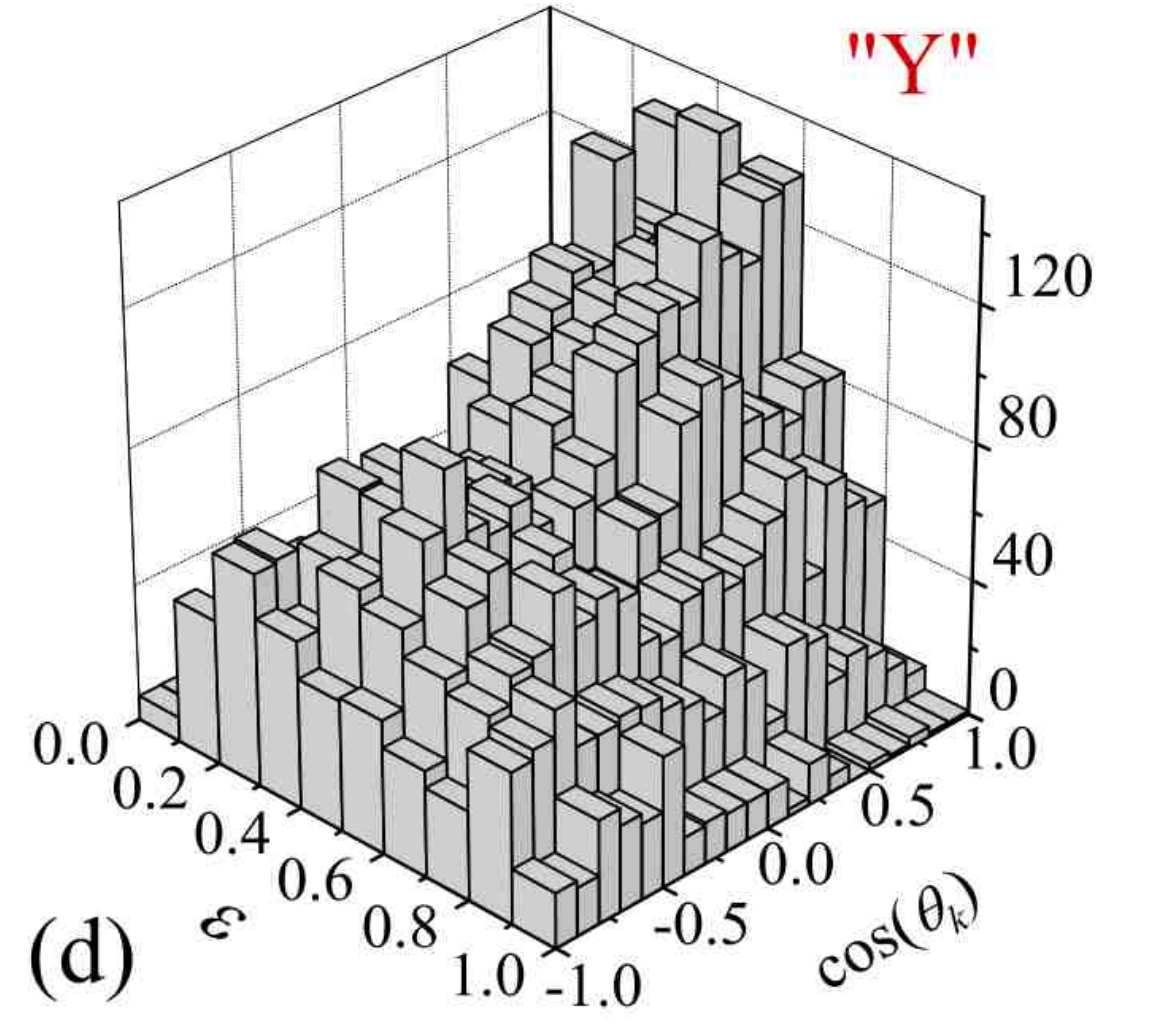}
}
\caption{(Colour online) Complete correlation picture for $^{6}$Be g.s.\ decay,
presented in ``T'' (a,c) and ``Y'' (b,d) Jacobi systems. The (a,b) is theory,
(c,d) is experimental data. Qualitative illustration of the meaning of different
kinematical regions is provided above the panels. Data and calculations are from
\cite{Grigorenko:2009}.}
\label{fig:VIIC1-corel-all-be}
\end{figure}

Recently, the full correlation picture for protons emitted by $^{6}$Be has been
experimentally established \cite{Grigorenko:2009} and was found to be in very
good agreement with the predictions of the three-body model which will be
discussed below. Both experimental and theoretical distributions in the ``T''
and ``Y'' Jacobi coordinates are presented in Fig.~\ref{fig:VIIC1-corel-all-be}.
Similarly, the new results of $p$-$p$ correlations in the decay of $^{16}$Ne
could be well described by the three-body model \cite{Mukha:2008}.

\subsubsection{Two-proton emission from excited states}
\label{sec:VIID2}

Several cases of $\beta$-delayed $2p$ emission are known, as mentioned in
Sec.~\ref{sec:IVD}. They are discussed in more detail in \textcite{Blank:2008}.
It is believed that in all cases studied, the mechanism of the emission is
sequential, i.e.\ it can be described as a sequence of two two-body decays.

In a few other cases, the $2p$ emitting excited states were populated by nuclear
reactions. Like in the case of ground-state resonances, the main motivation of
these studies was the search for \emph{p-p} correlations going beyond the
sequential mechanism. The $2^+$ state at 7.77~MeV in $^{14}$O was excited by the
two-proton transfer reaction from $^{3}$He impinging on a $^{12}$C target
\cite{Bain:1996}. The $2p$ emission from this state was found to proceed
sequentially through a $1/2^+$ state in $^{13}$N. The first excited state in
$^{17}$Ne ($3/2^-$) was reached by the Coulomb excitation of the radioactive
$^{17}$Ne beam to search for its $2p$ decay \cite{Chromik:2002}. Although the
conditions for the true $2p$ decay were fulfilled in this case, the
de-excitation by $\gamma$ radiation was found to dominate. In a different
approach, however, higher excited states in $^{17}$Ne were populated by
1n stripping reaction from the radioactive $^{18}$Ne beam
\cite{Zerguerras:2004}. The $2p$ angular correlation was found to be peaked at
small angles indicating a contribution from a seemingly non-sequential
mechanism. Due to small statistics and limited information on the identity of
the decaying states, however, no definite conclusions could be reached. The
$1^-$ resonance at 6.15 MeV in $^{18}$Ne was also thought to provide opportunity
for the true $2p$ emission, as no states are known in $^{17}$F to be located
within the decay energy window. The $2p$ emission from this state, populated in
the $^{17}$F+$^1$H reaction, was measured and a diproton-like contribution to
the \emph{p-p} correlation spectra was claimed \cite{Gomez:2001}. Similar
evidence was announced in another work, in which the same state was populated by
the Coulomb excitation of the $^{18}$Ne beam \cite{Raciti:2008}. The statistical
significance of both claims, however, is weak. The excited states of $^{19}$Ne,
populated in the inelastic scattering of $^{18}$Ne on hydrogen target, were
found to emit two protons sequentially \cite{Oliveira:2005}.

An interesting case is the claimed $2p$ emission from the high-spin $21^+$
isomeric state in $^{94}$Ag \cite{Mukha:2006}. The authors explain an
anomalously
high $2p$ decay rate and observed $p$-$p$ correlations by assuming a very high
deformation of the isomer. This work has caused a lot of controversies
\cite{Pechenaya:2007,Kankainen:2008,Mukha:2008a,Cerny:2009} which call for
further experimental investigations.

\subsubsection{Two-proton radioactivity}
\label{sec:VIID3}

\begin{figure}
\centerline{
\includegraphics[width=0.47\textwidth]{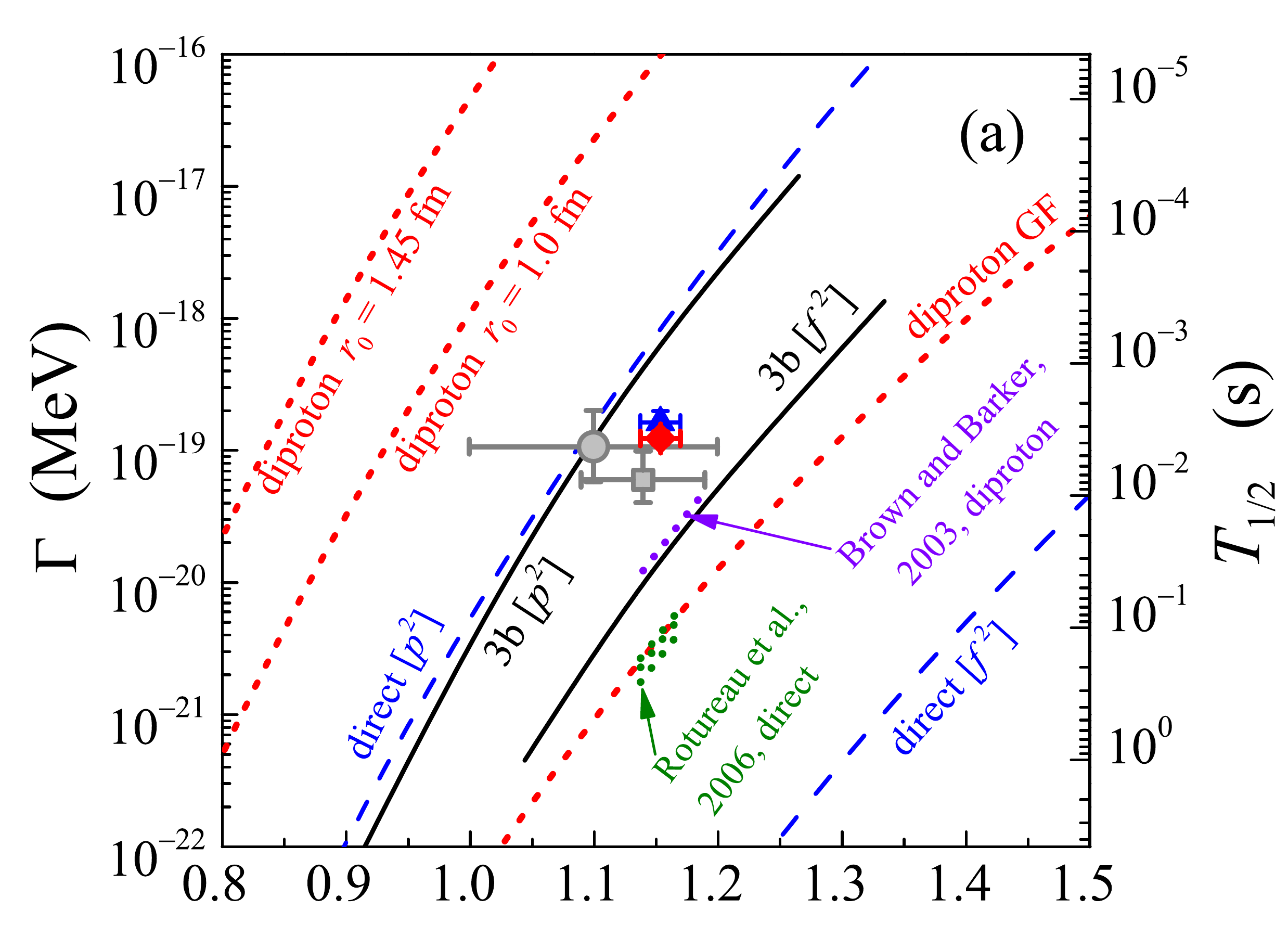}
}
\centerline{
\includegraphics[width=0.47\textwidth]{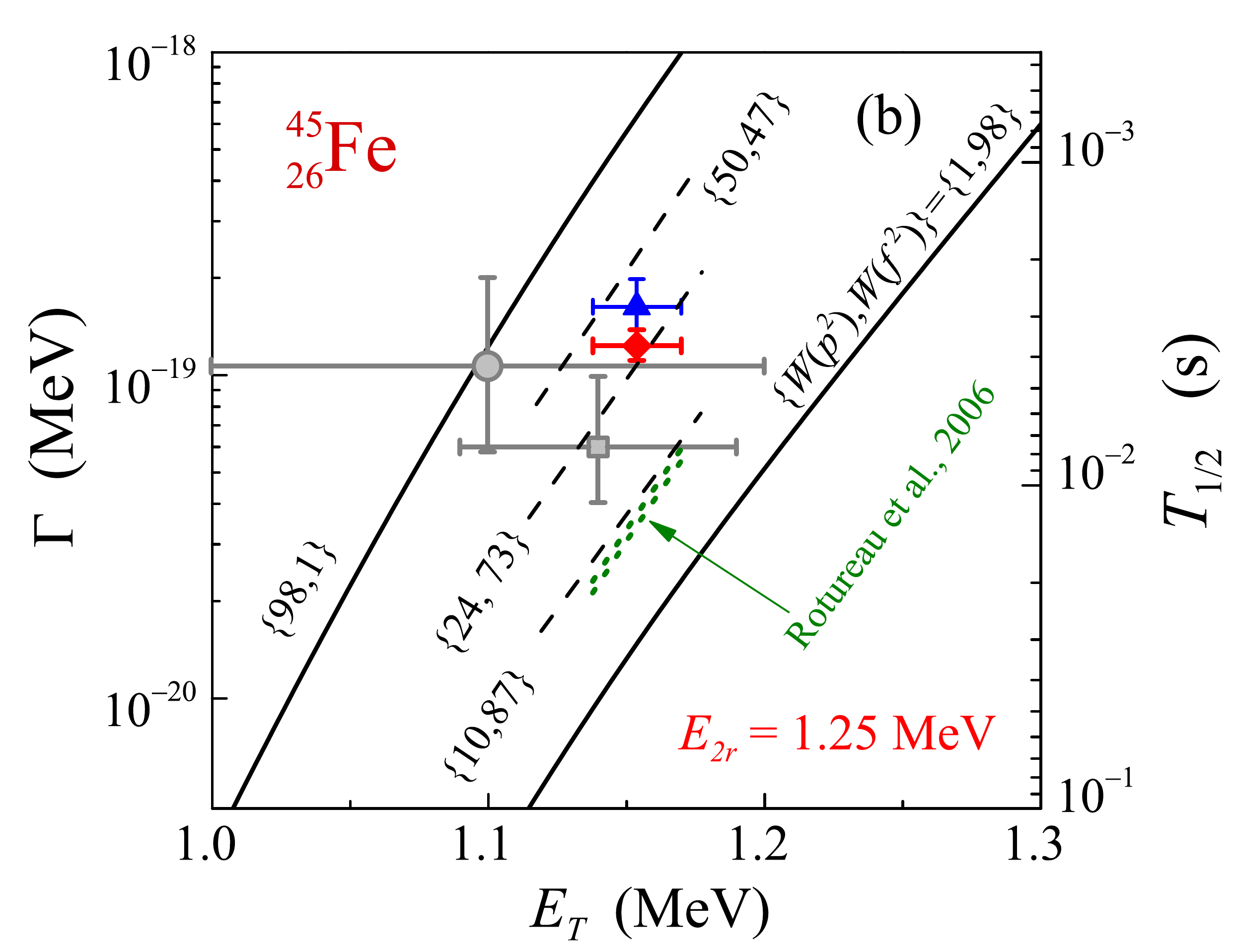}
}
\caption{(Color online) Lifetime of $^{45}$Fe vs.\ decay energy calculated in
different models.
(a) Simplified models of $2p$ decay. All spectroscopic factors are taken as
unity. (b) Three-body model
\cite{Grigorenko:2003,Grigorenko:2007} and continuum shell model results
\cite{Rotureau:2006}. The experimental points demonstrate the rapid improvement
of
the data: circle \cite{Pfutzner:2002}, square \cite{Giovinazzo:2002}, triangle
\cite{Dossat:2005}, and diamond \cite{Miernik:2007b}.}
\label{fig:VIIF1-lifetime-fe}
\end{figure}

The first case of $2p$ radioactivity was found in 2002 in the decay of $^{45}$Fe
measured at GSI Darmstadt \cite{Pfutzner:2002} and at GANIL
\cite{Giovinazzo:2002}. In both experiments ions of $^{45}$Fe were produced by
the fragmentation reaction of a $^{58}$Ni beam and separated using the
in-flight technique (Sec.~\ref{sec:IIIB1}). The selected ions were implanted
into silicon detectors and the only observables measured were the decay time and
energy. It sufficed to claim the observation of a new decay mode because only
the $2p$ emission hypothesis was consistent with the measured data. Later, in
another GANIL experiment the information on $^{45}$Fe was obtained with larger
statistics (30 atoms) and improved accuracy \cite{Dossat:2005}, see
Table~\ref{tab:VIIC3-2p} and Fig.~\ref{fig:VIIF1-lifetime-fe}. With the same
experimental technique, applied again at GANIL laboratory, $^{54}$Zn has been
identified to decay by the $2p$ radioactivity \cite{Blank:2005}. One decay event
of $^{48}$Ni was found to coincide with the $2p$ decay energy predicted for this
nucleus \cite{Dossat:2005}.

\begin{figure*}
\begin{center}
\includegraphics[width=0.49\textwidth]{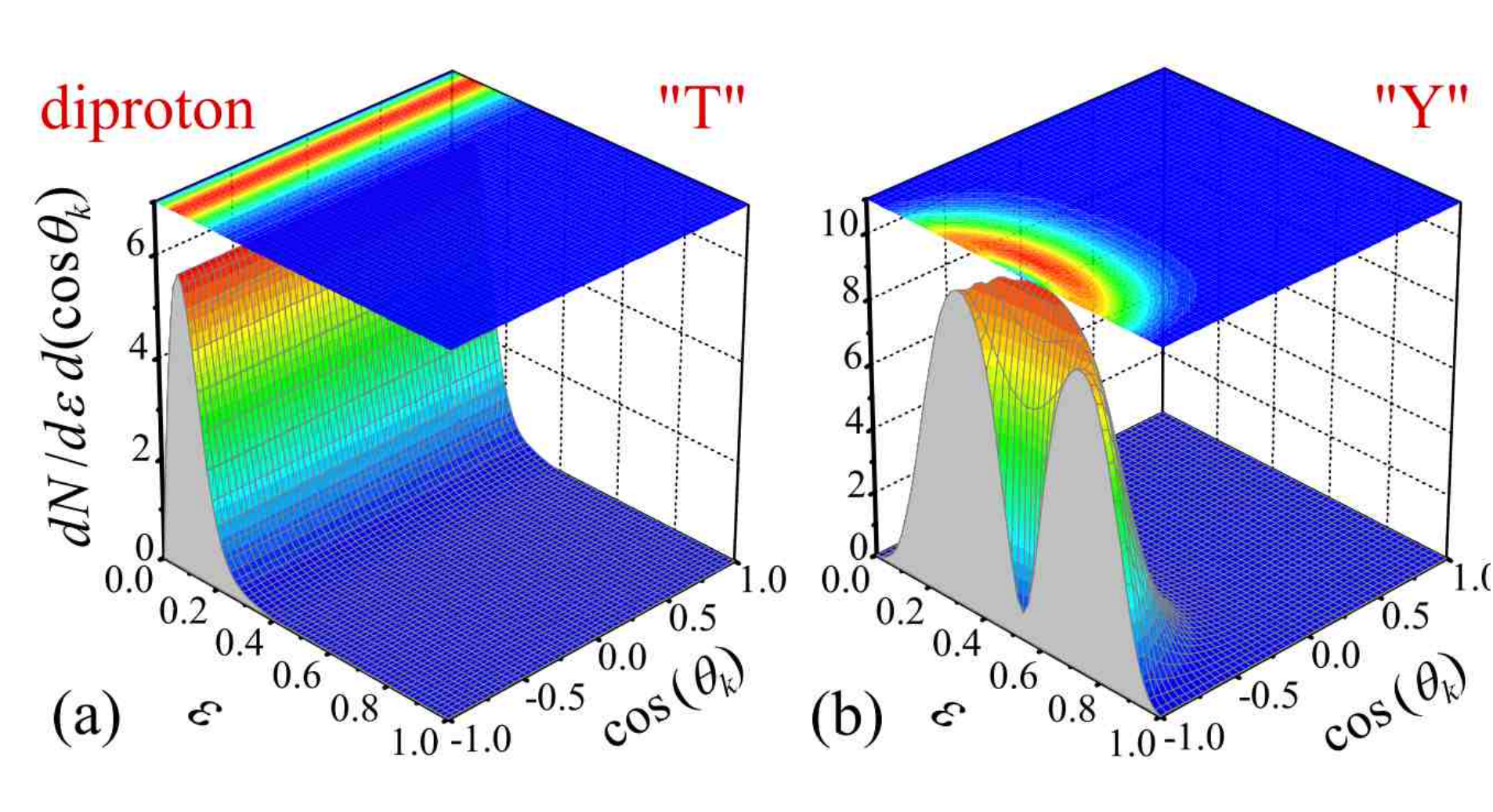}
\includegraphics[width=0.49\textwidth]{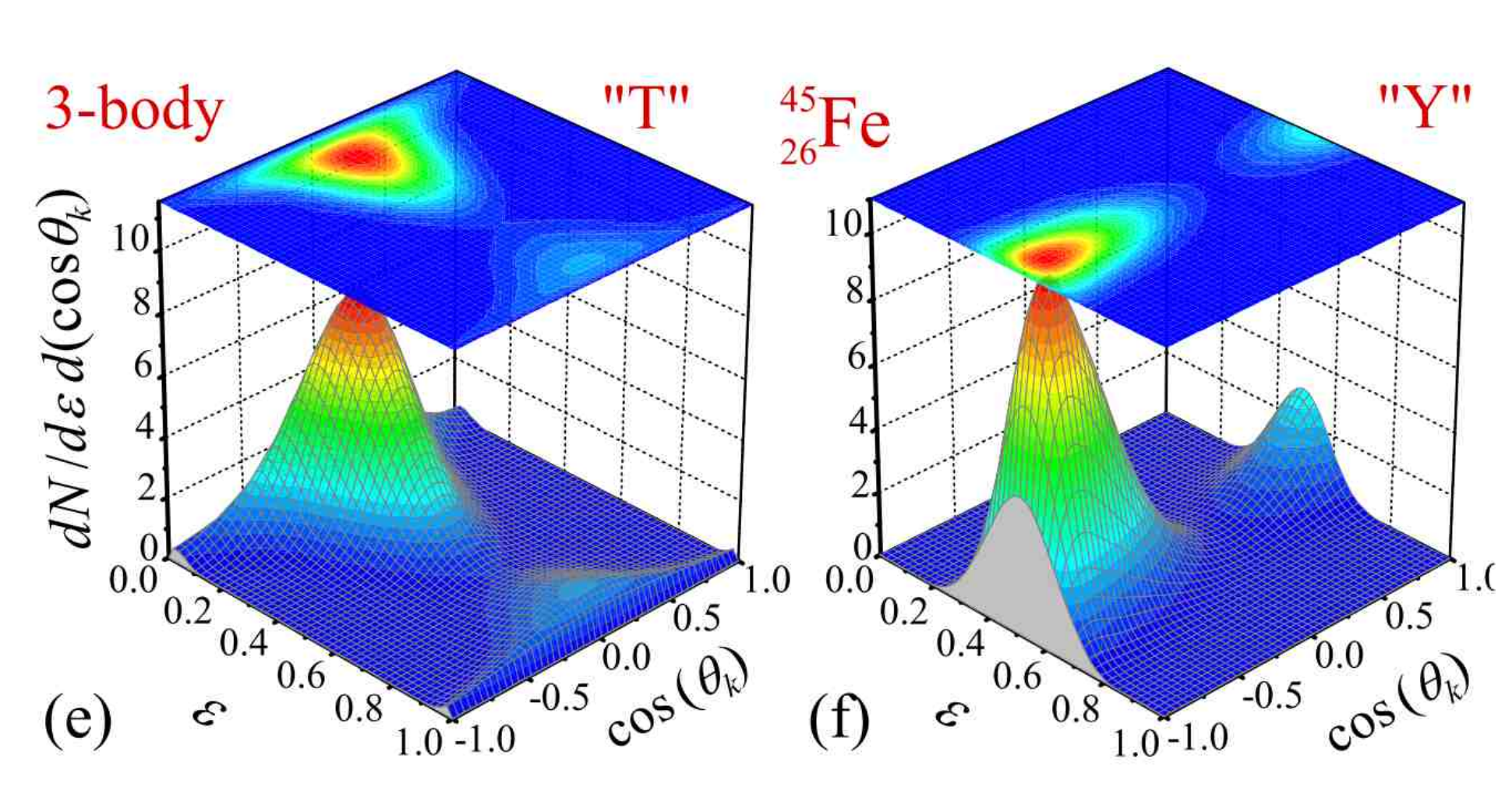}\\
\includegraphics[width=0.49\textwidth]{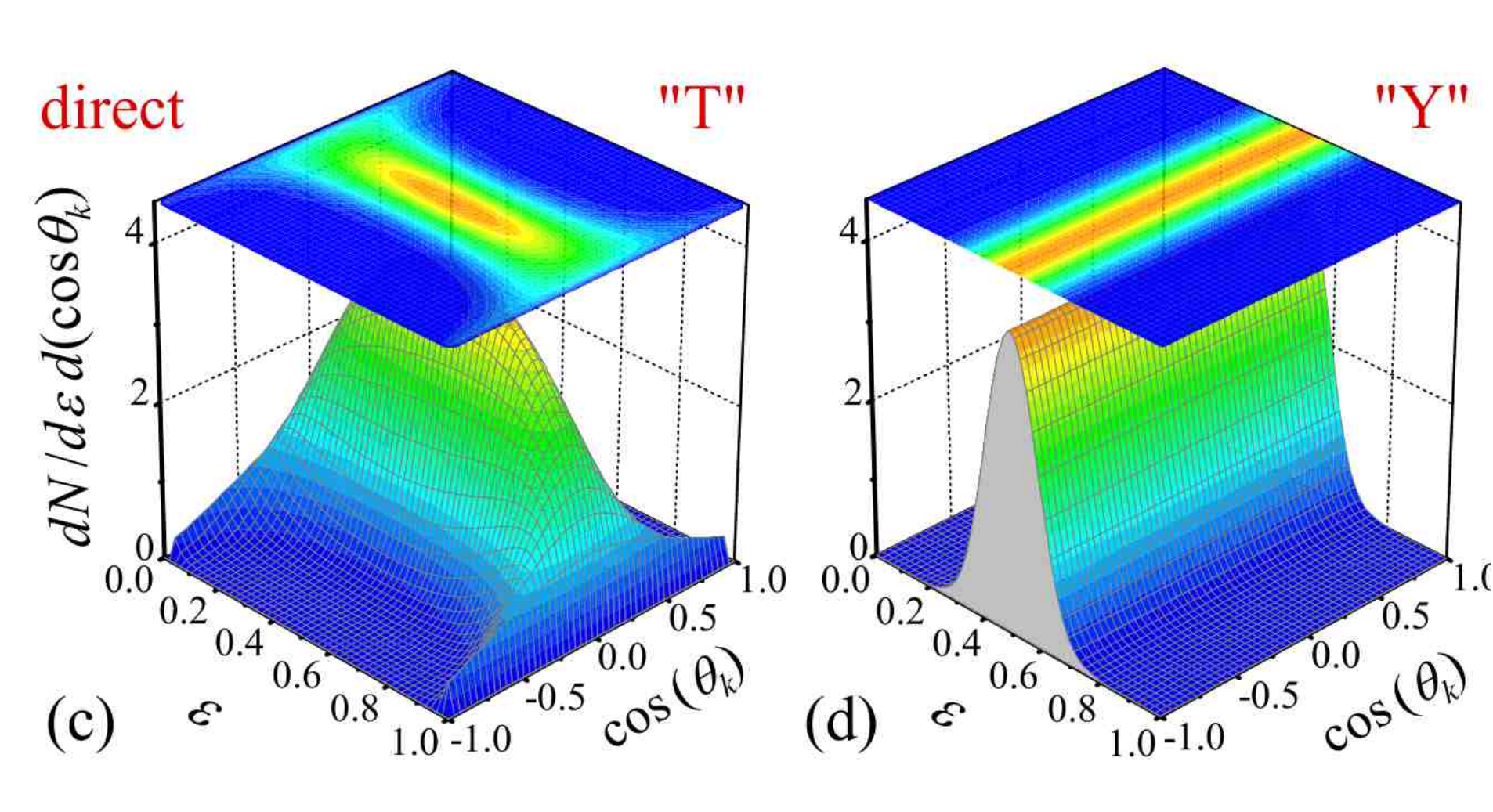}
\includegraphics[width=0.235\textwidth]{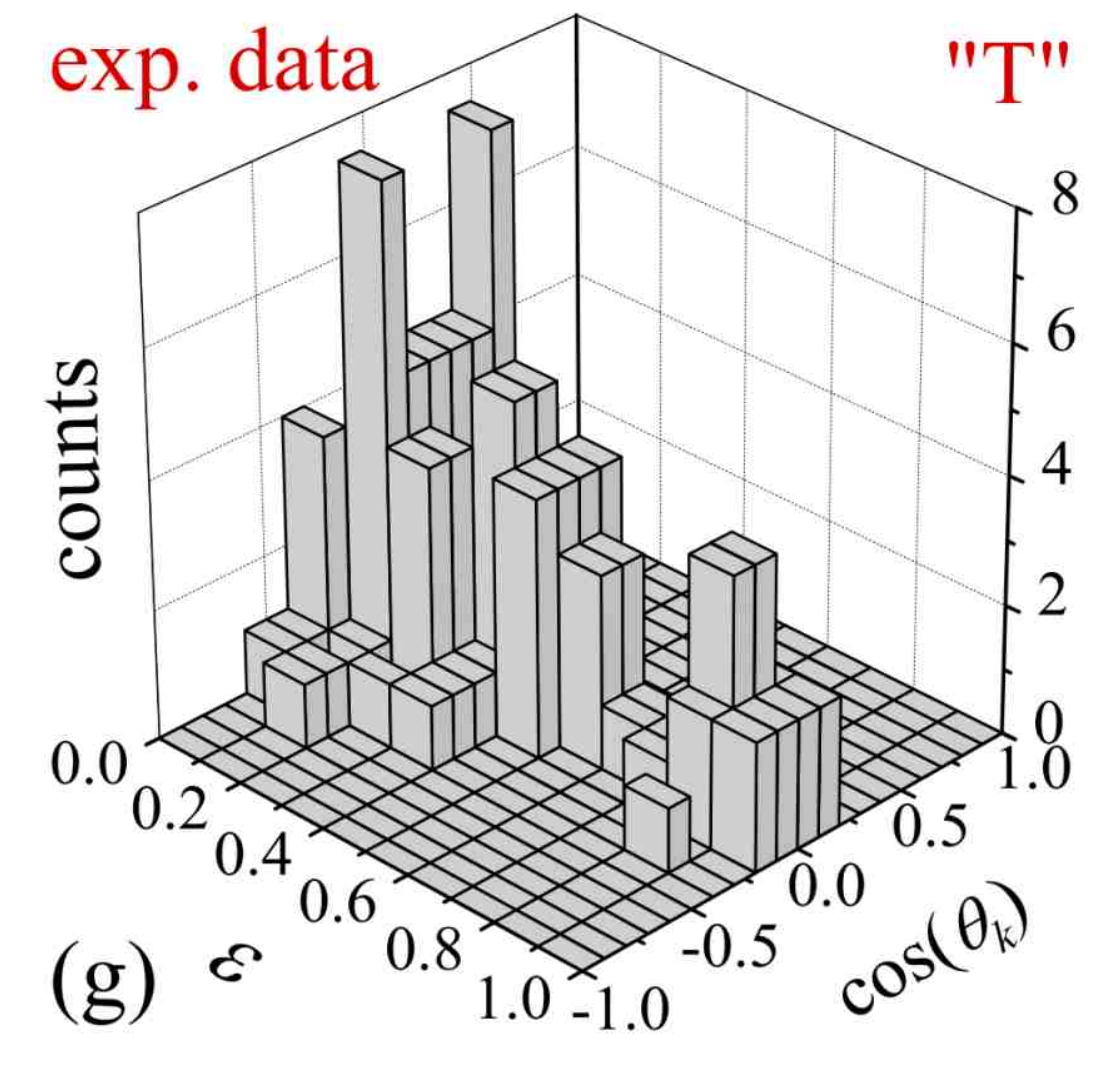}
\includegraphics[width=0.235\textwidth]{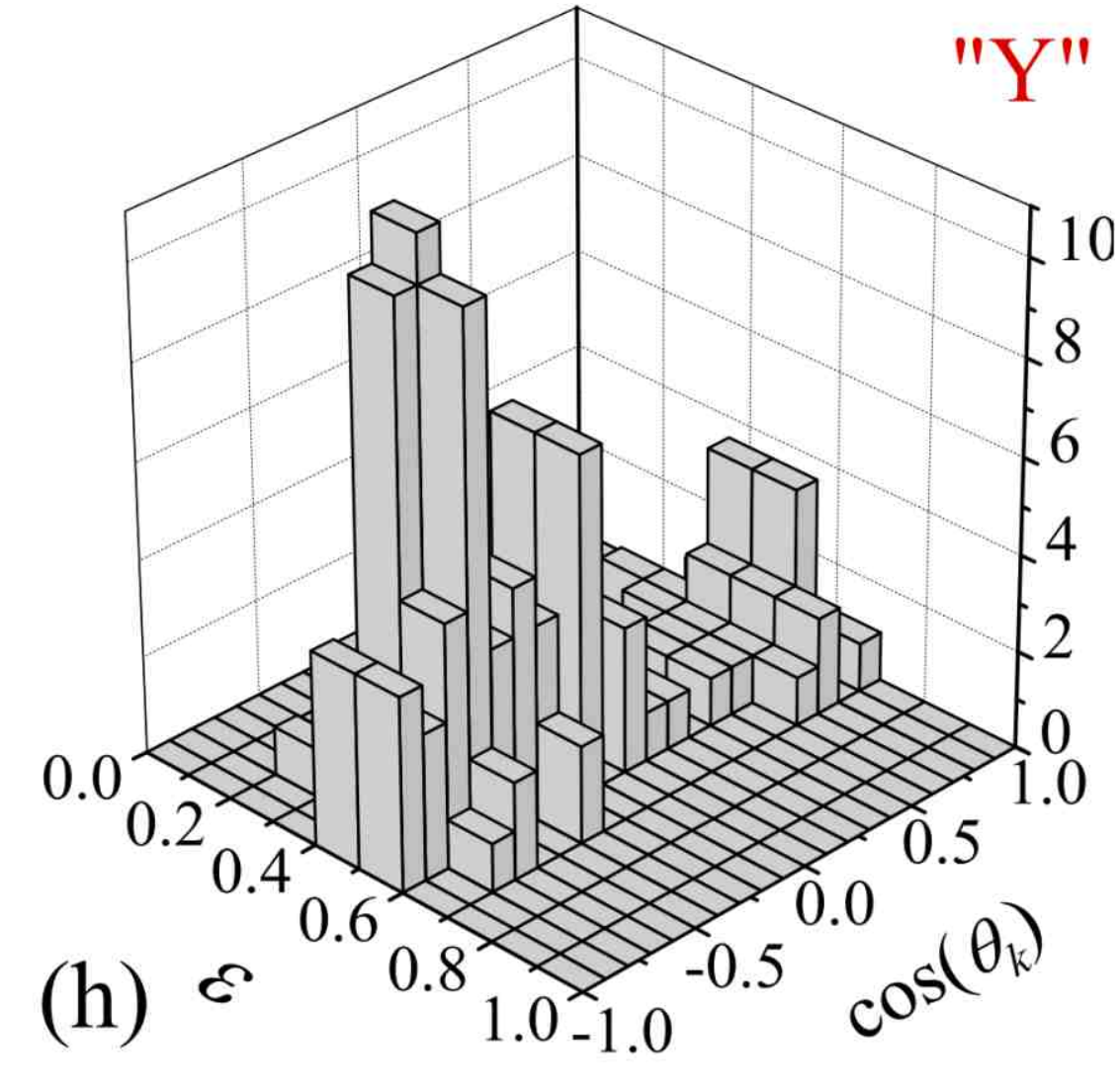}
\end{center}
\caption{(Color online) Momentum density distribution on the kinematical plane
$\{\varepsilon,\cos(\theta_k) \}$ for $^{45}$Fe. Diproton model (a,b), direct
decay model (c,d), three-body model(e,f), and experimental distribution (g,h).
Correlation patterns are provided in the ``T'' (a,c,e,g) and ``Y'' (c,d,f,h)
Jacobi systems. The calculations (e,f) are from \cite{Grigorenko:2010} and data
(g,h) are from \cite{Miernik:2007b}.}
\label{fig:VIIC1-corel-models}
\end{figure*}

\begin{figure}
\centerline{
\includegraphics[width=0.35\textwidth]{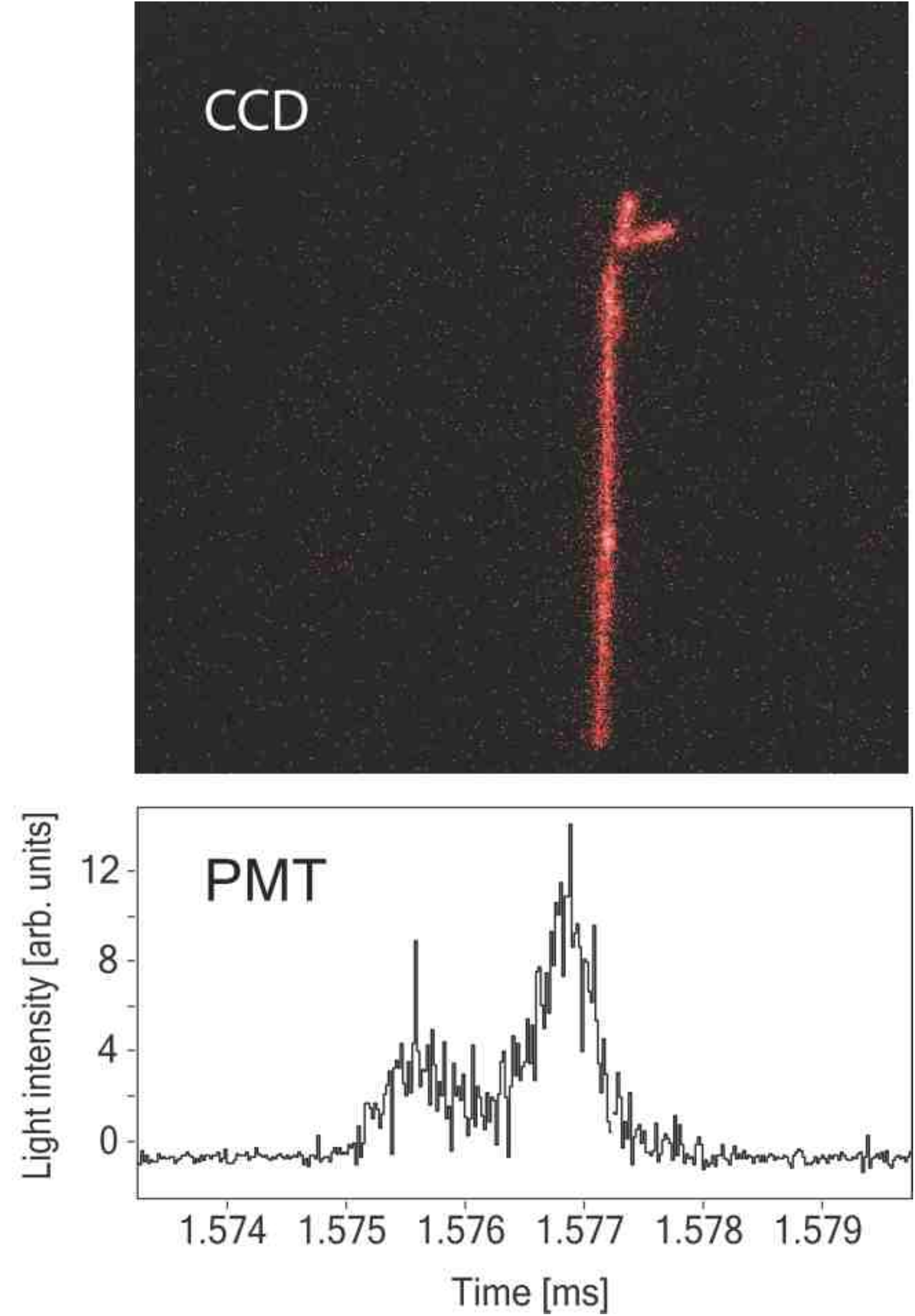}\mbox{\qquad}
}
\caption{An example of a two-proton decay event of $^{48}$Ni recorded
    with the optical time projection chamber described in
    section~\ref{sec:IIIC1}.
    Top: the image recorded by the CCD camera showing a long
    track of the $^{48}$Ni ion entering the chamber from below
    and the two bright, short tracks of protons emitted 1.576 ms
    after the implantation.
    Bottom: a part of the time profile of the total light intensity
    measured by the PMT showing in detail the \emph{2p} emission.
    From \cite{Pomorski:2011b}.}
\label{fig:VIID3_48Ni}
\end{figure}

In the next step, gaseous detectors, based on the principle of the time
projection chamber (TPC), were developed to directly record emitted protons and
to establish the correlations between them. The first direct observation of the
two protons ejected from $^{45}$Fe was achieved by \textcite{Giovinazzo:2007}
who recorded projections of protons' tracks on the anode plane of the TPC.
Later, this detector was used to directly demonstrate the two-protons emitted in
the decay of $^{54}$Zn \cite{Blank:2011}. \textcite{Miernik:2007b} applied a
novel type of detector, utilizing the optical readout of the TPC signals
(OTPC, see Sec.~\ref{sec:IIIC1}), to the detailed decay study of $^{45}$Fe at
the NSCL/MSU laboratory and succeeded to fully reconstruct tracks of emitted
protons in three dimensions. The full correlation picture for the $2p$ decay of
$^{45}$Fe established in this experiment is shown in
Fig.~\ref{fig:VIIC1-corel-models}. Recently, the OTPC detector was used to the
decay study of $^{48}$Ni at the NSCL/MSU laboratory and provided a direct
and unambiguous evidence for the $2p$ radioactivity of this nuclide
\cite{Pomorski:2011b}.
From the six decays recorded, four corresponded to the $2p$ emission
and two were interpreted as $\beta$-delayed proton emission.
An example of a $2p$ decay event of $^{48}$Ni is shown in
Fig.~\ref{fig:VIID3_48Ni}.

\begin{figure}
\centerline{
\includegraphics[width=0.48\textwidth]{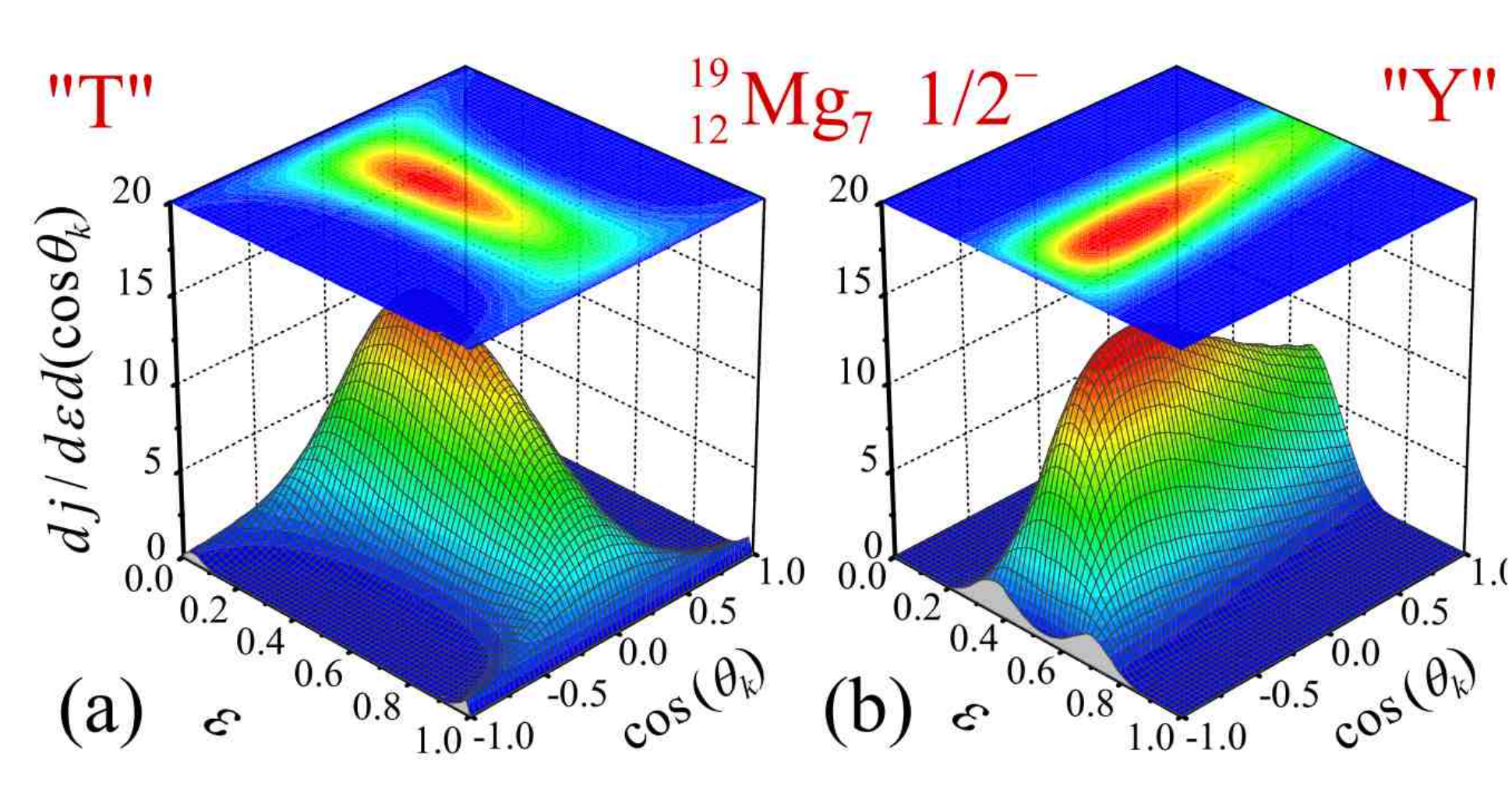}
}
\centerline{
\includegraphics[width=0.249\textwidth]{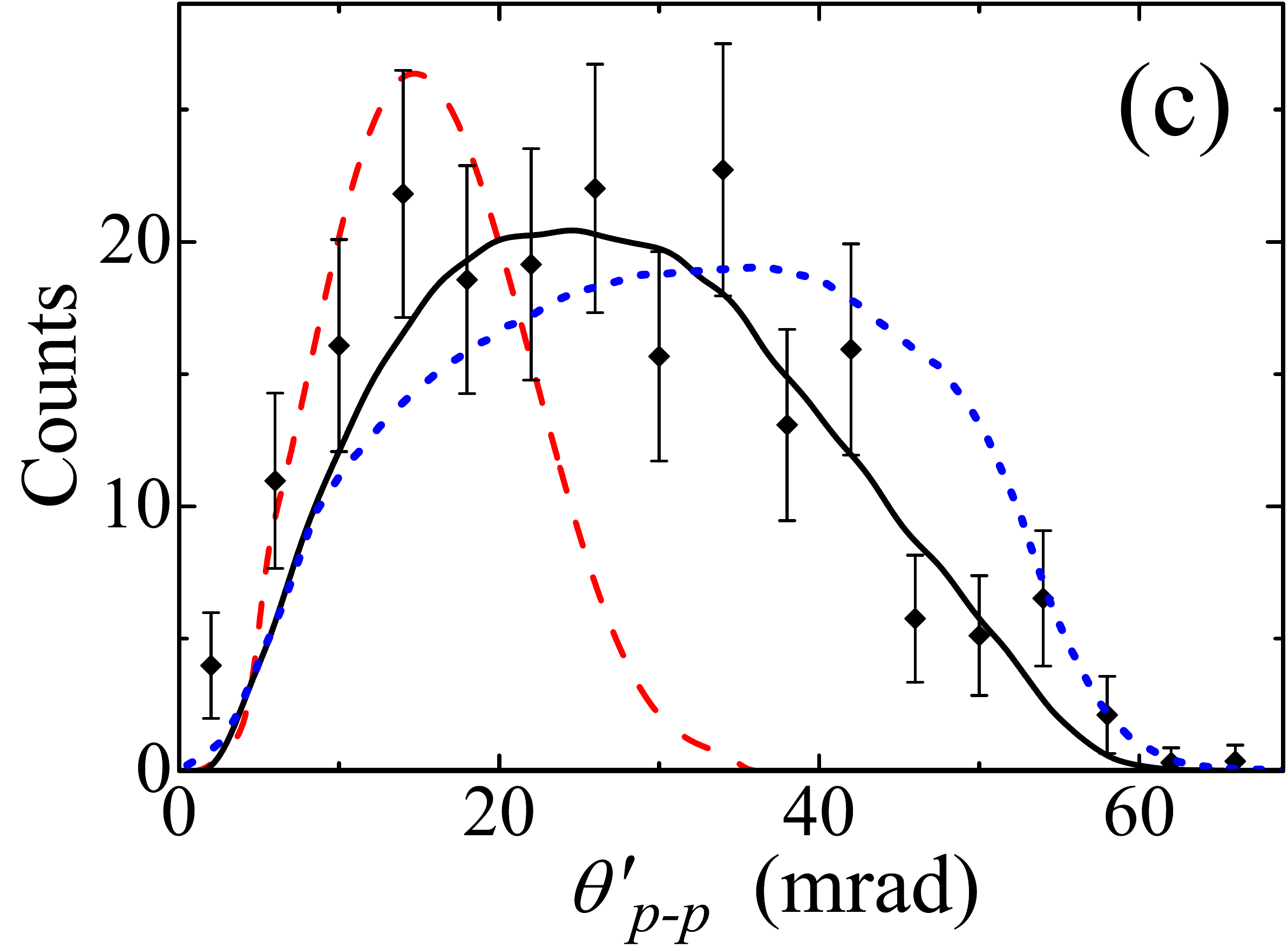}
\includegraphics[width=0.223\textwidth]{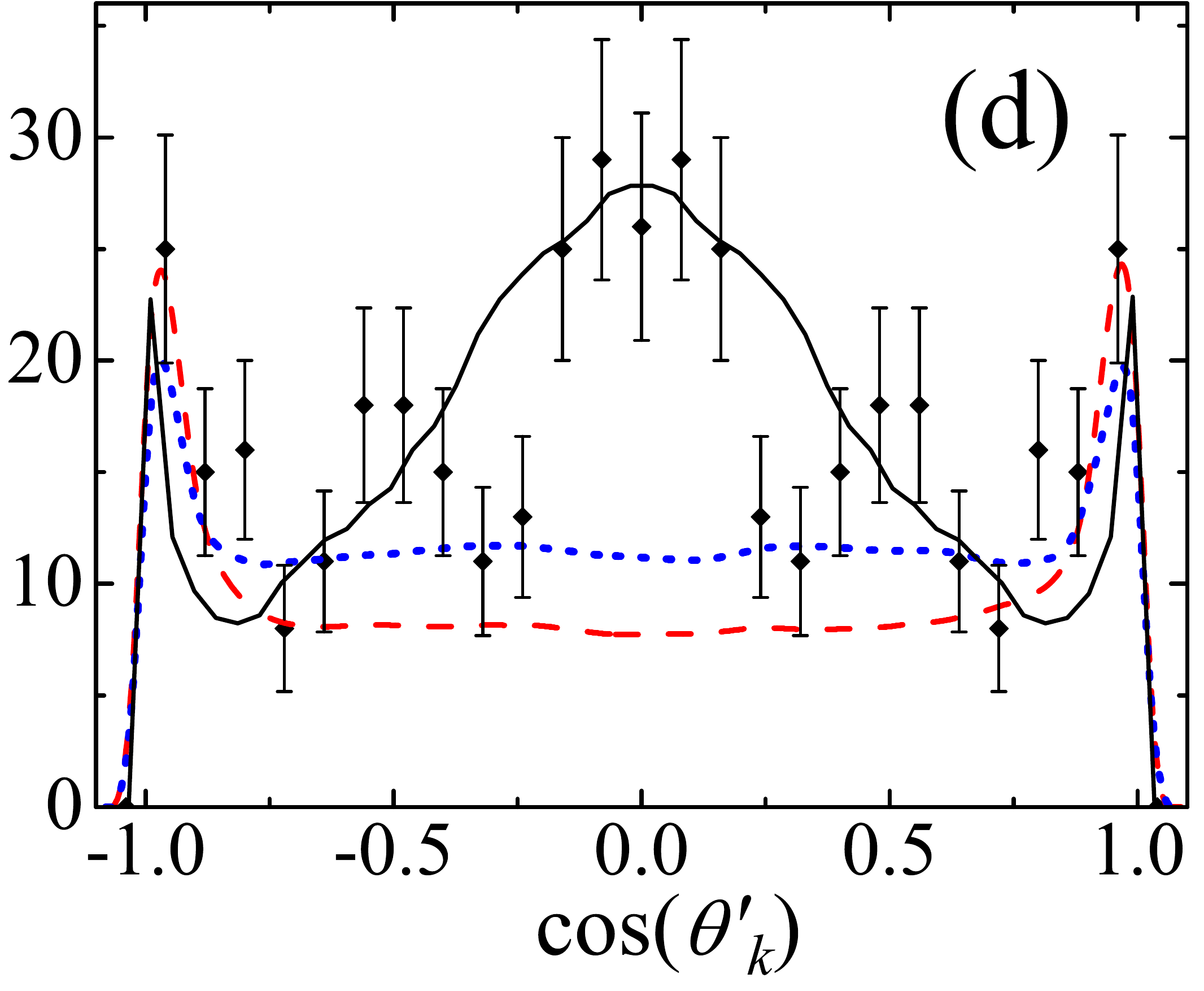}
}
\caption{(Colour online) Complete correlation picture for $^{19}$Mg g.s.\ decay,
presented in ``T'' (a) and ``Y'' (b) Jacobi systems. Comparison of two inclusive
distributions projected on a plane is given in the lower row \cite{Mukha:2008}.
The angles $\theta'_{p\text{-}p}$ and $\theta'_{k}$ are between projected
momenta of protons (c) and projected Jacobi momenta in the ``T'' system (d).
Solid, dashed, and doted curves correspond to three-body model, diproton model,
and phase space simulations. Data are from \cite{Mukha:2008} and calculations
are from \cite{Grigorenko:2010}.}
\label{fig:VIIC1-corel-all-mg}
\end{figure}

An application of the different technique based on decay in-flight and particle
tracking (Sec.~\ref{sec:IIIC1}) is necessary to extend decay studies to very
short half-lives in the subnanosecond range. This technique was used by
\textcite{Mukha:2007} to investigate the $2p$ radioactivity of $^{19}$Mg whose
half-life was found to be $4.0(15)$~ps. The $p$-$p$  momentum distributions,
projected on the transverse detector plane were obtained in \cite{Mukha:2008},
see Fig.~\ref{fig:VIIC1-corel-all-mg}.

\subsection{Simplified theoretical models}
\label{sec:VIIE}

Simplified theoretical models of three-body decays are typically based on the
factorization of the decay amplitude into the product of two-body terms. For
$2p$ emission this factorization can be done either in the ``T'' or in the ``Y''
Jacobi systems, see Fig.\ \ref{fig:VIIB-coord-syst}. This results in formulation
of \emph{diproton model} or  the \emph{direct decay model} respectively.
Factorization of amplitudes becomes possible only for specific forms of the
underlying Hamiltonians \cite{Grigorenko:2007}. Then, neglecting some
of the final state interactions (FSI), the three-body Green's function (GF)
can be constructed in an analytical form:
\begin{eqnarray}
\hat{G}^{(+)}_{3E}(\mathbf{XY},\mathbf{X}'\mathbf{Y}') = \frac{E}{2\pi i}
\int^{1}_0   d \varepsilon \,
\hat{G}^{(+)}_{\varepsilon E}(\mathbf{X},\mathbf{X}') \nonumber \\
\times \hat{G}^{(+)}_{(1-\varepsilon)E}(\mathbf{Y},\mathbf{Y}') \,.
\label{eq:VIID-gf3}
\end{eqnarray}
The operator $\hat{G}^{(+)}_{E}$ is the ordinary two-body GF for the respective
subsystem. This method allows to take into account \emph{one} (out of three
present) FSI \emph{exactly} in diproton model. In the direct decay model
\emph{two} FSIs can be treated exactly but only in the limit of infinitely heavy
core, see Figs.\ \ref{fig:VIIB-coord-syst} (b,c). For heavy $2p$ emitters this
is a good approximation.  The availability of approximate but analytic GF allows
to
determine decay properties without further approximations.

\subsubsection{Direct decay model}
\label{sec:VIIE1}

The decay amplitude can be factorized in the ``V'' coordinate system [Fig.\
\ref{fig:VIIB-coord-syst}(c)] if we neglect $p$-$p$ interaction and also assume
an
infinitely heavy core. The dominating contribution to the width can be obtained
in the single-pole approximation for two-body GFs where only the lowest states
in the subsystems core+$p$ with resonance energies $E_{p_1}$ and $E_{p_2}$ are
considered \cite{Galitsky:1964}:
\begin{eqnarray}
\Gamma_{\text{dir}}(E_T)  =  \frac{E_{T}\left \langle V_{3}\right \rangle
^{2}} {2\pi}
\int_0^{1} \!\! d \varepsilon \frac{ \Gamma_{p_1}(\varepsilon E_{T})}
{(\varepsilon E_{T}-E_{p_1})^{2}+\Gamma_{p_1}(\varepsilon E_{T})^{2}/4}
\nonumber \\
 \times   \frac{\Gamma_{p_2}((1-\varepsilon)E_{T})}
 {((1-\varepsilon)E_{T}-E_{p_2})^{2} + \Gamma_{p_2}
((1-\varepsilon)E_{T})^{2}/4}\;, \qquad
\label{eq:VIID1-rmat-wid-3}
\end{eqnarray}
It was shown in \cite{Grigorenko:2007a} that the matrix element $\left \langle
V_{3}\right \rangle$ can be well approximated as
\begin{equation}
\left \langle V_{3}\right \rangle ^{2}   =   D_3 (E_T-E_{p1}-E_{p2})^2 \; .
\label{eq:VIID1-v3}
\end{equation}
The parameter $D_3$ is a constant, $D_3 \approx 1.0-1.5$. The $\Gamma_{p_i}$ is
the width of the ground state resonance in the core+$p$ subsystem. It can be
expressed using the R-matrix approach to nuclear reactions \cite{Lane:1958}:
\begin{equation}
\Gamma_{p_i}(E) = 2 \gamma^{2} P_{l_i}(E,R,Z_1Z_2),
\label{eq:IB-r-mat-wid}
\end{equation}
The penetrability $P_{l}$ is defined via the
Coulomb functions $F$ and $G$ \cite{Abramowitz:1968},
regular and irregular at the origin:
\begin{equation}
P_{l}(E,R,Z_1Z_2)=\frac{kR}{F_{l}^{2} (\eta,kR)+G_{l}^{2}(\eta,kR)},
\label{eq:IB-r-mat-pen}
\end{equation}
In the above $\eta=Z_1Z_2 \alpha M/k$ is the Sommerfeld parameter, $\alpha$ is
the
fine structure constant. The $\gamma^2$ is the reduced width which is factorized
into the ``Wigner limit'' $\gamma_{WL}^{2}$:
\begin{equation}
\gamma^{2}=\gamma_{WL}^{2}\theta^{2}=\frac{1}{2MR^{2}}\theta^{2}\;,
\label{eq:IB-dimles-red-wid}
\end{equation}
and the dimensionless reduced width $\theta^{2}$ (spectroscopic factor).
The parameter $R$ is the ``channel radius'', typically taken as:
\begin{equation}
R=r_0 (A_{\text{core}}+1)^{1/3}\,,
\label{eq:IB-r-ch}
\end{equation}
where $r_0=1.4$ fm.

The results of calculations in the direct decay approximation,
Eq.~(\ref{eq:VIID1-rmat-wid-3}), for the case of $^{45}$Fe and for the values of
angular momentum $l=1$ and $l=3$ are presented in
Fig.~\ref{fig:VIIF1-lifetime-fe}. These are likely to provide upper and lower
width limits for given decay energy which are consistent with the limits
obtained in the complete three-body model, see Sec.\ \ref{sec:VIIF2}.

\begin{figure}
\centerline{
\includegraphics[width=0.42\textwidth]{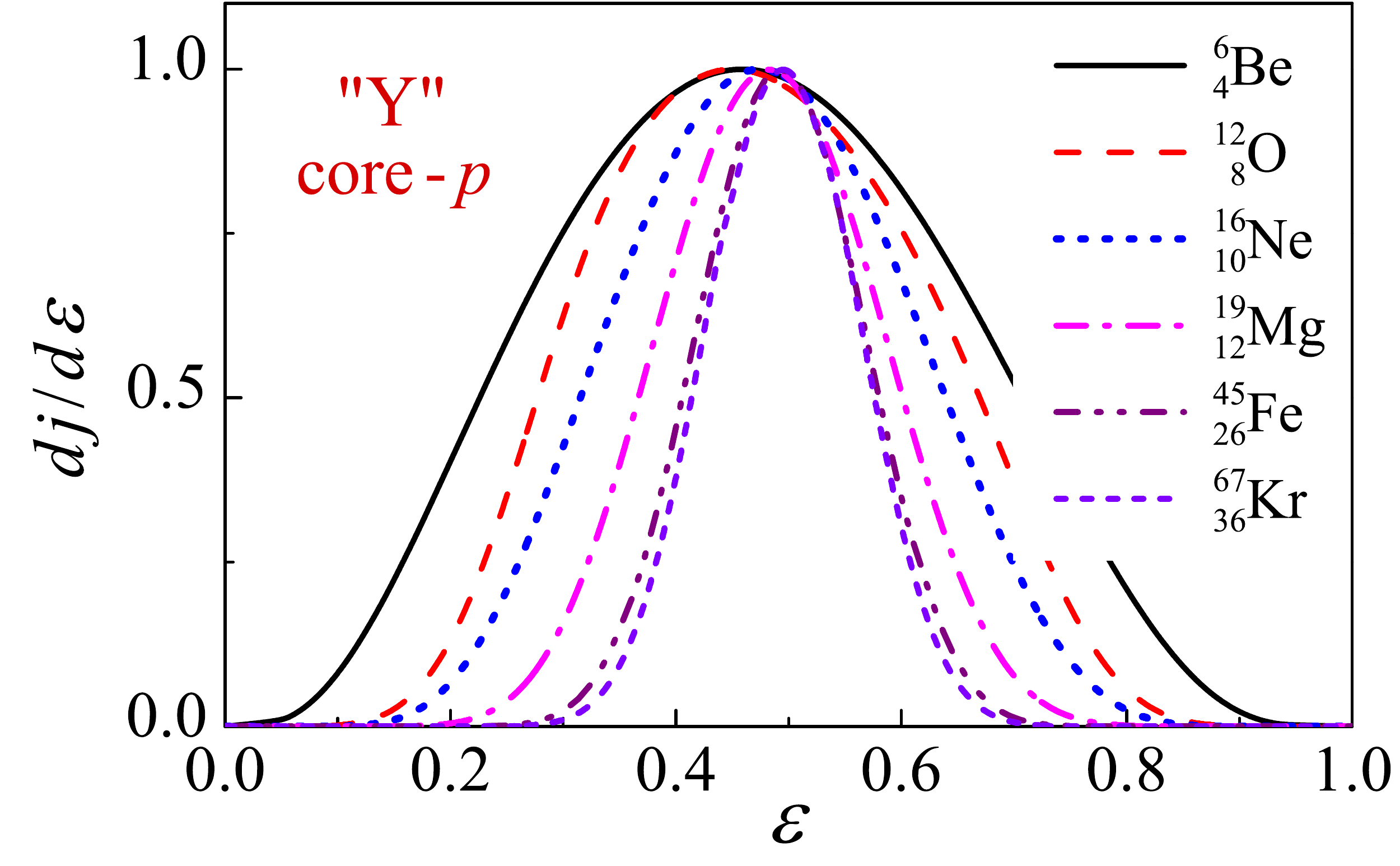}
}
\caption{(Color online) Systematics of energy distributions in the ``Y'' Jacobi
system (core-$p$ channel) calculated in the direct decay model of Eq.\
(\ref{eq:VIID1-rmat-wid-3}). All distributions are normalized to unity at the
maximum.}
\label{fig:VIID1-core-p-syst}
\end{figure}

Eq.~(\ref{eq:VIID1-rmat-wid-3}) has a similar structure and can be considered as
a
refined version of the original expression (\ref{eq:VIIB-w-goldan}) proposed by
Goldansky. Compared to the latter it provides quantitative results (the
preexponent is defined) and brings the explicit dependence on the property of
the core+$p$ subsystem. The energy distribution between the core and one of the
protons predicted by the direct decay approximation has a narrow bell-shape
peaked at $\varepsilon=1/2$, which reflects the symmetry between emitted
protons, see Fig.\ \ref{fig:VIID1-core-p-syst}. With increase of the atomic
number this distribution becomes narrower. The bell-shape  profile of this
inclusive distributions is by now well confirmed by experiment
\cite{Miernik:2007b,Grigorenko:2009,Blank:2011}. However, looking at the
complete distribution (see Fig.\ \ref{fig:VIIC1-corel-models} that shows
the distribution for the case of angular coupling $[p^2_{1/2}]_0$)
the deficiency of
the model becomes evident. Qualitatively, the direct decay model angular
distribution always has backward-forward symmetry, while there is a strong
angular asymmetry observed in experiment.

Using the fact that the numerator in Eq.\ (\ref{eq:VIID1-rmat-wid-3}) is sharply
peaked at $\varepsilon / 2$ and the denominator for the true $2p$ decays is a
smooth function within the decay window, it can be further approximated:
\begin{eqnarray}
\Gamma_{\text{dir}}(E_T) & \approx & \frac{E_{T}D_3 (E_T-E_{p1}-E_{p2})^2} {2\pi
(E_T/2-E_{p1})^2(E_T/2-E_{p2})^2}
\nonumber \\
& \times & \int_0^{1} \!\! d \varepsilon  \Gamma_{p_1}(\varepsilon
E_{T})\Gamma_{p_2}((1-\varepsilon)E_{T})\;,
\label{eq:VIID1-rmat-wid-3-app}
\end{eqnarray}
This approximation allows to estimate how the decay width depends on the
position of the lowest resonances in the core-$p$ subsystem. For the ground
state true $2p$ emitters $E_{p1}=E_{p1}=E_{2r}$ and we get :
\begin{equation}
\Gamma_{\text{dir}} \sim (E_T/2 - E_{2r})^{-2}\,.
\label{eq:VIID1-wid-3-ot-ep}
\end{equation}

We note that a somewhat different expression for the width, in the analogous
approximation, was introduced in Refs.\  \cite{Kryger:1995,Azhari:1998} for
$^{12}$O:
\begin{equation}
\Gamma_{\text{kr}}  =  \frac{E_{T}} {2\pi} \int_0^{1} \!\! d \varepsilon
\frac{ \Gamma_{p_1}(\varepsilon E_{T}) \Gamma_{p_2}((1-\varepsilon)E_{T})}
{(\varepsilon E_{T}-E_{p_1})^{2}+\Gamma_{p_1}(\varepsilon E_{T})^{2}/4}.
\label{eq:VIID1-rmat-wid-3-kr}
\end{equation}
The expression is given here in our notation. It was introduced by analogy with
two-body R-matrix expressions. Using the same approximation as in Eq.\
(\ref{eq:VIID1-rmat-wid-3-app}) we obtain
\begin{equation}
C=\frac{\Gamma_{\text{dir}}}{\Gamma_{\text{kr}}} \approx D_3
\frac{(E_T-E_{p_1}-E_{p_2})^2
(E_T/2-E_{p_1})^2}{(E_T/2-E_{p_1})^2(E_T/2-E_{p_2})^2}.
\end{equation}
For the ground state true $2p$ emitters $C \approx 4$. In the series of papers
\cite{Barker:1999,Barker:2001,Brown:2002,Barker:2002,Brown:2003,Barker:2003}
formulae equivalent to Eq.~(\ref{eq:VIID1-rmat-wid-3-kr}) were used. The results
obtained in these works should include the factor $C$ varying between 2 and 4 to
be consistent with the direct decay approximation.

\subsubsection{Simultaneous vs.\ sequential decay}
\label{sec:VIIE2}

\begin{figure}[tb]
\centerline{
\includegraphics[width=0.24\textwidth]{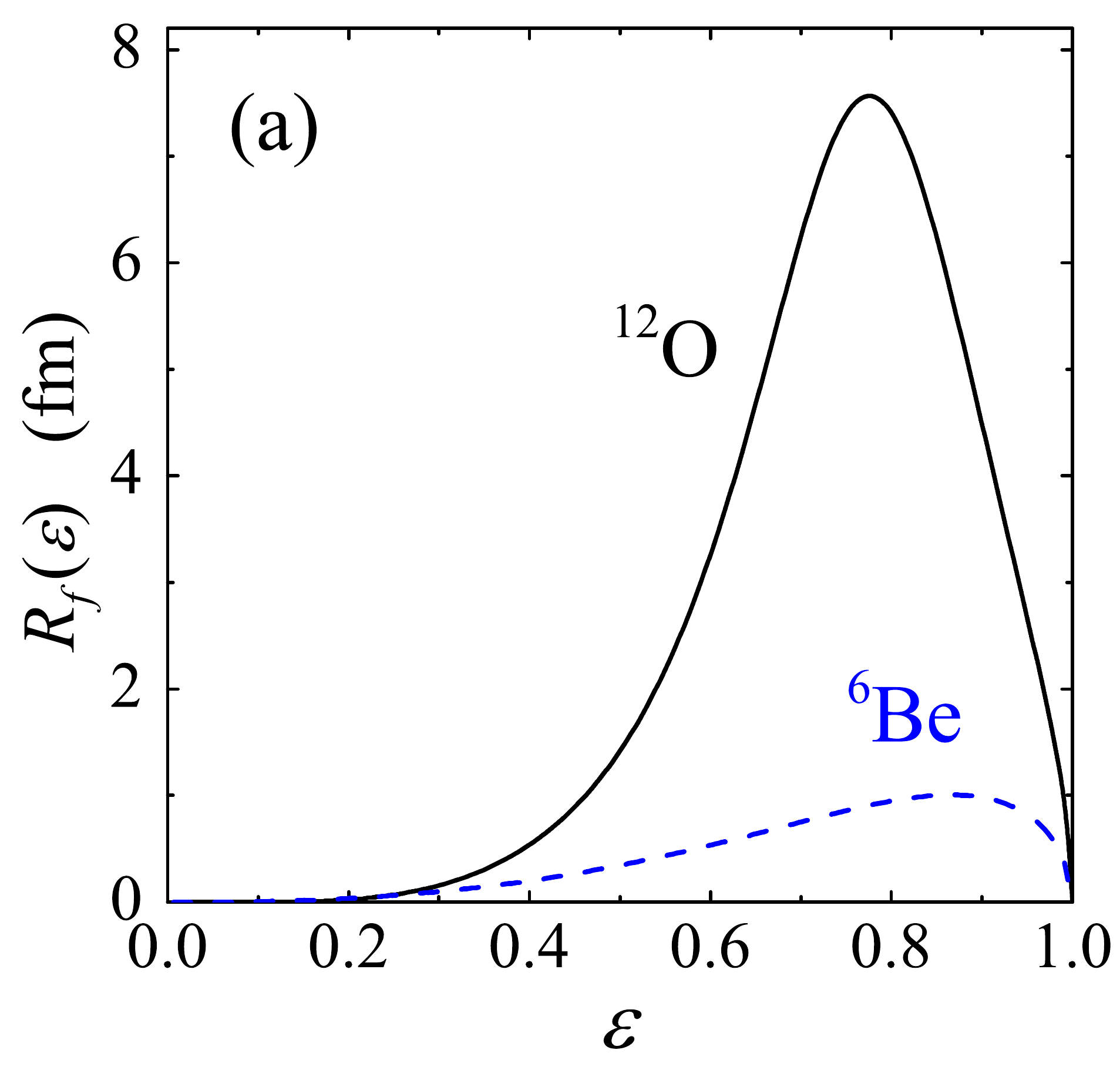}
\includegraphics[width=0.26\textwidth]{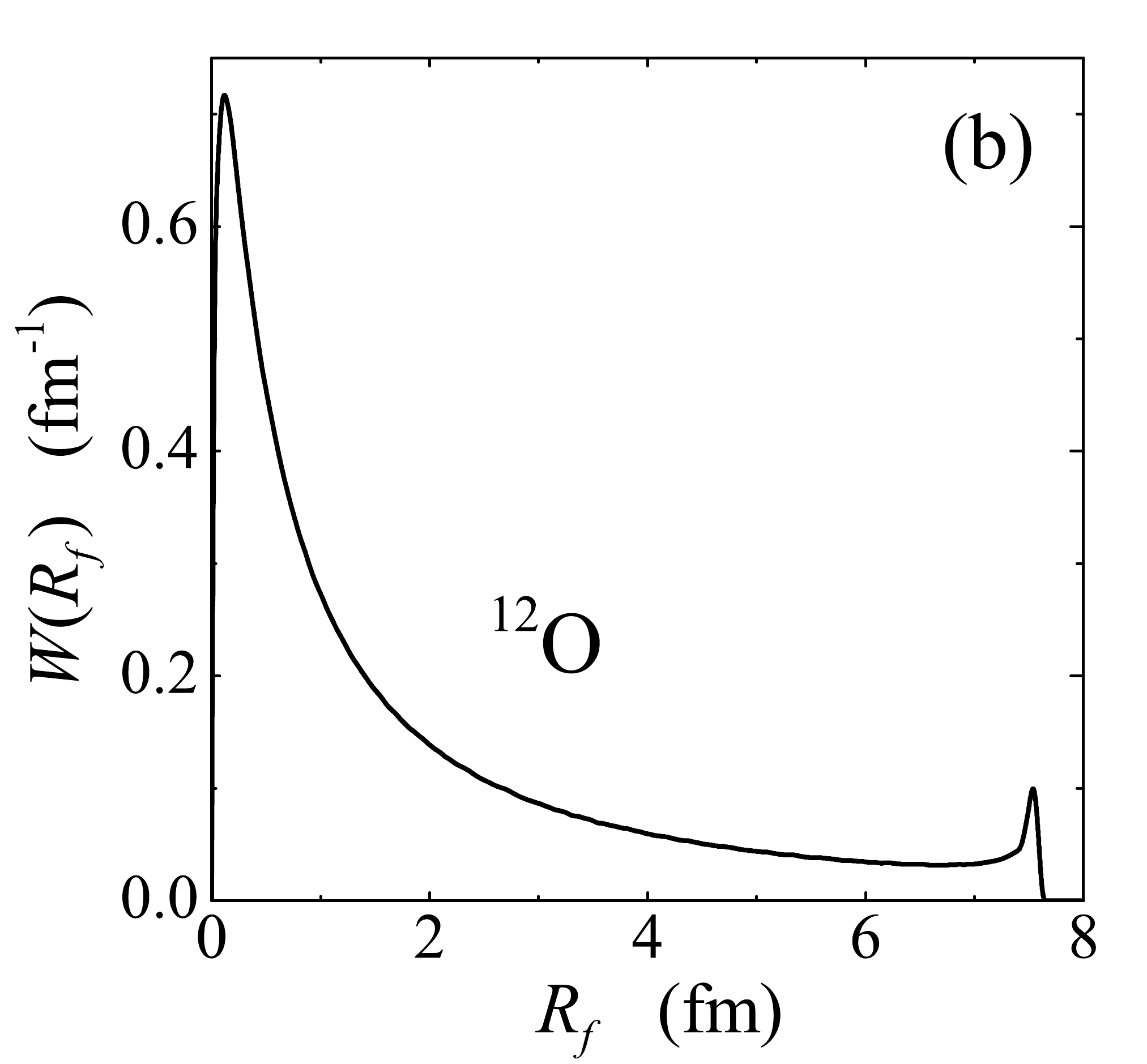}
}
\caption[Classical proton flight path.]{(Color online) (a) Classical one-proton
flight path $R_f$ for $^{12}$O and $^{6}$Be connected with the time delay in the
core+$p$ subsystem calculated by Eq.\ (\ref{eq:IB-time-delay-res}). (b)
Probability of different $R_f$ values for the decay of $^{12}$O estimated by
Eq.\ (\ref{eq:VIID1-rmat-wid-3}).}
\label{fig:VII:flight-dist}
\end{figure}

The true two-proton emission process is sometimes interpreted as ``sequential
decay via tails of higher-lying resonances''. Indeed, Eq.\
(\ref{eq:VIID1-wid-3-ot-ep}) shows that such resonances are very important for
the
true two-proton emission process. However, we shall argue that the sequential
interpretation may be misleading.
It can be shown that the resonance scattering can be interpreted in
terms of the \emph{time delay} $T_l(E,R)$ [e.g.\ \cite{Baz:1967}]:
\begin{equation}
T_l(E,R) \sim \frac{\Gamma(E)/4}{(E_r-E)^2+\Gamma(E)^2/4}.
\label{eq:IB-time-delay-res}
\end{equation}
Let us estimate the distance which the ``first'' emitted proton can travel while
the ``second'' is ``confined'' in the tail of the resonance. Among the known and
prospective two-proton emitters only the lightest $^{6}$Be and $^{12}$O have
such flight path comparable or exceeding $\sim 1$ fm. It can be seen in Fig.\
\ref{fig:VII:flight-dist} (a) that for $^{12}$O the estimated flight path can
achieve 8 fm, but the contribution of such situations to the total decay
probability
is minor, Fig.\ \ref{fig:VII:flight-dist} (b).  Furthermore,
having a typical scattering length in the nucleon-nucleon channel around 20 fm
even for $^{12}$O we can not maintain that a reliable spatial separation of
the core+$p$ subsystem is present.

So, the poles of the Greens' functions for two-body subsystems strongly increase
the probability of true two-proton decay when they come close to the three-body
decay window. However, only for one nuclear system ($^{12}$O) we can see
considerable sequential separation of subsystems in space and even in that
case the estimated contribution of sequential decay mechanism to the width is
minor.

\begin{figure*}[tb]
\centerline{
\includegraphics[width=0.33\textwidth]{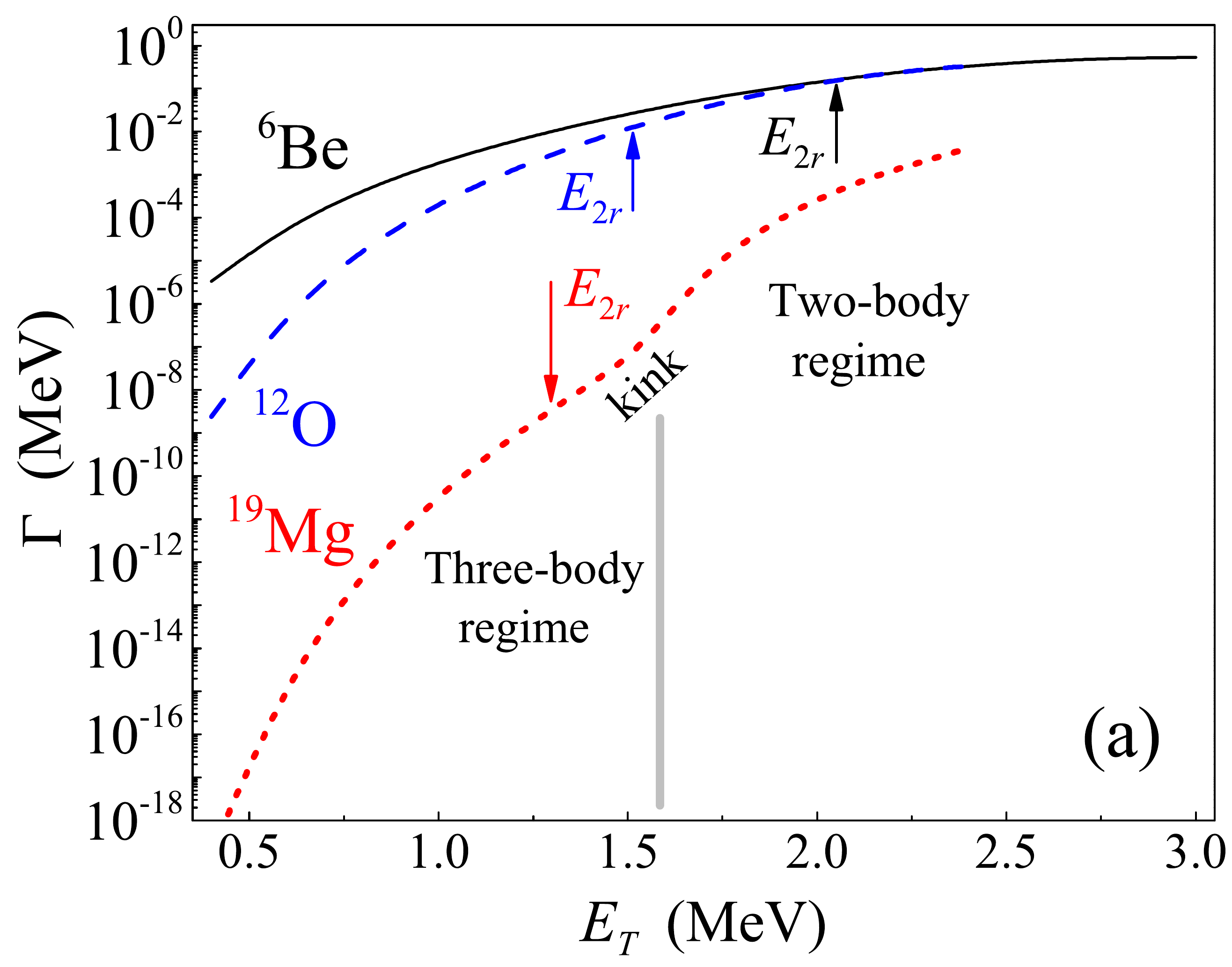}
\includegraphics[width=0.322\textwidth]{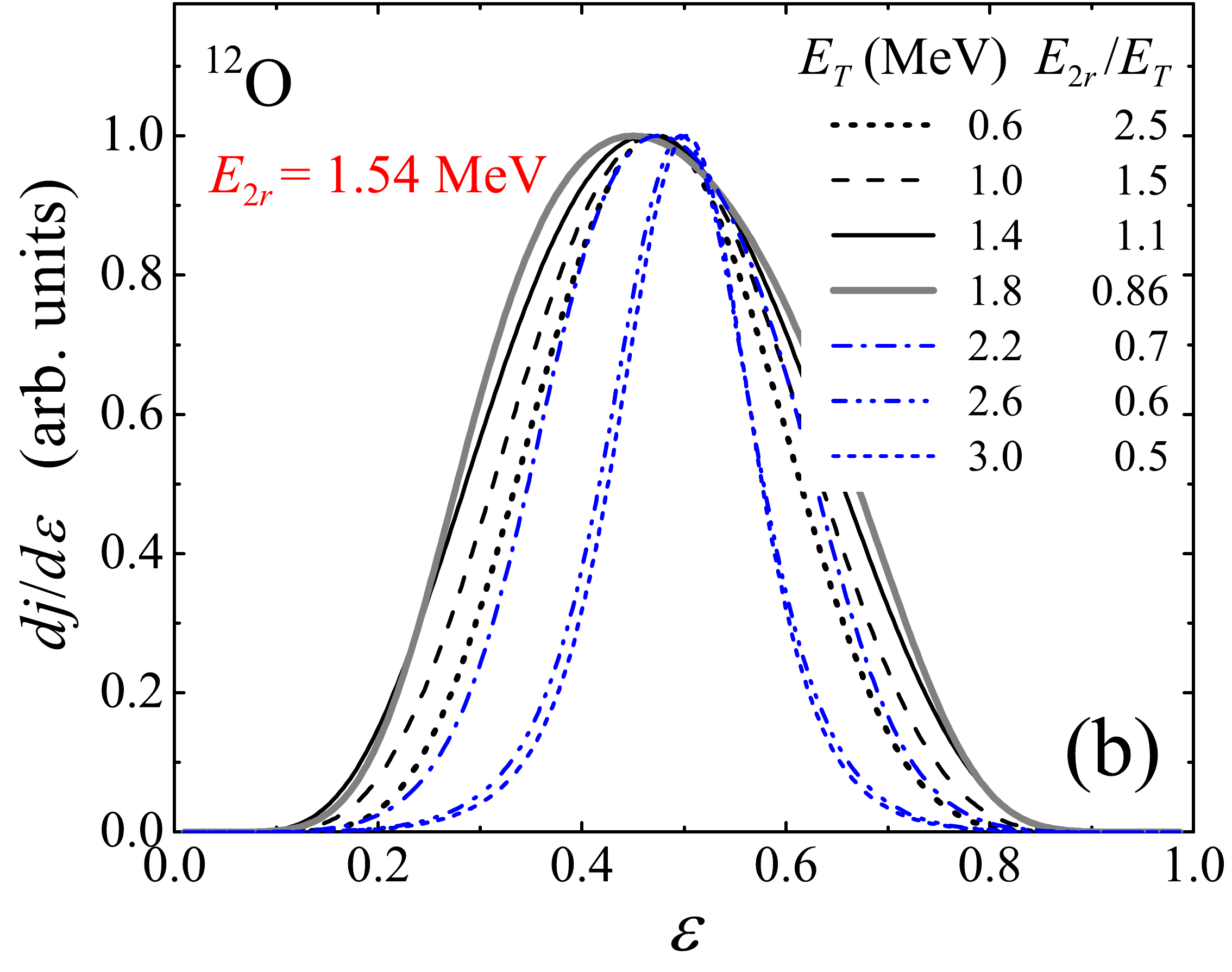}
\includegraphics[width=0.328\textwidth]{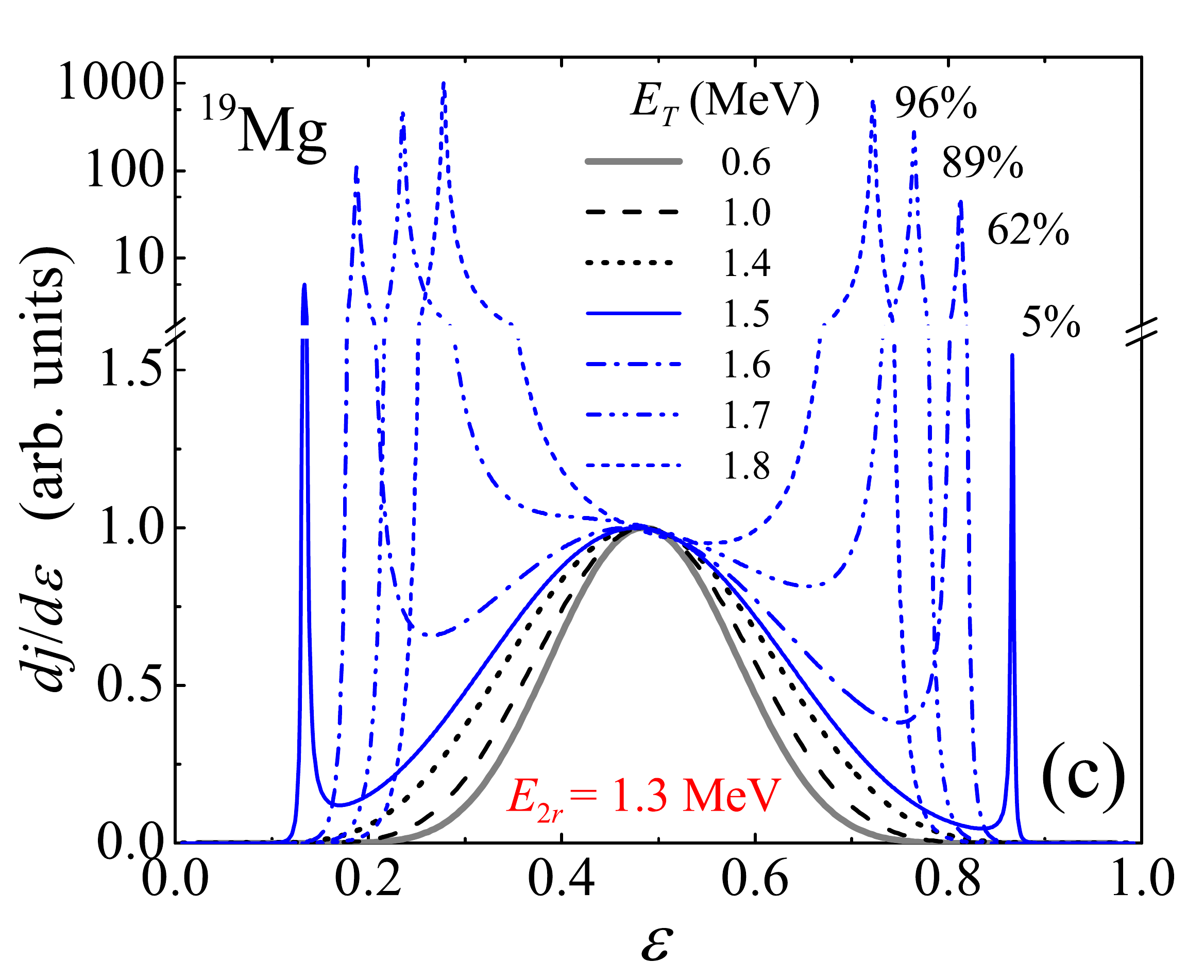}
}
\caption[Transition from true three-body to sequential decay.]{(Color online)
Transition from
the true three-body to the sequential decay for $^{6}$Be, $^{12}$O, and
$^{19}$Mg in the direct decay approximation Eq.\ (\ref{eq:VIID1-rmat-wid-3}).
The width vs.\ decay energy $E_T$ (a), evolution of energy distributions in the
``Y'' Jacobi system (core-$p$ channel) for the case of democratic decay (b), and
for radioactive decay (c). The curves in (b,c) are normalized to unity for
the maximum of the ``bell'' profile at $\varepsilon \approx 0.5$.}
\label{fig:VII:sim-seq}
\end{figure*}

Important insight about the transition from true three-body to sequential decay
can
be obtained using the direct decay model. In Fig.\ \ref{fig:VII:sim-seq} the
results of calculations for $^{6}$Be, $^{12}$O, and $^{19}$Mg with help of Eq.\
(\ref{eq:VIID1-rmat-wid-3}) are shown for a fixed energy $E_{2r}$ of the
resonance in the core+$p$ subsystem as a function of the three-body energy
$E_T$. We see a difference between the democratic decays (broad states in the
subsystems, overlapping with the decay window) and a radioactive decay (narrow
state in the subsystem). The curve for $^{19}$Mg in
Fig.~\ref{fig:VII:sim-seq}(a) shows two components: \emph{three-body regime}
(true three-body decay) and \emph{two-body regime} (sequential decay) with a
narrow transition zone in between (a kink in the line). In the cases of $^{6}$Be
and $^{12}$O (democratic decays) there is no sharp difference visible between
the two decay regimes and the transition is smooth.

The transition from the three-body to the two-body regime can be also
illustrated by the energy distribution in the core-$p$ channel.
Again, there is a qualitative difference between democratic and
radioactive decays. In the former case, when the two-body resonance
``enters'' the three-body energy window, the width of the distribution
suddenly becomes smaller, see Fig.\ \ref{fig:VII:sim-seq}(b).
On the other hand, if the state in the subsystem is narrow,
Fig.\ \ref{fig:VII:sim-seq}(c), the two ``horns'' appear in the energy
distribution corresponding to the proton energies defined by the intermediate
state. As long as the difference $E_{T}-E_{2r}$ is small, these horns do
not contribute significantly to the total width and the decay remains
effectively of the three-body type. Gradually the horns
become dominating, while the relative contribution of the three-body
``bell'' at $\varepsilon=E_T/2$ vanishes. It is reasonable to put the borderline
between true three-body decay and sequential decay as the moment when the
probabilities in the bell-like profile at $\varepsilon=E_T/2$ and in the
sequential horns become equal. From this one can derive a criterion
for the three-body decay mechanism:
%
%
\begin{equation}
\varepsilon_0 E_T < E_{2r},
\label{eq:VIIE2-cond-rad}
\end{equation}
where for radioactive decays $\varepsilon_0 \approx 0.76 - 0.84$ and for
democratic decays $\varepsilon_0 \approx 0.3 - 0.5$. Some uncertainty of the
estimated $\varepsilon_0$ values here is connected with particular selection of
parameter set for the system: $E_T$, $A$, $Z$, and $l$.

We note that the condition for the three-body decay proposed by
\textcite{Goldansky:1960}
was more restrictive:
\begin{equation}
E_T+\Gamma_{2r}/2<E_{2r}.
\label{eq:VIIE2-cond-gol}
\end{equation}

See also \cite{Alvarez:2008} for another formulation of how sequential and
direct
three-body decays may be distinguished.

\subsubsection{Diproton model}
\label{sec:VIIE3}

Historically the first idea of the diproton approximation appeared in the
original work of \textcite{Goldansky:1960}. Later, it was investigated in more
detail \cite{Goldansky:1961} and gradually it has become one of the major ideas
about the mechanism of the true $2p$ decay, frequently used as a synonym of it.
In this approach a single ``diproton'' particle (two protons in
$l=0$, $S=0$ state) is tunneling through the barrier along the
straight classical trajectory. The width in the ``T'' system is written as:
\begin{equation}
\Gamma_{\text{dp}}(E_T)  =  2 \gamma_{\text{dp}}^{2}  \int_0^{1} \!\! d
\varepsilon
\rho(\varepsilon E_{T}) \,P_{0}(E_{T}(1-\varepsilon)
,R_{\text{dp}},2Z_{\text{core}})\,,
\label{eq:VIIE3-dipro}
\end{equation}
where the typical definitions of the diproton channel radius \cite{Barker:2001}
is
\begin{equation}
R_{\text{dp}}=r_0(A_1^{1/3}+2^{1/3}),
\label{eq:VIIE3-dipro-rc}
\end{equation}
with $r_0=1.45$ \cite{Barker:2001}. Value $\gamma_{\text{dp}}$ is defined in an
ordinary R-matrix form (\ref{eq:IB-dimles-red-wid}). The function $\rho(E)$ is
the ``density of diproton states''  typically used in several forms:
\begin{eqnarray}
\rho(E) & = & E \, \delta(E-E_0) ,
\label{eq:VIIE3-dipro-delta} \\
\rho(E)  & \sim  & P_{0}(E,R_{\text{pp}},1) ,
\label{eq:VIIE3-dipro-pen} \\
\rho(E)  & \sim  & \sin^2[\delta_{0}(E)] .
\label{eq:VIIE3-dipro-mw}
\end{eqnarray}
Eq.\ (\ref{eq:VIIE3-dipro-delta}) corresponds to a ``fixed-energy diproton''.
It is evident that some energy should be contained in the internal motion of the
$p$-$p$ subsystem. In the early works it was taken $E_0 \approx 50-100$ keV
\cite{Goldansky:1961,Janecke:1965}. The diproton model was often used in a
trivial form with $E_0 = 0$ \cite{Brown:1991,Nazarewicz:1996,Ormand:1997}. The
results with $E_0 = 0$ for different $r_0$ values for $^{45}$Fe are given in
Fig.\ \ref{fig:VIIF1-lifetime-fe} (a). These results strongly overestimate $2p$
decay width and could be considered as its strict upper limit. Eq.\
(\ref{eq:VIIE3-dipro-pen}) corresponds to the  ``Coulomb-corrected phase
volume''. A radius of the $p$-$p$ channel of $R_{\text{pp}} \sim 1-2$ fm should
be
chosen here to reproduce reasonably the low-energy $p$-$p$ scattering
proporties. A treatment of the density in the spirit of the Migdal-Watson
approximation, Eq.\ (\ref{eq:VIIE3-dipro-mw}), was proposed by
\textcite{Barker:2001}. Here $\delta_0(E)$ is the phase shift for the $s$-wave
$p$-$p$ scattering. Results of the approaches (\ref{eq:VIIE3-dipro-pen}) and
(\ref{eq:VIIE3-dipro-mw}) can be well approximated by expression Eq.\
(\ref{eq:VIIE3-dipro-delta}) with an appropriate choice of $E_0$.

The diproton model in the form (\ref{eq:VIIE3-dipro}) has been \emph{inferred};
it
has never been \emph{derived}. To check the validity of this approach the
``dynamic'' diproton model was developed by \textcite{Grigorenko:2007} using the
analytical GF Eq.\ (\ref{eq:VIID-gf3}) of the simplified three-body Hamiltonian
with a diproton Ansatz. It was demonstrated that starting from
the Hamiltonian level, the results cannot be reduced to Eq.\
(\ref{eq:VIIE3-dipro}) with (\ref{eq:VIIE3-dipro-rc}).
Thus, the diproton model is typically used in an inconsistent way.
Correctly applied, the diproton model provides too small values of the width,
see Fig.\ \ref{fig:VIIF1-lifetime-fe}(a), ``diproton GF'' curve. Moreover,
Eq.~(\ref{eq:VIIE3-dipro}) leads to a single, narrow, low-energy peak in the
energy distribution in the ``T'' system, which differs from the experimental
distributions, see Figs.\ \ref{fig:VIIC1-corel-all-be}(c),
\ref{fig:VIIC1-corel-all-mg}(c), and \ref{fig:VIIC1-corel-models}(g). In
addition, Eq.~(\ref{eq:VIIE3-dipro}) provides an angle-independent decay
probability in the ``T'' system in contradiction to the experimental findings.


\subsection{Three-body model of $2p$ radioactivity}
\label{sec:VIIF}

In the three-cluster model of $2p$ radioactivity introduced by
\textcite{Grigorenko:2000b,Grigorenko:2002} the three-body Schr\"{o}dinger
equation with the complex energy
\begin{equation}
\left(  H_3-E +i\Gamma/2 \right)  \Psi_{3\varkappa}^{(+)}= 0
\label{eq:VIIF3-schred-1}
\end{equation}
is solved using hyperspherical harmonics (HH) method. The solution with outgoing
boundary condition is found in this method as
\begin{equation}
\Psi_{3\varkappa,JM}^{(+)}(\rho,\Omega_5)=\rho^{-5/2} \sum \nolimits _{K\gamma}
\chi_{\varkappa,K\gamma}^{(+)}(\rho)\,{\cal J}^{JM}_{K\gamma}(\Omega_5).
\label{eq:VIIF3-wf-3b-out}
\end{equation}
The hypermomentum $\varkappa=\sqrt{2ME}$ is the analogue
of the ordinary momentum in the HH three-body approach and the
hyperradius $\rho$ is a collective variable describing the ``breathing'' motion
of
the system
\begin{equation}
\rho=\frac{A_1A_2A_3}{A_1+A_2+A_3}\left( \frac{\mathbf{r}^2_{12}}{A_3} +
\frac{\mathbf{r}^2_{23}}{A_1} + \frac{\mathbf{r}^2_{31}}{A_2} \right).
\label{eq:VIIF3-rho}
\end{equation}
The hyperspherical harmonics $\mathcal{J}_{ K \gamma}$ are functions of the
5-dimensional ``solid angle'' $\Omega_5=\{\theta_{\rho},\Omega_x,
\Omega_y\}$. Here $\Omega_x$ and $\Omega_y$ are ordinary solid angles of the
Jacobi vectors $\mathbf{X}$ and $\mathbf{Y}$ and
$\tan(\theta_{\rho})=\sqrt{M_x/M_y} X/Y$.

For radioactivity problems with $|E_T| \gg \Gamma$ the following procedure was
found to be reliable. First, the discrete spectrum problem
\begin{equation}
\left(  H_3-E_{T} \right)  \tilde{\Psi}_{\text{3b}}=0
\end{equation}
is considered with \emph{some} box boundary conditions (e.g.\ zero or
quasistationary) yielding the ``box'' WF $\tilde{\Psi}_{\text{3b}}$ and the
value of the real resonant energy $E_{T}$. Then the inhomogeneous equation
\begin{equation}
\left(  H_3-E_{T} \right)  \Psi_{3\varkappa} ^{(+)}=-i\,\Gamma/2\,
\tilde{\Psi}_{\text{3b}}
\end{equation}
is solved with arbitrary $\Gamma$ and the actual width is defined afterwards
as the flux through a hypersphere of large radius divided by the normalization
within this radius:
\begin{equation}
\Gamma=\frac{j}{N}=\frac{\text{Im} \left. \left[ \int d\Omega_{5}\;
\Psi_{3\varkappa}^{(+)\ast}\rho
^{5/2}\frac{d}{d\rho}\rho^{5/2}\Psi_{3\varkappa}^{(+)} \right] \right\vert
_{\rho=\rho_{\max}}
}{M\;\int d \Omega_{5}\int_{0}^{\rho_{\max}}\rho^{5}d\rho\;\left\vert
\Psi_{3\varkappa}^{(+)}\right\vert ^{2}}\; .
\label{eq:VIIF3-wid-3b-def}
\end{equation}
Formula (\ref{eq:VIIF3-wid-3b-def}) is intuitive. Still, some formal
issues connected with its derivation can be found in Ref.\
\cite{Grigorenko:2007}

The asymptotic form of the three-body potentials in the hyperspherical harmonics
method is
\begin{equation}
V_{K \gamma, K' \gamma'}(\rho) = \frac{U_{K \gamma, K' \gamma'}}{\rho^{N_{K
\gamma, K' \gamma'}}} + \frac{C_K}{\rho^2}\delta_{K
\gamma, K' \gamma'}+ \frac{\varkappa \eta_{K \gamma, K' \gamma'}}{M \,\rho} \,,
\label{eq:VIIF3-pot-ass}
\end{equation}
where the multiindex $\{K \gamma\}=\{K,L,S,l_x, l_y,s_x\} $ is a complete set of
quantum numbers for three clusters. The matrix $U_{K \gamma, K' \gamma'}$ arises
due to contributions from the short-range nuclear forces, and $N_{K \gamma, K'
\gamma'} \geq 3$ are integers: the effective contribution of the
short-range forces decreases as $\rho^{-3}$ or faster in hypersherical space.
The diagonal centrifugal term $C_K=\mathcal{L}_K(\mathcal{L}_K+1)$ depends on
the ``effective angular momentum'' $\mathcal{L}_K=K+3/2$. Coulomb pairwise
potentials generate the long-range part of the hyperspherical potentials
behaving as $\rho^{-1}$. From the technical side, these three-body
hyperspherical ``Coulomb like'' potentials cause problems due to long-range
channel coupling (nondiagonal
``Sommerfeld parameters'' $\eta_{K \gamma, K' \gamma'}$) that does not allow one
to decouple the HH equations in the asymptotic region. To deal with this
problem, the finite-size potential matrix (in truncated hyperspherical basis)
can be diagonalized with respect to the long-range term by the orthogonal
transformation $\tilde{V}=A^T V A$:
\begin{equation}
\tilde{V}_{K \gamma, K' \gamma'}(\rho) = \frac{\tilde{U}_{K \gamma, K'
\gamma'}}{\rho^{3}} + \frac{C_{K \gamma, K' \gamma'}}{\rho^2} + \frac{\varkappa
\eta_{K
\gamma}}{M \, \rho}\delta_{K \gamma, K' \gamma'} \,.
\end{equation}
This potential includes nondiagonal ``centrifugal'' terms $C_{K \gamma, K'
\gamma'}$ and, to achieve the asymptotics in the diagonalized representation, we
still need to go very far in $\rho$, where the terms  $\sim \rho^{-2}$
become negligible compared to those with $\sim \rho^{-1}$. At such
values of $\rho$, the hyperradial part of the asymptotic solution with pure
outgoing
nature can be constructed using the Coulomb functions:
\begin{eqnarray}
\chi_{\varkappa,K \gamma}^{(+)}(\rho) & \sim & \sum \nolimits_{ K' \gamma'} \!
A_{K \gamma, K' \gamma'} \nonumber \\
 & \times  & \left[ G_{\mathcal{L}_0}(\eta_{K' \gamma'},\varkappa
 \rho)+iF_{\mathcal{L}_0}(\eta_{K' \gamma'},\varkappa \rho) \right]\,.\;
\label{eq:VIIF3-wf-ass}
\end{eqnarray}
These boundary conditions are exact in the \emph{truncated} HH basis at a
hypersphere of \emph{very large} radius, depending on energy and basis size. For
example, for $^{45}$Fe with $E_T=1.154$ MeV and $K_{\max}=20$, radii
$\rho_{\max}$ between 500 and 2000 fm are needed to get reasonable solutions.

The Coulomb three-body decays of nuclei have also been studied by other HH
based theoretical methods. The S-matrix approach was used in calculations of
$^{6}$Be g.s.\ in Refs.\ \cite{Danilin:1993,Vasilevsky:2001,Descouvemont:2006}
and $^{12}$O g.s.\ in Ref.\ \cite{Grigorenko:2002}. WKB calculations with
adiabatic HH potentials were done for the width of the first excited state of
$^{17}$Ne in \cite{Garrido:2008}. Having comparable quality for internal
structure, these methods may have deficiencies dealing specifically with
radioactivity problems: S-matrix calculations are difficult for very narrow
states and WKB can not provide momentum distributions.
We note that other HH based methods have been used to calculate the momentum
distributions from three-body decay of several excited states in light nuclei
\cite{Alvarez:2008a,Alvarez:2010}.

\subsubsection{Lifetimes}
\label{sec:VIIF2}

The Coulomb and $p$-$p$ potentials are the known ingredients of the
three-cluster calculations. In the light nuclei, the core-$p$ potentials are
fixed by fitting to the experimental single-particle spectra of the core-$p$
subsystem. In most cases this allows to obtain a reasonable description of
the energy and structure of the ground and lowest excited states. For heavier
nuclei, where the single particle description is not well justified and spectra
of subsystems are often not known experimentally, certain systematic guidelines
were proposed by \textcite{Grigorenko:2003}. Choosing the depths of potential
components with different $l$ values we can vary the structure of the $2p$
emitter for a fixed proton decay energy $E_{2r}$ in the core-$p$ subsystem.
A short-range (in the hyperradius) three-cluster potential is used to vary the
decay energy $E_T$.

For the majority of prospective $2p$ emitters the decay energies are not known.
Therefore, the lifetime predictions are provided in
terms of possible lifetime bands bound from above and below by calculations with
pure configurations. The calculations for the $p$-$f$ shell nucleus $^{45}$Fe
for different cases of configuration mixing are provided in Fig.\
\ref{fig:VIIF1-lifetime-fe}. It can be seen that within the three-body model the
precise lifetime data can be used to extract structural information about $2p$
emitters.

\begin{figure}[t]
\centerline{
\includegraphics[width=0.49\textwidth]{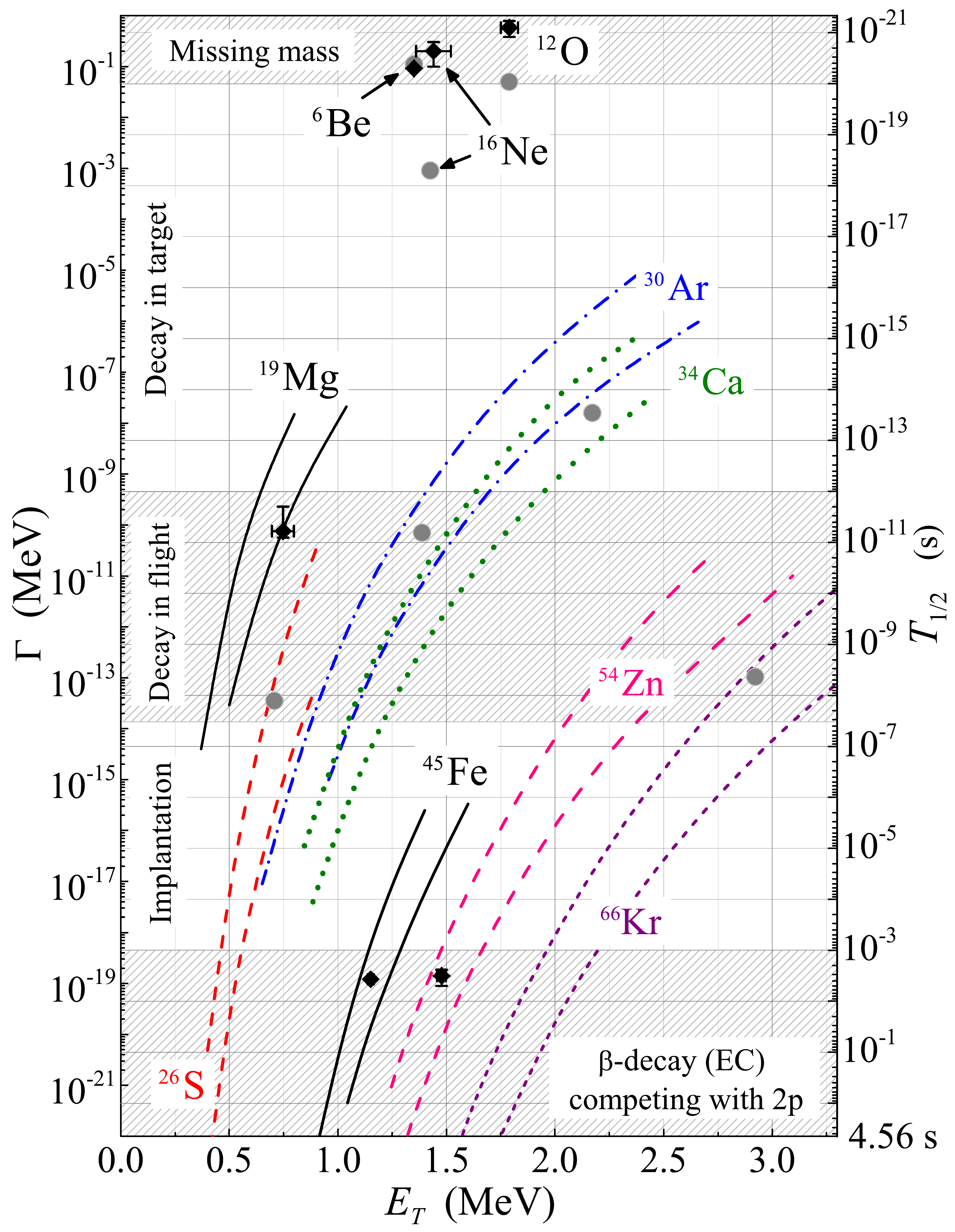}
}
\caption{(Color online) Lifetime vs.\ decay energy systematics for several known
and
prospective true $2p$ emitters calculated in three-body model. Hatching
indicates
lifetime ranges accessible to different experimental techniques. Experimental
results are shown by diamonds. Gray circles show specific predictions, where
available.}
\label{fig:VIIF1-lifetime-all}
\end{figure}

A broader view of the true $2p$ decay phenomenon is provided in Fig.\
\ref{fig:VIIF1-lifetime-all}. For the light $2p$ emitters specific lifetimes
are predicted (gray circles). For the heavier $2p$ emitters the predicted
lifetime ``bands'' are in a good agreement with the experimental data. The
lifetime range for the known true $2p$ emitters spans about 18 orders of the
magnitude. This plot emphasizes the complexity of the problem requiring a
variety of
experimental methods to cover the possible lifetime range.

\subsubsection{Spatial correlations}
\label{sec:VIIF3}

\begin{figure}[t]
\centerline{
\includegraphics[width=0.49\textwidth]{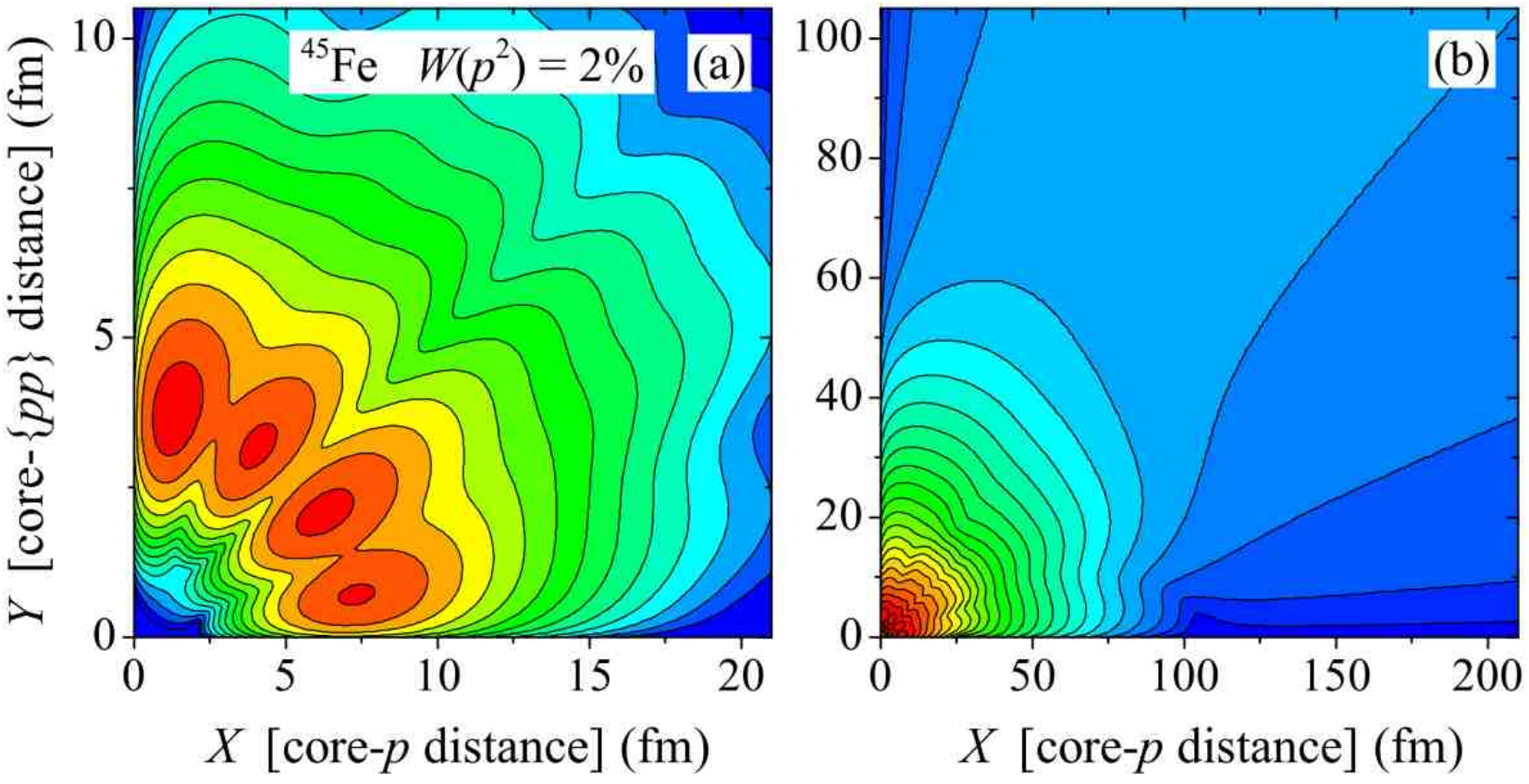}
}
\caption{(Color online) Spatial correlations in the $^{45}$Fe WF in the ``T''
system for
$W(p^2)=2\%$. Panels (a) and (b) shows different radial ranges. Logarithmic
scale: two (a) and one (b) contours per order of the magnitude.}
\label{fig:VIIF3-spatial-corel}
\end{figure}

The evolution of spatial correlations in the three-body model is illustrated by
the case of the $^{45}$Fe structure with $W(p^2)=98\%$, see Fig.\
\ref{fig:VIIF3-spatial-corel}. In the Jacobi ``T'' system, this case of almost
pure $[f^2]_0$ configuration has a very eminent correlation pattern with four
peaks in the internal region. Such correlations can be related to the so-called
Pauli focusing -- counting of excitation quanta in the ``T'' system.

It can be seen in Fig.\ \ref{fig:VIIF3-spatial-corel} that under the barrier the
four-hump structure is dissolved and a different correlation pattern is formed
while transition to the asymptotic region takes place. Finally, the momentum
distribution with one peak in energy between two protons is formed, see Fig.\
\ref{fig:VIIF4-pp-corel-structure}(b). This happens due to $p$-$p$ interaction
in the subbarrier region \cite{Grigorenko:2000b}. It is evident here that the
penetration process influences the distribution strongly and the information
about correlations in the nuclear interior is present in the final momentum
distributions in a very indirect way. Detailed theoretical calculations of the
subbarrier propagation process are required.

\subsubsection{Momentum correlations and nuclear structure}
\label{sec:VIIF4}

The correlation pictures calculated in the three-body model are provided in
Figs.\ \ref{fig:VIIC1-corel-all-be}(a,b), \ref{fig:VIIC1-corel-all-mg}(a,b), and
\ref{fig:VIIC1-corel-models}(e,f) for $^{6}$Be, $^{19}$Mg and $^{45}$Fe,
respectively. Despite the strong differences there are certain common features
in these distributions.

\noindent (i) There are Coulomb suppression regions due to $p$-$p$ repulsion at
$\varepsilon \sim 0$ in the ``T'' system and at $\varepsilon \sim 0.5$,
$\cos(\theta_k) \sim -1$ in the ``Y'' system. Analogous regions due to core-$p$
repulsion appear at $\varepsilon \sim 0.5$, $\cos(\theta_k) \sim \pm 1$ in the
``T'' system and at $\varepsilon \sim 0$, $\varepsilon \sim 1$  in the ``Y''
system. These regions grow drastically as we move from light to heavier $2p$
emitters.

\noindent (ii) In the ``Y'' system the particles are concentrated around
$\varepsilon \sim 0.5$. This effect becomes more pronounced as the
core charge increases. This particular aspect of the correlation picture is
consistent
with
the original predictions of Goldansky and with calculations performed in a
simplified
direct decay model, see Fig.\ \ref{fig:VIID1-core-p-syst}. In the ``T''
system this effect appears as a concentration of particles in the T-like
configuration with $\cos(\theta_k) \sim 0$.

\begin{figure}[t]
\centerline{
\includegraphics[width=0.4\textwidth]{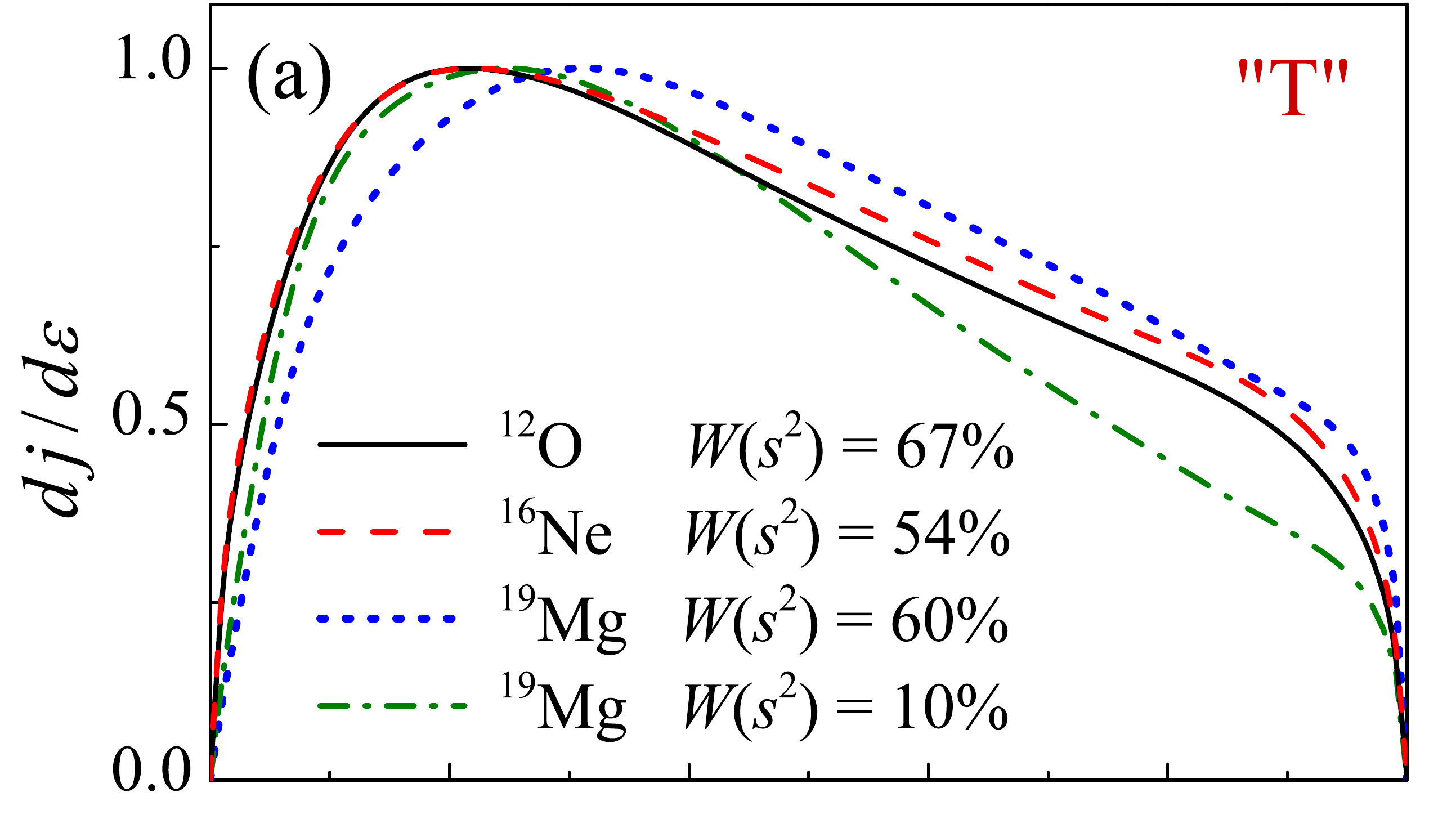}
}
\centerline{
\includegraphics[width=0.4\textwidth]{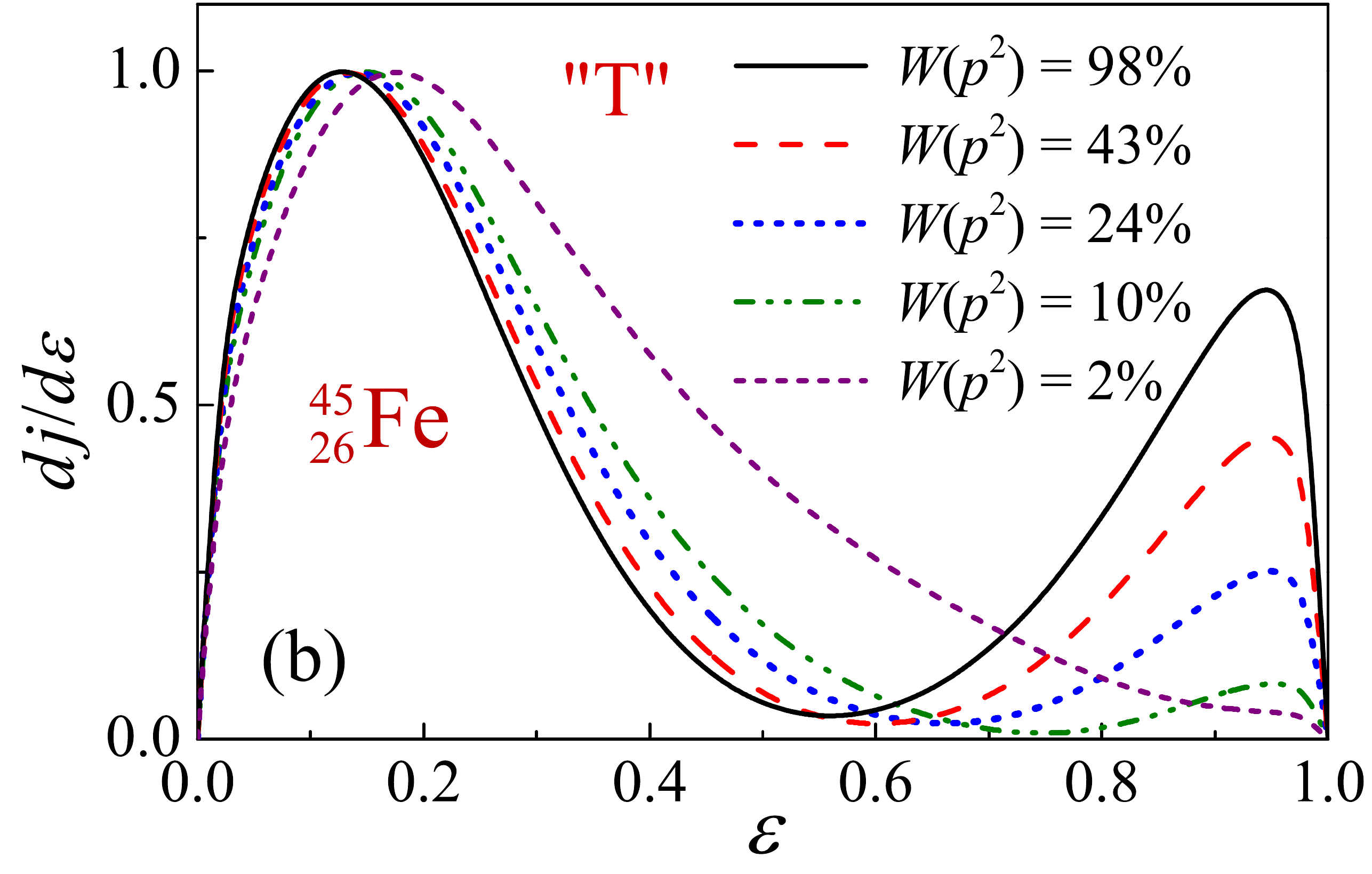}
}
\caption{(Color online) The energy distributions between two protons (``T''
Jacobi system) in
different $s$-$d$ shell nuclei (a) and with different assumptions about the
internal
structure of $^{45}$Fe (b).  $W(l^2)$  is  the  weight  of  the  $[l^2]$
configuration  in  the  nuclear interior. All distributions are normalized to
unity maximum value. From \cite{Grigorenko:2003a}.}
\label{fig:VIIF4-pp-corel-structure}
\end{figure}

\noindent (iii) The $p$-$p$ final-state interaction gives rise to enhancements
at small values of $\varepsilon$ in the ``T'' system. This enhancement varies
from minor to a very expressed one depending on the particular dynamics of the
system, see also Fig.\ \ref{fig:VIIF4-pp-corel-structure}.

An important question about $2p$ radioactivity asked from the earliest days of
this research is ``Which kind of information about internal structure can be
extracted from the $2p$ decay data?'' It was not clear in advance that
information about the nuclear interior would survive in the process of
penetration
of protons through the Coulomb barrier. The calculations in the three-body model
have demonstrated that correlations in the $p$-$p$ channel \emph{are
sufficiently sensitive} to nuclear structure to be discussed as prospective
instrument for nuclear spectroscopy in both  $s$-$d$ and $p$-$f$ shell nuclei,
see Fig.\ \ref{fig:VIIF4-pp-corel-structure}. These correlations are manifested
by the $\varepsilon$ variable in the ``T'' system and the $\cos(\theta_k)$
variable in
the ``Y'' system (Figs.\ \ref{fig:VIIC1-corel-all-be} and
\ref{fig:VIIC1-corel-models}).

\subsubsection{Long-range character of Coulomb interaction in the three-body
continuum}
\label{sec:VIIF5}

\begin{figure}[t]
\begin{center}
\includegraphics[width=0.25\textwidth]{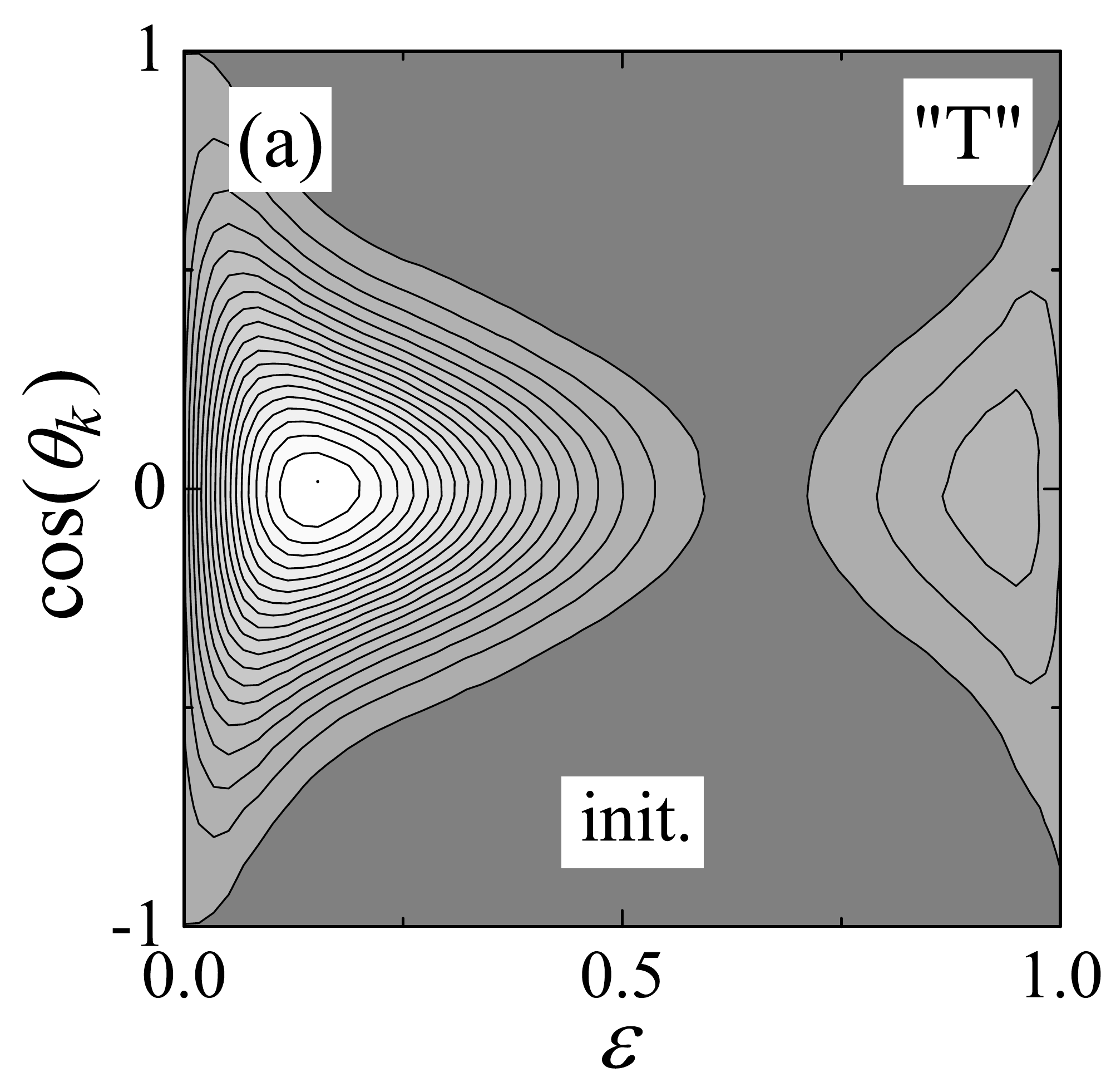}
\includegraphics[width=0.22\textwidth]{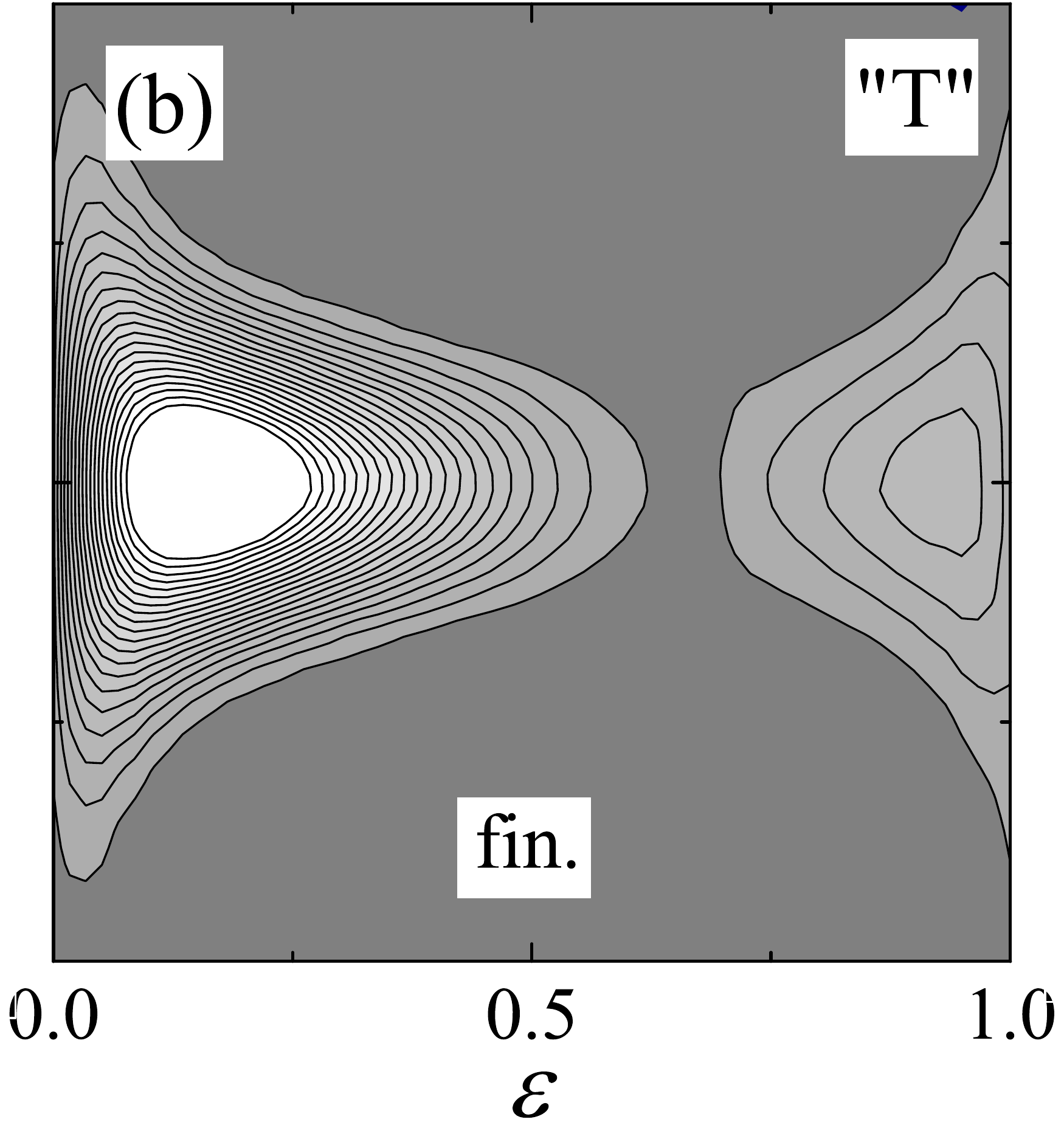} \\
\includegraphics[width=0.4\textwidth]{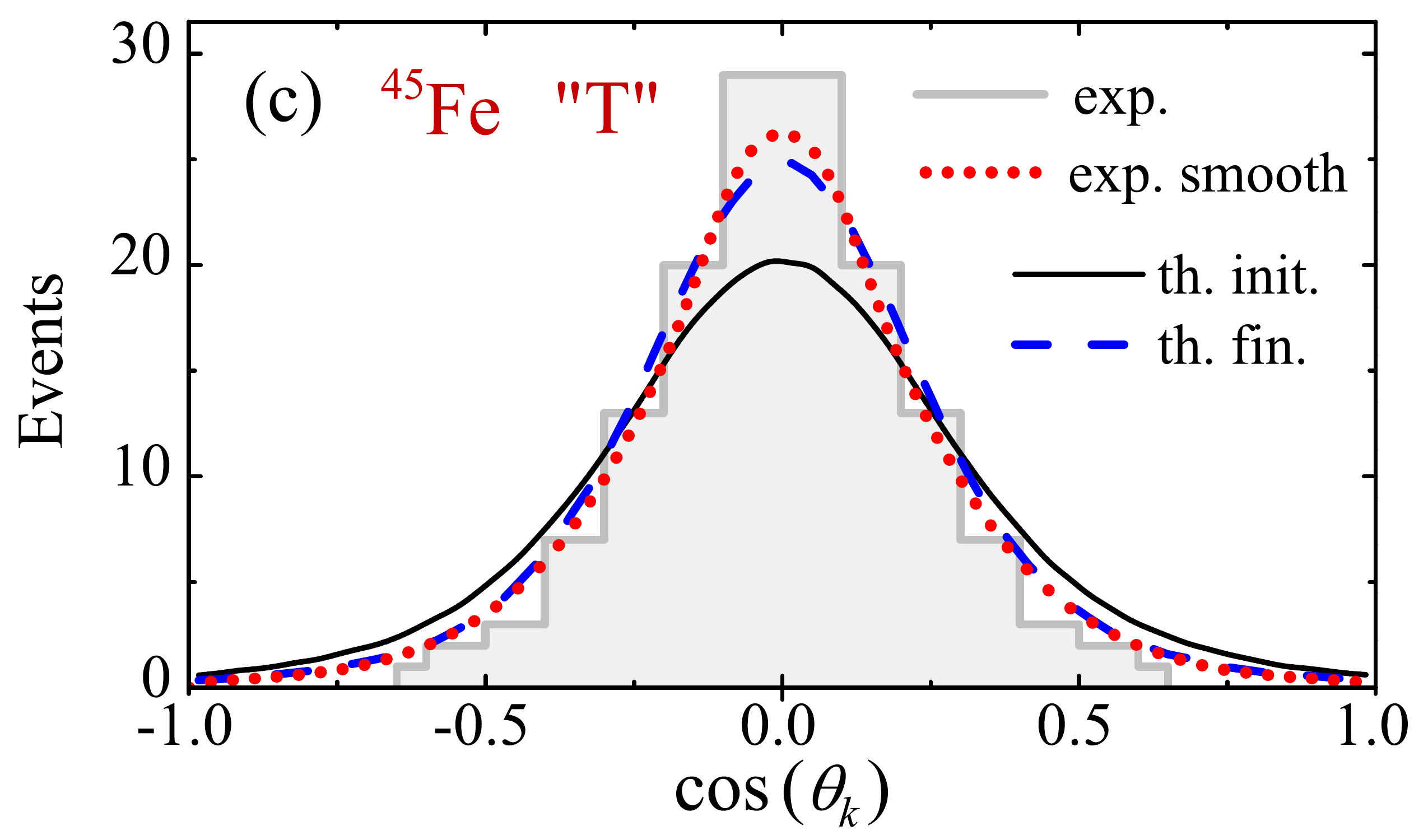}
\end{center}
\caption[Momentum density distribution for $^{45}$Fe without and with classical
extrapolation.]{Contour maps of the momentum density distribution on
the kinematical plane $\{\varepsilon,\cos(\theta_k) \}$ for $^{45}$Fe in the
``T'' Jacobi coordinate system without (a) and with (b) classical
extrapolation. Panel (c) shows comparison with experimental data of
\cite{Miernik:2007b}. Experimental and theoretical events are convoluted with
the experimental resolution. From \cite{Grigorenko:2010}.}
\label{fig:VIIF4-class-extrapol}
\end{figure}

The formulation of approximate boundary conditions Eq.\ (\ref{eq:VIIF3-wf-ass})
is satisfactory for distances around 10$^{3}$ fm. Such distances appear to be
insufficient to get converged momentum distributions for heavy $2p$ emitters.
Extrapolation by means of classical trajectories was used by
\textcite{Grigorenko:2010} to improve the momentum distributions. Classical
extrapolation modifies mainly the $\cos(\theta_k)$ distribution in ``T'' system
and
the $\varepsilon$ distribution in ``Y'' system, see Fig.\
\ref{fig:VIIF4-class-extrapol}(a,b). The effect of classical extrapolation is
important already on the current level of experimental precision, Fig.\
\ref{fig:VIIF4-class-extrapol}(c). A huge range is required both for the
extrapolation range ($\sim 10^{5}$ fm) and for the starting point  of  the
classical  procedure  ($\sim 10^{3}$ fm) under  typical $2p$ decay conditions.
The maximal extrapolation radius is comparable to atomic distances and
sensitivity of $2p$ momentum distributions to atomic screening of nuclear
Coulomb potential can be found. This brings the $2p$ decays to the borderline
with atomic phenomena.

\subsection{Two-proton radioactivity and many-body nuclear structure}
\label{sec:VIIG}

A very important aspect of $2p$ radioactivity studies is the understanding of
the
connection between specific asymptotic observables ($2p$ correlations)
and the WF structure in the internal region (nuclear spectroscopy). In Sec.\
\ref{sec:VIIF4} such a connection was demonstrated for the three-body model of
true $2p$ decay. However, the discussed three-body approach should be regarded
as a \emph{first approximation} model of this process.
By analogy to two-body decay, the exponential component of the width
is expected to be treated appropriately in this method.
The three-cluster model accounts well for single-particle degrees
of freedom in the core-$p$ channel. Therefore, calculation of the
``preexponent''
in three-cluster model is well justified only for the light closed-shell
(closed-subshell) systems or systems with closed-shell (closed-subshell) core.
With increasing mass number fewer systems can be found along the dripline whose
dynamics can be reasonably described in the three-cluster approximation. Taking
many-body effects into account in the $2p$ decay calculations is a natural next
step in the studies of the phenomenon. Here, we briefly sketch several possible
approaches to this problem which are being considered presently.

\subsubsection{Spectroscopic information in R-matrix approaches}
\label{sec:VIIG1}

The information about many body structure is embedded in the simplified models
of two proton radioactivity via spectroscopic factors. The single particle
proton factors $\theta^2$ Eq.\ (\ref{eq:IB-r-mat-wid}) were used for this
purpose
within the direct decay model Eq.\ (\ref{eq:VIID1-rmat-wid-3-kr})
\cite{Kryger:1995,Barker:1999,Brown:2002,Barker:2003}.

The diproton spectroscopic factors [for use within the diproton model Eq.\
(\ref{eq:VIIE3-dipro})] are obtained by projecting WFs of valence protons on the
``diproton quantum numbers'' for two protons (relative angular momentum zero and
total spin zero). It is also typically assumed that the relative motion of the
protons is represented by the lowest oscillator. Calculations of this type can
be found in Refs.\ \cite{Brown:1991,Brown:2002,Barker:2003}.

\subsubsection{Three-body model plus RMF amplitudes}
\label{sec:VIIG2}

A simple method of structure treatment in the three-cluster model was proposed
by
\textcite{Fomichev:2010}. This approach is equivalent to a standard method
used in the R-matrix description of two-body decays.  The major
components of the three-body cluster WFs with $J^{\pi}=0^+$ can be written in a
schematic spectroscopic notation as
\begin{equation}
\Psi_3^{(+)}   = \sum \nolimits _i X_i [l_i^2]_0 \,.
\label{eq:VIIG-wf3}
\end{equation}
The values which can be put in correspondence with the components of the
three-body WF Eq.\ (\ref{eq:VIIG-wf3}), to take into account the many-body
structure,
are overlaps of the many-body WFs of the precursor-daughter pair multiplied by a
combinatorial term. They can be written in the same spectroscopic notation:
\begin{equation}
\left( \frac{A!}{2!(A-2)!} \right) ^{1/2} \langle \Psi_{A} |
\Psi_{A-2} \rangle = \sum \nolimits_i \tilde{X}_i [l_i^2]_0 \,.
\label{eq:VIIG-psi-over}
\end{equation}
For WFs normalized to unity in the internal region the amplitudes of overlaps do
not sum to unity in the general case in contrast with amplitudes of components:
\begin{equation}
\sum \nolimits _i |X_i|^2 \equiv 1\;, \qquad N_{2p} = \sum \nolimits _i
|\tilde{X}_i|^2 \neq 1 \,.
\label{eq:VIIG-norm}
\end{equation}
Therefore the three-body width as a function of weights for all considered
single-particle configurations $\{ X^2_i \}=\{X^2_1, \ldots , X^2_n \}$ should
be renormalized as:
\begin{equation}
\Gamma(\{ X^2_i \}) \rightarrow  N_{2p} \Gamma (\{ \tilde{X}^2_i/N_{2p} \}).
\end{equation}
For systems which are not closed-shell/closed-subshell, the coefficient $N_{2p}$
is considerably different from unity already in the Hartree approximation.
The higher complexity of the three-body decays is reflected in the fact that
inclusion of structure information can not be reduced to multiplication by some
factor, but should precede three-body calculations.

An advantage of the method is the opportunity to use relatively standard
input from many-body theory, Eq.\ (\ref{eq:VIIG-psi-over}).
For example, \textcite{Fomichev:2010} calculated these amplitudes in the
relativistic
mean field model \cite{Litvinova:2008}. A disadvantage is the different
treatment of the internal structure and the decay process. The same problem,
however, exists also in the conventional two-body R-matrix phenomenology. The
center of mass is treated appropriately in the three-body decay calculations,
but only effectively in the many-body part.

\subsubsection{Shell model approaches}
\label{sec:VIIG3}

Several shell-model approaches have been developed for studies of two-nucleon
decays.

The Gamow shell model is based on the utilization of the so-called Berggren
basis \cite{Michel:2003}. This is a basis composed of Gamow states --
single-particle states with decay asymptotic $\sim e^{ikr}$. For homogeneous
Schr\"odinger equations such states should have complex energies. The
eigenenergy of
the shell-model Hamiltonian diagonalized in such a basis is naturally complex as
well, with the imaginary part interpreted as a half-width. Technicalities make
application of this method problematic for extremely narrow states. So far, it
has been applied to light neutron-rich systems \cite[and Refs.\
therein]{Michel:2010}.

The real-energy continuum shell model was employed for studies of narrow states
decaying by two-nucleon emission by \textcite{Rotureau:2005,Rotureau:2006} and
\textcite{Volya:2006}. In this approach the Hamiltonian is split in two parts
\begin{equation}
H = H_{\mathcal{QQ}} + H_{\mathcal{QP}} + H_{\mathcal{PQ}} + H_{\mathcal{PP}}.
\label{eq:ham-csm}
\end{equation}
Functional subspaces $\mathcal{Q}$ and $\mathcal{P}$ can be identified as
``internal region'' with discretized (shell-model) basis $\phi_i$ and ``external
region'' with continuum basis $\phi_E$ defined by projection operators:
\begin{equation}
\hat{\mathcal{Q}} = \sum_i \left| \phi_i \rangle \langle \phi_i \right|, \qquad
\hat{\mathcal{P}} = \int dE \left| \phi_E \rangle \langle \phi_E \right|.
\label{eq:csm-proj}
\end{equation}
Elimination of the subspace $\mathcal{P}$ can be done by the Green's function
methods leading to effective (energy dependent and non-Hermitian) Hamiltonian
$\mathcal{H}$ in $\mathcal{Q}$
\begin{equation}
\mathcal{H}_{\mathcal{QQ}}(E) = H_{\mathcal{QQ}} + H_{\mathcal{QP}}
G^{(+)}_P(E)H_{\mathcal{PQ}},
\label{eq:ham-csm-eff}
\end{equation}
which allows straightforward application of powerful shell-model methods. An
advantage of the continuum shell model is the use of extensive shell model
expertise and applicability to corresponding broad range of nuclear systems.

The continuum shell model method of \cite{Volya:2006} was applied in
extensive studies of the neutron-rich helium and oxygen isotopes providing good
description of the known cases of $2n$ decay and predictions for several
unknown.
In this approach the residual interaction in subspace  $\mathcal{P}$ is
neglected. Therefore, in the sense of the three-body continuum dynamics this
model is analogous to the direct decay model, Sec.\ \ref{sec:VIIE1}. It can be
shown that the expression for the $2n$ width from
\cite{Volya:2006} can be approximated by Eq.\ (\ref{eq:VIID1-rmat-wid-3-app}).

The shell model embedded into continuum (SMEC) was applied by
\textcite{Rotureau:2006} to studies of $2p$ radioactivity using either the
diproton or direct decay Ansatz. The later is called ``sequential 2p emission''
having in mind ``emission via tails of higher-lying states'', see also the
discussion of Sec.\ \ref{sec:VIIE2} to avoid confusion of terminology. The
calculated diproton lifetimes for $^{45}$Fe, $^{48}$Ni, and $^{54}$Zn are
typically exceeding the experimental values by a few times, see, e.g., Fig.\
\ref{fig:VIIF1-lifetime-fe} for $^{45}$Fe results. A possible reason could be
that
the diproton approximation tend to underestimate width, see the
discussion in Section \ref{sec:VIIE3}. The direct decay lifetimes calculated for
$^{45}$Fe are 20-50 times larger than experimental one, see Fig.\
\ref{fig:VIIF1-lifetime-fe}.

 A disadvantage of the shell-model methods is that no matching to the three-body
 boundary conditions was realized so far and calculations of momentum
 distributions for decay products are not possible. The necessity to incorporate
 three-body asymptotics in the SMEC calculations was emphasized in
 \cite{Rotureau:2006,Blank:2008a}. The formalism for this was proposed in
 \cite{Rotureau:2006} using hyperspherical decomposition
 (\ref{eq:VIIF3-wf-3b-out}) and boundary conditions (\ref{eq:VIIF3-wf-ass}). The
 obstacle, which  \textcite{Rotureau:2006} faced here is in the contact
 character of
 the residual interaction in SMEC. Delta-function interactions lead to collapse
 of
 three-body calculations. Use of finite-range residual interaction is
 technically complicated in SMEC \cite{Rotureau:2006}.

An important conceptual problem exists that should be common to
all shell-model approaches. The asymptotic interaction in the
$p$-$p$ channel (in the $\mathcal{P}$ subspace) should be the ``vacuum'' $p$-$p$
potential. This should be reconciled in the internal region with the shell-model
residual interaction which has effective character, reflecting many-body
dynamics of nuclear interior. This reconciliation should take place somewhere
close to the nuclear surface. The observables for $2p$ decay are especially
sensitive to this radial range, see Sec.\ \ref{sec:VIIF3}, making it an
important issue.

\subsubsection{Microscopic cluster models}
\label{sec:VIIG4}

Microscopic cluster models do not have the conceptual problem mentioned in the
end of the previous section utilizing the same (quasi) realistic nucleon-nucleon
interactions in all space. Microscopic resonating group methods (RGM) have been
applied to the lightest true $2p$ emitter $^{6}$Be.

The complex scaling method of \cite{Csoto:1994a} was the first example of such
studies. In this method the boundary conditions are not explicitly formulated
and therefore only resonance widths can be calculated.

Algebraic version of RGM \cite{Vasilevsky:2001,Nesterov:2010} for three-cluster
systems use WF in the form
\begin{equation}
\Psi(A) = \mathcal{A} \left[ \Psi(A_1)\Psi(A_2)\Psi(A_3)
\Psi_{3\varkappa}(\rho,\Omega_5) \right],
\label{eq:label}
\end{equation}
where $\Psi(A_i)$ are cluster WFs, $A=A_1+A_2+A_3$, and  $\Psi_{3\varkappa}$ is
three-body scattering WF in the ``S-matrix representation'' (three-body plane
wave plus three-body outgoing wave). The derived S-matrix is diagonalized and
the matrix elements with the most expressed resonance behavior (``eigenphases'')
are used to extract resonance energy and width;
\[
\left. d^2\delta(E)/dE^2 \right|_{E=E_T}=0, \qquad
\Gamma = 2\left. [d\delta(E)/dE]^{-1} \right|_{E=E_T}.
\]
A hyperspherical decomposition was used for WF $\Psi_{3\varkappa}$ as especially
suitable for imposing long-range boundary conditions for true three-body decays.
A somewhat simpler version of the boundary conditions than
(\ref{eq:VIIF3-wf-ass}) was
realized in \cite{Vasilevsky:2001} with diagonal terms only retained in the
Coulomb term of Eq.\ (\ref{eq:VIIF3-pot-ass}). The algebraic method employs
two different hyperspherical basis sets for the nuclear interior and asymptotic
providing a natural and controllable treatment of these two regions. The
disadvantage
is that the calculations already for $^{6}$Be seem to be not sufficiently
converged at the present level of sophistication compared to fully converged
pure three-body calculations \cite{Grigorenko:2009c}.

A general challenge for microscopic cluster models is that realistic
calculations can be performed only for the lightest systems with effective
nucleon-nucleon interactions typically employed. However, modern trend is to
move to heavier systems and realistic interactions.

\subsection{Three and four proton emission}
\label{sec:VIIJ}

The beta-delayed three-proton decay of $^{45}$Fe was reported by
\textcite{Miernik:2007}. Most likely this is a sequential decay and it has been
discussed already in Sec.\ \ref{sec:IV}. Three-proton decay of highly excited
states of $^{16}$Ne, populated in fragmentation reaction, via narrow
intermediate states in $^{15}$F and $^{14}$O, was observed by
\textcite{Mukha:2009}. Theoretical prospects of studies of this decay branch are
so far unclear.

Following the analogy with the true $2p$ emission ($1p$ emission is
energetically prohibited) we can define the true $4p$ emission (emission of $i$
protons is energetically prohibited for all $i<4$). The only candidate to
fulfil this condition studied so far is $^{8}$C. It was shown recently that it
is not a true $4p$ emitter \cite{Charity:2010}. This nucleus undergoes a
sequence of true $2p$ emissions decaying via the $^{6}$Be ground state. The next
candidate for the true $4p$ emitter is $^{21}$Si but no information is
available for this isotope yet.

\newpage

\section{EMISSION OF NEUTRONS}
\label{sec:VIII}


With the progress in reaching experimentally the neutron drip-line,
the interest to study nuclei beyond this limit is rising.
Emission of protons beyond the proton drip-line has an
analogue in neutron(s) emission beyond the neutron drip-line.
In this Section we want to comment on the possibility of neutron(s) emission
which may take the form of neutron/few-neutron radioactivity.
Some aspects of such processes were discussed by \textcite{Thoennessen:2004}.

For illustration we consider theoretical estimates for the two pairs:
$^{26}$S-$^{26}$O and $^{25}$P-$^{25}$O. They are isobaric partners but not
mirror nuclei (the $\{Z,N\}$ values are $\{16,10\}$-$\{8,18\}$ and
$\{15,10\}$-$\{8,17\}$). They have similar mass/charge ratios, opened $s$-$d$
shell, and should presumably have comparable structural and radial
characteristics. $^{26}$S is a candidate to be a true $2p$ emitter and
$^{26}$O could be a narrow two-neutron resonance.

\begin{figure*}[tb]
\begin{center}
\includegraphics[width=0.353\textwidth]{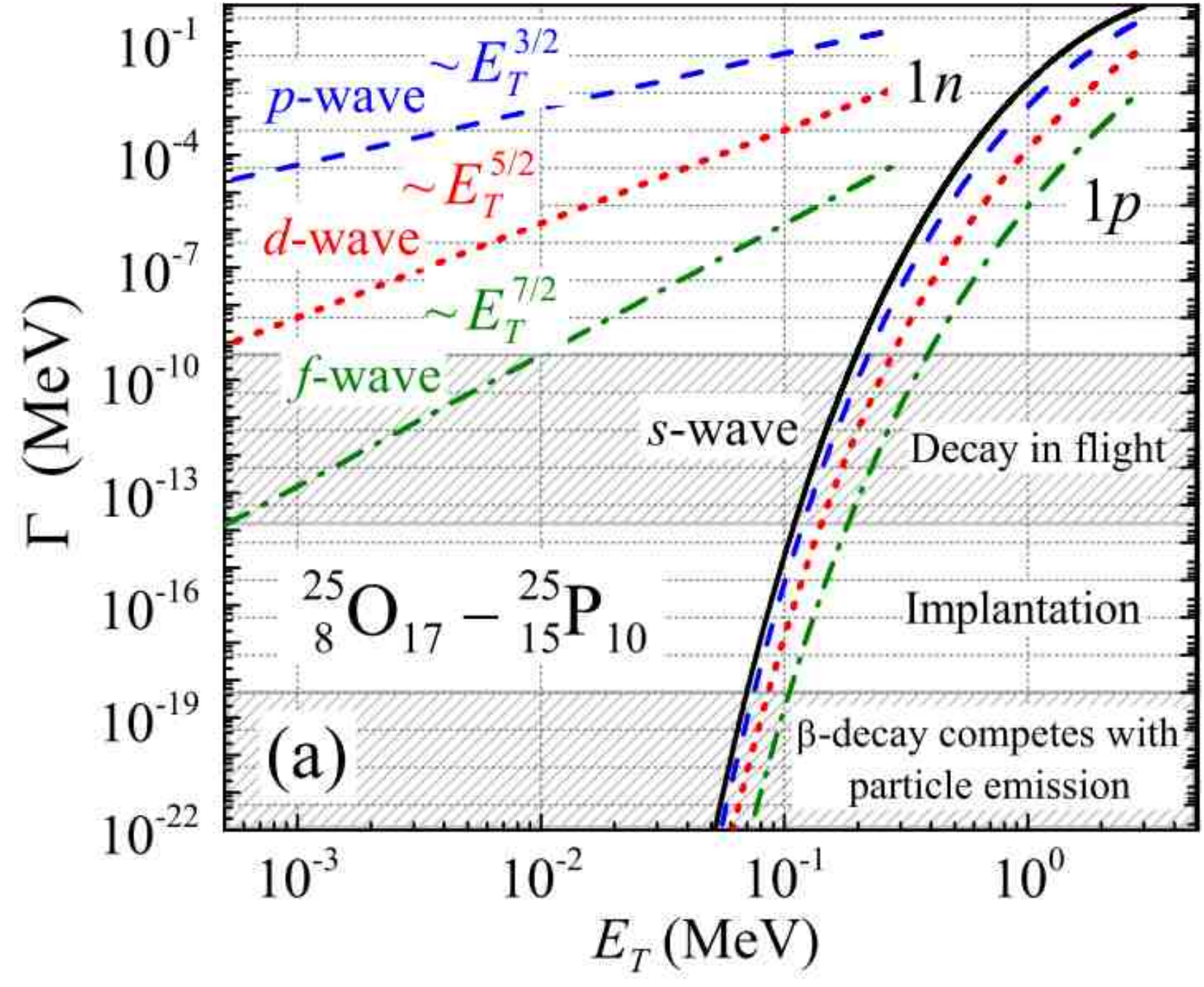}
\includegraphics[width=0.287\textwidth]{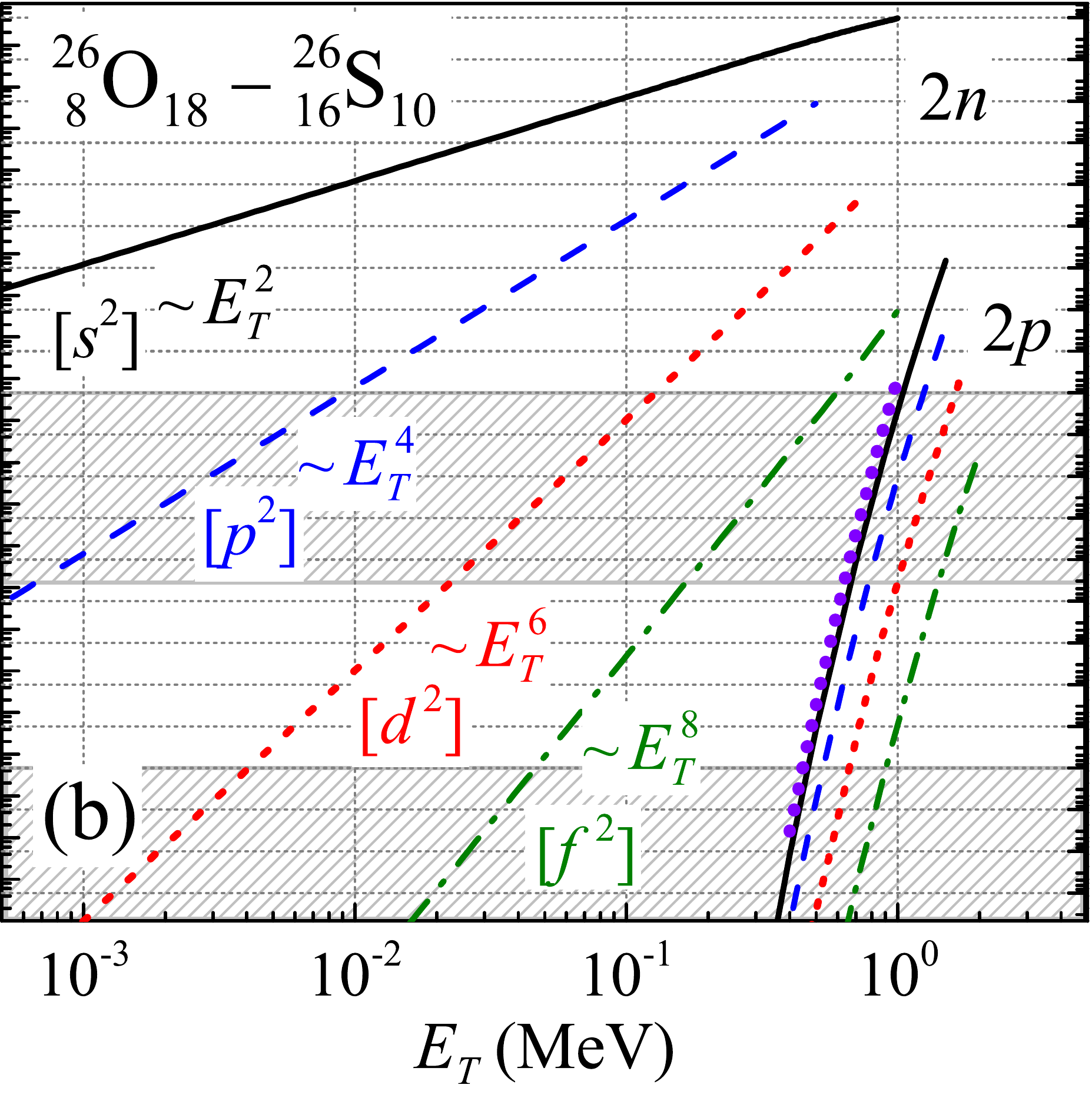}
\includegraphics[width=0.345\textwidth]{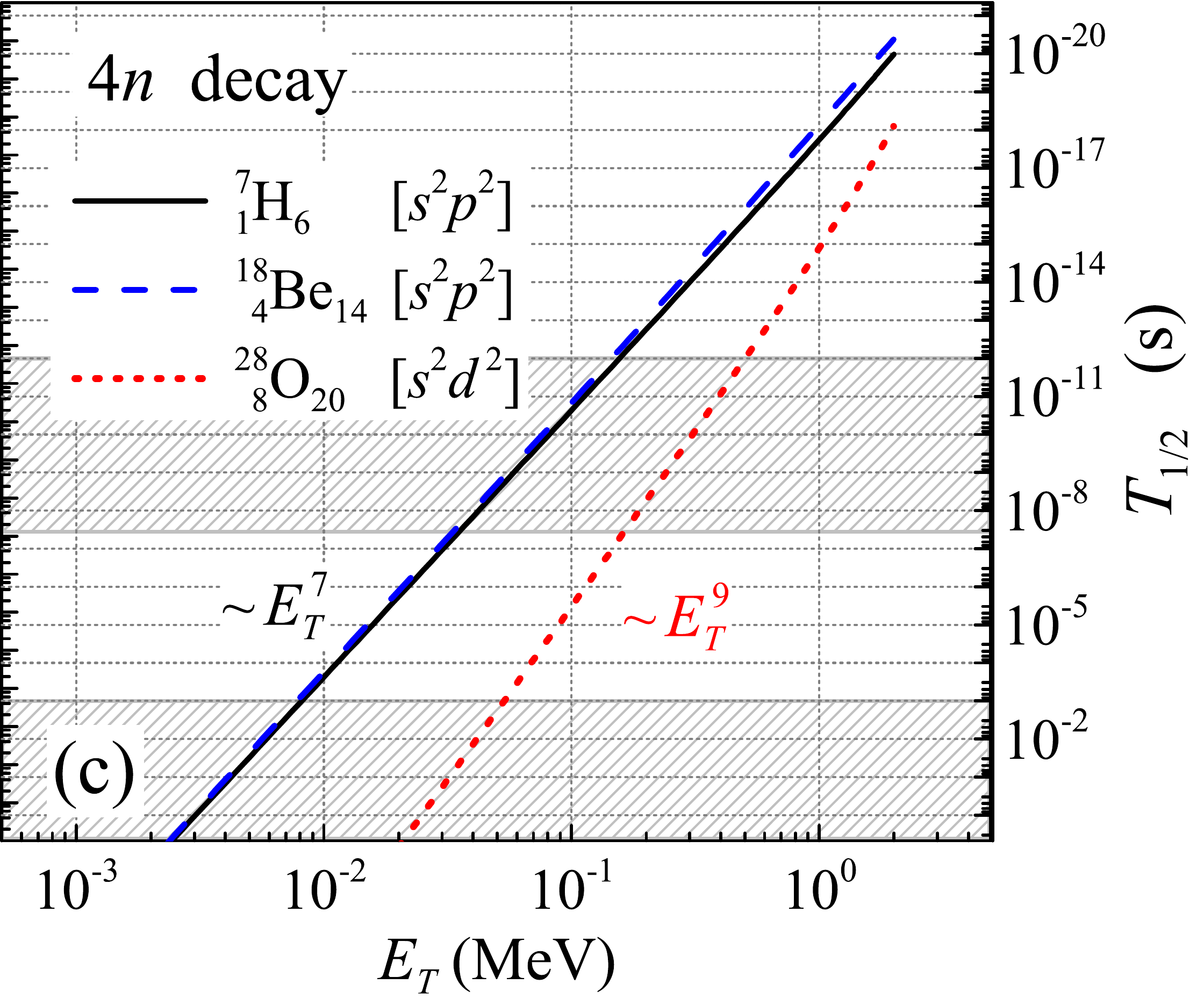}
\end{center}
\caption{(Color online) Estimated widths for the neutron (a) and the two-neutron
emission (b)
compared with widths for the proton and the two-proton emission. Panel (c) shows
estimates for the four-neutron emission. Hatched areas indicate lifetime ranges
accessible by different experimental techniques, see also Fig.\
\ref{fig:VIIF1-lifetime-all}.}
\label{fig:VIII-1n-2n}
\end{figure*}

For the one-neutron emission, simple estimates can be obtained by the standard
R-matrix expressions, Eqs.\ (\ref{eq:IB-r-mat-wid}) and (\ref{eq:IB-r-mat-pen}).
The results for protons and neutrons are shown in Fig.\ \ref{fig:VIII-1n-2n}(a).
Assuming a possible $s$- or $d$-wave ground state for $A=25$, one can see that
the decay energy window corresponding to proton radioactivity ranges from $\sim
50$ to $\sim 200$ keV. In contrast, for the neutron-emitting partner ($^{25}$O)
to be classified as radioactive, the decay energy, even for the $d$-wave, would
have to be smaller than 1 keV. It is highly improbable that such a fine-tuned
energy is actually found. It seems that a realistic chance to observe
one-neutron radioactivity may appear only for $f$-wave and higher-$l$ states. As
long as the heavier neutron drip-line nuclei are not known, long-lived neutron
emitters cannot be excluded. However, it is likely to happen beyond the $s$-$d$
shell.

The two-neutron radioactivity, or true two-neutron ($2n$) decay, is defined in
analogy to the true $2p$ decay (see Sec.\ \ref{sec:VIIA}) as a simultaneous
emission of two neutrons where one-neutron emission is energetically prohibited.
The possible $2n$ radioactivity, shown in Fig.\ \ref{fig:VIII-1n-2n}(b), has a
few important differences in comparison to the one-neutron radioactivity.

\noindent (i) Low-energy $s$-wave neutron emission could take place in the form
of a virtual state, which can not be interpreted in terms of width (the neutron
$s$-wave curve is missing in Fig.\ \ref{fig:VIII-1n-2n}(a)). For the $2n$
emission the phase space for a few-body final state creates additional effective
centrifugal barrier, so that a narrow resonance state is formed even for the
decay of a $[s^2]$ configuration. The possibility to observe narrow ``three-body
virtual states'' build on $[s^2]$ configurations was discussed by
\textcite{Grigorenko:2008} for the case of $^{10}$He. The search for the $2n$
ground-state decay has indicated only the existence of quite broad ($\Gamma
\gtrsim 1$ MeV) states in $^{5}$H, $^{10}$He, and $^{13}$Li
\cite{Golovkov:2005,Korsheninnikov:1994,Golovkov:2009,Johansson:2010}.

\noindent (ii) Similarly to the $p$-$2p$ decay comparison, the widths for the
true two-neutron emission are much smaller than for the one-neutron emission
with the same decay energy. The energy window for $2p$ radioactivity of $^{26}$S
extends up to 500-1700 keV, thus is about an order of magnitude larger than for
the proton radioactivity of $^{25}$P. The estimated relative increase of the
energy window for $2n$ radioactivity compared to $1n$ radioactivity is even
larger (about two orders of magnitude). For example, for the $[d^2]$ and the
$[f^2]$ configurations, the true $2n$  decays would be classified as radioactive
for decay energies ranging up to $\sim 200$ keV and $\sim 600$ keV,
respectively. Such broad ranges make the search for $2n$ radioactivity much more
promissing.

\noindent (iii) In contrast to the $1n$ situation, the $2n$ estimates in Fig.\
\ref{fig:VIII-1n-2n}(b) should be interpreted as \emph{lifetime limits} due to
the possibility of configuration mixing. The $[s^2]$ and $[p^2]$ curves are
likely to provide \emph{lower} lifetime limits for $s$-$d$ and $p$-$f$
configurations, respectively. The $[d^2]$ and $[f^2]$ curves provide
\emph{upper} lifetime limits for them. From the experience collected in the $2p$
decay studies we may argue that the realistic results are located between the
limiting curves, but closer to that for the lower $l$.

The systems decaying by a simultaneous emission of more than two
neutrons should have longer lifetimes for the given decay energy.
The estimates of the true four-neutron decay width can be performed in the
approximation generalizing the direct decay model of Eq.\
(\ref{eq:VIID1-rmat-wid-3}). For a true four-neutron decay this model gives
\cite{Grigorenko:2011}:
\begin{eqnarray}
\Gamma_{\text{dir}}(E_T) & = & \frac{E^3_{T}(E_T-\textstyle \sum
\nolimits^4_{i=1} E_{n_i})^2} {2\pi^3} \int_0^{1}  d \varepsilon_1
\int_0^{1-\varepsilon_1} d \varepsilon_2
\nonumber \\
 &  \times & \int_0^{1- \varepsilon_1- \varepsilon_2} \!\!\! d \varepsilon_3
 \prod^4_{i=1}  \frac{\Gamma_{n_i}(E_{i})}
 {(E_{i}-E_{n_i})^{2} + \Gamma_{n_i} (E_{i})^{2}/4}\,.\nonumber
\label{eq:VIID1-rmat-wid-5}
\end{eqnarray}
Here $E_i=\varepsilon_i E_T$ for $i=1,2,3$ and $E_4= (1- \varepsilon_1-
\varepsilon_2- \varepsilon_2)E_T$. The $E_{n_i}$ and $\Gamma_{n_i}$ are energy
and width of the lowest resonance in the core and $i$-th neutron subsystem.

The decay of $^{7}$H is strongly suspected to proceed by a true four-neutron
emission. The decay energy is uncertain, but seems to be smaller than
2 MeV \cite{Korsheninnikov:2003,Caamano:2008,Nikolskii:2010}. The ground-state
decay energies of the subsystems: $^{4}$H \cite{Tilley:1992} and $^{5}$H
\cite{Golovkov:2005} are around 2 MeV. There is only un upper limit of 1 ns for
the lifetime of $^{7}$H \cite{Golovkov:2004}. The nearest heavier candidates for
the true four-neutron emitters are $^{18}$Be and $^{28}$O.

In Fig.\ \ref{fig:VIII-1n-2n}(c) the estimates for the true four-neutron
emission lifetime are shown for $^{7}$H, $^{18}$Be, and $^{28}$O.
For simplicity we used $E_{n_i}=1.5$ MeV in all cases. The other
parameters are chosen according to guidelines from Sec.\ \ref{sec:VIIE1}. The
orbital configurations were chosen to yield results closer to
the upper limit for the width,
therefore, the provided estimates are conservative. These estimates
indicate that long-living states decaying by true four-neutron emission are
possible, with decay energies up to hundreds of keV, even for $p$ and $s$-$d$
shell nuclei.

In summary, the observation of neutron radioactivity in $s$-$d$ shell nuclei
seems unrealistic. It is more probable that this process occurs in heavier
($p$-$f$ shell)
systems. The discovery of $2n$ or $4n$ radioactivity is much more probable
since the energy windows corresponding to the radioactive timescale are much
broader. However, both $2n$ and $4n$ lifetimes depend strongly
on the nuclear structure and quantitative predictions require further
investigations.

\newpage

\section{CONCLUSIONS}
\label{sec:IX}

The exploration of the driplines of the nuclear chart have made great
progress during the last decade. Our knowledge on nuclei along and
beyond the proton dripline has increased significantly and studies of
the neutron dripline has now reached the Ne-Mg region. New radioactive
decay modes appear here, namely one-proton and two-proton emission
from long-lived states as well as many beta-delayed particle emission
branches. This review has given an overview of the present state of
the field.

Beta-delayed particle emission will be prominent at the driplines and
many modes will mainly occur there, the exception being beta-delayed
one- and multi-neutron emission that is expected to play a major role
for neutron-rich nuclei, in particular in the still unexplored part of
the nuclear chart.

The single-proton emission, be it as radioactivity or following a
beta-decay, is now well understood and established as a valuable probe
of nuclear structure evolution, in close analogy to the role that alpha
decay has played.

Two-proton radioactivity is the most recently discovered decay
mode. It is as well sensitive to nuclear structure, but there are
interesting physics questions in the emission process itself that are
not fully clarified and are intimately related to the more general
problem of three-body break-up. To some extent this also holds for
beta-delayed multi-particle decays, where only a few cases have been
thoroughly explored experimentally.

In all the cases mentioned theoretical and experimental activities are
ongoing in order to tackle the important questions that await
clarification. An essential ingredient in answering these questions is
the present rapid developments in the production of secondary
beams of near-dripline isotopes and the corresponding continuing
evolution of experimental techniques. Along with dedicated theoretical
efforts this should ensure that the progress continues.

\section*{Acknowledgments}

We are grateful to many colleagues for discussions and would like to
thank in particular H.O.U. Fynbo, G. Mart\'{\i}nez-Pinedo, I. Mukha,
G. Ter-Akopian, M.V. Zhukov, and J. \.Zylicz.
This work was supported in part by FAIR-Russia research center (LVG).

\newpage


\bibliographystyle{apsrmp}

\begin{thebibliography}{517}
\expandafter\ifx\csname natexlab\endcsname\relax\def\natexlab#1{#1}\fi
\expandafter\ifx\csname bibnamefont\endcsname\relax
  \def\bibnamefont#1{#1}\fi
\expandafter\ifx\csname bibfnamefont\endcsname\relax
  \def\bibfnamefont#1{#1}\fi
\expandafter\ifx\csname citenamefont\endcsname\relax
  \def\citenamefont#1{#1}\fi
\expandafter\ifx\csname url\endcsname\relax
  \def\url#1{\texttt{#1}}\fi
\expandafter\ifx\csname urlprefix\endcsname\relax\def\urlprefix{URL }\fi
\providecommand{\bibinfo}[2]{#2}
\providecommand{\eprint}[2][]{\url{#2}}

\bibitem[{\citenamefont{{\AA}berg} \emph{et~al.}(1997)\citenamefont{{\AA}berg,
  Semmes, and Nazarewicz}}]{Aberg:1997}
\bibinfo{author}{\bibnamefont{{\AA}berg}, \bibfnamefont{S.}},
  \bibinfo{author}{\bibfnamefont{P.}~\bibnamefont{Semmes}}, and
  \bibinfo{author}{\bibfnamefont{W.}~\bibnamefont{Nazarewicz}},
  \bibinfo{year}{1997}, \bibinfo{journal}{Phys.\ Rev.}
  \textbf{\bibinfo{volume}{C~56}}, \bibinfo{pages}{17762}.

\bibitem[{\citenamefont{Abramowitz and Stegun}(1968)}]{Abramowitz:1968}
\bibinfo{author}{\bibnamefont{Abramowitz}, \bibfnamefont{M.}}, and
  \bibinfo{author}{\bibfnamefont{I.~A.} \bibnamefont{Stegun}},
  \bibinfo{year}{1968}, \emph{\bibinfo{title}{Handbook of mathematical
  functions}} (\bibinfo{publisher}{Dover}, \bibinfo{address}{New-York}).

\bibitem[{\citenamefont{Achouri} \emph{et~al.}(2006)\citenamefont{Achouri,
  de~Oliveira~Santos, Lewitowicz, Blank, \"{A}yst\"{o}, Canchel, Czajkowski,
  Dendooven, Emsallem, Giovinazzo, Guillet, Jokinen}
  \emph{et~al.}}]{Achouri:2006}
\bibinfo{author}{\bibnamefont{Achouri}, \bibfnamefont{N.}},
  \bibinfo{author}{\bibfnamefont{F.}~\bibnamefont{de~Oliveira~Santos}},
  \bibinfo{author}{\bibfnamefont{M.}~\bibnamefont{Lewitowicz}},
  \bibinfo{author}{\bibfnamefont{B.}~\bibnamefont{Blank}},
  \bibinfo{author}{\bibfnamefont{J.}~\bibnamefont{\"{A}yst\"{o}}},
  \bibinfo{author}{\bibfnamefont{G.}~\bibnamefont{Canchel}},
  \bibinfo{author}{\bibfnamefont{S.}~\bibnamefont{Czajkowski}},
  \bibinfo{author}{\bibfnamefont{P.}~\bibnamefont{Dendooven}},
  \bibinfo{author}{\bibfnamefont{A.}~\bibnamefont{Emsallem}},
  \bibinfo{author}{\bibfnamefont{J.}~\bibnamefont{Giovinazzo}},
  \bibinfo{author}{\bibfnamefont{N.}~\bibnamefont{Guillet}},
  \bibinfo{author}{\bibfnamefont{A.}~\bibnamefont{Jokinen}}, \emph{et~al.},
  \bibinfo{year}{2006}, \bibinfo{journal}{The European Physical Journal A -
  Hadrons and Nuclei} \textbf{\bibinfo{volume}{27}}, \bibinfo{pages}{287}.

\bibitem[{\citenamefont{Adimi} \emph{et~al.}(2010)\citenamefont{Adimi,
  Dom\'\i{}nguez-Reyes, Alcorta, Bey, Blank, Borge, Santos, Dossat, Fynbo,
  Giovinazzo, Knudsen, Madurga} \emph{et~al.}}]{Adimi:2010}
\bibinfo{author}{\bibnamefont{Adimi}, \bibfnamefont{N.}},
  \bibinfo{author}{\bibfnamefont{R.}~\bibnamefont{Dom\'\i{}nguez-Reyes}},
  \bibinfo{author}{\bibfnamefont{M.}~\bibnamefont{Alcorta}},
  \bibinfo{author}{\bibfnamefont{A.}~\bibnamefont{Bey}},
  \bibinfo{author}{\bibfnamefont{B.}~\bibnamefont{Blank}},
  \bibinfo{author}{\bibfnamefont{M.~J.~G.} \bibnamefont{Borge}},
  \bibinfo{author}{\bibfnamefont{F.~d.~O.} \bibnamefont{Santos}},
  \bibinfo{author}{\bibfnamefont{C.}~\bibnamefont{Dossat}},
  \bibinfo{author}{\bibfnamefont{H.~O.~U.} \bibnamefont{Fynbo}},
  \bibinfo{author}{\bibfnamefont{J.}~\bibnamefont{Giovinazzo}},
  \bibinfo{author}{\bibfnamefont{H.~H.} \bibnamefont{Knudsen}},
  \bibinfo{author}{\bibfnamefont{M.}~\bibnamefont{Madurga}}, \emph{et~al.},
  \bibinfo{year}{2010}, \bibinfo{journal}{Phys. Rev. C}
  \textbf{\bibinfo{volume}{81}}(\bibinfo{number}{2}), \bibinfo{pages}{024311}.

\bibitem[{\citenamefont{Akovali}(1998)}]{Akovali:1998}
\bibinfo{author}{\bibnamefont{Akovali}, \bibfnamefont{Y.}},
  \bibinfo{year}{1998}, \bibinfo{journal}{Nucl.\ Data Sheets}
  \textbf{\bibinfo{volume}{84}}, \bibinfo{pages}{1}.

\bibitem[{\citenamefont{Algora} \emph{et~al.}(2010)\citenamefont{Algora,
  Jordan, Ta\'\i{}n, Rubio, Agramunt, Perez-Cerdan, Molina, Caballero,
  N\'acher, Krasznahorkay, Hunyadi, Guly\'as} \emph{et~al.}}]{Algora:2010}
\bibinfo{author}{\bibnamefont{Algora}, \bibfnamefont{A.}},
  \bibinfo{author}{\bibfnamefont{D.}~\bibnamefont{Jordan}},
  \bibinfo{author}{\bibfnamefont{J.~L.} \bibnamefont{Ta\'\i{}n}},
  \bibinfo{author}{\bibfnamefont{B.}~\bibnamefont{Rubio}},
  \bibinfo{author}{\bibfnamefont{J.}~\bibnamefont{Agramunt}},
  \bibinfo{author}{\bibfnamefont{A.~B.} \bibnamefont{Perez-Cerdan}},
  \bibinfo{author}{\bibfnamefont{F.}~\bibnamefont{Molina}},
  \bibinfo{author}{\bibfnamefont{L.}~\bibnamefont{Caballero}},
  \bibinfo{author}{\bibfnamefont{E.}~\bibnamefont{N\'acher}},
  \bibinfo{author}{\bibfnamefont{A.}~\bibnamefont{Krasznahorkay}},
  \bibinfo{author}{\bibfnamefont{M.~D.} \bibnamefont{Hunyadi}},
  \bibinfo{author}{\bibfnamefont{J.}~\bibnamefont{Guly\'as}}, \emph{et~al.},
  \bibinfo{year}{2010}, \bibinfo{journal}{Phys. Rev. Lett.}
  \textbf{\bibinfo{volume}{105}}(\bibinfo{number}{20}),
  \bibinfo{pages}{202501}.

\bibitem[{\citenamefont{Alvarez}(1937)}]{Alvarez:1937}
\bibinfo{author}{\bibnamefont{Alvarez}, \bibfnamefont{L.}},
  \bibinfo{year}{1937}, \bibinfo{journal}{Phys. Rev.}
  \textbf{\bibinfo{volume}{52}}(\bibinfo{number}{2}), \bibinfo{pages}{134}.

\bibitem[{\citenamefont{\'Alvarez-Rodr\'\i{}guez}
  \emph{et~al.}(2008{\natexlab{a}})\citenamefont{\'Alvarez-Rodr\'\i{}guez,
  Fynbo, Jensen, and Garrido}}]{Alvarez:2008}
\bibinfo{author}{\bibnamefont{\'Alvarez-Rodr\'\i{}guez}, \bibfnamefont{R.}},
  \bibinfo{author}{\bibfnamefont{H.~O.~U.} \bibnamefont{Fynbo}},
  \bibinfo{author}{\bibfnamefont{A.~S.} \bibnamefont{Jensen}}, and
  \bibinfo{author}{\bibfnamefont{E.}~\bibnamefont{Garrido}},
  \bibinfo{year}{2008}{\natexlab{a}}, \bibinfo{journal}{Phys. Rev. Lett.}
  \textbf{\bibinfo{volume}{100}}(\bibinfo{number}{19}),
  \bibinfo{pages}{192501}.

\bibitem[{\citenamefont{\'Alvarez-Rodr\'\i{}guez}
  \emph{et~al.}(2010)\citenamefont{\'Alvarez-Rodr\'\i{}guez, Jensen, Garrido,
  and Fedorov}}]{Alvarez:2010}
\bibinfo{author}{\bibnamefont{\'Alvarez-Rodr\'\i{}guez}, \bibfnamefont{R.}},
  \bibinfo{author}{\bibfnamefont{A.~S.} \bibnamefont{Jensen}},
  \bibinfo{author}{\bibfnamefont{E.}~\bibnamefont{Garrido}}, and
  \bibinfo{author}{\bibfnamefont{D.~V.} \bibnamefont{Fedorov}},
  \bibinfo{year}{2010}, \bibinfo{journal}{Phys. Rev. C}
  \textbf{\bibinfo{volume}{82}}(\bibinfo{number}{3}), \bibinfo{pages}{034001}.

\bibitem[{\citenamefont{\'Alvarez-Rodr\'\i{}guez}
  \emph{et~al.}(2008{\natexlab{b}})\citenamefont{\'Alvarez-Rodr\'\i{}guez,
  Jensen, Garrido, Fedorov, and Fynbo}}]{Alvarez:2008a}
\bibinfo{author}{\bibnamefont{\'Alvarez-Rodr\'\i{}guez}, \bibfnamefont{R.}},
  \bibinfo{author}{\bibfnamefont{A.~S.} \bibnamefont{Jensen}},
  \bibinfo{author}{\bibfnamefont{E.}~\bibnamefont{Garrido}},
  \bibinfo{author}{\bibfnamefont{D.~V.} \bibnamefont{Fedorov}}, and
  \bibinfo{author}{\bibfnamefont{H.~O.~U.} \bibnamefont{Fynbo}},
  \bibinfo{year}{2008}{\natexlab{b}}, \bibinfo{journal}{Phys. Rev. C}
  \textbf{\bibinfo{volume}{77}}(\bibinfo{number}{6}), \bibinfo{pages}{064305}.

\bibitem[{\citenamefont{Andreyev} \emph{et~al.}(2004)\citenamefont{Andreyev,
  Ackermann, He\ss{}berger, Heyde, Hofmann, Huyse, Karlgren, Kojouharov,
  Kindler, Lommel, M\"unzenberg, Page} \emph{et~al.}}]{Andreyev:2004}
\bibinfo{author}{\bibnamefont{Andreyev}, \bibfnamefont{A.}},
  \bibinfo{author}{\bibfnamefont{D.}~\bibnamefont{Ackermann}},
  \bibinfo{author}{\bibfnamefont{F.}~\bibnamefont{He\ss{}berger}},
  \bibinfo{author}{\bibfnamefont{K.}~\bibnamefont{Heyde}},
  \bibinfo{author}{\bibfnamefont{S.}~\bibnamefont{Hofmann}},
  \bibinfo{author}{\bibfnamefont{M.}~\bibnamefont{Huyse}},
  \bibinfo{author}{\bibfnamefont{D.}~\bibnamefont{Karlgren}},
  \bibinfo{author}{\bibfnamefont{I.}~\bibnamefont{Kojouharov}},
  \bibinfo{author}{\bibfnamefont{B.}~\bibnamefont{Kindler}},
  \bibinfo{author}{\bibfnamefont{B.}~\bibnamefont{Lommel}},
  \bibinfo{author}{\bibfnamefont{G.}~\bibnamefont{M\"unzenberg}},
  \bibinfo{author}{\bibfnamefont{R.}~\bibnamefont{Page}}, \emph{et~al.},
  \bibinfo{year}{2004}, \bibinfo{journal}{Phys.\ Rev.}
  \textbf{\bibinfo{volume}{C~69}}, \bibinfo{pages}{054308}.

\bibitem[{\citenamefont{Andreyev} \emph{et~al.}(2000)\citenamefont{Andreyev,
  Huyse, Duppen, Weissman, Ackermann, Gerl, He{\ss}berger, Hofmann,
  Kleinb\"{o}hl, M\"unzenberg, Reshitko, Schlegel}
  \emph{et~al.}}]{Andreyev:2000}
\bibinfo{author}{\bibnamefont{Andreyev}, \bibfnamefont{A.}},
  \bibinfo{author}{\bibfnamefont{M.}~\bibnamefont{Huyse}},
  \bibinfo{author}{\bibfnamefont{P.}~\bibnamefont{Duppen}},
  \bibinfo{author}{\bibfnamefont{L.}~\bibnamefont{Weissman}},
  \bibinfo{author}{\bibfnamefont{D.}~\bibnamefont{Ackermann}},
  \bibinfo{author}{\bibfnamefont{J.}~\bibnamefont{Gerl}},
  \bibinfo{author}{\bibfnamefont{F.}~\bibnamefont{He{\ss}berger}},
  \bibinfo{author}{\bibfnamefont{S.}~\bibnamefont{Hofmann}},
  \bibinfo{author}{\bibfnamefont{A.}~\bibnamefont{Kleinb\"{o}hl}},
  \bibinfo{author}{\bibfnamefont{G.}~\bibnamefont{M\"unzenberg}},
  \bibinfo{author}{\bibfnamefont{S.}~\bibnamefont{Reshitko}},
  \bibinfo{author}{\bibfnamefont{C.}~\bibnamefont{Schlegel}}, \emph{et~al.},
  \bibinfo{year}{2000}, \bibinfo{journal}{Nature}
  \textbf{\bibinfo{volume}{405}}, \bibinfo{pages}{430}.

\bibitem[{\citenamefont{Andreyev} \emph{et~al.}(2005)\citenamefont{Andreyev,
  Ackermann, Antalic, Darby, Franchoo, He\ss{}berger, Hofmann, Huyse,
  Kuusiniemi, Lommel, Kindler, Mann} \emph{et~al.}}]{Andreyev:2005}
\bibinfo{author}{\bibnamefont{Andreyev}, \bibfnamefont{A.~N.}},
  \bibinfo{author}{\bibfnamefont{D.}~\bibnamefont{Ackermann}},
  \bibinfo{author}{\bibfnamefont{S.}~\bibnamefont{Antalic}},
  \bibinfo{author}{\bibfnamefont{I.~G.} \bibnamefont{Darby}},
  \bibinfo{author}{\bibfnamefont{S.}~\bibnamefont{Franchoo}},
  \bibinfo{author}{\bibfnamefont{F.~P.} \bibnamefont{He\ss{}berger}},
  \bibinfo{author}{\bibfnamefont{S.}~\bibnamefont{Hofmann}},
  \bibinfo{author}{\bibfnamefont{M.}~\bibnamefont{Huyse}},
  \bibinfo{author}{\bibfnamefont{P.}~\bibnamefont{Kuusiniemi}},
  \bibinfo{author}{\bibfnamefont{B.}~\bibnamefont{Lommel}},
  \bibinfo{author}{\bibfnamefont{B.}~\bibnamefont{Kindler}},
  \bibinfo{author}{\bibfnamefont{R.}~\bibnamefont{Mann}}, \emph{et~al.},
  \bibinfo{year}{2005}, \bibinfo{journal}{Phys. Rev. C}
  \textbf{\bibinfo{volume}{72}}(\bibinfo{number}{1}), \bibinfo{pages}{014612}.

\bibitem[{\citenamefont{Andreyev} \emph{et~al.}(2010)\citenamefont{Andreyev,
  Elseviers, Huyse, Van~Duppen, Antalic, Barzakh, Bree, Cocolios, Comas,
  Diriken, Fedorov, Fedosseev} \emph{et~al.}}]{Andreyev:2010}
\bibinfo{author}{\bibnamefont{Andreyev}, \bibfnamefont{A.~N.}},
  \bibinfo{author}{\bibfnamefont{J.}~\bibnamefont{Elseviers}},
  \bibinfo{author}{\bibfnamefont{M.}~\bibnamefont{Huyse}},
  \bibinfo{author}{\bibfnamefont{P.}~\bibnamefont{Van~Duppen}},
  \bibinfo{author}{\bibfnamefont{S.}~\bibnamefont{Antalic}},
  \bibinfo{author}{\bibfnamefont{A.}~\bibnamefont{Barzakh}},
  \bibinfo{author}{\bibfnamefont{N.}~\bibnamefont{Bree}},
  \bibinfo{author}{\bibfnamefont{T.~E.} \bibnamefont{Cocolios}},
  \bibinfo{author}{\bibfnamefont{V.~F.} \bibnamefont{Comas}},
  \bibinfo{author}{\bibfnamefont{J.}~\bibnamefont{Diriken}},
  \bibinfo{author}{\bibfnamefont{D.}~\bibnamefont{Fedorov}},
  \bibinfo{author}{\bibfnamefont{V.}~\bibnamefont{Fedosseev}}, \emph{et~al.},
  \bibinfo{year}{2010}, \bibinfo{journal}{Phys. Rev. Lett.}
  \textbf{\bibinfo{volume}{105}}(\bibinfo{number}{25}),
  \bibinfo{pages}{252502}.

\bibitem[{\citenamefont{Ang\'{e}lique}
  \emph{et~al.}(2006)\citenamefont{Ang\'{e}lique, Timis, Pi\'{e}tri, Achouri,
  Baumann, Borcea, Buta, Catford, Courtin, Daugas, Dessagne, de~Oliveira}
  \emph{et~al.}}]{Angelique:2006}
\bibinfo{author}{\bibnamefont{Ang\'{e}lique}, \bibfnamefont{J.~C.}},
  \bibinfo{author}{\bibfnamefont{C.}~\bibnamefont{Timis}},
  \bibinfo{author}{\bibfnamefont{S.}~\bibnamefont{Pi\'{e}tri}},
  \bibinfo{author}{\bibfnamefont{N.~L.} \bibnamefont{Achouri}},
  \bibinfo{author}{\bibfnamefont{P.}~\bibnamefont{Baumann}},
  \bibinfo{author}{\bibfnamefont{C.}~\bibnamefont{Borcea}},
  \bibinfo{author}{\bibfnamefont{A.}~\bibnamefont{Buta}},
  \bibinfo{author}{\bibfnamefont{W.}~\bibnamefont{Catford}},
  \bibinfo{author}{\bibfnamefont{S.}~\bibnamefont{Courtin}},
  \bibinfo{author}{\bibfnamefont{J.~M.} \bibnamefont{Daugas}},
  \bibinfo{author}{\bibfnamefont{P.}~\bibnamefont{Dessagne}},
  \bibinfo{author}{\bibfnamefont{F.}~\bibnamefont{de~Oliveira}}, \emph{et~al.},
  \bibinfo{year}{2006}, \bibinfo{journal}{AIP Conference Proceedings}
  \textbf{\bibinfo{volume}{831}}(\bibinfo{number}{1}), \bibinfo{pages}{134}.

\bibitem[{\citenamefont{Anthony} \emph{et~al.}(2002)\citenamefont{Anthony,
  Buchmann, Bergbusch, D'Auria, Dombsky, Giesen, Jackson, King, Powell, and
  Barker}}]{Anthony:2002}
\bibinfo{author}{\bibnamefont{Anthony}, \bibfnamefont{D.}},
  \bibinfo{author}{\bibfnamefont{L.}~\bibnamefont{Buchmann}},
  \bibinfo{author}{\bibfnamefont{P.}~\bibnamefont{Bergbusch}},
  \bibinfo{author}{\bibfnamefont{J.~M.} \bibnamefont{D'Auria}},
  \bibinfo{author}{\bibfnamefont{M.}~\bibnamefont{Dombsky}},
  \bibinfo{author}{\bibfnamefont{U.}~\bibnamefont{Giesen}},
  \bibinfo{author}{\bibfnamefont{K.~P.} \bibnamefont{Jackson}},
  \bibinfo{author}{\bibfnamefont{J.~D.} \bibnamefont{King}},
  \bibinfo{author}{\bibfnamefont{J.}~\bibnamefont{Powell}}, and
  \bibinfo{author}{\bibfnamefont{F.~C.} \bibnamefont{Barker}},
  \bibinfo{year}{2002}, \bibinfo{journal}{Phys. Rev. C}
  \textbf{\bibinfo{volume}{65}}(\bibinfo{number}{3}), \bibinfo{pages}{034310}.

\bibitem[{\citenamefont{Antony} \emph{et~al.}(1997)\citenamefont{Antony, Pape,
  and Britz}}]{Antony:1997}
\bibinfo{author}{\bibnamefont{Antony}, \bibfnamefont{M.~S.}},
  \bibinfo{author}{\bibfnamefont{A.}~\bibnamefont{Pape}}, and
  \bibinfo{author}{\bibfnamefont{J.}~\bibnamefont{Britz}},
  \bibinfo{year}{1997}, \bibinfo{journal}{Atomic Data and Nuclear Data Tables}
  \textbf{\bibinfo{volume}{66}}(\bibinfo{number}{1}), \bibinfo{pages}{1}.

\bibitem[{\citenamefont{Arnould} \emph{et~al.}(2007)\citenamefont{Arnould,
  Goriely, and Takahashi}}]{Arnould:2007}
\bibinfo{author}{\bibnamefont{Arnould}, \bibfnamefont{M.}},
  \bibinfo{author}{\bibfnamefont{S.}~\bibnamefont{Goriely}}, and
  \bibinfo{author}{\bibfnamefont{K.}~\bibnamefont{Takahashi}},
  \bibinfo{year}{2007}, \bibinfo{journal}{Physics Reports}
  \textbf{\bibinfo{volume}{450}}(\bibinfo{number}{4-6}), \bibinfo{pages}{97}.

\bibitem[{\citenamefont{Arumugam} \emph{et~al.}(2008)\citenamefont{Arumugam,
  Ferreira, and Maglione}}]{Arumugam:2008}
\bibinfo{author}{\bibnamefont{Arumugam}, \bibfnamefont{P.}},
  \bibinfo{author}{\bibfnamefont{L.}~\bibnamefont{Ferreira}}, and
  \bibinfo{author}{\bibfnamefont{E.}~\bibnamefont{Maglione}},
  \bibinfo{year}{2008}, \bibinfo{journal}{Phys. Rev.}
  \textbf{\bibinfo{volume}{C~78}}, \bibinfo{pages}{041305(R)}.

\bibitem[{\citenamefont{Arumugam} \emph{et~al.}(2009)\citenamefont{Arumugam,
  Ferreira, and Maglione}}]{Arumugam:2009}
\bibinfo{author}{\bibnamefont{Arumugam}, \bibfnamefont{P.}},
  \bibinfo{author}{\bibfnamefont{L.}~\bibnamefont{Ferreira}}, and
  \bibinfo{author}{\bibfnamefont{E.}~\bibnamefont{Maglione}},
  \bibinfo{year}{2009}, \bibinfo{journal}{Phys.\ Lett.}
  \textbf{\bibinfo{volume}{B~680}}, \bibinfo{pages}{443}.

\bibitem[{\citenamefont{Arumugam} \emph{et~al.}(2007)\citenamefont{Arumugam,
  Maglione, and Ferreira}}]{Arumugam:2007}
\bibinfo{author}{\bibnamefont{Arumugam}, \bibfnamefont{P.}},
  \bibinfo{author}{\bibfnamefont{E.}~\bibnamefont{Maglione}}, and
  \bibinfo{author}{\bibfnamefont{L.~S.} \bibnamefont{Ferreira}},
  \bibinfo{year}{2007}, \bibinfo{journal}{Phys. Rev. C}
  \textbf{\bibinfo{volume}{76}}(\bibinfo{number}{4}), \bibinfo{pages}{044311}.

\bibitem[{\citenamefont{Audi}
  \emph{et~al.}(2003{\natexlab{a}})\citenamefont{Audi, Bersillon, Blachot, and
  Wapstra}}]{Audi:2003}
\bibinfo{author}{\bibnamefont{Audi}, \bibfnamefont{G.}},
  \bibinfo{author}{\bibfnamefont{O.}~\bibnamefont{Bersillon}},
  \bibinfo{author}{\bibfnamefont{J.}~\bibnamefont{Blachot}}, and
  \bibinfo{author}{\bibfnamefont{A.~H.} \bibnamefont{Wapstra}},
  \bibinfo{year}{2003}{\natexlab{a}}, \bibinfo{journal}{Nuclear Physics A}
  \textbf{\bibinfo{volume}{729}}(\bibinfo{number}{1}), \bibinfo{pages}{3}.

\bibitem[{\citenamefont{Audi}
  \emph{et~al.}(2003{\natexlab{b}})\citenamefont{Audi, Wapstra, and
  Thibault}}]{Audi:2003a}
\bibinfo{author}{\bibnamefont{Audi}, \bibfnamefont{G.}},
  \bibinfo{author}{\bibfnamefont{A.~H.} \bibnamefont{Wapstra}}, and
  \bibinfo{author}{\bibfnamefont{C.}~\bibnamefont{Thibault}},
  \bibinfo{year}{2003}{\natexlab{b}}, \bibinfo{journal}{Nuclear Physics A}
  \textbf{\bibinfo{volume}{729}}(\bibinfo{number}{1}), \bibinfo{pages}{337}.

\bibitem[{\citenamefont{Auerbach}(1983)}]{Auerbach:1983}
\bibinfo{author}{\bibnamefont{Auerbach}, \bibfnamefont{N.}},
  \bibinfo{year}{1983}, \bibinfo{journal}{Physics Reports}
  \textbf{\bibinfo{volume}{98}}(\bibinfo{number}{5}), \bibinfo{pages}{273}.

\bibitem[{\citenamefont{Aumann}(2005)}]{Aumann:2005}
\bibinfo{author}{\bibnamefont{Aumann}, \bibfnamefont{T.}},
  \bibinfo{year}{2005}, \bibinfo{journal}{Eur. Phys. J.}
  \textbf{\bibinfo{volume}{A26}}, \bibinfo{pages}{441}.

\bibitem[{\citenamefont{Axelsson} \emph{et~al.}(1998)\citenamefont{Axelsson,
  Äystö, Borge, Fraile, Fynbo, Honkanen, Hornsh{\o}j, Jokinen, Jonson, Lipas,
  Martel, Mukha} \emph{et~al.}}]{Axelsson:1998}
\bibinfo{author}{\bibnamefont{Axelsson}, \bibfnamefont{L.}},
  \bibinfo{author}{\bibfnamefont{J.}~\bibnamefont{Äystö}},
  \bibinfo{author}{\bibfnamefont{M.~J.~G.} \bibnamefont{Borge}},
  \bibinfo{author}{\bibfnamefont{L.~M.} \bibnamefont{Fraile}},
  \bibinfo{author}{\bibfnamefont{H.~O.~U.} \bibnamefont{Fynbo}},
  \bibinfo{author}{\bibfnamefont{A.}~\bibnamefont{Honkanen}},
  \bibinfo{author}{\bibfnamefont{P.}~\bibnamefont{Hornsh{\o}j}},
  \bibinfo{author}{\bibfnamefont{A.}~\bibnamefont{Jokinen}},
  \bibinfo{author}{\bibfnamefont{B.}~\bibnamefont{Jonson}},
  \bibinfo{author}{\bibfnamefont{P.~O.} \bibnamefont{Lipas}},
  \bibinfo{author}{\bibfnamefont{I.}~\bibnamefont{Martel}},
  \bibinfo{author}{\bibfnamefont{I.}~\bibnamefont{Mukha}}, \emph{et~al.},
  \bibinfo{year}{1998}, \bibinfo{journal}{Nuclear Physics A}
  \textbf{\bibinfo{volume}{634}}(\bibinfo{number}{4}), \bibinfo{pages}{475}.

\bibitem[{\citenamefont{Azhari} \emph{et~al.}(1998)\citenamefont{Azhari,
  Kryger, and Thoennessen}}]{Azhari:1998}
\bibinfo{author}{\bibnamefont{Azhari}, \bibfnamefont{A.}},
  \bibinfo{author}{\bibfnamefont{R.~A.} \bibnamefont{Kryger}}, and
  \bibinfo{author}{\bibfnamefont{M.}~\bibnamefont{Thoennessen}},
  \bibinfo{year}{1998}, \bibinfo{journal}{Phys. Rev. C}
  \textbf{\bibinfo{volume}{58}}, \bibinfo{pages}{2568}.

\bibitem[{\citenamefont{Azuma} \emph{et~al.}(1980)\citenamefont{Azuma,
  Bj\"{o}rnstad, Gustafsson, Hansen, Jonson, Mattsson, Nyman, Poskanzer, and
  Ravn}}]{Azuma:1980}
\bibinfo{author}{\bibnamefont{Azuma}, \bibfnamefont{R.~E.}},
  \bibinfo{author}{\bibfnamefont{T.}~\bibnamefont{Bj\"{o}rnstad}},
  \bibinfo{author}{\bibfnamefont{H.~{\AA}.} \bibnamefont{Gustafsson}},
  \bibinfo{author}{\bibfnamefont{P.~G.} \bibnamefont{Hansen}},
  \bibinfo{author}{\bibfnamefont{B.}~\bibnamefont{Jonson}},
  \bibinfo{author}{\bibfnamefont{S.}~\bibnamefont{Mattsson}},
  \bibinfo{author}{\bibfnamefont{G.}~\bibnamefont{Nyman}},
  \bibinfo{author}{\bibfnamefont{A.~M.} \bibnamefont{Poskanzer}}, and
  \bibinfo{author}{\bibfnamefont{H.~L.} \bibnamefont{Ravn}},
  \bibinfo{year}{1980}, \bibinfo{journal}{Physics Letters B}
  \textbf{\bibinfo{volume}{96}}(\bibinfo{number}{1-2}), \bibinfo{pages}{31}.

\bibitem[{\citenamefont{Azuma} \emph{et~al.}(1979)\citenamefont{Azuma, Carraz,
  Hansen, Jonson, Kratz, Mattsson, Nyman, Ohm, Ravn, Schr\"oder, and
  Ziegert}}]{Azuma:1979}
\bibinfo{author}{\bibnamefont{Azuma}, \bibfnamefont{R.~E.}},
  \bibinfo{author}{\bibfnamefont{L.~C.} \bibnamefont{Carraz}},
  \bibinfo{author}{\bibfnamefont{P.~G.} \bibnamefont{Hansen}},
  \bibinfo{author}{\bibfnamefont{B.}~\bibnamefont{Jonson}},
  \bibinfo{author}{\bibfnamefont{K.~L.} \bibnamefont{Kratz}},
  \bibinfo{author}{\bibfnamefont{S.}~\bibnamefont{Mattsson}},
  \bibinfo{author}{\bibfnamefont{G.}~\bibnamefont{Nyman}},
  \bibinfo{author}{\bibfnamefont{H.}~\bibnamefont{Ohm}},
  \bibinfo{author}{\bibfnamefont{H.~L.} \bibnamefont{Ravn}},
  \bibinfo{author}{\bibfnamefont{A.}~\bibnamefont{Schr\"oder}}, and
  \bibinfo{author}{\bibfnamefont{W.}~\bibnamefont{Ziegert}},
  \bibinfo{year}{1979}, \bibinfo{journal}{Phys. Rev. Lett.}
  \textbf{\bibinfo{volume}{43}}(\bibinfo{number}{22}), \bibinfo{pages}{1652}.

\bibitem[{\citenamefont{B\"ack} \emph{et~al.}(2003)\citenamefont{B\"ack,
  Cederwall, Lagergen, Wyss, Johnson, Karlgren, Greenlees, Jenkins, Jones,
  Joss, Julin, Juutinen} \emph{et~al.}}]{Back:2003}
\bibinfo{author}{\bibnamefont{B\"ack}, \bibfnamefont{T.}},
  \bibinfo{author}{\bibfnamefont{B.}~\bibnamefont{Cederwall}},
  \bibinfo{author}{\bibfnamefont{K.}~\bibnamefont{Lagergen}},
  \bibinfo{author}{\bibfnamefont{R.}~\bibnamefont{Wyss}},
  \bibinfo{author}{\bibfnamefont{A.}~\bibnamefont{Johnson}},
  \bibinfo{author}{\bibfnamefont{D.}~\bibnamefont{Karlgren}},
  \bibinfo{author}{\bibfnamefont{P.}~\bibnamefont{Greenlees}},
  \bibinfo{author}{\bibfnamefont{D.}~\bibnamefont{Jenkins}},
  \bibinfo{author}{\bibfnamefont{P.}~\bibnamefont{Jones}},
  \bibinfo{author}{\bibfnamefont{D.}~\bibnamefont{Joss}},
  \bibinfo{author}{\bibfnamefont{R.}~\bibnamefont{Julin}},
  \bibinfo{author}{\bibfnamefont{S.}~\bibnamefont{Juutinen}}, \emph{et~al.},
  \bibinfo{year}{2003}, \bibinfo{journal}{Eur.\ Phys.\ J.}
  \textbf{\bibinfo{volume}{A~16}}, \bibinfo{pages}{489}.

\bibitem[{\citenamefont{Bain} \emph{et~al.}(1996)\citenamefont{Bain, Woods,
  Coszach, Davinson, Decrock, Gaelens, Galster, Huyse, Irvine, Leleux, Lienard,
  Loiselet} \emph{et~al.}}]{Bain:1996}
\bibinfo{author}{\bibnamefont{Bain}, \bibfnamefont{C.}},
  \bibinfo{author}{\bibfnamefont{P.}~\bibnamefont{Woods}},
  \bibinfo{author}{\bibfnamefont{R.}~\bibnamefont{Coszach}},
  \bibinfo{author}{\bibfnamefont{T.}~\bibnamefont{Davinson}},
  \bibinfo{author}{\bibfnamefont{P.}~\bibnamefont{Decrock}},
  \bibinfo{author}{\bibfnamefont{M.}~\bibnamefont{Gaelens}},
  \bibinfo{author}{\bibfnamefont{W.}~\bibnamefont{Galster}},
  \bibinfo{author}{\bibfnamefont{M.}~\bibnamefont{Huyse}},
  \bibinfo{author}{\bibfnamefont{R.}~\bibnamefont{Irvine}},
  \bibinfo{author}{\bibfnamefont{P.}~\bibnamefont{Leleux}},
  \bibinfo{author}{\bibfnamefont{E.}~\bibnamefont{Lienard}},
  \bibinfo{author}{\bibfnamefont{M.}~\bibnamefont{Loiselet}}, \emph{et~al.},
  \bibinfo{year}{1996}, \bibinfo{journal}{Physics Letters B}
  \textbf{\bibinfo{volume}{373}}(\bibinfo{number}{1-3}), \bibinfo{pages}{35}.

\bibitem[{\citenamefont{Bambynek} \emph{et~al.}(1972)\citenamefont{Bambynek,
  Crasemann, Fink, Freund, Mark, Swift, Price, and Rao}}]{Bambynek:1972}
\bibinfo{author}{\bibnamefont{Bambynek}, \bibfnamefont{W.}},
  \bibinfo{author}{\bibfnamefont{B.}~\bibnamefont{Crasemann}},
  \bibinfo{author}{\bibfnamefont{R.~W.} \bibnamefont{Fink}},
  \bibinfo{author}{\bibfnamefont{H.~U.} \bibnamefont{Freund}},
  \bibinfo{author}{\bibfnamefont{H.}~\bibnamefont{Mark}},
  \bibinfo{author}{\bibfnamefont{C.~D.} \bibnamefont{Swift}},
  \bibinfo{author}{\bibfnamefont{R.~E.} \bibnamefont{Price}}, and
  \bibinfo{author}{\bibfnamefont{P.~V.} \bibnamefont{Rao}},
  \bibinfo{year}{1972}, \bibinfo{journal}{Rev. Mod. Phys.}
  \textbf{\bibinfo{volume}{44}}(\bibinfo{number}{4}), \bibinfo{pages}{716}.

\bibitem[{\citenamefont{Barker}(1989)}]{Barker:1989}
\bibinfo{author}{\bibnamefont{Barker}, \bibfnamefont{F.~C.}},
  \bibinfo{year}{1989}, \bibinfo{journal}{Aust. J. Phys.}
  \textbf{\bibinfo{volume}{42}}, \bibinfo{pages}{25}.

\bibitem[{\citenamefont{Barker}(1999)}]{Barker:1999}
\bibinfo{author}{\bibnamefont{Barker}, \bibfnamefont{F.~C.}},
  \bibinfo{year}{1999}, \bibinfo{journal}{Phys. Rev. C}
  \textbf{\bibinfo{volume}{59}}, \bibinfo{pages}{535}.

\bibitem[{\citenamefont{Barker}(2001)}]{Barker:2001}
\bibinfo{author}{\bibnamefont{Barker}, \bibfnamefont{F.~C.}},
  \bibinfo{year}{2001}, \bibinfo{journal}{Phys. Rev. C}
  \textbf{\bibinfo{volume}{63}}, \bibinfo{pages}{047303}.

\bibitem[{\citenamefont{Barker}(2002)}]{Barker:2002}
\bibinfo{author}{\bibnamefont{Barker}, \bibfnamefont{F.~C.}},
  \bibinfo{year}{2002}, \bibinfo{journal}{Phys. Rev. C}
  \textbf{\bibinfo{volume}{66}}, \bibinfo{pages}{047603},
  \bibinfo{note}{erratum Phys. Rev. C 67, 049902 (2003)}.

\bibitem[{\citenamefont{Barker}(2003)}]{Barker:2003}
\bibinfo{author}{\bibnamefont{Barker}, \bibfnamefont{F.~C.}},
  \bibinfo{year}{2003}, \bibinfo{journal}{Phys. Rev. C}
  \textbf{\bibinfo{volume}{68}}, \bibinfo{pages}{054602}.

\bibitem[{\citenamefont{Barker and Treacy}(1962)}]{Barker:1962}
\bibinfo{author}{\bibnamefont{Barker}, \bibfnamefont{F.~C.}}, and
  \bibinfo{author}{\bibfnamefont{P.~B.} \bibnamefont{Treacy}},
  \bibinfo{year}{1962}, \bibinfo{journal}{Nuclear Physics}
  \textbf{\bibinfo{volume}{38}}, \bibinfo{pages}{33}.

\bibitem[{\citenamefont{Barker and Warburton}(1988)}]{Barker:1988}
\bibinfo{author}{\bibnamefont{Barker}, \bibfnamefont{F.~C.}}, and
  \bibinfo{author}{\bibfnamefont{E.~K.} \bibnamefont{Warburton}},
  \bibinfo{year}{1988}, \bibinfo{journal}{Nuclear Physics A}
  \textbf{\bibinfo{volume}{487}}(\bibinfo{number}{2}), \bibinfo{pages}{269}.

\bibitem[{\citenamefont{Barmore} \emph{et~al.}(2000)\citenamefont{Barmore,
  Kruppa, Nazarewicz, and Vertse}}]{Barmore:2000}
\bibinfo{author}{\bibnamefont{Barmore}, \bibfnamefont{B.}},
  \bibinfo{author}{\bibfnamefont{A.}~\bibnamefont{Kruppa}},
  \bibinfo{author}{\bibfnamefont{W.}~\bibnamefont{Nazarewicz}}, and
  \bibinfo{author}{\bibfnamefont{T.}~\bibnamefont{Vertse}},
  \bibinfo{year}{2000}, \bibinfo{journal}{Phys.\ Rev.}
  \textbf{\bibinfo{volume}{C~62}}, \bibinfo{pages}{054315}.

\bibitem[{\citenamefont{Barton} \emph{et~al.}(1963)\citenamefont{Barton,
  McPherson, Bell, Frisken, Link, and Moore}}]{Burton:1963}
\bibinfo{author}{\bibnamefont{Barton}, \bibfnamefont{R.}},
  \bibinfo{author}{\bibfnamefont{R.}~\bibnamefont{McPherson}},
  \bibinfo{author}{\bibfnamefont{R.~E.} \bibnamefont{Bell}},
  \bibinfo{author}{\bibfnamefont{W.~R.} \bibnamefont{Frisken}},
  \bibinfo{author}{\bibfnamefont{W.~T.} \bibnamefont{Link}}, and
  \bibinfo{author}{\bibfnamefont{R.~B.} \bibnamefont{Moore}},
  \bibinfo{year}{1963}, \bibinfo{journal}{Can. J. Phys.}
  \textbf{\bibinfo{volume}{41}}, \bibinfo{pages}{2007}.

\bibitem[{\citenamefont{Batchelder}
  \emph{et~al.}(1998)\citenamefont{Batchelder, Bingham, Rykaczewski, Toth,
  Davinson, McKenzie, Woods, Ginter, Gross, McConnell, Zganjar, Hamilton}
  \emph{et~al.}}]{Batchelder:1998}
\bibinfo{author}{\bibnamefont{Batchelder}, \bibfnamefont{J.}},
  \bibinfo{author}{\bibfnamefont{C.}~\bibnamefont{Bingham}},
  \bibinfo{author}{\bibfnamefont{K.}~\bibnamefont{Rykaczewski}},
  \bibinfo{author}{\bibfnamefont{K.}~\bibnamefont{Toth}},
  \bibinfo{author}{\bibfnamefont{T.}~\bibnamefont{Davinson}},
  \bibinfo{author}{\bibfnamefont{J.}~\bibnamefont{McKenzie}},
  \bibinfo{author}{\bibfnamefont{P.}~\bibnamefont{Woods}},
  \bibinfo{author}{\bibfnamefont{T.}~\bibnamefont{Ginter}},
  \bibinfo{author}{\bibfnamefont{C.}~\bibnamefont{Gross}},
  \bibinfo{author}{\bibfnamefont{J.}~\bibnamefont{McConnell}},
  \bibinfo{author}{\bibfnamefont{E.}~\bibnamefont{Zganjar}},
  \bibinfo{author}{\bibfnamefont{J.}~\bibnamefont{Hamilton}}, \emph{et~al.},
  \bibinfo{year}{1998}, \bibinfo{journal}{Phys.\ Rev.}
  \textbf{\bibinfo{volume}{C~57}}, \bibinfo{pages}{R1042}.

\bibitem[{\citenamefont{Batist} \emph{et~al.}(2010)\citenamefont{Batist,
  G\'{o}rska, Grawe, Janas, Kavatsyuk, Karny, Kirchner, La~Commara, Mukha,
  Plochocki, and Roeckl}}]{Batist:2010}
\bibinfo{author}{\bibnamefont{Batist}, \bibfnamefont{L.}},
  \bibinfo{author}{\bibfnamefont{M.}~\bibnamefont{G\'{o}rska}},
  \bibinfo{author}{\bibfnamefont{H.}~\bibnamefont{Grawe}},
  \bibinfo{author}{\bibfnamefont{Z.}~\bibnamefont{Janas}},
  \bibinfo{author}{\bibfnamefont{M.}~\bibnamefont{Kavatsyuk}},
  \bibinfo{author}{\bibfnamefont{M.}~\bibnamefont{Karny}},
  \bibinfo{author}{\bibfnamefont{R.}~\bibnamefont{Kirchner}},
  \bibinfo{author}{\bibfnamefont{M.}~\bibnamefont{La~Commara}},
  \bibinfo{author}{\bibfnamefont{I.}~\bibnamefont{Mukha}},
  \bibinfo{author}{\bibfnamefont{A.}~\bibnamefont{Plochocki}}, and
  \bibinfo{author}{\bibfnamefont{E.}~\bibnamefont{Roeckl}},
  \bibinfo{year}{2010}, \bibinfo{journal}{The European Physical Journal A -
  Hadrons and Nuclei} \textbf{\bibinfo{volume}{46}}, \bibinfo{pages}{45}.

\bibitem[{\citenamefont{Baumann} \emph{et~al.}(2007)\citenamefont{Baumann,
  Amthor, Bazin, Brown, III, Gade, Ginter, Hausmann, Matos, Morrissey,
  Portillo, Schiller} \emph{et~al.}}]{Baumann:2007}
\bibinfo{author}{\bibnamefont{Baumann}, \bibfnamefont{T.}},
  \bibinfo{author}{\bibfnamefont{A.}~\bibnamefont{Amthor}},
  \bibinfo{author}{\bibfnamefont{D.}~\bibnamefont{Bazin}},
  \bibinfo{author}{\bibfnamefont{B.}~\bibnamefont{Brown}},
  \bibinfo{author}{\bibfnamefont{C.~F.} \bibnamefont{III}},
  \bibinfo{author}{\bibfnamefont{A.}~\bibnamefont{Gade}},
  \bibinfo{author}{\bibfnamefont{T.}~\bibnamefont{Ginter}},
  \bibinfo{author}{\bibfnamefont{M.}~\bibnamefont{Hausmann}},
  \bibinfo{author}{\bibfnamefont{M.}~\bibnamefont{Matos}},
  \bibinfo{author}{\bibfnamefont{D.}~\bibnamefont{Morrissey}},
  \bibinfo{author}{\bibfnamefont{M.}~\bibnamefont{Portillo}},
  \bibinfo{author}{\bibfnamefont{A.}~\bibnamefont{Schiller}}, \emph{et~al.},
  \bibinfo{year}{2007}, \bibinfo{journal}{Nature}
  \textbf{\bibinfo{volume}{449}}, \bibinfo{pages}{1022}.

\bibitem[{\citenamefont{Baye} \emph{et~al.}(2006)\citenamefont{Baye, Tursunov,
  and Descouvemont}}]{Baye:2006}
\bibinfo{author}{\bibnamefont{Baye}, \bibfnamefont{D.}},
  \bibinfo{author}{\bibfnamefont{E.~M.} \bibnamefont{Tursunov}}, and
  \bibinfo{author}{\bibfnamefont{P.}~\bibnamefont{Descouvemont}},
  \bibinfo{year}{2006}, \bibinfo{journal}{Phys. Rev. C}
  \textbf{\bibinfo{volume}{74}}(\bibinfo{number}{6}), \bibinfo{pages}{064302}.

\bibitem[{\citenamefont{Baz'}(1967)}]{Baz:1967}
\bibinfo{author}{\bibnamefont{Baz'}, \bibfnamefont{A.~I.}},
  \bibinfo{year}{1967}, \bibinfo{journal}{Yad. Fiz.}
  \textbf{\bibinfo{volume}{5}}, \bibinfo{pages}{229}.

\bibitem[{\citenamefont{Baz'} \emph{et~al.}(1972)\citenamefont{Baz', Goldansky,
  Goldberg, and Zeldovich}}]{Baz:1972}
\bibinfo{author}{\bibnamefont{Baz'}, \bibfnamefont{A.~I.}},
  \bibinfo{author}{\bibfnamefont{V.~I.} \bibnamefont{Goldansky}},
  \bibinfo{author}{\bibfnamefont{V.~Z.} \bibnamefont{Goldberg}}, and
  \bibinfo{author}{\bibfnamefont{Y.~B.} \bibnamefont{Zeldovich}},
  \bibinfo{year}{1972}, \emph{\bibinfo{title}{Light and intermediate nuclei
  near the border of nuclear stability}} (\bibinfo{publisher}{Nauka, Moscow}).

\bibitem[{\citenamefont{Bazin} \emph{et~al.}(2008)\citenamefont{Bazin, Montes,
  Becerril, Lorusso, Amthor, Baumann, Crawford, Estrade, Gade, Ginter, Guess,
  Hausmann} \emph{et~al.}}]{Bazin:2008}
\bibinfo{author}{\bibnamefont{Bazin}, \bibfnamefont{D.}},
  \bibinfo{author}{\bibfnamefont{F.}~\bibnamefont{Montes}},
  \bibinfo{author}{\bibfnamefont{A.}~\bibnamefont{Becerril}},
  \bibinfo{author}{\bibfnamefont{G.}~\bibnamefont{Lorusso}},
  \bibinfo{author}{\bibfnamefont{A.}~\bibnamefont{Amthor}},
  \bibinfo{author}{\bibfnamefont{T.}~\bibnamefont{Baumann}},
  \bibinfo{author}{\bibfnamefont{H.}~\bibnamefont{Crawford}},
  \bibinfo{author}{\bibfnamefont{A.}~\bibnamefont{Estrade}},
  \bibinfo{author}{\bibfnamefont{A.}~\bibnamefont{Gade}},
  \bibinfo{author}{\bibfnamefont{T.}~\bibnamefont{Ginter}},
  \bibinfo{author}{\bibfnamefont{C.~J.} \bibnamefont{Guess}},
  \bibinfo{author}{\bibfnamefont{M.}~\bibnamefont{Hausmann}}, \emph{et~al.},
  \bibinfo{year}{2008}, \bibinfo{journal}{Phys. Rev. Lett.}
  \textbf{\bibinfo{volume}{101}}(\bibinfo{number}{25}),
  \bibinfo{pages}{252501}.

\bibitem[{\citenamefont{Becchetti and Greenlees}(1969)}]{Becchetti:1969}
\bibinfo{author}{\bibnamefont{Becchetti}, \bibfnamefont{F.}}, and
  \bibinfo{author}{\bibfnamefont{G.}~\bibnamefont{Greenlees}},
  \bibinfo{year}{1969}, \bibinfo{journal}{Phys.\ Rev.}
  \textbf{\bibinfo{volume}{182}}, \bibinfo{pages}{1190}.

\bibitem[{\citenamefont{Becquerel}(1896)}]{Becquerel:1896}
\bibinfo{author}{\bibnamefont{Becquerel}, \bibfnamefont{H.}},
  \bibinfo{year}{1896}, \bibinfo{journal}{Comptes Rendus}
  \textbf{\bibinfo{volume}{122}}, \bibinfo{pages}{501}.

\bibitem[{\citenamefont{Behrens and B\"{u}ring}(1982)}]{Behrens:1982}
\bibinfo{author}{\bibnamefont{Behrens}, \bibfnamefont{H.}}, and
  \bibinfo{author}{\bibfnamefont{W.}~\bibnamefont{B\"{u}ring}},
  \bibinfo{year}{1982}, \emph{\bibinfo{title}{Electron radial wave functions
  and nuclear beta-decay}} (\bibinfo{publisher}{Oxford University Press}).

\bibitem[{\citenamefont{Benenson and Kashy}(1979)}]{Benenson:1979}
\bibinfo{author}{\bibnamefont{Benenson}, \bibfnamefont{W.}}, and
  \bibinfo{author}{\bibfnamefont{E.}~\bibnamefont{Kashy}},
  \bibinfo{year}{1979}, \bibinfo{journal}{Rev. Mod. Phys.}
  \textbf{\bibinfo{volume}{51}}(\bibinfo{number}{3}), \bibinfo{pages}{527}.

\bibitem[{\citenamefont{Benlliure} \emph{et~al.}(2008)\citenamefont{Benlliure,
  Fern\'andez-Ord\'o\~nez, Audouin, Boudard, Casarejos, Ducret, Enqvist, Heinz,
  Henzlova, Henzl, Kelic, Leray} \emph{et~al.}}]{Benlliure:2008}
\bibinfo{author}{\bibnamefont{Benlliure}, \bibfnamefont{J.}},
  \bibinfo{author}{\bibfnamefont{M.}~\bibnamefont{Fern\'andez-Ord\'o\~nez}},
  \bibinfo{author}{\bibfnamefont{L.}~\bibnamefont{Audouin}},
  \bibinfo{author}{\bibfnamefont{A.}~\bibnamefont{Boudard}},
  \bibinfo{author}{\bibfnamefont{E.}~\bibnamefont{Casarejos}},
  \bibinfo{author}{\bibfnamefont{J.~E.} \bibnamefont{Ducret}},
  \bibinfo{author}{\bibfnamefont{T.}~\bibnamefont{Enqvist}},
  \bibinfo{author}{\bibfnamefont{A.}~\bibnamefont{Heinz}},
  \bibinfo{author}{\bibfnamefont{D.}~\bibnamefont{Henzlova}},
  \bibinfo{author}{\bibfnamefont{V.}~\bibnamefont{Henzl}},
  \bibinfo{author}{\bibfnamefont{A.}~\bibnamefont{Kelic}},
  \bibinfo{author}{\bibfnamefont{S.}~\bibnamefont{Leray}}, \emph{et~al.},
  \bibinfo{year}{2008}, \bibinfo{journal}{Phys. Rev. C}
  \textbf{\bibinfo{volume}{78}}(\bibinfo{number}{5}), \bibinfo{pages}{054605}.

\bibitem[{\citenamefont{Benlliure} \emph{et~al.}(1999)\citenamefont{Benlliure,
  Schmidt, Cortina-Gil, Enqvist, Farget, Heinz, Junghans, Pereira, and
  Taieb}}]{Benlliure:1999}
\bibinfo{author}{\bibnamefont{Benlliure}, \bibfnamefont{J.}},
  \bibinfo{author}{\bibfnamefont{K.~H.} \bibnamefont{Schmidt}},
  \bibinfo{author}{\bibfnamefont{D.}~\bibnamefont{Cortina-Gil}},
  \bibinfo{author}{\bibfnamefont{T.}~\bibnamefont{Enqvist}},
  \bibinfo{author}{\bibfnamefont{F.}~\bibnamefont{Farget}},
  \bibinfo{author}{\bibfnamefont{A.}~\bibnamefont{Heinz}},
  \bibinfo{author}{\bibfnamefont{A.~R.} \bibnamefont{Junghans}},
  \bibinfo{author}{\bibfnamefont{J.}~\bibnamefont{Pereira}}, and
  \bibinfo{author}{\bibfnamefont{J.}~\bibnamefont{Taieb}},
  \bibinfo{year}{1999}, \bibinfo{journal}{Nuclear Physics A}
  \textbf{\bibinfo{volume}{660}}(\bibinfo{number}{1}), \bibinfo{pages}{87}.

\bibitem[{\citenamefont{Benlliure} \emph{et~al.}(2000)\citenamefont{Benlliure,
  Schmidt, Cortina-Gil, Enqvist, Farget, Heinz, Junghans, Pereira, and
  Taieb}}]{Benlliure:2000}
\bibinfo{author}{\bibnamefont{Benlliure}, \bibfnamefont{J.}},
  \bibinfo{author}{\bibfnamefont{K.~H.} \bibnamefont{Schmidt}},
  \bibinfo{author}{\bibfnamefont{D.}~\bibnamefont{Cortina-Gil}},
  \bibinfo{author}{\bibfnamefont{T.}~\bibnamefont{Enqvist}},
  \bibinfo{author}{\bibfnamefont{F.}~\bibnamefont{Farget}},
  \bibinfo{author}{\bibfnamefont{A.}~\bibnamefont{Heinz}},
  \bibinfo{author}{\bibfnamefont{A.~R.} \bibnamefont{Junghans}},
  \bibinfo{author}{\bibfnamefont{J.}~\bibnamefont{Pereira}}, and
  \bibinfo{author}{\bibfnamefont{J.}~\bibnamefont{Taieb}},
  \bibinfo{year}{2000}, \bibinfo{journal}{Nuclear Physics A}
  \textbf{\bibinfo{volume}{674}}(\bibinfo{number}{3-4}), \bibinfo{pages}{578}.

\bibitem[{\citenamefont{Bennaceur} \emph{et~al.}(1998)\citenamefont{Bennaceur,
  Nowacki, Oko{\l}owicz, and P{\l}oszajczak}}]{Bennaceur:1998}
\bibinfo{author}{\bibnamefont{Bennaceur}, \bibfnamefont{K.}},
  \bibinfo{author}{\bibfnamefont{F.}~\bibnamefont{Nowacki}},
  \bibinfo{author}{\bibfnamefont{J.}~\bibnamefont{Oko{\l}owicz}}, and
  \bibinfo{author}{\bibfnamefont{M.}~\bibnamefont{P{\l}oszajczak}},
  \bibinfo{year}{1998}, \bibinfo{journal}{Journal of Physics G: Nuclear and
  Particle Physics} \textbf{\bibinfo{volume}{24}}(\bibinfo{number}{8}),
  \bibinfo{pages}{1631}.

\bibitem[{\citenamefont{Bentley and Lenzi}(2007)}]{Bentley:2007}
\bibinfo{author}{\bibnamefont{Bentley}, \bibfnamefont{M.~A.}}, and
  \bibinfo{author}{\bibfnamefont{S.~M.} \bibnamefont{Lenzi}},
  \bibinfo{year}{2007}, \bibinfo{journal}{Progress in Particle and Nuclear
  Physics} \textbf{\bibinfo{volume}{59}}(\bibinfo{number}{2}),
  \bibinfo{pages}{497}.

\bibitem[{\citenamefont{Bergmann} \emph{et~al.}(1999)\citenamefont{Bergmann,
  Axelsson, Borge, Fedoseyev, Forss\'{e}n, Fynbo, Gr\'{e}vy, Hornsh{\o}j,
  Jading, Jonson, K\"{o}ster, Markenroth} \emph{et~al.}}]{Bergmann:1999}
\bibinfo{author}{\bibnamefont{Bergmann}, \bibfnamefont{U.~C.}},
  \bibinfo{author}{\bibfnamefont{L.}~\bibnamefont{Axelsson}},
  \bibinfo{author}{\bibfnamefont{M.~J.~G.} \bibnamefont{Borge}},
  \bibinfo{author}{\bibfnamefont{V.~N.} \bibnamefont{Fedoseyev}},
  \bibinfo{author}{\bibfnamefont{C.}~\bibnamefont{Forss\'{e}n}},
  \bibinfo{author}{\bibfnamefont{H.~O.~U.} \bibnamefont{Fynbo}},
  \bibinfo{author}{\bibfnamefont{S.}~\bibnamefont{Gr\'{e}vy}},
  \bibinfo{author}{\bibfnamefont{P.}~\bibnamefont{Hornsh{\o}j}},
  \bibinfo{author}{\bibfnamefont{Y.}~\bibnamefont{Jading}},
  \bibinfo{author}{\bibfnamefont{B.}~\bibnamefont{Jonson}},
  \bibinfo{author}{\bibfnamefont{U.}~\bibnamefont{K\"{o}ster}},
  \bibinfo{author}{\bibfnamefont{K.}~\bibnamefont{Markenroth}}, \emph{et~al.},
  \bibinfo{year}{1999}, \bibinfo{journal}{Nuclear Physics A}
  \textbf{\bibinfo{volume}{658}}(\bibinfo{number}{2}), \bibinfo{pages}{129}.

\bibitem[{\citenamefont{Bernas} \emph{et~al.}(2003)\citenamefont{Bernas,
  Armbruster, Benlliure, Boudard, Casarejos, Czajkowski, Enqvist, Legrain,
  Leray, Mustapha, Napolitani, Pereira} \emph{et~al.}}]{Bernas:2003}
\bibinfo{author}{\bibnamefont{Bernas}, \bibfnamefont{M.}},
  \bibinfo{author}{\bibfnamefont{P.}~\bibnamefont{Armbruster}},
  \bibinfo{author}{\bibfnamefont{J.}~\bibnamefont{Benlliure}},
  \bibinfo{author}{\bibfnamefont{A.}~\bibnamefont{Boudard}},
  \bibinfo{author}{\bibfnamefont{E.}~\bibnamefont{Casarejos}},
  \bibinfo{author}{\bibfnamefont{S.}~\bibnamefont{Czajkowski}},
  \bibinfo{author}{\bibfnamefont{T.}~\bibnamefont{Enqvist}},
  \bibinfo{author}{\bibfnamefont{R.}~\bibnamefont{Legrain}},
  \bibinfo{author}{\bibfnamefont{S.}~\bibnamefont{Leray}},
  \bibinfo{author}{\bibfnamefont{B.}~\bibnamefont{Mustapha}},
  \bibinfo{author}{\bibfnamefont{P.}~\bibnamefont{Napolitani}},
  \bibinfo{author}{\bibfnamefont{J.}~\bibnamefont{Pereira}}, \emph{et~al.},
  \bibinfo{year}{2003}, \bibinfo{journal}{Nuclear Physics A}
  \textbf{\bibinfo{volume}{725}}, \bibinfo{pages}{213}.

\bibitem[{\citenamefont{Bernas} \emph{et~al.}(1994)\citenamefont{Bernas,
  Czajkowski, Armbruster, Geissel, Dessagne, Donzaud, Faust, Hanelt, Heinz,
  Hesse, Kozhuharov, Miehe} \emph{et~al.}}]{Bernas:1994}
\bibinfo{author}{\bibnamefont{Bernas}, \bibfnamefont{M.}},
  \bibinfo{author}{\bibfnamefont{S.}~\bibnamefont{Czajkowski}},
  \bibinfo{author}{\bibfnamefont{P.}~\bibnamefont{Armbruster}},
  \bibinfo{author}{\bibfnamefont{H.}~\bibnamefont{Geissel}},
  \bibinfo{author}{\bibfnamefont{P.}~\bibnamefont{Dessagne}},
  \bibinfo{author}{\bibfnamefont{C.}~\bibnamefont{Donzaud}},
  \bibinfo{author}{\bibfnamefont{H.-R.} \bibnamefont{Faust}},
  \bibinfo{author}{\bibfnamefont{E.}~\bibnamefont{Hanelt}},
  \bibinfo{author}{\bibfnamefont{A.}~\bibnamefont{Heinz}},
  \bibinfo{author}{\bibfnamefont{M.}~\bibnamefont{Hesse}},
  \bibinfo{author}{\bibfnamefont{C.}~\bibnamefont{Kozhuharov}},
  \bibinfo{author}{\bibfnamefont{C.}~\bibnamefont{Miehe}}, \emph{et~al.},
  \bibinfo{year}{1994}, \bibinfo{journal}{Physics Letters B}
  \textbf{\bibinfo{volume}{331}}(\bibinfo{number}{1-2}), \bibinfo{pages}{19}.

\bibitem[{\citenamefont{Bernas} \emph{et~al.}(1997)\citenamefont{Bernas,
  Engelmann, Armbruster, Czajkowski, Ameil, Böckstiegel, Dessagne, Donzaud,
  Geissel, Heinz, Janas, Kozhuharov} \emph{et~al.}}]{Bernas:1997}
\bibinfo{author}{\bibnamefont{Bernas}, \bibfnamefont{M.}},
  \bibinfo{author}{\bibfnamefont{C.}~\bibnamefont{Engelmann}},
  \bibinfo{author}{\bibfnamefont{P.}~\bibnamefont{Armbruster}},
  \bibinfo{author}{\bibfnamefont{S.}~\bibnamefont{Czajkowski}},
  \bibinfo{author}{\bibfnamefont{F.}~\bibnamefont{Ameil}},
  \bibinfo{author}{\bibfnamefont{C.}~\bibnamefont{Böckstiegel}},
  \bibinfo{author}{\bibfnamefont{P.}~\bibnamefont{Dessagne}},
  \bibinfo{author}{\bibfnamefont{C.}~\bibnamefont{Donzaud}},
  \bibinfo{author}{\bibfnamefont{H.}~\bibnamefont{Geissel}},
  \bibinfo{author}{\bibfnamefont{A.}~\bibnamefont{Heinz}},
  \bibinfo{author}{\bibfnamefont{Z.}~\bibnamefont{Janas}},
  \bibinfo{author}{\bibfnamefont{C.}~\bibnamefont{Kozhuharov}}, \emph{et~al.},
  \bibinfo{year}{1997}, \bibinfo{journal}{Physics Letters B}
  \textbf{\bibinfo{volume}{415}}(\bibinfo{number}{2}), \bibinfo{pages}{111}.

\bibitem[{\citenamefont{Bertulani and Baur}(1986)}]{Bertulani:1986}
\bibinfo{author}{\bibnamefont{Bertulani}, \bibfnamefont{C.~A.}}, and
  \bibinfo{author}{\bibfnamefont{G.}~\bibnamefont{Baur}}, \bibinfo{year}{1986},
  \bibinfo{journal}{Nuclear Physics A}
  \textbf{\bibinfo{volume}{458}}(\bibinfo{number}{4}), \bibinfo{pages}{725}.

\bibitem[{\citenamefont{Bethe and Peierls}(1934)}]{Bethe:1934}
\bibinfo{author}{\bibnamefont{Bethe}, \bibfnamefont{H.}}, and
  \bibinfo{author}{\bibfnamefont{R.}~\bibnamefont{Peierls}},
  \bibinfo{year}{1934}, \bibinfo{journal}{Nature}
  \textbf{\bibinfo{volume}{133}}, \bibinfo{pages}{532}.

\bibitem[{\citenamefont{Bethe}(1937)}]{Bethe:1937}
\bibinfo{author}{\bibnamefont{Bethe}, \bibfnamefont{H.~A.}},
  \bibinfo{year}{1937}, \bibinfo{journal}{Rev. Mod. Phys.}
  \textbf{\bibinfo{volume}{9}}(\bibinfo{number}{2}), \bibinfo{pages}{69}.

\bibitem[{\citenamefont{Bhattacharya}
  \emph{et~al.}(2008)\citenamefont{Bhattacharya, Melconian, Komives, Triambak,
  Garc\'\i{}a, Adelberger, Brown, Cooper, Glasmacher, Guimaraes, Mantica,
  Oros-Peusquens} \emph{et~al.}}]{Bhattacharya:2008}
\bibinfo{author}{\bibnamefont{Bhattacharya}, \bibfnamefont{M.}},
  \bibinfo{author}{\bibfnamefont{D.}~\bibnamefont{Melconian}},
  \bibinfo{author}{\bibfnamefont{A.}~\bibnamefont{Komives}},
  \bibinfo{author}{\bibfnamefont{S.}~\bibnamefont{Triambak}},
  \bibinfo{author}{\bibfnamefont{A.}~\bibnamefont{Garc\'\i{}a}},
  \bibinfo{author}{\bibfnamefont{E.~G.} \bibnamefont{Adelberger}},
  \bibinfo{author}{\bibfnamefont{B.~A.} \bibnamefont{Brown}},
  \bibinfo{author}{\bibfnamefont{M.~W.} \bibnamefont{Cooper}},
  \bibinfo{author}{\bibfnamefont{T.}~\bibnamefont{Glasmacher}},
  \bibinfo{author}{\bibfnamefont{V.}~\bibnamefont{Guimaraes}},
  \bibinfo{author}{\bibfnamefont{P.~F.} \bibnamefont{Mantica}},
  \bibinfo{author}{\bibfnamefont{A.~M.} \bibnamefont{Oros-Peusquens}},
  \emph{et~al.}, \bibinfo{year}{2008}, \bibinfo{journal}{Phys. Rev. C}
  \textbf{\bibinfo{volume}{77}}(\bibinfo{number}{6}), \bibinfo{pages}{065503}.

\bibitem[{\citenamefont{Bilpuch} \emph{et~al.}(1976)\citenamefont{Bilpuch,
  Lane, Mitchell, and Moses}}]{Bilpuch:1976}
\bibinfo{author}{\bibnamefont{Bilpuch}, \bibfnamefont{E.~G.}},
  \bibinfo{author}{\bibfnamefont{A.~M.} \bibnamefont{Lane}},
  \bibinfo{author}{\bibfnamefont{G.~E.} \bibnamefont{Mitchell}}, and
  \bibinfo{author}{\bibfnamefont{J.~D.} \bibnamefont{Moses}},
  \bibinfo{year}{1976}, \bibinfo{journal}{Physics Reports}
  \textbf{\bibinfo{volume}{28}}(\bibinfo{number}{2}), \bibinfo{pages}{145}.

\bibitem[{\citenamefont{Bingham} \emph{et~al.}(1999)\citenamefont{Bingham,
  Batchelder, Rykaczewski, Toth, Yu, Ginter, Gross, Grzywacz, Karny, Kim,
  MacDonald, Mas} \emph{et~al.}}]{Bingham:1999}
\bibinfo{author}{\bibnamefont{Bingham}, \bibfnamefont{C.}},
  \bibinfo{author}{\bibfnamefont{J.}~\bibnamefont{Batchelder}},
  \bibinfo{author}{\bibfnamefont{K.}~\bibnamefont{Rykaczewski}},
  \bibinfo{author}{\bibfnamefont{K.}~\bibnamefont{Toth}},
  \bibinfo{author}{\bibfnamefont{C.-H.} \bibnamefont{Yu}},
  \bibinfo{author}{\bibfnamefont{T.}~\bibnamefont{Ginter}},
  \bibinfo{author}{\bibfnamefont{C.}~\bibnamefont{Gross}},
  \bibinfo{author}{\bibfnamefont{R.}~\bibnamefont{Grzywacz}},
  \bibinfo{author}{\bibfnamefont{M.}~\bibnamefont{Karny}},
  \bibinfo{author}{\bibfnamefont{S.}~\bibnamefont{Kim}},
  \bibinfo{author}{\bibfnamefont{B.}~\bibnamefont{MacDonald}},
  \bibinfo{author}{\bibfnamefont{J.}~\bibnamefont{Mas}}, \emph{et~al.},
  \bibinfo{year}{1999}, \bibinfo{journal}{Phys.\ Rev.}
  \textbf{\bibinfo{volume}{C~59}}, \bibinfo{pages}{R2984}.

\bibitem[{\citenamefont{Bingham} \emph{et~al.}(2005)\citenamefont{Bingham,
  Tantawy, Batchelder, Danchev, , Ginter, Gross, Fong, Grzywacz, Hagino,
  Hamilton, Karny} \emph{et~al.}}]{Bingham:2005}
\bibinfo{author}{\bibnamefont{Bingham}, \bibfnamefont{C.}},
  \bibinfo{author}{\bibfnamefont{M.}~\bibnamefont{Tantawy}},
  \bibinfo{author}{\bibfnamefont{J.}~\bibnamefont{Batchelder}},
  \bibinfo{author}{\bibfnamefont{M.}~\bibnamefont{Danchev}}, ,
  \bibinfo{author}{\bibfnamefont{T.}~\bibnamefont{Ginter}},
  \bibinfo{author}{\bibfnamefont{C.}~\bibnamefont{Gross}},
  \bibinfo{author}{\bibfnamefont{D.}~\bibnamefont{Fong}},
  \bibinfo{author}{\bibfnamefont{R.}~\bibnamefont{Grzywacz}},
  \bibinfo{author}{\bibfnamefont{K.}~\bibnamefont{Hagino}},
  \bibinfo{author}{\bibfnamefont{J.}~\bibnamefont{Hamilton}},
  \bibinfo{author}{\bibfnamefont{M.}~\bibnamefont{Karny}}, \emph{et~al.},
  \bibinfo{year}{2005}, \bibinfo{journal}{Nucl.\ Instr.\ Meth.\ Phys.\ Res.}
  \textbf{\bibinfo{volume}{B~241}}, \bibinfo{pages}{185}.

\bibitem[{\citenamefont{Blank} \emph{et~al.}(2011)\citenamefont{Blank, Ascher,
  Audirac, Canchel, Giovinazzo, Kurtukian-Nieto, de~Oliveira~Santos, Grévy,
  Thomas, Borcea, and Grigorenko}}]{Blank:2011}
\bibinfo{author}{\bibnamefont{Blank}, \bibfnamefont{B.}},
  \bibinfo{author}{\bibfnamefont{P.}~\bibnamefont{Ascher}},
  \bibinfo{author}{\bibfnamefont{L.}~\bibnamefont{Audirac}},
  \bibinfo{author}{\bibfnamefont{G.}~\bibnamefont{Canchel}},
  \bibinfo{author}{\bibfnamefont{J.}~\bibnamefont{Giovinazzo}},
  \bibinfo{author}{\bibfnamefont{T.}~\bibnamefont{Kurtukian-Nieto}},
  \bibinfo{author}{\bibfnamefont{F.}~\bibnamefont{de~Oliveira~Santos}},
  \bibinfo{author}{\bibfnamefont{S.}~\bibnamefont{Grévy}},
  \bibinfo{author}{\bibfnamefont{J.-C.} \bibnamefont{Thomas}},
  \bibinfo{author}{\bibfnamefont{C.}~\bibnamefont{Borcea}}, and
  \bibinfo{author}{\bibfnamefont{L.}~\bibnamefont{Grigorenko}},
  \bibinfo{year}{2011}, \bibinfo{journal}{Acta Physica Polonica B}
  \textbf{\bibinfo{volume}{42}}, \bibinfo{pages}{545}.

\bibitem[{\citenamefont{Blank} \emph{et~al.}(2005)\citenamefont{Blank, Bey,
  Canchel, Dossat, Fleury, Giovinazzo, Matea, Adimi, Oliveira, Stefan,
  Georgiev, Grevy} \emph{et~al.}}]{Blank:2005}
\bibinfo{author}{\bibnamefont{Blank}, \bibfnamefont{B.}},
  \bibinfo{author}{\bibfnamefont{A.}~\bibnamefont{Bey}},
  \bibinfo{author}{\bibfnamefont{G.}~\bibnamefont{Canchel}},
  \bibinfo{author}{\bibfnamefont{C.}~\bibnamefont{Dossat}},
  \bibinfo{author}{\bibfnamefont{A.}~\bibnamefont{Fleury}},
  \bibinfo{author}{\bibfnamefont{J.}~\bibnamefont{Giovinazzo}},
  \bibinfo{author}{\bibfnamefont{I.}~\bibnamefont{Matea}},
  \bibinfo{author}{\bibfnamefont{N.}~\bibnamefont{Adimi}},
  \bibinfo{author}{\bibfnamefont{F.~D.} \bibnamefont{Oliveira}},
  \bibinfo{author}{\bibfnamefont{I.}~\bibnamefont{Stefan}},
  \bibinfo{author}{\bibfnamefont{G.}~\bibnamefont{Georgiev}},
  \bibinfo{author}{\bibfnamefont{S.}~\bibnamefont{Grevy}}, \emph{et~al.},
  \bibinfo{year}{2005}, \bibinfo{journal}{Phys. Rev. Lett.}
  \textbf{\bibinfo{volume}{94}}, \bibinfo{pages}{232501}.

\bibitem[{\citenamefont{Blank and Borge}(2008)}]{Blank:2008}
\bibinfo{author}{\bibnamefont{Blank}, \bibfnamefont{B.}}, and
  \bibinfo{author}{\bibfnamefont{M.}~\bibnamefont{Borge}},
  \bibinfo{year}{2008}, \bibinfo{journal}{Progress in Particle and Nuclear
  Physics} \textbf{\bibinfo{volume}{60}}(\bibinfo{number}{2}),
  \bibinfo{pages}{403}.

\bibitem[{\citenamefont{Blank} \emph{et~al.}(2000)\citenamefont{Blank,
  Chartier, Czajkowski, Giovinazzo, Pravikoff, Thomas, de~France,
  de~Oliveira~Santos, Lewitowicz, Borcea, Grzywacz, Janas}
  \emph{et~al.}}]{Blank:2000}
\bibinfo{author}{\bibnamefont{Blank}, \bibfnamefont{B.}},
  \bibinfo{author}{\bibfnamefont{M.}~\bibnamefont{Chartier}},
  \bibinfo{author}{\bibfnamefont{S.}~\bibnamefont{Czajkowski}},
  \bibinfo{author}{\bibfnamefont{J.}~\bibnamefont{Giovinazzo}},
  \bibinfo{author}{\bibfnamefont{M.~S.} \bibnamefont{Pravikoff}},
  \bibinfo{author}{\bibfnamefont{J.-C.} \bibnamefont{Thomas}},
  \bibinfo{author}{\bibfnamefont{G.}~\bibnamefont{de~France}},
  \bibinfo{author}{\bibfnamefont{F.}~\bibnamefont{de~Oliveira~Santos}},
  \bibinfo{author}{\bibfnamefont{M.}~\bibnamefont{Lewitowicz}},
  \bibinfo{author}{\bibfnamefont{C.}~\bibnamefont{Borcea}},
  \bibinfo{author}{\bibfnamefont{R.}~\bibnamefont{Grzywacz}},
  \bibinfo{author}{\bibfnamefont{Z.}~\bibnamefont{Janas}}, \emph{et~al.},
  \bibinfo{year}{2000}, \bibinfo{journal}{Phys. Rev. Lett.}
  \textbf{\bibinfo{volume}{84}}(\bibinfo{number}{6}), \bibinfo{pages}{1116}.

\bibitem[{\citenamefont{Blank} \emph{et~al.}(2010)\citenamefont{Blank, Hay,
  Huikari, Leblanc, List, Pedroza, Ascher, Audirac, Borcea, Canchel, Delalee,
  Demonchy} \emph{et~al.}}]{Blank:2010}
\bibinfo{author}{\bibnamefont{Blank}, \bibfnamefont{B.}},
  \bibinfo{author}{\bibfnamefont{L.}~\bibnamefont{Hay}},
  \bibinfo{author}{\bibfnamefont{J.}~\bibnamefont{Huikari}},
  \bibinfo{author}{\bibfnamefont{S.}~\bibnamefont{Leblanc}},
  \bibinfo{author}{\bibfnamefont{S.}~\bibnamefont{List}},
  \bibinfo{author}{\bibfnamefont{J.-L.} \bibnamefont{Pedroza}},
  \bibinfo{author}{\bibfnamefont{P.}~\bibnamefont{Ascher}},
  \bibinfo{author}{\bibfnamefont{L.}~\bibnamefont{Audirac}},
  \bibinfo{author}{\bibfnamefont{C.}~\bibnamefont{Borcea}},
  \bibinfo{author}{\bibfnamefont{G.}~\bibnamefont{Canchel}},
  \bibinfo{author}{\bibfnamefont{F.}~\bibnamefont{Delalee}},
  \bibinfo{author}{\bibfnamefont{C.}~\bibnamefont{Demonchy}}, \emph{et~al.},
  \bibinfo{year}{2010}, \bibinfo{journal}{Nuclear Instruments and Methods in
  Physics Research Section A: Accelerators, Spectrometers, Detectors and
  Associated Equipment} \textbf{\bibinfo{volume}{613}}(\bibinfo{number}{1}),
  \bibinfo{pages}{65}.

\bibitem[{\citenamefont{Blank and P{\l}oszajczak}(2008)}]{Blank:2008a}
\bibinfo{author}{\bibnamefont{Blank}, \bibfnamefont{B.}}, and
  \bibinfo{author}{\bibfnamefont{M.}~\bibnamefont{P{\l}oszajczak}},
  \bibinfo{year}{2008}, \bibinfo{journal}{Rep. Prog. Phys.}
  \textbf{\bibinfo{volume}{71}}, \bibinfo{pages}{046301}.

\bibitem[{\citenamefont{{Blumenfeld}}
  \emph{et~al.}(2009)\citenamefont{{Blumenfeld}, {Butler}, {Cornell},
  {Fortuna}, and {Lindroos}}}]{Blumenfeld:2009}
\bibinfo{author}{\bibnamefont{{Blumenfeld}}, \bibfnamefont{Y.}},
  \bibinfo{author}{\bibfnamefont{P.}~\bibnamefont{{Butler}}},
  \bibinfo{author}{\bibfnamefont{J.}~\bibnamefont{{Cornell}}},
  \bibinfo{author}{\bibfnamefont{G.}~\bibnamefont{{Fortuna}}}, and
  \bibinfo{author}{\bibfnamefont{M.}~\bibnamefont{{Lindroos}}},
  \bibinfo{year}{2009}, \bibinfo{journal}{Int.J.Mod.Phys.}
  \textbf{\bibinfo{volume}{E18}}, \bibinfo{pages}{1960}.

\bibitem[{\citenamefont{Bochkarev} \emph{et~al.}(1984)\citenamefont{Bochkarev,
  Korsheninnikov, Kuz'min, Mukha, Ogloblin, Chulkov, and
  Yan'kov}}]{Bochkarev:1984}
\bibinfo{author}{\bibnamefont{Bochkarev}, \bibfnamefont{O.~V.}},
  \bibinfo{author}{\bibfnamefont{A.~A.} \bibnamefont{Korsheninnikov}},
  \bibinfo{author}{\bibfnamefont{E.~A.} \bibnamefont{Kuz'min}},
  \bibinfo{author}{\bibfnamefont{I.~G.} \bibnamefont{Mukha}},
  \bibinfo{author}{\bibfnamefont{A.~A.} \bibnamefont{Ogloblin}},
  \bibinfo{author}{\bibfnamefont{L.~V.} \bibnamefont{Chulkov}}, and
  \bibinfo{author}{\bibfnamefont{G.~B.} \bibnamefont{Yan'kov}},
  \bibinfo{year}{1984}, \bibinfo{journal}{JETP Lett.}
  \textbf{\bibinfo{volume}{40}}, \bibinfo{pages}{969}, \bibinfo{note}{[Zh.
  Eksp. Teor. Fiz. 40 (1984) 204-207]}.

\bibitem[{Bochkarev \emph{et~al.}(1989)\citenamefont{Bochkarev}
  \emph{et~al.}}]{Bochkarev:1989}
\bibinfo{author}{\bibnamefont{Bochkarev}, \bibfnamefont{O.~V.}}, \emph{et~al.},
  \bibinfo{year}{1989}, \bibinfo{journal}{Nucl. Phys. A}
  \textbf{\bibinfo{volume}{505}}, \bibinfo{pages}{215}.

\bibitem[{\citenamefont{Bogdanov} \emph{et~al.}(1973)\citenamefont{Bogdanov,
  Bochin, Karnaukhov, and Petrov}}]{Bogdanov:1973}
\bibinfo{author}{\bibnamefont{Bogdanov}, \bibfnamefont{D.}},
  \bibinfo{author}{\bibfnamefont{V.}~\bibnamefont{Bochin}},
  \bibinfo{author}{\bibfnamefont{V.}~\bibnamefont{Karnaukhov}}, and
  \bibinfo{author}{\bibfnamefont{L.}~\bibnamefont{Petrov}},
  \bibinfo{year}{1973}, \bibinfo{journal}{Sov.\ J.\ Nucl.\ Phys}
  \textbf{\bibinfo{volume}{16}}, \bibinfo{pages}{491}.

\bibitem[{\citenamefont{Bonetti and Guglielmetti}(2007)}]{Bonetti:2007}
\bibinfo{author}{\bibnamefont{Bonetti}, \bibfnamefont{R.}}, and
  \bibinfo{author}{\bibfnamefont{A.}~\bibnamefont{Guglielmetti}},
  \bibinfo{year}{2007}, \bibinfo{journal}{Romanian Reports on Physics}
  \textbf{\bibinfo{volume}{59}}(\bibinfo{number}{2}), \bibinfo{pages}{301}.

\bibitem[{\citenamefont{Borge} \emph{et~al.}(1988)\citenamefont{Borge,
  Cronberg, Cronqvist, Gabelmann, Hansen, Johannsen, Jonson, Mattsson, Nyman,
  Richter, Riisager, Tengblad} \emph{et~al.}}]{Borge:1988}
\bibinfo{author}{\bibnamefont{Borge}, \bibfnamefont{M.~J.~G.}},
  \bibinfo{author}{\bibfnamefont{H.}~\bibnamefont{Cronberg}},
  \bibinfo{author}{\bibfnamefont{M.}~\bibnamefont{Cronqvist}},
  \bibinfo{author}{\bibfnamefont{H.}~\bibnamefont{Gabelmann}},
  \bibinfo{author}{\bibfnamefont{P.~G.} \bibnamefont{Hansen}},
  \bibinfo{author}{\bibfnamefont{L.}~\bibnamefont{Johannsen}},
  \bibinfo{author}{\bibfnamefont{B.}~\bibnamefont{Jonson}},
  \bibinfo{author}{\bibfnamefont{S.}~\bibnamefont{Mattsson}},
  \bibinfo{author}{\bibfnamefont{G.}~\bibnamefont{Nyman}},
  \bibinfo{author}{\bibfnamefont{A.}~\bibnamefont{Richter}},
  \bibinfo{author}{\bibfnamefont{K.}~\bibnamefont{Riisager}},
  \bibinfo{author}{\bibfnamefont{O.}~\bibnamefont{Tengblad}}, \emph{et~al.},
  \bibinfo{year}{1988}, \bibinfo{journal}{Nuclear Physics A}
  \textbf{\bibinfo{volume}{490}}(\bibinfo{number}{2}), \bibinfo{pages}{287}.

\bibitem[{\citenamefont{Borge} \emph{et~al.}(1997)\citenamefont{Borge, Fynbo,
  Guillemaud-Mueller, Hornsh\o{}j, Humbert, Jonson, Leth, Mart\'\i{}nez-Pinedo,
  Nilsson, Nyman, Poves, Ramos-Lerate} \emph{et~al.}}]{Borge:1997}
\bibinfo{author}{\bibnamefont{Borge}, \bibfnamefont{M.~J.~G.}},
  \bibinfo{author}{\bibfnamefont{H.}~\bibnamefont{Fynbo}},
  \bibinfo{author}{\bibfnamefont{D.}~\bibnamefont{Guillemaud-Mueller}},
  \bibinfo{author}{\bibfnamefont{P.}~\bibnamefont{Hornsh\o{}j}},
  \bibinfo{author}{\bibfnamefont{F.}~\bibnamefont{Humbert}},
  \bibinfo{author}{\bibfnamefont{B.}~\bibnamefont{Jonson}},
  \bibinfo{author}{\bibfnamefont{T.~E.} \bibnamefont{Leth}},
  \bibinfo{author}{\bibfnamefont{G.}~\bibnamefont{Mart\'\i{}nez-Pinedo}},
  \bibinfo{author}{\bibfnamefont{T.}~\bibnamefont{Nilsson}},
  \bibinfo{author}{\bibfnamefont{G.}~\bibnamefont{Nyman}},
  \bibinfo{author}{\bibfnamefont{A.}~\bibnamefont{Poves}},
  \bibinfo{author}{\bibfnamefont{I.}~\bibnamefont{Ramos-Lerate}}, \emph{et~al.}
  (\bibinfo{collaboration}{the ISOLDE Collaboration}), \bibinfo{year}{1997},
  \bibinfo{journal}{Phys. Rev. C}
  \textbf{\bibinfo{volume}{55}}(\bibinfo{number}{1}), \bibinfo{pages}{R8}.

\bibitem[{\citenamefont{Borge} \emph{et~al.}(1989)\citenamefont{Borge, Hansen,
  Jonson, Mattsson, Nyman, Richter, and Riisager}}]{Borge:1989}
\bibinfo{author}{\bibnamefont{Borge}, \bibfnamefont{M.~J.~G.}},
  \bibinfo{author}{\bibfnamefont{P.~G.} \bibnamefont{Hansen}},
  \bibinfo{author}{\bibfnamefont{B.}~\bibnamefont{Jonson}},
  \bibinfo{author}{\bibfnamefont{S.}~\bibnamefont{Mattsson}},
  \bibinfo{author}{\bibfnamefont{G.}~\bibnamefont{Nyman}},
  \bibinfo{author}{\bibfnamefont{A.}~\bibnamefont{Richter}}, and
  \bibinfo{author}{\bibfnamefont{K.}~\bibnamefont{Riisager}},
  \bibinfo{year}{1989}, \bibinfo{journal}{Zeitschrift für Physik A Hadrons and
  Nuclei} \textbf{\bibinfo{volume}{332}}, \bibinfo{pages}{413}.

\bibitem[{\citenamefont{Borzov} \emph{et~al.}(2008)\citenamefont{Borzov,
  Cuenca-García, Langanke, Mart\'{\i}nez-Pinedo, and Montes}}]{Borzov:2008}
\bibinfo{author}{\bibnamefont{Borzov}, \bibfnamefont{I.}},
  \bibinfo{author}{\bibfnamefont{J.}~\bibnamefont{Cuenca-García}},
  \bibinfo{author}{\bibfnamefont{K.}~\bibnamefont{Langanke}},
  \bibinfo{author}{\bibfnamefont{G.}~\bibnamefont{Mart\'{\i}nez-Pinedo}}, and
  \bibinfo{author}{\bibfnamefont{F.}~\bibnamefont{Montes}},
  \bibinfo{year}{2008}, \bibinfo{journal}{Nuclear Physics A}
  \textbf{\bibinfo{volume}{814}}(\bibinfo{number}{1-4}), \bibinfo{pages}{159}.

\bibitem[{\citenamefont{Borzov}(2006)}]{Borzov:2006}
\bibinfo{author}{\bibnamefont{Borzov}, \bibfnamefont{I.~N.}},
  \bibinfo{year}{2006}, \bibinfo{journal}{Nuclear Physics A}
  \textbf{\bibinfo{volume}{777}}, \bibinfo{pages}{645}.

\bibitem[{\citenamefont{Borzov and Goriely}(2000)}]{Borzov:2000}
\bibinfo{author}{\bibnamefont{Borzov}, \bibfnamefont{I.~N.}}, and
  \bibinfo{author}{\bibfnamefont{S.}~\bibnamefont{Goriely}},
  \bibinfo{year}{2000}, \bibinfo{journal}{Phys. Rev. C}
  \textbf{\bibinfo{volume}{62}}(\bibinfo{number}{3}), \bibinfo{pages}{035501}.

\bibitem[{\citenamefont{Boudard} \emph{et~al.}(2002)\citenamefont{Boudard,
  Cugnon, Leray, and Volant}}]{Boudard:2002}
\bibinfo{author}{\bibnamefont{Boudard}, \bibfnamefont{A.}},
  \bibinfo{author}{\bibfnamefont{J.}~\bibnamefont{Cugnon}},
  \bibinfo{author}{\bibfnamefont{S.}~\bibnamefont{Leray}}, and
  \bibinfo{author}{\bibfnamefont{C.}~\bibnamefont{Volant}},
  \bibinfo{year}{2002}, \bibinfo{journal}{Phys. Rev. C}
  \textbf{\bibinfo{volume}{66}}(\bibinfo{number}{4}), \bibinfo{pages}{044615}.

\bibitem[{\citenamefont{Brink} \emph{et~al.}(1983)\citenamefont{Brink, Nemes,
  and Vautherin}}]{Brink:1983}
\bibinfo{author}{\bibnamefont{Brink}, \bibfnamefont{D.~M.}},
  \bibinfo{author}{\bibfnamefont{M.~C.} \bibnamefont{Nemes}}, and
  \bibinfo{author}{\bibfnamefont{D.}~\bibnamefont{Vautherin}},
  \bibinfo{year}{1983}, \bibinfo{journal}{Annals of Physics}
  \textbf{\bibinfo{volume}{147}}(\bibinfo{number}{1}), \bibinfo{pages}{171}.

\bibitem[{\citenamefont{Broda}(2006)}]{Broda:2006}
\bibinfo{author}{\bibnamefont{Broda}, \bibfnamefont{R.}}, \bibinfo{year}{2006},
  \bibinfo{journal}{Journal of Physics G: Nuclear and Particle Physics}
  \textbf{\bibinfo{volume}{32}}(\bibinfo{number}{6}), \bibinfo{pages}{R151}.

\bibitem[{\citenamefont{Brown}(1990)}]{Brown:1990}
\bibinfo{author}{\bibnamefont{Brown}, \bibfnamefont{B.~A.}},
  \bibinfo{year}{1990}, \bibinfo{journal}{Phys. Rev. Lett.}
  \textbf{\bibinfo{volume}{65}}(\bibinfo{number}{22}), \bibinfo{pages}{2753}.

\bibitem[{\citenamefont{Brown}(1991)}]{Brown:1991}
\bibinfo{author}{\bibnamefont{Brown}, \bibfnamefont{B.~A.}},
  \bibinfo{year}{1991}, \bibinfo{journal}{Phys. Rev. C}
  \textbf{\bibinfo{volume}{43}}, \bibinfo{pages}{R1513}.

\bibitem[{\citenamefont{Brown}(2001)}]{Brown:2001}
\bibinfo{author}{\bibnamefont{Brown}, \bibfnamefont{B.~A.}},
  \bibinfo{year}{2001}, \bibinfo{journal}{Progress in Particle and Nuclear
  Physics} \textbf{\bibinfo{volume}{47}}(\bibinfo{number}{2}),
  \bibinfo{pages}{517}.

\bibitem[{\citenamefont{Brown and Barker}(2003)}]{Brown:2003}
\bibinfo{author}{\bibnamefont{Brown}, \bibfnamefont{B.~A.}}, and
  \bibinfo{author}{\bibfnamefont{F.~C.} \bibnamefont{Barker}},
  \bibinfo{year}{2003}, \bibinfo{journal}{Phys.Rev. C}
  \textbf{\bibinfo{volume}{67}}, \bibinfo{pages}{041304}.

\bibitem[{\citenamefont{Brown} \emph{et~al.}(2002)\citenamefont{Brown, Barker,
  and Millener}}]{Brown:2002}
\bibinfo{author}{\bibnamefont{Brown}, \bibfnamefont{B.~A.}},
  \bibinfo{author}{\bibfnamefont{F.~C.} \bibnamefont{Barker}}, and
  \bibinfo{author}{\bibfnamefont{D.~J.} \bibnamefont{Millener}},
  \bibinfo{year}{2002}, \bibinfo{journal}{Phys. Rev. C}
  \textbf{\bibinfo{volume}{65}}, \bibinfo{pages}{051309}.

\bibitem[{\citenamefont{Büscher} \emph{et~al.}(2008)\citenamefont{Büscher,
  Ponsaers, Raabe, Huyse, Duppen, Aksouh, Smirnov, Fynbo, Hyldegaard, and
  Diget}}]{Buscher:2008}
\bibinfo{author}{\bibnamefont{Büscher}, \bibfnamefont{J.}},
  \bibinfo{author}{\bibfnamefont{J.}~\bibnamefont{Ponsaers}},
  \bibinfo{author}{\bibfnamefont{R.}~\bibnamefont{Raabe}},
  \bibinfo{author}{\bibfnamefont{M.}~\bibnamefont{Huyse}},
  \bibinfo{author}{\bibfnamefont{P.~V.} \bibnamefont{Duppen}},
  \bibinfo{author}{\bibfnamefont{F.}~\bibnamefont{Aksouh}},
  \bibinfo{author}{\bibfnamefont{D.}~\bibnamefont{Smirnov}},
  \bibinfo{author}{\bibfnamefont{H.}~\bibnamefont{Fynbo}},
  \bibinfo{author}{\bibfnamefont{S.}~\bibnamefont{Hyldegaard}}, and
  \bibinfo{author}{\bibfnamefont{C.}~\bibnamefont{Diget}},
  \bibinfo{year}{2008}, \bibinfo{journal}{Nuclear Instruments and Methods in
  Physics Research Section B: Beam Interactions with Materials and Atoms}
  \textbf{\bibinfo{volume}{266}}(\bibinfo{number}{19-20}),
  \bibinfo{pages}{4652}.

\bibitem[{\citenamefont{Buchmann} \emph{et~al.}(2007)\citenamefont{Buchmann,
  D'Auria, Dombsky, Giesen, Jackson, McNeely, Powell, and
  Volya}}]{Buchmann:2007}
\bibinfo{author}{\bibnamefont{Buchmann}, \bibfnamefont{L.}},
  \bibinfo{author}{\bibfnamefont{J.}~\bibnamefont{D'Auria}},
  \bibinfo{author}{\bibfnamefont{M.}~\bibnamefont{Dombsky}},
  \bibinfo{author}{\bibfnamefont{U.}~\bibnamefont{Giesen}},
  \bibinfo{author}{\bibfnamefont{K.~P.} \bibnamefont{Jackson}},
  \bibinfo{author}{\bibfnamefont{P.}~\bibnamefont{McNeely}},
  \bibinfo{author}{\bibfnamefont{J.}~\bibnamefont{Powell}}, and
  \bibinfo{author}{\bibfnamefont{A.}~\bibnamefont{Volya}},
  \bibinfo{year}{2007}, \bibinfo{journal}{Phys. Rev. C}
  \textbf{\bibinfo{volume}{75}}(\bibinfo{number}{1}), \bibinfo{pages}{012804}.

\bibitem[{\citenamefont{Buchmann} \emph{et~al.}(2009)\citenamefont{Buchmann,
  Ruprecht, and Ruiz}}]{Buchmann:2009}
\bibinfo{author}{\bibnamefont{Buchmann}, \bibfnamefont{L.}},
  \bibinfo{author}{\bibfnamefont{G.}~\bibnamefont{Ruprecht}}, and
  \bibinfo{author}{\bibfnamefont{C.}~\bibnamefont{Ruiz}}, \bibinfo{year}{2009},
  \bibinfo{journal}{Phys. Rev. C}
  \textbf{\bibinfo{volume}{80}}(\bibinfo{number}{4}), \bibinfo{pages}{045803}.

\bibitem[{\citenamefont{Buck} \emph{et~al.}(1992)\citenamefont{Buck, Merchant,
  and Perez}}]{Buck:1992}
\bibinfo{author}{\bibnamefont{Buck}, \bibfnamefont{B.}},
  \bibinfo{author}{\bibfnamefont{A.}~\bibnamefont{Merchant}}, and
  \bibinfo{author}{\bibfnamefont{S.}~\bibnamefont{Perez}},
  \bibinfo{year}{1992}, \bibinfo{journal}{Phys.\ Rev.}
  \textbf{\bibinfo{volume}{C~45}}, \bibinfo{pages}{1688}.

\bibitem[{\citenamefont{Burbidge} \emph{et~al.}(1957)\citenamefont{Burbidge,
  Burbidge, Fowler, and Hoyle}}]{Burbidge:1957}
\bibinfo{author}{\bibnamefont{Burbidge}, \bibfnamefont{E.~M.}},
  \bibinfo{author}{\bibfnamefont{G.~R.} \bibnamefont{Burbidge}},
  \bibinfo{author}{\bibfnamefont{W.~A.} \bibnamefont{Fowler}}, and
  \bibinfo{author}{\bibfnamefont{F.}~\bibnamefont{Hoyle}},
  \bibinfo{year}{1957}, \bibinfo{journal}{Rev. Mod. Phys.}
  \textbf{\bibinfo{volume}{29}}(\bibinfo{number}{4}), \bibinfo{pages}{547}.

\bibitem[{\citenamefont{Buta} \emph{et~al.}(2000)\citenamefont{Buta, Martin,
  Timis, Achouri, Angélique, Borcea, Cruceru, Genoux-Lubain, Grévy, Lewitowicz,
  Liénard, Marqués} \emph{et~al.}}]{Buta:2000}
\bibinfo{author}{\bibnamefont{Buta}, \bibfnamefont{A.}},
  \bibinfo{author}{\bibfnamefont{T.}~\bibnamefont{Martin}},
  \bibinfo{author}{\bibfnamefont{C.}~\bibnamefont{Timis}},
  \bibinfo{author}{\bibfnamefont{N.}~\bibnamefont{Achouri}},
  \bibinfo{author}{\bibfnamefont{J.~C.} \bibnamefont{Angélique}},
  \bibinfo{author}{\bibfnamefont{C.}~\bibnamefont{Borcea}},
  \bibinfo{author}{\bibfnamefont{I.}~\bibnamefont{Cruceru}},
  \bibinfo{author}{\bibfnamefont{A.}~\bibnamefont{Genoux-Lubain}},
  \bibinfo{author}{\bibfnamefont{S.}~\bibnamefont{Grévy}},
  \bibinfo{author}{\bibfnamefont{M.}~\bibnamefont{Lewitowicz}},
  \bibinfo{author}{\bibfnamefont{E.}~\bibnamefont{Liénard}},
  \bibinfo{author}{\bibfnamefont{F.~M.} \bibnamefont{Marqués}}, \emph{et~al.},
  \bibinfo{year}{2000}, \bibinfo{journal}{Nuclear Instruments and Methods in
  Physics Research Section A: Accelerators, Spectrometers, Detectors and
  Associated Equipment} \textbf{\bibinfo{volume}{455}}(\bibinfo{number}{2}),
  \bibinfo{pages}{412}.

\bibitem[{\citenamefont{Caamano} \emph{et~al.}(2008)\citenamefont{Caamano,
  Cortina-Gil, Mittig, Savajols, Chartier, Demonchy, Fernandez, Hornillos,
  Gillibert, Jurado, Kiselev, Lemmon} \emph{et~al.}}]{Caamano:2008}
\bibinfo{author}{\bibnamefont{Caamano}, \bibfnamefont{M.}},
  \bibinfo{author}{\bibfnamefont{D.}~\bibnamefont{Cortina-Gil}},
  \bibinfo{author}{\bibfnamefont{W.}~\bibnamefont{Mittig}},
  \bibinfo{author}{\bibfnamefont{H.}~\bibnamefont{Savajols}},
  \bibinfo{author}{\bibfnamefont{M.}~\bibnamefont{Chartier}},
  \bibinfo{author}{\bibfnamefont{C.}~\bibnamefont{Demonchy}},
  \bibinfo{author}{\bibfnamefont{B.}~\bibnamefont{Fernandez}},
  \bibinfo{author}{\bibfnamefont{M.~G.} \bibnamefont{Hornillos}},
  \bibinfo{author}{\bibfnamefont{A.}~\bibnamefont{Gillibert}},
  \bibinfo{author}{\bibfnamefont{B.}~\bibnamefont{Jurado}},
  \bibinfo{author}{\bibfnamefont{O.}~\bibnamefont{Kiselev}},
  \bibinfo{author}{\bibfnamefont{R.}~\bibnamefont{Lemmon}}, \emph{et~al.},
  \bibinfo{year}{2008}, \bibinfo{journal}{Phys. Rev. C}
  \textbf{\bibinfo{volume}{78}}, \bibinfo{pages}{044001}.

\bibitem[{\citenamefont{Cable} \emph{et~al.}(1983)\citenamefont{Cable,
  Honkanen, Parry, Zhou, Zhou, and Cerny}}]{Cable:1983}
\bibinfo{author}{\bibnamefont{Cable}, \bibfnamefont{M.~D.}},
  \bibinfo{author}{\bibfnamefont{J.}~\bibnamefont{Honkanen}},
  \bibinfo{author}{\bibfnamefont{R.~F.} \bibnamefont{Parry}},
  \bibinfo{author}{\bibfnamefont{S.~H.} \bibnamefont{Zhou}},
  \bibinfo{author}{\bibfnamefont{Z.~Y.} \bibnamefont{Zhou}}, and
  \bibinfo{author}{\bibfnamefont{J.}~\bibnamefont{Cerny}},
  \bibinfo{year}{1983}, \bibinfo{journal}{Phys. Rev. Lett.}
  \textbf{\bibinfo{volume}{50}}(\bibinfo{number}{6}), \bibinfo{pages}{404}.

\bibitem[{\citenamefont{del Campo} \emph{et~al.}(2001)\citenamefont{del Campo,
  Galindo-Uribarri, Beene, Gross, Liang, Halbert, Stracener, Shapira, Varner,
  Chavez-Lomeli, and Ortiz}}]{Gomez:2001}
\bibinfo{author}{\bibnamefont{del Campo}, \bibfnamefont{J.~G.}},
  \bibinfo{author}{\bibfnamefont{A.}~\bibnamefont{Galindo-Uribarri}},
  \bibinfo{author}{\bibfnamefont{J.~R.} \bibnamefont{Beene}},
  \bibinfo{author}{\bibfnamefont{C.~J.} \bibnamefont{Gross}},
  \bibinfo{author}{\bibfnamefont{J.~F.} \bibnamefont{Liang}},
  \bibinfo{author}{\bibfnamefont{M.~L.} \bibnamefont{Halbert}},
  \bibinfo{author}{\bibfnamefont{D.~W.} \bibnamefont{Stracener}},
  \bibinfo{author}{\bibfnamefont{D.}~\bibnamefont{Shapira}},
  \bibinfo{author}{\bibfnamefont{R.~L.} \bibnamefont{Varner}},
  \bibinfo{author}{\bibfnamefont{E.}~\bibnamefont{Chavez-Lomeli}}, and
  \bibinfo{author}{\bibfnamefont{M.~E.} \bibnamefont{Ortiz}},
  \bibinfo{year}{2001}, \bibinfo{journal}{Phys. Rev. Lett.}
  \textbf{\bibinfo{volume}{86}}, \bibinfo{pages}{43}.

\bibitem[{\citenamefont{Carpenter} \emph{et~al.}(1997)\citenamefont{Carpenter,
  Janssens, Amro, Blumenthal, Brown, Seweryniak, Woods, Ackermann, Ahmad,
  Davids, Fischer, Hackman} \emph{et~al.}}]{Carpenter:1997}
\bibinfo{author}{\bibnamefont{Carpenter}, \bibfnamefont{M.~P.}},
  \bibinfo{author}{\bibfnamefont{R.~V.~F.} \bibnamefont{Janssens}},
  \bibinfo{author}{\bibfnamefont{H.}~\bibnamefont{Amro}},
  \bibinfo{author}{\bibfnamefont{D.~J.} \bibnamefont{Blumenthal}},
  \bibinfo{author}{\bibfnamefont{L.~T.} \bibnamefont{Brown}},
  \bibinfo{author}{\bibfnamefont{D.}~\bibnamefont{Seweryniak}},
  \bibinfo{author}{\bibfnamefont{P.~J.} \bibnamefont{Woods}},
  \bibinfo{author}{\bibfnamefont{D.}~\bibnamefont{Ackermann}},
  \bibinfo{author}{\bibfnamefont{I.}~\bibnamefont{Ahmad}},
  \bibinfo{author}{\bibfnamefont{C.}~\bibnamefont{Davids}},
  \bibinfo{author}{\bibfnamefont{S.~M.} \bibnamefont{Fischer}},
  \bibinfo{author}{\bibfnamefont{G.}~\bibnamefont{Hackman}}, \emph{et~al.},
  \bibinfo{year}{1997}, \bibinfo{journal}{Phys. Rev. Lett.}
  \textbf{\bibinfo{volume}{78}}(\bibinfo{number}{19}), \bibinfo{pages}{3650}.

\bibitem[{\citenamefont{Casarejos} \emph{et~al.}(2006)\citenamefont{Casarejos,
  Benlliure, Pereira, Armbruster, Bernas, Boudard, Czajkowski, Enqvist,
  Legrain, Leray, Mustapha, Pravikoff} \emph{et~al.}}]{Casarejos:2006}
\bibinfo{author}{\bibnamefont{Casarejos}, \bibfnamefont{E.}},
  \bibinfo{author}{\bibfnamefont{J.}~\bibnamefont{Benlliure}},
  \bibinfo{author}{\bibfnamefont{J.}~\bibnamefont{Pereira}},
  \bibinfo{author}{\bibfnamefont{P.}~\bibnamefont{Armbruster}},
  \bibinfo{author}{\bibfnamefont{M.}~\bibnamefont{Bernas}},
  \bibinfo{author}{\bibfnamefont{A.}~\bibnamefont{Boudard}},
  \bibinfo{author}{\bibfnamefont{S.}~\bibnamefont{Czajkowski}},
  \bibinfo{author}{\bibfnamefont{T.}~\bibnamefont{Enqvist}},
  \bibinfo{author}{\bibfnamefont{R.}~\bibnamefont{Legrain}},
  \bibinfo{author}{\bibfnamefont{S.}~\bibnamefont{Leray}},
  \bibinfo{author}{\bibfnamefont{B.}~\bibnamefont{Mustapha}},
  \bibinfo{author}{\bibfnamefont{M.}~\bibnamefont{Pravikoff}}, \emph{et~al.},
  \bibinfo{year}{2006}, \bibinfo{journal}{Phys. Rev. C}
  \textbf{\bibinfo{volume}{74}}(\bibinfo{number}{4}), \bibinfo{pages}{044612}.

\bibitem[{\citenamefont{Celardo} \emph{et~al.}(2008)\citenamefont{Celardo,
  Izrailev, Zelevinsky, and Berman}}]{Celardo:2008}
\bibinfo{author}{\bibnamefont{Celardo}, \bibfnamefont{G.~L.}},
  \bibinfo{author}{\bibfnamefont{F.~M.} \bibnamefont{Izrailev}},
  \bibinfo{author}{\bibfnamefont{V.~G.} \bibnamefont{Zelevinsky}}, and
  \bibinfo{author}{\bibfnamefont{G.~P.} \bibnamefont{Berman}},
  \bibinfo{year}{2008}, \bibinfo{journal}{Physics Letters B}
  \textbf{\bibinfo{volume}{659}}(\bibinfo{number}{1-2}), \bibinfo{pages}{170}.

\bibitem[{\citenamefont{Cerny} \emph{et~al.}(1970)\citenamefont{Cerny, Esterl,
  R.A.Gough, and R.G.Sextro}}]{Cerny:1970}
\bibinfo{author}{\bibnamefont{Cerny}, \bibfnamefont{J.}},
  \bibinfo{author}{\bibfnamefont{J.}~\bibnamefont{Esterl}},
  \bibinfo{author}{\bibnamefont{R.A.Gough}}, and
  \bibinfo{author}{\bibnamefont{R.G.Sextro}}, \bibinfo{year}{1970},
  \bibinfo{journal}{Phys.\ Lett.} \textbf{\bibinfo{volume}{33~B}},
  \bibinfo{pages}{284}.

\bibitem[{\citenamefont{Cerny} \emph{et~al.}(2009)\citenamefont{Cerny, Moltz,
  Lee, Per\"aj\"arvi, Barquest, Grossman, Jeong, and Jewett}}]{Cerny:2009}
\bibinfo{author}{\bibnamefont{Cerny}, \bibfnamefont{J.}},
  \bibinfo{author}{\bibfnamefont{D.~M.} \bibnamefont{Moltz}},
  \bibinfo{author}{\bibfnamefont{D.~W.} \bibnamefont{Lee}},
  \bibinfo{author}{\bibfnamefont{K.}~\bibnamefont{Per\"aj\"arvi}},
  \bibinfo{author}{\bibfnamefont{B.~R.} \bibnamefont{Barquest}},
  \bibinfo{author}{\bibfnamefont{L.~E.} \bibnamefont{Grossman}},
  \bibinfo{author}{\bibfnamefont{W.}~\bibnamefont{Jeong}}, and
  \bibinfo{author}{\bibfnamefont{C.~C.} \bibnamefont{Jewett}},
  \bibinfo{year}{2009}, \bibinfo{journal}{Phys. Rev. Lett.}
  \textbf{\bibinfo{volume}{103}}(\bibinfo{number}{15}),
  \bibinfo{pages}{152502}.

\bibitem[{\citenamefont{Cetina} \emph{et~al.}(2002)\citenamefont{Cetina,
  Heimberg, Berman, Briscoe, Feldman, Murphy, Crannell, Longhi, Sober,
  Sanabria, and Kezerashvili}}]{Cetina:2002}
\bibinfo{author}{\bibnamefont{Cetina}, \bibfnamefont{C.}},
  \bibinfo{author}{\bibfnamefont{P.}~\bibnamefont{Heimberg}},
  \bibinfo{author}{\bibfnamefont{B.~L.} \bibnamefont{Berman}},
  \bibinfo{author}{\bibfnamefont{W.~J.} \bibnamefont{Briscoe}},
  \bibinfo{author}{\bibfnamefont{G.}~\bibnamefont{Feldman}},
  \bibinfo{author}{\bibfnamefont{L.~Y.} \bibnamefont{Murphy}},
  \bibinfo{author}{\bibfnamefont{H.}~\bibnamefont{Crannell}},
  \bibinfo{author}{\bibfnamefont{A.}~\bibnamefont{Longhi}},
  \bibinfo{author}{\bibfnamefont{D.~I.} \bibnamefont{Sober}},
  \bibinfo{author}{\bibfnamefont{J.~C.} \bibnamefont{Sanabria}}, and
  \bibinfo{author}{\bibfnamefont{G.~Y.} \bibnamefont{Kezerashvili}},
  \bibinfo{year}{2002}, \bibinfo{journal}{Phys. Rev. C}
  \textbf{\bibinfo{volume}{65}}(\bibinfo{number}{4}), \bibinfo{pages}{044622}.

\bibitem[{\citenamefont{Chadwick}(1932)}]{Chadwick:1932}
\bibinfo{author}{\bibnamefont{Chadwick}, \bibfnamefont{J.}},
  \bibinfo{year}{1932}, \bibinfo{journal}{Nature}
  \textbf{\bibinfo{volume}{129}}, \bibinfo{pages}{312}.

\bibitem[{\citenamefont{Charity} \emph{et~al.}(2010)\citenamefont{Charity,
  Elson, Manfredi, Shane, Sobotka, Chajecki, Coupland, Iwasaki, Kilburn, Lee,
  Lynch, Sanetullaev} \emph{et~al.}}]{Charity:2010}
\bibinfo{author}{\bibnamefont{Charity}, \bibfnamefont{R.~J.}},
  \bibinfo{author}{\bibfnamefont{J.~M.} \bibnamefont{Elson}},
  \bibinfo{author}{\bibfnamefont{J.}~\bibnamefont{Manfredi}},
  \bibinfo{author}{\bibfnamefont{R.}~\bibnamefont{Shane}},
  \bibinfo{author}{\bibfnamefont{L.~G.} \bibnamefont{Sobotka}},
  \bibinfo{author}{\bibfnamefont{Z.}~\bibnamefont{Chajecki}},
  \bibinfo{author}{\bibfnamefont{D.}~\bibnamefont{Coupland}},
  \bibinfo{author}{\bibfnamefont{H.}~\bibnamefont{Iwasaki}},
  \bibinfo{author}{\bibfnamefont{M.}~\bibnamefont{Kilburn}},
  \bibinfo{author}{\bibfnamefont{J.}~\bibnamefont{Lee}},
  \bibinfo{author}{\bibfnamefont{W.~G.} \bibnamefont{Lynch}},
  \bibinfo{author}{\bibfnamefont{A.}~\bibnamefont{Sanetullaev}}, \emph{et~al.},
  \bibinfo{year}{2010}, \bibinfo{journal}{Phys. Rev. C}
  \textbf{\bibinfo{volume}{82}}, \bibinfo{pages}{041304(R)}.

\bibitem[{\citenamefont{Charpak} \emph{et~al.}(1988)\citenamefont{Charpak,
  Dominik, Fabre, Gaudaen, Sauli, and Suzuki}}]{Charpak:1988}
\bibinfo{author}{\bibnamefont{Charpak}, \bibfnamefont{G.}},
  \bibinfo{author}{\bibfnamefont{W.}~\bibnamefont{Dominik}},
  \bibinfo{author}{\bibfnamefont{J.}~\bibnamefont{Fabre}},
  \bibinfo{author}{\bibfnamefont{J.}~\bibnamefont{Gaudaen}},
  \bibinfo{author}{\bibfnamefont{F.}~\bibnamefont{Sauli}}, and
  \bibinfo{author}{\bibfnamefont{M.}~\bibnamefont{Suzuki}},
  \bibinfo{year}{1988}, \bibinfo{journal}{Nuclear Instruments and Methods in
  Physics Research Section A: Accelerators, Spectrometers, Detectors and
  Associated Equipment} \textbf{\bibinfo{volume}{269}}(\bibinfo{number}{1}),
  \bibinfo{pages}{142}.

\bibitem[{\citenamefont{Cheal and Flanagan}(2010)}]{Cheal:2010}
\bibinfo{author}{\bibnamefont{Cheal}, \bibfnamefont{B.}}, and
  \bibinfo{author}{\bibfnamefont{K.~T.} \bibnamefont{Flanagan}},
  \bibinfo{year}{2010}, \bibinfo{journal}{Journal of Physics G: Nuclear and
  Particle Physics} \textbf{\bibinfo{volume}{37}}(\bibinfo{number}{11}),
  \bibinfo{pages}{113101}.

\bibitem[{\citenamefont{{Cheikh Mhamed}}
  \emph{et~al.}(2008)\citenamefont{{Cheikh Mhamed}, Essabaa, Lau, Lebois,
  Roussičre, Ducourtieux, Franchoo, Mueller, Ibrahim, LeDu, Lesrel, Mueller}
  \emph{et~al.}}]{CheikhMhamed:2008}
\bibinfo{author}{\bibnamefont{{Cheikh Mhamed}}, \bibfnamefont{M.}},
  \bibinfo{author}{\bibfnamefont{S.}~\bibnamefont{Essabaa}},
  \bibinfo{author}{\bibfnamefont{C.}~\bibnamefont{Lau}},
  \bibinfo{author}{\bibfnamefont{M.}~\bibnamefont{Lebois}},
  \bibinfo{author}{\bibfnamefont{B.}~\bibnamefont{Roussičre}},
  \bibinfo{author}{\bibfnamefont{M.}~\bibnamefont{Ducourtieux}},
  \bibinfo{author}{\bibfnamefont{S.}~\bibnamefont{Franchoo}},
  \bibinfo{author}{\bibfnamefont{D.~G.} \bibnamefont{Mueller}},
  \bibinfo{author}{\bibfnamefont{F.}~\bibnamefont{Ibrahim}},
  \bibinfo{author}{\bibfnamefont{J.}~\bibnamefont{LeDu}},
  \bibinfo{author}{\bibfnamefont{J.}~\bibnamefont{Lesrel}},
  \bibinfo{author}{\bibfnamefont{A.}~\bibnamefont{Mueller}}, \emph{et~al.},
  \bibinfo{year}{2008}, \bibinfo{journal}{Nuclear Instruments and Methods in
  Physics Research Section B: Beam Interactions with Materials and Atoms}
  \textbf{\bibinfo{volume}{266}}(\bibinfo{number}{19-20}),
  \bibinfo{pages}{4092}.

\bibitem[{\citenamefont{Chow} \emph{et~al.}(2002)\citenamefont{Chow, King,
  Bateman, Boyd, Buchmann, D'Auria, Davinson, Dombsky, Gete, Giesen, Iliadis,
  Jackson} \emph{et~al.}}]{Chow:2002}
\bibinfo{author}{\bibnamefont{Chow}, \bibfnamefont{J.~C.}},
  \bibinfo{author}{\bibfnamefont{J.~D.} \bibnamefont{King}},
  \bibinfo{author}{\bibfnamefont{N.~P.~T.} \bibnamefont{Bateman}},
  \bibinfo{author}{\bibfnamefont{R.~N.} \bibnamefont{Boyd}},
  \bibinfo{author}{\bibfnamefont{L.}~\bibnamefont{Buchmann}},
  \bibinfo{author}{\bibfnamefont{J.~M.} \bibnamefont{D'Auria}},
  \bibinfo{author}{\bibfnamefont{T.}~\bibnamefont{Davinson}},
  \bibinfo{author}{\bibfnamefont{M.}~\bibnamefont{Dombsky}},
  \bibinfo{author}{\bibfnamefont{E.}~\bibnamefont{Gete}},
  \bibinfo{author}{\bibfnamefont{U.}~\bibnamefont{Giesen}},
  \bibinfo{author}{\bibfnamefont{C.}~\bibnamefont{Iliadis}},
  \bibinfo{author}{\bibfnamefont{K.~P.} \bibnamefont{Jackson}}, \emph{et~al.},
  \bibinfo{year}{2002}, \bibinfo{journal}{Phys. Rev. C}
  \textbf{\bibinfo{volume}{66}}(\bibinfo{number}{6}), \bibinfo{pages}{064316}.

\bibitem[{\citenamefont{Chromik} \emph{et~al.}(2002)\citenamefont{Chromik,
  Thirolf, Thoennessen, Brown, Davinson, Gassmann, Heckman, Prisciandaro,
  Reiter, Tryggestad, and Woods}}]{Chromik:2002}
\bibinfo{author}{\bibnamefont{Chromik}, \bibfnamefont{M.~J.}},
  \bibinfo{author}{\bibfnamefont{P.~G.} \bibnamefont{Thirolf}},
  \bibinfo{author}{\bibfnamefont{M.}~\bibnamefont{Thoennessen}},
  \bibinfo{author}{\bibfnamefont{B.~A.} \bibnamefont{Brown}},
  \bibinfo{author}{\bibfnamefont{T.}~\bibnamefont{Davinson}},
  \bibinfo{author}{\bibfnamefont{D.}~\bibnamefont{Gassmann}},
  \bibinfo{author}{\bibfnamefont{P.}~\bibnamefont{Heckman}},
  \bibinfo{author}{\bibfnamefont{J.}~\bibnamefont{Prisciandaro}},
  \bibinfo{author}{\bibfnamefont{P.}~\bibnamefont{Reiter}},
  \bibinfo{author}{\bibfnamefont{E.}~\bibnamefont{Tryggestad}}, and
  \bibinfo{author}{\bibfnamefont{P.~J.} \bibnamefont{Woods}},
  \bibinfo{year}{2002}, \bibinfo{journal}{Phys. Rev. C}
  \textbf{\bibinfo{volume}{66}}(\bibinfo{number}{2}), \bibinfo{pages}{024313}.

\bibitem[{\citenamefont{Cinausero} \emph{et~al.}(2009)\citenamefont{Cinausero,
  Andrighetto, Biasetto, Calabretta, Esposito, Fagottia, Gramegna, Manzolaro,
  Mastinu, Lombardi, Pisent, and Prete}}]{Cinausero:2009}
\bibinfo{author}{\bibnamefont{Cinausero}, \bibfnamefont{M.}},
  \bibinfo{author}{\bibfnamefont{A.}~\bibnamefont{Andrighetto}},
  \bibinfo{author}{\bibfnamefont{L.}~\bibnamefont{Biasetto}},
  \bibinfo{author}{\bibfnamefont{L.}~\bibnamefont{Calabretta}},
  \bibinfo{author}{\bibfnamefont{J.}~\bibnamefont{Esposito}},
  \bibinfo{author}{\bibfnamefont{E.}~\bibnamefont{Fagottia}},
  \bibinfo{author}{\bibfnamefont{F.}~\bibnamefont{Gramegna}},
  \bibinfo{author}{\bibfnamefont{M.}~\bibnamefont{Manzolaro}},
  \bibinfo{author}{\bibfnamefont{P.}~\bibnamefont{Mastinu}},
  \bibinfo{author}{\bibfnamefont{A.}~\bibnamefont{Lombardi}},
  \bibinfo{author}{\bibfnamefont{A.}~\bibnamefont{Pisent}}, and
  \bibinfo{author}{\bibfnamefont{G.}~\bibnamefont{Prete}},
  \bibinfo{year}{2009}, \bibinfo{journal}{Acta Physica Polonica B}
  \textbf{\bibinfo{volume}{40}}, \bibinfo{pages}{821}.

\bibitem[{\citenamefont{Clifford} \emph{et~al.}(1989)\citenamefont{Clifford,
  Hagberg, Hardy, Schmeing, Azuma, Evans, Koslowsky, Schrewe, Sharma, and
  Towner}}]{Clifford:1989}
\bibinfo{author}{\bibnamefont{Clifford}, \bibfnamefont{E.~T.~H.}},
  \bibinfo{author}{\bibfnamefont{E.}~\bibnamefont{Hagberg}},
  \bibinfo{author}{\bibfnamefont{J.~C.} \bibnamefont{Hardy}},
  \bibinfo{author}{\bibfnamefont{H.}~\bibnamefont{Schmeing}},
  \bibinfo{author}{\bibfnamefont{R.~E.} \bibnamefont{Azuma}},
  \bibinfo{author}{\bibfnamefont{H.~C.} \bibnamefont{Evans}},
  \bibinfo{author}{\bibfnamefont{V.~T.} \bibnamefont{Koslowsky}},
  \bibinfo{author}{\bibfnamefont{U.~J.} \bibnamefont{Schrewe}},
  \bibinfo{author}{\bibfnamefont{K.~S.} \bibnamefont{Sharma}}, and
  \bibinfo{author}{\bibfnamefont{I.~S.} \bibnamefont{Towner}},
  \bibinfo{year}{1989}, \bibinfo{journal}{Nuclear Physics A}
  \textbf{\bibinfo{volume}{493}}(\bibinfo{number}{2}), \bibinfo{pages}{293}.

\bibitem[{\citenamefont{Cocks} \emph{et~al.}(2000)\citenamefont{Cocks, Butler,
  Cann, Greenlees, Jones, Smith, Jones, Julin, Juutinen, M\"uller, Piiparinen,
  Savelius} \emph{et~al.}}]{Cocks:2000}
\bibinfo{author}{\bibnamefont{Cocks}, \bibfnamefont{J.~F.~C.}},
  \bibinfo{author}{\bibfnamefont{P.~A.} \bibnamefont{Butler}},
  \bibinfo{author}{\bibfnamefont{K.~J.} \bibnamefont{Cann}},
  \bibinfo{author}{\bibfnamefont{P.~T.} \bibnamefont{Greenlees}},
  \bibinfo{author}{\bibfnamefont{G.~D.} \bibnamefont{Jones}},
  \bibinfo{author}{\bibfnamefont{J.~F.} \bibnamefont{Smith}},
  \bibinfo{author}{\bibfnamefont{P.~M.} \bibnamefont{Jones}},
  \bibinfo{author}{\bibfnamefont{R.}~\bibnamefont{Julin}},
  \bibinfo{author}{\bibfnamefont{S.}~\bibnamefont{Juutinen}},
  \bibinfo{author}{\bibfnamefont{D.}~\bibnamefont{M\"uller}},
  \bibinfo{author}{\bibfnamefont{M.}~\bibnamefont{Piiparinen}},
  \bibinfo{author}{\bibfnamefont{A.}~\bibnamefont{Savelius}}, \emph{et~al.},
  \bibinfo{year}{2000}, \bibinfo{journal}{Journal of Physics G: Nuclear and
  Particle Physics} \textbf{\bibinfo{volume}{26}}(\bibinfo{number}{1}),
  \bibinfo{pages}{23}.

\bibitem[{\citenamefont{Cole}(1996)}]{Cole:1996}
\bibinfo{author}{\bibnamefont{Cole}, \bibfnamefont{B.~J.}},
  \bibinfo{year}{1996}, \bibinfo{journal}{Phys. Rev. C}
  \textbf{\bibinfo{volume}{54}}, \bibinfo{pages}{1240}.

\bibitem[{\citenamefont{Corradi} \emph{et~al.}(2009)\citenamefont{Corradi,
  Pollarolo, and Szilner}}]{Corradi:2009}
\bibinfo{author}{\bibnamefont{Corradi}, \bibfnamefont{L.}},
  \bibinfo{author}{\bibfnamefont{G.}~\bibnamefont{Pollarolo}}, and
  \bibinfo{author}{\bibfnamefont{S.}~\bibnamefont{Szilner}},
  \bibinfo{year}{2009}, \bibinfo{journal}{Journal of Physics G: Nuclear and
  Particle Physics} \textbf{\bibinfo{volume}{36}}(\bibinfo{number}{11}),
  \bibinfo{pages}{113101}.

\bibitem[{\citenamefont{Crawford} \emph{et~al.}(2010)\citenamefont{Crawford,
  Janssens, Mantica, Berryman, Broda, Carpenter, Cieplicka, Fornal, Grinyer,
  Hoteling, Kay, Lauritsen} \emph{et~al.}}]{Crawford:2010}
\bibinfo{author}{\bibnamefont{Crawford}, \bibfnamefont{H.~L.}},
  \bibinfo{author}{\bibfnamefont{R.~V.~F.} \bibnamefont{Janssens}},
  \bibinfo{author}{\bibfnamefont{P.~F.} \bibnamefont{Mantica}},
  \bibinfo{author}{\bibfnamefont{J.~S.} \bibnamefont{Berryman}},
  \bibinfo{author}{\bibfnamefont{R.}~\bibnamefont{Broda}},
  \bibinfo{author}{\bibfnamefont{M.~P.} \bibnamefont{Carpenter}},
  \bibinfo{author}{\bibfnamefont{N.}~\bibnamefont{Cieplicka}},
  \bibinfo{author}{\bibfnamefont{B.}~\bibnamefont{Fornal}},
  \bibinfo{author}{\bibfnamefont{G.~F.} \bibnamefont{Grinyer}},
  \bibinfo{author}{\bibfnamefont{N.}~\bibnamefont{Hoteling}},
  \bibinfo{author}{\bibfnamefont{B.~P.} \bibnamefont{Kay}},
  \bibinfo{author}{\bibfnamefont{T.}~\bibnamefont{Lauritsen}}, \emph{et~al.},
  \bibinfo{year}{2010}, \bibinfo{journal}{Phys. Rev. C}
  \textbf{\bibinfo{volume}{82}}(\bibinfo{number}{1}), \bibinfo{pages}{014311}.

\bibitem[{\citenamefont{Crespi} \emph{et~al.}(2009)\citenamefont{Crespi,
  Camera, Bracco, Million, Wieland, Vandone, Recchia, Gadea, Kröll, Mengoni,
  Farnea, Ur} \emph{et~al.}}]{Crespi:2009}
\bibinfo{author}{\bibnamefont{Crespi}, \bibfnamefont{F.}},
  \bibinfo{author}{\bibfnamefont{F.}~\bibnamefont{Camera}},
  \bibinfo{author}{\bibfnamefont{A.}~\bibnamefont{Bracco}},
  \bibinfo{author}{\bibfnamefont{B.}~\bibnamefont{Million}},
  \bibinfo{author}{\bibfnamefont{O.}~\bibnamefont{Wieland}},
  \bibinfo{author}{\bibfnamefont{V.}~\bibnamefont{Vandone}},
  \bibinfo{author}{\bibfnamefont{F.}~\bibnamefont{Recchia}},
  \bibinfo{author}{\bibfnamefont{A.}~\bibnamefont{Gadea}},
  \bibinfo{author}{\bibfnamefont{T.}~\bibnamefont{Kröll}},
  \bibinfo{author}{\bibfnamefont{D.}~\bibnamefont{Mengoni}},
  \bibinfo{author}{\bibfnamefont{E.}~\bibnamefont{Farnea}},
  \bibinfo{author}{\bibfnamefont{C.}~\bibnamefont{Ur}}, \emph{et~al.},
  \bibinfo{year}{2009}, \bibinfo{journal}{Nuclear Instruments and Methods in
  Physics Research Section A: Accelerators, Spectrometers, Detectors and
  Associated Equipment} \textbf{\bibinfo{volume}{604}}(\bibinfo{number}{3}),
  \bibinfo{pages}{604}.

\bibitem[{\citenamefont{Cromaz} \emph{et~al.}(2008)\citenamefont{Cromaz, Riot,
  Fallon, Gros, Holmes, Lee, Macchiavelli, Vu, Yaver, and
  Zimmermann}}]{Cromaz:2008}
\bibinfo{author}{\bibnamefont{Cromaz}, \bibfnamefont{M.}},
  \bibinfo{author}{\bibfnamefont{V.}~\bibnamefont{Riot}},
  \bibinfo{author}{\bibfnamefont{P.}~\bibnamefont{Fallon}},
  \bibinfo{author}{\bibfnamefont{S.}~\bibnamefont{Gros}},
  \bibinfo{author}{\bibfnamefont{B.}~\bibnamefont{Holmes}},
  \bibinfo{author}{\bibfnamefont{I.}~\bibnamefont{Lee}},
  \bibinfo{author}{\bibfnamefont{A.}~\bibnamefont{Macchiavelli}},
  \bibinfo{author}{\bibfnamefont{C.}~\bibnamefont{Vu}},
  \bibinfo{author}{\bibfnamefont{H.}~\bibnamefont{Yaver}}, and
  \bibinfo{author}{\bibfnamefont{S.}~\bibnamefont{Zimmermann}},
  \bibinfo{year}{2008}, \bibinfo{journal}{Nuclear Instruments and Methods in
  Physics Research Section A: Accelerators, Spectrometers, Detectors and
  Associated Equipment} \textbf{\bibinfo{volume}{597}}(\bibinfo{number}{2-3}),
  \bibinfo{pages}{233}.

\bibitem[{\citenamefont{Cs\'{o}t\'{o}}(1994)}]{Csoto:1994a}
\bibinfo{author}{\bibnamefont{Cs\'{o}t\'{o}}, \bibfnamefont{A.}},
  \bibinfo{year}{1994}, \bibinfo{journal}{Phys. Rev. C}
  \textbf{\bibinfo{volume}{49}}, \bibinfo{pages}{3035}.

\bibitem[{\citenamefont{Curie and Joliot}(1934)}]{Curie:1934}
\bibinfo{author}{\bibnamefont{Curie}, \bibfnamefont{I.}}, and
  \bibinfo{author}{\bibfnamefont{F.}~\bibnamefont{Joliot}},
  \bibinfo{year}{1934}, \bibinfo{journal}{Comptes Rendus}
  \textbf{\bibinfo{volume}{198}}, \bibinfo{pages}{254}.

\bibitem[{\citenamefont{Curie and Curie}(1898)}]{Curie:1898}
\bibinfo{author}{\bibnamefont{Curie}, \bibfnamefont{P.}}, and
  \bibinfo{author}{\bibfnamefont{S.}~\bibnamefont{Curie}},
  \bibinfo{year}{1898}, \bibinfo{journal}{Comptes Rendus}
  \textbf{\bibinfo{volume}{127}}, \bibinfo{pages}{175}.

\bibitem[{\citenamefont{Curie} \emph{et~al.}(1898)\citenamefont{Curie, Curie,
  and B\'{e}mont}}]{Curie:1898b}
\bibinfo{author}{\bibnamefont{Curie}, \bibfnamefont{P.}},
  \bibinfo{author}{\bibfnamefont{S.}~\bibnamefont{Curie}}, and
  \bibinfo{author}{\bibfnamefont{G.}~\bibnamefont{B\'{e}mont}},
  \bibinfo{year}{1898}, \bibinfo{journal}{Comptes Rendus}
  \textbf{\bibinfo{volume}{127}}, \bibinfo{pages}{1215}.

\bibitem[{\citenamefont{Cuttone} \emph{et~al.}(2008)\citenamefont{Cuttone,
  Calabretta, Celona, Chines, Cosentino, Finocchiaro, Pappalardo, Re,
  Rifuggiato, and Rovelli}}]{Cuttone:2008}
\bibinfo{author}{\bibnamefont{Cuttone}, \bibfnamefont{G.}},
  \bibinfo{author}{\bibfnamefont{L.}~\bibnamefont{Calabretta}},
  \bibinfo{author}{\bibfnamefont{L.}~\bibnamefont{Celona}},
  \bibinfo{author}{\bibfnamefont{F.}~\bibnamefont{Chines}},
  \bibinfo{author}{\bibfnamefont{L.}~\bibnamefont{Cosentino}},
  \bibinfo{author}{\bibfnamefont{P.}~\bibnamefont{Finocchiaro}},
  \bibinfo{author}{\bibfnamefont{A.}~\bibnamefont{Pappalardo}},
  \bibinfo{author}{\bibfnamefont{M.}~\bibnamefont{Re}},
  \bibinfo{author}{\bibfnamefont{D.}~\bibnamefont{Rifuggiato}}, and
  \bibinfo{author}{\bibfnamefont{A.}~\bibnamefont{Rovelli}},
  \bibinfo{year}{2008}, \bibinfo{journal}{Nuclear Instruments and Methods in
  Physics Research Section B: Beam Interactions with Materials and Atoms}
  \textbf{\bibinfo{volume}{266}}(\bibinfo{number}{19-20}),
  \bibinfo{pages}{4108}.

\bibitem[{\citenamefont{Danilin and Zhukov}(1993)}]{Danilin:1993}
\bibinfo{author}{\bibnamefont{Danilin}, \bibfnamefont{B.~V.}}, and
  \bibinfo{author}{\bibfnamefont{M.~V.} \bibnamefont{Zhukov}},
  \bibinfo{year}{1993}, \bibinfo{journal}{Phys. At. Nucl.}
  \textbf{\bibinfo{volume}{56}}, \bibinfo{pages}{460}.

\bibitem[{\citenamefont{Darby} \emph{et~al.}(2010)\citenamefont{Darby,
  Grzywacz, Batchelder, Bingham, Cartegni, Gross, Hjorth-Jensen, Joss, Liddick,
  Nazarewicz, Padgett, Page} \emph{et~al.}}]{Darby:2010}
\bibinfo{author}{\bibnamefont{Darby}, \bibfnamefont{I.~G.}},
  \bibinfo{author}{\bibfnamefont{R.~K.} \bibnamefont{Grzywacz}},
  \bibinfo{author}{\bibfnamefont{J.~C.} \bibnamefont{Batchelder}},
  \bibinfo{author}{\bibfnamefont{C.~R.} \bibnamefont{Bingham}},
  \bibinfo{author}{\bibfnamefont{L.}~\bibnamefont{Cartegni}},
  \bibinfo{author}{\bibfnamefont{C.~J.} \bibnamefont{Gross}},
  \bibinfo{author}{\bibfnamefont{M.}~\bibnamefont{Hjorth-Jensen}},
  \bibinfo{author}{\bibfnamefont{D.~T.} \bibnamefont{Joss}},
  \bibinfo{author}{\bibfnamefont{S.~N.} \bibnamefont{Liddick}},
  \bibinfo{author}{\bibfnamefont{W.}~\bibnamefont{Nazarewicz}},
  \bibinfo{author}{\bibfnamefont{S.}~\bibnamefont{Padgett}},
  \bibinfo{author}{\bibfnamefont{R.~D.} \bibnamefont{Page}}, \emph{et~al.},
  \bibinfo{year}{2010}, \bibinfo{journal}{Phys. Rev. Lett.}
  \textbf{\bibinfo{volume}{105}}(\bibinfo{number}{16}),
  \bibinfo{pages}{162502}.

\bibitem[{\citenamefont{Datar} \emph{et~al.}(2005)\citenamefont{Datar, Kumar,
  Chakrabarty, Nanal, Mirgule, Mitra, and Oza}}]{Datar:2005}
\bibinfo{author}{\bibnamefont{Datar}, \bibfnamefont{V.~M.}},
  \bibinfo{author}{\bibfnamefont{S.}~\bibnamefont{Kumar}},
  \bibinfo{author}{\bibfnamefont{D.~R.} \bibnamefont{Chakrabarty}},
  \bibinfo{author}{\bibfnamefont{V.}~\bibnamefont{Nanal}},
  \bibinfo{author}{\bibfnamefont{E.~T.} \bibnamefont{Mirgule}},
  \bibinfo{author}{\bibfnamefont{A.}~\bibnamefont{Mitra}}, and
  \bibinfo{author}{\bibfnamefont{H.~H.} \bibnamefont{Oza}},
  \bibinfo{year}{2005}, \bibinfo{journal}{Phys. Rev. Lett.}
  \textbf{\bibinfo{volume}{94}}(\bibinfo{number}{12}), \bibinfo{pages}{122502}.

\bibitem[{\citenamefont{Davids} \emph{et~al.}(1992)\citenamefont{Davids, Back,
  Bindra, Henderson, Kutschera, Lauritsen, Nagame, Sugathan, Ramayya, and
  Walters}}]{Davids:1992}
\bibinfo{author}{\bibnamefont{Davids}, \bibfnamefont{C.}},
  \bibinfo{author}{\bibfnamefont{B.~B.} \bibnamefont{Back}},
  \bibinfo{author}{\bibfnamefont{K.}~\bibnamefont{Bindra}},
  \bibinfo{author}{\bibfnamefont{D.~J.} \bibnamefont{Henderson}},
  \bibinfo{author}{\bibfnamefont{W.}~\bibnamefont{Kutschera}},
  \bibinfo{author}{\bibfnamefont{T.}~\bibnamefont{Lauritsen}},
  \bibinfo{author}{\bibfnamefont{Y.}~\bibnamefont{Nagame}},
  \bibinfo{author}{\bibfnamefont{P.}~\bibnamefont{Sugathan}},
  \bibinfo{author}{\bibfnamefont{A.~V.} \bibnamefont{Ramayya}}, and
  \bibinfo{author}{\bibfnamefont{W.~B.} \bibnamefont{Walters}},
  \bibinfo{year}{1992}, \bibinfo{journal}{Nuclear Instruments and Methods in
  Physics Research Section B: Beam Interactions with Materials and Atoms}
  \textbf{\bibinfo{volume}{70}}(\bibinfo{number}{1-4}), \bibinfo{pages}{358}.

\bibitem[{\citenamefont{Davids and Esbensen}(2000)}]{Davids:2000}
\bibinfo{author}{\bibnamefont{Davids}, \bibfnamefont{C.}}, and
  \bibinfo{author}{\bibfnamefont{H.}~\bibnamefont{Esbensen}},
  \bibinfo{year}{2000}, \bibinfo{journal}{Phys.\ Rev.}
  \textbf{\bibinfo{volume}{C~61}}, \bibinfo{pages}{054302}.

\bibitem[{\citenamefont{Davids} \emph{et~al.}(1997)\citenamefont{Davids, Woods,
  Batchelder, Bingham, Blumenthal, Brown, Busse, Conticchio, Davinson, Freeman,
  Henderson, Irvine} \emph{et~al.}}]{Davids:1997}
\bibinfo{author}{\bibnamefont{Davids}, \bibfnamefont{C.}},
  \bibinfo{author}{\bibfnamefont{P.}~\bibnamefont{Woods}},
  \bibinfo{author}{\bibfnamefont{J.}~\bibnamefont{Batchelder}},
  \bibinfo{author}{\bibfnamefont{C.}~\bibnamefont{Bingham}},
  \bibinfo{author}{\bibfnamefont{D.}~\bibnamefont{Blumenthal}},
  \bibinfo{author}{\bibfnamefont{L.}~\bibnamefont{Brown}},
  \bibinfo{author}{\bibfnamefont{B.}~\bibnamefont{Busse}},
  \bibinfo{author}{\bibfnamefont{L.}~\bibnamefont{Conticchio}},
  \bibinfo{author}{\bibfnamefont{T.}~\bibnamefont{Davinson}},
  \bibinfo{author}{\bibfnamefont{S.}~\bibnamefont{Freeman}},
  \bibinfo{author}{\bibfnamefont{D.}~\bibnamefont{Henderson}},
  \bibinfo{author}{\bibfnamefont{R.}~\bibnamefont{Irvine}}, \emph{et~al.},
  \bibinfo{year}{1997}, \bibinfo{journal}{Phys.\ Rev.}
  \textbf{\bibinfo{volume}{C~55}}, \bibinfo{pages}{2255}.

\bibitem[{\citenamefont{Davids} \emph{et~al.}(2001)\citenamefont{Davids, Woods,
  J.C.~Batchelder, Brown, Busse, Carpenter, Conticchio, Davinson, DeBoer,
  Freeman, Hamada, Henderson} \emph{et~al.}}]{Davids:2001b}
\bibinfo{author}{\bibnamefont{Davids}, \bibfnamefont{C.}},
  \bibinfo{author}{\bibfnamefont{P.}~\bibnamefont{Woods}},
  \bibinfo{author}{\bibfnamefont{D.~B.} \bibnamefont{J.C.~Batchelder},
  \bibfnamefont{C.R.~Bingham}},
  \bibinfo{author}{\bibfnamefont{L.}~\bibnamefont{Brown}},
  \bibinfo{author}{\bibfnamefont{B.}~\bibnamefont{Busse}},
  \bibinfo{author}{\bibfnamefont{M.}~\bibnamefont{Carpenter}},
  \bibinfo{author}{\bibfnamefont{L.}~\bibnamefont{Conticchio}},
  \bibinfo{author}{\bibfnamefont{T.}~\bibnamefont{Davinson}},
  \bibinfo{author}{\bibfnamefont{J.}~\bibnamefont{DeBoer}},
  \bibinfo{author}{\bibfnamefont{S.}~\bibnamefont{Freeman}},
  \bibinfo{author}{\bibfnamefont{S.}~\bibnamefont{Hamada}},
  \bibinfo{author}{\bibfnamefont{D.}~\bibnamefont{Henderson}}, \emph{et~al.},
  \bibinfo{year}{2001}, \bibinfo{journal}{Hyperfine\ Inter.}
  \textbf{\bibinfo{volume}{132}}, \bibinfo{pages}{133}.

\bibitem[{\citenamefont{Davids} \emph{et~al.}(2004)\citenamefont{Davids, Woods,
  Mahmud, Davinson, Heinz, Ressler, Schmidt, Seweryniak, Shergur, Sonzogni, and
  Walters}}]{Davids:2004a}
\bibinfo{author}{\bibnamefont{Davids}, \bibfnamefont{C.}},
  \bibinfo{author}{\bibfnamefont{P.}~\bibnamefont{Woods}},
  \bibinfo{author}{\bibfnamefont{H.}~\bibnamefont{Mahmud}},
  \bibinfo{author}{\bibfnamefont{T.}~\bibnamefont{Davinson}},
  \bibinfo{author}{\bibfnamefont{A.}~\bibnamefont{Heinz}},
  \bibinfo{author}{\bibfnamefont{J.}~\bibnamefont{Ressler}},
  \bibinfo{author}{\bibfnamefont{K.}~\bibnamefont{Schmidt}},
  \bibinfo{author}{\bibfnamefont{D.}~\bibnamefont{Seweryniak}},
  \bibinfo{author}{\bibfnamefont{J.}~\bibnamefont{Shergur}},
  \bibinfo{author}{\bibfnamefont{A.}~\bibnamefont{Sonzogni}}, and
  \bibinfo{author}{\bibfnamefont{W.}~\bibnamefont{Walters}},
  \bibinfo{year}{2004}, \bibinfo{journal}{Phys.\ Rev.}
  \textbf{\bibinfo{volume}{C~69}}, \bibinfo{pages}{011302(R)}.

\bibitem[{\citenamefont{Davids} \emph{et~al.}(1996)\citenamefont{Davids, Woods,
  Penttil\"a, Batchelder, Bingham, Blumenthal, Brown, Busse, Conticchio,
  Davinson, Henderson, Irvine} \emph{et~al.}}]{Davids:1996}
\bibinfo{author}{\bibnamefont{Davids}, \bibfnamefont{C.}},
  \bibinfo{author}{\bibfnamefont{P.}~\bibnamefont{Woods}},
  \bibinfo{author}{\bibfnamefont{H.}~\bibnamefont{Penttil\"a}},
  \bibinfo{author}{\bibfnamefont{J.}~\bibnamefont{Batchelder}},
  \bibinfo{author}{\bibfnamefont{C.}~\bibnamefont{Bingham}},
  \bibinfo{author}{\bibfnamefont{D.}~\bibnamefont{Blumenthal}},
  \bibinfo{author}{\bibfnamefont{L.}~\bibnamefont{Brown}},
  \bibinfo{author}{\bibfnamefont{B.}~\bibnamefont{Busse}},
  \bibinfo{author}{\bibfnamefont{L.}~\bibnamefont{Conticchio}},
  \bibinfo{author}{\bibfnamefont{T.}~\bibnamefont{Davinson}},
  \bibinfo{author}{\bibfnamefont{D.}~\bibnamefont{Henderson}},
  \bibinfo{author}{\bibfnamefont{R.}~\bibnamefont{Irvine}}, \emph{et~al.},
  \bibinfo{year}{1996}, \bibinfo{journal}{Phys. Rev. Lett.}
  \textbf{\bibinfo{volume}{76}}(\bibinfo{number}{4}), \bibinfo{pages}{592}.

\bibitem[{\citenamefont{Davids} \emph{et~al.}(1998)\citenamefont{Davids, Woods,
  Seweryniak, Sonzogni, Batchelder, Bingham, Davinson, Henderson, Irvine, Poli,
  Uusitalo, and Walters}}]{Davids:1998}
\bibinfo{author}{\bibnamefont{Davids}, \bibfnamefont{C.}},
  \bibinfo{author}{\bibfnamefont{P.}~\bibnamefont{Woods}},
  \bibinfo{author}{\bibfnamefont{D.}~\bibnamefont{Seweryniak}},
  \bibinfo{author}{\bibfnamefont{A.}~\bibnamefont{Sonzogni}},
  \bibinfo{author}{\bibfnamefont{J.}~\bibnamefont{Batchelder}},
  \bibinfo{author}{\bibfnamefont{C.}~\bibnamefont{Bingham}},
  \bibinfo{author}{\bibfnamefont{T.}~\bibnamefont{Davinson}},
  \bibinfo{author}{\bibfnamefont{D.}~\bibnamefont{Henderson}},
  \bibinfo{author}{\bibfnamefont{R.}~\bibnamefont{Irvine}},
  \bibinfo{author}{\bibfnamefont{G.}~\bibnamefont{Poli}},
  \bibinfo{author}{\bibfnamefont{J.}~\bibnamefont{Uusitalo}}, and
  \bibinfo{author}{\bibfnamefont{W.}~\bibnamefont{Walters}},
  \bibinfo{year}{1998}, \bibinfo{journal}{Phys.\ Rev.\ Lett.}
  \textbf{\bibinfo{volume}{80}}, \bibinfo{pages}{1849}.

\bibitem[{\citenamefont{Davids}(2003)}]{Davids:2003}
\bibinfo{author}{\bibnamefont{Davids}, \bibfnamefont{C.~N.}},
  \bibinfo{year}{2003}, \bibinfo{journal}{Nuclear Instruments and Methods in
  Physics Research Section B: Beam Interactions with Materials and Atoms}
  \textbf{\bibinfo{volume}{204}}, \bibinfo{pages}{124}.

\bibitem[{\citenamefont{Davids and Esbensen}(2004)}]{Davids:2004}
\bibinfo{author}{\bibnamefont{Davids}, \bibfnamefont{C.~N.}}, and
  \bibinfo{author}{\bibfnamefont{H.}~\bibnamefont{Esbensen}},
  \bibinfo{year}{2004}, \bibinfo{journal}{Phys. Rev. C}
  \textbf{\bibinfo{volume}{69}}(\bibinfo{number}{3}), \bibinfo{pages}{034314}.

\bibitem[{\citenamefont{Davinson}(2010)}]{Davinson:2010}
\bibinfo{author}{\bibnamefont{Davinson}, \bibfnamefont{T.}},
  \bibinfo{year}{2010}, \bibinfo{title}{Advanced implantation detector array},
  \bibinfo{howpublished}{http://www.ph.ed.ac.uk/~td/AIDA}.

\bibitem[{\citenamefont{Delion}(2010)}]{Delion:2010}
\bibinfo{author}{\bibnamefont{Delion}, \bibfnamefont{D.}},
  \bibinfo{year}{2010}, \emph{\bibinfo{title}{Theory of Particle and Cluster
  Emission}}, volume \bibinfo{volume}{819} of \emph{\bibinfo{series}{Lecture
  Notes in Physics}} (\bibinfo{publisher}{Springer Berlin/Heidelberg}).

\bibitem[{\citenamefont{Delion}
  \emph{et~al.}(2006{\natexlab{a}})\citenamefont{Delion, Liotta, and
  Wyss}}]{Delion:2006}
\bibinfo{author}{\bibnamefont{Delion}, \bibfnamefont{D.~S.}},
  \bibinfo{author}{\bibfnamefont{R.~J.} \bibnamefont{Liotta}}, and
  \bibinfo{author}{\bibfnamefont{R.}~\bibnamefont{Wyss}},
  \bibinfo{year}{2006}{\natexlab{a}}, \bibinfo{journal}{Phys. Rev. Lett.}
  \textbf{\bibinfo{volume}{96}}(\bibinfo{number}{7}), \bibinfo{pages}{072501}.

\bibitem[{\citenamefont{Delion}
  \emph{et~al.}(2006{\natexlab{b}})\citenamefont{Delion, Liotta, and
  Wyss}}]{Delion:2006a}
\bibinfo{author}{\bibnamefont{Delion}, \bibfnamefont{D.~S.}},
  \bibinfo{author}{\bibfnamefont{R.~J.} \bibnamefont{Liotta}}, and
  \bibinfo{author}{\bibfnamefont{R.}~\bibnamefont{Wyss}},
  \bibinfo{year}{2006}{\natexlab{b}}, \bibinfo{journal}{Phys. Rep.}
  \textbf{\bibinfo{volume}{424}}(\bibinfo{number}{3}), \bibinfo{pages}{113}.

\bibitem[{\citenamefont{Descouvemont}
  \emph{et~al.}(2006)\citenamefont{Descouvemont, Tursunov, and
  Baye}}]{Descouvemont:2006}
\bibinfo{author}{\bibnamefont{Descouvemont}, \bibfnamefont{P.}},
  \bibinfo{author}{\bibfnamefont{E.}~\bibnamefont{Tursunov}}, and
  \bibinfo{author}{\bibfnamefont{D.}~\bibnamefont{Baye}}, \bibinfo{year}{2006},
  \bibinfo{journal}{Nucl. Phys.} \textbf{\bibinfo{volume}{A765}},
  \bibinfo{pages}{370}.

\bibitem[{\citenamefont{Diamond}(1999)}]{Diamond:1999}
\bibinfo{author}{\bibnamefont{Diamond}, \bibfnamefont{W.~T.}},
  \bibinfo{year}{1999}, \bibinfo{journal}{Nuclear Instruments and Methods in
  Physics Research Section A: Accelerators, Spectrometers, Detectors and
  Associated Equipment} \textbf{\bibinfo{volume}{432}}(\bibinfo{number}{2-3}),
  \bibinfo{pages}{471}.

\bibitem[{\citenamefont{Diget} \emph{et~al.}(2009)\citenamefont{Diget, Barker,
  Borge, Boutami, Dendooven, Eronen, Fox, Fulton, Fynbo, Huikari, Hyldegaard,
  Jeppesen} \emph{et~al.}}]{Diget:2009}
\bibinfo{author}{\bibnamefont{Diget}, \bibfnamefont{C.~A.}},
  \bibinfo{author}{\bibfnamefont{F.~C.} \bibnamefont{Barker}},
  \bibinfo{author}{\bibfnamefont{M.~J.~G.} \bibnamefont{Borge}},
  \bibinfo{author}{\bibfnamefont{R.}~\bibnamefont{Boutami}},
  \bibinfo{author}{\bibfnamefont{P.}~\bibnamefont{Dendooven}},
  \bibinfo{author}{\bibfnamefont{T.}~\bibnamefont{Eronen}},
  \bibinfo{author}{\bibfnamefont{S.~P.} \bibnamefont{Fox}},
  \bibinfo{author}{\bibfnamefont{B.~R.} \bibnamefont{Fulton}},
  \bibinfo{author}{\bibfnamefont{H.~O.~U.} \bibnamefont{Fynbo}},
  \bibinfo{author}{\bibfnamefont{J.}~\bibnamefont{Huikari}},
  \bibinfo{author}{\bibfnamefont{S.}~\bibnamefont{Hyldegaard}},
  \bibinfo{author}{\bibfnamefont{H.~B.} \bibnamefont{Jeppesen}}, \emph{et~al.},
  \bibinfo{year}{2009}, \bibinfo{journal}{Phys. Rev. C}
  \textbf{\bibinfo{volume}{80}}(\bibinfo{number}{3}), \bibinfo{pages}{034316}.

\bibitem[{\citenamefont{Dobaczewski}
  \emph{et~al.}(1994)\citenamefont{Dobaczewski, Hamamoto, Nazarewicz, and
  Sheikh}}]{Dobaczewski:1994}
\bibinfo{author}{\bibnamefont{Dobaczewski}, \bibfnamefont{J.}},
  \bibinfo{author}{\bibfnamefont{I.}~\bibnamefont{Hamamoto}},
  \bibinfo{author}{\bibfnamefont{W.}~\bibnamefont{Nazarewicz}}, and
  \bibinfo{author}{\bibfnamefont{J.~A.} \bibnamefont{Sheikh}},
  \bibinfo{year}{1994}, \bibinfo{journal}{Phys. Rev. Lett.}
  \textbf{\bibinfo{volume}{72}}(\bibinfo{number}{7}), \bibinfo{pages}{981}.

\bibitem[{\citenamefont{Dobaczewski}
  \emph{et~al.}(2007)\citenamefont{Dobaczewski, Michel, Nazarewicz,
  P{\l}oszajczak, and Rotureau}}]{Dobaczewski:2007}
\bibinfo{author}{\bibnamefont{Dobaczewski}, \bibfnamefont{J.}},
  \bibinfo{author}{\bibfnamefont{N.}~\bibnamefont{Michel}},
  \bibinfo{author}{\bibfnamefont{W.}~\bibnamefont{Nazarewicz}},
  \bibinfo{author}{\bibfnamefont{M.}~\bibnamefont{P{\l}oszajczak}}, and
  \bibinfo{author}{\bibfnamefont{J.}~\bibnamefont{Rotureau}},
  \bibinfo{year}{2007}, \bibinfo{journal}{Progress in Particle and Nuclear
  Physics} \textbf{\bibinfo{volume}{59}}(\bibinfo{number}{1}),
  \bibinfo{pages}{432}.

\bibitem[{\citenamefont{Dossat} \emph{et~al.}(2007)\citenamefont{Dossat, Adimi,
  Aksouh, Becker, Bey, Blank, Borcea, Borcea, Boston, Caamano, Canchel,
  Chartier} \emph{et~al.}}]{Dossat:2007}
\bibinfo{author}{\bibnamefont{Dossat}, \bibfnamefont{C.}},
  \bibinfo{author}{\bibfnamefont{N.}~\bibnamefont{Adimi}},
  \bibinfo{author}{\bibfnamefont{F.}~\bibnamefont{Aksouh}},
  \bibinfo{author}{\bibfnamefont{F.}~\bibnamefont{Becker}},
  \bibinfo{author}{\bibfnamefont{A.}~\bibnamefont{Bey}},
  \bibinfo{author}{\bibfnamefont{B.}~\bibnamefont{Blank}},
  \bibinfo{author}{\bibfnamefont{C.}~\bibnamefont{Borcea}},
  \bibinfo{author}{\bibfnamefont{R.}~\bibnamefont{Borcea}},
  \bibinfo{author}{\bibfnamefont{A.}~\bibnamefont{Boston}},
  \bibinfo{author}{\bibfnamefont{M.}~\bibnamefont{Caamano}},
  \bibinfo{author}{\bibfnamefont{G.}~\bibnamefont{Canchel}},
  \bibinfo{author}{\bibfnamefont{M.}~\bibnamefont{Chartier}}, \emph{et~al.},
  \bibinfo{year}{2007}, \bibinfo{journal}{Nuclear Physics A}
  \textbf{\bibinfo{volume}{792}}(\bibinfo{number}{1-2}), \bibinfo{pages}{18}.

\bibitem[{\citenamefont{Dossat} \emph{et~al.}(2005)\citenamefont{Dossat, Bey,
  Blank, Canchel, Fleury, Giovinazzo, Matea, de~Oliveira~Santos, Georgiev,
  Grevy, Stefan, Thomas} \emph{et~al.}}]{Dossat:2005}
\bibinfo{author}{\bibnamefont{Dossat}, \bibfnamefont{C.}},
  \bibinfo{author}{\bibfnamefont{A.}~\bibnamefont{Bey}},
  \bibinfo{author}{\bibfnamefont{B.}~\bibnamefont{Blank}},
  \bibinfo{author}{\bibfnamefont{G.}~\bibnamefont{Canchel}},
  \bibinfo{author}{\bibfnamefont{A.}~\bibnamefont{Fleury}},
  \bibinfo{author}{\bibfnamefont{J.}~\bibnamefont{Giovinazzo}},
  \bibinfo{author}{\bibfnamefont{I.}~\bibnamefont{Matea}},
  \bibinfo{author}{\bibfnamefont{F.}~\bibnamefont{de~Oliveira~Santos}},
  \bibinfo{author}{\bibfnamefont{G.}~\bibnamefont{Georgiev}},
  \bibinfo{author}{\bibfnamefont{S.}~\bibnamefont{Grevy}},
  \bibinfo{author}{\bibfnamefont{I.}~\bibnamefont{Stefan}},
  \bibinfo{author}{\bibfnamefont{J.~C.} \bibnamefont{Thomas}}, \emph{et~al.},
  \bibinfo{year}{2005}, \bibinfo{journal}{Phys. Rev. C}
  \textbf{\bibinfo{volume}{72}}, \bibinfo{pages}{054315}.

\bibitem[{\citenamefont{Dufour} \emph{et~al.}(1988)\citenamefont{Dufour, Moral,
  Hubert, Jean, Pravikoff, Fleury, Mueller, Schmidt, S\"{u}mmerer, Hanelt,
  Frehaut, Beau} \emph{et~al.}}]{Dufour:1988}
\bibinfo{author}{\bibnamefont{Dufour}, \bibfnamefont{J.~P.}},
  \bibinfo{author}{\bibfnamefont{R.~D.} \bibnamefont{Moral}},
  \bibinfo{author}{\bibfnamefont{F.}~\bibnamefont{Hubert}},
  \bibinfo{author}{\bibfnamefont{D.}~\bibnamefont{Jean}},
  \bibinfo{author}{\bibfnamefont{M.~S.} \bibnamefont{Pravikoff}},
  \bibinfo{author}{\bibfnamefont{A.}~\bibnamefont{Fleury}},
  \bibinfo{author}{\bibfnamefont{A.~C.} \bibnamefont{Mueller}},
  \bibinfo{author}{\bibfnamefont{K.~H.} \bibnamefont{Schmidt}},
  \bibinfo{author}{\bibfnamefont{K.}~\bibnamefont{S\"{u}mmerer}},
  \bibinfo{author}{\bibfnamefont{E.}~\bibnamefont{Hanelt}},
  \bibinfo{author}{\bibfnamefont{J.}~\bibnamefont{Frehaut}},
  \bibinfo{author}{\bibfnamefont{M.}~\bibnamefont{Beau}}, \emph{et~al.},
  \bibinfo{year}{1988}, \bibinfo{journal}{Physics Letters B}
  \textbf{\bibinfo{volume}{206}}(\bibinfo{number}{2}), \bibinfo{pages}{195}.

\bibitem[{\citenamefont{Eichler} \emph{et~al.}(2007)\citenamefont{Eichler,
  Aksenov, Belozerov, Bozhikov, Chepigin, Dmitriev, Dressler, Gäggeler,
  Gorshkov, Haenssler, Itkis, Laube} \emph{et~al.}}]{Eichler:2007}
\bibinfo{author}{\bibnamefont{Eichler}, \bibfnamefont{R.}},
  \bibinfo{author}{\bibfnamefont{N.~V.} \bibnamefont{Aksenov}},
  \bibinfo{author}{\bibfnamefont{A.~V.} \bibnamefont{Belozerov}},
  \bibinfo{author}{\bibfnamefont{G.~A.} \bibnamefont{Bozhikov}},
  \bibinfo{author}{\bibfnamefont{V.~I.} \bibnamefont{Chepigin}},
  \bibinfo{author}{\bibfnamefont{S.~N.} \bibnamefont{Dmitriev}},
  \bibinfo{author}{\bibfnamefont{R.}~\bibnamefont{Dressler}},
  \bibinfo{author}{\bibfnamefont{H.~W.} \bibnamefont{Gäggeler}},
  \bibinfo{author}{\bibfnamefont{V.~A.} \bibnamefont{Gorshkov}},
  \bibinfo{author}{\bibfnamefont{F.}~\bibnamefont{Haenssler}},
  \bibinfo{author}{\bibfnamefont{M.~G.} \bibnamefont{Itkis}},
  \bibinfo{author}{\bibfnamefont{A.}~\bibnamefont{Laube}}, \emph{et~al.},
  \bibinfo{year}{2007}, \bibinfo{journal}{Nature}
  \textbf{\bibinfo{volume}{447}}(\bibinfo{number}{3}), \bibinfo{pages}{72}.

\bibitem[{\citenamefont{Engel} \emph{et~al.}(2011)\citenamefont{Engel,
  Duggireddi, Vangapally, Elson, Sobotka, and Charity}}]{Engel:2011}
\bibinfo{author}{\bibnamefont{Engel}, \bibfnamefont{G.}},
  \bibinfo{author}{\bibfnamefont{N.}~\bibnamefont{Duggireddi}},
  \bibinfo{author}{\bibfnamefont{V.}~\bibnamefont{Vangapally}},
  \bibinfo{author}{\bibfnamefont{J.}~\bibnamefont{Elson}},
  \bibinfo{author}{\bibfnamefont{L.}~\bibnamefont{Sobotka}}, and
  \bibinfo{author}{\bibfnamefont{R.}~\bibnamefont{Charity}},
  \bibinfo{year}{2011}, \bibinfo{journal}{Nuclear Instruments and Methods in
  Physics Research Section A: Accelerators, Spectrometers, Detectors and
  Associated Equipment} \textbf{\bibinfo{volume}{In Press, Corrected Proof}}.

\bibitem[{\citenamefont{Engelmann} \emph{et~al.}(1995)\citenamefont{Engelmann,
  Ameil, Armbruster, Bernas, Czajkowski, Dessagne, Donzaud, Geissel, Heinz,
  Janas, Kozhuharov, Miehé} \emph{et~al.}}]{Engelmann:1995}
\bibinfo{author}{\bibnamefont{Engelmann}, \bibfnamefont{C.}},
  \bibinfo{author}{\bibfnamefont{F.}~\bibnamefont{Ameil}},
  \bibinfo{author}{\bibfnamefont{P.}~\bibnamefont{Armbruster}},
  \bibinfo{author}{\bibfnamefont{M.}~\bibnamefont{Bernas}},
  \bibinfo{author}{\bibfnamefont{S.}~\bibnamefont{Czajkowski}},
  \bibinfo{author}{\bibfnamefont{P.}~\bibnamefont{Dessagne}},
  \bibinfo{author}{\bibfnamefont{C.}~\bibnamefont{Donzaud}},
  \bibinfo{author}{\bibfnamefont{H.}~\bibnamefont{Geissel}},
  \bibinfo{author}{\bibfnamefont{A.}~\bibnamefont{Heinz}},
  \bibinfo{author}{\bibfnamefont{Z.}~\bibnamefont{Janas}},
  \bibinfo{author}{\bibfnamefont{C.}~\bibnamefont{Kozhuharov}},
  \bibinfo{author}{\bibfnamefont{C.}~\bibnamefont{Miehé}}, \emph{et~al.},
  \bibinfo{year}{1995}, \bibinfo{journal}{Zeitschrift für Physik A Hadrons and
  Nuclei} \textbf{\bibinfo{volume}{352}}, \bibinfo{pages}{351}.

\bibitem[{\citenamefont{Enqvist} \emph{et~al.}(1999)\citenamefont{Enqvist,
  Benlliure, Farget, Schmidt, Armbruster, Bernas, Tassan-Got, Boudard, Legrain,
  Volant, Böckstiegel, de~Jong} \emph{et~al.}}]{Enqvist:1999}
\bibinfo{author}{\bibnamefont{Enqvist}, \bibfnamefont{T.}},
  \bibinfo{author}{\bibfnamefont{J.}~\bibnamefont{Benlliure}},
  \bibinfo{author}{\bibfnamefont{F.}~\bibnamefont{Farget}},
  \bibinfo{author}{\bibfnamefont{K.~H.} \bibnamefont{Schmidt}},
  \bibinfo{author}{\bibfnamefont{P.}~\bibnamefont{Armbruster}},
  \bibinfo{author}{\bibfnamefont{M.}~\bibnamefont{Bernas}},
  \bibinfo{author}{\bibfnamefont{L.}~\bibnamefont{Tassan-Got}},
  \bibinfo{author}{\bibfnamefont{A.}~\bibnamefont{Boudard}},
  \bibinfo{author}{\bibfnamefont{R.}~\bibnamefont{Legrain}},
  \bibinfo{author}{\bibfnamefont{C.}~\bibnamefont{Volant}},
  \bibinfo{author}{\bibfnamefont{C.}~\bibnamefont{Böckstiegel}},
  \bibinfo{author}{\bibfnamefont{M.}~\bibnamefont{de~Jong}}, \emph{et~al.},
  \bibinfo{year}{1999}, \bibinfo{journal}{Nuclear Physics A}
  \textbf{\bibinfo{volume}{658}}(\bibinfo{number}{1}), \bibinfo{pages}{47}.

\bibitem[{\citenamefont{Eppinger} \emph{et~al.}(2009)\citenamefont{Eppinger,
  Hinke, Böhmer, Boutachkov, Faestermann, Geissel, Gernhäuser, G\'orska,
  Gottardo, Grebosz, Krücken, Kurz} \emph{et~al.}}]{Eppinger:2009}
\bibinfo{author}{\bibnamefont{Eppinger}, \bibfnamefont{K.}},
  \bibinfo{author}{\bibfnamefont{C.}~\bibnamefont{Hinke}},
  \bibinfo{author}{\bibfnamefont{M.}~\bibnamefont{Böhmer}},
  \bibinfo{author}{\bibfnamefont{P.}~\bibnamefont{Boutachkov}},
  \bibinfo{author}{\bibfnamefont{T.}~\bibnamefont{Faestermann}},
  \bibinfo{author}{\bibfnamefont{H.}~\bibnamefont{Geissel}},
  \bibinfo{author}{\bibfnamefont{R.}~\bibnamefont{Gernhäuser}},
  \bibinfo{author}{\bibfnamefont{M.}~\bibnamefont{G\'orska}},
  \bibinfo{author}{\bibfnamefont{A.}~\bibnamefont{Gottardo}},
  \bibinfo{author}{\bibfnamefont{J.}~\bibnamefont{Grebosz}},
  \bibinfo{author}{\bibfnamefont{R.}~\bibnamefont{Krücken}},
  \bibinfo{author}{\bibfnamefont{N.}~\bibnamefont{Kurz}}, \emph{et~al.},
  \bibinfo{year}{2009}, in \emph{\bibinfo{booktitle}{GSI Scientific Report
  2008}}, edited by \bibinfo{editor}{\bibfnamefont{K.}~\bibnamefont{Gro{\ss}e}}
  (\bibinfo{publisher}{GSI Helmholtzzentrum für Schwerionenforschung GmbH}), p.
  \bibinfo{pages}{147}.

\bibitem[{\citenamefont{Esbensen and Davids}(2001)}]{Esbensen:2001}
\bibinfo{author}{\bibnamefont{Esbensen}, \bibfnamefont{H.}}, and
  \bibinfo{author}{\bibfnamefont{C.}~\bibnamefont{Davids}},
  \bibinfo{year}{2001}, \bibinfo{journal}{Phys.\ Rev.}
  \textbf{\bibinfo{volume}{C~63}}, \bibinfo{pages}{014315}.

\bibitem[{\citenamefont{{F. de Oliveira Santos}}
  \emph{et~al.}(2005)\citenamefont{{F. de Oliveira Santos}, {P. Himpe}, {M.
  Lewitowicz}, {I. Stefan}, {N. Smirnova}, {N. L. Achouri}, {J. C.
  Ang\'elique}, {C. Angulo}, {L. Axelsson}, {D. Baiborodin}, {F. Becker}, {M.
  Bellegui}} \emph{et~al.}}]{Oliveira:2005}
\bibinfo{author}{\bibnamefont{{F. de Oliveira Santos}}},
  \bibinfo{author}{\bibnamefont{{P. Himpe}}}, \bibinfo{author}{\bibnamefont{{M.
  Lewitowicz}}}, \bibinfo{author}{\bibnamefont{{I. Stefan}}},
  \bibinfo{author}{\bibnamefont{{N. Smirnova}}},
  \bibinfo{author}{\bibnamefont{{N. L. Achouri}}},
  \bibinfo{author}{\bibnamefont{{J. C. Ang\'elique}}},
  \bibinfo{author}{\bibnamefont{{C. Angulo}}},
  \bibinfo{author}{\bibnamefont{{L. Axelsson}}},
  \bibinfo{author}{\bibnamefont{{D. Baiborodin}}},
  \bibinfo{author}{\bibnamefont{{F. Becker}}},
  \bibinfo{author}{\bibnamefont{{M. Bellegui}}}, \emph{et~al.},
  \bibinfo{year}{2005}, \bibinfo{journal}{Eur. Phys. J. A}
  \textbf{\bibinfo{volume}{24}}(\bibinfo{number}{2}), \bibinfo{pages}{237}.

\bibitem[{\citenamefont{Facina} \emph{et~al.}(2008)\citenamefont{Facina,
  Bachelet, Block, Bollen, Davies, {Folden~III}, Guenaut, Huikari, Kwan,
  Morrissey, Pang, Prinke} \emph{et~al.}}]{Facina:2008}
\bibinfo{author}{\bibnamefont{Facina}, \bibfnamefont{M.}},
  \bibinfo{author}{\bibfnamefont{C.}~\bibnamefont{Bachelet}},
  \bibinfo{author}{\bibfnamefont{M.}~\bibnamefont{Block}},
  \bibinfo{author}{\bibfnamefont{G.}~\bibnamefont{Bollen}},
  \bibinfo{author}{\bibfnamefont{D.}~\bibnamefont{Davies}},
  \bibinfo{author}{\bibfnamefont{C.}~\bibnamefont{{Folden~III}}},
  \bibinfo{author}{\bibfnamefont{C.}~\bibnamefont{Guenaut}},
  \bibinfo{author}{\bibfnamefont{J.}~\bibnamefont{Huikari}},
  \bibinfo{author}{\bibfnamefont{E.}~\bibnamefont{Kwan}},
  \bibinfo{author}{\bibfnamefont{D.}~\bibnamefont{Morrissey}},
  \bibinfo{author}{\bibfnamefont{G.}~\bibnamefont{Pang}},
  \bibinfo{author}{\bibfnamefont{A.}~\bibnamefont{Prinke}}, \emph{et~al.},
  \bibinfo{year}{2008}, \bibinfo{journal}{Nuclear Instruments and Methods in
  Physics Research Section B: Beam Interactions with Materials and Atoms}
  \textbf{\bibinfo{volume}{266}}(\bibinfo{number}{19-20}),
  \bibinfo{pages}{4471}.

\bibitem[{\citenamefont{Faesterman}
  \emph{et~al.}(1984)\citenamefont{Faesterman, Gillitzer, Hartel, Kleinle, and
  Nolte}}]{Faesterman:1984}
\bibinfo{author}{\bibnamefont{Faesterman}, \bibfnamefont{T.}},
  \bibinfo{author}{\bibfnamefont{A.}~\bibnamefont{Gillitzer}},
  \bibinfo{author}{\bibfnamefont{K.}~\bibnamefont{Hartel}},
  \bibinfo{author}{\bibfnamefont{P.}~\bibnamefont{Kleinle}}, and
  \bibinfo{author}{\bibfnamefont{E.}~\bibnamefont{Nolte}},
  \bibinfo{year}{1984}, \bibinfo{journal}{Phys.\ Lett.}
  \textbf{\bibinfo{volume}{137B}}, \bibinfo{pages}{23}.

\bibitem[{\citenamefont{Farnea} \emph{et~al.}(2010)\citenamefont{Farnea,
  Recchia, Bazzacco, Kröll, Podolyák, Quintana, and Gadea}}]{Farnea:2010}
\bibinfo{author}{\bibnamefont{Farnea}, \bibfnamefont{E.}},
  \bibinfo{author}{\bibfnamefont{F.}~\bibnamefont{Recchia}},
  \bibinfo{author}{\bibfnamefont{D.}~\bibnamefont{Bazzacco}},
  \bibinfo{author}{\bibfnamefont{T.}~\bibnamefont{Kröll}},
  \bibinfo{author}{\bibfnamefont{Z.}~\bibnamefont{Podolyák}},
  \bibinfo{author}{\bibfnamefont{B.}~\bibnamefont{Quintana}}, and
  \bibinfo{author}{\bibfnamefont{A.}~\bibnamefont{Gadea}},
  \bibinfo{year}{2010}, \bibinfo{journal}{Nuclear Instruments and Methods in
  Physics Research Section A: Accelerators, Spectrometers, Detectors and
  Associated Equipment} \textbf{\bibinfo{volume}{621}}(\bibinfo{number}{1-3}),
  \bibinfo{pages}{331}.

\bibitem[{\citenamefont{Fermi}(1934)}]{Fermi:1934}
\bibinfo{author}{\bibnamefont{Fermi}, \bibfnamefont{E.}}, \bibinfo{year}{1934},
  \bibinfo{journal}{Zeitschr. f. Phys.} \textbf{\bibinfo{volume}{88}},
  \bibinfo{pages}{161}.

\bibitem[{\citenamefont{Fernández-Domínguez}
  \emph{et~al.}(2005)\citenamefont{Fernández-Domínguez, Armbruster, Audouin,
  Benlliure, Bernas, Boudard, Casarejos, Czajkowski, Ducret, Enqvist, Jurado,
  Legrain} \emph{et~al.}}]{FernandezDominguez:2005}
\bibinfo{author}{\bibnamefont{Fernández-Domínguez}, \bibfnamefont{B.}},
  \bibinfo{author}{\bibfnamefont{P.}~\bibnamefont{Armbruster}},
  \bibinfo{author}{\bibfnamefont{L.}~\bibnamefont{Audouin}},
  \bibinfo{author}{\bibfnamefont{J.}~\bibnamefont{Benlliure}},
  \bibinfo{author}{\bibfnamefont{M.}~\bibnamefont{Bernas}},
  \bibinfo{author}{\bibfnamefont{A.}~\bibnamefont{Boudard}},
  \bibinfo{author}{\bibfnamefont{E.}~\bibnamefont{Casarejos}},
  \bibinfo{author}{\bibfnamefont{S.}~\bibnamefont{Czajkowski}},
  \bibinfo{author}{\bibfnamefont{J.}~\bibnamefont{Ducret}},
  \bibinfo{author}{\bibfnamefont{T.}~\bibnamefont{Enqvist}},
  \bibinfo{author}{\bibfnamefont{B.}~\bibnamefont{Jurado}},
  \bibinfo{author}{\bibfnamefont{R.}~\bibnamefont{Legrain}}, \emph{et~al.},
  \bibinfo{year}{2005}, \bibinfo{journal}{Nuclear Physics A}
  \textbf{\bibinfo{volume}{747}}(\bibinfo{number}{2-4}), \bibinfo{pages}{227}.

\bibitem[{\citenamefont{Ferreira and Maglione}(2000)}]{Ferreira:2000}
\bibinfo{author}{\bibnamefont{Ferreira}, \bibfnamefont{L.}}, and
  \bibinfo{author}{\bibfnamefont{E.}~\bibnamefont{Maglione}},
  \bibinfo{year}{2000}, \bibinfo{journal}{Phys.\ Rev.}
  \textbf{\bibinfo{volume}{C~61}}, \bibinfo{pages}{021304(R)}.

\bibitem[{\citenamefont{Ferreira and Maglione}(2001)}]{Ferreira:2001}
\bibinfo{author}{\bibnamefont{Ferreira}, \bibfnamefont{L.}}, and
  \bibinfo{author}{\bibfnamefont{E.}~\bibnamefont{Maglione}},
  \bibinfo{year}{2001}, \bibinfo{journal}{Phys.\ Rev.\ Lett.}
  \textbf{\bibinfo{volume}{86}}, \bibinfo{pages}{1721}.

\bibitem[{\citenamefont{Ferreira and Maglione}(2005)}]{Ferreira:2005}
\bibinfo{author}{\bibnamefont{Ferreira}, \bibfnamefont{L.}}, and
  \bibinfo{author}{\bibfnamefont{E.}~\bibnamefont{Maglione}},
  \bibinfo{year}{2005}, \bibinfo{journal}{J.\ Phys.}
  \textbf{\bibinfo{volume}{G~31}}, \bibinfo{pages}{S1569}.

\bibitem[{\citenamefont{Ferreira} \emph{et~al.}(2002)\citenamefont{Ferreira,
  Maglione, and Fernandes}}]{Ferreira:2002}
\bibinfo{author}{\bibnamefont{Ferreira}, \bibfnamefont{L.}},
  \bibinfo{author}{\bibfnamefont{E.}~\bibnamefont{Maglione}}, and
  \bibinfo{author}{\bibfnamefont{D.}~\bibnamefont{Fernandes}},
  \bibinfo{year}{2002}, \bibinfo{journal}{Phys.\ Rev.}
  \textbf{\bibinfo{volume}{C~65}}, \bibinfo{pages}{024323}.

\bibitem[{\citenamefont{Ferreira and Arumugan}(2007)}]{Ferreira:2007}
\bibinfo{editor}{\bibnamefont{Ferreira}, \bibfnamefont{L.~S.}}, and
  \bibinfo{editor}{\bibfnamefont{P.}~\bibnamefont{Arumugan}} (eds.),
  \bibinfo{year}{2007}, \emph{\bibinfo{title}{Proton Emitting Nuclei and
  Related Topics}}, number \bibinfo{number}{961} in \bibinfo{series}{AIP
  Conference Proceedings} (\bibinfo{publisher}{AIP},
  \bibinfo{address}{Melville, New York}).

\bibitem[{\citenamefont{Fiorin} \emph{et~al.}(2003)\citenamefont{Fiorin,
  Maglione, and Ferreira}}]{Fiorin:2003}
\bibinfo{author}{\bibnamefont{Fiorin}, \bibfnamefont{G.}},
  \bibinfo{author}{\bibfnamefont{E.}~\bibnamefont{Maglione}}, and
  \bibinfo{author}{\bibfnamefont{L.~S.} \bibnamefont{Ferreira}},
  \bibinfo{year}{2003}, \bibinfo{journal}{Phys. Rev. C}
  \textbf{\bibinfo{volume}{67}}(\bibinfo{number}{5}), \bibinfo{pages}{054302}.

\bibitem[{\citenamefont{Flanagan} \emph{et~al.}(2009)\citenamefont{Flanagan,
  Vingerhoets, Avgoulea, Billowes, Bissell, Blaum, Cheal, De~Rydt, Fedosseev,
  Forest, Geppert, K\"oster} \emph{et~al.}}]{Flanagan:2009}
\bibinfo{author}{\bibnamefont{Flanagan}, \bibfnamefont{K.~T.}},
  \bibinfo{author}{\bibfnamefont{P.}~\bibnamefont{Vingerhoets}},
  \bibinfo{author}{\bibfnamefont{M.}~\bibnamefont{Avgoulea}},
  \bibinfo{author}{\bibfnamefont{J.}~\bibnamefont{Billowes}},
  \bibinfo{author}{\bibfnamefont{M.~L.} \bibnamefont{Bissell}},
  \bibinfo{author}{\bibfnamefont{K.}~\bibnamefont{Blaum}},
  \bibinfo{author}{\bibfnamefont{B.}~\bibnamefont{Cheal}},
  \bibinfo{author}{\bibfnamefont{M.}~\bibnamefont{De~Rydt}},
  \bibinfo{author}{\bibfnamefont{V.~N.} \bibnamefont{Fedosseev}},
  \bibinfo{author}{\bibfnamefont{D.~H.} \bibnamefont{Forest}},
  \bibinfo{author}{\bibfnamefont{C.}~\bibnamefont{Geppert}},
  \bibinfo{author}{\bibfnamefont{U.}~\bibnamefont{K\"oster}}, \emph{et~al.},
  \bibinfo{year}{2009}, \bibinfo{journal}{Phys. Rev. Lett.}
  \textbf{\bibinfo{volume}{103}}(\bibinfo{number}{14}),
  \bibinfo{pages}{142501}.

\bibitem[{\citenamefont{Flerov and Petrzhak}(1940)}]{Flerov:1940}
\bibinfo{author}{\bibnamefont{Flerov}, \bibfnamefont{G.}}, and
  \bibinfo{author}{\bibfnamefont{K.}~\bibnamefont{Petrzhak}},
  \bibinfo{year}{1940}, \bibinfo{journal}{J. Phys. U.S.S.R.}
  \textbf{\bibinfo{volume}{3}}, \bibinfo{pages}{275}.

\bibitem[{\citenamefont{{Fomichev}}
  \emph{et~al.}(2011)\citenamefont{{Fomichev}, {Mukha}, {Stepantsov},
  {Grigorenko}, {Litvinova}, {Chudoba}, {Egorova}, {Golovkov}, {Gorshkov},
  {Gorshkov}, {Kaminski}, {Krupko}} \emph{et~al.}}]{Fomichev:2010}
\bibinfo{author}{\bibnamefont{{Fomichev}}, \bibfnamefont{A.~S.}},
  \bibinfo{author}{\bibfnamefont{I.~G.} \bibnamefont{{Mukha}}},
  \bibinfo{author}{\bibfnamefont{S.~V.} \bibnamefont{{Stepantsov}}},
  \bibinfo{author}{\bibfnamefont{L.~V.} \bibnamefont{{Grigorenko}}},
  \bibinfo{author}{\bibfnamefont{E.~V.} \bibnamefont{{Litvinova}}},
  \bibinfo{author}{\bibfnamefont{V.}~\bibnamefont{{Chudoba}}},
  \bibinfo{author}{\bibfnamefont{I.~A.} \bibnamefont{{Egorova}}},
  \bibinfo{author}{\bibfnamefont{M.~S.} \bibnamefont{{Golovkov}}},
  \bibinfo{author}{\bibfnamefont{A.~V.} \bibnamefont{{Gorshkov}}},
  \bibinfo{author}{\bibfnamefont{V.~A.} \bibnamefont{{Gorshkov}}},
  \bibinfo{author}{\bibfnamefont{G.}~\bibnamefont{{Kaminski}}},
  \bibinfo{author}{\bibfnamefont{S.~A.} \bibnamefont{{Krupko}}}, \emph{et~al.},
  \bibinfo{year}{2011}, \bibinfo{journal}{Int. J. Mod. Phys.}
  \textbf{\bibinfo{volume}{E20}}, \bibinfo{pages}{1491}.

\bibitem[{\citenamefont{Fraile and Äystö}(2003)}]{Fraile:2003}
\bibinfo{author}{\bibnamefont{Fraile}, \bibfnamefont{L.~M.}}, and
  \bibinfo{author}{\bibfnamefont{J.}~\bibnamefont{Äystö}},
  \bibinfo{year}{2003}, \bibinfo{journal}{Nuclear Instruments and Methods in
  Physics Research Section A: Accelerators, Spectrometers, Detectors and
  Associated Equipment} \textbf{\bibinfo{volume}{513}}(\bibinfo{number}{1-2}),
  \bibinfo{pages}{287}.

\bibitem[{\citenamefont{Fynbo} \emph{et~al.}(2000)\citenamefont{Fynbo, Borge,
  Axelsson, \"{A}yst\"{o}, Bergmann, Fraile, Honkanen, Hornsh{\o}j, Jading,
  Jokinen, Jonson, Martel} \emph{et~al.}}]{Fynbo:2000}
\bibinfo{author}{\bibnamefont{Fynbo}, \bibfnamefont{H.~O.~U.}},
  \bibinfo{author}{\bibfnamefont{M.~J.~G.} \bibnamefont{Borge}},
  \bibinfo{author}{\bibfnamefont{L.}~\bibnamefont{Axelsson}},
  \bibinfo{author}{\bibfnamefont{J.}~\bibnamefont{\"{A}yst\"{o}}},
  \bibinfo{author}{\bibfnamefont{U.~C.} \bibnamefont{Bergmann}},
  \bibinfo{author}{\bibfnamefont{L.~M.} \bibnamefont{Fraile}},
  \bibinfo{author}{\bibfnamefont{A.}~\bibnamefont{Honkanen}},
  \bibinfo{author}{\bibfnamefont{P.}~\bibnamefont{Hornsh{\o}j}},
  \bibinfo{author}{\bibfnamefont{Y.}~\bibnamefont{Jading}},
  \bibinfo{author}{\bibfnamefont{A.}~\bibnamefont{Jokinen}},
  \bibinfo{author}{\bibfnamefont{B.}~\bibnamefont{Jonson}},
  \bibinfo{author}{\bibfnamefont{I.}~\bibnamefont{Martel}}, \emph{et~al.},
  \bibinfo{year}{2000}, \bibinfo{journal}{Nuclear Physics A}
  \textbf{\bibinfo{volume}{677}}(\bibinfo{number}{1-4}), \bibinfo{pages}{38}.

\bibitem[{\citenamefont{Fynbo} \emph{et~al.}(2004)\citenamefont{Fynbo, Borge,
  Cederk\"{a}ll, Courtin, Dessagne, Jonson, Scornet, Nilsson, Nyman, Poirier,
  Riisager, Tengblad} \emph{et~al.}}]{Fynbo:2004}
\bibinfo{author}{\bibnamefont{Fynbo}, \bibfnamefont{H.~O.~U.}},
  \bibinfo{author}{\bibfnamefont{M.~J.~G.} \bibnamefont{Borge}},
  \bibinfo{author}{\bibfnamefont{J.}~\bibnamefont{Cederk\"{a}ll}},
  \bibinfo{author}{\bibfnamefont{S.}~\bibnamefont{Courtin}},
  \bibinfo{author}{\bibfnamefont{P.}~\bibnamefont{Dessagne}},
  \bibinfo{author}{\bibfnamefont{B.}~\bibnamefont{Jonson}},
  \bibinfo{author}{\bibfnamefont{G.~L.} \bibnamefont{Scornet}},
  \bibinfo{author}{\bibfnamefont{T.}~\bibnamefont{Nilsson}},
  \bibinfo{author}{\bibfnamefont{G.}~\bibnamefont{Nyman}},
  \bibinfo{author}{\bibfnamefont{E.}~\bibnamefont{Poirier}},
  \bibinfo{author}{\bibfnamefont{K.}~\bibnamefont{Riisager}},
  \bibinfo{author}{\bibfnamefont{O.}~\bibnamefont{Tengblad}}, \emph{et~al.},
  \bibinfo{year}{2004}, \bibinfo{journal}{Nuclear Physics A}
  \textbf{\bibinfo{volume}{736}}, \bibinfo{pages}{39}.

\bibitem[{\citenamefont{Fynbo} \emph{et~al.}(2005)\citenamefont{Fynbo, Diget,
  Bergmann, Borge, Cederk\"{a}ll, Dendooven, Fraile, Franchoo, Fedosseev,
  Fulton, Huang, Huikari} \emph{et~al.}}]{Fynbo:2005}
\bibinfo{author}{\bibnamefont{Fynbo}, \bibfnamefont{H.~O.~U.}},
  \bibinfo{author}{\bibfnamefont{C.~A.} \bibnamefont{Diget}},
  \bibinfo{author}{\bibfnamefont{U.~C.} \bibnamefont{Bergmann}},
  \bibinfo{author}{\bibfnamefont{M.~J.~G.} \bibnamefont{Borge}},
  \bibinfo{author}{\bibfnamefont{J.}~\bibnamefont{Cederk\"{a}ll}},
  \bibinfo{author}{\bibfnamefont{P.}~\bibnamefont{Dendooven}},
  \bibinfo{author}{\bibfnamefont{L.~M.} \bibnamefont{Fraile}},
  \bibinfo{author}{\bibfnamefont{S.}~\bibnamefont{Franchoo}},
  \bibinfo{author}{\bibfnamefont{V.~N.} \bibnamefont{Fedosseev}},
  \bibinfo{author}{\bibfnamefont{B.~R.} \bibnamefont{Fulton}},
  \bibinfo{author}{\bibfnamefont{W.}~\bibnamefont{Huang}},
  \bibinfo{author}{\bibfnamefont{J.}~\bibnamefont{Huikari}}, \emph{et~al.},
  \bibinfo{year}{2005}, \bibinfo{journal}{Nature}
  \textbf{\bibinfo{volume}{433}}, \bibinfo{pages}{136}.

\bibitem[{\citenamefont{Gaimard and Schmidt}(1991)}]{Gaimard:1991}
\bibinfo{author}{\bibnamefont{Gaimard}, \bibfnamefont{J.~J.}}, and
  \bibinfo{author}{\bibfnamefont{K.~H.} \bibnamefont{Schmidt}},
  \bibinfo{year}{1991}, \bibinfo{journal}{Nuclear Physics A}
  \textbf{\bibinfo{volume}{531}}(\bibinfo{number}{3-4}), \bibinfo{pages}{709}.

\bibitem[{\citenamefont{Galitsky and Cheltsov}(1964)}]{Galitsky:1964}
\bibinfo{author}{\bibnamefont{Galitsky}, \bibfnamefont{V.~M.}}, and
  \bibinfo{author}{\bibfnamefont{V.~F.} \bibnamefont{Cheltsov}},
  \bibinfo{year}{1964}, \bibinfo{journal}{Nucl. Phys.}
  \textbf{\bibinfo{volume}{56}}, \bibinfo{pages}{86}.

\bibitem[{\citenamefont{Gamow}(1928)}]{Gamow:1928}
\bibinfo{author}{\bibnamefont{Gamow}, \bibfnamefont{G.}}, \bibinfo{year}{1928},
  \bibinfo{journal}{Zeitschr. f. Phys.} \textbf{\bibinfo{volume}{51}},
  \bibinfo{pages}{204}.

\bibitem[{\citenamefont{Garrido} \emph{et~al.}(2007)\citenamefont{Garrido,
  Fedorov, Fynbo, and Jensen}}]{Garrido:2007}
\bibinfo{author}{\bibnamefont{Garrido}, \bibfnamefont{E.}},
  \bibinfo{author}{\bibfnamefont{D.~V.} \bibnamefont{Fedorov}},
  \bibinfo{author}{\bibfnamefont{H.~O.~U.} \bibnamefont{Fynbo}}, and
  \bibinfo{author}{\bibfnamefont{A.~S.} \bibnamefont{Jensen}},
  \bibinfo{year}{2007}, \bibinfo{journal}{Physics Letters B}
  \textbf{\bibinfo{volume}{648}}(\bibinfo{number}{4}), \bibinfo{pages}{274}.

\bibitem[{\citenamefont{Garrido} \emph{et~al.}(2008)\citenamefont{Garrido,
  Jensen, and Fedorov}}]{Garrido:2008}
\bibinfo{author}{\bibnamefont{Garrido}, \bibfnamefont{E.}},
  \bibinfo{author}{\bibfnamefont{A.}~\bibnamefont{Jensen}}, and
  \bibinfo{author}{\bibfnamefont{D.}~\bibnamefont{Fedorov}},
  \bibinfo{year}{2008}, \bibinfo{journal}{Phys. Rev. C}
  \textbf{\bibinfo{volume}{78}}, \bibinfo{pages}{034004}.

\bibitem[{\citenamefont{Gavron}(1980)}]{Gavron:1980}
\bibinfo{author}{\bibnamefont{Gavron}, \bibfnamefont{A.}},
  \bibinfo{year}{1980}, \bibinfo{journal}{Phys. Rev. C}
  \textbf{\bibinfo{volume}{21}}(\bibinfo{number}{1}), \bibinfo{pages}{230}.

\bibitem[{\citenamefont{Geesaman} \emph{et~al.}(1977)\citenamefont{Geesaman,
  McGrath, Lesser, Urone, and VerWest}}]{Geesaman:1977}
\bibinfo{author}{\bibnamefont{Geesaman}, \bibfnamefont{D.~F.}},
  \bibinfo{author}{\bibfnamefont{R.~L.} \bibnamefont{McGrath}},
  \bibinfo{author}{\bibfnamefont{P.~M.~S.} \bibnamefont{Lesser}},
  \bibinfo{author}{\bibfnamefont{P.~P.} \bibnamefont{Urone}}, and
  \bibinfo{author}{\bibfnamefont{B.}~\bibnamefont{VerWest}},
  \bibinfo{year}{1977}, \bibinfo{journal}{Phys. Rev. C}
  \textbf{\bibinfo{volume}{15}}, \bibinfo{pages}{1835}.

\bibitem[{\citenamefont{Geiger and Nutall}(1912)}]{Geiger:1912}
\bibinfo{author}{\bibnamefont{Geiger}, \bibfnamefont{H.}}, and
  \bibinfo{author}{\bibfnamefont{J.}~\bibnamefont{Nutall}},
  \bibinfo{year}{1912}, \bibinfo{journal}{Phil. Mag.}
  \textbf{\bibinfo{volume}{23}}, \bibinfo{pages}{439}.

\bibitem[{\citenamefont{Geissel} \emph{et~al.}(1992)\citenamefont{Geissel,
  Armbruster, Behr, Brünle, Burkard, Chen, Folger, Franczak, Keller, Klepper,
  Langenbeck, Nickel} \emph{et~al.}}]{Geissel:1992}
\bibinfo{author}{\bibnamefont{Geissel}, \bibfnamefont{H.}},
  \bibinfo{author}{\bibfnamefont{P.}~\bibnamefont{Armbruster}},
  \bibinfo{author}{\bibfnamefont{K.~H.} \bibnamefont{Behr}},
  \bibinfo{author}{\bibfnamefont{A.}~\bibnamefont{Brünle}},
  \bibinfo{author}{\bibfnamefont{K.}~\bibnamefont{Burkard}},
  \bibinfo{author}{\bibfnamefont{M.}~\bibnamefont{Chen}},
  \bibinfo{author}{\bibfnamefont{H.}~\bibnamefont{Folger}},
  \bibinfo{author}{\bibfnamefont{B.}~\bibnamefont{Franczak}},
  \bibinfo{author}{\bibfnamefont{H.}~\bibnamefont{Keller}},
  \bibinfo{author}{\bibfnamefont{O.}~\bibnamefont{Klepper}},
  \bibinfo{author}{\bibfnamefont{B.}~\bibnamefont{Langenbeck}},
  \bibinfo{author}{\bibfnamefont{F.}~\bibnamefont{Nickel}}, \emph{et~al.},
  \bibinfo{year}{1992}, \bibinfo{journal}{Nuclear Instruments and Methods in
  Physics Research Section B: Beam Interactions with Materials and Atoms}
  \textbf{\bibinfo{volume}{70}}(\bibinfo{number}{1-4}), \bibinfo{pages}{286}.

\bibitem[{\citenamefont{Geissel} \emph{et~al.}(1995)\citenamefont{Geissel,
  M\"{u}nzenberg, and Riisager}}]{Geissel:1995}
\bibinfo{author}{\bibnamefont{Geissel}, \bibfnamefont{H.}},
  \bibinfo{author}{\bibfnamefont{G.}~\bibnamefont{M\"{u}nzenberg}}, and
  \bibinfo{author}{\bibfnamefont{K.}~\bibnamefont{Riisager}},
  \bibinfo{year}{1995}, \bibinfo{journal}{Annual Review of Nuclear and Particle
  Science} \textbf{\bibinfo{volume}{45}}(\bibinfo{number}{1}),
  \bibinfo{pages}{163}.

\bibitem[{\citenamefont{Gelletly and Eberth}(2006)}]{Gelletly:2006}
\bibinfo{author}{\bibnamefont{Gelletly}, \bibfnamefont{W.}}, and
  \bibinfo{author}{\bibfnamefont{J.}~\bibnamefont{Eberth}},
  \bibinfo{year}{2006}, in \emph{\bibinfo{booktitle}{The Euroschool Lectures on
  Physics With Exotic Beams, Vol.II, Lect. Notes Phys. 700}}, edited by
  \bibinfo{editor}{\bibfnamefont{J.}~\bibnamefont{Al-Khalili}} and
  \bibinfo{editor}{\bibfnamefont{E.}~\bibnamefont{Roeckl}}
  (\bibinfo{publisher}{Springer, Berlin Heidelberg}), p.~\bibinfo{pages}{79}.

\bibitem[{\citenamefont{Gillitzer} \emph{et~al.}(1987)\citenamefont{Gillitzer,
  Faesterman, Hartel, Kleinle, and Nolte}}]{Gillitzer:1987}
\bibinfo{author}{\bibnamefont{Gillitzer}, \bibfnamefont{A.}},
  \bibinfo{author}{\bibfnamefont{T.}~\bibnamefont{Faesterman}},
  \bibinfo{author}{\bibfnamefont{K.}~\bibnamefont{Hartel}},
  \bibinfo{author}{\bibfnamefont{P.}~\bibnamefont{Kleinle}}, and
  \bibinfo{author}{\bibfnamefont{E.}~\bibnamefont{Nolte}},
  \bibinfo{year}{1987}, \bibinfo{journal}{Z.\ Phys.}
  \textbf{\bibinfo{volume}{A~326}}, \bibinfo{pages}{107}.

\bibitem[{\citenamefont{Ginter} \emph{et~al.}(2000)\citenamefont{Ginter,
  Batchelder, Bingham, Gross, Grzywacz, Hamilton, Janas, Karny, Kim, Mas,
  McConnel, Piechaczek} \emph{et~al.}}]{Ginter:2000}
\bibinfo{author}{\bibnamefont{Ginter}, \bibfnamefont{T.}},
  \bibinfo{author}{\bibfnamefont{J.}~\bibnamefont{Batchelder}},
  \bibinfo{author}{\bibfnamefont{C.}~\bibnamefont{Bingham}},
  \bibinfo{author}{\bibfnamefont{C.}~\bibnamefont{Gross}},
  \bibinfo{author}{\bibfnamefont{R.}~\bibnamefont{Grzywacz}},
  \bibinfo{author}{\bibfnamefont{J.}~\bibnamefont{Hamilton}},
  \bibinfo{author}{\bibfnamefont{Z.}~\bibnamefont{Janas}},
  \bibinfo{author}{\bibfnamefont{M.}~\bibnamefont{Karny}},
  \bibinfo{author}{\bibfnamefont{S.}~\bibnamefont{Kim}},
  \bibinfo{author}{\bibfnamefont{J.}~\bibnamefont{Mas}},
  \bibinfo{author}{\bibfnamefont{J.}~\bibnamefont{McConnel}},
  \bibinfo{author}{\bibfnamefont{A.}~\bibnamefont{Piechaczek}}, \emph{et~al.},
  \bibinfo{year}{2000}, \bibinfo{journal}{Phys.\ Rev.}
  \textbf{\bibinfo{volume}{C~61}}, \bibinfo{pages}{014308}.

\bibitem[{\citenamefont{Ginter} \emph{et~al.}(2003)\citenamefont{Ginter,
  Batchelder, Bingham, Gross, Grzywacz, Hamilton, Janas, Karny, Piechaczek,
  Ramayya, Rykaczewski, Walters} \emph{et~al.}}]{Ginter:2003}
\bibinfo{author}{\bibnamefont{Ginter}, \bibfnamefont{T.}},
  \bibinfo{author}{\bibfnamefont{J.}~\bibnamefont{Batchelder}},
  \bibinfo{author}{\bibfnamefont{C.}~\bibnamefont{Bingham}},
  \bibinfo{author}{\bibfnamefont{C.}~\bibnamefont{Gross}},
  \bibinfo{author}{\bibfnamefont{R.}~\bibnamefont{Grzywacz}},
  \bibinfo{author}{\bibfnamefont{J.}~\bibnamefont{Hamilton}},
  \bibinfo{author}{\bibfnamefont{Z.}~\bibnamefont{Janas}},
  \bibinfo{author}{\bibfnamefont{M.}~\bibnamefont{Karny}},
  \bibinfo{author}{\bibfnamefont{A.}~\bibnamefont{Piechaczek}},
  \bibinfo{author}{\bibfnamefont{A.}~\bibnamefont{Ramayya}},
  \bibinfo{author}{\bibfnamefont{K.}~\bibnamefont{Rykaczewski}},
  \bibinfo{author}{\bibfnamefont{W.}~\bibnamefont{Walters}}, \emph{et~al.},
  \bibinfo{year}{2003}, \bibinfo{journal}{Phys.\ Rev.}
  \textbf{\bibinfo{volume}{C~68}}, \bibinfo{pages}{032330}.

\bibitem[{\citenamefont{Giovinazzo}
  \emph{et~al.}(2007)\citenamefont{Giovinazzo, Blank, Borcea, Canchel, Dalouzy,
  Demonchy, de~Oliveira~Santos, Dossat, Grevy, Hay, Huikari, Leblanc}
  \emph{et~al.}}]{Giovinazzo:2007}
\bibinfo{author}{\bibnamefont{Giovinazzo}, \bibfnamefont{J.}},
  \bibinfo{author}{\bibfnamefont{B.}~\bibnamefont{Blank}},
  \bibinfo{author}{\bibfnamefont{C.}~\bibnamefont{Borcea}},
  \bibinfo{author}{\bibfnamefont{G.}~\bibnamefont{Canchel}},
  \bibinfo{author}{\bibfnamefont{J.~C.} \bibnamefont{Dalouzy}},
  \bibinfo{author}{\bibfnamefont{C.~E.} \bibnamefont{Demonchy}},
  \bibinfo{author}{\bibfnamefont{F.}~\bibnamefont{de~Oliveira~Santos}},
  \bibinfo{author}{\bibfnamefont{C.}~\bibnamefont{Dossat}},
  \bibinfo{author}{\bibfnamefont{S.}~\bibnamefont{Grevy}},
  \bibinfo{author}{\bibfnamefont{L.}~\bibnamefont{Hay}},
  \bibinfo{author}{\bibfnamefont{J.}~\bibnamefont{Huikari}},
  \bibinfo{author}{\bibfnamefont{S.}~\bibnamefont{Leblanc}}, \emph{et~al.},
  \bibinfo{year}{2007}, \bibinfo{journal}{Phys. Rev. Lett.}
  \textbf{\bibinfo{volume}{99}}, \bibinfo{pages}{102501}.

\bibitem[{\citenamefont{Giovinazzo}
  \emph{et~al.}(2002)\citenamefont{Giovinazzo, Blank, Chartier, Czajkowski,
  Fleury, Lopez~Jimenez, Pravikoff, Thomas, de~Oliveira~Santos, Lewitowicz,
  Maslov, Stanoiu} \emph{et~al.}}]{Giovinazzo:2002}
\bibinfo{author}{\bibnamefont{Giovinazzo}, \bibfnamefont{J.}},
  \bibinfo{author}{\bibfnamefont{B.}~\bibnamefont{Blank}},
  \bibinfo{author}{\bibfnamefont{M.}~\bibnamefont{Chartier}},
  \bibinfo{author}{\bibfnamefont{S.}~\bibnamefont{Czajkowski}},
  \bibinfo{author}{\bibfnamefont{A.}~\bibnamefont{Fleury}},
  \bibinfo{author}{\bibfnamefont{M.~J.} \bibnamefont{Lopez~Jimenez}},
  \bibinfo{author}{\bibfnamefont{M.~S.} \bibnamefont{Pravikoff}},
  \bibinfo{author}{\bibfnamefont{J.-C.} \bibnamefont{Thomas}},
  \bibinfo{author}{\bibfnamefont{F.}~\bibnamefont{de~Oliveira~Santos}},
  \bibinfo{author}{\bibfnamefont{M.}~\bibnamefont{Lewitowicz}},
  \bibinfo{author}{\bibfnamefont{V.}~\bibnamefont{Maslov}},
  \bibinfo{author}{\bibfnamefont{M.}~\bibnamefont{Stanoiu}}, \emph{et~al.},
  \bibinfo{year}{2002}, \bibinfo{journal}{Phys. Rev. Lett.}
  \textbf{\bibinfo{volume}{89}}(\bibinfo{number}{10}), \bibinfo{pages}{102501}.

\bibitem[{\citenamefont{Giovinazzo}
  \emph{et~al.}(2000)\citenamefont{Giovinazzo, Dessagne, and
  Mieh\'{e}}}]{Giovinazzo:2000}
\bibinfo{author}{\bibnamefont{Giovinazzo}, \bibfnamefont{J.}},
  \bibinfo{author}{\bibfnamefont{P.}~\bibnamefont{Dessagne}}, and
  \bibinfo{author}{\bibfnamefont{C.}~\bibnamefont{Mieh\'{e}}},
  \bibinfo{year}{2000}, \bibinfo{journal}{Nuclear Physics A}
  \textbf{\bibinfo{volume}{674}}(\bibinfo{number}{3-4}), \bibinfo{pages}{394}.

\bibitem[{\citenamefont{Goldansky}(1960)}]{Goldansky:1960}
\bibinfo{author}{\bibnamefont{Goldansky}, \bibfnamefont{V.}},
  \bibinfo{year}{1960}, \bibinfo{journal}{Nucl. Phys.}
  \textbf{\bibinfo{volume}{19}}, \bibinfo{pages}{482}.

\bibitem[{\citenamefont{Goldansky}(1961)}]{Goldansky:1961}
\bibinfo{author}{\bibnamefont{Goldansky}, \bibfnamefont{V.~I.}},
  \bibinfo{year}{1961}, \bibinfo{journal}{Nucl. Phys.}
  \textbf{\bibinfo{volume}{27}}, \bibinfo{pages}{648}.

\bibitem[{\citenamefont{Golovkov} \emph{et~al.}(2005)\citenamefont{Golovkov,
  Grigorenko, Fomichev, Krupko, Oganessian, Rodin, Sidorchuk, Slepnev,
  Stepantsov, Ter-Akopian, Wolski, Itkis} \emph{et~al.}}]{Golovkov:2005}
\bibinfo{author}{\bibnamefont{Golovkov}, \bibfnamefont{M.~S.}},
  \bibinfo{author}{\bibfnamefont{L.~V.} \bibnamefont{Grigorenko}},
  \bibinfo{author}{\bibfnamefont{A.~S.} \bibnamefont{Fomichev}},
  \bibinfo{author}{\bibfnamefont{S.~A.} \bibnamefont{Krupko}},
  \bibinfo{author}{\bibfnamefont{Y.~T.} \bibnamefont{Oganessian}},
  \bibinfo{author}{\bibfnamefont{A.~M.} \bibnamefont{Rodin}},
  \bibinfo{author}{\bibfnamefont{S.~I.} \bibnamefont{Sidorchuk}},
  \bibinfo{author}{\bibfnamefont{R.~S.} \bibnamefont{Slepnev}},
  \bibinfo{author}{\bibfnamefont{S.~V.} \bibnamefont{Stepantsov}},
  \bibinfo{author}{\bibfnamefont{G.~M.} \bibnamefont{Ter-Akopian}},
  \bibinfo{author}{\bibfnamefont{R.}~\bibnamefont{Wolski}},
  \bibinfo{author}{\bibfnamefont{M.~G.} \bibnamefont{Itkis}}, \emph{et~al.},
  \bibinfo{year}{2005}, \bibinfo{journal}{Phys. Rev. C}
  \textbf{\bibinfo{volume}{72}}, \bibinfo{pages}{064612}.

\bibitem[{\citenamefont{Golovkov} \emph{et~al.}(2004)\citenamefont{Golovkov,
  Grigorenko, Fomichev, Oganessian, Orlov, Rodin, Sidorchuk, Slepnev,
  Stepantsov, Ter-Akopian, and Wolski}}]{Golovkov:2004}
\bibinfo{author}{\bibnamefont{Golovkov}, \bibfnamefont{M.~S.}},
  \bibinfo{author}{\bibfnamefont{L.~V.} \bibnamefont{Grigorenko}},
  \bibinfo{author}{\bibfnamefont{A.~S.} \bibnamefont{Fomichev}},
  \bibinfo{author}{\bibfnamefont{Y.~T.} \bibnamefont{Oganessian}},
  \bibinfo{author}{\bibfnamefont{Y.~I.} \bibnamefont{Orlov}},
  \bibinfo{author}{\bibfnamefont{A.~M.} \bibnamefont{Rodin}},
  \bibinfo{author}{\bibfnamefont{S.~I.} \bibnamefont{Sidorchuk}},
  \bibinfo{author}{\bibfnamefont{R.~S.} \bibnamefont{Slepnev}},
  \bibinfo{author}{\bibfnamefont{S.~V.} \bibnamefont{Stepantsov}},
  \bibinfo{author}{\bibfnamefont{G.~M.} \bibnamefont{Ter-Akopian}}, and
  \bibinfo{author}{\bibfnamefont{R.}~\bibnamefont{Wolski}},
  \bibinfo{year}{2004}, \bibinfo{journal}{Phys. Lett. B}
  \textbf{\bibinfo{volume}{588}}, \bibinfo{pages}{163}.

\bibitem[{\citenamefont{Golovkov} \emph{et~al.}(2009)\citenamefont{Golovkov,
  Grigorenko, Ter-Akopian, Fomichev, Oganessian, Gorshkov, Krupko, Rodin,
  Sidorchuk, Slepnev, Stepantsov, Wolski} \emph{et~al.}}]{Golovkov:2009}
\bibinfo{author}{\bibnamefont{Golovkov}, \bibfnamefont{M.~S.}},
  \bibinfo{author}{\bibfnamefont{L.~V.} \bibnamefont{Grigorenko}},
  \bibinfo{author}{\bibfnamefont{G.~M.} \bibnamefont{Ter-Akopian}},
  \bibinfo{author}{\bibfnamefont{A.~S.} \bibnamefont{Fomichev}},
  \bibinfo{author}{\bibfnamefont{Y.~T.} \bibnamefont{Oganessian}},
  \bibinfo{author}{\bibfnamefont{V.~A.} \bibnamefont{Gorshkov}},
  \bibinfo{author}{\bibfnamefont{S.~A.} \bibnamefont{Krupko}},
  \bibinfo{author}{\bibfnamefont{A.~M.} \bibnamefont{Rodin}},
  \bibinfo{author}{\bibfnamefont{S.~I.} \bibnamefont{Sidorchuk}},
  \bibinfo{author}{\bibfnamefont{R.~S.} \bibnamefont{Slepnev}},
  \bibinfo{author}{\bibfnamefont{S.~V.} \bibnamefont{Stepantsov}},
  \bibinfo{author}{\bibfnamefont{R.}~\bibnamefont{Wolski}}, \emph{et~al.},
  \bibinfo{year}{2009}, \bibinfo{journal}{Phys. Lett. B}
  \textbf{\bibinfo{volume}{672}}, \bibinfo{pages}{22}.

\bibitem[{\citenamefont{Goriely} \emph{et~al.}(2009)\citenamefont{Goriely,
  Chamel, and Pearson}}]{Goriely:2009}
\bibinfo{author}{\bibnamefont{Goriely}, \bibfnamefont{S.}},
  \bibinfo{author}{\bibfnamefont{N.}~\bibnamefont{Chamel}}, and
  \bibinfo{author}{\bibfnamefont{J.~M.} \bibnamefont{Pearson}},
  \bibinfo{year}{2009}, \bibinfo{journal}{Phys. Rev. Lett.}
  \textbf{\bibinfo{volume}{102}}(\bibinfo{number}{15}),
  \bibinfo{pages}{152503}.

\bibitem[{\citenamefont{Goriely} \emph{et~al.}(2010)\citenamefont{Goriely,
  Chamel, and Pearson}}]{Goriely:2010}
\bibinfo{author}{\bibnamefont{Goriely}, \bibfnamefont{S.}},
  \bibinfo{author}{\bibfnamefont{N.}~\bibnamefont{Chamel}}, and
  \bibinfo{author}{\bibfnamefont{J.~M.} \bibnamefont{Pearson}},
  \bibinfo{year}{2010}, \bibinfo{journal}{Phys. Rev. C}
  \textbf{\bibinfo{volume}{82}}(\bibinfo{number}{3}), \bibinfo{pages}{035804}.

\bibitem[{\citenamefont{Grawe} \emph{et~al.}(2007)\citenamefont{Grawe,
  Langanke, and Mart\'\i{}nez-Pinedo}}]{Grawe:2007}
\bibinfo{author}{\bibnamefont{Grawe}, \bibfnamefont{H.}},
  \bibinfo{author}{\bibfnamefont{K.}~\bibnamefont{Langanke}}, and
  \bibinfo{author}{\bibfnamefont{G.}~\bibnamefont{Mart\'\i{}nez-Pinedo}},
  \bibinfo{year}{2007}, \bibinfo{journal}{Reports on Progress in Physics}
  \textbf{\bibinfo{volume}{70}}(\bibinfo{number}{9}), \bibinfo{pages}{1525}.

\bibitem[{\citenamefont{Gr\'{e}vy} \emph{et~al.}(2004)\citenamefont{Gr\'{e}vy,
  Ang\'{e}lique, Baumann, Borcea, Buta, Canchel, Catford, Courtin, Daugas,
  de~Oliveira, Dessagne, Dlouhy} \emph{et~al.}}]{Grevy:2004}
\bibinfo{author}{\bibnamefont{Gr\'{e}vy}, \bibfnamefont{S.}},
  \bibinfo{author}{\bibfnamefont{J.~C.} \bibnamefont{Ang\'{e}lique}},
  \bibinfo{author}{\bibfnamefont{P.}~\bibnamefont{Baumann}},
  \bibinfo{author}{\bibfnamefont{C.}~\bibnamefont{Borcea}},
  \bibinfo{author}{\bibfnamefont{A.}~\bibnamefont{Buta}},
  \bibinfo{author}{\bibfnamefont{G.}~\bibnamefont{Canchel}},
  \bibinfo{author}{\bibfnamefont{W.~N.} \bibnamefont{Catford}},
  \bibinfo{author}{\bibfnamefont{S.}~\bibnamefont{Courtin}},
  \bibinfo{author}{\bibfnamefont{J.~M.} \bibnamefont{Daugas}},
  \bibinfo{author}{\bibfnamefont{F.}~\bibnamefont{de~Oliveira}},
  \bibinfo{author}{\bibfnamefont{P.}~\bibnamefont{Dessagne}},
  \bibinfo{author}{\bibfnamefont{Z.}~\bibnamefont{Dlouhy}}, \emph{et~al.},
  \bibinfo{year}{2004}, \bibinfo{journal}{Physics Letters B}
  \textbf{\bibinfo{volume}{594}}(\bibinfo{number}{3-4}), \bibinfo{pages}{252}.

\bibitem[{\citenamefont{Grigorenko}(2009)}]{Grigorenko:2009b}
\bibinfo{author}{\bibnamefont{Grigorenko}, \bibfnamefont{L.~V.}},
  \bibinfo{year}{2009}, \bibinfo{journal}{Physics of Particles and Nuclei}
  \textbf{\bibinfo{volume}{40}}, \bibinfo{pages}{674}.

\bibitem[{\citenamefont{Grigorenko}
  \emph{et~al.}(2010)\citenamefont{Grigorenko, Egorova, Zhukov, Charity, and
  Miernik}}]{Grigorenko:2010}
\bibinfo{author}{\bibnamefont{Grigorenko}, \bibfnamefont{L.~V.}},
  \bibinfo{author}{\bibfnamefont{I.~A.} \bibnamefont{Egorova}},
  \bibinfo{author}{\bibfnamefont{M.~V.} \bibnamefont{Zhukov}},
  \bibinfo{author}{\bibfnamefont{R.~J.} \bibnamefont{Charity}}, and
  \bibinfo{author}{\bibfnamefont{K.}~\bibnamefont{Miernik}},
  \bibinfo{year}{2010}, \bibinfo{journal}{Phys. Rev. C}
  \textbf{\bibinfo{volume}{82}}, \bibinfo{pages}{014615}, \bibinfo{note}{12
  pages}.

\bibitem[{\citenamefont{Grigorenko}
  \emph{et~al.}(2000)\citenamefont{Grigorenko, Johnson, Mukha, Thompson, and
  Zhukov}}]{Grigorenko:2000b}
\bibinfo{author}{\bibnamefont{Grigorenko}, \bibfnamefont{L.~V.}},
  \bibinfo{author}{\bibfnamefont{R.~C.} \bibnamefont{Johnson}},
  \bibinfo{author}{\bibfnamefont{I.~G.} \bibnamefont{Mukha}},
  \bibinfo{author}{\bibfnamefont{I.~J.} \bibnamefont{Thompson}}, and
  \bibinfo{author}{\bibfnamefont{M.~V.} \bibnamefont{Zhukov}},
  \bibinfo{year}{2000}, \bibinfo{journal}{Phys. Rev. Lett.}
  \textbf{\bibinfo{volume}{85}}, \bibinfo{pages}{22}.

\bibitem[{\citenamefont{Grigorenko}
  \emph{et~al.}(2002)\citenamefont{Grigorenko, Mukha, Thompson, and
  Zhukov}}]{Grigorenko:2002}
\bibinfo{author}{\bibnamefont{Grigorenko}, \bibfnamefont{L.~V.}},
  \bibinfo{author}{\bibfnamefont{I.~G.} \bibnamefont{Mukha}},
  \bibinfo{author}{\bibfnamefont{I.~J.} \bibnamefont{Thompson}}, and
  \bibinfo{author}{\bibfnamefont{M.~V.} \bibnamefont{Zhukov}},
  \bibinfo{year}{2002}, \bibinfo{journal}{Phys. Rev. Lett.}
  \textbf{\bibinfo{volume}{88}}, \bibinfo{pages}{042502}.

\bibitem[{\citenamefont{Grigorenko}
  \emph{et~al.}(2003{\natexlab{a}})\citenamefont{Grigorenko, Mukha, and
  Zhukov}}]{Grigorenko:2003}
\bibinfo{author}{\bibnamefont{Grigorenko}, \bibfnamefont{L.~V.}},
  \bibinfo{author}{\bibfnamefont{I.~G.} \bibnamefont{Mukha}}, and
  \bibinfo{author}{\bibfnamefont{M.~V.} \bibnamefont{Zhukov}},
  \bibinfo{year}{2003}{\natexlab{a}}, \bibinfo{journal}{Nucl. Phys. A}
  \textbf{\bibinfo{volume}{713}}, \bibinfo{pages}{372}, \bibinfo{note}{erratum
  Nucl. Phys. A740 (2004) 401}.

\bibitem[{\citenamefont{Grigorenko}
  \emph{et~al.}(2009{\natexlab{a}})\citenamefont{Grigorenko, Wiser, Mercurio,
  Charity, Shane, Sobotka, Elson, Wuosmaa, Banu, McCleskey, Trache, Tribble}
  \emph{et~al.}}]{Grigorenko:2009c}
\bibinfo{author}{\bibnamefont{Grigorenko}, \bibfnamefont{L.~V.}},
  \bibinfo{author}{\bibfnamefont{T.~D.} \bibnamefont{Wiser}},
  \bibinfo{author}{\bibfnamefont{K.}~\bibnamefont{Mercurio}},
  \bibinfo{author}{\bibfnamefont{R.~J.} \bibnamefont{Charity}},
  \bibinfo{author}{\bibfnamefont{R.}~\bibnamefont{Shane}},
  \bibinfo{author}{\bibfnamefont{L.~G.} \bibnamefont{Sobotka}},
  \bibinfo{author}{\bibfnamefont{J.~M.} \bibnamefont{Elson}},
  \bibinfo{author}{\bibfnamefont{A.~H.} \bibnamefont{Wuosmaa}},
  \bibinfo{author}{\bibfnamefont{A.}~\bibnamefont{Banu}},
  \bibinfo{author}{\bibfnamefont{M.}~\bibnamefont{McCleskey}},
  \bibinfo{author}{\bibfnamefont{L.}~\bibnamefont{Trache}},
  \bibinfo{author}{\bibfnamefont{R.~E.} \bibnamefont{Tribble}}, \emph{et~al.},
  \bibinfo{year}{2009}{\natexlab{a}}, \bibinfo{journal}{Phys. Rev. C}
  \textbf{\bibinfo{volume}{80}}, \bibinfo{pages}{034602}.

\bibitem[{\citenamefont{Grigorenko}
  \emph{et~al.}(2009{\natexlab{b}})\citenamefont{Grigorenko, Wiser, Miernik,
  Charity, Pf\"utzner, Banu, Bingham, \'{C}wiok, Darby, Dominik, Elson, Ginter}
  \emph{et~al.}}]{Grigorenko:2009}
\bibinfo{author}{\bibnamefont{Grigorenko}, \bibfnamefont{L.~V.}},
  \bibinfo{author}{\bibfnamefont{T.~D.} \bibnamefont{Wiser}},
  \bibinfo{author}{\bibfnamefont{K.}~\bibnamefont{Miernik}},
  \bibinfo{author}{\bibfnamefont{R.~J.} \bibnamefont{Charity}},
  \bibinfo{author}{\bibfnamefont{M.}~\bibnamefont{Pf\"utzner}},
  \bibinfo{author}{\bibfnamefont{A.}~\bibnamefont{Banu}},
  \bibinfo{author}{\bibfnamefont{C.~R.} \bibnamefont{Bingham}},
  \bibinfo{author}{\bibfnamefont{M.}~\bibnamefont{\'{C}wiok}},
  \bibinfo{author}{\bibfnamefont{I.~G.} \bibnamefont{Darby}},
  \bibinfo{author}{\bibfnamefont{W.}~\bibnamefont{Dominik}},
  \bibinfo{author}{\bibfnamefont{J.~M.} \bibnamefont{Elson}},
  \bibinfo{author}{\bibfnamefont{T.}~\bibnamefont{Ginter}}, \emph{et~al.},
  \bibinfo{year}{2009}{\natexlab{b}}, \bibinfo{journal}{Phys. Lett. B}
  \textbf{\bibinfo{volume}{677}}, \bibinfo{pages}{30}.

\bibitem[{\citenamefont{Grigorenko and
  Zhukov}(2007{\natexlab{a}})}]{Grigorenko:2007}
\bibinfo{author}{\bibnamefont{Grigorenko}, \bibfnamefont{L.~V.}}, and
  \bibinfo{author}{\bibfnamefont{M.~V.} \bibnamefont{Zhukov}},
  \bibinfo{year}{2007}{\natexlab{a}}, \bibinfo{journal}{Phys. Rev. C}
  \textbf{\bibinfo{volume}{76}}, \bibinfo{pages}{014008}.

\bibitem[{\citenamefont{Grigorenko and
  Zhukov}(2007{\natexlab{b}})}]{Grigorenko:2007a}
\bibinfo{author}{\bibnamefont{Grigorenko}, \bibfnamefont{L.~V.}}, and
  \bibinfo{author}{\bibfnamefont{M.~V.} \bibnamefont{Zhukov}},
  \bibinfo{year}{2007}{\natexlab{b}}, \bibinfo{journal}{Phys. Rev. C}
  \textbf{\bibinfo{volume}{76}}, \bibinfo{pages}{014009}.

\bibitem[{\citenamefont{Grigorenko and Zhukov}(2008)}]{Grigorenko:2008}
\bibinfo{author}{\bibnamefont{Grigorenko}, \bibfnamefont{L.~V.}}, and
  \bibinfo{author}{\bibfnamefont{M.~V.} \bibnamefont{Zhukov}},
  \bibinfo{year}{2008}, \bibinfo{journal}{Phys. Rev. C}
  \textbf{\bibinfo{volume}{77}}, \bibinfo{pages}{034611}.

\bibitem[{\citenamefont{Grigorenko and
  Zhukov}(2003{\natexlab{b}})}]{Grigorenko:2003a}
\bibinfo{author}{\bibnamefont{Grigorenko}, \bibfnamefont{L.~V.}}, and
  \bibinfo{author}{\bibfnamefont{M.~V.} \bibnamefont{Zhukov}},
  \bibinfo{year}{2003}{\natexlab{b}}, \bibinfo{journal}{Phys. Rev. C}
  \textbf{\bibinfo{volume}{68}}, \bibinfo{pages}{054005}.

\bibitem[{\citenamefont{Grigorenko}
  \emph{et~al.}(2011)\citenamefont{Grigorenko, Mukha, Scheidenberger, and
  Zhukov}}]{Grigorenko:2011}
\bibinfo{author}{\bibnamefont{Grigorenko}, \bibfnamefont{L.~V.}},
  \bibinfo{author}{\bibfnamefont{I.~G.} \bibnamefont{Mukha}},
  \bibinfo{author}{\bibfnamefont{C.} \bibnamefont{Scheidenberger}}, and
  \bibinfo{author}{\bibfnamefont{M.~V.} \bibnamefont{Zhukov}},
  \bibinfo{year}{2011}, \bibinfo{journal}{Phys. Rev. C}
  \textbf{\bibinfo{volume}{84}}, \bibinfo{pages}{021303(R)}.

\bibitem[{\citenamefont{Grinyer} \emph{et~al.}(2010)\citenamefont{Grinyer,
  Svensson, and Brown}}]{Grinyer:2010}
\bibinfo{author}{\bibnamefont{Grinyer}, \bibfnamefont{G.~F.}},
  \bibinfo{author}{\bibfnamefont{C.~E.} \bibnamefont{Svensson}}, and
  \bibinfo{author}{\bibfnamefont{B.~A.} \bibnamefont{Brown}},
  \bibinfo{year}{2010}, \bibinfo{journal}{Nuclear Instruments and Methods in
  Physics Research Section A: Accelerators, Spectrometers, Detectors and
  Associated Equipment} \textbf{\bibinfo{volume}{622}}(\bibinfo{number}{1}),
  \bibinfo{pages}{236}.

\bibitem[{\citenamefont{Gross} \emph{et~al.}(2000)\citenamefont{Gross, Ginter,
  Shapira, Milner, McConnell, James, Johnson, Mas, Mantica, Auble, Das,
  Blankenship} \emph{et~al.}}]{Gross:2000}
\bibinfo{author}{\bibnamefont{Gross}, \bibfnamefont{C.~J.}},
  \bibinfo{author}{\bibfnamefont{T.~N.} \bibnamefont{Ginter}},
  \bibinfo{author}{\bibfnamefont{D.}~\bibnamefont{Shapira}},
  \bibinfo{author}{\bibfnamefont{W.~T.} \bibnamefont{Milner}},
  \bibinfo{author}{\bibfnamefont{J.~W.} \bibnamefont{McConnell}},
  \bibinfo{author}{\bibfnamefont{A.~N.} \bibnamefont{James}},
  \bibinfo{author}{\bibfnamefont{J.~W.} \bibnamefont{Johnson}},
  \bibinfo{author}{\bibfnamefont{J.}~\bibnamefont{Mas}},
  \bibinfo{author}{\bibfnamefont{P.~F.} \bibnamefont{Mantica}},
  \bibinfo{author}{\bibfnamefont{R.~L.} \bibnamefont{Auble}},
  \bibinfo{author}{\bibfnamefont{J.~J.} \bibnamefont{Das}},
  \bibinfo{author}{\bibfnamefont{J.~L.} \bibnamefont{Blankenship}},
  \emph{et~al.}, \bibinfo{year}{2000}, \bibinfo{journal}{Nuclear Instruments
  and Methods in Physics Research Section A: Accelerators, Spectrometers,
  Detectors and Associated Equipment}
  \textbf{\bibinfo{volume}{450}}(\bibinfo{number}{1}), \bibinfo{pages}{12}.

\bibitem[{\citenamefont{Grotz and Klapdor}(1990)}]{Grotz:1990}
\bibinfo{author}{\bibnamefont{Grotz}, \bibfnamefont{K.}}, and
  \bibinfo{author}{\bibfnamefont{H.~V.} \bibnamefont{Klapdor}},
  \bibinfo{year}{1990}, \emph{\bibinfo{title}{The weak interaction in nuclear,
  particle and astrophysics}} (\bibinfo{publisher}{Adam Hilger}).

\bibitem[{\citenamefont{Grzywacz}(2003)}]{Grzywacz:2003}
\bibinfo{author}{\bibnamefont{Grzywacz}, \bibfnamefont{R.}},
  \bibinfo{year}{2003}, \bibinfo{journal}{Nuclear Instruments and Methods in
  Physics Research Section B: Beam Interactions with Materials and Atoms}
  \textbf{\bibinfo{volume}{204}}, \bibinfo{pages}{649}.

\bibitem[{\citenamefont{{Grzywacz}}
  \emph{et~al.}(1998)\citenamefont{{Grzywacz}, {Beraud}, {Borcea}, {Emsallem},
  {Glogowski}, {Grawe}, {Guillemaud-Mueller}, {Hjorth-Jensen}, {Houry},
  {Lewitowicz}, {Mueller}, {Nowak}} \emph{et~al.}}]{Grzywacz:1998}
\bibinfo{author}{\bibnamefont{{Grzywacz}}, \bibfnamefont{R.}},
  \bibinfo{author}{\bibfnamefont{R.}~\bibnamefont{{Beraud}}},
  \bibinfo{author}{\bibfnamefont{C.}~\bibnamefont{{Borcea}}},
  \bibinfo{author}{\bibfnamefont{A.}~\bibnamefont{{Emsallem}}},
  \bibinfo{author}{\bibfnamefont{M.}~\bibnamefont{{Glogowski}}},
  \bibinfo{author}{\bibfnamefont{H.}~\bibnamefont{{Grawe}}},
  \bibinfo{author}{\bibfnamefont{D.}~\bibnamefont{{Guillemaud-Mueller}}},
  \bibinfo{author}{\bibfnamefont{M.}~\bibnamefont{{Hjorth-Jensen}}},
  \bibinfo{author}{\bibfnamefont{M.}~\bibnamefont{{Houry}}},
  \bibinfo{author}{\bibfnamefont{M.}~\bibnamefont{{Lewitowicz}}},
  \bibinfo{author}{\bibfnamefont{A.~C.} \bibnamefont{{Mueller}}},
  \bibinfo{author}{\bibfnamefont{A.}~\bibnamefont{{Nowak}}}, \emph{et~al.},
  \bibinfo{year}{1998}, \bibinfo{journal}{Phys.Rev.Lett.}
  \textbf{\bibinfo{volume}{81}}, \bibinfo{pages}{766}.

\bibitem[{\citenamefont{Grzywacz} \emph{et~al.}(2005)\citenamefont{Grzywacz,
  Karny, Rykaczewski, Batchelder, Bingham, Fong, Gross, Kr\'olas, Mazzocchi,
  Piechaczek, Tantawy, Winger} \emph{et~al.}}]{Grzywacz:2005}
\bibinfo{author}{\bibnamefont{Grzywacz}, \bibfnamefont{R.}},
  \bibinfo{author}{\bibfnamefont{M.}~\bibnamefont{Karny}},
  \bibinfo{author}{\bibfnamefont{K.}~\bibnamefont{Rykaczewski}},
  \bibinfo{author}{\bibfnamefont{J.}~\bibnamefont{Batchelder}},
  \bibinfo{author}{\bibfnamefont{C.}~\bibnamefont{Bingham}},
  \bibinfo{author}{\bibfnamefont{D.}~\bibnamefont{Fong}},
  \bibinfo{author}{\bibfnamefont{C.}~\bibnamefont{Gross}},
  \bibinfo{author}{\bibfnamefont{W.}~\bibnamefont{Kr\'olas}},
  \bibinfo{author}{\bibfnamefont{C.}~\bibnamefont{Mazzocchi}},
  \bibinfo{author}{\bibfnamefont{A.}~\bibnamefont{Piechaczek}},
  \bibinfo{author}{\bibfnamefont{M.}~\bibnamefont{Tantawy}},
  \bibinfo{author}{\bibfnamefont{J.}~\bibnamefont{Winger}}, \emph{et~al.},
  \bibinfo{year}{2005}, \bibinfo{journal}{Eur.\ Phys.\ J.}
  \textbf{\bibinfo{volume}{A~25}}, \bibinfo{pages}{145}.

\bibitem[{\citenamefont{Gurney and Condon}(1928)}]{Gurney:1928}
\bibinfo{author}{\bibnamefont{Gurney}, \bibfnamefont{R.}}, and
  \bibinfo{author}{\bibfnamefont{E.}~\bibnamefont{Condon}},
  \bibinfo{year}{1928}, \bibinfo{journal}{Nature}
  \textbf{\bibinfo{volume}{122}}, \bibinfo{pages}{439}.

\bibitem[{\citenamefont{Gurvitz and Kalbermann}(1987)}]{Gurvitz:1987}
\bibinfo{author}{\bibnamefont{Gurvitz}, \bibfnamefont{S.~A.}}, and
  \bibinfo{author}{\bibfnamefont{G.}~\bibnamefont{Kalbermann}},
  \bibinfo{year}{1987}, \bibinfo{journal}{Phys. Rev. Lett.}
  \textbf{\bibinfo{volume}{59}}(\bibinfo{number}{3}), \bibinfo{pages}{262}.

\bibitem[{\citenamefont{Hagino}(2001)}]{Hagino:2001}
\bibinfo{author}{\bibnamefont{Hagino}, \bibfnamefont{K.}},
  \bibinfo{year}{2001}, \bibinfo{journal}{Phys.\ Rev.}
  \textbf{\bibinfo{volume}{C~64}}, \bibinfo{pages}{04104(R)}.

\bibitem[{\citenamefont{Hall and Hoffman}(1992)}]{Hall:1992}
\bibinfo{author}{\bibnamefont{Hall}, \bibfnamefont{H.~L.}}, and
  \bibinfo{author}{\bibfnamefont{D.~C.} \bibnamefont{Hoffman}},
  \bibinfo{year}{1992}, \bibinfo{journal}{Annual Review of Nuclear and Particle
  Science} \textbf{\bibinfo{volume}{42}}(\bibinfo{number}{1}),
  \bibinfo{pages}{147}.

\bibitem[{\citenamefont{Hamamoto and Sagawa}(1993)}]{Hamamoto:1993}
\bibinfo{author}{\bibnamefont{Hamamoto}, \bibfnamefont{I.}}, and
  \bibinfo{author}{\bibfnamefont{H.}~\bibnamefont{Sagawa}},
  \bibinfo{year}{1993}, \bibinfo{journal}{Phys. Rev. C}
  \textbf{\bibinfo{volume}{48}}(\bibinfo{number}{3}), \bibinfo{pages}{R960}.

\bibitem[{\citenamefont{Hansen and Jonson}(1988)}]{Hansen:1988}
\bibinfo{author}{\bibnamefont{Hansen}, \bibfnamefont{P.~G.}}, and
  \bibinfo{author}{\bibfnamefont{B.}~\bibnamefont{Jonson}},
  \bibinfo{year}{1988}, in \emph{\bibinfo{booktitle}{Particle Emission from
  Nuclei}}, edited by \bibinfo{editor}{\bibfnamefont{D.}~\bibnamefont{Poenaru}}
  and \bibinfo{editor}{\bibfnamefont{M.}~\bibnamefont{Ivascu}}
  (\bibinfo{publisher}{CRC}), volume \bibinfo{volume}{III}, p.
  \bibinfo{pages}{157}.

\bibitem[{\citenamefont{Hansen} \emph{et~al.}(1990)\citenamefont{Hansen,
  Jonson, and Richter}}]{Hansen:1990}
\bibinfo{author}{\bibnamefont{Hansen}, \bibfnamefont{P.~G.}},
  \bibinfo{author}{\bibfnamefont{B.}~\bibnamefont{Jonson}}, and
  \bibinfo{author}{\bibfnamefont{A.}~\bibnamefont{Richter}},
  \bibinfo{year}{1990}, \bibinfo{journal}{Nuclear Physics A}
  \textbf{\bibinfo{volume}{518}}(\bibinfo{number}{1-2}), \bibinfo{pages}{13}.

\bibitem[{\citenamefont{Hardy and Hagberg}(1988)}]{Hardy:1988}
\bibinfo{author}{\bibnamefont{Hardy}, \bibfnamefont{J.~C.}}, and
  \bibinfo{author}{\bibfnamefont{E.}~\bibnamefont{Hagberg}},
  \bibinfo{year}{1988}, in \emph{\bibinfo{booktitle}{Particle Emission from
  Nuclei}}, edited by \bibinfo{editor}{\bibfnamefont{D.}~\bibnamefont{Poenaru}}
  and \bibinfo{editor}{\bibfnamefont{M.}~\bibnamefont{Ivascu}}
  (\bibinfo{publisher}{CRC}), volume \bibinfo{volume}{III},
  p.~\bibinfo{pages}{99}.

\bibitem[{\citenamefont{Hardy} \emph{et~al.}(1978)\citenamefont{Hardy, Jonson,
  and Hansen}}]{Hardy:1978}
\bibinfo{author}{\bibnamefont{Hardy}, \bibfnamefont{J.~C.}},
  \bibinfo{author}{\bibfnamefont{B.}~\bibnamefont{Jonson}}, and
  \bibinfo{author}{\bibfnamefont{P.~G.} \bibnamefont{Hansen}},
  \bibinfo{year}{1978}, \bibinfo{journal}{Nuclear Physics A}
  \textbf{\bibinfo{volume}{305}}(\bibinfo{number}{1}), \bibinfo{pages}{15}.

\bibitem[{\citenamefont{Harney} \emph{et~al.}(1986)\citenamefont{Harney,
  Richter, and Weidenm\"uller}}]{Harney:1986}
\bibinfo{author}{\bibnamefont{Harney}, \bibfnamefont{H.~L.}},
  \bibinfo{author}{\bibfnamefont{A.}~\bibnamefont{Richter}}, and
  \bibinfo{author}{\bibfnamefont{H.~A.} \bibnamefont{Weidenm\"uller}},
  \bibinfo{year}{1986}, \bibinfo{journal}{Rev. Mod. Phys.}
  \textbf{\bibinfo{volume}{58}}(\bibinfo{number}{3}), \bibinfo{pages}{607}.

\bibitem[{\citenamefont{Heine} \emph{et~al.}(1991)\citenamefont{Heine,
  Faesterman, Gillitzer, Hamolka, Kopf, and Wagner}}]{Heine:1991}
\bibinfo{author}{\bibnamefont{Heine}, \bibfnamefont{A.}},
  \bibinfo{author}{\bibfnamefont{T.}~\bibnamefont{Faesterman}},
  \bibinfo{author}{\bibfnamefont{A.}~\bibnamefont{Gillitzer}},
  \bibinfo{author}{\bibfnamefont{J.}~\bibnamefont{Hamolka}},
  \bibinfo{author}{\bibfnamefont{M.}~\bibnamefont{Kopf}}, and
  \bibinfo{author}{\bibfnamefont{W.}~\bibnamefont{Wagner}},
  \bibinfo{year}{1991}, \bibinfo{journal}{Z.\ Phys.}
  \textbf{\bibinfo{volume}{A~340}}, \bibinfo{pages}{225}.

\bibitem[{\citenamefont{Henzlova} \emph{et~al.}(2008)\citenamefont{Henzlova,
  Schmidt, Ricciardi, Keli\ifmmode~\acute{c}\else \'{c}\fi{}, Henzl,
  Napolitani, Audouin, Benlliure, Boudard, Casarejos, Ducret, Enqvist}
  \emph{et~al.}}]{Henzlova:2008}
\bibinfo{author}{\bibnamefont{Henzlova}, \bibfnamefont{D.}},
  \bibinfo{author}{\bibfnamefont{K.-H.} \bibnamefont{Schmidt}},
  \bibinfo{author}{\bibfnamefont{M.~V.} \bibnamefont{Ricciardi}},
  \bibinfo{author}{\bibfnamefont{A.}~\bibnamefont{Keli\ifmmode~\acute{c}\else
  \'{c}\fi{}}}, \bibinfo{author}{\bibfnamefont{V.}~\bibnamefont{Henzl}},
  \bibinfo{author}{\bibfnamefont{P.}~\bibnamefont{Napolitani}},
  \bibinfo{author}{\bibfnamefont{L.}~\bibnamefont{Audouin}},
  \bibinfo{author}{\bibfnamefont{J.}~\bibnamefont{Benlliure}},
  \bibinfo{author}{\bibfnamefont{A.}~\bibnamefont{Boudard}},
  \bibinfo{author}{\bibfnamefont{E.}~\bibnamefont{Casarejos}},
  \bibinfo{author}{\bibfnamefont{J.~E.} \bibnamefont{Ducret}},
  \bibinfo{author}{\bibfnamefont{T.}~\bibnamefont{Enqvist}}, \emph{et~al.},
  \bibinfo{year}{2008}, \bibinfo{journal}{Phys. Rev. C}
  \textbf{\bibinfo{volume}{78}}(\bibinfo{number}{4}), \bibinfo{pages}{044616}.

\bibitem[{\citenamefont{Hirayama} \emph{et~al.}(2005)\citenamefont{Hirayama,
  Shimoda, Izumi, Hatakeyama, Jackson, Levy, Miyatake, Yagi, and
  Yano}}]{Hirayama:2005}
\bibinfo{author}{\bibnamefont{Hirayama}, \bibfnamefont{Y.}},
  \bibinfo{author}{\bibfnamefont{T.}~\bibnamefont{Shimoda}},
  \bibinfo{author}{\bibfnamefont{H.}~\bibnamefont{Izumi}},
  \bibinfo{author}{\bibfnamefont{A.}~\bibnamefont{Hatakeyama}},
  \bibinfo{author}{\bibfnamefont{K.~P.} \bibnamefont{Jackson}},
  \bibinfo{author}{\bibfnamefont{C.~D.~P.} \bibnamefont{Levy}},
  \bibinfo{author}{\bibfnamefont{H.}~\bibnamefont{Miyatake}},
  \bibinfo{author}{\bibfnamefont{M.}~\bibnamefont{Yagi}}, and
  \bibinfo{author}{\bibfnamefont{H.}~\bibnamefont{Yano}}, \bibinfo{year}{2005},
  \bibinfo{journal}{Physics Letters B}
  \textbf{\bibinfo{volume}{611}}(\bibinfo{number}{3-4}), \bibinfo{pages}{239}.

\bibitem[{\citenamefont{Hofmann}(1995)}]{Hofmann:1995}
\bibinfo{author}{\bibnamefont{Hofmann}, \bibfnamefont{S.}},
  \bibinfo{year}{1995}, \bibinfo{journal}{Radiochim.\ Acta}
  \textbf{\bibinfo{volume}{70/71}}, \bibinfo{pages}{93}.

\bibitem[{\citenamefont{Hofmann}(1996)}]{Hofmann:1996}
\bibinfo{author}{\bibnamefont{Hofmann}, \bibfnamefont{S.}},
  \bibinfo{year}{1996}, \emph{\bibinfo{title}{Nuclear Decay Modes.}}
  (\bibinfo{publisher}{Institute of Physics Publishing}), chapter
  \bibinfo{chapter}{3.Proton Radioactivity}, pp. \bibinfo{pages}{143--203}.

\bibitem[{\citenamefont{Hofmann}(2009{\natexlab{a}})}]{Hofmann:2009a}
\bibinfo{author}{\bibnamefont{Hofmann}, \bibfnamefont{S.}},
  \bibinfo{year}{2009}{\natexlab{a}}, in \emph{\bibinfo{booktitle}{The
  Euroschool Lectures on Physics with Exotic Beams, Vol. III, Lect. Notes Phys.
  764}}, edited by \bibinfo{editor}{\bibfnamefont{J.}~\bibnamefont{Al-Khalili}}
  and \bibinfo{editor}{\bibfnamefont{E.}~\bibnamefont{Roeckl}}
  (\bibinfo{publisher}{Springer, Berlin Heidelberg}), p. \bibinfo{pages}{203}.

\bibitem[{\citenamefont{Hofmann}(2009{\natexlab{b}})}]{Hofmann:2009}
\bibinfo{author}{\bibnamefont{Hofmann}, \bibfnamefont{S.}},
  \bibinfo{year}{2009}{\natexlab{b}}, \bibinfo{journal}{Russian Chemical
  Reviews} \textbf{\bibinfo{volume}{78}}(\bibinfo{number}{12}),
  \bibinfo{pages}{1123}.

\bibitem[{\citenamefont{Hofmann and M\"unzenberg}(2000)}]{Hofmann:2000}
\bibinfo{author}{\bibnamefont{Hofmann}, \bibfnamefont{S.}}, and
  \bibinfo{author}{\bibfnamefont{G.}~\bibnamefont{M\"unzenberg}},
  \bibinfo{year}{2000}, \bibinfo{journal}{Rev. Mod. Phys.}
  \textbf{\bibinfo{volume}{72}}(\bibinfo{number}{3}), \bibinfo{pages}{733}.

\bibitem[{\citenamefont{{Hofmann}} \emph{et~al.}(1982)\citenamefont{{Hofmann},
  {Reisdorf}, {M\"{u}nzenberg}, {Hessberger}, {Schneider}, and
  {Armbruster}}}]{Hofmann:1982}
\bibinfo{author}{\bibnamefont{{Hofmann}}, \bibfnamefont{S.}},
  \bibinfo{author}{\bibfnamefont{W.}~\bibnamefont{{Reisdorf}}},
  \bibinfo{author}{\bibfnamefont{G.}~\bibnamefont{{M\"{u}nzenberg}}},
  \bibinfo{author}{\bibfnamefont{F.~P.} \bibnamefont{{Hessberger}}},
  \bibinfo{author}{\bibfnamefont{J.~R.~H.} \bibnamefont{{Schneider}}}, and
  \bibinfo{author}{\bibfnamefont{P.}~\bibnamefont{{Armbruster}}},
  \bibinfo{year}{1982}, \bibinfo{journal}{Z. Phys.}
  \textbf{\bibinfo{volume}{A305}}, \bibinfo{pages}{111}.

\bibitem[{\citenamefont{Holstein}(1974)}]{Holstein:1974}
\bibinfo{author}{\bibnamefont{Holstein}, \bibfnamefont{B.~R.}},
  \bibinfo{year}{1974}, \bibinfo{journal}{Rev. Mod. Phys.}
  \textbf{\bibinfo{volume}{46}}(\bibinfo{number}{4}), \bibinfo{pages}{789}.

\bibitem[{\citenamefont{Homma} \emph{et~al.}(1996)\citenamefont{Homma, Bender,
  Hirsch, Muto, Klapdor-Kleingrothaus, and Oda}}]{Honma:1996}
\bibinfo{author}{\bibnamefont{Homma}, \bibfnamefont{H.}},
  \bibinfo{author}{\bibfnamefont{E.}~\bibnamefont{Bender}},
  \bibinfo{author}{\bibfnamefont{M.}~\bibnamefont{Hirsch}},
  \bibinfo{author}{\bibfnamefont{K.}~\bibnamefont{Muto}},
  \bibinfo{author}{\bibfnamefont{H.~V.} \bibnamefont{Klapdor-Kleingrothaus}},
  and \bibinfo{author}{\bibfnamefont{T.}~\bibnamefont{Oda}},
  \bibinfo{year}{1996}, \bibinfo{journal}{Phys. Rev. C}
  \textbf{\bibinfo{volume}{54}}(\bibinfo{number}{6}), \bibinfo{pages}{2972}.

\bibitem[{\citenamefont{Hosmer} \emph{et~al.}(2010)\citenamefont{Hosmer,
  Schatz, Aprahamian, Arndt, Clement, Estrade, Farouqi, Kratz, Liddick,
  Lisetskiy, Mantica, M\"{o}ller} \emph{et~al.}}]{Hosmer:2010}
\bibinfo{author}{\bibnamefont{Hosmer}, \bibfnamefont{P.}},
  \bibinfo{author}{\bibfnamefont{H.}~\bibnamefont{Schatz}},
  \bibinfo{author}{\bibfnamefont{A.}~\bibnamefont{Aprahamian}},
  \bibinfo{author}{\bibfnamefont{O.}~\bibnamefont{Arndt}},
  \bibinfo{author}{\bibfnamefont{R.~R.~C.} \bibnamefont{Clement}},
  \bibinfo{author}{\bibfnamefont{A.}~\bibnamefont{Estrade}},
  \bibinfo{author}{\bibfnamefont{K.}~\bibnamefont{Farouqi}},
  \bibinfo{author}{\bibfnamefont{K.-L.} \bibnamefont{Kratz}},
  \bibinfo{author}{\bibfnamefont{S.~N.} \bibnamefont{Liddick}},
  \bibinfo{author}{\bibfnamefont{A.~F.} \bibnamefont{Lisetskiy}},
  \bibinfo{author}{\bibfnamefont{P.~F.} \bibnamefont{Mantica}},
  \bibinfo{author}{\bibfnamefont{P.}~\bibnamefont{M\"{o}ller}}, \emph{et~al.},
  \bibinfo{year}{2010}, \bibinfo{journal}{Phys. Rev. C}
  \textbf{\bibinfo{volume}{82}}(\bibinfo{number}{2}), \bibinfo{pages}{025806}.

\bibitem[{\citenamefont{{Hosmer}} \emph{et~al.}(2005)\citenamefont{{Hosmer},
  {Schatz}, {Aprahamian}, {Arndt}, {Clement}, {Estrade}, {Kratz}, {Liddick},
  {Mantica}, {Mueller}, {Montes}, {Morton}} \emph{et~al.}}]{Hosmer:2005}
\bibinfo{author}{\bibnamefont{{Hosmer}}, \bibfnamefont{P.~T.}},
  \bibinfo{author}{\bibfnamefont{H.}~\bibnamefont{{Schatz}}},
  \bibinfo{author}{\bibfnamefont{A.}~\bibnamefont{{Aprahamian}}},
  \bibinfo{author}{\bibfnamefont{O.}~\bibnamefont{{Arndt}}},
  \bibinfo{author}{\bibfnamefont{R.~R.~C.} \bibnamefont{{Clement}}},
  \bibinfo{author}{\bibfnamefont{A.}~\bibnamefont{{Estrade}}},
  \bibinfo{author}{\bibfnamefont{K.~L.} \bibnamefont{{Kratz}}},
  \bibinfo{author}{\bibfnamefont{S.~N.} \bibnamefont{{Liddick}}},
  \bibinfo{author}{\bibfnamefont{P.~F.} \bibnamefont{{Mantica}}},
  \bibinfo{author}{\bibfnamefont{W.~F.} \bibnamefont{{Mueller}}},
  \bibinfo{author}{\bibfnamefont{F.}~\bibnamefont{{Montes}}},
  \bibinfo{author}{\bibfnamefont{A.~C.} \bibnamefont{{Morton}}}, \emph{et~al.},
  \bibinfo{year}{2005}, \bibinfo{journal}{Phys.Rev.Lett.}
  \textbf{\bibinfo{volume}{94}}, \bibinfo{pages}{112501}.

\bibitem[{\citenamefont{Huang} \emph{et~al.}(1976)\citenamefont{Huang, Auyangi,
  Chen, Crasemann, and Mark}}]{Huang:1976}
\bibinfo{author}{\bibnamefont{Huang}, \bibfnamefont{K.-N.}},
  \bibinfo{author}{\bibfnamefont{M.}~\bibnamefont{Auyangi}},
  \bibinfo{author}{\bibfnamefont{M.}~\bibnamefont{Chen}},
  \bibinfo{author}{\bibfnamefont{B.}~\bibnamefont{Crasemann}}, and
  \bibinfo{author}{\bibfnamefont{H.}~\bibnamefont{Mark}}, \bibinfo{year}{1976},
  \bibinfo{journal}{Atom.\ Data\ Nucl.\ Data\ Tabl.}
  \textbf{\bibinfo{volume}{18}}, \bibinfo{pages}{243}.

\bibitem[{\citenamefont{Hubbard-Nelson}
  \emph{et~al.}(1999)\citenamefont{Hubbard-Nelson, Momayezi, and
  Warburton}}]{Hubbard:1999}
\bibinfo{author}{\bibnamefont{Hubbard-Nelson}, \bibfnamefont{B.}},
  \bibinfo{author}{\bibfnamefont{M.}~\bibnamefont{Momayezi}}, and
  \bibinfo{author}{\bibfnamefont{W.~K.} \bibnamefont{Warburton}},
  \bibinfo{year}{1999}, \bibinfo{journal}{Nuclear Instruments and Methods in
  Physics Research Section A: Accelerators, Spectrometers, Detectors and
  Associated Equipment} \textbf{\bibinfo{volume}{422}}(\bibinfo{number}{1-3}),
  \bibinfo{pages}{411}.

\bibitem[{\citenamefont{Hyldegaard}(2010)}]{Hyldegaard:2010a}
\bibinfo{author}{\bibnamefont{Hyldegaard}, \bibfnamefont{S.}},
  \bibinfo{year}{2010}, \emph{\bibinfo{title}{Beta-decay studies of $^8$Be and
  $^{12}$C}}, Ph.D. thesis, \bibinfo{school}{Aarhus University}.

\bibitem[{\citenamefont{Hyldegaard}
  \emph{et~al.}(2010)\citenamefont{Hyldegaard, Alcorta, Bastin, Borge, Boutami,
  Brandenburg, B\"uscher, Dendooven, Diget, Van~Duppen, Eronen, Fox}
  \emph{et~al.}}]{Hyldegaard:2010}
\bibinfo{author}{\bibnamefont{Hyldegaard}, \bibfnamefont{S.}},
  \bibinfo{author}{\bibfnamefont{M.}~\bibnamefont{Alcorta}},
  \bibinfo{author}{\bibfnamefont{B.}~\bibnamefont{Bastin}},
  \bibinfo{author}{\bibfnamefont{M.~J.~G.} \bibnamefont{Borge}},
  \bibinfo{author}{\bibfnamefont{R.}~\bibnamefont{Boutami}},
  \bibinfo{author}{\bibfnamefont{S.}~\bibnamefont{Brandenburg}},
  \bibinfo{author}{\bibfnamefont{J.}~\bibnamefont{B\"uscher}},
  \bibinfo{author}{\bibfnamefont{P.}~\bibnamefont{Dendooven}},
  \bibinfo{author}{\bibfnamefont{C.~A.} \bibnamefont{Diget}},
  \bibinfo{author}{\bibfnamefont{P.}~\bibnamefont{Van~Duppen}},
  \bibinfo{author}{\bibfnamefont{T.}~\bibnamefont{Eronen}},
  \bibinfo{author}{\bibfnamefont{S.~P.} \bibnamefont{Fox}}, \emph{et~al.},
  \bibinfo{year}{2010}, \bibinfo{journal}{Phys. Rev. C}
  \textbf{\bibinfo{volume}{81}}(\bibinfo{number}{2}), \bibinfo{pages}{024303}.

\bibitem[{\citenamefont{Hyldegaard}
  \emph{et~al.}(2009)\citenamefont{Hyldegaard, Forss\'{e}n, Diget, Alcorta,
  Barker, Bastin, Borge, Boutami, Brandenburg, B\"{u}scher, Dendooven, Duppen}
  \emph{et~al.}}]{Hyldegaard:2009}
\bibinfo{author}{\bibnamefont{Hyldegaard}, \bibfnamefont{S.}},
  \bibinfo{author}{\bibfnamefont{C.}~\bibnamefont{Forss\'{e}n}},
  \bibinfo{author}{\bibfnamefont{C.~A.} \bibnamefont{Diget}},
  \bibinfo{author}{\bibfnamefont{M.}~\bibnamefont{Alcorta}},
  \bibinfo{author}{\bibfnamefont{F.~C.} \bibnamefont{Barker}},
  \bibinfo{author}{\bibfnamefont{B.}~\bibnamefont{Bastin}},
  \bibinfo{author}{\bibfnamefont{M.~J.~G.} \bibnamefont{Borge}},
  \bibinfo{author}{\bibfnamefont{R.}~\bibnamefont{Boutami}},
  \bibinfo{author}{\bibfnamefont{S.}~\bibnamefont{Brandenburg}},
  \bibinfo{author}{\bibfnamefont{J.}~\bibnamefont{B\"{u}scher}},
  \bibinfo{author}{\bibfnamefont{P.}~\bibnamefont{Dendooven}},
  \bibinfo{author}{\bibfnamefont{P.~V.} \bibnamefont{Duppen}}, \emph{et~al.},
  \bibinfo{year}{2009}, \bibinfo{journal}{Physics Letters B}
  \textbf{\bibinfo{volume}{678}}(\bibinfo{number}{5}), \bibinfo{pages}{459}.

\bibitem[{\citenamefont{Ichimura} \emph{et~al.}(2006)\citenamefont{Ichimura,
  Sakai, and Wakasa}}]{Ichimura:2006}
\bibinfo{author}{\bibnamefont{Ichimura}, \bibfnamefont{M.}},
  \bibinfo{author}{\bibfnamefont{H.}~\bibnamefont{Sakai}}, and
  \bibinfo{author}{\bibfnamefont{T.}~\bibnamefont{Wakasa}},
  \bibinfo{year}{2006}, \bibinfo{journal}{Progress in Particle and Nuclear
  Physics} \textbf{\bibinfo{volume}{56}}(\bibinfo{number}{2}),
  \bibinfo{pages}{446}.

\bibitem[{\citenamefont{Irvine} \emph{et~al.}(1997)\citenamefont{Irvine,
  Davids, Woods, Blumenthal, Brown, Conticchio, Davinson, Henderson, Mackenzie,
  Penttil\"a, Seweryniak, and Walters}}]{Irvine:1997}
\bibinfo{author}{\bibnamefont{Irvine}, \bibfnamefont{R.}},
  \bibinfo{author}{\bibfnamefont{C.}~\bibnamefont{Davids}},
  \bibinfo{author}{\bibfnamefont{P.}~\bibnamefont{Woods}},
  \bibinfo{author}{\bibfnamefont{D.}~\bibnamefont{Blumenthal}},
  \bibinfo{author}{\bibfnamefont{L.}~\bibnamefont{Brown}},
  \bibinfo{author}{\bibfnamefont{L.}~\bibnamefont{Conticchio}},
  \bibinfo{author}{\bibfnamefont{T.}~\bibnamefont{Davinson}},
  \bibinfo{author}{\bibfnamefont{D.}~\bibnamefont{Henderson}},
  \bibinfo{author}{\bibfnamefont{J.}~\bibnamefont{Mackenzie}},
  \bibinfo{author}{\bibfnamefont{H.}~\bibnamefont{Penttil\"a}},
  \bibinfo{author}{\bibfnamefont{D.}~\bibnamefont{Seweryniak}}, and
  \bibinfo{author}{\bibfnamefont{W.}~\bibnamefont{Walters}},
  \bibinfo{year}{1997}, \bibinfo{journal}{Phys.\ Rev.}
  \textbf{\bibinfo{volume}{C~55}}, \bibinfo{pages}{R1621}.

\bibitem[{\citenamefont{Jackson} \emph{et~al.}(1970)\citenamefont{Jackson,
  Cardinal, Evans, Jelley, and Cerny}}]{Jackson:1970}
\bibinfo{author}{\bibnamefont{Jackson}, \bibfnamefont{K.}},
  \bibinfo{author}{\bibfnamefont{C.}~\bibnamefont{Cardinal}},
  \bibinfo{author}{\bibfnamefont{H.}~\bibnamefont{Evans}},
  \bibinfo{author}{\bibfnamefont{N.}~\bibnamefont{Jelley}}, and
  \bibinfo{author}{\bibfnamefont{J.}~\bibnamefont{Cerny}},
  \bibinfo{year}{1970}, \bibinfo{journal}{Phys.\ Lett.}
  \textbf{\bibinfo{volume}{33~B}}, \bibinfo{pages}{281}.

\bibitem[{\citenamefont{Janas} \emph{et~al.}(2005)\citenamefont{Janas, Batist,
  Döring, Gierlik, Kirchner, Kurcewicz, Mahmud, Mazzocchi, Plochocki, Roeckl,
  Schmidt, Woods} \emph{et~al.}}]{Janas:2005}
\bibinfo{author}{\bibnamefont{Janas}, \bibfnamefont{Z.}},
  \bibinfo{author}{\bibfnamefont{L.}~\bibnamefont{Batist}},
  \bibinfo{author}{\bibfnamefont{J.}~\bibnamefont{Döring}},
  \bibinfo{author}{\bibfnamefont{M.}~\bibnamefont{Gierlik}},
  \bibinfo{author}{\bibfnamefont{R.}~\bibnamefont{Kirchner}},
  \bibinfo{author}{\bibfnamefont{J.}~\bibnamefont{Kurcewicz}},
  \bibinfo{author}{\bibfnamefont{H.}~\bibnamefont{Mahmud}},
  \bibinfo{author}{\bibfnamefont{C.}~\bibnamefont{Mazzocchi}},
  \bibinfo{author}{\bibfnamefont{A.}~\bibnamefont{Plochocki}},
  \bibinfo{author}{\bibfnamefont{E.}~\bibnamefont{Roeckl}},
  \bibinfo{author}{\bibfnamefont{K.}~\bibnamefont{Schmidt}},
  \bibinfo{author}{\bibfnamefont{P.~J.} \bibnamefont{Woods}}, \emph{et~al.},
  \bibinfo{year}{2005}, \bibinfo{journal}{The European Physical Journal A -
  Hadrons and Nuclei} \textbf{\bibinfo{volume}{23}}, \bibinfo{pages}{401}.

\bibitem[{\citenamefont{J\"{a}necke}(1965)}]{Janecke:1965}
\bibinfo{author}{\bibnamefont{J\"{a}necke}, \bibfnamefont{J.}},
  \bibinfo{year}{1965}, \bibinfo{journal}{Nucl. Phys.}
  \textbf{\bibinfo{volume}{61}}, \bibinfo{pages}{326}.

\bibitem[{\citenamefont{Jenkins} \emph{et~al.}(2000)\citenamefont{Jenkins,
  Muikku, Greenlees, Hauschild, Helariutta, Jones, Julin, Juutinen,
  Kankaanp\"a\"a, Kelsall, Kettunen, Kuusiniemi} \emph{et~al.}}]{Jenkins:2000}
\bibinfo{author}{\bibnamefont{Jenkins}, \bibfnamefont{D.~G.}},
  \bibinfo{author}{\bibfnamefont{M.}~\bibnamefont{Muikku}},
  \bibinfo{author}{\bibfnamefont{P.~T.} \bibnamefont{Greenlees}},
  \bibinfo{author}{\bibfnamefont{K.}~\bibnamefont{Hauschild}},
  \bibinfo{author}{\bibfnamefont{K.}~\bibnamefont{Helariutta}},
  \bibinfo{author}{\bibfnamefont{P.~M.} \bibnamefont{Jones}},
  \bibinfo{author}{\bibfnamefont{R.}~\bibnamefont{Julin}},
  \bibinfo{author}{\bibfnamefont{S.}~\bibnamefont{Juutinen}},
  \bibinfo{author}{\bibfnamefont{H.}~\bibnamefont{Kankaanp\"a\"a}},
  \bibinfo{author}{\bibfnamefont{N.~S.} \bibnamefont{Kelsall}},
  \bibinfo{author}{\bibfnamefont{H.}~\bibnamefont{Kettunen}},
  \bibinfo{author}{\bibfnamefont{P.}~\bibnamefont{Kuusiniemi}}, \emph{et~al.},
  \bibinfo{year}{2000}, \bibinfo{journal}{Phys. Rev. C}
  \textbf{\bibinfo{volume}{62}}(\bibinfo{number}{2}), \bibinfo{pages}{021302}.

\bibitem[{\citenamefont{Jensen} \emph{et~al.}(2004)\citenamefont{Jensen,
  Riisager, Fedorov, and Garrido}}]{Jensen:2004}
\bibinfo{author}{\bibnamefont{Jensen}, \bibfnamefont{A.~S.}},
  \bibinfo{author}{\bibfnamefont{K.}~\bibnamefont{Riisager}},
  \bibinfo{author}{\bibfnamefont{D.~V.} \bibnamefont{Fedorov}}, and
  \bibinfo{author}{\bibfnamefont{E.}~\bibnamefont{Garrido}},
  \bibinfo{year}{2004}, \bibinfo{journal}{Rev. Mod. Phys.}
  \textbf{\bibinfo{volume}{76}}(\bibinfo{number}{1}), \bibinfo{pages}{215}.

\bibitem[{\citenamefont{Jeppesen} \emph{et~al.}(2002)\citenamefont{Jeppesen,
  Bergmann, Borge, Cederk\"{a}ll, Fedoseyev, Fynbo, Hansper, Jonson,
  Markenroth, Mishin, Nilsson, Nyman} \emph{et~al.}}]{Jeppesen:2002}
\bibinfo{author}{\bibnamefont{Jeppesen}, \bibfnamefont{H.}},
  \bibinfo{author}{\bibfnamefont{U.~C.} \bibnamefont{Bergmann}},
  \bibinfo{author}{\bibfnamefont{M.~J.~G.} \bibnamefont{Borge}},
  \bibinfo{author}{\bibfnamefont{J.}~\bibnamefont{Cederk\"{a}ll}},
  \bibinfo{author}{\bibfnamefont{V.~N.} \bibnamefont{Fedoseyev}},
  \bibinfo{author}{\bibfnamefont{H.~O.~U.} \bibnamefont{Fynbo}},
  \bibinfo{author}{\bibfnamefont{V.~Y.} \bibnamefont{Hansper}},
  \bibinfo{author}{\bibfnamefont{B.}~\bibnamefont{Jonson}},
  \bibinfo{author}{\bibfnamefont{K.}~\bibnamefont{Markenroth}},
  \bibinfo{author}{\bibfnamefont{V.~I.} \bibnamefont{Mishin}},
  \bibinfo{author}{\bibfnamefont{T.}~\bibnamefont{Nilsson}},
  \bibinfo{author}{\bibfnamefont{G.}~\bibnamefont{Nyman}}, \emph{et~al.},
  \bibinfo{year}{2002}, \bibinfo{journal}{Nuclear Physics A}
  \textbf{\bibinfo{volume}{709}}(\bibinfo{number}{1-4}), \bibinfo{pages}{119}.

\bibitem[{\citenamefont{Johansson} \emph{et~al.}(2010)\citenamefont{Johansson,
  Aksyutina, Aumann, Boretzky, Borge, Chatillon, Chulkov, Cortina-Gil,
  Pramanik, Emling, Forssen, Fynbo} \emph{et~al.}}]{Johansson:2010}
\bibinfo{author}{\bibnamefont{Johansson}, \bibfnamefont{H.}},
  \bibinfo{author}{\bibfnamefont{Y.}~\bibnamefont{Aksyutina}},
  \bibinfo{author}{\bibfnamefont{T.}~\bibnamefont{Aumann}},
  \bibinfo{author}{\bibfnamefont{K.}~\bibnamefont{Boretzky}},
  \bibinfo{author}{\bibfnamefont{M.}~\bibnamefont{Borge}},
  \bibinfo{author}{\bibfnamefont{A.}~\bibnamefont{Chatillon}},
  \bibinfo{author}{\bibfnamefont{L.}~\bibnamefont{Chulkov}},
  \bibinfo{author}{\bibfnamefont{D.}~\bibnamefont{Cortina-Gil}},
  \bibinfo{author}{\bibfnamefont{U.~D.} \bibnamefont{Pramanik}},
  \bibinfo{author}{\bibfnamefont{H.}~\bibnamefont{Emling}},
  \bibinfo{author}{\bibfnamefont{C.}~\bibnamefont{Forssen}},
  \bibinfo{author}{\bibfnamefont{H.}~\bibnamefont{Fynbo}}, \emph{et~al.},
  \bibinfo{year}{2010}, \bibinfo{journal}{Nucl. Phys.}
  \textbf{\bibinfo{volume}{A847}}, \bibinfo{pages}{66}.

\bibitem[{\citenamefont{Jonson and Nyman}(1996)}]{Jonson:1996}
\bibinfo{author}{\bibnamefont{Jonson}, \bibfnamefont{B.}}, and
  \bibinfo{author}{\bibfnamefont{G.}~\bibnamefont{Nyman}},
  \bibinfo{year}{1996}, in \emph{\bibinfo{booktitle}{Nuclear Decay Modes}},
  edited by \bibinfo{editor}{\bibfnamefont{D.}~\bibnamefont{Poenaru}}
  (\bibinfo{publisher}{IOP publishing}), p. \bibinfo{pages}{102}.

\bibitem[{\citenamefont{Jonson and Riisager}(2001)}]{Jonson:2001}
\bibinfo{author}{\bibnamefont{Jonson}, \bibfnamefont{B.}}, and
  \bibinfo{author}{\bibfnamefont{K.}~\bibnamefont{Riisager}},
  \bibinfo{year}{2001}, \bibinfo{journal}{Nuclear Physics A}
  \textbf{\bibinfo{volume}{693}}(\bibinfo{number}{1-2}), \bibinfo{pages}{77}.

\bibitem[{\citenamefont{{Joss}} \emph{et~al.}(2006)\citenamefont{{Joss},
  {Darby}, {Page}, {Uusitalo}, {Eeckhaudt}, {Grahn}, {Greenlees}, {Jones},
  {Julin}, {Juutinen}, {Ketelhut}, {Leino}} \emph{et~al.}}]{Joss:2006}
\bibinfo{author}{\bibnamefont{{Joss}}, \bibfnamefont{D.~T.}},
  \bibinfo{author}{\bibfnamefont{I.~G.} \bibnamefont{{Darby}}},
  \bibinfo{author}{\bibfnamefont{R.~D.} \bibnamefont{{Page}}},
  \bibinfo{author}{\bibfnamefont{J.}~\bibnamefont{{Uusitalo}}},
  \bibinfo{author}{\bibfnamefont{S.}~\bibnamefont{{Eeckhaudt}}},
  \bibinfo{author}{\bibfnamefont{T.}~\bibnamefont{{Grahn}}},
  \bibinfo{author}{\bibfnamefont{P.~T.} \bibnamefont{{Greenlees}}},
  \bibinfo{author}{\bibfnamefont{P.~M.} \bibnamefont{{Jones}}},
  \bibinfo{author}{\bibfnamefont{R.}~\bibnamefont{{Julin}}},
  \bibinfo{author}{\bibfnamefont{S.}~\bibnamefont{{Juutinen}}},
  \bibinfo{author}{\bibfnamefont{S.}~\bibnamefont{{Ketelhut}}},
  \bibinfo{author}{\bibfnamefont{M.}~\bibnamefont{{Leino}}}, \emph{et~al.},
  \bibinfo{year}{2006}, \bibinfo{journal}{Phys.\ Lett.}
  \textbf{\bibinfo{volume}{B~641}}, \bibinfo{pages}{34}.

\bibitem[{\citenamefont{JPG}(2011)}]{ISOLbeams:2011}
\bibinfo{author}{\bibnamefont{JPG}}, \bibinfo{year}{2011},
  \bibinfo{title}{Focus section: "physics with reaccelerated isol beams"},
  \bibinfo{howpublished}{Journal of Physics G: Nuclear and particle Physics,
  Volume 38, Number 2}.

\bibitem[{\citenamefont{Julin}(2010)}]{Julin:2010}
\bibinfo{author}{\bibnamefont{Julin}, \bibfnamefont{R.}}, \bibinfo{year}{2010},
  \bibinfo{journal}{Nucl.\ Phys.} \textbf{\bibinfo{volume}{A~834}},
  \bibinfo{pages}{15c}.

\bibitem[{\citenamefont{Kadmensky and Sonzogni}(2000)}]{Kadmensky:2000}
\bibinfo{author}{\bibnamefont{Kadmensky}, \bibfnamefont{S.}}, and
  \bibinfo{author}{\bibfnamefont{A.}~\bibnamefont{Sonzogni}},
  \bibinfo{year}{2000}, \bibinfo{journal}{Phys.\ Rev.}
  \textbf{\bibinfo{volume}{C~62}}, \bibinfo{pages}{054601}.

\bibitem[{\citenamefont{Kanada-En'yo}(2010)}]{Kanada-Enyo:2010}
\bibinfo{author}{\bibnamefont{Kanada-En'yo}, \bibfnamefont{Y.}},
  \bibinfo{year}{2010}, \bibinfo{journal}{Phys. Rev. C}
  \textbf{\bibinfo{volume}{81}}(\bibinfo{number}{3}), \bibinfo{pages}{034321}.

\bibitem[{\citenamefont{Kankainen} \emph{et~al.}(2008)\citenamefont{Kankainen,
  Elomaa, Batist, Eliseev, Eronen, Hager, Hakala, Jokinen, Moore, Novikov,
  Penttil\"{a}, Popov} \emph{et~al.}}]{Kankainen:2008}
\bibinfo{author}{\bibnamefont{Kankainen}, \bibfnamefont{A.}},
  \bibinfo{author}{\bibfnamefont{V.-V.} \bibnamefont{Elomaa}},
  \bibinfo{author}{\bibfnamefont{L.}~\bibnamefont{Batist}},
  \bibinfo{author}{\bibfnamefont{S.}~\bibnamefont{Eliseev}},
  \bibinfo{author}{\bibfnamefont{T.}~\bibnamefont{Eronen}},
  \bibinfo{author}{\bibfnamefont{U.}~\bibnamefont{Hager}},
  \bibinfo{author}{\bibfnamefont{J.}~\bibnamefont{Hakala}},
  \bibinfo{author}{\bibfnamefont{A.}~\bibnamefont{Jokinen}},
  \bibinfo{author}{\bibfnamefont{I.~D.} \bibnamefont{Moore}},
  \bibinfo{author}{\bibfnamefont{Y.~N.} \bibnamefont{Novikov}},
  \bibinfo{author}{\bibfnamefont{H.}~\bibnamefont{Penttil\"{a}}},
  \bibinfo{author}{\bibfnamefont{A.}~\bibnamefont{Popov}}, \emph{et~al.},
  \bibinfo{year}{2008}, \bibinfo{journal}{Phys. Rev. Lett.}
  \textbf{\bibinfo{volume}{101}}, \bibinfo{pages}{142503}.

\bibitem[{\citenamefont{Karnaukhov}
  \emph{et~al.}(1963)\citenamefont{Karnaukhov, Ter-Akopian, and
  Subbotin}}]{Karnaukhov:1963}
\bibinfo{author}{\bibnamefont{Karnaukhov}, \bibfnamefont{V.}},
  \bibinfo{author}{\bibfnamefont{G.}~\bibnamefont{Ter-Akopian}}, and
  \bibinfo{author}{\bibfnamefont{V.}~\bibnamefont{Subbotin}},
  \bibinfo{year}{1963}, in \emph{\bibinfo{booktitle}{Proc. Asilomar Conf. on
  Reactions Between Complex Nuclei}}, edited by
  \bibinfo{editor}{\bibfnamefont{e.~a.} \bibnamefont{A.~Ghiorso}}
  (\bibinfo{publisher}{University of California Press}), p.
  \bibinfo{pages}{434}.

\bibitem[{\citenamefont{Karny} \emph{et~al.}(2003)\citenamefont{Karny,
  Grzywacz, Batchelder, Bingham, Gross, Hagino, Hamilton, Janas, Kulp,
  McConnell, Momayezi, Piechaczek} \emph{et~al.}}]{Karny:2003}
\bibinfo{author}{\bibnamefont{Karny}, \bibfnamefont{M.}},
  \bibinfo{author}{\bibfnamefont{R.~K.} \bibnamefont{Grzywacz}},
  \bibinfo{author}{\bibfnamefont{J.~C.} \bibnamefont{Batchelder}},
  \bibinfo{author}{\bibfnamefont{C.~R.} \bibnamefont{Bingham}},
  \bibinfo{author}{\bibfnamefont{C.~J.} \bibnamefont{Gross}},
  \bibinfo{author}{\bibfnamefont{K.}~\bibnamefont{Hagino}},
  \bibinfo{author}{\bibfnamefont{J.~H.} \bibnamefont{Hamilton}},
  \bibinfo{author}{\bibfnamefont{Z.}~\bibnamefont{Janas}},
  \bibinfo{author}{\bibfnamefont{W.~D.} \bibnamefont{Kulp}},
  \bibinfo{author}{\bibfnamefont{J.~W.} \bibnamefont{McConnell}},
  \bibinfo{author}{\bibfnamefont{M.}~\bibnamefont{Momayezi}},
  \bibinfo{author}{\bibfnamefont{A.}~\bibnamefont{Piechaczek}}, \emph{et~al.},
  \bibinfo{year}{2003}, \bibinfo{journal}{Phys. Rev. Lett.}
  \textbf{\bibinfo{volume}{90}}(\bibinfo{number}{1}), \bibinfo{pages}{012502}.

\bibitem[{\citenamefont{Karny} \emph{et~al.}(2008)\citenamefont{Karny,
  Rykaczewski, Grzywacz, Batchelder, Bingham, Goodin, Gross, Hamilton, Korgul,
  Kr{\'o}las, Liddick, Li} \emph{et~al.}}]{Karny:2008}
\bibinfo{author}{\bibnamefont{Karny}, \bibfnamefont{M.}},
  \bibinfo{author}{\bibfnamefont{K.}~\bibnamefont{Rykaczewski}},
  \bibinfo{author}{\bibfnamefont{R.}~\bibnamefont{Grzywacz}},
  \bibinfo{author}{\bibfnamefont{J.}~\bibnamefont{Batchelder}},
  \bibinfo{author}{\bibfnamefont{C.}~\bibnamefont{Bingham}},
  \bibinfo{author}{\bibfnamefont{C.}~\bibnamefont{Goodin}},
  \bibinfo{author}{\bibfnamefont{C.}~\bibnamefont{Gross}},
  \bibinfo{author}{\bibfnamefont{J.}~\bibnamefont{Hamilton}},
  \bibinfo{author}{\bibfnamefont{A.}~\bibnamefont{Korgul}},
  \bibinfo{author}{\bibfnamefont{W.}~\bibnamefont{Kr{\'o}las}},
  \bibinfo{author}{\bibfnamefont{S.}~\bibnamefont{Liddick}},
  \bibinfo{author}{\bibfnamefont{K.}~\bibnamefont{Li}}, \emph{et~al.},
  \bibinfo{year}{2008}, \bibinfo{journal}{Phys.\ Lett.}
  \textbf{\bibinfo{volume}{B~664}}, \bibinfo{pages}{52}.

\bibitem[{\citenamefont{Kaufmann}(1902)}]{Kaufmann:1902}
\bibinfo{author}{\bibnamefont{Kaufmann}, \bibfnamefont{W.}},
  \bibinfo{year}{1902}, \bibinfo{journal}{Phys. Zeitschr.}
  \textbf{\bibinfo{volume}{4}}, \bibinfo{pages}{54}.

\bibitem[{\citenamefont{KeKelis} \emph{et~al.}(1978)\citenamefont{KeKelis,
  Zisman, Scott, Jahn, Vieira, Cerny, and Ajzenberg-Selove}}]{KeKelis:1978}
\bibinfo{author}{\bibnamefont{KeKelis}, \bibfnamefont{G.~J.}},
  \bibinfo{author}{\bibfnamefont{M.~S.} \bibnamefont{Zisman}},
  \bibinfo{author}{\bibfnamefont{D.~K.} \bibnamefont{Scott}},
  \bibinfo{author}{\bibfnamefont{R.}~\bibnamefont{Jahn}},
  \bibinfo{author}{\bibfnamefont{D.~J.} \bibnamefont{Vieira}},
  \bibinfo{author}{\bibfnamefont{J.}~\bibnamefont{Cerny}}, and
  \bibinfo{author}{\bibfnamefont{F.}~\bibnamefont{Ajzenberg-Selove}},
  \bibinfo{year}{1978}, \bibinfo{journal}{Phys. Rev. C}
  \textbf{\bibinfo{volume}{17}}, \bibinfo{pages}{1929}.

\bibitem[{\citenamefont{Keli\'{c}} \emph{et~al.}(2008)\citenamefont{Keli\'{c},
  Ricciardi, and Schmidt}}]{Kelic:2008}
\bibinfo{author}{\bibnamefont{Keli\'{c}}, \bibfnamefont{A.}},
  \bibinfo{author}{\bibfnamefont{M.}~\bibnamefont{Ricciardi}}, and
  \bibinfo{author}{\bibfnamefont{K.-H.} \bibnamefont{Schmidt}},
  \bibinfo{year}{2008}, in \emph{\bibinfo{booktitle}{Proceedings of the Joint
  ICTP-IAEA Advanced Workshop on Model Codes for Spallation Reactions}}, edited
  by \bibinfo{editor}{\bibfnamefont{D.}~\bibnamefont{Filges}},
  \bibinfo{editor}{\bibfnamefont{S.}~\bibnamefont{Leray}},
  \bibinfo{editor}{\bibfnamefont{Y.}~\bibnamefont{Yariv}},
  \bibinfo{editor}{\bibfnamefont{A.}~\bibnamefont{Mengoni}},
  \bibinfo{editor}{\bibfnamefont{A.}~\bibnamefont{Stanculescu}}, and
  \bibinfo{editor}{\bibfnamefont{G.}~\bibnamefont{Mank}}
  (\bibinfo{publisher}{IAEA INDC(NDS)-530, Vienna}), pp. \bibinfo{pages}{181 --
  221}.

\bibitem[{\citenamefont{Kettunen} \emph{et~al.}(2004)\citenamefont{Kettunen,
  Enqvist, Grahn, Greenlees, Jones, Julin, Juutinen, Keenan, Kuusiniemi, Leino,
  A.-P.Lepp\"anen, Nieminen} \emph{et~al.}}]{Kettunen:2004}
\bibinfo{author}{\bibnamefont{Kettunen}, \bibfnamefont{H.}},
  \bibinfo{author}{\bibfnamefont{T.}~\bibnamefont{Enqvist}},
  \bibinfo{author}{\bibfnamefont{T.}~\bibnamefont{Grahn}},
  \bibinfo{author}{\bibfnamefont{P.}~\bibnamefont{Greenlees}},
  \bibinfo{author}{\bibfnamefont{P.}~\bibnamefont{Jones}},
  \bibinfo{author}{\bibfnamefont{R.}~\bibnamefont{Julin}},
  \bibinfo{author}{\bibfnamefont{S.}~\bibnamefont{Juutinen}},
  \bibinfo{author}{\bibfnamefont{A.}~\bibnamefont{Keenan}},
  \bibinfo{author}{\bibfnamefont{P.}~\bibnamefont{Kuusiniemi}},
  \bibinfo{author}{\bibfnamefont{M.}~\bibnamefont{Leino}},
  \bibinfo{author}{\bibnamefont{A.-P.Lepp\"anen}},
  \bibinfo{author}{\bibfnamefont{P.}~\bibnamefont{Nieminen}}, \emph{et~al.},
  \bibinfo{year}{2004}, \bibinfo{journal}{Phys.\ Rev.}
  \textbf{\bibinfo{volume}{C~69}}, \bibinfo{pages}{054323}.

\bibitem[{\citenamefont{Kettunen} \emph{et~al.}(2001)\citenamefont{Kettunen,
  Greenlees, Helariutta, Jones, Julin, Juutinen, Kuusiniemi, Leino, Muikku,
  Nieminen, and Uusitalo}}]{Kettunen:2001}
\bibinfo{author}{\bibnamefont{Kettunen}, \bibfnamefont{H.}},
  \bibinfo{author}{\bibfnamefont{P.}~\bibnamefont{Greenlees}},
  \bibinfo{author}{\bibfnamefont{K.}~\bibnamefont{Helariutta}},
  \bibinfo{author}{\bibfnamefont{P.}~\bibnamefont{Jones}},
  \bibinfo{author}{\bibfnamefont{R.}~\bibnamefont{Julin}},
  \bibinfo{author}{\bibfnamefont{S.}~\bibnamefont{Juutinen}},
  \bibinfo{author}{\bibfnamefont{P.}~\bibnamefont{Kuusiniemi}},
  \bibinfo{author}{\bibfnamefont{M.}~\bibnamefont{Leino}},
  \bibinfo{author}{\bibfnamefont{M.}~\bibnamefont{Muikku}},
  \bibinfo{author}{\bibfnamefont{P.}~\bibnamefont{Nieminen}}, and
  \bibinfo{author}{\bibfnamefont{J.}~\bibnamefont{Uusitalo}},
  \bibinfo{year}{2001}, \bibinfo{journal}{Acta.\ Phys.\ Pol.}
  \textbf{\bibinfo{volume}{B~32}}, \bibinfo{pages}{989}.

\bibitem[{\citenamefont{Kirsebom}(2010)}]{Kirsebom:2010}
\bibinfo{author}{\bibnamefont{Kirsebom}, \bibfnamefont{O.~S.}},
  \bibinfo{year}{2010}, \emph{\bibinfo{title}{$^8$B Neutrinos and $^{12}$C
  Resonances}}, Ph.D. thesis, \bibinfo{school}{Aarhus University}.

\bibitem[{\citenamefont{Kirsebom} \emph{et~al.}(2009)\citenamefont{Kirsebom,
  Alcorta, Borge, Cubero, Diget, Dominguez-Reyes, Fraile, Fulton, Fynbo,
  Galaviz, Garcia, Hyldegaard} \emph{et~al.}}]{Kirsebom:2009}
\bibinfo{author}{\bibnamefont{Kirsebom}, \bibfnamefont{O.~S.}},
  \bibinfo{author}{\bibfnamefont{M.}~\bibnamefont{Alcorta}},
  \bibinfo{author}{\bibfnamefont{M.~J.~G.} \bibnamefont{Borge}},
  \bibinfo{author}{\bibfnamefont{M.}~\bibnamefont{Cubero}},
  \bibinfo{author}{\bibfnamefont{C.~A.} \bibnamefont{Diget}},
  \bibinfo{author}{\bibfnamefont{R.}~\bibnamefont{Dominguez-Reyes}},
  \bibinfo{author}{\bibfnamefont{L.}~\bibnamefont{Fraile}},
  \bibinfo{author}{\bibfnamefont{B.~R.} \bibnamefont{Fulton}},
  \bibinfo{author}{\bibfnamefont{H.}~\bibnamefont{Fynbo}},
  \bibinfo{author}{\bibfnamefont{D.}~\bibnamefont{Galaviz}},
  \bibinfo{author}{\bibfnamefont{G.}~\bibnamefont{Garcia}},
  \bibinfo{author}{\bibfnamefont{S.}~\bibnamefont{Hyldegaard}}, \emph{et~al.},
  \bibinfo{year}{2009}, \bibinfo{journal}{Physics Letters B}
  \textbf{\bibinfo{volume}{680}}(\bibinfo{number}{1}), \bibinfo{pages}{44}.

\bibitem[{\citenamefont{{Klepper}} \emph{et~al.}(1982)\citenamefont{{Klepper},
  {Batsch}, {Hofmann}, {Kirchner}, {Kurcewicz}, {Reisdorf}, {Roeckl},
  {Schardt}, and {Nyman}}}]{Klepper:1982}
\bibinfo{author}{\bibnamefont{{Klepper}}, \bibfnamefont{O.}},
  \bibinfo{author}{\bibfnamefont{T.}~\bibnamefont{{Batsch}}},
  \bibinfo{author}{\bibfnamefont{S.}~\bibnamefont{{Hofmann}}},
  \bibinfo{author}{\bibfnamefont{R.}~\bibnamefont{{Kirchner}}},
  \bibinfo{author}{\bibfnamefont{W.}~\bibnamefont{{Kurcewicz}}},
  \bibinfo{author}{\bibfnamefont{W.}~\bibnamefont{{Reisdorf}}},
  \bibinfo{author}{\bibfnamefont{E.}~\bibnamefont{{Roeckl}}},
  \bibinfo{author}{\bibfnamefont{D.}~\bibnamefont{{Schardt}}}, and
  \bibinfo{author}{\bibfnamefont{G.}~\bibnamefont{{Nyman}}},
  \bibinfo{year}{1982}, \bibinfo{journal}{Z. Phys.}
  \textbf{\bibinfo{volume}{A305}}, \bibinfo{pages}{125}.

\bibitem[{\citenamefont{Kluge} \emph{et~al.}(2008)\citenamefont{Kluge,
  Herfurth, Kester, Nörtershäuser, and Quint}}]{Kluge:2008}
\bibinfo{author}{\bibnamefont{Kluge}, \bibfnamefont{H.-J.}},
  \bibinfo{author}{\bibfnamefont{F.}~\bibnamefont{Herfurth}},
  \bibinfo{author}{\bibfnamefont{O.}~\bibnamefont{Kester}},
  \bibinfo{author}{\bibfnamefont{W.}~\bibnamefont{Nörtershäuser}}, and
  \bibinfo{author}{\bibfnamefont{W.}~\bibnamefont{Quint}},
  \bibinfo{year}{2008}, \bibinfo{journal}{Nuclear Instruments and Methods in
  Physics Research Section B: Beam Interactions with Materials and Atoms}
  \textbf{\bibinfo{volume}{266}}(\bibinfo{number}{19-20}),
  \bibinfo{pages}{4542}.

\bibitem[{\citenamefont{Korgul} \emph{et~al.}(2008)\citenamefont{Korgul,
  Rykaczewski, Gross, Grzywacz, Liddick, Mazzocchi, Batchelder, Bingham, Darby,
  Goodin, Hamilton, Hwang} \emph{et~al.}}]{Korgul:2008}
\bibinfo{author}{\bibnamefont{Korgul}, \bibfnamefont{A.}},
  \bibinfo{author}{\bibfnamefont{K.~P.} \bibnamefont{Rykaczewski}},
  \bibinfo{author}{\bibfnamefont{C.~J.} \bibnamefont{Gross}},
  \bibinfo{author}{\bibfnamefont{R.~K.} \bibnamefont{Grzywacz}},
  \bibinfo{author}{\bibfnamefont{S.~N.} \bibnamefont{Liddick}},
  \bibinfo{author}{\bibfnamefont{C.}~\bibnamefont{Mazzocchi}},
  \bibinfo{author}{\bibfnamefont{J.~C.} \bibnamefont{Batchelder}},
  \bibinfo{author}{\bibfnamefont{C.~R.} \bibnamefont{Bingham}},
  \bibinfo{author}{\bibfnamefont{I.~G.} \bibnamefont{Darby}},
  \bibinfo{author}{\bibfnamefont{C.}~\bibnamefont{Goodin}},
  \bibinfo{author}{\bibfnamefont{J.~H.} \bibnamefont{Hamilton}},
  \bibinfo{author}{\bibfnamefont{J.~K.} \bibnamefont{Hwang}}, \emph{et~al.},
  \bibinfo{year}{2008}, \bibinfo{journal}{Phys. Rev. C}
  \textbf{\bibinfo{volume}{77}}(\bibinfo{number}{3}), \bibinfo{pages}{034301}.

\bibitem[{\citenamefont{Korsheninnikov}
  \emph{et~al.}(2003)\citenamefont{Korsheninnikov, Nikolskii, Kuzmin, Ozawa,
  Morimoto, Tokanai, Kanungo, Tanihata, Timofeyuk, Golovkov, Fomichev, Rodin}
  \emph{et~al.}}]{Korsheninnikov:2003}
\bibinfo{author}{\bibnamefont{Korsheninnikov}, \bibfnamefont{A.}},
  \bibinfo{author}{\bibfnamefont{E.}~\bibnamefont{Nikolskii}},
  \bibinfo{author}{\bibfnamefont{E.}~\bibnamefont{Kuzmin}},
  \bibinfo{author}{\bibfnamefont{A.}~\bibnamefont{Ozawa}},
  \bibinfo{author}{\bibfnamefont{K.}~\bibnamefont{Morimoto}},
  \bibinfo{author}{\bibfnamefont{F.}~\bibnamefont{Tokanai}},
  \bibinfo{author}{\bibfnamefont{R.}~\bibnamefont{Kanungo}},
  \bibinfo{author}{\bibfnamefont{I.}~\bibnamefont{Tanihata}},
  \bibinfo{author}{\bibfnamefont{N.}~\bibnamefont{Timofeyuk}},
  \bibinfo{author}{\bibfnamefont{M.}~\bibnamefont{Golovkov}},
  \bibinfo{author}{\bibfnamefont{A.}~\bibnamefont{Fomichev}},
  \bibinfo{author}{\bibfnamefont{A.}~\bibnamefont{Rodin}}, \emph{et~al.},
  \bibinfo{year}{2003}, \bibinfo{journal}{Phys. Rev. Lett.}
  \textbf{\bibinfo{volume}{90}}, \bibinfo{pages}{082501}.

\bibitem[{\citenamefont{Korsheninnikov}
  \emph{et~al.}(1994)\citenamefont{Korsheninnikov, Yoshida, Aleksandrov, Aoi,
  Doki, Inabe, Fujimaki, Kobayashi, Kumagai, Moon, Nikolsky, Obuti}
  \emph{et~al.}}]{Korsheninnikov:1994}
\bibinfo{author}{\bibnamefont{Korsheninnikov}, \bibfnamefont{A.}},
  \bibinfo{author}{\bibfnamefont{K.}~\bibnamefont{Yoshida}},
  \bibinfo{author}{\bibfnamefont{D.}~\bibnamefont{Aleksandrov}},
  \bibinfo{author}{\bibfnamefont{N.}~\bibnamefont{Aoi}},
  \bibinfo{author}{\bibfnamefont{Y.}~\bibnamefont{Doki}},
  \bibinfo{author}{\bibfnamefont{N.}~\bibnamefont{Inabe}},
  \bibinfo{author}{\bibfnamefont{M.}~\bibnamefont{Fujimaki}},
  \bibinfo{author}{\bibfnamefont{T.}~\bibnamefont{Kobayashi}},
  \bibinfo{author}{\bibfnamefont{H.}~\bibnamefont{Kumagai}},
  \bibinfo{author}{\bibfnamefont{C.-B.} \bibnamefont{Moon}},
  \bibinfo{author}{\bibfnamefont{E.}~\bibnamefont{Nikolsky}},
  \bibinfo{author}{\bibfnamefont{M.}~\bibnamefont{Obuti}}, \emph{et~al.},
  \bibinfo{year}{1994}, \bibinfo{journal}{Phys. Lett.}
  \textbf{\bibinfo{volume}{B326}}, \bibinfo{pages}{31}.

\bibitem[{\citenamefont{Kox} \emph{et~al.}(1987)\citenamefont{Kox, Gamp,
  Perrin, Arvieux, Bertholet, Bruandet, Buenerd, Cherkaoui, Cole, El-Masri,
  Longequeue, Menet} \emph{et~al.}}]{Kox:1987}
\bibinfo{author}{\bibnamefont{Kox}, \bibfnamefont{S.}},
  \bibinfo{author}{\bibfnamefont{A.}~\bibnamefont{Gamp}},
  \bibinfo{author}{\bibfnamefont{C.}~\bibnamefont{Perrin}},
  \bibinfo{author}{\bibfnamefont{J.}~\bibnamefont{Arvieux}},
  \bibinfo{author}{\bibfnamefont{R.}~\bibnamefont{Bertholet}},
  \bibinfo{author}{\bibfnamefont{J.~F.} \bibnamefont{Bruandet}},
  \bibinfo{author}{\bibfnamefont{M.}~\bibnamefont{Buenerd}},
  \bibinfo{author}{\bibfnamefont{R.}~\bibnamefont{Cherkaoui}},
  \bibinfo{author}{\bibfnamefont{A.~J.} \bibnamefont{Cole}},
  \bibinfo{author}{\bibfnamefont{Y.}~\bibnamefont{El-Masri}},
  \bibinfo{author}{\bibfnamefont{N.}~\bibnamefont{Longequeue}},
  \bibinfo{author}{\bibfnamefont{J.}~\bibnamefont{Menet}}, \emph{et~al.},
  \bibinfo{year}{1987}, \bibinfo{journal}{Phys. Rev. C}
  \textbf{\bibinfo{volume}{35}}(\bibinfo{number}{5}), \bibinfo{pages}{1678}.

\bibitem[{\citenamefont{Kratz} \emph{et~al.}(2007)\citenamefont{Kratz, Farouqi,
  and Pfeiffer}}]{Kratz:2007}
\bibinfo{author}{\bibnamefont{Kratz}, \bibfnamefont{K.-L.}},
  \bibinfo{author}{\bibfnamefont{K.}~\bibnamefont{Farouqi}}, and
  \bibinfo{author}{\bibfnamefont{B.}~\bibnamefont{Pfeiffer}},
  \bibinfo{year}{2007}, \bibinfo{journal}{Progress in Particle and Nuclear
  Physics} \textbf{\bibinfo{volume}{59}}(\bibinfo{number}{1}),
  \bibinfo{pages}{147}.

\bibitem[{\citenamefont{Kruppa} \emph{et~al.}(2000)\citenamefont{Kruppa,
  Barmore, Nazarewicz, and Vertse}}]{Kruppa:2000}
\bibinfo{author}{\bibnamefont{Kruppa}, \bibfnamefont{A.}},
  \bibinfo{author}{\bibfnamefont{B.}~\bibnamefont{Barmore}},
  \bibinfo{author}{\bibfnamefont{W.}~\bibnamefont{Nazarewicz}}, and
  \bibinfo{author}{\bibfnamefont{T.}~\bibnamefont{Vertse}},
  \bibinfo{year}{2000}, \bibinfo{journal}{Phys.\ Rev.\ Lett.}
  \textbf{\bibinfo{volume}{84}}, \bibinfo{pages}{4549}.

\bibitem[{\citenamefont{Kruppa and Nazarewicz}(2004)}]{Kruppa:2004}
\bibinfo{author}{\bibnamefont{Kruppa}, \bibfnamefont{A.~T.}}, and
  \bibinfo{author}{\bibfnamefont{W.}~\bibnamefont{Nazarewicz}},
  \bibinfo{year}{2004}, \bibinfo{journal}{Phys. Rev. C}
  \textbf{\bibinfo{volume}{69}}(\bibinfo{number}{5}), \bibinfo{pages}{054311}.

\bibitem[{\citenamefont{Kryger} \emph{et~al.}(1995)\citenamefont{Kryger,
  Azhari, Hellstr\"{o}m, Kelley, Kubo, Pfaff, Ramakrishnan, Sherrill,
  Thoennessen, Yokoyama, Charity, Dempsey} \emph{et~al.}}]{Kryger:1995}
\bibinfo{author}{\bibnamefont{Kryger}, \bibfnamefont{R.~A.}},
  \bibinfo{author}{\bibfnamefont{A.}~\bibnamefont{Azhari}},
  \bibinfo{author}{\bibfnamefont{M.}~\bibnamefont{Hellstr\"{o}m}},
  \bibinfo{author}{\bibfnamefont{J.~H.} \bibnamefont{Kelley}},
  \bibinfo{author}{\bibfnamefont{T.}~\bibnamefont{Kubo}},
  \bibinfo{author}{\bibfnamefont{R.}~\bibnamefont{Pfaff}},
  \bibinfo{author}{\bibfnamefont{E.}~\bibnamefont{Ramakrishnan}},
  \bibinfo{author}{\bibfnamefont{B.~M.} \bibnamefont{Sherrill}},
  \bibinfo{author}{\bibfnamefont{M.}~\bibnamefont{Thoennessen}},
  \bibinfo{author}{\bibfnamefont{S.}~\bibnamefont{Yokoyama}},
  \bibinfo{author}{\bibfnamefont{R.~J.} \bibnamefont{Charity}},
  \bibinfo{author}{\bibfnamefont{J.}~\bibnamefont{Dempsey}}, \emph{et~al.},
  \bibinfo{year}{1995}, \bibinfo{journal}{Phys. Rev. Lett.}
  \textbf{\bibinfo{volume}{74}}, \bibinfo{pages}{860}.

\bibitem[{\citenamefont{Kudryavtsev}
  \emph{et~al.}(2008)\citenamefont{Kudryavtsev, Cocolios, Gentens, Ivanov,
  Huyse, Pauwels, Sawicka, Sonoda, den Bergh, and Duppen}}]{Kudryavtsev:2008}
\bibinfo{author}{\bibnamefont{Kudryavtsev}, \bibfnamefont{Y.}},
  \bibinfo{author}{\bibfnamefont{T.}~\bibnamefont{Cocolios}},
  \bibinfo{author}{\bibfnamefont{J.}~\bibnamefont{Gentens}},
  \bibinfo{author}{\bibfnamefont{O.}~\bibnamefont{Ivanov}},
  \bibinfo{author}{\bibfnamefont{M.}~\bibnamefont{Huyse}},
  \bibinfo{author}{\bibfnamefont{D.}~\bibnamefont{Pauwels}},
  \bibinfo{author}{\bibfnamefont{M.}~\bibnamefont{Sawicka}},
  \bibinfo{author}{\bibfnamefont{T.}~\bibnamefont{Sonoda}},
  \bibinfo{author}{\bibfnamefont{P.~V.} \bibnamefont{den Bergh}}, and
  \bibinfo{author}{\bibfnamefont{P.~V.} \bibnamefont{Duppen}},
  \bibinfo{year}{2008}, \bibinfo{journal}{Nuclear Instruments and Methods in
  Physics Research Section B: Beam Interactions with Materials and Atoms}
  \textbf{\bibinfo{volume}{266}}(\bibinfo{number}{19-20}),
  \bibinfo{pages}{4368}.

\bibitem[{\citenamefont{Kumar} \emph{et~al.}(2009)\citenamefont{Kumar, Molina,
  Pietri, Casarejos, Algora, Benlliure, Doornenbal, Gerl, G\'{o}rska,
  Kojouharov, Podolyák, Prokopowicz} \emph{et~al.}}]{Kumar:2009}
\bibinfo{author}{\bibnamefont{Kumar}, \bibfnamefont{R.}},
  \bibinfo{author}{\bibfnamefont{F.}~\bibnamefont{Molina}},
  \bibinfo{author}{\bibfnamefont{S.}~\bibnamefont{Pietri}},
  \bibinfo{author}{\bibfnamefont{E.}~\bibnamefont{Casarejos}},
  \bibinfo{author}{\bibfnamefont{A.}~\bibnamefont{Algora}},
  \bibinfo{author}{\bibfnamefont{J.}~\bibnamefont{Benlliure}},
  \bibinfo{author}{\bibfnamefont{P.}~\bibnamefont{Doornenbal}},
  \bibinfo{author}{\bibfnamefont{J.}~\bibnamefont{Gerl}},
  \bibinfo{author}{\bibfnamefont{M.}~\bibnamefont{G\'{o}rska}},
  \bibinfo{author}{\bibfnamefont{I.}~\bibnamefont{Kojouharov}},
  \bibinfo{author}{\bibfnamefont{Z.}~\bibnamefont{Podolyák}},
  \bibinfo{author}{\bibfnamefont{W.}~\bibnamefont{Prokopowicz}}, \emph{et~al.},
  \bibinfo{year}{2009}, \bibinfo{journal}{Nuclear Instruments and Methods in
  Physics Research Section A: Accelerators, Spectrometers, Detectors and
  Associated Equipment} \textbf{\bibinfo{volume}{598}}(\bibinfo{number}{3}),
  \bibinfo{pages}{754}.

\bibitem[{\citenamefont{Kurtukian-Nieto}
  \emph{et~al.}(2008)\citenamefont{Kurtukian-Nieto, Benlliure, and
  Schmidt}}]{KurtukianNieto:2008}
\bibinfo{author}{\bibnamefont{Kurtukian-Nieto}, \bibfnamefont{T.}},
  \bibinfo{author}{\bibfnamefont{J.}~\bibnamefont{Benlliure}}, and
  \bibinfo{author}{\bibfnamefont{K.-H.} \bibnamefont{Schmidt}},
  \bibinfo{year}{2008}, \bibinfo{journal}{Nuclear Instruments and Methods in
  Physics Research Section A: Accelerators, Spectrometers, Detectors and
  Associated Equipment} \textbf{\bibinfo{volume}{589}}(\bibinfo{number}{3}),
  \bibinfo{pages}{472}.

\bibitem[{\citenamefont{Kuznetsov and Skobelev}(1999)}]{Kuznetsov:1999}
\bibinfo{author}{\bibnamefont{Kuznetsov}, \bibfnamefont{V.}}, and
  \bibinfo{author}{\bibfnamefont{N.}~\bibnamefont{Skobelev}},
  \bibinfo{year}{1999}, \bibinfo{journal}{Physics of Particles and Nuclei}
  \textbf{\bibinfo{volume}{30}}(\bibinfo{number}{6}), \bibinfo{pages}{666}.

\bibitem[{\citenamefont{Lagergren} \emph{et~al.}(2006)\citenamefont{Lagergren,
  Joss, Wyss, Cederwall, Barton, Eeckhaudt, Grahn, Greenlees, Hadinia, Jones,
  Julin, Juutinen} \emph{et~al.}}]{Lagergren:2006}
\bibinfo{author}{\bibnamefont{Lagergren}, \bibfnamefont{K.}},
  \bibinfo{author}{\bibfnamefont{D.}~\bibnamefont{Joss}},
  \bibinfo{author}{\bibfnamefont{R.}~\bibnamefont{Wyss}},
  \bibinfo{author}{\bibfnamefont{B.}~\bibnamefont{Cederwall}},
  \bibinfo{author}{\bibfnamefont{C.}~\bibnamefont{Barton}},
  \bibinfo{author}{\bibfnamefont{S.}~\bibnamefont{Eeckhaudt}},
  \bibinfo{author}{\bibfnamefont{T.}~\bibnamefont{Grahn}},
  \bibinfo{author}{\bibfnamefont{P.}~\bibnamefont{Greenlees}},
  \bibinfo{author}{\bibfnamefont{B.}~\bibnamefont{Hadinia}},
  \bibinfo{author}{\bibfnamefont{P.}~\bibnamefont{Jones}},
  \bibinfo{author}{\bibfnamefont{R.}~\bibnamefont{Julin}},
  \bibinfo{author}{\bibfnamefont{S.}~\bibnamefont{Juutinen}}, \emph{et~al.},
  \bibinfo{year}{2006}, \bibinfo{journal}{Phys.\ Rev.}
  \textbf{\bibinfo{volume}{C~74}}, \bibinfo{pages}{024316}.

\bibitem[{\citenamefont{Lane and Thomas}(1958)}]{Lane:1958}
\bibinfo{author}{\bibnamefont{Lane}, \bibfnamefont{A.~M.}}, and
  \bibinfo{author}{\bibfnamefont{R.~G.} \bibnamefont{Thomas}},
  \bibinfo{year}{1958}, \bibinfo{journal}{Rev. Mod. Phys.}
  \textbf{\bibinfo{volume}{30}}, \bibinfo{pages}{257}.

\bibitem[{\citenamefont{Langanke and
  Mart\'{\i}nez-Pinedo}(2000)}]{Langanke:2000}
\bibinfo{author}{\bibnamefont{Langanke}, \bibfnamefont{K.}}, and
  \bibinfo{author}{\bibfnamefont{G.}~\bibnamefont{Mart\'{\i}nez-Pinedo}},
  \bibinfo{year}{2000}, \bibinfo{journal}{Nuclear Physics A}
  \textbf{\bibinfo{volume}{673}}(\bibinfo{number}{1-4}), \bibinfo{pages}{481}.

\bibitem[{\citenamefont{Langanke and
  Mart\'\i{}nez-Pinedo}(2003)}]{Langanke:2003}
\bibinfo{author}{\bibnamefont{Langanke}, \bibfnamefont{K.}}, and
  \bibinfo{author}{\bibfnamefont{G.}~\bibnamefont{Mart\'\i{}nez-Pinedo}},
  \bibinfo{year}{2003}, \bibinfo{journal}{Rev. Mod. Phys.}
  \textbf{\bibinfo{volume}{75}}(\bibinfo{number}{3}), \bibinfo{pages}{819}.

\bibitem[{\citenamefont{{Larsson}} \emph{et~al.}(1983)\citenamefont{{Larsson},
  {Batsch}, {Kirchner}, {Klepper}, {Kurcewicz}, {Roeckl}, {Schardt}, {Feix},
  {Nyman}, and {Tidemand-Petersson}}}]{Larsson:1983}
\bibinfo{author}{\bibnamefont{{Larsson}}, \bibfnamefont{P.~O.}},
  \bibinfo{author}{\bibfnamefont{T.}~\bibnamefont{{Batsch}}},
  \bibinfo{author}{\bibfnamefont{R.}~\bibnamefont{{Kirchner}}},
  \bibinfo{author}{\bibfnamefont{O.}~\bibnamefont{{Klepper}}},
  \bibinfo{author}{\bibfnamefont{W.}~\bibnamefont{{Kurcewicz}}},
  \bibinfo{author}{\bibfnamefont{E.}~\bibnamefont{{Roeckl}}},
  \bibinfo{author}{\bibfnamefont{D.}~\bibnamefont{{Schardt}}},
  \bibinfo{author}{\bibfnamefont{W.~F.} \bibnamefont{{Feix}}},
  \bibinfo{author}{\bibfnamefont{G.}~\bibnamefont{{Nyman}}}, and
  \bibinfo{author}{\bibfnamefont{P.}~\bibnamefont{{Tidemand-Petersson}}},
  \bibinfo{year}{1983}, \bibinfo{journal}{Z.Phys.}
  \textbf{\bibinfo{volume}{A314}}, \bibinfo{pages}{9}.

\bibitem[{\citenamefont{Le~Gentil} \emph{et~al.}(2008)\citenamefont{Le~Gentil,
  Aumann, Bacri, Benlliure, Bianchin, B\"ohmer, Boudard, Brzychczyk, Casarejos,
  Combet, Donadille, Ducret} \emph{et~al.}}]{Gentil:2008}
\bibinfo{author}{\bibnamefont{Le~Gentil}, \bibfnamefont{E.}},
  \bibinfo{author}{\bibfnamefont{T.}~\bibnamefont{Aumann}},
  \bibinfo{author}{\bibfnamefont{C.~O.} \bibnamefont{Bacri}},
  \bibinfo{author}{\bibfnamefont{J.}~\bibnamefont{Benlliure}},
  \bibinfo{author}{\bibfnamefont{S.}~\bibnamefont{Bianchin}},
  \bibinfo{author}{\bibfnamefont{M.}~\bibnamefont{B\"ohmer}},
  \bibinfo{author}{\bibfnamefont{A.}~\bibnamefont{Boudard}},
  \bibinfo{author}{\bibfnamefont{J.}~\bibnamefont{Brzychczyk}},
  \bibinfo{author}{\bibfnamefont{E.}~\bibnamefont{Casarejos}},
  \bibinfo{author}{\bibfnamefont{M.}~\bibnamefont{Combet}},
  \bibinfo{author}{\bibfnamefont{L.}~\bibnamefont{Donadille}},
  \bibinfo{author}{\bibfnamefont{J.~E.} \bibnamefont{Ducret}}, \emph{et~al.},
  \bibinfo{year}{2008}, \bibinfo{journal}{Phys. Rev. Lett.}
  \textbf{\bibinfo{volume}{100}}(\bibinfo{number}{2}), \bibinfo{pages}{022701}.

\bibitem[{\citenamefont{Lecesne}(2008)}]{Lecesne:2008}
\bibinfo{author}{\bibnamefont{Lecesne}, \bibfnamefont{N.}},
  \bibinfo{year}{2008}, \bibinfo{journal}{Nuclear Instruments and Methods in
  Physics Research Section B: Beam Interactions with Materials and Atoms}
  \textbf{\bibinfo{volume}{266}}(\bibinfo{number}{19-20}),
  \bibinfo{pages}{4338}.

\bibitem[{\citenamefont{Lee} \emph{et~al.}(2003)\citenamefont{Lee, Deleplanque,
  and Vetter}}]{Lee:2003}
\bibinfo{author}{\bibnamefont{Lee}, \bibfnamefont{I.~Y.}},
  \bibinfo{author}{\bibfnamefont{M.~A.} \bibnamefont{Deleplanque}}, and
  \bibinfo{author}{\bibfnamefont{K.}~\bibnamefont{Vetter}},
  \bibinfo{year}{2003}, \bibinfo{journal}{Reports on Progress in Physics}
  \textbf{\bibinfo{volume}{66}}(\bibinfo{number}{7}), \bibinfo{pages}{1095}.

\bibitem[{\citenamefont{Leino} \emph{et~al.}(1995)\citenamefont{Leino, Äystö,
  Enqvist, Heikkinen, Jokinen, Nurmia, Ostrowski, Trzaska, Uusitalo, Eskola,
  Armbruster, and Ninov}}]{Leino:1995}
\bibinfo{author}{\bibnamefont{Leino}, \bibfnamefont{M.}},
  \bibinfo{author}{\bibfnamefont{J.}~\bibnamefont{Äystö}},
  \bibinfo{author}{\bibfnamefont{T.}~\bibnamefont{Enqvist}},
  \bibinfo{author}{\bibfnamefont{P.}~\bibnamefont{Heikkinen}},
  \bibinfo{author}{\bibfnamefont{A.}~\bibnamefont{Jokinen}},
  \bibinfo{author}{\bibfnamefont{M.}~\bibnamefont{Nurmia}},
  \bibinfo{author}{\bibfnamefont{A.}~\bibnamefont{Ostrowski}},
  \bibinfo{author}{\bibfnamefont{W.~H.} \bibnamefont{Trzaska}},
  \bibinfo{author}{\bibfnamefont{J.}~\bibnamefont{Uusitalo}},
  \bibinfo{author}{\bibfnamefont{K.}~\bibnamefont{Eskola}},
  \bibinfo{author}{\bibfnamefont{P.}~\bibnamefont{Armbruster}}, and
  \bibinfo{author}{\bibfnamefont{V.}~\bibnamefont{Ninov}},
  \bibinfo{year}{1995}, \bibinfo{journal}{Nuclear Instruments and Methods in
  Physics Research Section B: Beam Interactions with Materials and Atoms}
  \textbf{\bibinfo{volume}{99}}(\bibinfo{number}{1-4}), \bibinfo{pages}{653}.

\bibitem[{\citenamefont{Lewitowicz}(2008)}]{Lewitowicz:2008}
\bibinfo{author}{\bibnamefont{Lewitowicz}, \bibfnamefont{M.}},
  \bibinfo{year}{2008}, \bibinfo{journal}{Nuclear Physics A}
  \textbf{\bibinfo{volume}{805}}(\bibinfo{number}{1-4}), \bibinfo{pages}{519c}.

\bibitem[{\citenamefont{Lewitowicz}
  \emph{et~al.}(1994)\citenamefont{Lewitowicz, Anne, Auger, Bazin, Borcea,
  Borrel, Corre, Dörfler, Fomichov, Grzywacz, Guillemaud-Mueller, Hue}
  \emph{et~al.}}]{Lewitowicz:1994}
\bibinfo{author}{\bibnamefont{Lewitowicz}, \bibfnamefont{M.}},
  \bibinfo{author}{\bibfnamefont{R.}~\bibnamefont{Anne}},
  \bibinfo{author}{\bibfnamefont{G.}~\bibnamefont{Auger}},
  \bibinfo{author}{\bibfnamefont{D.}~\bibnamefont{Bazin}},
  \bibinfo{author}{\bibfnamefont{C.}~\bibnamefont{Borcea}},
  \bibinfo{author}{\bibfnamefont{V.}~\bibnamefont{Borrel}},
  \bibinfo{author}{\bibfnamefont{J.}~\bibnamefont{Corre}},
  \bibinfo{author}{\bibfnamefont{T.}~\bibnamefont{Dörfler}},
  \bibinfo{author}{\bibfnamefont{A.}~\bibnamefont{Fomichov}},
  \bibinfo{author}{\bibfnamefont{R.}~\bibnamefont{Grzywacz}},
  \bibinfo{author}{\bibfnamefont{D.}~\bibnamefont{Guillemaud-Mueller}},
  \bibinfo{author}{\bibfnamefont{R.}~\bibnamefont{Hue}}, \emph{et~al.},
  \bibinfo{year}{1994}, \bibinfo{journal}{Physics Letters B}
  \textbf{\bibinfo{volume}{332}}(\bibinfo{number}{1-2}), \bibinfo{pages}{20}.

\bibitem[{\citenamefont{Li} \emph{et~al.}(2009)\citenamefont{Li, Lou, Ye, Hua,
  Jiang, Li, Zhang, Zheng, Ge, Kong, Lv, Li} \emph{et~al.}}]{Li:2009}
\bibinfo{author}{\bibnamefont{Li}, \bibfnamefont{Z.~H.}},
  \bibinfo{author}{\bibfnamefont{J.~L.} \bibnamefont{Lou}},
  \bibinfo{author}{\bibfnamefont{Y.~L.} \bibnamefont{Ye}},
  \bibinfo{author}{\bibfnamefont{H.}~\bibnamefont{Hua}},
  \bibinfo{author}{\bibfnamefont{D.~X.} \bibnamefont{Jiang}},
  \bibinfo{author}{\bibfnamefont{X.~Q.} \bibnamefont{Li}},
  \bibinfo{author}{\bibfnamefont{S.~Q.} \bibnamefont{Zhang}},
  \bibinfo{author}{\bibfnamefont{T.}~\bibnamefont{Zheng}},
  \bibinfo{author}{\bibfnamefont{Y.~C.} \bibnamefont{Ge}},
  \bibinfo{author}{\bibfnamefont{Z.}~\bibnamefont{Kong}},
  \bibinfo{author}{\bibfnamefont{L.~H.} \bibnamefont{Lv}},
  \bibinfo{author}{\bibfnamefont{C.}~\bibnamefont{Li}}, \emph{et~al.},
  \bibinfo{year}{2009}, \bibinfo{journal}{Phys. Rev. C}
  \textbf{\bibinfo{volume}{80}}(\bibinfo{number}{5}), \bibinfo{pages}{054315}.

\bibitem[{\citenamefont{Liddick} \emph{et~al.}(2006)\citenamefont{Liddick,
  Grzywacz, Mazzocchi, Page, Rykaczewski, Batchelder, Bingham, Darby, Drafta,
  Goodin, Gross, Hamilton} \emph{et~al.}}]{Liddick:2006}
\bibinfo{author}{\bibnamefont{Liddick}, \bibfnamefont{S.~N.}},
  \bibinfo{author}{\bibfnamefont{R.}~\bibnamefont{Grzywacz}},
  \bibinfo{author}{\bibfnamefont{C.}~\bibnamefont{Mazzocchi}},
  \bibinfo{author}{\bibfnamefont{R.~D.} \bibnamefont{Page}},
  \bibinfo{author}{\bibfnamefont{K.~P.} \bibnamefont{Rykaczewski}},
  \bibinfo{author}{\bibfnamefont{J.~C.} \bibnamefont{Batchelder}},
  \bibinfo{author}{\bibfnamefont{C.~R.} \bibnamefont{Bingham}},
  \bibinfo{author}{\bibfnamefont{I.~G.} \bibnamefont{Darby}},
  \bibinfo{author}{\bibfnamefont{G.}~\bibnamefont{Drafta}},
  \bibinfo{author}{\bibfnamefont{C.}~\bibnamefont{Goodin}},
  \bibinfo{author}{\bibfnamefont{C.~J.} \bibnamefont{Gross}},
  \bibinfo{author}{\bibfnamefont{J.~H.} \bibnamefont{Hamilton}}, \emph{et~al.},
  \bibinfo{year}{2006}, \bibinfo{journal}{Phys. Rev. Lett.}
  \textbf{\bibinfo{volume}{97}}(\bibinfo{number}{8}), \bibinfo{pages}{082501}.

\bibitem[{\citenamefont{Litvinov and Bosch}(2011)}]{Litvinov:2011}
\bibinfo{author}{\bibnamefont{Litvinov}, \bibfnamefont{Y.~A.}}, and
  \bibinfo{author}{\bibfnamefont{F.}~\bibnamefont{Bosch}},
  \bibinfo{year}{2011}, \bibinfo{journal}{Reports on Progress in Physics}
  \textbf{\bibinfo{volume}{74}}(\bibinfo{number}{1}), \bibinfo{pages}{016301}.

\bibitem[{\citenamefont{Litvinova} \emph{et~al.}(2008)\citenamefont{Litvinova,
  Ring, and Tselyaev}}]{Litvinova:2008}
\bibinfo{author}{\bibnamefont{Litvinova}, \bibfnamefont{E.}},
  \bibinfo{author}{\bibfnamefont{P.}~\bibnamefont{Ring}}, and
  \bibinfo{author}{\bibfnamefont{V.}~\bibnamefont{Tselyaev}},
  \bibinfo{year}{2008}, \bibinfo{journal}{Phys. Rev. C}
  \textbf{\bibinfo{volume}{78}}, \bibinfo{pages}{014312}.

\bibitem[{\citenamefont{Livingston}
  \emph{et~al.}(1993)\citenamefont{Livingston, Woods, Davinson, Davis, Hofmann,
  James, Page, Selin, and Shotter}}]{Livingston:1993}
\bibinfo{author}{\bibnamefont{Livingston}, \bibfnamefont{K.}},
  \bibinfo{author}{\bibfnamefont{P.}~\bibnamefont{Woods}},
  \bibinfo{author}{\bibfnamefont{T.}~\bibnamefont{Davinson}},
  \bibinfo{author}{\bibfnamefont{N.}~\bibnamefont{Davis}},
  \bibinfo{author}{\bibfnamefont{S.}~\bibnamefont{Hofmann}},
  \bibinfo{author}{\bibfnamefont{A.}~\bibnamefont{James}},
  \bibinfo{author}{\bibfnamefont{R.}~\bibnamefont{Page}},
  \bibinfo{author}{\bibfnamefont{P.}~\bibnamefont{Selin}}, and
  \bibinfo{author}{\bibfnamefont{A.}~\bibnamefont{Shotter}},
  \bibinfo{year}{1993}, \bibinfo{journal}{Phys.\ Lett.}
  \textbf{\bibinfo{volume}{B~312}}, \bibinfo{pages}{46}.

\bibitem[{\citenamefont{Lunney} \emph{et~al.}(2003)\citenamefont{Lunney,
  Pearson, and Thibault}}]{Lunney:2003}
\bibinfo{author}{\bibnamefont{Lunney}, \bibfnamefont{D.}},
  \bibinfo{author}{\bibfnamefont{J.~M.} \bibnamefont{Pearson}}, and
  \bibinfo{author}{\bibfnamefont{C.}~\bibnamefont{Thibault}},
  \bibinfo{year}{2003}, \bibinfo{journal}{Rev. Mod. Phys.}
  \textbf{\bibinfo{volume}{75}}(\bibinfo{number}{3}), \bibinfo{pages}{1021}.

\bibitem[{\citenamefont{Macfarlane and Siivola}(1965)}]{Macfarlane:1965}
\bibinfo{author}{\bibnamefont{Macfarlane}, \bibfnamefont{R.~D.}}, and
  \bibinfo{author}{\bibfnamefont{A.}~\bibnamefont{Siivola}},
  \bibinfo{year}{1965}, \bibinfo{journal}{Phys. Rev. Lett.}
  \textbf{\bibinfo{volume}{14}}(\bibinfo{number}{4}), \bibinfo{pages}{114}.

\bibitem[{\citenamefont{Madurga} \emph{et~al.}(2009)\citenamefont{Madurga,
  Borge, Alcorta, Fraile, Fynbo, Jonson, Kirsebom, Nilsson, Nyman, Perea,
  Riisager, Tengblad} \emph{et~al.}}]{Madurga:2009}
\bibinfo{author}{\bibnamefont{Madurga}, \bibfnamefont{M.}},
  \bibinfo{author}{\bibfnamefont{M.~J.~G.} \bibnamefont{Borge}},
  \bibinfo{author}{\bibfnamefont{M.}~\bibnamefont{Alcorta}},
  \bibinfo{author}{\bibfnamefont{L.~M.} \bibnamefont{Fraile}},
  \bibinfo{author}{\bibfnamefont{H.~O.~U.} \bibnamefont{Fynbo}},
  \bibinfo{author}{\bibfnamefont{B.}~\bibnamefont{Jonson}},
  \bibinfo{author}{\bibfnamefont{O.}~\bibnamefont{Kirsebom}},
  \bibinfo{author}{\bibfnamefont{T.}~\bibnamefont{Nilsson}},
  \bibinfo{author}{\bibfnamefont{G.}~\bibnamefont{Nyman}},
  \bibinfo{author}{\bibfnamefont{A.}~\bibnamefont{Perea}},
  \bibinfo{author}{\bibfnamefont{K.}~\bibnamefont{Riisager}},
  \bibinfo{author}{\bibfnamefont{O.}~\bibnamefont{Tengblad}}, \emph{et~al.},
  \bibinfo{year}{2009}, \bibinfo{journal}{The European Physical Journal A -
  Hadrons and Nuclei} \textbf{\bibinfo{volume}{42}}, \bibinfo{pages}{415}.

\bibitem[{\citenamefont{Madurga} \emph{et~al.}(2008)\citenamefont{Madurga,
  Borge, Angelique, Bao, Bergmann, Buta, Cederk\"{a}ll, Diget, Fraile, Fynbo,
  Jeppesen, Jonson} \emph{et~al.}}]{Madurga:2008}
\bibinfo{author}{\bibnamefont{Madurga}, \bibfnamefont{M.}},
  \bibinfo{author}{\bibfnamefont{M.~J.~G.} \bibnamefont{Borge}},
  \bibinfo{author}{\bibfnamefont{J.~C.} \bibnamefont{Angelique}},
  \bibinfo{author}{\bibfnamefont{L.}~\bibnamefont{Bao}},
  \bibinfo{author}{\bibfnamefont{U.}~\bibnamefont{Bergmann}},
  \bibinfo{author}{\bibfnamefont{A.}~\bibnamefont{Buta}},
  \bibinfo{author}{\bibfnamefont{J.}~\bibnamefont{Cederk\"{a}ll}},
  \bibinfo{author}{\bibfnamefont{C.~A.} \bibnamefont{Diget}},
  \bibinfo{author}{\bibfnamefont{L.~M.} \bibnamefont{Fraile}},
  \bibinfo{author}{\bibfnamefont{H.~O.~U.} \bibnamefont{Fynbo}},
  \bibinfo{author}{\bibfnamefont{H.~B.} \bibnamefont{Jeppesen}},
  \bibinfo{author}{\bibfnamefont{B.}~\bibnamefont{Jonson}}, \emph{et~al.},
  \bibinfo{year}{2008}, \bibinfo{journal}{Nuclear Physics A}
  \textbf{\bibinfo{volume}{810}}(\bibinfo{number}{1-4}), \bibinfo{pages}{1}.

\bibitem[{\citenamefont{Magill} \emph{et~al.}(2009)\citenamefont{Magill,
  Pfennig, and Galy}}]{Magill:2009}
\bibinfo{author}{\bibnamefont{Magill}, \bibfnamefont{J.}},
  \bibinfo{author}{\bibfnamefont{G.}~\bibnamefont{Pfennig}}, and
  \bibinfo{author}{\bibfnamefont{J.}~\bibnamefont{Galy}}, \bibinfo{year}{2009},
  \emph{\bibinfo{title}{Karlsruher Nuklidkarte}} (\bibinfo{publisher}{European
  Communities 2006-2009}), \bibinfo{edition}{7th edition, revised printing
  2009} edition.

\bibitem[{\citenamefont{Maglione and Ferreira}(2000)}]{Maglione:2000}
\bibinfo{author}{\bibnamefont{Maglione}, \bibfnamefont{E.}}, and
  \bibinfo{author}{\bibfnamefont{L.}~\bibnamefont{Ferreira}},
  \bibinfo{year}{2000}, \bibinfo{journal}{Phys.\ Reev.}
  \textbf{\bibinfo{volume}{C~61}}, \bibinfo{pages}{047307}.

\bibitem[{\citenamefont{Maglione and Ferreira}(2002)}]{Maglione:2002}
\bibinfo{author}{\bibnamefont{Maglione}, \bibfnamefont{E.}}, and
  \bibinfo{author}{\bibfnamefont{L.}~\bibnamefont{Ferreira}},
  \bibinfo{year}{2002}, \bibinfo{journal}{Eur.\ Phys.\ J.}
  \textbf{\bibinfo{volume}{A~15}}, \bibinfo{pages}{89}.

\bibitem[{\citenamefont{Maglione} \emph{et~al.}(1998)\citenamefont{Maglione,
  Ferreira, and Liotta}}]{Maglione:1998}
\bibinfo{author}{\bibnamefont{Maglione}, \bibfnamefont{E.}},
  \bibinfo{author}{\bibfnamefont{L.}~\bibnamefont{Ferreira}}, and
  \bibinfo{author}{\bibfnamefont{R.}~\bibnamefont{Liotta}},
  \bibinfo{year}{1998}, \bibinfo{journal}{Phys.\ Rev.\ Lett.}
  \textbf{\bibinfo{volume}{81}}, \bibinfo{pages}{538}.

\bibitem[{\citenamefont{Maglione} \emph{et~al.}(1999)\citenamefont{Maglione,
  Ferreira, and Liotta}}]{Maglione:1999}
\bibinfo{author}{\bibnamefont{Maglione}, \bibfnamefont{E.}},
  \bibinfo{author}{\bibfnamefont{S.}~\bibnamefont{Ferreira}}, and
  \bibinfo{author}{\bibfnamefont{R.}~\bibnamefont{Liotta}},
  \bibinfo{year}{1999}, \bibinfo{journal}{Phys.\ Rev.}
  \textbf{\bibinfo{volume}{C~59}}, \bibinfo{pages}{R589}.

\bibitem[{\citenamefont{Mahmud} \emph{et~al.}(2002)\citenamefont{Mahmud,
  Davids, Woods, Davidson, Heinz, Ressler, Schmidt, Seweryniak, Shergur,
  Sonsogni, and Walters}}]{Mahmud:2002}
\bibinfo{author}{\bibnamefont{Mahmud}, \bibfnamefont{H.}},
  \bibinfo{author}{\bibfnamefont{C.}~\bibnamefont{Davids}},
  \bibinfo{author}{\bibfnamefont{P.}~\bibnamefont{Woods}},
  \bibinfo{author}{\bibfnamefont{T.}~\bibnamefont{Davidson}},
  \bibinfo{author}{\bibfnamefont{A.}~\bibnamefont{Heinz}},
  \bibinfo{author}{\bibfnamefont{J.}~\bibnamefont{Ressler}},
  \bibinfo{author}{\bibfnamefont{K.}~\bibnamefont{Schmidt}},
  \bibinfo{author}{\bibfnamefont{D.}~\bibnamefont{Seweryniak}},
  \bibinfo{author}{\bibfnamefont{J.}~\bibnamefont{Shergur}},
  \bibinfo{author}{\bibfnamefont{A.}~\bibnamefont{Sonsogni}}, and
  \bibinfo{author}{\bibfnamefont{W.}~\bibnamefont{Walters}},
  \bibinfo{year}{2002}, \bibinfo{journal}{Eur.\ Phys.\ J.}
  \textbf{\bibinfo{volume}{A~15}}, \bibinfo{pages}{85}.

\bibitem[{\citenamefont{Mahmud} \emph{et~al.}(2001)\citenamefont{Mahmud,
  Davids, Woods, Davinson, Heinz, Poli, Ressler, Schmidt, Seweryniak, Smith,
  Sonzodni, Uusitalo} \emph{et~al.}}]{Mahmud:2001}
\bibinfo{author}{\bibnamefont{Mahmud}, \bibfnamefont{H.}},
  \bibinfo{author}{\bibfnamefont{C.}~\bibnamefont{Davids}},
  \bibinfo{author}{\bibfnamefont{P.}~\bibnamefont{Woods}},
  \bibinfo{author}{\bibfnamefont{T.}~\bibnamefont{Davinson}},
  \bibinfo{author}{\bibfnamefont{A.}~\bibnamefont{Heinz}},
  \bibinfo{author}{\bibfnamefont{G.}~\bibnamefont{Poli}},
  \bibinfo{author}{\bibfnamefont{J.}~\bibnamefont{Ressler}},
  \bibinfo{author}{\bibfnamefont{K.}~\bibnamefont{Schmidt}},
  \bibinfo{author}{\bibfnamefont{D.}~\bibnamefont{Seweryniak}},
  \bibinfo{author}{\bibfnamefont{M.}~\bibnamefont{Smith}},
  \bibinfo{author}{\bibfnamefont{A.}~\bibnamefont{Sonzodni}},
  \bibinfo{author}{\bibfnamefont{J.}~\bibnamefont{Uusitalo}}, \emph{et~al.},
  \bibinfo{year}{2001}, \bibinfo{journal}{Phys.\ Rev.}
  \textbf{\bibinfo{volume}{C~64}}, \bibinfo{pages}{031303}.

\bibitem[{\citenamefont{Mart\'{\i}nez-Pinedo}
  \emph{et~al.}(2007)\citenamefont{Mart\'{\i}nez-Pinedo, Mocelj, Zinner, Kelic,
  Langanke, Panov, Pfeiffer, Rauscher, Schmidt, and
  Thielemann}}]{Martinez:2007}
\bibinfo{author}{\bibnamefont{Mart\'{\i}nez-Pinedo}, \bibfnamefont{G.}},
  \bibinfo{author}{\bibfnamefont{D.}~\bibnamefont{Mocelj}},
  \bibinfo{author}{\bibfnamefont{N.~T.} \bibnamefont{Zinner}},
  \bibinfo{author}{\bibfnamefont{A.}~\bibnamefont{Kelic}},
  \bibinfo{author}{\bibfnamefont{K.}~\bibnamefont{Langanke}},
  \bibinfo{author}{\bibfnamefont{I.}~\bibnamefont{Panov}},
  \bibinfo{author}{\bibfnamefont{B.}~\bibnamefont{Pfeiffer}},
  \bibinfo{author}{\bibfnamefont{T.}~\bibnamefont{Rauscher}},
  \bibinfo{author}{\bibfnamefont{K.-H.} \bibnamefont{Schmidt}}, and
  \bibinfo{author}{\bibfnamefont{F.-K.} \bibnamefont{Thielemann}},
  \bibinfo{year}{2007}, \bibinfo{journal}{Progress in Particle and Nuclear
  Physics} \textbf{\bibinfo{volume}{59}}(\bibinfo{number}{1}),
  \bibinfo{pages}{199}.

\bibitem[{\citenamefont{Mattoon} \emph{et~al.}(2009)\citenamefont{Mattoon,
  Sarazin, Andreoiu, Andreyev, Austin, Ball, Chakrawarthy, Cross, Cunningham,
  Daoud, Garrett, Grinyer} \emph{et~al.}}]{Mattoon:2009}
\bibinfo{author}{\bibnamefont{Mattoon}, \bibfnamefont{C.~M.}},
  \bibinfo{author}{\bibfnamefont{F.}~\bibnamefont{Sarazin}},
  \bibinfo{author}{\bibfnamefont{C.}~\bibnamefont{Andreoiu}},
  \bibinfo{author}{\bibfnamefont{A.~N.} \bibnamefont{Andreyev}},
  \bibinfo{author}{\bibfnamefont{R.~A.~E.} \bibnamefont{Austin}},
  \bibinfo{author}{\bibfnamefont{G.~C.} \bibnamefont{Ball}},
  \bibinfo{author}{\bibfnamefont{R.~S.} \bibnamefont{Chakrawarthy}},
  \bibinfo{author}{\bibfnamefont{D.}~\bibnamefont{Cross}},
  \bibinfo{author}{\bibfnamefont{E.~S.} \bibnamefont{Cunningham}},
  \bibinfo{author}{\bibfnamefont{J.}~\bibnamefont{Daoud}},
  \bibinfo{author}{\bibfnamefont{P.~E.} \bibnamefont{Garrett}},
  \bibinfo{author}{\bibfnamefont{G.~F.} \bibnamefont{Grinyer}}, \emph{et~al.},
  \bibinfo{year}{2009}, \bibinfo{journal}{Phys. Rev. C}
  \textbf{\bibinfo{volume}{80}}(\bibinfo{number}{3}), \bibinfo{pages}{034318}.

\bibitem[{\citenamefont{Mazzocchi} \emph{et~al.}(2007)\citenamefont{Mazzocchi,
  Grzywacz, Liddick, Rykaczewski, Schatz, Batchelder, Bingham, Gross, Hamilton,
  Hwang, Ilyushkin, Korgul} \emph{et~al.}}]{Mazzocchi:2007}
\bibinfo{author}{\bibnamefont{Mazzocchi}, \bibfnamefont{C.}},
  \bibinfo{author}{\bibfnamefont{R.}~\bibnamefont{Grzywacz}},
  \bibinfo{author}{\bibfnamefont{S.}~\bibnamefont{Liddick}},
  \bibinfo{author}{\bibfnamefont{K.}~\bibnamefont{Rykaczewski}},
  \bibinfo{author}{\bibfnamefont{H.}~\bibnamefont{Schatz}},
  \bibinfo{author}{\bibfnamefont{J.}~\bibnamefont{Batchelder}},
  \bibinfo{author}{\bibfnamefont{C.}~\bibnamefont{Bingham}},
  \bibinfo{author}{\bibfnamefont{C.}~\bibnamefont{Gross}},
  \bibinfo{author}{\bibfnamefont{J.}~\bibnamefont{Hamilton}},
  \bibinfo{author}{\bibfnamefont{J.}~\bibnamefont{Hwang}},
  \bibinfo{author}{\bibfnamefont{S.}~\bibnamefont{Ilyushkin}},
  \bibinfo{author}{\bibfnamefont{A.}~\bibnamefont{Korgul}}, \emph{et~al.},
  \bibinfo{year}{2007}, \bibinfo{journal}{Phys. Rev. Lett.}
  \textbf{\bibinfo{volume}{98}}, \bibinfo{pages}{212501}.

\bibitem[{\citenamefont{Mehren} \emph{et~al.}(1996)\citenamefont{Mehren,
  Pfeiffer, Schoedder, Kratz, Huhta, Dendooven, Honkanen, Lhersonneau, Oinonen,
  Parmonen, Penttil\"a, Popov} \emph{et~al.}}]{Mehren:1996}
\bibinfo{author}{\bibnamefont{Mehren}, \bibfnamefont{T.}},
  \bibinfo{author}{\bibfnamefont{B.}~\bibnamefont{Pfeiffer}},
  \bibinfo{author}{\bibfnamefont{S.}~\bibnamefont{Schoedder}},
  \bibinfo{author}{\bibfnamefont{K.-L.} \bibnamefont{Kratz}},
  \bibinfo{author}{\bibfnamefont{M.}~\bibnamefont{Huhta}},
  \bibinfo{author}{\bibfnamefont{P.}~\bibnamefont{Dendooven}},
  \bibinfo{author}{\bibfnamefont{A.}~\bibnamefont{Honkanen}},
  \bibinfo{author}{\bibfnamefont{G.}~\bibnamefont{Lhersonneau}},
  \bibinfo{author}{\bibfnamefont{M.}~\bibnamefont{Oinonen}},
  \bibinfo{author}{\bibfnamefont{J.-M.} \bibnamefont{Parmonen}},
  \bibinfo{author}{\bibfnamefont{H.}~\bibnamefont{Penttil\"a}},
  \bibinfo{author}{\bibfnamefont{A.}~\bibnamefont{Popov}}, \emph{et~al.},
  \bibinfo{year}{1996}, \bibinfo{journal}{Phys. Rev. Lett.}
  \textbf{\bibinfo{volume}{77}}(\bibinfo{number}{3}), \bibinfo{pages}{458}.

\bibitem[{\citenamefont{Mianowski} \emph{et~al.}(2010)\citenamefont{Mianowski,
  Czyrkowski, D\c{a}browski, Dominik, Janas, Miernik, Pf\"{u}tzner, Fomichev,
  Golovkov, Grigorenko, Krupko, Sidorchuk} \emph{et~al.}}]{Mianowski:2010}
\bibinfo{author}{\bibnamefont{Mianowski}, \bibfnamefont{S.}},
  \bibinfo{author}{\bibfnamefont{H.}~\bibnamefont{Czyrkowski}},
  \bibinfo{author}{\bibfnamefont{R.}~\bibnamefont{D\c{a}browski}},
  \bibinfo{author}{\bibfnamefont{W.}~\bibnamefont{Dominik}},
  \bibinfo{author}{\bibfnamefont{Z.}~\bibnamefont{Janas}},
  \bibinfo{author}{\bibfnamefont{K.}~\bibnamefont{Miernik}},
  \bibinfo{author}{\bibfnamefont{M.}~\bibnamefont{Pf\"{u}tzner}},
  \bibinfo{author}{\bibfnamefont{A.~S.} \bibnamefont{Fomichev}},
  \bibinfo{author}{\bibfnamefont{M.~S.} \bibnamefont{Golovkov}},
  \bibinfo{author}{\bibfnamefont{L.~V.} \bibnamefont{Grigorenko}},
  \bibinfo{author}{\bibfnamefont{S.~A.} \bibnamefont{Krupko}},
  \bibinfo{author}{\bibfnamefont{S.~I.} \bibnamefont{Sidorchuk}},
  \emph{et~al.}, \bibinfo{year}{2010}, \bibinfo{journal}{Acta Physica Polonica
  B} \textbf{\bibinfo{volume}{41}}, \bibinfo{pages}{449}.

\bibitem[{\citenamefont{Michel}
  \emph{et~al.}(2010{\natexlab{a}})\citenamefont{Michel, Nazarewicz, Okołowicz,
  and Płoszajczak}}]{Michel:2010}
\bibinfo{author}{\bibnamefont{Michel}, \bibfnamefont{N.}},
  \bibinfo{author}{\bibfnamefont{W.}~\bibnamefont{Nazarewicz}},
  \bibinfo{author}{\bibfnamefont{J.}~\bibnamefont{Okołowicz}}, and
  \bibinfo{author}{\bibfnamefont{M.}~\bibnamefont{Płoszajczak}},
  \bibinfo{year}{2010}{\natexlab{a}}, \bibinfo{journal}{Journal of Physics G:
  Nuclear and Particle Physics}
  \textbf{\bibinfo{volume}{37}}(\bibinfo{number}{6}), \bibinfo{pages}{064042}.

\bibitem[{\citenamefont{Michel}
  \emph{et~al.}(2010{\natexlab{b}})\citenamefont{Michel, Nazarewicz, and
  P\l{}oszajczak}}]{Michel:2010a}
\bibinfo{author}{\bibnamefont{Michel}, \bibfnamefont{N.}},
  \bibinfo{author}{\bibfnamefont{W.}~\bibnamefont{Nazarewicz}}, and
  \bibinfo{author}{\bibfnamefont{M.}~\bibnamefont{P\l{}oszajczak}},
  \bibinfo{year}{2010}{\natexlab{b}}, \bibinfo{journal}{Phys. Rev. C}
  \textbf{\bibinfo{volume}{82}}(\bibinfo{number}{4}), \bibinfo{pages}{044315}.

\bibitem[{\citenamefont{Michel} \emph{et~al.}(2002)\citenamefont{Michel,
  Nazarewicz, P\l{}oszajczak, and Bennaceur}}]{Michel:2002}
\bibinfo{author}{\bibnamefont{Michel}, \bibfnamefont{N.}},
  \bibinfo{author}{\bibfnamefont{W.}~\bibnamefont{Nazarewicz}},
  \bibinfo{author}{\bibfnamefont{M.}~\bibnamefont{P\l{}oszajczak}}, and
  \bibinfo{author}{\bibfnamefont{K.}~\bibnamefont{Bennaceur}},
  \bibinfo{year}{2002}, \bibinfo{journal}{Phys. Rev. Lett.}
  \textbf{\bibinfo{volume}{89}}(\bibinfo{number}{4}), \bibinfo{pages}{042502}.

\bibitem[{\citenamefont{Michel} \emph{et~al.}(2003)\citenamefont{Michel,
  Nazarewicz, P{\l}oszajczak, and Oko{\l}owicz}}]{Michel:2003}
\bibinfo{author}{\bibnamefont{Michel}, \bibfnamefont{N.}},
  \bibinfo{author}{\bibfnamefont{W.}~\bibnamefont{Nazarewicz}},
  \bibinfo{author}{\bibfnamefont{M.}~\bibnamefont{P{\l}oszajczak}}, and
  \bibinfo{author}{\bibfnamefont{J.}~\bibnamefont{Oko{\l}owicz}},
  \bibinfo{year}{2003}, \bibinfo{journal}{Phys. Rev. C}
  \textbf{\bibinfo{volume}{67}}, \bibinfo{pages}{054311}.

\bibitem[{\citenamefont{Michel} \emph{et~al.}(2009)\citenamefont{Michel,
  Nazarewicz, Płoszajczak, and Vertse}}]{Michel:2009}
\bibinfo{author}{\bibnamefont{Michel}, \bibfnamefont{N.}},
  \bibinfo{author}{\bibfnamefont{W.}~\bibnamefont{Nazarewicz}},
  \bibinfo{author}{\bibfnamefont{M.}~\bibnamefont{Płoszajczak}}, and
  \bibinfo{author}{\bibfnamefont{T.}~\bibnamefont{Vertse}},
  \bibinfo{year}{2009}, \bibinfo{journal}{Journal of Physics G: Nuclear and
  Particle Physics} \textbf{\bibinfo{volume}{36}}(\bibinfo{number}{1}),
  \bibinfo{pages}{013101}.

\bibitem[{\citenamefont{Miernik}
  \emph{et~al.}(2007{\natexlab{a}})\citenamefont{Miernik, Dominik, Czyrkowski,
  D\k{a}browski, Fomitchev, Golovkov, Janas, Ku\'smierz, Pfützner, Rodin,
  Stepantsov, Slepniev} \emph{et~al.}}]{Miernik:2007c}
\bibinfo{author}{\bibnamefont{Miernik}, \bibfnamefont{K.}},
  \bibinfo{author}{\bibfnamefont{W.}~\bibnamefont{Dominik}},
  \bibinfo{author}{\bibfnamefont{H.}~\bibnamefont{Czyrkowski}},
  \bibinfo{author}{\bibfnamefont{R.}~\bibnamefont{D\k{a}browski}},
  \bibinfo{author}{\bibfnamefont{A.}~\bibnamefont{Fomitchev}},
  \bibinfo{author}{\bibfnamefont{M.}~\bibnamefont{Golovkov}},
  \bibinfo{author}{\bibfnamefont{Z.}~\bibnamefont{Janas}},
  \bibinfo{author}{\bibfnamefont{W.}~\bibnamefont{Ku\'smierz}},
  \bibinfo{author}{\bibfnamefont{M.}~\bibnamefont{Pfützner}},
  \bibinfo{author}{\bibfnamefont{A.}~\bibnamefont{Rodin}},
  \bibinfo{author}{\bibfnamefont{S.}~\bibnamefont{Stepantsov}},
  \bibinfo{author}{\bibfnamefont{R.}~\bibnamefont{Slepniev}}, \emph{et~al.},
  \bibinfo{year}{2007}{\natexlab{a}}, \bibinfo{journal}{Nuclear Instruments and
  Methods in Physics Research Section A: Accelerators, Spectrometers, Detectors
  and Associated Equipment}
  \textbf{\bibinfo{volume}{581}}(\bibinfo{number}{1-2}), \bibinfo{pages}{194}.

\bibitem[{\citenamefont{Miernik}
  \emph{et~al.}(2007{\natexlab{b}})\citenamefont{Miernik, Dominik, Janas,
  Pf\"utzner, Bingham, Czyrkowski, \ifmmode~\acute{C}\else \'{C}\fi{}wiok,
  Darby, D\k{a}browski, Ginter, Grzywacz, Karny} \emph{et~al.}}]{Miernik:2007}
\bibinfo{author}{\bibnamefont{Miernik}, \bibfnamefont{K.}},
  \bibinfo{author}{\bibfnamefont{W.}~\bibnamefont{Dominik}},
  \bibinfo{author}{\bibfnamefont{Z.}~\bibnamefont{Janas}},
  \bibinfo{author}{\bibfnamefont{M.}~\bibnamefont{Pf\"utzner}},
  \bibinfo{author}{\bibfnamefont{C.~R.} \bibnamefont{Bingham}},
  \bibinfo{author}{\bibfnamefont{H.}~\bibnamefont{Czyrkowski}},
  \bibinfo{author}{\bibfnamefont{M.}~\bibnamefont{\ifmmode~\acute{C}\else
  \'{C}\fi{}wiok}}, \bibinfo{author}{\bibfnamefont{I.~G.} \bibnamefont{Darby}},
  \bibinfo{author}{\bibfnamefont{R.}~\bibnamefont{D\k{a}browski}},
  \bibinfo{author}{\bibfnamefont{T.}~\bibnamefont{Ginter}},
  \bibinfo{author}{\bibfnamefont{R.}~\bibnamefont{Grzywacz}},
  \bibinfo{author}{\bibfnamefont{M.}~\bibnamefont{Karny}}, \emph{et~al.},
  \bibinfo{year}{2007}{\natexlab{b}}, \bibinfo{journal}{Phys. Rev. C}
  \textbf{\bibinfo{volume}{76}}(\bibinfo{number}{4}), \bibinfo{pages}{041304}.

\bibitem[{\citenamefont{Miernik}
  \emph{et~al.}(2007{\natexlab{c}})\citenamefont{Miernik, Dominik, Janas,
  Pf\"utzner, Grigorenko, Bingham, Czyrkowski, \'Cwiok, Darby, D\k{a}browski,
  Ginter, Grzywacz} \emph{et~al.}}]{Miernik:2007b}
\bibinfo{author}{\bibnamefont{Miernik}, \bibfnamefont{K.}},
  \bibinfo{author}{\bibfnamefont{W.}~\bibnamefont{Dominik}},
  \bibinfo{author}{\bibfnamefont{Z.}~\bibnamefont{Janas}},
  \bibinfo{author}{\bibfnamefont{M.}~\bibnamefont{Pf\"utzner}},
  \bibinfo{author}{\bibfnamefont{L.}~\bibnamefont{Grigorenko}},
  \bibinfo{author}{\bibfnamefont{C.~R.} \bibnamefont{Bingham}},
  \bibinfo{author}{\bibfnamefont{H.}~\bibnamefont{Czyrkowski}},
  \bibinfo{author}{\bibfnamefont{M.}~\bibnamefont{\'Cwiok}},
  \bibinfo{author}{\bibfnamefont{I.~G.} \bibnamefont{Darby}},
  \bibinfo{author}{\bibfnamefont{R.}~\bibnamefont{D\k{a}browski}},
  \bibinfo{author}{\bibfnamefont{T.}~\bibnamefont{Ginter}},
  \bibinfo{author}{\bibfnamefont{R.}~\bibnamefont{Grzywacz}}, \emph{et~al.},
  \bibinfo{year}{2007}{\natexlab{c}}, \bibinfo{journal}{Phys. Rev. Lett.}
  \textbf{\bibinfo{volume}{99}}, \bibinfo{pages}{192501}.

\bibitem[{\citenamefont{Miernik} \emph{et~al.}(2009)\citenamefont{Miernik,
  Dominik, Janas, Pf\"utzner, Grigorenko, Bingham, Czyrkowski, \'Cwiok, Darby,
  D\k{a}browski, Ginter, Grzywacz} \emph{et~al.}}]{Miernik:2009}
\bibinfo{author}{\bibnamefont{Miernik}, \bibfnamefont{K.}},
  \bibinfo{author}{\bibfnamefont{W.}~\bibnamefont{Dominik}},
  \bibinfo{author}{\bibfnamefont{Z.}~\bibnamefont{Janas}},
  \bibinfo{author}{\bibfnamefont{M.}~\bibnamefont{Pf\"utzner}},
  \bibinfo{author}{\bibfnamefont{L.~V.} \bibnamefont{Grigorenko}},
  \bibinfo{author}{\bibfnamefont{C.}~\bibnamefont{Bingham}},
  \bibinfo{author}{\bibfnamefont{H.}~\bibnamefont{Czyrkowski}},
  \bibinfo{author}{\bibfnamefont{M.}~\bibnamefont{\'Cwiok}},
  \bibinfo{author}{\bibfnamefont{I.~G.} \bibnamefont{Darby}},
  \bibinfo{author}{\bibfnamefont{R.}~\bibnamefont{D\k{a}browski}},
  \bibinfo{author}{\bibfnamefont{T.}~\bibnamefont{Ginter}},
  \bibinfo{author}{\bibfnamefont{R.}~\bibnamefont{Grzywacz}}, \emph{et~al.},
  \bibinfo{year}{2009}, \bibinfo{journal}{Eur. Phys. J. A}
  \textbf{\bibinfo{volume}{42}}, \bibinfo{pages}{431}.

\bibitem[{\citenamefont{Millener}(2005)}]{Millener:2005}
\bibinfo{author}{\bibnamefont{Millener}, \bibfnamefont{D.}},
  \bibinfo{year}{2005}, \bibinfo{journal}{The European Physical Journal A -
  Hadrons and Nuclei} \textbf{\bibinfo{volume}{25}}, \bibinfo{pages}{97}.

\bibitem[{\citenamefont{Mitchell} \emph{et~al.}(2010)\citenamefont{Mitchell,
  Richter, and Weidenm\"uller}}]{Mitchell:2010}
\bibinfo{author}{\bibnamefont{Mitchell}, \bibfnamefont{G.~E.}},
  \bibinfo{author}{\bibfnamefont{A.}~\bibnamefont{Richter}}, and
  \bibinfo{author}{\bibfnamefont{H.~A.} \bibnamefont{Weidenm\"uller}},
  \bibinfo{year}{2010}, \bibinfo{journal}{Rev. Mod. Phys.}
  \textbf{\bibinfo{volume}{82}}(\bibinfo{number}{4}), \bibinfo{pages}{2845}.

\bibitem[{\citenamefont{Möller} \emph{et~al.}(1995)\citenamefont{Möller, Nix,
  Myers, and Swištecki}}]{Moller:1995}
\bibinfo{author}{\bibnamefont{Möller}, \bibfnamefont{P.}},
  \bibinfo{author}{\bibfnamefont{J.~R.} \bibnamefont{Nix}},
  \bibinfo{author}{\bibfnamefont{W.~D.} \bibnamefont{Myers}}, and
  \bibinfo{author}{\bibfnamefont{W.~J.} \bibnamefont{Swištecki}},
  \bibinfo{year}{1995}, \bibinfo{journal}{Atomic Data and Nuclear Data Tables}
  \textbf{\bibinfo{volume}{59}}(\bibinfo{number}{2}), \bibinfo{pages}{185}.

\bibitem[{\citenamefont{Münzenberg}
  \emph{et~al.}(1979)\citenamefont{Münzenberg, Faust, Hofmann, Armbruster,
  Güttner, and Ewald}}]{Muenzenberg:1979}
\bibinfo{author}{\bibnamefont{Münzenberg}, \bibfnamefont{G.}},
  \bibinfo{author}{\bibfnamefont{W.}~\bibnamefont{Faust}},
  \bibinfo{author}{\bibfnamefont{S.}~\bibnamefont{Hofmann}},
  \bibinfo{author}{\bibfnamefont{P.}~\bibnamefont{Armbruster}},
  \bibinfo{author}{\bibfnamefont{K.}~\bibnamefont{Güttner}}, and
  \bibinfo{author}{\bibfnamefont{H.}~\bibnamefont{Ewald}},
  \bibinfo{year}{1979}, \bibinfo{journal}{Nucl. Inst. Meth.}
  \textbf{\bibinfo{volume}{161}}, \bibinfo{pages}{65}.

\bibitem[{\citenamefont{Mohr}(2007)}]{Mohr:2007}
\bibinfo{author}{\bibnamefont{Mohr}, \bibfnamefont{P.}}, \bibinfo{year}{2007},
  \bibinfo{journal}{Eur. Phys. J. A} \textbf{\bibinfo{volume}{31}},
  \bibinfo{pages}{23}.

\bibitem[{\citenamefont{M\"{o}ller}
  \emph{et~al.}(1997)\citenamefont{M\"{o}ller, Nix, and Kratz}}]{Moller:1997}
\bibinfo{author}{\bibnamefont{M\"{o}ller}, \bibfnamefont{P.}},
  \bibinfo{author}{\bibfnamefont{J.~R.} \bibnamefont{Nix}}, and
  \bibinfo{author}{\bibfnamefont{K.-L.} \bibnamefont{Kratz}},
  \bibinfo{year}{1997}, \bibinfo{journal}{Atomic Data and Nuclear Data Tables}
  \textbf{\bibinfo{volume}{66}}(\bibinfo{number}{2}), \bibinfo{pages}{131}.

\bibitem[{\citenamefont{M\"{o}ller}
  \emph{et~al.}(2003)\citenamefont{M\"{o}ller, Pfeiffer, and
  Kratz}}]{Moller:2003}
\bibinfo{author}{\bibnamefont{M\"{o}ller}, \bibfnamefont{P.}},
  \bibinfo{author}{\bibfnamefont{B.}~\bibnamefont{Pfeiffer}}, and
  \bibinfo{author}{\bibfnamefont{K.-L.} \bibnamefont{Kratz}},
  \bibinfo{year}{2003}, \bibinfo{journal}{Phys. Rev. C}
  \textbf{\bibinfo{volume}{67}}(\bibinfo{number}{5}), \bibinfo{pages}{055802}.

\bibitem[{\citenamefont{M\"{o}ller}
  \emph{et~al.}(2009)\citenamefont{M\"{o}ller, Sierk, Ichikawa, Iwamoto,
  Bengtsson, Uhrenholt, and \AA{}berg}}]{Moller:2009}
\bibinfo{author}{\bibnamefont{M\"{o}ller}, \bibfnamefont{P.}},
  \bibinfo{author}{\bibfnamefont{A.~J.} \bibnamefont{Sierk}},
  \bibinfo{author}{\bibfnamefont{T.}~\bibnamefont{Ichikawa}},
  \bibinfo{author}{\bibfnamefont{A.}~\bibnamefont{Iwamoto}},
  \bibinfo{author}{\bibfnamefont{R.}~\bibnamefont{Bengtsson}},
  \bibinfo{author}{\bibfnamefont{H.}~\bibnamefont{Uhrenholt}}, and
  \bibinfo{author}{\bibfnamefont{S.}~\bibnamefont{\AA{}berg}},
  \bibinfo{year}{2009}, \bibinfo{journal}{Phys. Rev. C}
  \textbf{\bibinfo{volume}{79}}(\bibinfo{number}{6}), \bibinfo{pages}{064304}.

\bibitem[{\citenamefont{Moltz} \emph{et~al.}(1994)\citenamefont{Moltz,
  Robertson, Batchelder, and Cerny}}]{Moltz:1994}
\bibinfo{author}{\bibnamefont{Moltz}, \bibfnamefont{D.}},
  \bibinfo{author}{\bibfnamefont{J.}~\bibnamefont{Robertson}},
  \bibinfo{author}{\bibfnamefont{J.}~\bibnamefont{Batchelder}}, and
  \bibinfo{author}{\bibfnamefont{J.}~\bibnamefont{Cerny}},
  \bibinfo{year}{1994}, \bibinfo{journal}{Nuclear Instruments and Methods in
  Physics Research Section A: Accelerators, Spectrometers, Detectors and
  Associated Equipment} \textbf{\bibinfo{volume}{349}}(\bibinfo{number}{1}),
  \bibinfo{pages}{210}.

\bibitem[{\citenamefont{Montanari} \emph{et~al.}(2011)\citenamefont{Montanari,
  Leoni, Mengoni, Benzoni, Blasi, Bocchi, Bortignon, Bracco, Camera, Colo,
  Corsi, Crespi} \emph{et~al.}}]{Montanari:2011}
\bibinfo{author}{\bibnamefont{Montanari}, \bibfnamefont{D.}},
  \bibinfo{author}{\bibfnamefont{S.}~\bibnamefont{Leoni}},
  \bibinfo{author}{\bibfnamefont{D.}~\bibnamefont{Mengoni}},
  \bibinfo{author}{\bibfnamefont{G.}~\bibnamefont{Benzoni}},
  \bibinfo{author}{\bibfnamefont{N.}~\bibnamefont{Blasi}},
  \bibinfo{author}{\bibfnamefont{G.}~\bibnamefont{Bocchi}},
  \bibinfo{author}{\bibfnamefont{P.}~\bibnamefont{Bortignon}},
  \bibinfo{author}{\bibfnamefont{A.}~\bibnamefont{Bracco}},
  \bibinfo{author}{\bibfnamefont{F.}~\bibnamefont{Camera}},
  \bibinfo{author}{\bibfnamefont{G.}~\bibnamefont{Colo}},
  \bibinfo{author}{\bibfnamefont{A.}~\bibnamefont{Corsi}},
  \bibinfo{author}{\bibfnamefont{F.}~\bibnamefont{Crespi}}, \emph{et~al.},
  \bibinfo{year}{2011}, \bibinfo{journal}{Physics Letters B}
  \textbf{\bibinfo{volume}{697}}(\bibinfo{number}{4}), \bibinfo{pages}{288}.

\bibitem[{\citenamefont{Morrissey} \emph{et~al.}(2003)\citenamefont{Morrissey,
  Sherrill, Steiner, Stolz, and Wiedenhoever}}]{Morrissey:2003}
\bibinfo{author}{\bibnamefont{Morrissey}, \bibfnamefont{D.~J.}},
  \bibinfo{author}{\bibfnamefont{B.~M.} \bibnamefont{Sherrill}},
  \bibinfo{author}{\bibfnamefont{M.}~\bibnamefont{Steiner}},
  \bibinfo{author}{\bibfnamefont{A.}~\bibnamefont{Stolz}}, and
  \bibinfo{author}{\bibfnamefont{I.}~\bibnamefont{Wiedenhoever}},
  \bibinfo{year}{2003}, \bibinfo{journal}{Nuclear Instruments and Methods in
  Physics Research Section B: Beam Interactions with Materials and Atoms}
  \textbf{\bibinfo{volume}{204}}, \bibinfo{pages}{90}.

\bibitem[{\citenamefont{Mueller and Anne}(1991)}]{Mueller:1991}
\bibinfo{author}{\bibnamefont{Mueller}, \bibfnamefont{A.~C.}}, and
  \bibinfo{author}{\bibfnamefont{R.}~\bibnamefont{Anne}}, \bibinfo{year}{1991},
  \bibinfo{journal}{Nuclear Instruments and Methods in Physics Research Section
  B: Beam Interactions with Materials and Atoms}
  \textbf{\bibinfo{volume}{56-57}}(\bibinfo{number}{Part 1}),
  \bibinfo{pages}{559}.

\bibitem[{\citenamefont{Mukha}
  \emph{et~al.}(2008{\natexlab{a}})\citenamefont{Mukha, Grawe, Roeckl, and
  Tabor}}]{Mukha:2008a}
\bibinfo{author}{\bibnamefont{Mukha}, \bibfnamefont{I.}},
  \bibinfo{author}{\bibfnamefont{H.}~\bibnamefont{Grawe}},
  \bibinfo{author}{\bibfnamefont{E.}~\bibnamefont{Roeckl}}, and
  \bibinfo{author}{\bibfnamefont{S.}~\bibnamefont{Tabor}},
  \bibinfo{year}{2008}{\natexlab{a}}, \bibinfo{journal}{Phys. Rev. C}
  \textbf{\bibinfo{volume}{78}}, \bibinfo{pages}{039803}.

\bibitem[{\citenamefont{Mukha}
  \emph{et~al.}(2008{\natexlab{b}})\citenamefont{Mukha, Grigorenko, S\"ummerer,
  Acosta, Alvarez, Casarejos, Chatillon, Cortina-Gil, Espino, Fomichev,
  Garcia-Ramos, Geissel} \emph{et~al.}}]{Mukha:2008}
\bibinfo{author}{\bibnamefont{Mukha}, \bibfnamefont{I.}},
  \bibinfo{author}{\bibfnamefont{L.}~\bibnamefont{Grigorenko}},
  \bibinfo{author}{\bibfnamefont{K.}~\bibnamefont{S\"ummerer}},
  \bibinfo{author}{\bibfnamefont{L.}~\bibnamefont{Acosta}},
  \bibinfo{author}{\bibfnamefont{M.~A.~G.} \bibnamefont{Alvarez}},
  \bibinfo{author}{\bibfnamefont{E.}~\bibnamefont{Casarejos}},
  \bibinfo{author}{\bibfnamefont{A.}~\bibnamefont{Chatillon}},
  \bibinfo{author}{\bibfnamefont{D.}~\bibnamefont{Cortina-Gil}},
  \bibinfo{author}{\bibfnamefont{J.}~\bibnamefont{Espino}},
  \bibinfo{author}{\bibfnamefont{A.}~\bibnamefont{Fomichev}},
  \bibinfo{author}{\bibfnamefont{J.~E.} \bibnamefont{Garcia-Ramos}},
  \bibinfo{author}{\bibfnamefont{H.}~\bibnamefont{Geissel}}, \emph{et~al.},
  \bibinfo{year}{2008}{\natexlab{b}}, \bibinfo{journal}{Phys. Rev. C}
  \textbf{\bibinfo{volume}{77}}, \bibinfo{pages}{061303(R)}.

\bibitem[{\citenamefont{Mukha} \emph{et~al.}(2006)\citenamefont{Mukha, Roeckl,
  Batist, Blazhev, D\"{o}ring, Grawe, Grigorenko, Huyse, Janas, Kirchner,
  Commara, Mazzocchi} \emph{et~al.}}]{Mukha:2006}
\bibinfo{author}{\bibnamefont{Mukha}, \bibfnamefont{I.}},
  \bibinfo{author}{\bibfnamefont{E.}~\bibnamefont{Roeckl}},
  \bibinfo{author}{\bibfnamefont{L.}~\bibnamefont{Batist}},
  \bibinfo{author}{\bibfnamefont{A.}~\bibnamefont{Blazhev}},
  \bibinfo{author}{\bibfnamefont{J.}~\bibnamefont{D\"{o}ring}},
  \bibinfo{author}{\bibfnamefont{H.}~\bibnamefont{Grawe}},
  \bibinfo{author}{\bibfnamefont{L.}~\bibnamefont{Grigorenko}},
  \bibinfo{author}{\bibfnamefont{M.}~\bibnamefont{Huyse}},
  \bibinfo{author}{\bibfnamefont{Z.}~\bibnamefont{Janas}},
  \bibinfo{author}{\bibfnamefont{R.}~\bibnamefont{Kirchner}},
  \bibinfo{author}{\bibfnamefont{M.~L.} \bibnamefont{Commara}},
  \bibinfo{author}{\bibfnamefont{C.}~\bibnamefont{Mazzocchi}}, \emph{et~al.},
  \bibinfo{year}{2006}, \bibinfo{journal}{Nature}
  \textbf{\bibinfo{volume}{439}}, \bibinfo{pages}{298}.

\bibitem[{\citenamefont{Mukha} \emph{et~al.}(2005)\citenamefont{Mukha, Roeckl,
  D\"oring, Batist, Blazhev, Grawe, Hoffman, Huyse, Janas, Kirchner,
  La~Commara, Mazzocchi} \emph{et~al.}}]{Mukha:2005b}
\bibinfo{author}{\bibnamefont{Mukha}, \bibfnamefont{I.}},
  \bibinfo{author}{\bibfnamefont{E.}~\bibnamefont{Roeckl}},
  \bibinfo{author}{\bibfnamefont{J.}~\bibnamefont{D\"oring}},
  \bibinfo{author}{\bibfnamefont{L.}~\bibnamefont{Batist}},
  \bibinfo{author}{\bibfnamefont{A.}~\bibnamefont{Blazhev}},
  \bibinfo{author}{\bibfnamefont{H.}~\bibnamefont{Grawe}},
  \bibinfo{author}{\bibfnamefont{C.~R.} \bibnamefont{Hoffman}},
  \bibinfo{author}{\bibfnamefont{M.}~\bibnamefont{Huyse}},
  \bibinfo{author}{\bibfnamefont{Z.}~\bibnamefont{Janas}},
  \bibinfo{author}{\bibfnamefont{R.}~\bibnamefont{Kirchner}},
  \bibinfo{author}{\bibfnamefont{M.}~\bibnamefont{La~Commara}},
  \bibinfo{author}{\bibfnamefont{C.}~\bibnamefont{Mazzocchi}}, \emph{et~al.},
  \bibinfo{year}{2005}, \bibinfo{journal}{Phys. Rev. Lett.}
  \textbf{\bibinfo{volume}{95}}(\bibinfo{number}{2}), \bibinfo{pages}{022501}.

\bibitem[{\citenamefont{Mukha} \emph{et~al.}(2010)\citenamefont{Mukha,
  S\"ummerer, Acosta, Alvarez, Casarejos, Chatillon, Cortina-Gil, Egorova,
  Espino, Fomichev, Garc\'\i{}a-Ramos, Geissel} \emph{et~al.}}]{Mukha:2010}
\bibinfo{author}{\bibnamefont{Mukha}, \bibfnamefont{I.}},
  \bibinfo{author}{\bibfnamefont{K.}~\bibnamefont{S\"ummerer}},
  \bibinfo{author}{\bibfnamefont{L.}~\bibnamefont{Acosta}},
  \bibinfo{author}{\bibfnamefont{M.~A.~G.} \bibnamefont{Alvarez}},
  \bibinfo{author}{\bibfnamefont{E.}~\bibnamefont{Casarejos}},
  \bibinfo{author}{\bibfnamefont{A.}~\bibnamefont{Chatillon}},
  \bibinfo{author}{\bibfnamefont{D.}~\bibnamefont{Cortina-Gil}},
  \bibinfo{author}{\bibfnamefont{I.~A.} \bibnamefont{Egorova}},
  \bibinfo{author}{\bibfnamefont{J.~M.} \bibnamefont{Espino}},
  \bibinfo{author}{\bibfnamefont{A.}~\bibnamefont{Fomichev}},
  \bibinfo{author}{\bibfnamefont{J.~E.} \bibnamefont{Garc\'\i{}a-Ramos}},
  \bibinfo{author}{\bibfnamefont{H.}~\bibnamefont{Geissel}}, \emph{et~al.},
  \bibinfo{year}{2010}, \bibinfo{journal}{Phys. Rev. C}
  \textbf{\bibinfo{volume}{82}}(\bibinfo{number}{5}), \bibinfo{pages}{054315}.

\bibitem[{\citenamefont{Mukha} \emph{et~al.}(2007)\citenamefont{Mukha,
  S\"ummerer, Acosta, Alvarez, Casarejos, Chatillon, Cortina-Gil, Espino,
  Fomichev, Garcia-Ramos, Geissel, Gomez-Camacho} \emph{et~al.}}]{Mukha:2007}
\bibinfo{author}{\bibnamefont{Mukha}, \bibfnamefont{I.}},
  \bibinfo{author}{\bibfnamefont{K.}~\bibnamefont{S\"ummerer}},
  \bibinfo{author}{\bibfnamefont{L.}~\bibnamefont{Acosta}},
  \bibinfo{author}{\bibfnamefont{M.~A.~G.} \bibnamefont{Alvarez}},
  \bibinfo{author}{\bibfnamefont{E.}~\bibnamefont{Casarejos}},
  \bibinfo{author}{\bibfnamefont{A.}~\bibnamefont{Chatillon}},
  \bibinfo{author}{\bibfnamefont{D.}~\bibnamefont{Cortina-Gil}},
  \bibinfo{author}{\bibfnamefont{J.}~\bibnamefont{Espino}},
  \bibinfo{author}{\bibfnamefont{A.}~\bibnamefont{Fomichev}},
  \bibinfo{author}{\bibfnamefont{J.~E.} \bibnamefont{Garcia-Ramos}},
  \bibinfo{author}{\bibfnamefont{H.}~\bibnamefont{Geissel}},
  \bibinfo{author}{\bibfnamefont{J.}~\bibnamefont{Gomez-Camacho}},
  \emph{et~al.}, \bibinfo{year}{2007}, \bibinfo{journal}{Phys. Rev. Lett.}
  \textbf{\bibinfo{volume}{99}}, \bibinfo{pages}{182501}.

\bibitem[{\citenamefont{Mukha} \emph{et~al.}(2009)\citenamefont{Mukha,
  Timofeyuk, S\"ummerer, Acosta, Alvarez, Casarejos, Chatillon, Cortina-Gil,
  Espino, Fomichev, Garcia-Ramos, Geissel} \emph{et~al.}}]{Mukha:2009}
\bibinfo{author}{\bibnamefont{Mukha}, \bibfnamefont{I.}},
  \bibinfo{author}{\bibfnamefont{N.~K.} \bibnamefont{Timofeyuk}},
  \bibinfo{author}{\bibfnamefont{K.}~\bibnamefont{S\"ummerer}},
  \bibinfo{author}{\bibfnamefont{L.}~\bibnamefont{Acosta}},
  \bibinfo{author}{\bibfnamefont{M.~A.~G.} \bibnamefont{Alvarez}},
  \bibinfo{author}{\bibfnamefont{E.}~\bibnamefont{Casarejos}},
  \bibinfo{author}{\bibfnamefont{A.}~\bibnamefont{Chatillon}},
  \bibinfo{author}{\bibfnamefont{D.}~\bibnamefont{Cortina-Gil}},
  \bibinfo{author}{\bibfnamefont{J.~M.} \bibnamefont{Espino}},
  \bibinfo{author}{\bibfnamefont{A.}~\bibnamefont{Fomichev}},
  \bibinfo{author}{\bibfnamefont{J.~E.} \bibnamefont{Garcia-Ramos}},
  \bibinfo{author}{\bibfnamefont{H.}~\bibnamefont{Geissel}}, \emph{et~al.},
  \bibinfo{year}{2009}, \bibinfo{journal}{Phys. Rev. C}
  \textbf{\bibinfo{volume}{79}}, \bibinfo{pages}{061301(R)}.

\bibitem[{\citenamefont{Nakata} \emph{et~al.}(1997)\citenamefont{Nakata,
  Tachibana, and Yamada}}]{Nakata:1997}
\bibinfo{author}{\bibnamefont{Nakata}, \bibfnamefont{H.}},
  \bibinfo{author}{\bibfnamefont{T.}~\bibnamefont{Tachibana}}, and
  \bibinfo{author}{\bibfnamefont{M.}~\bibnamefont{Yamada}},
  \bibinfo{year}{1997}, \bibinfo{journal}{Nuclear Physics A}
  \textbf{\bibinfo{volume}{625}}(\bibinfo{number}{3}), \bibinfo{pages}{521}.

\bibitem[{\citenamefont{Napolitani}
  \emph{et~al.}(2007)\citenamefont{Napolitani, Schmidt, Tassan-Got, Armbruster,
  Enqvist, Heinz, Henzl, Henzlova, Keli\ifmmode~\acute{c}\else \'{c}\fi{},
  Pleska\ifmmode~\check{c}\else \v{c}\fi{}, Ricciardi, Schmitt}
  \emph{et~al.}}]{Napolitani:2007}
\bibinfo{author}{\bibnamefont{Napolitani}, \bibfnamefont{P.}},
  \bibinfo{author}{\bibfnamefont{K.-H.} \bibnamefont{Schmidt}},
  \bibinfo{author}{\bibfnamefont{L.}~\bibnamefont{Tassan-Got}},
  \bibinfo{author}{\bibfnamefont{P.}~\bibnamefont{Armbruster}},
  \bibinfo{author}{\bibfnamefont{T.}~\bibnamefont{Enqvist}},
  \bibinfo{author}{\bibfnamefont{A.}~\bibnamefont{Heinz}},
  \bibinfo{author}{\bibfnamefont{V.}~\bibnamefont{Henzl}},
  \bibinfo{author}{\bibfnamefont{D.}~\bibnamefont{Henzlova}},
  \bibinfo{author}{\bibfnamefont{A.}~\bibnamefont{Keli\ifmmode~\acute{c}\else
  \'{c}\fi{}}},
  \bibinfo{author}{\bibfnamefont{R.}~\bibnamefont{Pleska\ifmmode~\check{c}\else
  \v{c}\fi{}}}, \bibinfo{author}{\bibfnamefont{M.~V.} \bibnamefont{Ricciardi}},
  \bibinfo{author}{\bibfnamefont{C.}~\bibnamefont{Schmitt}}, \emph{et~al.},
  \bibinfo{year}{2007}, \bibinfo{journal}{Phys. Rev. C}
  \textbf{\bibinfo{volume}{76}}(\bibinfo{number}{6}), \bibinfo{pages}{064609}.

\bibitem[{\citenamefont{Navr\'{a}til}
  \emph{et~al.}(2007)\citenamefont{Navr\'{a}til, Gueorguiev, Vary, Ormand, and
  Nogga}}]{Navratil:2007}
\bibinfo{author}{\bibnamefont{Navr\'{a}til}, \bibfnamefont{P.}},
  \bibinfo{author}{\bibfnamefont{V.~G.} \bibnamefont{Gueorguiev}},
  \bibinfo{author}{\bibfnamefont{J.~P.} \bibnamefont{Vary}},
  \bibinfo{author}{\bibfnamefont{W.~E.} \bibnamefont{Ormand}}, and
  \bibinfo{author}{\bibfnamefont{A.}~\bibnamefont{Nogga}},
  \bibinfo{year}{2007}, \bibinfo{journal}{Phys. Rev. Lett.}
  \textbf{\bibinfo{volume}{99}}(\bibinfo{number}{4}), \bibinfo{pages}{042501}.

\bibitem[{\citenamefont{Nazarewicz}
  \emph{et~al.}(1996)\citenamefont{Nazarewicz, Dobaczewski, Werner, Maruhn,
  Reinhard, Rutz, Chinn, Umar, and Strayer}}]{Nazarewicz:1996}
\bibinfo{author}{\bibnamefont{Nazarewicz}, \bibfnamefont{W.}},
  \bibinfo{author}{\bibfnamefont{J.}~\bibnamefont{Dobaczewski}},
  \bibinfo{author}{\bibfnamefont{T.~R.} \bibnamefont{Werner}},
  \bibinfo{author}{\bibfnamefont{J.~A.} \bibnamefont{Maruhn}},
  \bibinfo{author}{\bibfnamefont{P.-G.} \bibnamefont{Reinhard}},
  \bibinfo{author}{\bibfnamefont{K.}~\bibnamefont{Rutz}},
  \bibinfo{author}{\bibfnamefont{C.~R.} \bibnamefont{Chinn}},
  \bibinfo{author}{\bibfnamefont{A.~S.} \bibnamefont{Umar}}, and
  \bibinfo{author}{\bibfnamefont{M.~R.} \bibnamefont{Strayer}},
  \bibinfo{year}{1996}, \bibinfo{journal}{Phys. Rev. C}
  \textbf{\bibinfo{volume}{53}}(\bibinfo{number}{2}), \bibinfo{pages}{740}.

\bibitem[{\citenamefont{Nesterov} \emph{et~al.}(2010)\citenamefont{Nesterov,
  Arickx, Broeckhove, and Vasilevsky}}]{Nesterov:2010}
\bibinfo{author}{\bibnamefont{Nesterov}, \bibfnamefont{A.}},
  \bibinfo{author}{\bibfnamefont{F.}~\bibnamefont{Arickx}},
  \bibinfo{author}{\bibfnamefont{J.}~\bibnamefont{Broeckhove}}, and
  \bibinfo{author}{\bibfnamefont{V.}~\bibnamefont{Vasilevsky}},
  \bibinfo{year}{2010}, \bibinfo{journal}{Physics of Particles and Nuclei}
  \textbf{\bibinfo{volume}{41}}, \bibinfo{pages}{716}.

\bibitem[{\citenamefont{Nikolskii} \emph{et~al.}(2010)\citenamefont{Nikolskii,
  Korsheninnikov, Otsu, Suzuki, Yoneda, Baba, Yamada, Kondo, Aoi, Denikin,
  Golovkov, Fomichev} \emph{et~al.}}]{Nikolskii:2010}
\bibinfo{author}{\bibnamefont{Nikolskii}, \bibfnamefont{E.}},
  \bibinfo{author}{\bibfnamefont{A.}~\bibnamefont{Korsheninnikov}},
  \bibinfo{author}{\bibfnamefont{H.}~\bibnamefont{Otsu}},
  \bibinfo{author}{\bibfnamefont{H.}~\bibnamefont{Suzuki}},
  \bibinfo{author}{\bibfnamefont{K.}~\bibnamefont{Yoneda}},
  \bibinfo{author}{\bibfnamefont{H.}~\bibnamefont{Baba}},
  \bibinfo{author}{\bibfnamefont{K.}~\bibnamefont{Yamada}},
  \bibinfo{author}{\bibfnamefont{Y.}~\bibnamefont{Kondo}},
  \bibinfo{author}{\bibfnamefont{N.}~\bibnamefont{Aoi}},
  \bibinfo{author}{\bibfnamefont{A.}~\bibnamefont{Denikin}},
  \bibinfo{author}{\bibfnamefont{M.}~\bibnamefont{Golovkov}},
  \bibinfo{author}{\bibfnamefont{A.}~\bibnamefont{Fomichev}}, \emph{et~al.},
  \bibinfo{year}{2010}, \bibinfo{journal}{Phys.Rev. C}
  \textbf{\bibinfo{volume}{81}}, \bibinfo{pages}{064606}.

\bibitem[{\citenamefont{Nilsson} \emph{et~al.}(2000)\citenamefont{Nilsson,
  Nyman, and Riisager}}]{Nilsson:2000}
\bibinfo{author}{\bibnamefont{Nilsson}, \bibfnamefont{T.}},
  \bibinfo{author}{\bibfnamefont{G.}~\bibnamefont{Nyman}}, and
  \bibinfo{author}{\bibfnamefont{K.}~\bibnamefont{Riisager}},
  \bibinfo{year}{2000}, \bibinfo{journal}{Hyperfine Interactions}
  \textbf{\bibinfo{volume}{129}}(\bibinfo{number}{1}), \bibinfo{pages}{67}.

\bibitem[{\citenamefont{Nolden} \emph{et~al.}(2008)\citenamefont{Nolden,
  Dimopoulou, Dolinskii, and Steck}}]{Nolden:2008}
\bibinfo{author}{\bibnamefont{Nolden}, \bibfnamefont{F.}},
  \bibinfo{author}{\bibfnamefont{C.}~\bibnamefont{Dimopoulou}},
  \bibinfo{author}{\bibfnamefont{A.}~\bibnamefont{Dolinskii}}, and
  \bibinfo{author}{\bibfnamefont{M.}~\bibnamefont{Steck}},
  \bibinfo{year}{2008}, \bibinfo{journal}{Nuclear Instruments and Methods in
  Physics Research Section B: Beam Interactions with Materials and Atoms}
  \textbf{\bibinfo{volume}{266}}(\bibinfo{number}{19-20}),
  \bibinfo{pages}{4569}.

\bibitem[{\citenamefont{Nummela} \emph{et~al.}(2001)\citenamefont{Nummela,
  Nowacki, Baumann, Caurier, Cederk\"all, Courtin, Dessagne, Jokinen, Knipper,
  Le~Scornet, Lyapin, Mieh\'e} \emph{et~al.}}]{Nummela:2001}
\bibinfo{author}{\bibnamefont{Nummela}, \bibfnamefont{S.}},
  \bibinfo{author}{\bibfnamefont{F.}~\bibnamefont{Nowacki}},
  \bibinfo{author}{\bibfnamefont{P.}~\bibnamefont{Baumann}},
  \bibinfo{author}{\bibfnamefont{E.}~\bibnamefont{Caurier}},
  \bibinfo{author}{\bibfnamefont{J.}~\bibnamefont{Cederk\"all}},
  \bibinfo{author}{\bibfnamefont{S.}~\bibnamefont{Courtin}},
  \bibinfo{author}{\bibfnamefont{P.}~\bibnamefont{Dessagne}},
  \bibinfo{author}{\bibfnamefont{A.}~\bibnamefont{Jokinen}},
  \bibinfo{author}{\bibfnamefont{A.}~\bibnamefont{Knipper}},
  \bibinfo{author}{\bibfnamefont{G.}~\bibnamefont{Le~Scornet}},
  \bibinfo{author}{\bibfnamefont{L.~G.} \bibnamefont{Lyapin}},
  \bibinfo{author}{\bibfnamefont{C.}~\bibnamefont{Mieh\'e}}, \emph{et~al.},
  \bibinfo{year}{2001}, \bibinfo{journal}{Phys. Rev. C}
  \textbf{\bibinfo{volume}{64}}(\bibinfo{number}{5}), \bibinfo{pages}{054313}.

\bibitem[{\citenamefont{Nyman} \emph{et~al.}(1990)\citenamefont{Nyman, Azuma,
  Hansen, Jonson, Larsson, Mattsson, Richter, Riisager, Tengblad, and
  Wilhelmsen}}]{Nyman:1990}
\bibinfo{author}{\bibnamefont{Nyman}, \bibfnamefont{G.}},
  \bibinfo{author}{\bibfnamefont{R.~E.} \bibnamefont{Azuma}},
  \bibinfo{author}{\bibfnamefont{P.~G.} \bibnamefont{Hansen}},
  \bibinfo{author}{\bibfnamefont{B.}~\bibnamefont{Jonson}},
  \bibinfo{author}{\bibfnamefont{P.~O.} \bibnamefont{Larsson}},
  \bibinfo{author}{\bibfnamefont{S.}~\bibnamefont{Mattsson}},
  \bibinfo{author}{\bibfnamefont{A.}~\bibnamefont{Richter}},
  \bibinfo{author}{\bibfnamefont{K.}~\bibnamefont{Riisager}},
  \bibinfo{author}{\bibfnamefont{O.}~\bibnamefont{Tengblad}}, and
  \bibinfo{author}{\bibfnamefont{K.}~\bibnamefont{Wilhelmsen}},
  \bibinfo{year}{1990}, \bibinfo{journal}{Nuclear Physics A}
  \textbf{\bibinfo{volume}{510}}(\bibinfo{number}{2}), \bibinfo{pages}{189}.

\bibitem[{\citenamefont{Oganessian}(2007)}]{Oganessian:2007}
\bibinfo{author}{\bibnamefont{Oganessian}, \bibfnamefont{Y.}},
  \bibinfo{year}{2007}, \bibinfo{journal}{Journal of Physics G: Nuclear and
  Particle Physics} \textbf{\bibinfo{volume}{34}}(\bibinfo{number}{4}),
  \bibinfo{pages}{R165}.

\bibitem[{\citenamefont{Oganessian}
  \emph{et~al.}(2010)\citenamefont{Oganessian, Abdullin, Bailey, Benker,
  Bennett, Dmitriev, Ezold, Hamilton, Henderson, Itkis, Lobanov, Mezentsev}
  \emph{et~al.}}]{Oganessian:2010}
\bibinfo{author}{\bibnamefont{Oganessian}, \bibfnamefont{Y.~T.}},
  \bibinfo{author}{\bibfnamefont{F.~S.} \bibnamefont{Abdullin}},
  \bibinfo{author}{\bibfnamefont{P.~D.} \bibnamefont{Bailey}},
  \bibinfo{author}{\bibfnamefont{D.~E.} \bibnamefont{Benker}},
  \bibinfo{author}{\bibfnamefont{M.~E.} \bibnamefont{Bennett}},
  \bibinfo{author}{\bibfnamefont{S.~N.} \bibnamefont{Dmitriev}},
  \bibinfo{author}{\bibfnamefont{J.~G.} \bibnamefont{Ezold}},
  \bibinfo{author}{\bibfnamefont{J.~H.} \bibnamefont{Hamilton}},
  \bibinfo{author}{\bibfnamefont{R.~A.} \bibnamefont{Henderson}},
  \bibinfo{author}{\bibfnamefont{M.~G.} \bibnamefont{Itkis}},
  \bibinfo{author}{\bibfnamefont{Y.~V.} \bibnamefont{Lobanov}},
  \bibinfo{author}{\bibfnamefont{A.~N.} \bibnamefont{Mezentsev}},
  \emph{et~al.}, \bibinfo{year}{2010}, \bibinfo{journal}{Phys. Rev. Lett.}
  \textbf{\bibinfo{volume}{104}}(\bibinfo{number}{14}),
  \bibinfo{pages}{142502}.

\bibitem[{\citenamefont{Ohnishi} \emph{et~al.}(2010)\citenamefont{Ohnishi,
  Kubo, Kusaka, Yoshida, Yoshida, Ohtake, Fukuda, Takeda, Kameda, Tanaka,
  Inabe, Yanagisawa} \emph{et~al.}}]{Ohnishi:2010}
\bibinfo{author}{\bibnamefont{Ohnishi}, \bibfnamefont{T.}},
  \bibinfo{author}{\bibfnamefont{T.}~\bibnamefont{Kubo}},
  \bibinfo{author}{\bibfnamefont{K.}~\bibnamefont{Kusaka}},
  \bibinfo{author}{\bibfnamefont{A.}~\bibnamefont{Yoshida}},
  \bibinfo{author}{\bibfnamefont{K.}~\bibnamefont{Yoshida}},
  \bibinfo{author}{\bibfnamefont{M.}~\bibnamefont{Ohtake}},
  \bibinfo{author}{\bibfnamefont{N.}~\bibnamefont{Fukuda}},
  \bibinfo{author}{\bibfnamefont{H.}~\bibnamefont{Takeda}},
  \bibinfo{author}{\bibfnamefont{D.}~\bibnamefont{Kameda}},
  \bibinfo{author}{\bibfnamefont{K.}~\bibnamefont{Tanaka}},
  \bibinfo{author}{\bibfnamefont{N.}~\bibnamefont{Inabe}},
  \bibinfo{author}{\bibfnamefont{Y.}~\bibnamefont{Yanagisawa}}, \emph{et~al.},
  \bibinfo{year}{2010}, \bibinfo{journal}{Journal of the Physical Society of
  Japan} \textbf{\bibinfo{volume}{79}}(\bibinfo{number}{7}),
  \bibinfo{pages}{073201}.

\bibitem[{\citenamefont{Oko{\l}owicz}
  \emph{et~al.}(2003)\citenamefont{Oko{\l}owicz, P{\l}oszajczak, and
  Rotter}}]{Okolowicz:2003}
\bibinfo{author}{\bibnamefont{Oko{\l}owicz}, \bibfnamefont{J.}},
  \bibinfo{author}{\bibfnamefont{M.}~\bibnamefont{P{\l}oszajczak}}, and
  \bibinfo{author}{\bibfnamefont{I.}~\bibnamefont{Rotter}},
  \bibinfo{year}{2003}, \bibinfo{journal}{Physics Reports}
  \textbf{\bibinfo{volume}{374}}(\bibinfo{number}{4-5}), \bibinfo{pages}{271}.

\bibitem[{\citenamefont{Ormand}(1996)}]{Ormand:1996}
\bibinfo{author}{\bibnamefont{Ormand}, \bibfnamefont{W.~E.}},
  \bibinfo{year}{1996}, \bibinfo{journal}{Phys. Rev. C}
  \textbf{\bibinfo{volume}{53}}(\bibinfo{number}{1}), \bibinfo{pages}{214}.

\bibitem[{\citenamefont{Ormand}(1997)}]{Ormand:1997}
\bibinfo{author}{\bibnamefont{Ormand}, \bibfnamefont{W.~E.}},
  \bibinfo{year}{1997}, \bibinfo{journal}{Phys. Rev. C}
  \textbf{\bibinfo{volume}{55}}, \bibinfo{pages}{2407}.

\bibitem[{\citenamefont{Osterfeld}(1992)}]{Osterfeld:1992}
\bibinfo{author}{\bibnamefont{Osterfeld}, \bibfnamefont{F.}},
  \bibinfo{year}{1992}, \bibinfo{journal}{Rev. Mod. Phys.}
  \textbf{\bibinfo{volume}{64}}(\bibinfo{number}{2}), \bibinfo{pages}{491}.

\bibitem[{\citenamefont{Ozawa} \emph{et~al.}(1995)\citenamefont{Ozawa, Raimann,
  Boyd, Chloupek, Fujimaki, Kimura, Kitagawa, Kobayashi, Kolata, Kubono,
  Tanihata, Watanabe} \emph{et~al.}}]{Ozawa:1995}
\bibinfo{author}{\bibnamefont{Ozawa}, \bibfnamefont{A.}},
  \bibinfo{author}{\bibfnamefont{G.}~\bibnamefont{Raimann}},
  \bibinfo{author}{\bibfnamefont{R.~N.} \bibnamefont{Boyd}},
  \bibinfo{author}{\bibfnamefont{F.~R.} \bibnamefont{Chloupek}},
  \bibinfo{author}{\bibfnamefont{M.}~\bibnamefont{Fujimaki}},
  \bibinfo{author}{\bibfnamefont{K.}~\bibnamefont{Kimura}},
  \bibinfo{author}{\bibfnamefont{H.}~\bibnamefont{Kitagawa}},
  \bibinfo{author}{\bibfnamefont{T.}~\bibnamefont{Kobayashi}},
  \bibinfo{author}{\bibfnamefont{J.~J.} \bibnamefont{Kolata}},
  \bibinfo{author}{\bibfnamefont{S.}~\bibnamefont{Kubono}},
  \bibinfo{author}{\bibfnamefont{I.}~\bibnamefont{Tanihata}},
  \bibinfo{author}{\bibfnamefont{Y.}~\bibnamefont{Watanabe}}, \emph{et~al.},
  \bibinfo{year}{1995}, \bibinfo{journal}{Nuclear Physics A}
  \textbf{\bibinfo{volume}{592}}(\bibinfo{number}{2}), \bibinfo{pages}{244}.

\bibitem[{\citenamefont{Page} \emph{et~al.}(2003)\citenamefont{Page, Andreyev,
  Appelbe, Butler, Freeman, Greenlees, Herzberg, Jenkins, Jones, Jones, Joss,
  Julin} \emph{et~al.}}]{Page:2003}
\bibinfo{author}{\bibnamefont{Page}, \bibfnamefont{R.}},
  \bibinfo{author}{\bibfnamefont{A.~N.} \bibnamefont{Andreyev}},
  \bibinfo{author}{\bibfnamefont{D.~E.} \bibnamefont{Appelbe}},
  \bibinfo{author}{\bibfnamefont{P.~A.} \bibnamefont{Butler}},
  \bibinfo{author}{\bibfnamefont{S.~J.} \bibnamefont{Freeman}},
  \bibinfo{author}{\bibfnamefont{P.~T.} \bibnamefont{Greenlees}},
  \bibinfo{author}{\bibfnamefont{R.~D.} \bibnamefont{Herzberg}},
  \bibinfo{author}{\bibfnamefont{D.~G.} \bibnamefont{Jenkins}},
  \bibinfo{author}{\bibfnamefont{G.~D.} \bibnamefont{Jones}},
  \bibinfo{author}{\bibfnamefont{P.}~\bibnamefont{Jones}},
  \bibinfo{author}{\bibfnamefont{D.~T.} \bibnamefont{Joss}},
  \bibinfo{author}{\bibfnamefont{R.}~\bibnamefont{Julin}}, \emph{et~al.},
  \bibinfo{year}{2003}, \bibinfo{journal}{Nuclear Instruments and Methods in
  Physics Research Section B: Beam Interactions with Materials and Atoms}
  \textbf{\bibinfo{volume}{204}}, \bibinfo{pages}{634}.

\bibitem[{\citenamefont{Page} \emph{et~al.}(2007)\citenamefont{Page, Bianco,
  Darby, Uusitalo, Joss, Grahn, Herzberg, Pakarinen, Thomson, Eeckhaudt,
  Greenlees, Jones} \emph{et~al.}}]{Page:2007}
\bibinfo{author}{\bibnamefont{Page}, \bibfnamefont{R.}},
  \bibinfo{author}{\bibfnamefont{L.}~\bibnamefont{Bianco}},
  \bibinfo{author}{\bibfnamefont{I.}~\bibnamefont{Darby}},
  \bibinfo{author}{\bibfnamefont{J.}~\bibnamefont{Uusitalo}},
  \bibinfo{author}{\bibfnamefont{D.}~\bibnamefont{Joss}},
  \bibinfo{author}{\bibfnamefont{T.}~\bibnamefont{Grahn}},
  \bibinfo{author}{\bibfnamefont{R.-D.} \bibnamefont{Herzberg}},
  \bibinfo{author}{\bibfnamefont{J.}~\bibnamefont{Pakarinen}},
  \bibinfo{author}{\bibfnamefont{J.}~\bibnamefont{Thomson}},
  \bibinfo{author}{\bibfnamefont{S.}~\bibnamefont{Eeckhaudt}},
  \bibinfo{author}{\bibfnamefont{P.}~\bibnamefont{Greenlees}},
  \bibinfo{author}{\bibfnamefont{P.}~\bibnamefont{Jones}}, \emph{et~al.},
  \bibinfo{year}{2007}, \bibinfo{journal}{Phys.\ Rev.}
  \textbf{\bibinfo{volume}{C~75}}, \bibinfo{pages}{061302}.

\bibitem[{\citenamefont{Page} \emph{et~al.}(1992)\citenamefont{Page, Woods,
  Cunningham, Davinson, Davis, Hofmann, James, Livingston, Sellin, and
  Shotter}}]{Page:1992}
\bibinfo{author}{\bibnamefont{Page}, \bibfnamefont{R.}},
  \bibinfo{author}{\bibfnamefont{P.}~\bibnamefont{Woods}},
  \bibinfo{author}{\bibfnamefont{R.}~\bibnamefont{Cunningham}},
  \bibinfo{author}{\bibfnamefont{T.}~\bibnamefont{Davinson}},
  \bibinfo{author}{\bibfnamefont{N.}~\bibnamefont{Davis}},
  \bibinfo{author}{\bibfnamefont{S.}~\bibnamefont{Hofmann}},
  \bibinfo{author}{\bibfnamefont{A.}~\bibnamefont{James}},
  \bibinfo{author}{\bibfnamefont{K.}~\bibnamefont{Livingston}},
  \bibinfo{author}{\bibfnamefont{P.}~\bibnamefont{Sellin}}, and
  \bibinfo{author}{\bibfnamefont{A.}~\bibnamefont{Shotter}},
  \bibinfo{year}{1992}, \bibinfo{journal}{Phys.\ Rev.\ Lett.}
  \textbf{\bibinfo{volume}{68}}, \bibinfo{pages}{1287}.

\bibitem[{\citenamefont{Page} \emph{et~al.}(1994)\citenamefont{Page, Woods,
  Cunningham, Davinson, Davis, James, Livingston, Sellin, and
  Shotter}}]{Page:1994}
\bibinfo{author}{\bibnamefont{Page}, \bibfnamefont{R.}},
  \bibinfo{author}{\bibfnamefont{P.}~\bibnamefont{Woods}},
  \bibinfo{author}{\bibfnamefont{R.}~\bibnamefont{Cunningham}},
  \bibinfo{author}{\bibfnamefont{T.}~\bibnamefont{Davinson}},
  \bibinfo{author}{\bibfnamefont{N.}~\bibnamefont{Davis}},
  \bibinfo{author}{\bibfnamefont{A.}~\bibnamefont{James}},
  \bibinfo{author}{\bibfnamefont{K.}~\bibnamefont{Livingston}},
  \bibinfo{author}{\bibfnamefont{P.}~\bibnamefont{Sellin}}, and
  \bibinfo{author}{\bibfnamefont{A.}~\bibnamefont{Shotter}},
  \bibinfo{year}{1994}, \bibinfo{journal}{Phys.\ Rev.\ Lett.}
  \textbf{\bibinfo{volume}{72}}, \bibinfo{pages}{1178}.

\bibitem[{\citenamefont{Page} \emph{et~al.}(1996)\citenamefont{Page, Woods,
  Cunningham, Davinson, Davis, James, Livingston, Sellin, and
  Shotter}}]{Page:1996}
\bibinfo{author}{\bibnamefont{Page}, \bibfnamefont{R.}},
  \bibinfo{author}{\bibfnamefont{P.}~\bibnamefont{Woods}},
  \bibinfo{author}{\bibfnamefont{R.}~\bibnamefont{Cunningham}},
  \bibinfo{author}{\bibfnamefont{T.}~\bibnamefont{Davinson}},
  \bibinfo{author}{\bibfnamefont{N.}~\bibnamefont{Davis}},
  \bibinfo{author}{\bibfnamefont{A.}~\bibnamefont{James}},
  \bibinfo{author}{\bibfnamefont{K.}~\bibnamefont{Livingston}},
  \bibinfo{author}{\bibfnamefont{P.}~\bibnamefont{Sellin}}, and
  \bibinfo{author}{\bibfnamefont{A.}~\bibnamefont{Shotter}},
  \bibinfo{year}{1996}, \bibinfo{journal}{Phys.\ Rev.}
  \textbf{\bibinfo{volume}{C~53}}, \bibinfo{pages}{660}.

\bibitem[{\citenamefont{Page}(2011)}]{Page:2011}
\bibinfo{author}{\bibnamefont{Page}, \bibfnamefont{R.~D.}},
  \bibinfo{year}{2011}, \bibinfo{journal}{Phys. Rev. C}
  \textbf{\bibinfo{volume}{83}}(\bibinfo{number}{1}), \bibinfo{pages}{014305}.

\bibitem[{\citenamefont{Pais}(1986)}]{Pais:1986}
\bibinfo{author}{\bibnamefont{Pais}, \bibfnamefont{A.}}, \bibinfo{year}{1986},
  \emph{\bibinfo{title}{Inward bound: of matter and forces in the physical
  world}} (\bibinfo{publisher}{Oxford University Press}).

\bibitem[{\citenamefont{Paul} \emph{et~al.}(1995)\citenamefont{Paul, Woods,
  Davinson, Page, Sellin, Beausang, Clark, Cunningham, Forbes, Fossan, Gizon,
  Gizon} \emph{et~al.}}]{Paul:1995}
\bibinfo{author}{\bibnamefont{Paul}, \bibfnamefont{E.~S.}},
  \bibinfo{author}{\bibfnamefont{P.~J.} \bibnamefont{Woods}},
  \bibinfo{author}{\bibfnamefont{T.}~\bibnamefont{Davinson}},
  \bibinfo{author}{\bibfnamefont{R.~D.} \bibnamefont{Page}},
  \bibinfo{author}{\bibfnamefont{P.~J.} \bibnamefont{Sellin}},
  \bibinfo{author}{\bibfnamefont{C.~W.} \bibnamefont{Beausang}},
  \bibinfo{author}{\bibfnamefont{R.~M.} \bibnamefont{Clark}},
  \bibinfo{author}{\bibfnamefont{R.~A.} \bibnamefont{Cunningham}},
  \bibinfo{author}{\bibfnamefont{S.~A.} \bibnamefont{Forbes}},
  \bibinfo{author}{\bibfnamefont{D.~B.} \bibnamefont{Fossan}},
  \bibinfo{author}{\bibfnamefont{A.}~\bibnamefont{Gizon}},
  \bibinfo{author}{\bibfnamefont{J.}~\bibnamefont{Gizon}}, \emph{et~al.},
  \bibinfo{year}{1995}, \bibinfo{journal}{Phys. Rev. C}
  \textbf{\bibinfo{volume}{51}}(\bibinfo{number}{1}), \bibinfo{pages}{78}.

\bibitem[{\citenamefont{Pauli}(1930)}]{Pauli:1930}
\bibinfo{author}{\bibnamefont{Pauli}, \bibfnamefont{W.}}, \bibinfo{year}{1930},
  in \emph{\bibinfo{booktitle}{W. Pauli, collected scientific papers}}, edited
  by \bibinfo{editor}{\bibfnamefont{R.}~\bibnamefont{Kronig}} and
  \bibinfo{editor}{\bibfnamefont{V.}~\bibnamefont{Weisskopf}}
  (\bibinfo{publisher}{Interscience, New York, 1964}),
  volume~\bibinfo{volume}{2}, p. \bibinfo{pages}{1313}.

\bibitem[{\citenamefont{Pechenaya} \emph{et~al.}(2007)\citenamefont{Pechenaya,
  Chiara, Sarantites, Reviol, Charity, Carpenter, Janssens, Lauritsen, Lister,
  Seweryniak, Zhu, Andersson} \emph{et~al.}}]{Pechenaya:2007}
\bibinfo{author}{\bibnamefont{Pechenaya}, \bibfnamefont{O.~L.}},
  \bibinfo{author}{\bibfnamefont{C.~J.} \bibnamefont{Chiara}},
  \bibinfo{author}{\bibfnamefont{D.~G.} \bibnamefont{Sarantites}},
  \bibinfo{author}{\bibfnamefont{W.}~\bibnamefont{Reviol}},
  \bibinfo{author}{\bibfnamefont{R.~J.} \bibnamefont{Charity}},
  \bibinfo{author}{\bibfnamefont{M.~P.} \bibnamefont{Carpenter}},
  \bibinfo{author}{\bibfnamefont{R.~V.~F.} \bibnamefont{Janssens}},
  \bibinfo{author}{\bibfnamefont{T.}~\bibnamefont{Lauritsen}},
  \bibinfo{author}{\bibfnamefont{C.~J.} \bibnamefont{Lister}},
  \bibinfo{author}{\bibfnamefont{D.}~\bibnamefont{Seweryniak}},
  \bibinfo{author}{\bibfnamefont{S.}~\bibnamefont{Zhu}},
  \bibinfo{author}{\bibfnamefont{L.-L.} \bibnamefont{Andersson}},
  \emph{et~al.}, \bibinfo{year}{2007}, \bibinfo{journal}{Phys. Rev. C}
  \textbf{\bibinfo{volume}{76}}, \bibinfo{pages}{011304(R)}.

\bibitem[{\citenamefont{Penttilä} \emph{et~al.}(2010)\citenamefont{Penttilä,
  Karvonen, Eronen, Elomaa, Hager, Hakala, Jokinen, Kankainen, Moore,
  Peräjärvi, Rahaman, Rinta-Antila} \emph{et~al.}}]{Penttila:2010}
\bibinfo{author}{\bibnamefont{Penttilä}, \bibfnamefont{H.}},
  \bibinfo{author}{\bibfnamefont{P.}~\bibnamefont{Karvonen}},
  \bibinfo{author}{\bibfnamefont{T.}~\bibnamefont{Eronen}},
  \bibinfo{author}{\bibfnamefont{V.}~\bibnamefont{Elomaa}},
  \bibinfo{author}{\bibfnamefont{U.}~\bibnamefont{Hager}},
  \bibinfo{author}{\bibfnamefont{J.}~\bibnamefont{Hakala}},
  \bibinfo{author}{\bibfnamefont{A.}~\bibnamefont{Jokinen}},
  \bibinfo{author}{\bibfnamefont{A.}~\bibnamefont{Kankainen}},
  \bibinfo{author}{\bibfnamefont{I.}~\bibnamefont{Moore}},
  \bibinfo{author}{\bibfnamefont{K.}~\bibnamefont{Peräjärvi}},
  \bibinfo{author}{\bibfnamefont{S.}~\bibnamefont{Rahaman}},
  \bibinfo{author}{\bibfnamefont{S.}~\bibnamefont{Rinta-Antila}},
  \emph{et~al.}, \bibinfo{year}{2010}, \bibinfo{journal}{The European Physical
  Journal A - Hadrons and Nuclei} \textbf{\bibinfo{volume}{44}},
  \bibinfo{pages}{147}.

\bibitem[{\citenamefont{Pereira} \emph{et~al.}(2007)\citenamefont{Pereira,
  Benlliure, Casarejos, Armbruster, Bernas, Boudard, Czajkowski, Enqvist,
  Legrain, Leray, Mustapha, Pravikoff} \emph{et~al.}}]{Pereira:2007}
\bibinfo{author}{\bibnamefont{Pereira}, \bibfnamefont{J.}},
  \bibinfo{author}{\bibfnamefont{J.}~\bibnamefont{Benlliure}},
  \bibinfo{author}{\bibfnamefont{E.}~\bibnamefont{Casarejos}},
  \bibinfo{author}{\bibfnamefont{P.}~\bibnamefont{Armbruster}},
  \bibinfo{author}{\bibfnamefont{M.}~\bibnamefont{Bernas}},
  \bibinfo{author}{\bibfnamefont{A.}~\bibnamefont{Boudard}},
  \bibinfo{author}{\bibfnamefont{S.}~\bibnamefont{Czajkowski}},
  \bibinfo{author}{\bibfnamefont{T.}~\bibnamefont{Enqvist}},
  \bibinfo{author}{\bibfnamefont{R.}~\bibnamefont{Legrain}},
  \bibinfo{author}{\bibfnamefont{S.}~\bibnamefont{Leray}},
  \bibinfo{author}{\bibfnamefont{B.}~\bibnamefont{Mustapha}},
  \bibinfo{author}{\bibfnamefont{M.}~\bibnamefont{Pravikoff}}, \emph{et~al.},
  \bibinfo{year}{2007}, \bibinfo{journal}{Phys. Rev. C}
  \textbf{\bibinfo{volume}{75}}(\bibinfo{number}{1}), \bibinfo{pages}{014602}.

\bibitem[{\citenamefont{Pereira} \emph{et~al.}(2010)\citenamefont{Pereira,
  Hosmer, Lorusso, Santi, Couture, Daly, Santo, Elliot, Görres, Herlitzius,
  Kratz, Lamm} \emph{et~al.}}]{Pereira:2010}
\bibinfo{author}{\bibnamefont{Pereira}, \bibfnamefont{J.}},
  \bibinfo{author}{\bibfnamefont{P.}~\bibnamefont{Hosmer}},
  \bibinfo{author}{\bibfnamefont{G.}~\bibnamefont{Lorusso}},
  \bibinfo{author}{\bibfnamefont{P.}~\bibnamefont{Santi}},
  \bibinfo{author}{\bibfnamefont{A.}~\bibnamefont{Couture}},
  \bibinfo{author}{\bibfnamefont{J.}~\bibnamefont{Daly}},
  \bibinfo{author}{\bibfnamefont{M.~D.} \bibnamefont{Santo}},
  \bibinfo{author}{\bibfnamefont{T.}~\bibnamefont{Elliot}},
  \bibinfo{author}{\bibfnamefont{J.}~\bibnamefont{Görres}},
  \bibinfo{author}{\bibfnamefont{C.}~\bibnamefont{Herlitzius}},
  \bibinfo{author}{\bibfnamefont{K.-L.} \bibnamefont{Kratz}},
  \bibinfo{author}{\bibfnamefont{L.}~\bibnamefont{Lamm}}, \emph{et~al.},
  \bibinfo{year}{2010}, \bibinfo{journal}{Nuclear Instruments and Methods in
  Physics Research Section A: Accelerators, Spectrometers, Detectors and
  Associated Equipment} \textbf{\bibinfo{volume}{618}}(\bibinfo{number}{1-3}),
  \bibinfo{pages}{275}.

\bibitem[{\citenamefont{Pervin} \emph{et~al.}(2007)\citenamefont{Pervin,
  Pieper, and Wiringa}}]{Pervin:2007}
\bibinfo{author}{\bibnamefont{Pervin}, \bibfnamefont{M.}},
  \bibinfo{author}{\bibfnamefont{S.~C.} \bibnamefont{Pieper}}, and
  \bibinfo{author}{\bibfnamefont{R.~B.} \bibnamefont{Wiringa}},
  \bibinfo{year}{2007}, \bibinfo{journal}{Phys. Rev. C}
  \textbf{\bibinfo{volume}{76}}(\bibinfo{number}{6}), \bibinfo{pages}{064319}.

\bibitem[{\citenamefont{Pfeiffer} \emph{et~al.}(2001)\citenamefont{Pfeiffer,
  Kratz, Thielemann, and Walters}}]{Pfeiffer:2001}
\bibinfo{author}{\bibnamefont{Pfeiffer}, \bibfnamefont{B.}},
  \bibinfo{author}{\bibfnamefont{K.~L.} \bibnamefont{Kratz}},
  \bibinfo{author}{\bibfnamefont{F.~K.} \bibnamefont{Thielemann}}, and
  \bibinfo{author}{\bibfnamefont{W.~B.} \bibnamefont{Walters}},
  \bibinfo{year}{2001}, \bibinfo{journal}{Nuclear Physics A}
  \textbf{\bibinfo{volume}{693}}(\bibinfo{number}{1-2}), \bibinfo{pages}{282}.

\bibitem[{\citenamefont{Pfennig} \emph{et~al.}(2008)\citenamefont{Pfennig,
  Normand, Magill, and Fangh\"anel}}]{Pfennig:2008}
\bibinfo{editor}{\bibnamefont{Pfennig}, \bibfnamefont{G.}},
  \bibinfo{editor}{\bibfnamefont{C.}~\bibnamefont{Normand}},
  \bibinfo{editor}{\bibfnamefont{J.}~\bibnamefont{Magill}}, and
  \bibinfo{editor}{\bibfnamefont{T.}~\bibnamefont{Fangh\"anel}} (eds.),
  \bibinfo{year}{2008}, \emph{\bibinfo{title}{Karlsruher Nuklidkarte.
  Commemoration of the 50th Anniversary}} (\bibinfo{publisher}{Luxembourg:
  Office for Official Publications of the European Communities}).

\bibitem[{\citenamefont{Pf\"utzner}
  \emph{et~al.}(2002)\citenamefont{Pf\"utzner, Badura, Bingham, Blank,
  Chartier, Geissel, Giovinazzo, Grigorenko, Grzywacz, Hellstr\"om, Janas,
  Kurcewicz} \emph{et~al.}}]{Pfutzner:2002}
\bibinfo{author}{\bibnamefont{Pf\"utzner}, \bibfnamefont{M.}},
  \bibinfo{author}{\bibfnamefont{E.}~\bibnamefont{Badura}},
  \bibinfo{author}{\bibfnamefont{C.}~\bibnamefont{Bingham}},
  \bibinfo{author}{\bibfnamefont{B.}~\bibnamefont{Blank}},
  \bibinfo{author}{\bibfnamefont{M.}~\bibnamefont{Chartier}},
  \bibinfo{author}{\bibfnamefont{H.}~\bibnamefont{Geissel}},
  \bibinfo{author}{\bibfnamefont{J.}~\bibnamefont{Giovinazzo}},
  \bibinfo{author}{\bibfnamefont{L.}~\bibnamefont{Grigorenko}},
  \bibinfo{author}{\bibfnamefont{R.}~\bibnamefont{Grzywacz}},
  \bibinfo{author}{\bibfnamefont{M.}~\bibnamefont{Hellstr\"om}},
  \bibinfo{author}{\bibfnamefont{Z.}~\bibnamefont{Janas}},
  \bibinfo{author}{\bibfnamefont{J.}~\bibnamefont{Kurcewicz}}, \emph{et~al.},
  \bibinfo{year}{2002}, \bibinfo{journal}{Eur. Phys. J.}
  \textbf{\bibinfo{volume}{A 14}}, \bibinfo{pages}{279}.

\bibitem[{\citenamefont{Pühlhofer}(1977)}]{Puhlhofer:1977}
\bibinfo{author}{\bibnamefont{Pühlhofer}, \bibfnamefont{F.}},
  \bibinfo{year}{1977}, \bibinfo{journal}{Nuclear Physics A}
  \textbf{\bibinfo{volume}{280}}(\bibinfo{number}{1}), \bibinfo{pages}{267}.

\bibitem[{\citenamefont{Pietri} \emph{et~al.}(2007)\citenamefont{Pietri, Regan,
  Podolyák, Rudolph, Steer, Garnsworthy, Werner-Malento, Hoischen, Górska,
  Gerl, Wollersheim, Kojouharov} \emph{et~al.}}]{Pietri:2007}
\bibinfo{author}{\bibnamefont{Pietri}, \bibfnamefont{S.}},
  \bibinfo{author}{\bibfnamefont{P.}~\bibnamefont{Regan}},
  \bibinfo{author}{\bibfnamefont{Z.}~\bibnamefont{Podolyák}},
  \bibinfo{author}{\bibfnamefont{D.}~\bibnamefont{Rudolph}},
  \bibinfo{author}{\bibfnamefont{S.}~\bibnamefont{Steer}},
  \bibinfo{author}{\bibfnamefont{A.}~\bibnamefont{Garnsworthy}},
  \bibinfo{author}{\bibfnamefont{E.}~\bibnamefont{Werner-Malento}},
  \bibinfo{author}{\bibfnamefont{R.}~\bibnamefont{Hoischen}},
  \bibinfo{author}{\bibfnamefont{M.}~\bibnamefont{Górska}},
  \bibinfo{author}{\bibfnamefont{J.}~\bibnamefont{Gerl}},
  \bibinfo{author}{\bibfnamefont{H.}~\bibnamefont{Wollersheim}},
  \bibinfo{author}{\bibfnamefont{I.}~\bibnamefont{Kojouharov}}, \emph{et~al.},
  \bibinfo{year}{2007}, \bibinfo{journal}{Nuclear Instruments and Methods in
  Physics Research Section B: Beam Interactions with Materials and Atoms}
  \textbf{\bibinfo{volume}{261}}(\bibinfo{number}{1-2}), \bibinfo{pages}{1079}.

\bibitem[{\citenamefont{Poenaru} \emph{et~al.}(2002)\citenamefont{Poenaru,
  Nagame, Gherghescu, and Greiner}}]{Poenaru:2002}
\bibinfo{author}{\bibnamefont{Poenaru}, \bibfnamefont{D.~N.}},
  \bibinfo{author}{\bibfnamefont{Y.}~\bibnamefont{Nagame}},
  \bibinfo{author}{\bibfnamefont{R.~A.} \bibnamefont{Gherghescu}}, and
  \bibinfo{author}{\bibfnamefont{W.}~\bibnamefont{Greiner}},
  \bibinfo{year}{2002}, \bibinfo{journal}{Phys. Rev. C}
  \textbf{\bibinfo{volume}{65}}(\bibinfo{number}{5}), \bibinfo{pages}{054308}.

\bibitem[{\citenamefont{Poli} \emph{et~al.}(1999)\citenamefont{Poli, Davids,
  Woods, Seweryniak, Batchelder, Brown, Bingham, Carpenter, Conticchio,
  Davinson, DeBoer, Hamada} \emph{et~al.}}]{Poli:1999}
\bibinfo{author}{\bibnamefont{Poli}, \bibfnamefont{G.}},
  \bibinfo{author}{\bibfnamefont{C.}~\bibnamefont{Davids}},
  \bibinfo{author}{\bibfnamefont{P.}~\bibnamefont{Woods}},
  \bibinfo{author}{\bibfnamefont{D.}~\bibnamefont{Seweryniak}},
  \bibinfo{author}{\bibfnamefont{J.}~\bibnamefont{Batchelder}},
  \bibinfo{author}{\bibfnamefont{L.}~\bibnamefont{Brown}},
  \bibinfo{author}{\bibfnamefont{C.}~\bibnamefont{Bingham}},
  \bibinfo{author}{\bibfnamefont{M.}~\bibnamefont{Carpenter}},
  \bibinfo{author}{\bibfnamefont{L.}~\bibnamefont{Conticchio}},
  \bibinfo{author}{\bibfnamefont{T.}~\bibnamefont{Davinson}},
  \bibinfo{author}{\bibfnamefont{J.}~\bibnamefont{DeBoer}},
  \bibinfo{author}{\bibfnamefont{S.}~\bibnamefont{Hamada}}, \emph{et~al.},
  \bibinfo{year}{1999}, \bibinfo{journal}{Phys.\ Rev.}
  \textbf{\bibinfo{volume}{C~59}}, \bibinfo{pages}{R2979}.

\bibitem[{\citenamefont{Poli} \emph{et~al.}(2001)\citenamefont{Poli, Davids,
  Woods, Seweryniak, Carpenter, Ci\.zewski, Davinson, Heinz, Janssens, Lister,
  Ressler, Sonzogni} \emph{et~al.}}]{Poli:2001}
\bibinfo{author}{\bibnamefont{Poli}, \bibfnamefont{G.}},
  \bibinfo{author}{\bibfnamefont{C.}~\bibnamefont{Davids}},
  \bibinfo{author}{\bibfnamefont{P.}~\bibnamefont{Woods}},
  \bibinfo{author}{\bibfnamefont{D.}~\bibnamefont{Seweryniak}},
  \bibinfo{author}{\bibfnamefont{M.}~\bibnamefont{Carpenter}},
  \bibinfo{author}{\bibfnamefont{J.}~\bibnamefont{Ci\.zewski}},
  \bibinfo{author}{\bibfnamefont{T.}~\bibnamefont{Davinson}},
  \bibinfo{author}{\bibfnamefont{A.}~\bibnamefont{Heinz}},
  \bibinfo{author}{\bibfnamefont{R.}~\bibnamefont{Janssens}},
  \bibinfo{author}{\bibfnamefont{C.}~\bibnamefont{Lister}},
  \bibinfo{author}{\bibfnamefont{J.}~\bibnamefont{Ressler}},
  \bibinfo{author}{\bibfnamefont{A.}~\bibnamefont{Sonzogni}}, \emph{et~al.},
  \bibinfo{year}{2001}, \bibinfo{journal}{Phys.\ Rev.}
  \textbf{\bibinfo{volume}{C~63}}(\bibinfo{number}{4}),
  \bibinfo{pages}{044304}.

\bibitem[{\citenamefont{Pomorski}
  \emph{et~al.}(2011{\natexlab{a}})\citenamefont{Pomorski, Miernik, Dominik,
  Janas, Pf\"utzner, Bingham, Czyrkowski, \ifmmode~\acute{C}\else
  \'{C}\fi{}wiok, Darby, D\k{a}browski, Ginter, Grzywacz}
  \emph{et~al.}}]{Pomorski:2011}
\bibinfo{author}{\bibnamefont{Pomorski}, \bibfnamefont{M.}},
  \bibinfo{author}{\bibfnamefont{K.}~\bibnamefont{Miernik}},
  \bibinfo{author}{\bibfnamefont{W.}~\bibnamefont{Dominik}},
  \bibinfo{author}{\bibfnamefont{Z.}~\bibnamefont{Janas}},
  \bibinfo{author}{\bibfnamefont{M.}~\bibnamefont{Pf\"utzner}},
  \bibinfo{author}{\bibfnamefont{C.~R.} \bibnamefont{Bingham}},
  \bibinfo{author}{\bibfnamefont{H.}~\bibnamefont{Czyrkowski}},
  \bibinfo{author}{\bibfnamefont{M.}~\bibnamefont{\ifmmode~\acute{C}\else
  \'{C}\fi{}wiok}}, \bibinfo{author}{\bibfnamefont{I.~G.} \bibnamefont{Darby}},
  \bibinfo{author}{\bibfnamefont{R.}~\bibnamefont{D\k{a}browski}},
  \bibinfo{author}{\bibfnamefont{T.}~\bibnamefont{Ginter}},
  \bibinfo{author}{\bibfnamefont{R.}~\bibnamefont{Grzywacz}}, \emph{et~al.},
  \bibinfo{year}{2011}{\natexlab{a}}, \bibinfo{journal}{Phys. Rev. C}
  \textbf{\bibinfo{volume}{83}}(\bibinfo{number}{1}), \bibinfo{pages}{014306}.

\bibitem[{\citenamefont{Pomorski}
  \emph{et~al.}(2011{\natexlab{b}})\citenamefont{Pomorski, Pf\"utzner, Dominik,
  Grzywacz, Baumann, Berryman, Czyrkowski, D\k{a}browski, Ginter, Johnson,
  Kami\ifmmode~\acute{n}\else \'{n}\fi{}ski, Ku\ifmmode~\acute{z}\else
  \'{z}\fi{}niak} \emph{et~al.}}]{Pomorski:2011b}
\bibinfo{author}{\bibnamefont{Pomorski}, \bibfnamefont{M.}},
  \bibinfo{author}{\bibfnamefont{M.}~\bibnamefont{Pf\"utzner}},
  \bibinfo{author}{\bibfnamefont{W.}~\bibnamefont{Dominik}},
  \bibinfo{author}{\bibfnamefont{R.}~\bibnamefont{Grzywacz}},
  \bibinfo{author}{\bibfnamefont{T.}~\bibnamefont{Baumann}},
  \bibinfo{author}{\bibfnamefont{J.~S.} \bibnamefont{Berryman}},
  \bibinfo{author}{\bibfnamefont{H.}~\bibnamefont{Czyrkowski}},
  \bibinfo{author}{\bibfnamefont{R.}~\bibnamefont{D\k{a}browski}},
  \bibinfo{author}{\bibfnamefont{T.}~\bibnamefont{Ginter}},
  \bibinfo{author}{\bibfnamefont{J.}~\bibnamefont{Johnson}},
  \bibinfo{author}{\bibfnamefont{G.}~\bibnamefont{Kami\ifmmode~\acute{n}\else
  \'{n}\fi{}ski}},
  \bibinfo{author}{\bibfnamefont{A.}~\bibnamefont{Ku\ifmmode~\acute{z}\else
  \'{z}\fi{}niak}}, \emph{et~al.}, \bibinfo{year}{2011}{\natexlab{b}},
  \bibinfo{journal}{Phys. Rev. C}
  \textbf{\bibinfo{volume}{83}}(\bibinfo{number}{6}), \bibinfo{pages}{061303}.

\bibitem[{\citenamefont{Prezado} \emph{et~al.}(2003)\citenamefont{Prezado,
  Bergmann, Borge, Cederk\"{a}ll, Diget, Fraile, Fynbo, Jeppesen, Jonson,
  Meister, Nilsson, Nyman} \emph{et~al.}}]{Prezado:2003}
\bibinfo{author}{\bibnamefont{Prezado}, \bibfnamefont{Y.}},
  \bibinfo{author}{\bibfnamefont{U.~C.} \bibnamefont{Bergmann}},
  \bibinfo{author}{\bibfnamefont{M.~J.~G.} \bibnamefont{Borge}},
  \bibinfo{author}{\bibfnamefont{J.}~\bibnamefont{Cederk\"{a}ll}},
  \bibinfo{author}{\bibfnamefont{C.~A.} \bibnamefont{Diget}},
  \bibinfo{author}{\bibfnamefont{L.~M.} \bibnamefont{Fraile}},
  \bibinfo{author}{\bibfnamefont{H.~O.~U.} \bibnamefont{Fynbo}},
  \bibinfo{author}{\bibfnamefont{H.}~\bibnamefont{Jeppesen}},
  \bibinfo{author}{\bibfnamefont{B.}~\bibnamefont{Jonson}},
  \bibinfo{author}{\bibfnamefont{M.}~\bibnamefont{Meister}},
  \bibinfo{author}{\bibfnamefont{T.}~\bibnamefont{Nilsson}},
  \bibinfo{author}{\bibfnamefont{G.}~\bibnamefont{Nyman}}, \emph{et~al.},
  \bibinfo{year}{2003}, \bibinfo{journal}{Physics Letters B}
  \textbf{\bibinfo{volume}{576}}(\bibinfo{number}{1-2}), \bibinfo{pages}{55}.

\bibitem[{\citenamefont{Prezado} \emph{et~al.}(2005)\citenamefont{Prezado,
  Borge, Diget, Fraile, Fulton, Fynbo, Jeppesen, Jonson, Meister, Nilsson,
  Nyman, Riisager} \emph{et~al.}}]{Prezado:2005}
\bibinfo{author}{\bibnamefont{Prezado}, \bibfnamefont{Y.}},
  \bibinfo{author}{\bibfnamefont{M.~J.~G.} \bibnamefont{Borge}},
  \bibinfo{author}{\bibfnamefont{C.~A.} \bibnamefont{Diget}},
  \bibinfo{author}{\bibfnamefont{L.~M.} \bibnamefont{Fraile}},
  \bibinfo{author}{\bibfnamefont{B.~R.} \bibnamefont{Fulton}},
  \bibinfo{author}{\bibfnamefont{H.~O.~U.} \bibnamefont{Fynbo}},
  \bibinfo{author}{\bibfnamefont{H.~B.} \bibnamefont{Jeppesen}},
  \bibinfo{author}{\bibfnamefont{B.}~\bibnamefont{Jonson}},
  \bibinfo{author}{\bibfnamefont{M.}~\bibnamefont{Meister}},
  \bibinfo{author}{\bibfnamefont{T.}~\bibnamefont{Nilsson}},
  \bibinfo{author}{\bibfnamefont{G.}~\bibnamefont{Nyman}},
  \bibinfo{author}{\bibfnamefont{K.}~\bibnamefont{Riisager}}, \emph{et~al.},
  \bibinfo{year}{2005}, \bibinfo{journal}{Physics Letters B}
  \textbf{\bibinfo{volume}{618}}(\bibinfo{number}{1-4}), \bibinfo{pages}{43}.

\bibitem[{\citenamefont{Qi} \emph{et~al.}(2009{\natexlab{a}})\citenamefont{Qi,
  Xu, Liotta, and Wyss}}]{Qi:2009}
\bibinfo{author}{\bibnamefont{Qi}, \bibfnamefont{C.}},
  \bibinfo{author}{\bibfnamefont{F.~R.} \bibnamefont{Xu}},
  \bibinfo{author}{\bibfnamefont{R.~J.} \bibnamefont{Liotta}}, and
  \bibinfo{author}{\bibfnamefont{R.}~\bibnamefont{Wyss}},
  \bibinfo{year}{2009}{\natexlab{a}}, \bibinfo{journal}{Phys. Rev. Lett.}
  \textbf{\bibinfo{volume}{103}}(\bibinfo{number}{7}), \bibinfo{pages}{072501}.

\bibitem[{\citenamefont{Qi} \emph{et~al.}(2009{\natexlab{b}})\citenamefont{Qi,
  Xu, Liotta, Wyss, Zhang, Asawatangtrakuldee, and Hu}}]{Qi:2009a}
\bibinfo{author}{\bibnamefont{Qi}, \bibfnamefont{C.}},
  \bibinfo{author}{\bibfnamefont{F.~R.} \bibnamefont{Xu}},
  \bibinfo{author}{\bibfnamefont{R.~J.} \bibnamefont{Liotta}},
  \bibinfo{author}{\bibfnamefont{R.}~\bibnamefont{Wyss}},
  \bibinfo{author}{\bibfnamefont{M.~Y.} \bibnamefont{Zhang}},
  \bibinfo{author}{\bibfnamefont{C.}~\bibnamefont{Asawatangtrakuldee}}, and
  \bibinfo{author}{\bibfnamefont{D.}~\bibnamefont{Hu}},
  \bibinfo{year}{2009}{\natexlab{b}}, \bibinfo{journal}{Phys. Rev. C}
  \textbf{\bibinfo{volume}{80}}(\bibinfo{number}{4}), \bibinfo{pages}{044326}.

\bibitem[{\citenamefont{Raabe} \emph{et~al.}(2008)\citenamefont{Raabe,
  Andreyev, Borge, Buchmann, Capel, Fynbo, Huyse, Kanungo, Kirchner, Mattoon,
  Morton, Mukha} \emph{et~al.}}]{Raabe:2008}
\bibinfo{author}{\bibnamefont{Raabe}, \bibfnamefont{R.}},
  \bibinfo{author}{\bibfnamefont{A.}~\bibnamefont{Andreyev}},
  \bibinfo{author}{\bibfnamefont{M.~J.~G.} \bibnamefont{Borge}},
  \bibinfo{author}{\bibfnamefont{L.}~\bibnamefont{Buchmann}},
  \bibinfo{author}{\bibfnamefont{P.}~\bibnamefont{Capel}},
  \bibinfo{author}{\bibfnamefont{H.~O.~U.} \bibnamefont{Fynbo}},
  \bibinfo{author}{\bibfnamefont{M.}~\bibnamefont{Huyse}},
  \bibinfo{author}{\bibfnamefont{R.}~\bibnamefont{Kanungo}},
  \bibinfo{author}{\bibfnamefont{T.}~\bibnamefont{Kirchner}},
  \bibinfo{author}{\bibfnamefont{C.}~\bibnamefont{Mattoon}},
  \bibinfo{author}{\bibfnamefont{A.~C.} \bibnamefont{Morton}},
  \bibinfo{author}{\bibfnamefont{I.}~\bibnamefont{Mukha}}, \emph{et~al.},
  \bibinfo{year}{2008}, \bibinfo{journal}{Phys. Rev. Lett.}
  \textbf{\bibinfo{volume}{101}}(\bibinfo{number}{21}),
  \bibinfo{pages}{212501}.

\bibitem[{\citenamefont{Raabe} \emph{et~al.}(2009)\citenamefont{Raabe,
  B\"uscher, Ponsaers, Aksouh, Huyse, Ivanov, Lesher, Mukha, Pauwels, Sawicka,
  Smirnov, Stefanescu} \emph{et~al.}}]{Raabe:2009}
\bibinfo{author}{\bibnamefont{Raabe}, \bibfnamefont{R.}},
  \bibinfo{author}{\bibfnamefont{J.}~\bibnamefont{B\"uscher}},
  \bibinfo{author}{\bibfnamefont{J.}~\bibnamefont{Ponsaers}},
  \bibinfo{author}{\bibfnamefont{F.}~\bibnamefont{Aksouh}},
  \bibinfo{author}{\bibfnamefont{M.}~\bibnamefont{Huyse}},
  \bibinfo{author}{\bibfnamefont{O.}~\bibnamefont{Ivanov}},
  \bibinfo{author}{\bibfnamefont{S.~R.} \bibnamefont{Lesher}},
  \bibinfo{author}{\bibfnamefont{I.}~\bibnamefont{Mukha}},
  \bibinfo{author}{\bibfnamefont{D.}~\bibnamefont{Pauwels}},
  \bibinfo{author}{\bibfnamefont{M.}~\bibnamefont{Sawicka}},
  \bibinfo{author}{\bibfnamefont{D.}~\bibnamefont{Smirnov}},
  \bibinfo{author}{\bibfnamefont{I.}~\bibnamefont{Stefanescu}}, \emph{et~al.},
  \bibinfo{year}{2009}, \bibinfo{journal}{Phys. Rev. C}
  \textbf{\bibinfo{volume}{80}}(\bibinfo{number}{5}), \bibinfo{pages}{054307}.

\bibitem[{\citenamefont{Raciti} \emph{et~al.}(2008)\citenamefont{Raciti,
  Cardella, De~Napoli, Rapisarda, Amorini, and Sfienti}}]{Raciti:2008}
\bibinfo{author}{\bibnamefont{Raciti}, \bibfnamefont{G.}},
  \bibinfo{author}{\bibfnamefont{G.}~\bibnamefont{Cardella}},
  \bibinfo{author}{\bibfnamefont{M.}~\bibnamefont{De~Napoli}},
  \bibinfo{author}{\bibfnamefont{E.}~\bibnamefont{Rapisarda}},
  \bibinfo{author}{\bibfnamefont{F.}~\bibnamefont{Amorini}}, and
  \bibinfo{author}{\bibfnamefont{C.}~\bibnamefont{Sfienti}},
  \bibinfo{year}{2008}, \bibinfo{journal}{Phys. Rev. Lett.}
  \textbf{\bibinfo{volume}{100}}(\bibinfo{number}{19}),
  \bibinfo{pages}{192503}.

\bibitem[{\citenamefont{Radivojevic}
  \emph{et~al.}(2002)\citenamefont{Radivojevic, Baumann, Caurier,
  Cederk\"{a}ll, Courtin, Dessagne, Jokinen, Knipper, Scornet, Lyapin,
  Mieh\'{e}, Nowacki} \emph{et~al.}}]{Radivojevic:2002}
\bibinfo{author}{\bibnamefont{Radivojevic}, \bibfnamefont{Z.}},
  \bibinfo{author}{\bibfnamefont{P.}~\bibnamefont{Baumann}},
  \bibinfo{author}{\bibfnamefont{E.}~\bibnamefont{Caurier}},
  \bibinfo{author}{\bibfnamefont{J.}~\bibnamefont{Cederk\"{a}ll}},
  \bibinfo{author}{\bibfnamefont{S.}~\bibnamefont{Courtin}},
  \bibinfo{author}{\bibfnamefont{P.}~\bibnamefont{Dessagne}},
  \bibinfo{author}{\bibfnamefont{A.}~\bibnamefont{Jokinen}},
  \bibinfo{author}{\bibfnamefont{A.}~\bibnamefont{Knipper}},
  \bibinfo{author}{\bibfnamefont{G.~L.} \bibnamefont{Scornet}},
  \bibinfo{author}{\bibfnamefont{V.}~\bibnamefont{Lyapin}},
  \bibinfo{author}{\bibfnamefont{C.}~\bibnamefont{Mieh\'{e}}},
  \bibinfo{author}{\bibfnamefont{F.}~\bibnamefont{Nowacki}}, \emph{et~al.},
  \bibinfo{year}{2002}, \bibinfo{journal}{Nuclear Instruments and Methods in
  Physics Research Section A: Accelerators, Spectrometers, Detectors and
  Associated Equipment} \textbf{\bibinfo{volume}{481}}(\bibinfo{number}{1-3}),
  \bibinfo{pages}{464}.

\bibitem[{\citenamefont{Raimann} \emph{et~al.}(1996)\citenamefont{Raimann,
  Ozawa, Boyd, Chloupek, Fujimaki, Kimura, Kobayashi, Kolata, Kubono, Tanihata,
  Watanabe, and Yoshida}}]{Raimann:1996}
\bibinfo{author}{\bibnamefont{Raimann}, \bibfnamefont{G.}},
  \bibinfo{author}{\bibfnamefont{A.}~\bibnamefont{Ozawa}},
  \bibinfo{author}{\bibfnamefont{R.~N.} \bibnamefont{Boyd}},
  \bibinfo{author}{\bibfnamefont{F.~R.} \bibnamefont{Chloupek}},
  \bibinfo{author}{\bibfnamefont{M.}~\bibnamefont{Fujimaki}},
  \bibinfo{author}{\bibfnamefont{K.}~\bibnamefont{Kimura}},
  \bibinfo{author}{\bibfnamefont{T.}~\bibnamefont{Kobayashi}},
  \bibinfo{author}{\bibfnamefont{J.~J.} \bibnamefont{Kolata}},
  \bibinfo{author}{\bibfnamefont{S.}~\bibnamefont{Kubono}},
  \bibinfo{author}{\bibfnamefont{I.}~\bibnamefont{Tanihata}},
  \bibinfo{author}{\bibfnamefont{Y.}~\bibnamefont{Watanabe}}, and
  \bibinfo{author}{\bibfnamefont{K.}~\bibnamefont{Yoshida}},
  \bibinfo{year}{1996}, \bibinfo{journal}{Phys. Rev. C}
  \textbf{\bibinfo{volume}{53}}(\bibinfo{number}{1}), \bibinfo{pages}{453}.

\bibitem[{\citenamefont{Rasmussen}(1966)}]{Rasmussen:1966}
\bibinfo{author}{\bibnamefont{Rasmussen}, \bibfnamefont{J.}},
  \bibinfo{year}{1966}, \emph{\bibinfo{title}{Alpha-, Beta-, and Gamma-Ray
  Spectroscopy}} (\bibinfo{publisher}{North-Holland}), p. \bibinfo{pages}{701}.

\bibitem[{\citenamefont{Reed} \emph{et~al.}(1999)\citenamefont{Reed, Tarasov,
  Page, Guillemaud-Mueller, Penionzhkevich, Allatt, Ang\'elique, Anne, Borcea,
  Burjan, Catford, Dlouh\'y} \emph{et~al.}}]{Reed:1999}
\bibinfo{author}{\bibnamefont{Reed}, \bibfnamefont{A.~T.}},
  \bibinfo{author}{\bibfnamefont{O.}~\bibnamefont{Tarasov}},
  \bibinfo{author}{\bibfnamefont{R.~D.} \bibnamefont{Page}},
  \bibinfo{author}{\bibfnamefont{D.}~\bibnamefont{Guillemaud-Mueller}},
  \bibinfo{author}{\bibfnamefont{Y.~E.} \bibnamefont{Penionzhkevich}},
  \bibinfo{author}{\bibfnamefont{R.~G.} \bibnamefont{Allatt}},
  \bibinfo{author}{\bibfnamefont{J.~C.} \bibnamefont{Ang\'elique}},
  \bibinfo{author}{\bibfnamefont{R.}~\bibnamefont{Anne}},
  \bibinfo{author}{\bibfnamefont{C.}~\bibnamefont{Borcea}},
  \bibinfo{author}{\bibfnamefont{V.}~\bibnamefont{Burjan}},
  \bibinfo{author}{\bibfnamefont{W.~N.} \bibnamefont{Catford}},
  \bibinfo{author}{\bibfnamefont{Z.}~\bibnamefont{Dlouh\'y}}, \emph{et~al.},
  \bibinfo{year}{1999}, \bibinfo{journal}{Phys. Rev. C}
  \textbf{\bibinfo{volume}{60}}(\bibinfo{number}{2}), \bibinfo{pages}{024311}.

\bibitem[{\citenamefont{Reisdorf}(1981)}]{Reisdorf:1981}
\bibinfo{author}{\bibnamefont{Reisdorf}, \bibfnamefont{W.}},
  \bibinfo{year}{1981}, \bibinfo{journal}{Zeitschrift für Physik A Hadrons and
  Nuclei} \textbf{\bibinfo{volume}{300}}, \bibinfo{pages}{227}.

\bibitem[{\citenamefont{Reisdorf and Schädel}(1992)}]{Reisdorf:1992}
\bibinfo{author}{\bibnamefont{Reisdorf}, \bibfnamefont{W.}}, and
  \bibinfo{author}{\bibfnamefont{M.}~\bibnamefont{Schädel}},
  \bibinfo{year}{1992}, \bibinfo{journal}{Zeitschrift für Physik A Hadrons and
  Nuclei} \textbf{\bibinfo{volume}{343}}, \bibinfo{pages}{47}.

\bibitem[{\citenamefont{Reiter} \emph{et~al.}(1999)\citenamefont{Reiter, Khoo,
  Lister, Seweryniak, Ahmad, Alcorta, Carpenter, Cizewski, Davids, Gervais,
  Greene, Henning} \emph{et~al.}}]{Reiter:1999}
\bibinfo{author}{\bibnamefont{Reiter}, \bibfnamefont{P.}},
  \bibinfo{author}{\bibfnamefont{T.~L.} \bibnamefont{Khoo}},
  \bibinfo{author}{\bibfnamefont{C.~J.} \bibnamefont{Lister}},
  \bibinfo{author}{\bibfnamefont{D.}~\bibnamefont{Seweryniak}},
  \bibinfo{author}{\bibfnamefont{I.}~\bibnamefont{Ahmad}},
  \bibinfo{author}{\bibfnamefont{M.}~\bibnamefont{Alcorta}},
  \bibinfo{author}{\bibfnamefont{M.~P.} \bibnamefont{Carpenter}},
  \bibinfo{author}{\bibfnamefont{J.~A.} \bibnamefont{Cizewski}},
  \bibinfo{author}{\bibfnamefont{C.~N.} \bibnamefont{Davids}},
  \bibinfo{author}{\bibfnamefont{G.}~\bibnamefont{Gervais}},
  \bibinfo{author}{\bibfnamefont{J.~P.} \bibnamefont{Greene}},
  \bibinfo{author}{\bibfnamefont{W.~F.} \bibnamefont{Henning}}, \emph{et~al.},
  \bibinfo{year}{1999}, \bibinfo{journal}{Phys. Rev. Lett.}
  \textbf{\bibinfo{volume}{82}}(\bibinfo{number}{3}), \bibinfo{pages}{509}.

\bibitem[{\citenamefont{Ricciardi} \emph{et~al.}(2006)\citenamefont{Ricciardi,
  Armbruster, Benlliure, Bernas, Boudard, Czajkowski, Enqvist,
  Keli\ifmmode~\acute{c}\else \'{c}\fi{}, Leray, Legrain, Mustapha, Pereira}
  \emph{et~al.}}]{Ricciardi:2006}
\bibinfo{author}{\bibnamefont{Ricciardi}, \bibfnamefont{M.~V.}},
  \bibinfo{author}{\bibfnamefont{P.}~\bibnamefont{Armbruster}},
  \bibinfo{author}{\bibfnamefont{J.}~\bibnamefont{Benlliure}},
  \bibinfo{author}{\bibfnamefont{M.}~\bibnamefont{Bernas}},
  \bibinfo{author}{\bibfnamefont{A.}~\bibnamefont{Boudard}},
  \bibinfo{author}{\bibfnamefont{S.}~\bibnamefont{Czajkowski}},
  \bibinfo{author}{\bibfnamefont{T.}~\bibnamefont{Enqvist}},
  \bibinfo{author}{\bibfnamefont{A.}~\bibnamefont{Keli\ifmmode~\acute{c}\else
  \'{c}\fi{}}}, \bibinfo{author}{\bibfnamefont{S.}~\bibnamefont{Leray}},
  \bibinfo{author}{\bibfnamefont{R.}~\bibnamefont{Legrain}},
  \bibinfo{author}{\bibfnamefont{B.}~\bibnamefont{Mustapha}},
  \bibinfo{author}{\bibfnamefont{J.}~\bibnamefont{Pereira}}, \emph{et~al.},
  \bibinfo{year}{2006}, \bibinfo{journal}{Phys. Rev. C}
  \textbf{\bibinfo{volume}{73}}(\bibinfo{number}{1}), \bibinfo{pages}{014607}.

\bibitem[{\citenamefont{Robinson} \emph{et~al.}(2003)\citenamefont{Robinson,
  Davids, Mukherjee, Seweryniak, Sinha, Wilt, and Woods}}]{Robinson:2003}
\bibinfo{author}{\bibnamefont{Robinson}, \bibfnamefont{A.}},
  \bibinfo{author}{\bibfnamefont{C.}~\bibnamefont{Davids}},
  \bibinfo{author}{\bibfnamefont{G.}~\bibnamefont{Mukherjee}},
  \bibinfo{author}{\bibfnamefont{D.}~\bibnamefont{Seweryniak}},
  \bibinfo{author}{\bibfnamefont{S.}~\bibnamefont{Sinha}},
  \bibinfo{author}{\bibfnamefont{P.}~\bibnamefont{Wilt}}, and
  \bibinfo{author}{\bibfnamefont{P.}~\bibnamefont{Woods}},
  \bibinfo{year}{2003}, \bibinfo{journal}{Phys.\ Rev.}
  \textbf{\bibinfo{volume}{C~68}}, \bibinfo{pages}{054301}.

\bibitem[{\citenamefont{Robinson} \emph{et~al.}(2005)\citenamefont{Robinson,
  Woods, Seweryniak, Davids, Carpenter, Hecht, Peterson, Sinha, Walters, and
  Zhu}}]{Robinson:2005}
\bibinfo{author}{\bibnamefont{Robinson}, \bibfnamefont{A.}},
  \bibinfo{author}{\bibfnamefont{P.}~\bibnamefont{Woods}},
  \bibinfo{author}{\bibfnamefont{D.}~\bibnamefont{Seweryniak}},
  \bibinfo{author}{\bibfnamefont{C.}~\bibnamefont{Davids}},
  \bibinfo{author}{\bibfnamefont{M.}~\bibnamefont{Carpenter}},
  \bibinfo{author}{\bibfnamefont{A.}~\bibnamefont{Hecht}},
  \bibinfo{author}{\bibfnamefont{D.}~\bibnamefont{Peterson}},
  \bibinfo{author}{\bibfnamefont{S.}~\bibnamefont{Sinha}},
  \bibinfo{author}{\bibfnamefont{W.}~\bibnamefont{Walters}}, and
  \bibinfo{author}{\bibfnamefont{S.}~\bibnamefont{Zhu}}, \bibinfo{year}{2005},
  \bibinfo{journal}{Phys.\ Rev.\ Lett.} \textbf{\bibinfo{volume}{95}},
  \bibinfo{pages}{032502}.

\bibitem[{\citenamefont{Robson}(1975)}]{Robson:1975}
\bibinfo{author}{\bibnamefont{Robson}, \bibfnamefont{D.}},
  \bibinfo{year}{1975}, in \emph{\bibinfo{booktitle}{Nuclear Spectroscopy and
  Reactions}}, edited by
  \bibinfo{editor}{\bibfnamefont{J.}~\bibnamefont{Cerny}}
  (\bibinfo{publisher}{Academic Press}), volume~\bibinfo{volume}{D}, pp.
  \bibinfo{pages}{179--248}.

\bibitem[{\citenamefont{Rochman} \emph{et~al.}(2004)\citenamefont{Rochman,
  Tsekhanovich, Gönnenwein, Sokolov, Storrer, Simpson, and
  Serot}}]{Rochman:2004}
\bibinfo{author}{\bibnamefont{Rochman}, \bibfnamefont{D.}},
  \bibinfo{author}{\bibfnamefont{I.}~\bibnamefont{Tsekhanovich}},
  \bibinfo{author}{\bibfnamefont{F.}~\bibnamefont{Gönnenwein}},
  \bibinfo{author}{\bibfnamefont{V.}~\bibnamefont{Sokolov}},
  \bibinfo{author}{\bibfnamefont{F.}~\bibnamefont{Storrer}},
  \bibinfo{author}{\bibfnamefont{G.}~\bibnamefont{Simpson}}, and
  \bibinfo{author}{\bibfnamefont{O.}~\bibnamefont{Serot}},
  \bibinfo{year}{2004}, \bibinfo{journal}{Nuclear Physics A}
  \textbf{\bibinfo{volume}{735}}(\bibinfo{number}{1-2}), \bibinfo{pages}{3}.

\bibitem[{\citenamefont{Rodin} \emph{et~al.}(2003)\citenamefont{Rodin,
  Stepantsov, Bogdanov, Golovkov, Fomichev, Sidorchuk, Slepnev, Wolski,
  Ter-Akopian, Oganessian, Yukhimchuk, Perevozchikov}
  \emph{et~al.}}]{Rodin:2003}
\bibinfo{author}{\bibnamefont{Rodin}, \bibfnamefont{A.~M.}},
  \bibinfo{author}{\bibfnamefont{S.~V.} \bibnamefont{Stepantsov}},
  \bibinfo{author}{\bibfnamefont{D.~D.} \bibnamefont{Bogdanov}},
  \bibinfo{author}{\bibfnamefont{M.~S.} \bibnamefont{Golovkov}},
  \bibinfo{author}{\bibfnamefont{A.~S.} \bibnamefont{Fomichev}},
  \bibinfo{author}{\bibfnamefont{S.~I.} \bibnamefont{Sidorchuk}},
  \bibinfo{author}{\bibfnamefont{R.~S.} \bibnamefont{Slepnev}},
  \bibinfo{author}{\bibfnamefont{R.}~\bibnamefont{Wolski}},
  \bibinfo{author}{\bibfnamefont{G.~M.} \bibnamefont{Ter-Akopian}},
  \bibinfo{author}{\bibfnamefont{Y.~T.} \bibnamefont{Oganessian}},
  \bibinfo{author}{\bibfnamefont{A.~A.} \bibnamefont{Yukhimchuk}},
  \bibinfo{author}{\bibfnamefont{V.~V.} \bibnamefont{Perevozchikov}},
  \emph{et~al.}, \bibinfo{year}{2003}, \bibinfo{journal}{Nuclear Instruments
  and Methods in Physics Research Section B: Beam Interactions with Materials
  and Atoms} \textbf{\bibinfo{volume}{204}}, \bibinfo{pages}{114}.

\bibitem[{\citenamefont{Roeckl}(1995)}]{Roeckl:1995}
\bibinfo{author}{\bibnamefont{Roeckl}, \bibfnamefont{E.}},
  \bibinfo{year}{1995}, \bibinfo{journal}{Radiochimica Acta}
  \textbf{\bibinfo{volume}{70/71}}, \bibinfo{pages}{107}.

\bibitem[{\citenamefont{Roeckl}(1996)}]{Roeckl:1996}
\bibinfo{author}{\bibnamefont{Roeckl}, \bibfnamefont{E.}},
  \bibinfo{year}{1996}, \emph{\bibinfo{title}{Nuclear Decay Modes.}}
  (\bibinfo{publisher}{Institute of Physics Publishing}), chapter
  \bibinfo{chapter}{Alpha Decay}, pp. \bibinfo{pages}{237--274}.

\bibitem[{\citenamefont{Rogers} \emph{et~al.}(2011)\citenamefont{Rogers,
  Famiano, Lynch, Wallace, Amorini, Bazin, Charity, Delaunay, de~Souza, Elson,
  Gade, Galaviz} \emph{et~al.}}]{Rogers:2011}
\bibinfo{author}{\bibnamefont{Rogers}, \bibfnamefont{A.~M.}},
  \bibinfo{author}{\bibfnamefont{M.~A.} \bibnamefont{Famiano}},
  \bibinfo{author}{\bibfnamefont{W.~G.} \bibnamefont{Lynch}},
  \bibinfo{author}{\bibfnamefont{M.~S.} \bibnamefont{Wallace}},
  \bibinfo{author}{\bibfnamefont{F.}~\bibnamefont{Amorini}},
  \bibinfo{author}{\bibfnamefont{D.}~\bibnamefont{Bazin}},
  \bibinfo{author}{\bibfnamefont{R.~J.} \bibnamefont{Charity}},
  \bibinfo{author}{\bibfnamefont{F.}~\bibnamefont{Delaunay}},
  \bibinfo{author}{\bibfnamefont{R.~T.} \bibnamefont{de~Souza}},
  \bibinfo{author}{\bibfnamefont{J.}~\bibnamefont{Elson}},
  \bibinfo{author}{\bibfnamefont{A.}~\bibnamefont{Gade}},
  \bibinfo{author}{\bibfnamefont{D.}~\bibnamefont{Galaviz}}, \emph{et~al.},
  \bibinfo{year}{2011}, \bibinfo{journal}{Phys. Rev. Lett.}
  \textbf{\bibinfo{volume}{106}}(\bibinfo{number}{25}),
  \bibinfo{pages}{252503}.

\bibitem[{\citenamefont{Rose and Jones}(1984)}]{Rose:1984}
\bibinfo{author}{\bibnamefont{Rose}, \bibfnamefont{H.}}, and
  \bibinfo{author}{\bibfnamefont{G.}~\bibnamefont{Jones}},
  \bibinfo{year}{1984}, \bibinfo{journal}{Nature}
  \textbf{\bibinfo{volume}{307}}, \bibinfo{pages}{245}.

\bibitem[{\citenamefont{Rotureau} \emph{et~al.}(2005)\citenamefont{Rotureau,
  Oko{\l}owicz, and P{\l}oszajczak}}]{Rotureau:2005}
\bibinfo{author}{\bibnamefont{Rotureau}, \bibfnamefont{J.}},
  \bibinfo{author}{\bibfnamefont{J.}~\bibnamefont{Oko{\l}owicz}}, and
  \bibinfo{author}{\bibfnamefont{M.}~\bibnamefont{P{\l}oszajczak}},
  \bibinfo{year}{2005}, \bibinfo{journal}{Phys. Rev. Lett.}
  \textbf{\bibinfo{volume}{95}}, \bibinfo{pages}{042503}, \bibinfo{note}{4
  pages}.

\bibitem[{\citenamefont{Rotureau} \emph{et~al.}(2006)\citenamefont{Rotureau,
  Okołowicz, and Płoszajczak}}]{Rotureau:2006}
\bibinfo{author}{\bibnamefont{Rotureau}, \bibfnamefont{J.}},
  \bibinfo{author}{\bibfnamefont{J.}~\bibnamefont{Okołowicz}}, and
  \bibinfo{author}{\bibfnamefont{M.}~\bibnamefont{Płoszajczak}},
  \bibinfo{year}{2006}, \bibinfo{journal}{Nuclear Physics A}
  \textbf{\bibinfo{volume}{767}}, \bibinfo{pages}{13 }, ISSN
  \bibinfo{issn}{0375-9474}.

\bibitem[{\citenamefont{Rubio} \emph{et~al.}(2005)\citenamefont{Rubio,
  Gelletly, N\'{a}cher, Algora, Ta\'{\i}n, P\'{e}rez, and
  Caballero}}]{Rubio:2005}
\bibinfo{author}{\bibnamefont{Rubio}, \bibfnamefont{B.}},
  \bibinfo{author}{\bibfnamefont{W.}~\bibnamefont{Gelletly}},
  \bibinfo{author}{\bibfnamefont{E.}~\bibnamefont{N\'{a}cher}},
  \bibinfo{author}{\bibfnamefont{A.}~\bibnamefont{Algora}},
  \bibinfo{author}{\bibfnamefont{J.~L.} \bibnamefont{Ta\'{\i}n}},
  \bibinfo{author}{\bibfnamefont{A.}~\bibnamefont{P\'{e}rez}}, and
  \bibinfo{author}{\bibfnamefont{L.}~\bibnamefont{Caballero}},
  \bibinfo{year}{2005}, \bibinfo{journal}{Journal of Physics G: Nuclear and
  Particle Physics} \textbf{\bibinfo{volume}{31}}(\bibinfo{number}{10}),
  \bibinfo{pages}{S1477}.

\bibitem[{\citenamefont{Rudolph}(2002)}]{Rudolph:2002}
\bibinfo{author}{\bibnamefont{Rudolph}, \bibfnamefont{D.}},
  \bibinfo{year}{2002}, \bibinfo{journal}{The European Physical Journal A -
  Hadrons and Nuclei} \textbf{\bibinfo{volume}{15}}, \bibinfo{pages}{161}.

\bibitem[{\citenamefont{Rudolph} \emph{et~al.}(2008)\citenamefont{Rudolph,
  Hoischen, Hellstr\"om, Pietri, Podoly\'ak, Regan, Garnsworthy, Steer, Becker,
  Bednarczyk, C\'aceres, Doornenbal} \emph{et~al.}}]{Rudolph:2008}
\bibinfo{author}{\bibnamefont{Rudolph}, \bibfnamefont{D.}},
  \bibinfo{author}{\bibfnamefont{R.}~\bibnamefont{Hoischen}},
  \bibinfo{author}{\bibfnamefont{M.}~\bibnamefont{Hellstr\"om}},
  \bibinfo{author}{\bibfnamefont{S.}~\bibnamefont{Pietri}},
  \bibinfo{author}{\bibfnamefont{Z.}~\bibnamefont{Podoly\'ak}},
  \bibinfo{author}{\bibfnamefont{P.~H.} \bibnamefont{Regan}},
  \bibinfo{author}{\bibfnamefont{A.~B.} \bibnamefont{Garnsworthy}},
  \bibinfo{author}{\bibfnamefont{S.~J.} \bibnamefont{Steer}},
  \bibinfo{author}{\bibfnamefont{F.}~\bibnamefont{Becker}},
  \bibinfo{author}{\bibfnamefont{P.}~\bibnamefont{Bednarczyk}},
  \bibinfo{author}{\bibfnamefont{L.}~\bibnamefont{C\'aceres}},
  \bibinfo{author}{\bibfnamefont{P.}~\bibnamefont{Doornenbal}}, \emph{et~al.},
  \bibinfo{year}{2008}, \bibinfo{journal}{Phys. Rev. C}
  \textbf{\bibinfo{volume}{78}}(\bibinfo{number}{2}), \bibinfo{pages}{021301}.

\bibitem[{\citenamefont{Rutherford}(1899)}]{Rutherford:1899}
\bibinfo{author}{\bibnamefont{Rutherford}, \bibfnamefont{E.}},
  \bibinfo{year}{1899}, \bibinfo{journal}{Phil. Mag.}
  \textbf{\bibinfo{volume}{47}}, \bibinfo{pages}{109}.

\bibitem[{\citenamefont{Rutherford}(1911)}]{Rutherford:1911}
\bibinfo{author}{\bibnamefont{Rutherford}, \bibfnamefont{E.}},
  \bibinfo{year}{1911}, \bibinfo{journal}{Phil. Mag.}
  \textbf{\bibinfo{volume}{21}}, \bibinfo{pages}{669}.

\bibitem[{\citenamefont{Rutherford and da~C.~Andrade}(1914)}]{Rutherford:1914}
\bibinfo{author}{\bibnamefont{Rutherford}, \bibfnamefont{E.}}, and
  \bibinfo{author}{\bibfnamefont{E.}~\bibnamefont{da~C.~Andrade}},
  \bibinfo{year}{1914}, \bibinfo{journal}{Phil. Mag.}
  \textbf{\bibinfo{volume}{27}}, \bibinfo{pages}{854}.

\bibitem[{\citenamefont{Rutherford and Geiger}(1908)}]{Rutherford:1908}
\bibinfo{author}{\bibnamefont{Rutherford}, \bibfnamefont{E.}}, and
  \bibinfo{author}{\bibfnamefont{H.}~\bibnamefont{Geiger}},
  \bibinfo{year}{1908}, \bibinfo{journal}{Proc. Roy. Soc.}
  \textbf{\bibinfo{volume}{A81}}, \bibinfo{pages}{162}.

\bibitem[{\citenamefont{Rykaczewski}(2002)}]{Rykaczewski:2002}
\bibinfo{author}{\bibnamefont{Rykaczewski}, \bibfnamefont{K.}},
  \bibinfo{year}{2002}, \bibinfo{journal}{Eur.\ Phys.\ J.}
  \textbf{\bibinfo{volume}{A~15}}, \bibinfo{pages}{81}.

\bibitem[{\citenamefont{Rykaczewski}
  \emph{et~al.}(1999)\citenamefont{Rykaczewski, Batchelder, Bingham, Davinson,
  Ginter, Gross, Grzywacz, Karny, MacDonald, Mas, McConnell, Piechaczek}
  \emph{et~al.}}]{Rykaczewski:1999}
\bibinfo{author}{\bibnamefont{Rykaczewski}, \bibfnamefont{K.}},
  \bibinfo{author}{\bibfnamefont{J.}~\bibnamefont{Batchelder}},
  \bibinfo{author}{\bibfnamefont{C.}~\bibnamefont{Bingham}},
  \bibinfo{author}{\bibfnamefont{T.}~\bibnamefont{Davinson}},
  \bibinfo{author}{\bibfnamefont{T.}~\bibnamefont{Ginter}},
  \bibinfo{author}{\bibfnamefont{C.}~\bibnamefont{Gross}},
  \bibinfo{author}{\bibfnamefont{R.}~\bibnamefont{Grzywacz}},
  \bibinfo{author}{\bibfnamefont{M.}~\bibnamefont{Karny}},
  \bibinfo{author}{\bibfnamefont{B.}~\bibnamefont{MacDonald}},
  \bibinfo{author}{\bibfnamefont{J.}~\bibnamefont{Mas}},
  \bibinfo{author}{\bibfnamefont{J.}~\bibnamefont{McConnell}},
  \bibinfo{author}{\bibfnamefont{A.}~\bibnamefont{Piechaczek}}, \emph{et~al.},
  \bibinfo{year}{1999}, \bibinfo{journal}{Phys.\ Rev.}
  \textbf{\bibinfo{volume}{C~60}}, \bibinfo{pages}{011301}.

\bibitem[{\citenamefont{Rykaczewski}
  \emph{et~al.}(2001{\natexlab{a}})\citenamefont{Rykaczewski, Batchelder,
  Bingham, Ginter, Gross, Grzywacz, Hamilton, Hartley, Janas, Karny, Kulp,
  Lipoglavsek} \emph{et~al.}}]{Rykaczewski:2001b}
\bibinfo{author}{\bibnamefont{Rykaczewski}, \bibfnamefont{K.}},
  \bibinfo{author}{\bibfnamefont{J.}~\bibnamefont{Batchelder}},
  \bibinfo{author}{\bibfnamefont{C.}~\bibnamefont{Bingham}},
  \bibinfo{author}{\bibfnamefont{T.}~\bibnamefont{Ginter}},
  \bibinfo{author}{\bibfnamefont{C.}~\bibnamefont{Gross}},
  \bibinfo{author}{\bibfnamefont{R.}~\bibnamefont{Grzywacz}},
  \bibinfo{author}{\bibfnamefont{J.}~\bibnamefont{Hamilton}},
  \bibinfo{author}{\bibfnamefont{D.}~\bibnamefont{Hartley}},
  \bibinfo{author}{\bibfnamefont{Z.}~\bibnamefont{Janas}},
  \bibinfo{author}{\bibfnamefont{M.}~\bibnamefont{Karny}},
  \bibinfo{author}{\bibfnamefont{W.}~\bibnamefont{Kulp}},
  \bibinfo{author}{\bibfnamefont{M.}~\bibnamefont{Lipoglavsek}}, \emph{et~al.},
  \bibinfo{year}{2001}{\natexlab{a}}, \bibinfo{journal}{Acta Phys.\ Pol.}
  \textbf{\bibinfo{volume}{B~32}}, \bibinfo{pages}{971}.

\bibitem[{\citenamefont{Rykaczewski}
  \emph{et~al.}(2001{\natexlab{b}})\citenamefont{Rykaczewski, Grzywacz, Karny,
  McConnell, Momayezi, Wahl, Janas, Batchelder, Hartley, Tantawy, Gross,
  Ginter} \emph{et~al.}}]{Rykaczewski:2001}
\bibinfo{author}{\bibnamefont{Rykaczewski}, \bibfnamefont{K.}},
  \bibinfo{author}{\bibfnamefont{R.}~\bibnamefont{Grzywacz}},
  \bibinfo{author}{\bibfnamefont{M.}~\bibnamefont{Karny}},
  \bibinfo{author}{\bibfnamefont{J.}~\bibnamefont{McConnell}},
  \bibinfo{author}{\bibfnamefont{M.}~\bibnamefont{Momayezi}},
  \bibinfo{author}{\bibfnamefont{J.}~\bibnamefont{Wahl}},
  \bibinfo{author}{\bibfnamefont{Z.}~\bibnamefont{Janas}},
  \bibinfo{author}{\bibfnamefont{J.}~\bibnamefont{Batchelder}},
  \bibinfo{author}{\bibfnamefont{C.~B.~D.} \bibnamefont{Hartley}},
  \bibinfo{author}{\bibfnamefont{M.}~\bibnamefont{Tantawy}},
  \bibinfo{author}{\bibfnamefont{C.}~\bibnamefont{Gross}},
  \bibinfo{author}{\bibfnamefont{T.}~\bibnamefont{Ginter}}, \emph{et~al.},
  \bibinfo{year}{2001}{\natexlab{b}}, \bibinfo{journal}{Nucl.\ Phys.}
  \textbf{\bibinfo{volume}{A~682}}, \bibinfo{pages}{270c}.

\bibitem[{\citenamefont{Sagawa} \emph{et~al.}(1993)\citenamefont{Sagawa,
  Hamamoto, and Ishihara}}]{Sagawa:1993}
\bibinfo{author}{\bibnamefont{Sagawa}, \bibfnamefont{H.}},
  \bibinfo{author}{\bibfnamefont{I.}~\bibnamefont{Hamamoto}}, and
  \bibinfo{author}{\bibfnamefont{M.}~\bibnamefont{Ishihara}},
  \bibinfo{year}{1993}, \bibinfo{journal}{Physics Letters B}
  \textbf{\bibinfo{volume}{303}}(\bibinfo{number}{3-4}), \bibinfo{pages}{215}.

\bibitem[{\citenamefont{Sakurai}(2008)}]{Sakurai:2008}
\bibinfo{author}{\bibnamefont{Sakurai}, \bibfnamefont{H.}},
  \bibinfo{year}{2008}, \bibinfo{journal}{Nuclear Physics A}
  \textbf{\bibinfo{volume}{805}}(\bibinfo{number}{1-4}), \bibinfo{pages}{526c}.

\bibitem[{\citenamefont{Schardt and Riisager}(1993)}]{Schardt:1993}
\bibinfo{author}{\bibnamefont{Schardt}, \bibfnamefont{D.}}, and
  \bibinfo{author}{\bibfnamefont{K.}~\bibnamefont{Riisager}},
  \bibinfo{year}{1993}, \bibinfo{journal}{Zeitschrift für Physik A Hadrons and
  Nuclei} \textbf{\bibinfo{volume}{345}}, \bibinfo{pages}{265}.

\bibitem[{\citenamefont{Schatz} \emph{et~al.}(1998)\citenamefont{Schatz,
  Aprahamian, Görres, Wiescher, Rauscher, Rembges, Thielemann, Pfeiffer,
  Möller, Kratz, Herndl, Brown} \emph{et~al.}}]{Schatz:1998}
\bibinfo{author}{\bibnamefont{Schatz}, \bibfnamefont{H.}},
  \bibinfo{author}{\bibfnamefont{A.}~\bibnamefont{Aprahamian}},
  \bibinfo{author}{\bibfnamefont{J.}~\bibnamefont{Görres}},
  \bibinfo{author}{\bibfnamefont{M.}~\bibnamefont{Wiescher}},
  \bibinfo{author}{\bibfnamefont{T.}~\bibnamefont{Rauscher}},
  \bibinfo{author}{\bibfnamefont{J.~F.} \bibnamefont{Rembges}},
  \bibinfo{author}{\bibfnamefont{F.~K.} \bibnamefont{Thielemann}},
  \bibinfo{author}{\bibfnamefont{B.}~\bibnamefont{Pfeiffer}},
  \bibinfo{author}{\bibfnamefont{P.}~\bibnamefont{Möller}},
  \bibinfo{author}{\bibfnamefont{K.~L.} \bibnamefont{Kratz}},
  \bibinfo{author}{\bibfnamefont{H.}~\bibnamefont{Herndl}},
  \bibinfo{author}{\bibfnamefont{B.~A.} \bibnamefont{Brown}}, \emph{et~al.},
  \bibinfo{year}{1998}, \bibinfo{journal}{Physics Reports}
  \textbf{\bibinfo{volume}{294}}(\bibinfo{number}{4}), \bibinfo{pages}{167}.

\bibitem[{\citenamefont{Schmidt} \emph{et~al.}(2002)\citenamefont{Schmidt,
  Benlliure, Enqvist, Junghans, Rejmund, and Ricciardi}}]{Schmidt:2002}
\bibinfo{author}{\bibnamefont{Schmidt}, \bibfnamefont{K.~H.}},
  \bibinfo{author}{\bibfnamefont{J.}~\bibnamefont{Benlliure}},
  \bibinfo{author}{\bibfnamefont{T.}~\bibnamefont{Enqvist}},
  \bibinfo{author}{\bibfnamefont{A.~R.} \bibnamefont{Junghans}},
  \bibinfo{author}{\bibfnamefont{F.}~\bibnamefont{Rejmund}}, and
  \bibinfo{author}{\bibfnamefont{M.~V.} \bibnamefont{Ricciardi}},
  \bibinfo{year}{2002}, \bibinfo{journal}{Nuclear Physics A}
  \textbf{\bibinfo{volume}{701}}(\bibinfo{number}{1-4}), \bibinfo{pages}{115}.

\bibitem[{\citenamefont{Schmidt} \emph{et~al.}(1993)\citenamefont{Schmidt,
  Brohm, Clerc, Dornik, Fauerbach, Geissel, Grewe, Hanelt, Junghans, Magel,
  Morawek, Münzenberg} \emph{et~al.}}]{Schmidt:1993}
\bibinfo{author}{\bibnamefont{Schmidt}, \bibfnamefont{K.~H.}},
  \bibinfo{author}{\bibfnamefont{T.}~\bibnamefont{Brohm}},
  \bibinfo{author}{\bibfnamefont{H.~G.} \bibnamefont{Clerc}},
  \bibinfo{author}{\bibfnamefont{M.}~\bibnamefont{Dornik}},
  \bibinfo{author}{\bibfnamefont{M.}~\bibnamefont{Fauerbach}},
  \bibinfo{author}{\bibfnamefont{H.}~\bibnamefont{Geissel}},
  \bibinfo{author}{\bibfnamefont{A.}~\bibnamefont{Grewe}},
  \bibinfo{author}{\bibfnamefont{E.}~\bibnamefont{Hanelt}},
  \bibinfo{author}{\bibfnamefont{A.}~\bibnamefont{Junghans}},
  \bibinfo{author}{\bibfnamefont{A.}~\bibnamefont{Magel}},
  \bibinfo{author}{\bibfnamefont{W.}~\bibnamefont{Morawek}},
  \bibinfo{author}{\bibfnamefont{G.}~\bibnamefont{Münzenberg}}, \emph{et~al.},
  \bibinfo{year}{1993}, \bibinfo{journal}{Physics Letters B}
  \textbf{\bibinfo{volume}{300}}(\bibinfo{number}{4}), \bibinfo{pages}{313}.

\bibitem[{\citenamefont{Schmidt and Morawek}(1991)}]{Schmidt:1991}
\bibinfo{author}{\bibnamefont{Schmidt}, \bibfnamefont{K.~H.}}, and
  \bibinfo{author}{\bibfnamefont{W.}~\bibnamefont{Morawek}},
  \bibinfo{year}{1991}, \bibinfo{journal}{Reports on Progress in Physics}
  \textbf{\bibinfo{volume}{54}}(\bibinfo{number}{7}), \bibinfo{pages}{949}.

\bibitem[{\citenamefont{Schneider} \emph{et~al.}(1994)\citenamefont{Schneider,
  Friese, Reinhold, Zeitelhack, Faestermann, Gernhäuser, Gilg, Heine, Homolka,
  Kienle, Körner, Geissel} \emph{et~al.}}]{Schneider:1994}
\bibinfo{author}{\bibnamefont{Schneider}, \bibfnamefont{R.}},
  \bibinfo{author}{\bibfnamefont{J.}~\bibnamefont{Friese}},
  \bibinfo{author}{\bibfnamefont{J.}~\bibnamefont{Reinhold}},
  \bibinfo{author}{\bibfnamefont{K.}~\bibnamefont{Zeitelhack}},
  \bibinfo{author}{\bibfnamefont{T.}~\bibnamefont{Faestermann}},
  \bibinfo{author}{\bibfnamefont{R.}~\bibnamefont{Gernhäuser}},
  \bibinfo{author}{\bibfnamefont{H.}~\bibnamefont{Gilg}},
  \bibinfo{author}{\bibfnamefont{F.}~\bibnamefont{Heine}},
  \bibinfo{author}{\bibfnamefont{J.}~\bibnamefont{Homolka}},
  \bibinfo{author}{\bibfnamefont{P.}~\bibnamefont{Kienle}},
  \bibinfo{author}{\bibfnamefont{H.~J.} \bibnamefont{Körner}},
  \bibinfo{author}{\bibfnamefont{H.}~\bibnamefont{Geissel}}, \emph{et~al.},
  \bibinfo{year}{1994}, \bibinfo{journal}{Zeitschrift für Physik A Hadrons and
  Nuclei} \textbf{\bibinfo{volume}{348}}, \bibinfo{pages}{241}.

\bibitem[{\citenamefont{Scholey} \emph{et~al.}(2005)\citenamefont{Scholey,
  Sandzelius, Eeckhaudt, Grahn, Greenlees, Jones, Julin, Juutinen, Leino,
  Lepp\"anen, Nieminen, Nyman} \emph{et~al.}}]{Scholey:2005}
\bibinfo{author}{\bibnamefont{Scholey}, \bibfnamefont{C.}},
  \bibinfo{author}{\bibfnamefont{M.}~\bibnamefont{Sandzelius}},
  \bibinfo{author}{\bibfnamefont{S.}~\bibnamefont{Eeckhaudt}},
  \bibinfo{author}{\bibfnamefont{T.}~\bibnamefont{Grahn}},
  \bibinfo{author}{\bibfnamefont{P.}~\bibnamefont{Greenlees}},
  \bibinfo{author}{\bibfnamefont{P.}~\bibnamefont{Jones}},
  \bibinfo{author}{\bibfnamefont{R.}~\bibnamefont{Julin}},
  \bibinfo{author}{\bibfnamefont{S.}~\bibnamefont{Juutinen}},
  \bibinfo{author}{\bibfnamefont{M.}~\bibnamefont{Leino}},
  \bibinfo{author}{\bibfnamefont{A.-P.} \bibnamefont{Lepp\"anen}},
  \bibinfo{author}{\bibfnamefont{P.}~\bibnamefont{Nieminen}},
  \bibinfo{author}{\bibfnamefont{M.}~\bibnamefont{Nyman}}, \emph{et~al.},
  \bibinfo{year}{2005}, \bibinfo{journal}{J.\ Phys.}
  \textbf{\bibinfo{volume}{G~31}}, \bibinfo{pages}{S1719}.

\bibitem[{\citenamefont{Sellin} \emph{et~al.}(1993)\citenamefont{Sellin, Woods,
  Davinson, Davis, Livingston, Page, Shotter, Hofmann, and
  James}}]{Sellin:1993}
\bibinfo{author}{\bibnamefont{Sellin}, \bibfnamefont{P.}},
  \bibinfo{author}{\bibfnamefont{P.}~\bibnamefont{Woods}},
  \bibinfo{author}{\bibfnamefont{T.}~\bibnamefont{Davinson}},
  \bibinfo{author}{\bibfnamefont{N.}~\bibnamefont{Davis}},
  \bibinfo{author}{\bibfnamefont{K.}~\bibnamefont{Livingston}},
  \bibinfo{author}{\bibfnamefont{R.}~\bibnamefont{Page}},
  \bibinfo{author}{\bibfnamefont{A.}~\bibnamefont{Shotter}},
  \bibinfo{author}{\bibfnamefont{S.}~\bibnamefont{Hofmann}}, and
  \bibinfo{author}{\bibfnamefont{A.}~\bibnamefont{James}},
  \bibinfo{year}{1993}, \bibinfo{journal}{Phys.\ Rev.}
  \textbf{\bibinfo{volume}{C~47}}, \bibinfo{pages}{1933}.

\bibitem[{\citenamefont{Sellin} \emph{et~al.}(1992)\citenamefont{Sellin, Woods,
  Branford, Davinson, Davis, Ireland, Livingston, Page, Shotter, Hofmann, Hunt,
  James} \emph{et~al.}}]{Sellin:1992}
\bibinfo{author}{\bibnamefont{Sellin}, \bibfnamefont{P.~J.}},
  \bibinfo{author}{\bibfnamefont{P.~J.} \bibnamefont{Woods}},
  \bibinfo{author}{\bibfnamefont{D.}~\bibnamefont{Branford}},
  \bibinfo{author}{\bibfnamefont{T.}~\bibnamefont{Davinson}},
  \bibinfo{author}{\bibfnamefont{N.~J.} \bibnamefont{Davis}},
  \bibinfo{author}{\bibfnamefont{D.~G.} \bibnamefont{Ireland}},
  \bibinfo{author}{\bibfnamefont{K.}~\bibnamefont{Livingston}},
  \bibinfo{author}{\bibfnamefont{R.~D.} \bibnamefont{Page}},
  \bibinfo{author}{\bibfnamefont{A.~C.} \bibnamefont{Shotter}},
  \bibinfo{author}{\bibfnamefont{S.}~\bibnamefont{Hofmann}},
  \bibinfo{author}{\bibfnamefont{R.~A.} \bibnamefont{Hunt}},
  \bibinfo{author}{\bibfnamefont{A.~N.} \bibnamefont{James}}, \emph{et~al.},
  \bibinfo{year}{1992}, \bibinfo{journal}{Nuclear Instruments and Methods in
  Physics Research Section A: Accelerators, Spectrometers, Detectors and
  Associated Equipment} \textbf{\bibinfo{volume}{311}}(\bibinfo{number}{1-2}),
  \bibinfo{pages}{217}.

\bibitem[{\citenamefont{Severijns} \emph{et~al.}(2006)\citenamefont{Severijns,
  Beck, and Naviliat-Cuncic}}]{Severijns:2006}
\bibinfo{author}{\bibnamefont{Severijns}, \bibfnamefont{N.}},
  \bibinfo{author}{\bibfnamefont{M.}~\bibnamefont{Beck}}, and
  \bibinfo{author}{\bibfnamefont{O.}~\bibnamefont{Naviliat-Cuncic}},
  \bibinfo{year}{2006}, \bibinfo{journal}{Rev. Mod. Phys.}
  \textbf{\bibinfo{volume}{78}}(\bibinfo{number}{3}), \bibinfo{pages}{991}.

\bibitem[{\citenamefont{Seweryniak}
  \emph{et~al.}(1998)\citenamefont{Seweryniak, Ackermann, Amro, Brown,
  Carpenter, Conticchio, Davids, Fischer, Hackman, Hamada, Henderson, Janssens}
  \emph{et~al.}}]{Seweryniak:1998}
\bibinfo{author}{\bibnamefont{Seweryniak}, \bibfnamefont{D.}},
  \bibinfo{author}{\bibfnamefont{D.}~\bibnamefont{Ackermann}},
  \bibinfo{author}{\bibfnamefont{H.}~\bibnamefont{Amro}},
  \bibinfo{author}{\bibfnamefont{L.~T.} \bibnamefont{Brown}},
  \bibinfo{author}{\bibfnamefont{M.~P.} \bibnamefont{Carpenter}},
  \bibinfo{author}{\bibfnamefont{L.}~\bibnamefont{Conticchio}},
  \bibinfo{author}{\bibfnamefont{C.~N.} \bibnamefont{Davids}},
  \bibinfo{author}{\bibfnamefont{S.~M.} \bibnamefont{Fischer}},
  \bibinfo{author}{\bibfnamefont{G.}~\bibnamefont{Hackman}},
  \bibinfo{author}{\bibfnamefont{S.}~\bibnamefont{Hamada}},
  \bibinfo{author}{\bibfnamefont{D.~J.} \bibnamefont{Henderson}},
  \bibinfo{author}{\bibfnamefont{R.~V.~F.} \bibnamefont{Janssens}},
  \emph{et~al.}, \bibinfo{year}{1998}, \bibinfo{journal}{Phys. Rev. C}
  \textbf{\bibinfo{volume}{58}}(\bibinfo{number}{5}), \bibinfo{pages}{2710}.

\bibitem[{\citenamefont{Seweryniak}
  \emph{et~al.}(2007{\natexlab{a}})\citenamefont{Seweryniak, Blank, Carpenter,
  Davids, Davinson, Freeman, Hammond, Hoteling, Janssens, Khoo, Liu, Mukherjee}
  \emph{et~al.}}]{Seweryniak:2007a}
\bibinfo{author}{\bibnamefont{Seweryniak}, \bibfnamefont{D.}},
  \bibinfo{author}{\bibfnamefont{B.}~\bibnamefont{Blank}},
  \bibinfo{author}{\bibfnamefont{M.}~\bibnamefont{Carpenter}},
  \bibinfo{author}{\bibfnamefont{C.}~\bibnamefont{Davids}},
  \bibinfo{author}{\bibfnamefont{T.}~\bibnamefont{Davinson}},
  \bibinfo{author}{\bibfnamefont{S.}~\bibnamefont{Freeman}},
  \bibinfo{author}{\bibfnamefont{N.}~\bibnamefont{Hammond}},
  \bibinfo{author}{\bibfnamefont{N.}~\bibnamefont{Hoteling}},
  \bibinfo{author}{\bibfnamefont{R.}~\bibnamefont{Janssens}},
  \bibinfo{author}{\bibfnamefont{T.}~\bibnamefont{Khoo}},
  \bibinfo{author}{\bibfnamefont{Z.}~\bibnamefont{Liu}},
  \bibinfo{author}{\bibfnamefont{G.}~\bibnamefont{Mukherjee}}, \emph{et~al.},
  \bibinfo{year}{2007}{\natexlab{a}}, \bibinfo{journal}{Phys.\ Rev. Lett.}
  \textbf{\bibinfo{volume}{99}}, \bibinfo{pages}{082502}.

\bibitem[{\citenamefont{Seweryniak}
  \emph{et~al.}(2007{\natexlab{b}})\citenamefont{Seweryniak, Carpenter, Gros,
  Hecht, Hoteling, Janssens, Khoo, Lauritsen, Lister, Lotay, Peterson,
  Robinson} \emph{et~al.}}]{Seweryniak:2007}
\bibinfo{author}{\bibnamefont{Seweryniak}, \bibfnamefont{D.}},
  \bibinfo{author}{\bibfnamefont{M.~P.} \bibnamefont{Carpenter}},
  \bibinfo{author}{\bibfnamefont{S.}~\bibnamefont{Gros}},
  \bibinfo{author}{\bibfnamefont{A.~A.} \bibnamefont{Hecht}},
  \bibinfo{author}{\bibfnamefont{N.}~\bibnamefont{Hoteling}},
  \bibinfo{author}{\bibfnamefont{R.~V.~F.} \bibnamefont{Janssens}},
  \bibinfo{author}{\bibfnamefont{T.~L.} \bibnamefont{Khoo}},
  \bibinfo{author}{\bibfnamefont{T.}~\bibnamefont{Lauritsen}},
  \bibinfo{author}{\bibfnamefont{C.~J.} \bibnamefont{Lister}},
  \bibinfo{author}{\bibfnamefont{G.}~\bibnamefont{Lotay}},
  \bibinfo{author}{\bibfnamefont{D.}~\bibnamefont{Peterson}},
  \bibinfo{author}{\bibfnamefont{A.~P.} \bibnamefont{Robinson}}, \emph{et~al.},
  \bibinfo{year}{2007}{\natexlab{b}}, \bibinfo{journal}{Phys. Rev. Lett.}
  \textbf{\bibinfo{volume}{99}}(\bibinfo{number}{2}), \bibinfo{pages}{022504}.

\bibitem[{\citenamefont{Seweryniak}
  \emph{et~al.}(2005{\natexlab{a}})\citenamefont{Seweryniak, Davids, Robinson,
  Woods, Blank, Carpenter, Davinson, Freeman, Hammond, Hoteling, Janssens,
  Khoo} \emph{et~al.}}]{Seweryniak:2005}
\bibinfo{author}{\bibnamefont{Seweryniak}, \bibfnamefont{D.}},
  \bibinfo{author}{\bibfnamefont{C.}~\bibnamefont{Davids}},
  \bibinfo{author}{\bibfnamefont{A.}~\bibnamefont{Robinson}},
  \bibinfo{author}{\bibfnamefont{P.}~\bibnamefont{Woods}},
  \bibinfo{author}{\bibfnamefont{B.}~\bibnamefont{Blank}},
  \bibinfo{author}{\bibfnamefont{M.}~\bibnamefont{Carpenter}},
  \bibinfo{author}{\bibfnamefont{T.}~\bibnamefont{Davinson}},
  \bibinfo{author}{\bibfnamefont{S.}~\bibnamefont{Freeman}},
  \bibinfo{author}{\bibfnamefont{N.}~\bibnamefont{Hammond}},
  \bibinfo{author}{\bibfnamefont{N.}~\bibnamefont{Hoteling}},
  \bibinfo{author}{\bibfnamefont{R.}~\bibnamefont{Janssens}},
  \bibinfo{author}{\bibfnamefont{T.}~\bibnamefont{Khoo}}, \emph{et~al.},
  \bibinfo{year}{2005}{\natexlab{a}}, \bibinfo{journal}{Eur.\ Phys.\ J.}
  \textbf{\bibinfo{volume}{A~25}}, \bibinfo{pages}{159}.

\bibitem[{\citenamefont{Seweryniak}
  \emph{et~al.}(1997)\citenamefont{Seweryniak, Davids, Walters, Woods, Ahmad,
  Amro, Blumenthal, Brown, Carpenter, Davinson, Fischer, Henderson}
  \emph{et~al.}}]{Seweryniak:1997}
\bibinfo{author}{\bibnamefont{Seweryniak}, \bibfnamefont{D.}},
  \bibinfo{author}{\bibfnamefont{C.~N.} \bibnamefont{Davids}},
  \bibinfo{author}{\bibfnamefont{W.~B.} \bibnamefont{Walters}},
  \bibinfo{author}{\bibfnamefont{P.~J.} \bibnamefont{Woods}},
  \bibinfo{author}{\bibfnamefont{I.}~\bibnamefont{Ahmad}},
  \bibinfo{author}{\bibfnamefont{H.}~\bibnamefont{Amro}},
  \bibinfo{author}{\bibfnamefont{D.~J.} \bibnamefont{Blumenthal}},
  \bibinfo{author}{\bibfnamefont{L.~T.} \bibnamefont{Brown}},
  \bibinfo{author}{\bibfnamefont{M.~P.} \bibnamefont{Carpenter}},
  \bibinfo{author}{\bibfnamefont{T.}~\bibnamefont{Davinson}},
  \bibinfo{author}{\bibfnamefont{S.~M.} \bibnamefont{Fischer}},
  \bibinfo{author}{\bibfnamefont{D.~J.} \bibnamefont{Henderson}},
  \emph{et~al.}, \bibinfo{year}{1997}, \bibinfo{journal}{Phys. Rev. C}
  \textbf{\bibinfo{volume}{55}}(\bibinfo{number}{5}), \bibinfo{pages}{R2137}.

\bibitem[{\citenamefont{Seweryniak}
  \emph{et~al.}(2006)\citenamefont{Seweryniak, Starosta, Davids, Gros, Hecht,
  Hoteling, Khoo, Lagergren, Lotay, Peterson, Robinson, Vaman}
  \emph{et~al.}}]{Seweryniak:2006}
\bibinfo{author}{\bibnamefont{Seweryniak}, \bibfnamefont{D.}},
  \bibinfo{author}{\bibfnamefont{K.}~\bibnamefont{Starosta}},
  \bibinfo{author}{\bibfnamefont{C.~N.} \bibnamefont{Davids}},
  \bibinfo{author}{\bibfnamefont{S.}~\bibnamefont{Gros}},
  \bibinfo{author}{\bibfnamefont{A.~A.} \bibnamefont{Hecht}},
  \bibinfo{author}{\bibfnamefont{N.}~\bibnamefont{Hoteling}},
  \bibinfo{author}{\bibfnamefont{T.~L.} \bibnamefont{Khoo}},
  \bibinfo{author}{\bibfnamefont{K.}~\bibnamefont{Lagergren}},
  \bibinfo{author}{\bibfnamefont{G.}~\bibnamefont{Lotay}},
  \bibinfo{author}{\bibfnamefont{D.}~\bibnamefont{Peterson}},
  \bibinfo{author}{\bibfnamefont{A.}~\bibnamefont{Robinson}},
  \bibinfo{author}{\bibfnamefont{C.}~\bibnamefont{Vaman}}, \emph{et~al.},
  \bibinfo{year}{2006}, \bibinfo{journal}{Phys. Rev. C}
  \textbf{\bibinfo{volume}{73}}(\bibinfo{number}{6}), \bibinfo{pages}{061301}.

\bibitem[{\citenamefont{Seweryniak}
  \emph{et~al.}(2005{\natexlab{b}})\citenamefont{Seweryniak, Uusitalo,
  Bhattacharyya, Carpenter, Cizewski, Ding, Davids, Fotiades, Janssens,
  Lauritsen, Lister, Macchiavelli} \emph{et~al.}}]{Seweryniak:2005a}
\bibinfo{author}{\bibnamefont{Seweryniak}, \bibfnamefont{D.}},
  \bibinfo{author}{\bibfnamefont{J.}~\bibnamefont{Uusitalo}},
  \bibinfo{author}{\bibfnamefont{P.}~\bibnamefont{Bhattacharyya}},
  \bibinfo{author}{\bibfnamefont{M.~P.} \bibnamefont{Carpenter}},
  \bibinfo{author}{\bibfnamefont{J.~A.} \bibnamefont{Cizewski}},
  \bibinfo{author}{\bibfnamefont{K.~Y.} \bibnamefont{Ding}},
  \bibinfo{author}{\bibfnamefont{C.~N.} \bibnamefont{Davids}},
  \bibinfo{author}{\bibfnamefont{N.}~\bibnamefont{Fotiades}},
  \bibinfo{author}{\bibfnamefont{R.~V.~F.} \bibnamefont{Janssens}},
  \bibinfo{author}{\bibfnamefont{T.}~\bibnamefont{Lauritsen}},
  \bibinfo{author}{\bibfnamefont{C.~J.} \bibnamefont{Lister}},
  \bibinfo{author}{\bibfnamefont{A.~O.} \bibnamefont{Macchiavelli}},
  \emph{et~al.}, \bibinfo{year}{2005}{\natexlab{b}}, \bibinfo{journal}{Phys.
  Rev. C} \textbf{\bibinfo{volume}{71}}(\bibinfo{number}{5}),
  \bibinfo{pages}{054319}.

\bibitem[{\citenamefont{Seweryniak}
  \emph{et~al.}(2001)\citenamefont{Seweryniak, Woods, Ressler, Davids, Heinz,
  Uusitalo, Walters, Caggiano, Carpenter, Cizewski, Davinson, Ding}
  \emph{et~al.}}]{Seweryniak:2001}
\bibinfo{author}{\bibnamefont{Seweryniak}, \bibfnamefont{D.}},
  \bibinfo{author}{\bibfnamefont{P.}~\bibnamefont{Woods}},
  \bibinfo{author}{\bibfnamefont{J.}~\bibnamefont{Ressler}},
  \bibinfo{author}{\bibfnamefont{C.}~\bibnamefont{Davids}},
  \bibinfo{author}{\bibfnamefont{A.}~\bibnamefont{Heinz}},
  \bibinfo{author}{\bibfnamefont{J.}~\bibnamefont{Uusitalo}},
  \bibinfo{author}{\bibfnamefont{W.}~\bibnamefont{Walters}},
  \bibinfo{author}{\bibfnamefont{J.}~\bibnamefont{Caggiano}},
  \bibinfo{author}{\bibfnamefont{M.}~\bibnamefont{Carpenter}},
  \bibinfo{author}{\bibfnamefont{J.}~\bibnamefont{Cizewski}},
  \bibinfo{author}{\bibfnamefont{T.}~\bibnamefont{Davinson}},
  \bibinfo{author}{\bibfnamefont{K.}~\bibnamefont{Ding}}, \emph{et~al.},
  \bibinfo{year}{2001}, \bibinfo{journal}{Phys.\ Rev.\ Lett.}
  \textbf{\bibinfo{volume}{86}}, \bibinfo{pages}{1458}.

\bibitem[{\citenamefont{Shaughnessy}
  \emph{et~al.}(2002)\citenamefont{Shaughnessy, Gregorich, Adams, Lane, Laue,
  Lee, McGrath, Ninov, Patin, Strellis, Sylwester, Wilk}
  \emph{et~al.}}]{Shaughnessy:2002}
\bibinfo{author}{\bibnamefont{Shaughnessy}, \bibfnamefont{D.~A.}},
  \bibinfo{author}{\bibfnamefont{K.~E.} \bibnamefont{Gregorich}},
  \bibinfo{author}{\bibfnamefont{J.~L.} \bibnamefont{Adams}},
  \bibinfo{author}{\bibfnamefont{M.~R.} \bibnamefont{Lane}},
  \bibinfo{author}{\bibfnamefont{C.~A.} \bibnamefont{Laue}},
  \bibinfo{author}{\bibfnamefont{D.~M.} \bibnamefont{Lee}},
  \bibinfo{author}{\bibfnamefont{C.~A.} \bibnamefont{McGrath}},
  \bibinfo{author}{\bibfnamefont{V.}~\bibnamefont{Ninov}},
  \bibinfo{author}{\bibfnamefont{J.~B.} \bibnamefont{Patin}},
  \bibinfo{author}{\bibfnamefont{D.~A.} \bibnamefont{Strellis}},
  \bibinfo{author}{\bibfnamefont{E.~R.} \bibnamefont{Sylwester}},
  \bibinfo{author}{\bibfnamefont{P.~A.} \bibnamefont{Wilk}}, \emph{et~al.},
  \bibinfo{year}{2002}, \bibinfo{journal}{Phys. Rev. C}
  \textbf{\bibinfo{volume}{65}}(\bibinfo{number}{2}), \bibinfo{pages}{024612}.

\bibitem[{\citenamefont{Shotter}(2003)}]{Shotter:2003}
\bibinfo{author}{\bibnamefont{Shotter}, \bibfnamefont{A.}},
  \bibinfo{year}{2003}, \bibinfo{journal}{Nuclear Instruments and Methods in
  Physics Research Section B: Beam Interactions with Materials and Atoms}
  \textbf{\bibinfo{volume}{204}}, \bibinfo{pages}{17}.

\bibitem[{\citenamefont{Skeppstedt}
  \emph{et~al.}(1999)\citenamefont{Skeppstedt, Roth, Lindström, Wadsworth,
  Hibbert, Kelsall, Jenkins, Grawe, Górska, Moszynski, Sujkowski, Wolski}
  \emph{et~al.}}]{Skeppstedt:1999}
\bibinfo{author}{\bibnamefont{Skeppstedt}, \bibfnamefont{O.}},
  \bibinfo{author}{\bibfnamefont{H.~A.} \bibnamefont{Roth}},
  \bibinfo{author}{\bibfnamefont{L.}~\bibnamefont{Lindström}},
  \bibinfo{author}{\bibfnamefont{R.}~\bibnamefont{Wadsworth}},
  \bibinfo{author}{\bibfnamefont{I.}~\bibnamefont{Hibbert}},
  \bibinfo{author}{\bibfnamefont{N.}~\bibnamefont{Kelsall}},
  \bibinfo{author}{\bibfnamefont{D.}~\bibnamefont{Jenkins}},
  \bibinfo{author}{\bibfnamefont{H.}~\bibnamefont{Grawe}},
  \bibinfo{author}{\bibfnamefont{M.}~\bibnamefont{Górska}},
  \bibinfo{author}{\bibfnamefont{M.}~\bibnamefont{Moszynski}},
  \bibinfo{author}{\bibfnamefont{Z.}~\bibnamefont{Sujkowski}},
  \bibinfo{author}{\bibfnamefont{D.}~\bibnamefont{Wolski}}, \emph{et~al.},
  \bibinfo{year}{1999}, \bibinfo{journal}{Nuclear Instruments and Methods in
  Physics Research Section A: Accelerators, Spectrometers, Detectors and
  Associated Equipment} \textbf{\bibinfo{volume}{421}}(\bibinfo{number}{3}),
  \bibinfo{pages}{531}.

\bibitem[{\citenamefont{Sonzogni}(2002)}]{Sonzogni:2002}
\bibinfo{author}{\bibnamefont{Sonzogni}, \bibfnamefont{A.}},
  \bibinfo{year}{2002}, \bibinfo{journal}{Nuclear Data Sheets}
  \textbf{\bibinfo{volume}{95}}, \bibinfo{pages}{1}.

\bibitem[{\citenamefont{Sonzogni} \emph{et~al.}(1999)\citenamefont{Sonzogni,
  Davids, Woods, Seweryniak, Carpenter, Ressler, Schwartz, Uusitalo, and
  Walters}}]{Sonzogni:1999}
\bibinfo{author}{\bibnamefont{Sonzogni}, \bibfnamefont{A.}},
  \bibinfo{author}{\bibfnamefont{C.}~\bibnamefont{Davids}},
  \bibinfo{author}{\bibfnamefont{P.}~\bibnamefont{Woods}},
  \bibinfo{author}{\bibfnamefont{D.}~\bibnamefont{Seweryniak}},
  \bibinfo{author}{\bibfnamefont{M.}~\bibnamefont{Carpenter}},
  \bibinfo{author}{\bibfnamefont{J.}~\bibnamefont{Ressler}},
  \bibinfo{author}{\bibfnamefont{J.}~\bibnamefont{Schwartz}},
  \bibinfo{author}{\bibfnamefont{J.}~\bibnamefont{Uusitalo}}, and
  \bibinfo{author}{\bibfnamefont{W.}~\bibnamefont{Walters}},
  \bibinfo{year}{1999}, \bibinfo{journal}{Phys.\ Rev.\ Lett.}
  \textbf{\bibinfo{volume}{83}}, \bibinfo{pages}{1116}.

\bibitem[{\citenamefont{Soramel} \emph{et~al.}(2001)\citenamefont{Soramel,
  Guglielmetti, Stroe, Muller, Bonetti, Poli, Malerba, Bianchi, Andrighetto,
  Guo, Li, Maglione} \emph{et~al.}}]{Soramel:2001}
\bibinfo{author}{\bibnamefont{Soramel}, \bibfnamefont{F.}},
  \bibinfo{author}{\bibfnamefont{A.}~\bibnamefont{Guglielmetti}},
  \bibinfo{author}{\bibfnamefont{L.}~\bibnamefont{Stroe}},
  \bibinfo{author}{\bibfnamefont{L.}~\bibnamefont{Muller}},
  \bibinfo{author}{\bibfnamefont{R.}~\bibnamefont{Bonetti}},
  \bibinfo{author}{\bibfnamefont{G.}~\bibnamefont{Poli}},
  \bibinfo{author}{\bibfnamefont{F.}~\bibnamefont{Malerba}},
  \bibinfo{author}{\bibfnamefont{E.}~\bibnamefont{Bianchi}},
  \bibinfo{author}{\bibfnamefont{A.}~\bibnamefont{Andrighetto}},
  \bibinfo{author}{\bibfnamefont{J.}~\bibnamefont{Guo}},
  \bibinfo{author}{\bibfnamefont{Z.}~\bibnamefont{Li}},
  \bibinfo{author}{\bibfnamefont{E.}~\bibnamefont{Maglione}}, \emph{et~al.},
  \bibinfo{year}{2001}, \bibinfo{journal}{Phys.\ Rev.}
  \textbf{\bibinfo{volume}{C~63}}, \bibinfo{pages}{031304}.

\bibitem[{\citenamefont{Sorlin} \emph{et~al.}(2003)\citenamefont{Sorlin,
  Donzaud, Azaiez, Bourgeois, Gaudefroy, Ibrahim, Guillemaud-Mueller, Pougheon,
  Lewitowicz, de~Oliveira~Santos, Saint-Laurent, Stanoiu}
  \emph{et~al.}}]{Sorlin:2003}
\bibinfo{author}{\bibnamefont{Sorlin}, \bibfnamefont{O.}},
  \bibinfo{author}{\bibfnamefont{C.}~\bibnamefont{Donzaud}},
  \bibinfo{author}{\bibfnamefont{F.}~\bibnamefont{Azaiez}},
  \bibinfo{author}{\bibfnamefont{C.}~\bibnamefont{Bourgeois}},
  \bibinfo{author}{\bibfnamefont{L.}~\bibnamefont{Gaudefroy}},
  \bibinfo{author}{\bibfnamefont{F.}~\bibnamefont{Ibrahim}},
  \bibinfo{author}{\bibfnamefont{D.}~\bibnamefont{Guillemaud-Mueller}},
  \bibinfo{author}{\bibfnamefont{F.}~\bibnamefont{Pougheon}},
  \bibinfo{author}{\bibfnamefont{M.}~\bibnamefont{Lewitowicz}},
  \bibinfo{author}{\bibfnamefont{F.}~\bibnamefont{de~Oliveira~Santos}},
  \bibinfo{author}{\bibfnamefont{M.~G.} \bibnamefont{Saint-Laurent}},
  \bibinfo{author}{\bibfnamefont{M.}~\bibnamefont{Stanoiu}}, \emph{et~al.},
  \bibinfo{year}{2003}, \bibinfo{journal}{Nuclear Physics A}
  \textbf{\bibinfo{volume}{719}}, \bibinfo{pages}{C193}.

\bibitem[{\citenamefont{{Sorlin} and {Porquet}}(2008)}]{Sorlin:2008}
\bibinfo{author}{\bibnamefont{{Sorlin}}, \bibfnamefont{O.}}, and
  \bibinfo{author}{\bibfnamefont{M.~G.} \bibnamefont{{Porquet}}},
  \bibinfo{year}{2008}, \bibinfo{journal}{Prog.Part.Nucl.Phys.}
  \textbf{\bibinfo{volume}{61}}, \bibinfo{pages}{602}.

\bibitem[{\citenamefont{Stanoiu} \emph{et~al.}(2008)\citenamefont{Stanoiu,
  Sümmerer, Mukha, Chatillon, Gil, Heil, Hoffman, Kiselev, Kurz, and
  Ott}}]{Stanoiu:2008}
\bibinfo{author}{\bibnamefont{Stanoiu}, \bibfnamefont{M.}},
  \bibinfo{author}{\bibfnamefont{K.}~\bibnamefont{Sümmerer}},
  \bibinfo{author}{\bibfnamefont{I.}~\bibnamefont{Mukha}},
  \bibinfo{author}{\bibfnamefont{A.}~\bibnamefont{Chatillon}},
  \bibinfo{author}{\bibfnamefont{E.~C.} \bibnamefont{Gil}},
  \bibinfo{author}{\bibfnamefont{M.}~\bibnamefont{Heil}},
  \bibinfo{author}{\bibfnamefont{J.}~\bibnamefont{Hoffman}},
  \bibinfo{author}{\bibfnamefont{O.}~\bibnamefont{Kiselev}},
  \bibinfo{author}{\bibfnamefont{N.}~\bibnamefont{Kurz}}, and
  \bibinfo{author}{\bibfnamefont{W.}~\bibnamefont{Ott}}, \bibinfo{year}{2008},
  \bibinfo{journal}{Nuclear Instruments and Methods in Physics Research Section
  B: Beam Interactions with Materials and Atoms}
  \textbf{\bibinfo{volume}{266}}(\bibinfo{number}{19-20}),
  \bibinfo{pages}{4625}.

\bibitem[{\citenamefont{Starosta} \emph{et~al.}(2009)\citenamefont{Starosta,
  Vaman, Miller, Voss, Bazin, Glasmacher, Crawford, Mantica, Tan, Hennig,
  Walby, Fallu-Labruyere} \emph{et~al.}}]{Starosta:2009}
\bibinfo{author}{\bibnamefont{Starosta}, \bibfnamefont{K.}},
  \bibinfo{author}{\bibfnamefont{C.}~\bibnamefont{Vaman}},
  \bibinfo{author}{\bibfnamefont{D.}~\bibnamefont{Miller}},
  \bibinfo{author}{\bibfnamefont{P.}~\bibnamefont{Voss}},
  \bibinfo{author}{\bibfnamefont{D.}~\bibnamefont{Bazin}},
  \bibinfo{author}{\bibfnamefont{T.}~\bibnamefont{Glasmacher}},
  \bibinfo{author}{\bibfnamefont{H.}~\bibnamefont{Crawford}},
  \bibinfo{author}{\bibfnamefont{P.}~\bibnamefont{Mantica}},
  \bibinfo{author}{\bibfnamefont{H.}~\bibnamefont{Tan}},
  \bibinfo{author}{\bibfnamefont{W.}~\bibnamefont{Hennig}},
  \bibinfo{author}{\bibfnamefont{M.}~\bibnamefont{Walby}},
  \bibinfo{author}{\bibfnamefont{A.}~\bibnamefont{Fallu-Labruyere}},
  \emph{et~al.}, \bibinfo{year}{2009}, \bibinfo{journal}{Nuclear Instruments
  and Methods in Physics Research Section A: Accelerators, Spectrometers,
  Detectors and Associated Equipment}
  \textbf{\bibinfo{volume}{610}}(\bibinfo{number}{3}), \bibinfo{pages}{700}.

\bibitem[{\citenamefont{Stracener}(2003)}]{Stracener:2003}
\bibinfo{author}{\bibnamefont{Stracener}, \bibfnamefont{D.~W.}},
  \bibinfo{year}{2003}, \bibinfo{journal}{Nuclear Instruments and Methods in
  Physics Research Section B: Beam Interactions with Materials and Atoms}
  \textbf{\bibinfo{volume}{204}}, \bibinfo{pages}{42}.

\bibitem[{\citenamefont{Sumithrarachchi}
  \emph{et~al.}(2010)\citenamefont{Sumithrarachchi, Morrissey, Davies, Davies,
  Facina, Kwan, Mantica, Portillo, Shimbara, Stoker, and
  Weerasiri}}]{Sumithrarachchi:2010}
\bibinfo{author}{\bibnamefont{Sumithrarachchi}, \bibfnamefont{C.~S.}},
  \bibinfo{author}{\bibfnamefont{D.~J.} \bibnamefont{Morrissey}},
  \bibinfo{author}{\bibfnamefont{A.~D.} \bibnamefont{Davies}},
  \bibinfo{author}{\bibfnamefont{D.~A.} \bibnamefont{Davies}},
  \bibinfo{author}{\bibfnamefont{M.}~\bibnamefont{Facina}},
  \bibinfo{author}{\bibfnamefont{E.}~\bibnamefont{Kwan}},
  \bibinfo{author}{\bibfnamefont{P.~F.} \bibnamefont{Mantica}},
  \bibinfo{author}{\bibfnamefont{M.}~\bibnamefont{Portillo}},
  \bibinfo{author}{\bibfnamefont{Y.}~\bibnamefont{Shimbara}},
  \bibinfo{author}{\bibfnamefont{J.}~\bibnamefont{Stoker}}, and
  \bibinfo{author}{\bibfnamefont{R.~R.} \bibnamefont{Weerasiri}},
  \bibinfo{year}{2010}, \bibinfo{journal}{Phys. Rev. C}
  \textbf{\bibinfo{volume}{81}}(\bibinfo{number}{1}), \bibinfo{pages}{014302}.

\bibitem[{\citenamefont{S\"ummerer and Blank}(2000)}]{Summerer:2000}
\bibinfo{author}{\bibnamefont{S\"ummerer}, \bibfnamefont{K.}}, and
  \bibinfo{author}{\bibfnamefont{B.}~\bibnamefont{Blank}},
  \bibinfo{year}{2000}, \bibinfo{journal}{Phys. Rev. C}
  \textbf{\bibinfo{volume}{61}}(\bibinfo{number}{3}), \bibinfo{pages}{034607}.

\bibitem[{\citenamefont{Sun} \emph{et~al.}(2003)\citenamefont{Sun, Zhan, Guo,
  Xiao, and Li}}]{Sun:2003}
\bibinfo{author}{\bibnamefont{Sun}, \bibfnamefont{Z.}},
  \bibinfo{author}{\bibfnamefont{W.~L.} \bibnamefont{Zhan}},
  \bibinfo{author}{\bibfnamefont{Z.~Y.} \bibnamefont{Guo}},
  \bibinfo{author}{\bibfnamefont{G.}~\bibnamefont{Xiao}}, and
  \bibinfo{author}{\bibfnamefont{J.~X.} \bibnamefont{Li}},
  \bibinfo{year}{2003}, \bibinfo{journal}{Nuclear Instruments and Methods in
  Physics Research Section A: Accelerators, Spectrometers, Detectors and
  Associated Equipment} \textbf{\bibinfo{volume}{503}}(\bibinfo{number}{3}),
  \bibinfo{pages}{496}.

\bibitem[{\citenamefont{Suzuki} \emph{et~al.}(2009)\citenamefont{Suzuki,
  Iwasaki, Beaumel, Nalpas, Pollacco, Assie, Baba, Blumenfeld, Sereville,
  Drouart, Franchoo, Gillibert} \emph{et~al.}}]{Suzuki:2009}
\bibinfo{author}{\bibnamefont{Suzuki}, \bibfnamefont{D.}},
  \bibinfo{author}{\bibfnamefont{H.}~\bibnamefont{Iwasaki}},
  \bibinfo{author}{\bibfnamefont{D.}~\bibnamefont{Beaumel}},
  \bibinfo{author}{\bibfnamefont{L.}~\bibnamefont{Nalpas}},
  \bibinfo{author}{\bibfnamefont{E.}~\bibnamefont{Pollacco}},
  \bibinfo{author}{\bibfnamefont{M.}~\bibnamefont{Assie}},
  \bibinfo{author}{\bibfnamefont{H.}~\bibnamefont{Baba}},
  \bibinfo{author}{\bibfnamefont{Y.}~\bibnamefont{Blumenfeld}},
  \bibinfo{author}{\bibfnamefont{N.~D.} \bibnamefont{Sereville}},
  \bibinfo{author}{\bibfnamefont{A.}~\bibnamefont{Drouart}},
  \bibinfo{author}{\bibfnamefont{S.}~\bibnamefont{Franchoo}},
  \bibinfo{author}{\bibfnamefont{A.}~\bibnamefont{Gillibert}}, \emph{et~al.},
  \bibinfo{year}{2009}, \bibinfo{journal}{Phys. Rev. Lett.}
  \textbf{\bibinfo{volume}{103}}, \bibinfo{pages}{152503}.

\bibitem[{\citenamefont{Symons} \emph{et~al.}(1979)\citenamefont{Symons,
  Viyogi, Westfall, Doll, Greiner, Faraggi, Lindstrom, Scott, Crawford, and
  McParland}}]{Symons:1979}
\bibinfo{author}{\bibnamefont{Symons}, \bibfnamefont{T.~J.~M.}},
  \bibinfo{author}{\bibfnamefont{Y.~P.} \bibnamefont{Viyogi}},
  \bibinfo{author}{\bibfnamefont{G.~D.} \bibnamefont{Westfall}},
  \bibinfo{author}{\bibfnamefont{P.}~\bibnamefont{Doll}},
  \bibinfo{author}{\bibfnamefont{D.~E.} \bibnamefont{Greiner}},
  \bibinfo{author}{\bibfnamefont{H.}~\bibnamefont{Faraggi}},
  \bibinfo{author}{\bibfnamefont{P.~J.} \bibnamefont{Lindstrom}},
  \bibinfo{author}{\bibfnamefont{D.~K.} \bibnamefont{Scott}},
  \bibinfo{author}{\bibfnamefont{H.~J.} \bibnamefont{Crawford}}, and
  \bibinfo{author}{\bibfnamefont{C.}~\bibnamefont{McParland}},
  \bibinfo{year}{1979}, \bibinfo{journal}{Phys. Rev. Lett.}
  \textbf{\bibinfo{volume}{42}}(\bibinfo{number}{1}), \bibinfo{pages}{40}.

\bibitem[{\citenamefont{Taieb} \emph{et~al.}(2003)\citenamefont{Taieb, Schmidt,
  Tassan-Got, Armbruster, Benlliure, Bernas, Boudard, Casarejos, Czajkowski,
  Enqvist, Legrain, Leray} \emph{et~al.}}]{Taieb:2003}
\bibinfo{author}{\bibnamefont{Taieb}, \bibfnamefont{J.}},
  \bibinfo{author}{\bibfnamefont{K.~H.} \bibnamefont{Schmidt}},
  \bibinfo{author}{\bibfnamefont{L.}~\bibnamefont{Tassan-Got}},
  \bibinfo{author}{\bibfnamefont{P.}~\bibnamefont{Armbruster}},
  \bibinfo{author}{\bibfnamefont{J.}~\bibnamefont{Benlliure}},
  \bibinfo{author}{\bibfnamefont{M.}~\bibnamefont{Bernas}},
  \bibinfo{author}{\bibfnamefont{A.}~\bibnamefont{Boudard}},
  \bibinfo{author}{\bibfnamefont{E.}~\bibnamefont{Casarejos}},
  \bibinfo{author}{\bibfnamefont{S.}~\bibnamefont{Czajkowski}},
  \bibinfo{author}{\bibfnamefont{T.}~\bibnamefont{Enqvist}},
  \bibinfo{author}{\bibfnamefont{R.}~\bibnamefont{Legrain}},
  \bibinfo{author}{\bibfnamefont{S.}~\bibnamefont{Leray}}, \emph{et~al.},
  \bibinfo{year}{2003}, \bibinfo{journal}{Nuclear Physics A}
  \textbf{\bibinfo{volume}{724}}(\bibinfo{number}{3-4}), \bibinfo{pages}{413}.

\bibitem[{\citenamefont{Tanihata}(2008)}]{Tanihata:2008}
\bibinfo{author}{\bibnamefont{Tanihata}, \bibfnamefont{I.}},
  \bibinfo{year}{2008}, \bibinfo{journal}{Nuclear Instruments and Methods in
  Physics Research Section B: Beam Interactions with Materials and Atoms}
  \textbf{\bibinfo{volume}{266}}(\bibinfo{number}{19-20}),
  \bibinfo{pages}{4067}.

\bibitem[{\citenamefont{Tantawy} \emph{et~al.}(2006)\citenamefont{Tantawy,
  Bingham, Rykaczewski, Batchelder, Kr{\'o}las, Danchev, Fong, Ginter, Gross,
  Grzywacz, Hagino, Hamilton} \emph{et~al.}}]{Tantawy:2006}
\bibinfo{author}{\bibnamefont{Tantawy}, \bibfnamefont{M.}},
  \bibinfo{author}{\bibfnamefont{C.}~\bibnamefont{Bingham}},
  \bibinfo{author}{\bibfnamefont{K.}~\bibnamefont{Rykaczewski}},
  \bibinfo{author}{\bibfnamefont{J.}~\bibnamefont{Batchelder}},
  \bibinfo{author}{\bibfnamefont{W.}~\bibnamefont{Kr{\'o}las}},
  \bibinfo{author}{\bibfnamefont{M.}~\bibnamefont{Danchev}},
  \bibinfo{author}{\bibfnamefont{D.}~\bibnamefont{Fong}},
  \bibinfo{author}{\bibfnamefont{T.}~\bibnamefont{Ginter}},
  \bibinfo{author}{\bibfnamefont{C.}~\bibnamefont{Gross}},
  \bibinfo{author}{\bibfnamefont{R.}~\bibnamefont{Grzywacz}},
  \bibinfo{author}{\bibfnamefont{K.}~\bibnamefont{Hagino}},
  \bibinfo{author}{\bibfnamefont{J.}~\bibnamefont{Hamilton}}, \emph{et~al.},
  \bibinfo{year}{2006}, \bibinfo{journal}{Phys.\ Rev.}
  \textbf{\bibinfo{volume}{C~73}}, \bibinfo{pages}{024316}.

\bibitem[{\citenamefont{Tarasov} \emph{et~al.}(2009)\citenamefont{Tarasov,
  Morrissey, Amthor, Baumann, Bazin, Gade, Ginter, Hausmann, Inabe, Kubo,
  Nettleton, Pereira} \emph{et~al.}}]{Tarasov:2009}
\bibinfo{author}{\bibnamefont{Tarasov}, \bibfnamefont{O.~B.}},
  \bibinfo{author}{\bibfnamefont{D.~J.} \bibnamefont{Morrissey}},
  \bibinfo{author}{\bibfnamefont{A.~M.} \bibnamefont{Amthor}},
  \bibinfo{author}{\bibfnamefont{T.}~\bibnamefont{Baumann}},
  \bibinfo{author}{\bibfnamefont{D.}~\bibnamefont{Bazin}},
  \bibinfo{author}{\bibfnamefont{A.}~\bibnamefont{Gade}},
  \bibinfo{author}{\bibfnamefont{T.~N.} \bibnamefont{Ginter}},
  \bibinfo{author}{\bibfnamefont{M.}~\bibnamefont{Hausmann}},
  \bibinfo{author}{\bibfnamefont{N.}~\bibnamefont{Inabe}},
  \bibinfo{author}{\bibfnamefont{T.}~\bibnamefont{Kubo}},
  \bibinfo{author}{\bibfnamefont{A.}~\bibnamefont{Nettleton}},
  \bibinfo{author}{\bibfnamefont{J.}~\bibnamefont{Pereira}}, \emph{et~al.},
  \bibinfo{year}{2009}, \bibinfo{journal}{Phys. Rev. Lett.}
  \textbf{\bibinfo{volume}{102}}(\bibinfo{number}{14}),
  \bibinfo{pages}{142501}.

\bibitem[{\citenamefont{Tengblad} \emph{et~al.}(2004)\citenamefont{Tengblad,
  Bergmann, Fraile, Fynbo, and Walsh}}]{Tengblad:2004}
\bibinfo{author}{\bibnamefont{Tengblad}, \bibfnamefont{O.}},
  \bibinfo{author}{\bibfnamefont{U.~C.} \bibnamefont{Bergmann}},
  \bibinfo{author}{\bibfnamefont{L.~M.} \bibnamefont{Fraile}},
  \bibinfo{author}{\bibfnamefont{H.~O.~U.} \bibnamefont{Fynbo}}, and
  \bibinfo{author}{\bibfnamefont{S.}~\bibnamefont{Walsh}},
  \bibinfo{year}{2004}, \bibinfo{journal}{Nuclear Instruments and Methods in
  Physics Research Section A: Accelerators, Spectrometers, Detectors and
  Associated Equipment} \textbf{\bibinfo{volume}{525}}(\bibinfo{number}{3}),
  \bibinfo{pages}{458}.

\bibitem[{\citenamefont{Thaysen} \emph{et~al.}(1999)\citenamefont{Thaysen,
  Axelsson, \"{A}yst\"{o}, Borge, Fraile, Fynbo, Honkanen, Hornsh{\o}j, Jading,
  Jokinen, Jonson, Martel} \emph{et~al.}}]{Thaysen:1999}
\bibinfo{author}{\bibnamefont{Thaysen}, \bibfnamefont{J.}},
  \bibinfo{author}{\bibfnamefont{L.}~\bibnamefont{Axelsson}},
  \bibinfo{author}{\bibfnamefont{J.}~\bibnamefont{\"{A}yst\"{o}}},
  \bibinfo{author}{\bibfnamefont{M.~J.~G.} \bibnamefont{Borge}},
  \bibinfo{author}{\bibfnamefont{L.~M.} \bibnamefont{Fraile}},
  \bibinfo{author}{\bibfnamefont{H.~O.~U.} \bibnamefont{Fynbo}},
  \bibinfo{author}{\bibfnamefont{A.}~\bibnamefont{Honkanen}},
  \bibinfo{author}{\bibfnamefont{P.}~\bibnamefont{Hornsh{\o}j}},
  \bibinfo{author}{\bibfnamefont{Y.}~\bibnamefont{Jading}},
  \bibinfo{author}{\bibfnamefont{A.}~\bibnamefont{Jokinen}},
  \bibinfo{author}{\bibfnamefont{B.}~\bibnamefont{Jonson}},
  \bibinfo{author}{\bibfnamefont{I.}~\bibnamefont{Martel}}, \emph{et~al.},
  \bibinfo{year}{1999}, \bibinfo{journal}{Physics Letters B}
  \textbf{\bibinfo{volume}{467}}(\bibinfo{number}{3-4}), \bibinfo{pages}{194}.

\bibitem[{\citenamefont{Thibault} \emph{et~al.}(1975)\citenamefont{Thibault,
  Klapisch, Rigaud, Poskanzer, Prieels, Lessard, and Reisdorf}}]{Thibault:1975}
\bibinfo{author}{\bibnamefont{Thibault}, \bibfnamefont{C.}},
  \bibinfo{author}{\bibfnamefont{R.}~\bibnamefont{Klapisch}},
  \bibinfo{author}{\bibfnamefont{C.}~\bibnamefont{Rigaud}},
  \bibinfo{author}{\bibfnamefont{A.~M.} \bibnamefont{Poskanzer}},
  \bibinfo{author}{\bibfnamefont{R.}~\bibnamefont{Prieels}},
  \bibinfo{author}{\bibfnamefont{L.}~\bibnamefont{Lessard}}, and
  \bibinfo{author}{\bibfnamefont{W.}~\bibnamefont{Reisdorf}},
  \bibinfo{year}{1975}, \bibinfo{journal}{Phys. Rev. C}
  \textbf{\bibinfo{volume}{12}}(\bibinfo{number}{2}), \bibinfo{pages}{644}.

\bibitem[{\citenamefont{Thoennessen}(2004)}]{Thoennessen:2004}
\bibinfo{author}{\bibnamefont{Thoennessen}, \bibfnamefont{M.}},
  \bibinfo{year}{2004}, \bibinfo{journal}{Reports on Progress in Physics}
  \textbf{\bibinfo{volume}{67}}(\bibinfo{number}{7}), \bibinfo{pages}{1187}.

\bibitem[{\citenamefont{Thoennessen}(2010)}]{Thoennessen:2010}
\bibinfo{author}{\bibnamefont{Thoennessen}, \bibfnamefont{M.}},
  \bibinfo{year}{2010}, \bibinfo{journal}{Nuclear Physics A}
  \textbf{\bibinfo{volume}{834}}(\bibinfo{number}{1-4}), \bibinfo{pages}{688c}.

\bibitem[{\citenamefont{Tilley} \emph{et~al.}(1992)\citenamefont{Tilley,
  Weller, and Hale}}]{Tilley:1992}
\bibinfo{author}{\bibnamefont{Tilley}, \bibfnamefont{D.}},
  \bibinfo{author}{\bibfnamefont{H.}~\bibnamefont{Weller}}, and
  \bibinfo{author}{\bibfnamefont{G.}~\bibnamefont{Hale}}, \bibinfo{year}{1992},
  \bibinfo{journal}{Nucl. Phys.} \textbf{\bibinfo{volume}{A541}},
  \bibinfo{pages}{1}.

\bibitem[{\citenamefont{Toivanen} \emph{et~al.}(2010)\citenamefont{Toivanen,
  Carlsson, Dobaczewski, Mizuyama, Rodr\'\i{}guez-Guzm\'an, Toivanen, and
  Vesel\'y}}]{Toivanen:2010}
\bibinfo{author}{\bibnamefont{Toivanen}, \bibfnamefont{J.}},
  \bibinfo{author}{\bibfnamefont{B.~G.} \bibnamefont{Carlsson}},
  \bibinfo{author}{\bibfnamefont{J.}~\bibnamefont{Dobaczewski}},
  \bibinfo{author}{\bibfnamefont{K.}~\bibnamefont{Mizuyama}},
  \bibinfo{author}{\bibfnamefont{R.~R.} \bibnamefont{Rodr\'\i{}guez-Guzm\'an}},
  \bibinfo{author}{\bibfnamefont{P.}~\bibnamefont{Toivanen}}, and
  \bibinfo{author}{\bibfnamefont{P.}~\bibnamefont{Vesel\'y}},
  \bibinfo{year}{2010}, \bibinfo{journal}{Phys. Rev. C}
  \textbf{\bibinfo{volume}{81}}(\bibinfo{number}{3}), \bibinfo{pages}{034312}.

\bibitem[{\citenamefont{Toth} \emph{et~al.}(1993)\citenamefont{Toth, Sousa,
  Wilmarth, Nitschke, and Vierinen}}]{Toth:1993}
\bibinfo{author}{\bibnamefont{Toth}, \bibfnamefont{K.}},
  \bibinfo{author}{\bibfnamefont{D.}~\bibnamefont{Sousa}},
  \bibinfo{author}{\bibfnamefont{P.}~\bibnamefont{Wilmarth}},
  \bibinfo{author}{\bibfnamefont{J.}~\bibnamefont{Nitschke}}, and
  \bibinfo{author}{\bibfnamefont{K.}~\bibnamefont{Vierinen}},
  \bibinfo{year}{1993}, \bibinfo{journal}{Phys.\ Rev.}
  \textbf{\bibinfo{volume}{C~47}}, \bibinfo{pages}{1804}.

\bibitem[{\citenamefont{Towner and Hardy}(2010)}]{Towner:2010}
\bibinfo{author}{\bibnamefont{Towner}, \bibfnamefont{I.~S.}}, and
  \bibinfo{author}{\bibfnamefont{J.~C.} \bibnamefont{Hardy}},
  \bibinfo{year}{2010}, \bibinfo{journal}{Reports on Progress in Physics}
  \textbf{\bibinfo{volume}{73}}(\bibinfo{number}{4}), \bibinfo{pages}{046301}.

\bibitem[{\citenamefont{Trinder} \emph{et~al.}(1997)\citenamefont{Trinder,
  Adelberger, Brown, Janas, Keller, Krumbholz, Kunze, Magnus, Meissner,
  Piechaczek, Pf\"{u}tzner, Roeckl} \emph{et~al.}}]{Trinder:1997}
\bibinfo{author}{\bibnamefont{Trinder}, \bibfnamefont{W.}},
  \bibinfo{author}{\bibfnamefont{E.~G.} \bibnamefont{Adelberger}},
  \bibinfo{author}{\bibfnamefont{B.~A.} \bibnamefont{Brown}},
  \bibinfo{author}{\bibfnamefont{Z.}~\bibnamefont{Janas}},
  \bibinfo{author}{\bibfnamefont{H.}~\bibnamefont{Keller}},
  \bibinfo{author}{\bibfnamefont{K.}~\bibnamefont{Krumbholz}},
  \bibinfo{author}{\bibfnamefont{V.}~\bibnamefont{Kunze}},
  \bibinfo{author}{\bibfnamefont{P.}~\bibnamefont{Magnus}},
  \bibinfo{author}{\bibfnamefont{F.}~\bibnamefont{Meissner}},
  \bibinfo{author}{\bibfnamefont{A.}~\bibnamefont{Piechaczek}},
  \bibinfo{author}{\bibfnamefont{M.}~\bibnamefont{Pf\"{u}tzner}},
  \bibinfo{author}{\bibfnamefont{E.}~\bibnamefont{Roeckl}}, \emph{et~al.},
  \bibinfo{year}{1997}, \bibinfo{journal}{Nuclear Physics A}
  \textbf{\bibinfo{volume}{620}}(\bibinfo{number}{2}), \bibinfo{pages}{191}.

\bibitem[{\citenamefont{Tripathi} \emph{et~al.}(2006)\citenamefont{Tripathi,
  Tabor, Hoffman, Wiedeking, Volya, Mantica, Davies, Liddick, Mueller, Stolz,
  Tomlin, Otsuka} \emph{et~al.}}]{Tripathi:2006}
\bibinfo{author}{\bibnamefont{Tripathi}, \bibfnamefont{V.}},
  \bibinfo{author}{\bibfnamefont{S.~L.} \bibnamefont{Tabor}},
  \bibinfo{author}{\bibfnamefont{C.~R.} \bibnamefont{Hoffman}},
  \bibinfo{author}{\bibfnamefont{M.}~\bibnamefont{Wiedeking}},
  \bibinfo{author}{\bibfnamefont{A.}~\bibnamefont{Volya}},
  \bibinfo{author}{\bibfnamefont{P.~F.} \bibnamefont{Mantica}},
  \bibinfo{author}{\bibfnamefont{A.~D.} \bibnamefont{Davies}},
  \bibinfo{author}{\bibfnamefont{S.~N.} \bibnamefont{Liddick}},
  \bibinfo{author}{\bibfnamefont{W.~F.} \bibnamefont{Mueller}},
  \bibinfo{author}{\bibfnamefont{A.}~\bibnamefont{Stolz}},
  \bibinfo{author}{\bibfnamefont{B.~E.} \bibnamefont{Tomlin}},
  \bibinfo{author}{\bibfnamefont{T.}~\bibnamefont{Otsuka}}, \emph{et~al.},
  \bibinfo{year}{2006}, \bibinfo{journal}{Phys. Rev. C}
  \textbf{\bibinfo{volume}{73}}(\bibinfo{number}{5}), \bibinfo{pages}{054303}.

\bibitem[{\citenamefont{Tursunov}
  \emph{et~al.}(2006{\natexlab{a}})\citenamefont{Tursunov, Baye, and
  Descouvemont}}]{Tursonov:2006}
\bibinfo{author}{\bibnamefont{Tursunov}, \bibfnamefont{E.~M.}},
  \bibinfo{author}{\bibfnamefont{D.}~\bibnamefont{Baye}}, and
  \bibinfo{author}{\bibfnamefont{P.}~\bibnamefont{Descouvemont}},
  \bibinfo{year}{2006}{\natexlab{a}}, \bibinfo{journal}{Phys. Rev. C}
  \textbf{\bibinfo{volume}{73}}(\bibinfo{number}{1}), \bibinfo{pages}{014303}.

\bibitem[{\citenamefont{Tursunov}
  \emph{et~al.}(2006{\natexlab{b}})\citenamefont{Tursunov, Baye, and
  Descouvemont}}]{Tursonov:2006E}
\bibinfo{author}{\bibnamefont{Tursunov}, \bibfnamefont{E.~M.}},
  \bibinfo{author}{\bibfnamefont{D.}~\bibnamefont{Baye}}, and
  \bibinfo{author}{\bibfnamefont{P.}~\bibnamefont{Descouvemont}},
  \bibinfo{year}{2006}{\natexlab{b}}, \bibinfo{journal}{Phys. Rev. C}
  \textbf{\bibinfo{volume}{74}}(\bibinfo{number}{6}), \bibinfo{pages}{069904}.

\bibitem[{\citenamefont{Uusitalo} \emph{et~al.}(1999)\citenamefont{Uusitalo,
  Davids, Woods, Seweryniak, Sonzogni, Batchelder, Bingham, Davinson, de~Boer,
  Henderson, Maier, Ressler} \emph{et~al.}}]{Uusitalo:1999}
\bibinfo{author}{\bibnamefont{Uusitalo}, \bibfnamefont{J.}},
  \bibinfo{author}{\bibfnamefont{C.}~\bibnamefont{Davids}},
  \bibinfo{author}{\bibfnamefont{P.}~\bibnamefont{Woods}},
  \bibinfo{author}{\bibfnamefont{D.}~\bibnamefont{Seweryniak}},
  \bibinfo{author}{\bibfnamefont{A.}~\bibnamefont{Sonzogni}},
  \bibinfo{author}{\bibfnamefont{J.}~\bibnamefont{Batchelder}},
  \bibinfo{author}{\bibfnamefont{C.}~\bibnamefont{Bingham}},
  \bibinfo{author}{\bibfnamefont{T.}~\bibnamefont{Davinson}},
  \bibinfo{author}{\bibfnamefont{J.}~\bibnamefont{de~Boer}},
  \bibinfo{author}{\bibfnamefont{D.}~\bibnamefont{Henderson}},
  \bibinfo{author}{\bibfnamefont{H.}~\bibnamefont{Maier}},
  \bibinfo{author}{\bibfnamefont{J.}~\bibnamefont{Ressler}}, \emph{et~al.},
  \bibinfo{year}{1999}, \bibinfo{journal}{Phys.\ Rev.}
  \textbf{\bibinfo{volume}{C~59}}(\bibinfo{number}{6}), \bibinfo{pages}{R2975}.

\bibitem[{\citenamefont{Varga} \emph{et~al.}(1992)\citenamefont{Varga, Lovas,
  and Liotta}}]{Varga:1992}
\bibinfo{author}{\bibnamefont{Varga}, \bibfnamefont{K.}},
  \bibinfo{author}{\bibfnamefont{R.~G.} \bibnamefont{Lovas}}, and
  \bibinfo{author}{\bibfnamefont{R.~J.} \bibnamefont{Liotta}},
  \bibinfo{year}{1992}, \bibinfo{journal}{Phys. Rev. Lett.}
  \textbf{\bibinfo{volume}{69}}(\bibinfo{number}{1}), \bibinfo{pages}{37}.

\bibitem[{\citenamefont{Vasilevsky}
  \emph{et~al.}(2001)\citenamefont{Vasilevsky, Nesterov, Arickx, and
  Broeckhove}}]{Vasilevsky:2001}
\bibinfo{author}{\bibnamefont{Vasilevsky}, \bibfnamefont{V.}},
  \bibinfo{author}{\bibfnamefont{A.}~\bibnamefont{Nesterov}},
  \bibinfo{author}{\bibfnamefont{F.}~\bibnamefont{Arickx}}, and
  \bibinfo{author}{\bibfnamefont{J.}~\bibnamefont{Broeckhove}},
  \bibinfo{year}{2001}, \bibinfo{journal}{Phys. Rev. C}
  \textbf{\bibinfo{volume}{63}}, \bibinfo{pages}{034607}.

\bibitem[{\citenamefont{Villard}(1900)}]{Villard:1900}
\bibinfo{author}{\bibnamefont{Villard}, \bibfnamefont{P.}},
  \bibinfo{year}{1900}, \bibinfo{journal}{Comptes Rendus}
  \textbf{\bibinfo{volume}{130}}, \bibinfo{pages}{1010,1178}.

\bibitem[{\citenamefont{Villari}(2003)}]{Villari:2003}
\bibinfo{author}{\bibnamefont{Villari}, \bibfnamefont{A.~C.~C.}},
  \bibinfo{year}{2003}, \bibinfo{journal}{Nuclear Instruments and Methods in
  Physics Research Section B: Beam Interactions with Materials and Atoms}
  \textbf{\bibinfo{volume}{204}}, \bibinfo{pages}{31}.

\bibitem[{\citenamefont{Volya and Zelevinsky}(2006)}]{Volya:2006}
\bibinfo{author}{\bibnamefont{Volya}, \bibfnamefont{A.}}, and
  \bibinfo{author}{\bibfnamefont{V.}~\bibnamefont{Zelevinsky}},
  \bibinfo{year}{2006}, \bibinfo{journal}{Phys. Rev. C}
  \textbf{\bibinfo{volume}{74}}, \bibinfo{pages}{064314}.

\bibitem[{\citenamefont{Voulot} \emph{et~al.}(2008)\citenamefont{Voulot,
  Wenander, Piselli, Scrivens, Lindroos, Jeppesen, Fraile, Sturm, and
  Delahaye}}]{Voulot:2008}
\bibinfo{author}{\bibnamefont{Voulot}, \bibfnamefont{D.}},
  \bibinfo{author}{\bibfnamefont{F.}~\bibnamefont{Wenander}},
  \bibinfo{author}{\bibfnamefont{E.}~\bibnamefont{Piselli}},
  \bibinfo{author}{\bibfnamefont{R.}~\bibnamefont{Scrivens}},
  \bibinfo{author}{\bibfnamefont{M.}~\bibnamefont{Lindroos}},
  \bibinfo{author}{\bibfnamefont{H.}~\bibnamefont{Jeppesen}},
  \bibinfo{author}{\bibfnamefont{L.}~\bibnamefont{Fraile}},
  \bibinfo{author}{\bibfnamefont{S.}~\bibnamefont{Sturm}}, and
  \bibinfo{author}{\bibfnamefont{P.}~\bibnamefont{Delahaye}},
  \bibinfo{year}{2008}, \bibinfo{journal}{Nuclear Instruments and Methods in
  Physics Research Section B: Beam Interactions with Materials and Atoms}
  \textbf{\bibinfo{volume}{266}}(\bibinfo{number}{19-20}),
  \bibinfo{pages}{4103}.

\bibitem[{\citenamefont{Wahl}(1988)}]{Wahl:1988}
\bibinfo{author}{\bibnamefont{Wahl}, \bibfnamefont{A.~C.}},
  \bibinfo{year}{1988}, \bibinfo{journal}{Atomic Data and Nuclear Data Tables}
  \textbf{\bibinfo{volume}{39}}(\bibinfo{number}{1}), \bibinfo{pages}{1}.

\bibitem[{\citenamefont{Wallace} \emph{et~al.}(2007)\citenamefont{Wallace,
  Famiano, van Goethem, Rogers, Lynch, Clifford, Delaunay, Lee, Labostov,
  Mocko, Morris, Moroni} \emph{et~al.}}]{Wallace:2007}
\bibinfo{author}{\bibnamefont{Wallace}, \bibfnamefont{M.}},
  \bibinfo{author}{\bibfnamefont{M.}~\bibnamefont{Famiano}},
  \bibinfo{author}{\bibfnamefont{M.-J.} \bibnamefont{van Goethem}},
  \bibinfo{author}{\bibfnamefont{A.}~\bibnamefont{Rogers}},
  \bibinfo{author}{\bibfnamefont{W.}~\bibnamefont{Lynch}},
  \bibinfo{author}{\bibfnamefont{J.}~\bibnamefont{Clifford}},
  \bibinfo{author}{\bibfnamefont{F.}~\bibnamefont{Delaunay}},
  \bibinfo{author}{\bibfnamefont{J.}~\bibnamefont{Lee}},
  \bibinfo{author}{\bibfnamefont{S.}~\bibnamefont{Labostov}},
  \bibinfo{author}{\bibfnamefont{M.}~\bibnamefont{Mocko}},
  \bibinfo{author}{\bibfnamefont{L.}~\bibnamefont{Morris}},
  \bibinfo{author}{\bibfnamefont{A.}~\bibnamefont{Moroni}}, \emph{et~al.},
  \bibinfo{year}{2007}, \bibinfo{journal}{Nuclear Instruments and Methods in
  Physics Research Section A: Accelerators, Spectrometers, Detectors and
  Associated Equipment} \textbf{\bibinfo{volume}{583}}(\bibinfo{number}{2-3}),
  \bibinfo{pages}{302}.

\bibitem[{\citenamefont{Warburton and Grudberg}(2006)}]{Warburton:2006}
\bibinfo{author}{\bibnamefont{Warburton}, \bibfnamefont{W.}}, and
  \bibinfo{author}{\bibfnamefont{P.}~\bibnamefont{Grudberg}},
  \bibinfo{year}{2006}, \bibinfo{journal}{Nuclear Instruments and Methods in
  Physics Research Section A: Accelerators, Spectrometers, Detectors and
  Associated Equipment} \textbf{\bibinfo{volume}{568}}(\bibinfo{number}{1}),
  \bibinfo{pages}{350}.

\bibitem[{\citenamefont{Webber} \emph{et~al.}(1990)\citenamefont{Webber, Kish,
  and Schrier}}]{Webber:1990}
\bibinfo{author}{\bibnamefont{Webber}, \bibfnamefont{W.~R.}},
  \bibinfo{author}{\bibfnamefont{J.~C.} \bibnamefont{Kish}}, and
  \bibinfo{author}{\bibfnamefont{D.~A.} \bibnamefont{Schrier}},
  \bibinfo{year}{1990}, \bibinfo{journal}{Phys. Rev. C}
  \textbf{\bibinfo{volume}{41}}(\bibinfo{number}{2}), \bibinfo{pages}{520}.

\bibitem[{\citenamefont{Weidenm\"uller and Mitchell}(2009)}]{Weidenmuller:2009}
\bibinfo{author}{\bibnamefont{Weidenm\"uller}, \bibfnamefont{H.~A.}}, and
  \bibinfo{author}{\bibfnamefont{G.~E.} \bibnamefont{Mitchell}},
  \bibinfo{year}{2009}, \bibinfo{journal}{Rev. Mod. Phys.}
  \textbf{\bibinfo{volume}{81}}(\bibinfo{number}{2}), \bibinfo{pages}{539}.

\bibitem[{\citenamefont{Weissman} \emph{et~al.}(2003)\citenamefont{Weissman,
  Bergmann, Cederkall, Fraile, Franchoo, Fynbo, Gausemel, Jeppesen, K\"oster,
  Kratz, Nilsson, Pfeiffer} \emph{et~al.}}]{Weissman:2003}
\bibinfo{author}{\bibnamefont{Weissman}, \bibfnamefont{L.}},
  \bibinfo{author}{\bibfnamefont{U.}~\bibnamefont{Bergmann}},
  \bibinfo{author}{\bibfnamefont{J.}~\bibnamefont{Cederkall}},
  \bibinfo{author}{\bibfnamefont{L.~M.} \bibnamefont{Fraile}},
  \bibinfo{author}{\bibfnamefont{S.}~\bibnamefont{Franchoo}},
  \bibinfo{author}{\bibfnamefont{H.}~\bibnamefont{Fynbo}},
  \bibinfo{author}{\bibfnamefont{H.}~\bibnamefont{Gausemel}},
  \bibinfo{author}{\bibfnamefont{H.}~\bibnamefont{Jeppesen}},
  \bibinfo{author}{\bibfnamefont{U.}~\bibnamefont{K\"oster}},
  \bibinfo{author}{\bibfnamefont{K.-L.} \bibnamefont{Kratz}},
  \bibinfo{author}{\bibfnamefont{T.}~\bibnamefont{Nilsson}},
  \bibinfo{author}{\bibfnamefont{B.}~\bibnamefont{Pfeiffer}}, \emph{et~al.},
  \bibinfo{year}{2003}, \bibinfo{journal}{Phys. Rev. C}
  \textbf{\bibinfo{volume}{67}}(\bibinfo{number}{5}), \bibinfo{pages}{054314}.

\bibitem[{\citenamefont{Wenander}(2008)}]{Wenander:2008}
\bibinfo{author}{\bibnamefont{Wenander}, \bibfnamefont{F.}},
  \bibinfo{year}{2008}, \bibinfo{journal}{Nuclear Instruments and Methods in
  Physics Research Section B: Beam Interactions with Materials and Atoms}
  \textbf{\bibinfo{volume}{266}}(\bibinfo{number}{19-20}),
  \bibinfo{pages}{4346}.

\bibitem[{\citenamefont{Westfall} \emph{et~al.}(1979)\citenamefont{Westfall,
  Symons, Greiner, Heckman, Lindstrom, Mahoney, Shotter, Scott, Crawford,
  McParland, Awes, Gelbke} \emph{et~al.}}]{Westfall:1979}
\bibinfo{author}{\bibnamefont{Westfall}, \bibfnamefont{G.~D.}},
  \bibinfo{author}{\bibfnamefont{T.~J.~M.} \bibnamefont{Symons}},
  \bibinfo{author}{\bibfnamefont{D.~E.} \bibnamefont{Greiner}},
  \bibinfo{author}{\bibfnamefont{H.~H.} \bibnamefont{Heckman}},
  \bibinfo{author}{\bibfnamefont{P.~J.} \bibnamefont{Lindstrom}},
  \bibinfo{author}{\bibfnamefont{J.}~\bibnamefont{Mahoney}},
  \bibinfo{author}{\bibfnamefont{A.~C.} \bibnamefont{Shotter}},
  \bibinfo{author}{\bibfnamefont{D.~K.} \bibnamefont{Scott}},
  \bibinfo{author}{\bibfnamefont{H.~J.} \bibnamefont{Crawford}},
  \bibinfo{author}{\bibfnamefont{C.}~\bibnamefont{McParland}},
  \bibinfo{author}{\bibfnamefont{T.~C.} \bibnamefont{Awes}},
  \bibinfo{author}{\bibfnamefont{C.~K.} \bibnamefont{Gelbke}}, \emph{et~al.},
  \bibinfo{year}{1979}, \bibinfo{journal}{Phys. Rev. Lett.}
  \textbf{\bibinfo{volume}{43}}(\bibinfo{number}{25}), \bibinfo{pages}{1859}.

\bibitem[{\citenamefont{Whaling}(1966)}]{Whaling:1966}
\bibinfo{author}{\bibnamefont{Whaling}, \bibfnamefont{W.}},
  \bibinfo{year}{1966}, \bibinfo{journal}{Phys. Rev.}
  \textbf{\bibinfo{volume}{150}}, \bibinfo{pages}{836}.

\bibitem[{\citenamefont{Wick}(1934)}]{Wick:1934}
\bibinfo{author}{\bibnamefont{Wick}, \bibfnamefont{G.-C.}},
  \bibinfo{year}{1934}, \bibinfo{journal}{Rend. Acc. Lincei}
  \textbf{\bibinfo{volume}{19}}, \bibinfo{pages}{319}.

\bibitem[{\citenamefont{Wilkinson}(1995)}]{Wilkinson:1995}
\bibinfo{author}{\bibnamefont{Wilkinson}, \bibfnamefont{D.~H.}},
  \bibinfo{year}{1995}, \bibinfo{journal}{Nuclear Instruments and Methods in
  Physics Research Section A: Accelerators, Spectrometers, Detectors and
  Associated Equipment} \textbf{\bibinfo{volume}{365}}(\bibinfo{number}{2-3}),
  \bibinfo{pages}{497}.

\bibitem[{\citenamefont{Winger} \emph{et~al.}(2009)\citenamefont{Winger,
  Ilyushkin, Rykaczewski, Gross, Batchelder, Goodin, Grzywacz, Hamilton,
  Korgul, Kr\'olas, Liddick, Mazzocchi} \emph{et~al.}}]{Winger:2009}
\bibinfo{author}{\bibnamefont{Winger}, \bibfnamefont{J.~A.}},
  \bibinfo{author}{\bibfnamefont{S.~V.} \bibnamefont{Ilyushkin}},
  \bibinfo{author}{\bibfnamefont{K.~P.} \bibnamefont{Rykaczewski}},
  \bibinfo{author}{\bibfnamefont{C.~J.} \bibnamefont{Gross}},
  \bibinfo{author}{\bibfnamefont{J.~C.} \bibnamefont{Batchelder}},
  \bibinfo{author}{\bibfnamefont{C.}~\bibnamefont{Goodin}},
  \bibinfo{author}{\bibfnamefont{R.}~\bibnamefont{Grzywacz}},
  \bibinfo{author}{\bibfnamefont{J.~H.} \bibnamefont{Hamilton}},
  \bibinfo{author}{\bibfnamefont{A.}~\bibnamefont{Korgul}},
  \bibinfo{author}{\bibfnamefont{W.}~\bibnamefont{Kr\'olas}},
  \bibinfo{author}{\bibfnamefont{S.~N.} \bibnamefont{Liddick}},
  \bibinfo{author}{\bibfnamefont{C.}~\bibnamefont{Mazzocchi}}, \emph{et~al.},
  \bibinfo{year}{2009}, \bibinfo{journal}{Phys. Rev. Lett.}
  \textbf{\bibinfo{volume}{102}}(\bibinfo{number}{14}),
  \bibinfo{pages}{142502}.

\bibitem[{\citenamefont{Winkler} \emph{et~al.}(2008)\citenamefont{Winkler,
  Geissel, Weick, Achenbach, Behr, Boutin, Brünle, Gleim, Hüller, Karagiannis,
  Kelic, Kindler} \emph{et~al.}}]{Winkler:2008}
\bibinfo{author}{\bibnamefont{Winkler}, \bibfnamefont{M.}},
  \bibinfo{author}{\bibfnamefont{H.}~\bibnamefont{Geissel}},
  \bibinfo{author}{\bibfnamefont{H.}~\bibnamefont{Weick}},
  \bibinfo{author}{\bibfnamefont{B.}~\bibnamefont{Achenbach}},
  \bibinfo{author}{\bibfnamefont{K.-H.} \bibnamefont{Behr}},
  \bibinfo{author}{\bibfnamefont{D.}~\bibnamefont{Boutin}},
  \bibinfo{author}{\bibfnamefont{A.}~\bibnamefont{Brünle}},
  \bibinfo{author}{\bibfnamefont{M.}~\bibnamefont{Gleim}},
  \bibinfo{author}{\bibfnamefont{W.}~\bibnamefont{Hüller}},
  \bibinfo{author}{\bibfnamefont{C.}~\bibnamefont{Karagiannis}},
  \bibinfo{author}{\bibfnamefont{A.}~\bibnamefont{Kelic}},
  \bibinfo{author}{\bibfnamefont{B.}~\bibnamefont{Kindler}}, \emph{et~al.},
  \bibinfo{year}{2008}, \bibinfo{journal}{Nuclear Instruments and Methods in
  Physics Research Section B: Beam Interactions with Materials and Atoms}
  \textbf{\bibinfo{volume}{266}}(\bibinfo{number}{19-20}),
  \bibinfo{pages}{4183}.

\bibitem[{\citenamefont{Woods} \emph{et~al.}(1993)\citenamefont{Woods,
  Davinson, Davis, Hofmann, James, Livingston, Page, Selin, and
  Shotter}}]{Woods:1993}
\bibinfo{author}{\bibnamefont{Woods}, \bibfnamefont{P.}},
  \bibinfo{author}{\bibfnamefont{T.}~\bibnamefont{Davinson}},
  \bibinfo{author}{\bibfnamefont{N.}~\bibnamefont{Davis}},
  \bibinfo{author}{\bibfnamefont{S.}~\bibnamefont{Hofmann}},
  \bibinfo{author}{\bibfnamefont{A.}~\bibnamefont{James}},
  \bibinfo{author}{\bibfnamefont{K.}~\bibnamefont{Livingston}},
  \bibinfo{author}{\bibfnamefont{R.}~\bibnamefont{Page}},
  \bibinfo{author}{\bibfnamefont{P.}~\bibnamefont{Selin}}, and
  \bibinfo{author}{\bibfnamefont{A.}~\bibnamefont{Shotter}},
  \bibinfo{year}{1993}, \bibinfo{journal}{Nucl.\ Phys.}
  \textbf{\bibinfo{volume}{A~553}}, \bibinfo{pages}{485c}.

\bibitem[{\citenamefont{Woods and Davis}(1997)}]{Woods:1997}
\bibinfo{author}{\bibnamefont{Woods}, \bibfnamefont{P.}}, and
  \bibinfo{author}{\bibfnamefont{C.}~\bibnamefont{Davis}},
  \bibinfo{year}{1997}, \bibinfo{journal}{Annu.\ Rev.\ Nucl.\ Part.\ Sci.}
  \textbf{\bibinfo{volume}{47}}, \bibinfo{pages}{541}.

\bibitem[{\citenamefont{Woods} \emph{et~al.}(2004)\citenamefont{Woods, Munro,
  Seweryniak, Davids, Heinz, Mahud, Sarazin, Shergur, Walters, and
  Woehr}}]{Woods:2004}
\bibinfo{author}{\bibnamefont{Woods}, \bibfnamefont{P.}},
  \bibinfo{author}{\bibfnamefont{P.}~\bibnamefont{Munro}},
  \bibinfo{author}{\bibfnamefont{D.}~\bibnamefont{Seweryniak}},
  \bibinfo{author}{\bibfnamefont{C.}~\bibnamefont{Davids}},
  \bibinfo{author}{\bibfnamefont{T.~D.~A.} \bibnamefont{Heinz}},
  \bibinfo{author}{\bibfnamefont{H.}~\bibnamefont{Mahud}},
  \bibinfo{author}{\bibfnamefont{F.}~\bibnamefont{Sarazin}},
  \bibinfo{author}{\bibfnamefont{J.}~\bibnamefont{Shergur}},
  \bibinfo{author}{\bibfnamefont{W.}~\bibnamefont{Walters}}, and
  \bibinfo{author}{\bibfnamefont{A.}~\bibnamefont{Woehr}},
  \bibinfo{year}{2004}, \bibinfo{journal}{Phys.\ Rev.}
  \textbf{\bibinfo{volume}{C~69}}, \bibinfo{pages}{051302(R)}.

\bibitem[{\citenamefont{Woodward} \emph{et~al.}(1983)\citenamefont{Woodward,
  Tribble, and Tanner}}]{Woodward:1983}
\bibinfo{author}{\bibnamefont{Woodward}, \bibfnamefont{C.}},
  \bibinfo{author}{\bibfnamefont{R.}~\bibnamefont{Tribble}}, and
  \bibinfo{author}{\bibfnamefont{D.}~\bibnamefont{Tanner}},
  \bibinfo{year}{1983}, \bibinfo{journal}{Phys. Rev. C}
  \textbf{\bibinfo{volume}{27}}, \bibinfo{pages}{27}.

\bibitem[{\citenamefont{Wormer} \emph{et~al.}(1994)\citenamefont{Wormer,
  G\"{o}rres, Iliadis, Wiescher, and Thielemann}}]{Wormer:1994}
\bibinfo{author}{\bibnamefont{Wormer}, \bibfnamefont{L.~V.}},
  \bibinfo{author}{\bibfnamefont{J.}~\bibnamefont{G\"{o}rres}},
  \bibinfo{author}{\bibfnamefont{C.}~\bibnamefont{Iliadis}},
  \bibinfo{author}{\bibfnamefont{M.}~\bibnamefont{Wiescher}}, and
  \bibinfo{author}{\bibfnamefont{F.-K.} \bibnamefont{Thielemann}},
  \bibinfo{year}{1994}, \bibinfo{journal}{The Astrophysical Journal}
  \textbf{\bibinfo{volume}{432}}, \bibinfo{pages}{326}.

\bibitem[{\citenamefont{Wrede} \emph{et~al.}(2009)\citenamefont{Wrede,
  Caggiano, Clark, Deibel, Parikh, and Parker}}]{Wrede:2009}
\bibinfo{author}{\bibnamefont{Wrede}, \bibfnamefont{C.}},
  \bibinfo{author}{\bibfnamefont{J.~A.} \bibnamefont{Caggiano}},
  \bibinfo{author}{\bibfnamefont{J.~A.} \bibnamefont{Clark}},
  \bibinfo{author}{\bibfnamefont{C.~M.} \bibnamefont{Deibel}},
  \bibinfo{author}{\bibfnamefont{A.}~\bibnamefont{Parikh}}, and
  \bibinfo{author}{\bibfnamefont{P.~D.} \bibnamefont{Parker}},
  \bibinfo{year}{2009}, \bibinfo{journal}{Phys. Rev. C}
  \textbf{\bibinfo{volume}{79}}(\bibinfo{number}{4}), \bibinfo{pages}{045808}.

\bibitem[{\citenamefont{Wu} \emph{et~al.}(1957)\citenamefont{Wu, Ambler,
  Hayward, Hoppes, and Hudson}}]{Wu:1957}
\bibinfo{author}{\bibnamefont{Wu}, \bibfnamefont{C.~S.}},
  \bibinfo{author}{\bibfnamefont{E.}~\bibnamefont{Ambler}},
  \bibinfo{author}{\bibfnamefont{R.~W.} \bibnamefont{Hayward}},
  \bibinfo{author}{\bibfnamefont{D.~D.} \bibnamefont{Hoppes}}, and
  \bibinfo{author}{\bibfnamefont{R.~P.} \bibnamefont{Hudson}},
  \bibinfo{year}{1957}, \bibinfo{journal}{Phys. Rev.}
  \textbf{\bibinfo{volume}{105}}(\bibinfo{number}{4}), \bibinfo{pages}{1413}.

\bibitem[{\citenamefont{Yoneda} \emph{et~al.}(2003)\citenamefont{Yoneda, Aoi,
  Iwasaki, Sakurai, Ogawa, Nakamura, Schmidt-Ott, Sch\"afer, Notani, Fukuda,
  Ideguchi, Kishida} \emph{et~al.}}]{Yoneda:2003}
\bibinfo{author}{\bibnamefont{Yoneda}, \bibfnamefont{K.}},
  \bibinfo{author}{\bibfnamefont{N.}~\bibnamefont{Aoi}},
  \bibinfo{author}{\bibfnamefont{H.}~\bibnamefont{Iwasaki}},
  \bibinfo{author}{\bibfnamefont{H.}~\bibnamefont{Sakurai}},
  \bibinfo{author}{\bibfnamefont{H.}~\bibnamefont{Ogawa}},
  \bibinfo{author}{\bibfnamefont{T.}~\bibnamefont{Nakamura}},
  \bibinfo{author}{\bibfnamefont{W.-D.} \bibnamefont{Schmidt-Ott}},
  \bibinfo{author}{\bibfnamefont{M.}~\bibnamefont{Sch\"afer}},
  \bibinfo{author}{\bibfnamefont{M.}~\bibnamefont{Notani}},
  \bibinfo{author}{\bibfnamefont{N.}~\bibnamefont{Fukuda}},
  \bibinfo{author}{\bibfnamefont{E.}~\bibnamefont{Ideguchi}},
  \bibinfo{author}{\bibfnamefont{T.}~\bibnamefont{Kishida}}, \emph{et~al.},
  \bibinfo{year}{2003}, \bibinfo{journal}{Phys. Rev. C}
  \textbf{\bibinfo{volume}{67}}(\bibinfo{number}{1}), \bibinfo{pages}{014316}.

\bibitem[{\citenamefont{Yordanov} \emph{et~al.}(2007)\citenamefont{Yordanov,
  Kowalska, Blaum, De~Rydt, Flanagan, Lievens, Neugart, Neyens, and
  Stroke}}]{Yordanov:2007}
\bibinfo{author}{\bibnamefont{Yordanov}, \bibfnamefont{D.~T.}},
  \bibinfo{author}{\bibfnamefont{M.}~\bibnamefont{Kowalska}},
  \bibinfo{author}{\bibfnamefont{K.}~\bibnamefont{Blaum}},
  \bibinfo{author}{\bibfnamefont{M.}~\bibnamefont{De~Rydt}},
  \bibinfo{author}{\bibfnamefont{K.~T.} \bibnamefont{Flanagan}},
  \bibinfo{author}{\bibfnamefont{P.}~\bibnamefont{Lievens}},
  \bibinfo{author}{\bibfnamefont{R.}~\bibnamefont{Neugart}},
  \bibinfo{author}{\bibfnamefont{G.}~\bibnamefont{Neyens}}, and
  \bibinfo{author}{\bibfnamefont{H.~H.} \bibnamefont{Stroke}},
  \bibinfo{year}{2007}, \bibinfo{journal}{Phys. Rev. Lett.}
  \textbf{\bibinfo{volume}{99}}(\bibinfo{number}{21}), \bibinfo{pages}{212501}.

\bibitem[{\citenamefont{Äystö}(2001)}]{Aysto:2001}
\bibinfo{author}{\bibnamefont{Äystö}, \bibfnamefont{J.}}, \bibinfo{year}{2001},
  \bibinfo{journal}{Nuclear Physics A}
  \textbf{\bibinfo{volume}{693}}(\bibinfo{number}{1-2}), \bibinfo{pages}{477}.

\bibitem[{\citenamefont{Yu} \emph{et~al.}(1998)\citenamefont{Yu, Batchelder,
  Bingham, Grzywacz, Rykaczewski, Toth, Akovali, Baktash, Galindo-Uribarri,
  Ginter, Gross, Karny} \emph{et~al.}}]{Yu:1998}
\bibinfo{author}{\bibnamefont{Yu}, \bibfnamefont{C.-H.}},
  \bibinfo{author}{\bibfnamefont{J.}~\bibnamefont{Batchelder}},
  \bibinfo{author}{\bibfnamefont{C.}~\bibnamefont{Bingham}},
  \bibinfo{author}{\bibfnamefont{R.}~\bibnamefont{Grzywacz}},
  \bibinfo{author}{\bibfnamefont{K.}~\bibnamefont{Rykaczewski}},
  \bibinfo{author}{\bibfnamefont{K.}~\bibnamefont{Toth}},
  \bibinfo{author}{\bibfnamefont{Y.}~\bibnamefont{Akovali}},
  \bibinfo{author}{\bibfnamefont{C.}~\bibnamefont{Baktash}},
  \bibinfo{author}{\bibfnamefont{A.}~\bibnamefont{Galindo-Uribarri}},
  \bibinfo{author}{\bibfnamefont{T.}~\bibnamefont{Ginter}},
  \bibinfo{author}{\bibfnamefont{C.}~\bibnamefont{Gross}},
  \bibinfo{author}{\bibfnamefont{M.}~\bibnamefont{Karny}}, \emph{et~al.},
  \bibinfo{year}{1998}, \bibinfo{journal}{Phys.\ Rev.}
  \textbf{\bibinfo{volume}{C~58}}, \bibinfo{pages}{R3042}.

\bibitem[{\citenamefont{Zagrebaev and Greiner}(2008)}]{Zagrebaev:2008}
\bibinfo{author}{\bibnamefont{Zagrebaev}, \bibfnamefont{V.}}, and
  \bibinfo{author}{\bibfnamefont{W.}~\bibnamefont{Greiner}},
  \bibinfo{year}{2008}, \bibinfo{journal}{Phys. Rev. C}
  \textbf{\bibinfo{volume}{78}}(\bibinfo{number}{3}), \bibinfo{pages}{034610}.

\bibitem[{\citenamefont{Zeldovich}(1960)}]{Zeldovich:1960}
\bibinfo{author}{\bibnamefont{Zeldovich}, \bibfnamefont{Y.~B.}},
  \bibinfo{year}{1960}, \bibinfo{journal}{JETP} \textbf{\bibinfo{volume}{38}},
  \bibinfo{pages}{1123}.

\bibitem[{\citenamefont{Zelevinsky}(1996)}]{Zelevinsky:1996}
\bibinfo{author}{\bibnamefont{Zelevinsky}, \bibfnamefont{V.}},
  \bibinfo{year}{1996}, \bibinfo{journal}{Annual Review of Nuclear and Particle
  Science} \textbf{\bibinfo{volume}{46}}, \bibinfo{pages}{237}.

\bibitem[{\citenamefont{Zerguerras}
  \emph{et~al.}(2004)\citenamefont{Zerguerras, Blank, Blumenfeld, Suomij\"arvi,
  Beaumel, Brown, Chartier, Fallot, Giovinazzo, Jouanne, Lapoux, Lhenry-Yvon}
  \emph{et~al.}}]{Zerguerras:2004}
\bibinfo{author}{\bibnamefont{Zerguerras}, \bibfnamefont{T.}},
  \bibinfo{author}{\bibfnamefont{B.}~\bibnamefont{Blank}},
  \bibinfo{author}{\bibfnamefont{Y.}~\bibnamefont{Blumenfeld}},
  \bibinfo{author}{\bibfnamefont{T.}~\bibnamefont{Suomij\"arvi}},
  \bibinfo{author}{\bibfnamefont{D.}~\bibnamefont{Beaumel}},
  \bibinfo{author}{\bibfnamefont{B.}~\bibnamefont{Brown}},
  \bibinfo{author}{\bibfnamefont{M.}~\bibnamefont{Chartier}},
  \bibinfo{author}{\bibfnamefont{M.}~\bibnamefont{Fallot}},
  \bibinfo{author}{\bibfnamefont{J.}~\bibnamefont{Giovinazzo}},
  \bibinfo{author}{\bibfnamefont{C.}~\bibnamefont{Jouanne}},
  \bibinfo{author}{\bibfnamefont{V.}~\bibnamefont{Lapoux}},
  \bibinfo{author}{\bibfnamefont{I.}~\bibnamefont{Lhenry-Yvon}}, \emph{et~al.},
  \bibinfo{year}{2004}, \bibinfo{journal}{Eur. Phys. J. A}
  \textbf{\bibinfo{volume}{20}}(\bibinfo{number}{3}), \bibinfo{pages}{389}.

\bibitem[{\citenamefont{Zhang} \emph{et~al.}(2007)\citenamefont{Zhang, Ren,
  Zhi, and Zheng}}]{Zhang:2007}
\bibinfo{author}{\bibnamefont{Zhang}, \bibfnamefont{X.}},
  \bibinfo{author}{\bibfnamefont{Z.}~\bibnamefont{Ren}},
  \bibinfo{author}{\bibfnamefont{Q.}~\bibnamefont{Zhi}}, and
  \bibinfo{author}{\bibfnamefont{Q.}~\bibnamefont{Zheng}},
  \bibinfo{year}{2007}, \bibinfo{journal}{Journal of Physics G: Nuclear and
  Particle Physics} \textbf{\bibinfo{volume}{34}}(\bibinfo{number}{12}),
  \bibinfo{pages}{2611}.

\bibitem[{\citenamefont{Zhukov} \emph{et~al.}(1993)\citenamefont{Zhukov,
  Danilin, Fedorov, Bang, Thompson, and Vaagen}}]{Zhukov:1993}
\bibinfo{author}{\bibnamefont{Zhukov}, \bibfnamefont{M.~V.}},
  \bibinfo{author}{\bibfnamefont{B.~V.} \bibnamefont{Danilin}},
  \bibinfo{author}{\bibfnamefont{D.~V.} \bibnamefont{Fedorov}},
  \bibinfo{author}{\bibfnamefont{J.~M.} \bibnamefont{Bang}},
  \bibinfo{author}{\bibfnamefont{I.~J.} \bibnamefont{Thompson}}, and
  \bibinfo{author}{\bibfnamefont{J.~S.} \bibnamefont{Vaagen}},
  \bibinfo{year}{1993}, \bibinfo{journal}{Physics Reports}
  \textbf{\bibinfo{volume}{231}}(\bibinfo{number}{4}), \bibinfo{pages}{151}.

\end{thebibliography}

\providecommand{\noopsort}[1]{}\providecommand{\singleletter}[1]{#1}%

\end{document}